\documentclass[12pt, uga]{report}

\usepackage[english]{babel}
 
\usepackage[margin=1in]{geometry}
 
\usepackage{amsmath}
\usepackage{amssymb}
\usepackage{amsfonts}
\usepackage{amsthm}
\usepackage{mathtools}
\usepackage{latexsym}
 
\usepackage{bbm}
\usepackage{bbold}
\usepackage{bm}
\usepackage{braket}
\usepackage{physics}              
\DeclareMathAlphabet{\mathpzc}{OT1}{pzc}{m}{it}
\usepackage[mathscr]{euscript}
 
\usepackage{array}
\usepackage{booktabs}
\usepackage{multirow}
\usepackage{tabularx}
\usepackage{longtable}
 
\usepackage{graphicx}
\usepackage{float}
\usepackage{caption}
\usepackage{placeins}
 
\usepackage{enumitem}
 
\usepackage{xcolor}
\usepackage{tcolorbox}
 
\usepackage[all]{xy}
\usepackage{tikz}
\usepackage{qcircuit}

\usetikzlibrary{calc}
 
\usepackage{algorithm}
\usepackage{algpseudocode}
 
\makeatletter
\newenvironment{breakablealgorithm}
  {
   \begin{center}
     \refstepcounter{algorithm}%
     \hrule height.8pt depth0pt \kern2pt%
     \renewcommand{\caption}[2][\relax]{%
       {\raggedright\textbf{\fname@algorithm~\thealgorithm} ##2\par}%
       \vspace{2pt}\hrule\vspace{2pt}%
     }
  }{
     \kern2pt\hrule\relax%
   \end{center}
  }
\makeatother
 
\usepackage[colorlinks=true,
            linkcolor=blue!60!black,
            citecolor=green!50!black,
            urlcolor=blue!70!black]{hyperref}
\usepackage{cleveref}
\usepackage[square, numbers, sort&compress]{natbib}
\numberwithin{equation}{section}
 
\newtheorem{theorem}{Theorem}[section]

\theoremstyle{definition}
\newtheorem{definition}[theorem]{Definition}
\newtheorem{example}[theorem]{Example}
 
\theoremstyle{remark}
\newtheorem{remark}[theorem]{Remark}
 
\newcommand{\Nvib}{N_{\mathrm{vib}}}
\newcommand{\Nel}{N_{\mathrm{el}}}

\newcommand{\nK}{n_{\mathrm{K}}}
\newcommand{\Tprop}{T_{\mathrm{prop}}}
\newcommand{\Twall}{T_{\mathrm{wall}}}
\newcommand{\Tref}{T_{\mathrm{ref}}}
\newcommand{\NSOP}{N_{\mathrm{SOP}}}
\newcommand{\NA}{N_{\!A}}
\newcommand{\Nnodes}{N_{\mathrm{nodes}}}
\newcommand{\Nleaf}{N_{\mathrm{leaf}}}
\newcommand{\grouporder}{\lvert G \rvert}
\newcommand{\Mspec}{M_{\mathrm{spec}}}
\newcommand{\Order}{\mathcal{O}}
 
\newcommand{\Tgate}{\ensuremath{T}}
\newcommand{\Tcount}{\ensuremath{T\text{-count}}}
\newcommand{\Rz}[1]{\ensuremath{R_z\!\left(#1\right)}}
\newcommand{\Rx}[1]{\ensuremath{R_x\!\left(#1\right)}}
\newcommand{\BigO}{\ensuremath{\mathcal{O}}}
\newcommand{\eps}{\ensuremath{\varepsilon}}
\newcommand{\epssynth}{\ensuremath{\varepsilon_{\mathrm{synth}}}}
 
\newcommand{\ceil}[1]{\lceil #1 \rceil}
\newcommand{\floor}[1]{\lfloor #1 \rfloor}
\newcommand{\e}{\mathrm{e}}
\newcommand{\bvec}[1]{\boldsymbol{#1}}
\newcommand{\Reg}[1]{\mathcal{#1}}

\usepackage{uga}

\title{An Oracle-Free Quantum Algorithm for Nonadiabatic Quantum Molecular Dynamics}

\author{Joshua Michael Courtney}

\previousdegrees{B.S., The University of Georgia, 2022}

\thisdegree{Doctor of Philosophy}

\thesistype{Dissertation} 

\thesismonth{May}
\thesisyear{2026}

\professor{Phillip C. Stancil}

\fulltitle{An Oracle-Free Quantum Algorithm for Nonadiabatic Quantum Molecular Dynamics\ugastyversion}

\indexwords{Quantum Simulation, Quantum Circuit, Nonadiabatic Dynamics,
            Real-Time Evolution, Quantum Chemistry, Multi-channel, Many-body}

\dean{Ronald Walcott}

\memberi{Michael Geller}
\memberii{Yohannes Abate}

\begin{document}
\pagenumbering{roman}
\begin{abstract}
Quantum computation is an attractive front for many problems that are intractable for computers today. 
One such problem is nonadiabatic quantum molecular dynamics, where quantized internal states coupling to parameterized modes result in a Hamiltonian resistant to oracle-based models and spectral decomposition. 
This dissertation applies diabatic Hamiltonian operators directly to the computational basis as first-quantized split-operator propagators, validated with dynamic observables including absorption and recurrence spectra, scattering cross-sections, population dynamics, and quantum scars. 
Circuits are derived and specified, with focused circuit optimization in multi-mode and multi-channel extensions, including multivariate potential energy terms and graph theoretic optimization from molecular symmetry. 
Resource estimation shows circuit depth advantage against QROM-loading architectures on a fault-tolerant scale, and a quantitative comparison against quantum signal processing variants confirms that a Trotter-based architecture retains a scalable $\Tgate$-gate advantage.
Expanding beyond electronic states demonstrates that duality between finite basis and discrete variable representations permits congruent structural decompositions into quantum circuits, expanding the use of multi-channel dynamics far beyond chemistry.
\end{abstract}

\maketitle
\pagestyle{plain}
\chapter*{Dedication}
\centerline{
\textit{A steed is not praised for his strength, but praised for his mettle.}}
\centerline{\textit{-- The Analects Book XIV, Confucius}}

\pseudochapter{Acknowledgments}

I thank my siblings, Ya'el, Elya, and especially my brother Elijah who have protected me, helped me learn, and loved me. 
I would have dropped out if not for the friends who were patient and blunt with me.
Thank you Manthan Sonawane, Preston Jamieson, and Jon Cooper for helping me find my voice. 

Melanie Reber and Nicholas Cooper brought patience while I worked to help build part of an ultrafast laser for time-resolved spectroscopy during my undergraduate studies, and Steven Hancock showed me balance in performing computational work and staying healthy.
Conversations with David Landau and Kanzo Nakayama helped me find my footing while I was still aimless and without passion. 
Loris Magnani grounded my perspective and made physics real as a study and as a profession, making me realize I needed a project that I could attach myself to.
Valuable conversations with Stephen Thomas made it clear I was not the only one striving.

My PI (Phillip Stancil) invested in me since my junior year in undergrad, and in graduate school gave me my own project. 
I am grateful for his patience as I learned and built what is now the topic of this dissertation.
My committee members Michael Geller and Yohannes Abate directed me to find state of the art in quantum algorithms and to make the goal to publish when I was confident I had good science.

Thank you, reader, in advance for working through the immense number of initialisms and acronyms.
A list of all abbreviations used in this dissertation can be found in Appendix~\ref{app:acronyms}, with a glossary of terms in Appendix~\ref{app:glossary}.

Finally I would like to thank all quantum computing experimentalists. 
I hope we continue to build toward a future where high-overhead algorithms like this can be applied and new breakthroughs can be realized.

\tableofcontents

\listoffigures
\listoftables   

\newpage
\pagenumbering{arabic}

\chapter{Introduction}\label{ch:Introduction}

Nonadiabatic quantum molecular dynamics (NA-QMD) encompasses a set of processes where molecular degrees of freedom become inseparably coupled to electronic or angular energy channels. 
The NA-QMD problem falls under the class of finding-all-paths (or number of shortest paths) problems, reducible to counting the number of shortest orbits taken by an initial state in a manifold, with an accurate map of their trajectories. 
This problem therefore has a classical \#P-hard complexity \cite{valiant1979complexity}, counting the solutions to an NP problem, being verifiable in polynomial time. 
An initial state $\ket{\psi_0}$ propagates according to its initialization and its Hamiltonian; tracking the evolution of the subsequent wavefunction, extending to its distribution between regimes and dictating time-dependent transitions, excitations, and state coherence, is a finding-all-paths problem \cite{kassal2011simulating}.
Dynamic observables, including reaction rates and absorption spectra, follow naturally from calculating the wavefunction in this way.
This thesis develops and numerically validates an algorithm to solve NA-QMD problems efficiently (using polynomially scaling time and resources) using quantum circuits.

Quantum advantage is loosely defined by a quantum computer's ability to outperform a classical computer for a particular problem, typically measured by end-to-end computation time, total energy expenditure, or accuracy of the result.
The proverbial ``goal posts" for quantum advantage continue to move, with advances in machine learning \cite{westermayr2019machine, dral2018nonadiabatic, akimov2021extending, zhang2025advancing, muller2025machine, axelrod2022excited} and GPU acceleration \cite{kondratyuk2021gpu, seritan2021terachem, peters2019nonadiabatic} pushing the limits of classical computation. 
Dynamic chemistry simulation is rapidly advancing, with fundamental application in early universe research \cite{stancil1998deuterium, fortenberry2024quantum, miyake2011rovibrationally} and photon-excited systems present in photovoltaics \cite{wibowo2019nonadiabatic}, drug discovery \cite{kobauri2023rational, yu2020nonadiabatic}, cancer research \cite{liang2022effects}, biological systems \cite{nelson2020non, wang2016recent}, materials development \cite{liu2024breaking, zheng2019ab}, and singlet fission chromophores \cite{smith2010singlet, akimov2014nonadiabatic, tao2014electronically}.
Methods are developed and refined for specific problems, allowing tractability for accurate computations on classical hardware. 
These methods typically make fair assumptions regarding Hamiltonian structure, with efficient solutions found when entropy is localized \cite{freitag2020density}, crossing behavior is well-approximated \cite{tully1990molecular}, or when the system is favorable to stochastic methods \cite{herman2005numerical, gorshkov2013semiclassical}.
As systems become more complex, these solutions falter and accurate calculation must rely upon first principles \cite{zheng2019ab}.
The presence of system topologies that induce chaos, nonlocal entropy, and the introduction of many-body terms enacts a curse of dimensionality~\cite{bellman1966dynamic} on nonadiabatic systems, leaving the field of dynamic chemistry open to quantum computation \cite{sangiogo2025nonadiabatic}.

Translating to quantum complexity, nonadiabatic dynamics is in the bounded-error quantum polynomial (BQP) complexity class \cite{lloyd1996universal, low2019hamiltonian, schleich2026chemically}, meaning the problem is solvable efficiently using quantum computers up to a given error $\eps$.
Beyond proof, the gap between ``it is possible" and ``how does it solve?" has rapidly narrowed in the last ten years. 
Advances in signal processing and block encoding allow for query-optimal Hamiltonian simulation, treating the Hamiltonian as a ``black-box" and approximating using a minimal number of queries to an oracle to load structured or unstructured data \cite{low2017optimal, low2019hamiltonian, gilyen2019quantum, motlagh2024generalized}.
Second quantization perspectives have found early success in analog quantum computation, with conical intersection dynamics being successfully simulated on a single trapped ion \cite{navickas2025experimental}.

With an accurate assessment in scaling difficulty for second-quantized approaches \cite{navickas2025experimental, whitlow2023quantum}, first-quantized approaches have received increased attention.
Early implementations focused on variational algorithms, in hopes of a paradigmatic shift in NISQ architectures \cite{ollitrault2021solving}. 
Ease of applying arbitrary phase rotations on physical qubits is an attractive feature, but noise suppression and scaling have yet to find breakthroughs necessary for the long-time simulations required for chemistry \cite{bidart2025quantum}.
In contrast, the interaction picture \cite{georges2025quantum} allows efficient Hamiltonian encoding for toy models, and oracle-based approaches for the Hamiltonian are favorable to fault-tolerant architectures, with computational limits dictated by coherence times and magic state distillation of T-gates. 
Other recent approaches focus on a block-diagonalization through a XOR-class decomposition, applying diabatic potential surfaces with a phase gradient arithmetic scheme, loading coefficients from quantum read-only memory (QROM) \cite{motlagh2025quantum}. 
Despite these advances, present first-quantized quantum algorithms have not yet provided numerical validation for scaled systems, instead relying on resource estimations and optimistic projection.
NISQ-era methods showed promise with validated reaction rates \cite{ollitrault2020nonadiabatic} but are not extended to multiple dimensions or multiple electronic channels. 
Current fault-tolerant era schemes \cite{motlagh2025quantum, georges2025quantum} do not present converged, accurate observables or extension beyond harmonic potential surfaces with sparse electronic coupling. 
With limiting factors in quantum hardware scale, fidelity, and coherence time, only toy models with sparse Hamiltonians are reasonable for simulation. 
For strongly interacting systems with dense Hamiltonians and chaotic dynamics, classical emulation can give a ``proof-of-principle," with resource estimations projecting feasibility for future quantum hardware. 
Quantum advantage for these problems is proven in theory, but given that proposed algorithms have rarely been emulated and validated, a concrete system size remains elusive.

This thesis first considers classical methods designed for NA-QMD, identifying precise computational bottlenecks and motivating the use of quantum computers.
Hamiltonian simulation is well-studied in quantum computation, so an alternate proposal for nonadiabatic simulation requires commensurate motivation. 
Current Hamiltonian simulation techniques are considered from an information theoretic perspective, narrowing their ability and assessing practicality in solving chaotic, dynamical systems. 
Once a new algorithm for simulating nonadiabatic systems is motivated, constraints are imposed on a quantum algorithm that may be able to handle NA-QMD efficiently, considering circuit complexity and error accumulation.
The current direction of quantum computation is noted to frame the feasibility of such an algorithm.
At the end of this chapter, the structure to the thesis is presented, providing a product-formula based algorithm to accurately simulate NA-QMD.

\section{Classical Computational Chemistry}
Chemists are invested in solving NA-QMD problems using classical computational methods. 
Four classical methods are discussed here to motivate the use and application of quantum algorithms: Multi-layer multiconfigurational time-dependent Hartree (ML-MCTDH), time-dependent density matrix renormalization group (TD-DMRG), time-dependent coupled cluster (TD-CC), and ab-initio multiple spawning (AIMS). 
In noting the strengths and computational bottleneck of each method, this background identifies areas of nonadiabatic chemistry where quantum algorithms may give advantage. 

Other common methods in computational chemistry are not discussed, including fewest switches surface hopping (FSSH) \cite{tully1990molecular}, semiclassical initial value representation (SC-IVR) \cite{sun1997semiclassical}, and coupled-coherent states (CCS) \cite{shalashilin2004real}.
These methods have similar limitations to examples described here and are actively developing in chemistry simulation.
The four examples described here present the major characteristics of classical NA-QMD simulation with similar limits in simulating strongly-correlated, low-symmetry molecules.
Treated as archetypes, these methods show where a proposed quantum algorithm needs to provide efficiency for future practicality. 

\subsection{Multi-Layer Multiconfigurational \\ Time-Dependent Hartree (ML-MCTDH)}

ML-MCTDH is a classical algorithm designed to efficiently simulate quantum dynamics for highly symmetric many-body systems \cite{meyer1990multi, manthe1992wave}. 
Advantages beyond standard MCTDH theory arise in a hierarchical correlation structure, where degrees of freedom are ``grouped" into ``logical particles." 
These terms are presented in quotations to distinguish the parlance used in chemistry literature from the formal semantics of `group' and `logical' as used in computation and mathematics. 
A thirty-thousand foot view shows how the curse of dimensionality limits this method, primarily in the breakdown of correlation hierarchy.

MCTDH theory decomposes a time-dependent wavefunction $\ket{\Psi(t)}$ into an orthogonal set of single-particle functions (SPFs) $\ket{\varphi_J^{(\kappa)}(t)}$ . 
\begin{equation}\label{eq:MCTDH_Orthogonal_Decomposition}
    \ket{\Psi(t)} = \sum_{i=1}^p\sum_{j_i}A_{j_1j_2\cdots j_p}(t)\prod_{\kappa=1}^p\ket{\varphi_J^{(\kappa)}(t)}
\end{equation}
SPFs are given for $\kappa=1,2,\ldots,p$ single particle ``groups" \cite{wang2015multilayer}. $J$ represents each combination of indices $\{j_1, j_2,\ldots,j_p\}$ constructing an order-$p$ tensor, decomposed by the Tucker form. 
This form, seen in Eq.~\ref{eq:MCTDH_Orthogonal_Decomposition}, asserts that the full wavefunction $\ket{\Psi(t)}$ can be represented by an orthonormal basis $\ket{\varphi_J^{(\kappa)}(t)}$ weighted by a `core tensor' $A_{j_1j_2\cdots j_p}$.
The wavefunction can then be stored in $\mathcal{O}(N_\kappa^p+pN_\kappa n_{\kappa,q}^{F(\kappa)})$ instead of $\mathcal{O}(n_{\kappa,q}^{pF(\kappa)})$, where $N_\kappa$ is the total number of SPFs in group $\kappa$, $p$ is the number of SP groups, $F(\kappa)$ is the degrees of freedom per group, and $n_{\kappa,q}$ is the number of primitive basis functions for the $q$th coordinate of a group $\kappa$.
This form is still exponential in in the degrees of freedom and number of SP groups, though the problem is no longer exponential in total dimension $pF(\kappa)$.

Each SPF is decomposed into primitive basis functions via multidimensional expansion in an SP subspace:
\begin{equation}\label{Primitive_Basis_MCTDH}
    \begin{split}
        \ket{\varphi_n^{(\kappa, n)}(t)} & = \sum_IB_I^{\kappa, n}(t)\ket{u_I^{\kappa}} \\
        & \equiv \sum_{\alpha=1}^{F(\kappa)}\sum_{i_\alpha}B_{i_1i_2\ldots i_{F(\kappa)}}^{\kappa,n}(t)\prod_{q=1}^{F(\kappa)}\ket{\varphi_{i_q}^{\kappa, q}}
    \end{split}
\end{equation}
Here, $F(\kappa)$ are the Cartesian degrees of freedom in the $\kappa$th SP group, with primitive basis functions for each $q$th coordinate $\ket{\varphi_{i_q}^{\kappa, q}}$. 

By design, the number of SP groups $p$ is less than the full dimensionality $f$, allowing for numerical convergence for large molecular system dynamics. 
MCTDH is efficient when each SP group is uncorrelated to other SP groups. 
A full configurational model leads to an exponential complexity in MCTDH with no multi-layer advantage, giving an upper-bound of a few tens of degrees of freedom (up to 100 with basis set contraction) \cite{wang2015multilayer}.

ML-MCTDH gives basis function contractions dynamically, advantaging tensor-network architecture to allow primitive basis functions in different SP groups to interact according to their proximity in basis encoding \cite{vendrell2011multilayer}.
Adding time-dependence gives
\begin{equation}\label{TD_Primitive_Basis_MCTDH}
    \begin{split}
        \ket{\varphi_n^{(\kappa)}(t)} & = \sum_IB_I^{\kappa, n}(t)\ket{u_I^{\kappa}(t)} \\
        & \equiv \sum_{\alpha=1}^{Q(\kappa)}\sum_{i_\alpha}B_{i_1i_2\ldots i_{Q(\kappa)}}^{\kappa,n}(t)\prod_{q=1}^{Q(\kappa)}\ket{v_{i_q}^{\kappa, q}(t)}.
    \end{split}
\end{equation}
The ``multi-layer" aspect of ML-MCTDH is the construction of SPs out of multiple other SPs. This can be performed until the primitive basis is reached.
An example of a two-layer ML-MCTDH wavefunction is given in \cite{wang2015multilayer}, iterated here
\begin{equation}\label{Two-Layer_ML-MCTDH}
    \begin{split}
        \ket{\Psi(t)} & = \sum_{k=1}^p\sum_{j_k}A_{j_1j_2\ldots j_p}(t)\times \\
        & \prod_{\kappa=1}^p \left[\sum_{\alpha=1}^{Q(\kappa)}\sum_{i_\alpha}B_{i_1i_2\ldots i_{Q(\kappa)}}^{\kappa,n}(t)\prod_{q=1}^{Q(\kappa)}\ket{v_{i_q}^{\kappa, q}(t)}\right].
    \end{split}
\end{equation}

Wang \cite{wang2015multilayer} provides ML-MCTDH complexity in terms of the number of layers, assuming that the number of SPFs is the same for all SP groups and for all layers, the number of SP groups are the same for all layers, and the number of primitive basis functions are the same for all degrees of freedom, as
\begin{equation}
    C(f) \propto \exp\left[A+B\times f^{1/(L+1)}\right],
\end{equation}
for constants $A$ and $B$, number of layers $L$, and number of degrees of freedom $f$. 

MCTDH (L=1) shows clear benefit relative to classical wavepacket propagation, with further improvement through the hierarchical assumption $L>1$.
Exponential complexity is seen in the full-configurational limit $L=0$, when either the hierarchical assumption cannot be made, or when non-hierarchical correlations require dynamic decomposition and recomposition of SPFs. 
From this result, if quantum computation is to be used in conjunction with or as an alternative to ML-MCTDH, it must at least allow for efficient (polynomial scaling) application of non-hierarchical correlations between degrees of freedom and SPFs. 
ML-MCTDH further uses a mean-field approximation, introducing some limitation in the localization of the wavefunction. The L1 mean-field operator $\langle \hat{H}\rangle ^\kappa(t)$ and L1-SPFs depend on the primitive basis functions. 
While a quantum computational method may leverage wavefunction coherence to make mean-field approximations, a basal construction may prefer a more precise method.

\subsection{Time-Dependent Density Matrix \\ Renormalization Group (TD-DMRG)}
Matrix product states (MPS) are relevant when considering qubitization and quantum signal processing (QSP), so a brief background is given here for context.

The electronic Hamiltonian can be written as 
\begin{equation}\label{MPS_Hamiltonian}
\begin{split}
    \hat{H} & = E_0 + \sum_{ij}t_{ij}\sum_\sigma \hat{a}_{i\sigma}^\dagger \hat{a}_{j\sigma} \\
    & + \frac{1}{2}\sum_{ijkl}v_{ij;kl}\sum_{\sigma\tau}\hat{a}_{i\sigma}^\dagger \hat{a}_{j\tau}^{\dagger}\hat{a}_{l\tau}\hat{a}_{k\sigma},
\end{split}
\end{equation}
with Latin indices representing spatial orbitals and Greek indices denote electron spin projections. 
Here, $t_{ij}$ is the one-electron integral and $v_{ij;kl}$ is the two-electron integral.
For fixed particle number $N$, this can be rewritten as
\begin{equation}\label{MPS_Hamiltonian_rewritten}
\begin{split}
    \hat{H} & = E_0 + \sum_{ijkl}h_{ij;kl}\sum_{\sigma\tau}\hat{a}_{i\sigma}^\dagger \hat{a}_{j\tau}^{\dagger}\hat{a}_{l\tau}\hat{a}_{k\sigma}, \\
    h_{ij;kl} &= v_{ij;kl} + \frac{1}{N-1}(t_{ik}\delta_{jl} + t_{jl}\delta_{ik}),
\end{split}
\end{equation}
where $\delta_{ab}$ is the Kronecker delta.

In the occupation number representation, Hilbert space basis states are
\begin{equation}\label{Hilbert_space_basis_MPS}
    \begin{split}
        \ket{n_{1\uparrow} n_{1\downarrow}\ldots n_{L\uparrow}n_{L\downarrow}} & = \left(\hat{a}_{1\uparrow}^\dagger\right)^{n_{1\uparrow}}\left(\hat{a}_{1\downarrow}^\dagger\right)^{n_{1\downarrow}} \cdots \\
        & \times \left(\hat{a}_{L\uparrow}^\dagger\right)^{n_{L\uparrow}}\left(\hat{a}_{L\downarrow}^\dagger\right)^{n_{L\downarrow}}\ket{0}.
    \end{split}
\end{equation}

A linear combination of unitaries (LCU) can be constructed, transforming according to the chosen irrep of the molecular point group.  
An eigenstate of the Hamiltonian is written as
\begin{equation}
\begin{split}
    \ket{\Psi} &= \sum_{\{n_{j\sigma}\}}C^{n_{1\uparrow}n_{1\downarrow}\ldots n_{L\uparrow}n_{L\downarrow}} \ket{n_{1\uparrow}n_{1\downarrow}\ldots n_{L\uparrow}n_{L\downarrow}}.
\end{split}
\end{equation}

The full configuration interaction (FCI) tensor grows as $4^L$ for $L$ spatial orbitals, just as the decomposition of an arbitrary Hamiltonian gives a sum of $4^L$ Pauli operators $\{I, X, Y, Z\}^{\otimes 2L}$.

Singular value decomposition (SVD) allows decomposition into matrix products
\begin{equation}
    \begin{split}
        C^{n_{1\uparrow}n_{1\downarrow}\ldots n_{L\uparrow}n_{L\downarrow}} & = \sum_{\{\alpha_k\}}A[1]_{\alpha_1}^{n_{1\uparrow}n_{1\downarrow}}A[2]_{\alpha_1\alpha_2}^{n_{2\uparrow}n_{2\downarrow}} \\
        & \times A[3]_{\alpha_2\alpha_3}^{n_{3\uparrow}n_{3\downarrow}}\cdots A[L]_{\alpha_{L-1}}^{n_{L\uparrow}n_{L\downarrow}},
    \end{split}
\end{equation}

So,
\begin{equation}
    \ket{\Psi} = \sum_{\{\sigma_k\}, \{\alpha_k\}} A[1]_{\alpha_1}^{\sigma_1}A[2]_{\alpha_1\alpha_2}^{\sigma_2}\cdots A[L]_{\alpha_{L-1}}^{\sigma_L}\ket{\sigma_1\sigma_2 \ldots \sigma_L}.
\end{equation}

Bond dimension $\chi_k \equiv \dim(\alpha_k)$ grows exponentially with $\chi_k = \min(4^k, 4^{L-k})$ but can be truncated to $\chi_k = \min(4^k, 4^{L-k}, D)$ for maximum bond dimension $D$ by partitioning into left and right blocks, retaining the $D$ largest singular values.

DMRG variationally optimizes MPS tensors to approximate ground states \cite{white1992density, white1993density}.
Partitioning into ``system'' and ``environment'' blocks $S$ and $E$ and tracing over the environment yields a reduced density matrix
\begin{equation}
    \hat{\rho}_S = \text{Tr}_E \ket{\Psi}\bra{\Psi}.
\end{equation}
States with the largest eigenvalues capture the most important correlations with the environment, prompting truncation to $D$ largest eigenvalues for computational tractability \cite{schollwock2005density, schollwock2011density}.

Ground-state DMRG scales as $\mathcal{O}(L D^3 d^2)$ for a chain of $L$ sites with local Hilbert space dimension $d$ and bond dimension $D$.
One-dimensional systems with short-range interactions follow an area law, accurately represented with a sufficiently small bond dimension \cite{schollwock2005density}.
Higher dimensional systems consequently suffer from the area law relationship.

Time-dependent extensions of DMRG solve the time-dependent Schr\"odinger equation (TDSE) while maintaining the MPS structure.
Several algorithms have been developed, each with distinct error characteristics \cite{paeckel2019time}.

Time evolution via block decimation (TEBD) yields
\begin{equation}
    e^{-i\hat{H}\delta t} \approx \prod_{\langle i,j\rangle} e^{-i\hat{h}_{ij}\delta t} + \mathcal{O}(\delta t^2),
\end{equation}
where $\hat{h}_{ij}$ are nearest-neighbor terms in the Hamiltonian \cite{zwolak2004mixed}.
MPS form is restored by sequentially applying SVD-truncated two-site gates $e^{-i\hat{h}_{ij}\delta t}$ \cite{daley2004time}.
An alternative is the time-dependent variational principle (TDVP), which conserves energy and unitarity by construction \cite{haegeman2011time, vanderstraeten2016gradient}.
TDVP projects the TDSE onto the MPS manifold's tangent space for a fixed bond dimension
\begin{equation}
    \frac{\partial}{\partial t}\ket{\Psi(t)} = -i\hat{P}_{T_\Psi}\hat{H}\ket{\Psi(t)},
\end{equation}
where $\hat{P}_{T_\Psi}$ is the projector onto the tangent space.
Sweeping through the chain of product states and integrating local effective Schr\"odinger equations give the equations of motion \cite{paeckel2019time}.

The primary limitation of TD-DMRG is entanglement entropy growth \cite{calabrese2005evolution}.
The linear-until-saturated growth of entanglement entropy for generic Hamiltonians increases bond dimension exponentially for a fixed truncation error $D(t) \sim e^{S(t)} \sim e^{v_E t}$ for an effective velocity $v_E$ (bounded by the Lieb-Robinson velocity). 
A quantum alternative to TD-DMRG must be able to handle high entanglement entropies natively and provide accurate results over long times without exponential resource scaling.

\subsection{Time-Dependent Coupled Cluster (TD-CC)}
Throughout this document, the initialism `CC' is reserved to represent `coupled cluster,' as opposed to `close coupling.' 
The coupled cluster wavefunction is written as
\begin{equation}\label{CC_Ansatz}
    \ket{\Psi(t)} = e^{\hat{T}(t)}\ket{\Phi_0},
\end{equation}
where $\ket{\Phi_0}$ is a reference determinant (typically Hartree-Fock) and $\hat{T}(t)$ is the cluster operator decomposed as $\hat{T} = \hat{T}_1 + \hat{T}_2 + \cdots$, with single and double excitations
\begin{equation}
    \hat{T}_1 = \sum_{i}^{\text{occ}}\sum_{a}^{\text{virt}} t_i^a(t)\, \hat{a}^\dagger_a \hat{a}_i\text{, }\quad\hat{T}_2 = \frac{1}{4}\sum_{ij}^{\text{occ}}\sum_{ab}^{\text{virt}} t_{ij}^{ab}(t)\, \hat{a}^\dagger_a \hat{a}^\dagger_b \hat{a}_j \hat{a}_i,
\end{equation}
where $i,j$ index occupied orbitals and $a,b$ index virtual orbitals \cite{vcivzek1969cluster, bartlett2007coupled}. 
In practice, perturbative triples, quadruples, and higher-order terms are feasible and actively incorporated \cite{koch1997cc3, bomble2005coupled}.
Time-dependent equations of motion derive from the bivariational principle \cite{kvaal2025time}.

The similarity-transformed Hamiltonian $\bar{H} = e^{-\hat{T}}\hat{H}e^{\hat{T}}$ is non-Hermitian, creating fundamental challenges for dynamics.
Biorthogonal left and right eigenvectors result, along with non-unitary time evolution and complex eigenvalues near conical intersections between same-symmetry excited states \cite{kohn2007can}. 
Standard CC methods yield unphysical complex excitation energies at conical intersections, requiring specialized formulations like similarity-constrained CC (SCC) \cite{kjonstad2019orbital} that enforce state orthogonality to recover correct intersection topology.

CC method complexity scales polynomially with the number of molecular orbitals $N_{\text{orb}}$, where typically $N_{\text{orb}} = o + v$ with $o$ occupied and $v$ virtual orbitals. Since $v \gg o$ in most applications, scaling is dominated by the virtual space. For singles and doubles (TD-CCSD) and singles, doubles, and triples (TD-CCSDT), complexity scales as
\begin{itemize}
    \item TD-CCSD: $\mathcal{O}(o^2 v^4) \sim \mathcal{O}(N_{\text{orb}}^6)$
    \item TD-CCSDT: $\mathcal{O}(o^3 v^5) \sim \mathcal{O}(N_{\text{orb}}^8)$
    \item TD-CCSD(T): $\mathcal{O}(o^3 v^4) \sim \mathcal{O}(N_{\text{orb}}^7)$ (perturbative triples)
\end{itemize}

CC techniques have evolved for specific problems, including TD-CCSD for driven systems \cite{huber2011explicitly}, equation-of-motion CC (EOM-CC) for excitation energies and electron affinities \cite{krylov2008equation, bomble2005equation}, orbital adaptive techniques to reach the full configuration interation (FCI) limit with size consistency \cite{kvaal2012ab, sato2018communication, pathak2022time}, and GPU-enhanced methods giving $17\times$ speedup with polarizability errors under $0.1\%$ compared to linear response \cite{pandey2024ab}.
Being so customizable, an overview of CC emphasizes that prior to using a quantum computer for simulational chemistry, a battery of comparisons must be made against relevant, state-of-the-art classical methods with heuristics for total wall time and energy expenditure. 

Limitations of CC for nonadiabatic dynamics vary by optimization technique. 
Truncation errors miss higher order contributions (triples, quadruples, etc.), though computational power sometimes allows higher-order terms exceeding sextuples. 
Standard TD-CC produces incorrect branching plane dimensionality, with recent corrections for similarity-constrained systems \cite{kjonstad2024coupled}.
Single-reference CC breaks down with comparably-weighted configurations \cite{bartlett2020index}, and numerical instability frequently affects larger systems, presently handled by dynamic biorthonormal references \cite{kristiansen2020numerical, kjonstad2020biorthonormal}. 
Recent advances including time-dependent multireference CC theory introduce extended CC operators for propagating general initial states including superposition, being required for simulating wavepacket dynamics through a conical intersection \cite{mosquera2025time}. 
Following these recent improvements, the moving target for quantum advantage grows apparent in CC methods, where a broad ``scaling up" limitation is compensated with problem-specific optimization. 

\subsection{Ab Initio Multiple Spawning (AIMS)}

The first three methods were chosen for their broad applicability, structure, and flexibility. 
AIMS is chosen as the final classical example due to its first-principles approach.
The molecular wavefunction is given using the Born-Huang expansion with nuclear wavefunctions decomposed into trajectory basis functions (TBFs) \cite{mignolet2018walk}
\begin{equation}\label{AIMS_Wavefunction}
    \Psi(\mathbf{r},\mathbf{R},t) = \sum_J \sum_{i=1}^{N_J(t)} C_i^{(J)}(t)\, \chi_i^{(J)}(\mathbf{R};t)\, \Phi_J(\mathbf{r};\mathbf{R}),
\end{equation}
with complex amplitudes $C_i^{(J)}(t)$, trajectory basis functions $\chi_i^{(J)}$, adiabatic electronic states $\Phi_J(\mathbf{r};\mathbf{R})$ , and time-dependent TBFs $N_J(t)$ for electronic state $J$.
Each TBF is a `frozen' Gaussian
\begin{equation}\label{Frozen_Gaussian}
    \chi_i^{(J)}(\mathbf{R};t) = e^{i\gamma_i^{(J)}(t)} \prod_{\rho=1}^{3N_{\text{nuc}}} \left(\frac{2\alpha_\rho}{\pi}\right)^{1/4} \exp\left[-\alpha_\rho(R_\rho - \bar{R}_{i\rho}^{(J)})^2 + i\bar{P}_{i\rho}^{(J)}(R_\rho - \bar{R}_{i\rho}^{(J)})\right],
\end{equation}
where $\bar{R}_{i\rho}^{(J)}(t)$ and $\bar{P}_{i\rho}^{(J)}(t)$ are the time-dependent position and momentum centers, $\alpha_\rho$ are frozen width parameters, and $\gamma_i^{(J)}(t)$ is a phase factor.
The `frozen' quality is a lack of broadening of the wavepacket, an assumption made for computational convenience.
While convenient in problems handled by AIMS, this behavior gives a critical limit to simulation accuracy.

The TBF centers evolve classically via Hamilton's equations, while the complex amplitudes satisfy the TDSE in the nonorthogonal TBF basis
\begin{equation}\label{Amplitude_EOM}
    i\hbar\dot{\mathbf{C}} = \mathbf{S}^{-1}\left[\mathbf{H} - i\hbar\dot{\mathbf{S}}\right]\mathbf{C},
\end{equation}
with overlap matrix $\mathbf{S}$ and $\mathbf{H}$ containing kinetic, potential, and nonadiabatic coupling contributions.

A saddle-point approximation transforms the problem from requiring global potential energy surface knowledge to local electronic structure calculations. 
Potential energy integrals are evaluated by Taylor expansion about the centroid position $\bar{\mathbf{R}}_{ki} = (\bar{\mathbf{R}}_k + \bar{\mathbf{R}}_i)/2$, enabling on-the-fly dynamics with quantum chemistry calculations only at TBF centers and centroids \cite{levine2008implementation}.

Spawning expands the TBF basis when trajectories enter regions of strong nonadiabatic coupling \cite{ben2000ab}. Each TBF monitors the effective coupling strength
\begin{equation}
    \Lambda_{IJ}^{\text{eff}} = \mathbf{d}_{IJ}(\mathbf{R}) \cdot \dot{\mathbf{R}},
\end{equation}
where $\mathbf{d}_{IJ} = \bra{\Phi_I}\nabla_{\mathbf{R}}\ket{\Phi_J}$ is the first-order nonadiabatic coupling vector.
When $|\Lambda_{IJ}^{\text{eff}}|$ exceeds a threshold, the TBF creates a child TBF on the coupled electronic state, rescaling momentum for energy conservation.
The child is initialized with zero amplitude, with population transfer occurring through the coupled amplitude equations in Eq.~\ref{Amplitude_EOM}, preserving total norm.

AIMSWISS (Ab Initio Multiple Spawning with Informed Stochastic Selections) addresses the exponential growth of TBFs using a parameter-free adaptive criterion \cite{lassmann2021aimswiss}.
A decay time $\tau_D$ estimates when parent-child TBF groups become dynamically uncoupled
\begin{equation}\label{Decay_Time}
    \tau_D = \sqrt{\frac{\pi}{(\mathbf{F}_0^P - \mathbf{F}_0^C)^T \boldsymbol{\alpha}^{-1} (\mathbf{F}_0^P - \mathbf{F}_0^C)}},
\end{equation}
where $\mathbf{F}_0^P$ and $\mathbf{F}_0^C$ are nuclear gradients at spawning time on parent and child TBFs, and $\boldsymbol{\alpha}$ is the diagonal matrix of Gaussian width parameters.

After coupled dynamics for time $\tau_D$, stochastic selection weighted by nuclear population retains one TBF group while discarding others, with an additional step of amplitude renormalization. 

Spawning explosion is the primary computational bottleneck of this method. 
Each nonadiabatic region spawns new TBFs with pathological growth in systems with dense state manifolds or multiple sequential crossings as $N_{\text{TBF}} \propto 2^{N_{\text{crossings}}}$.
Additional bottlenecks include linear dependence when TBFs converge in phase space ($\det(\mathbf{S}) \rightarrow 0$). 
New energy, gradient, and coupling calculations are then required at every timestep. 
Progresses in GPU acceleration (TeraChem) give $10$-$100\times$ improvement through raw matrix multiplication speedup, enabling recent applications to photoswitches and DNA photostability \cite{seritan2021terachem}, though small system sizes ($\lesssim 40$ degrees of freedom) limits the application of AIMS/AIMSWISS in practice.

\subsection{Where Quantum Computation Fits}
Classical approaches occupy niches in optimizing accuracy, cost, and applicability.
ML-MCTDH is cost-effective and applicable to highly symmetric molecules, sacrificing accuracy when a tree topology poorly approximates mode entanglement. 
TD-DMRG is accurate for short times and limited to systems with area law entanglement scaling.
TD-CC methods provide systematically improvable electronic structure accuracy with $\mathcal{O}(N_{\text{orb}}^6)-\mathcal{O}(N_{\text{orb}}^8)$ scaling, now capable of treating conical intersections through similarity-constrained formulations, but remain limited to relatively small systems and challenged by strong correlation.
AIMS enables fully quantum nuclear dynamics on high-dimensional potential surfaces through adaptive Gaussian bases, with costs dominated by on-the-fly electronic structure. 
The addition of informed stochastic selection in AIMSWISS controls trajectory proliferation while preserving coherence.
The common thread among these classical bottlenecks is the curse of dimensionality manifesting through different mechanisms. 

Quantum algorithms offer exponentially large Hilbert spaces on qubit registers, where superpositions of multiple configurations or paths overcome the limits of classical Complete Active Space Self-Consistent Field (CASSCF) and Multireference Configuration Interaction (MRCI) methods \cite{sangiogo2025nonadiabatic}. 
The opportunity to exhibit advantage is readily present and discussed, with complexity calculations and early implementations promising efficiency \cite{kassal2008polynomial, kassal2011simulating}.

\section{Quantum Simulation Overview}
NA-QMD simulation presents a set of challenges apart from the goals of estimating ground states and phases.
Quantum Signal Processing (QSP) and qubitization offer asymptotically optimal query complexity for Hamiltonian simulation, the computational overhead dependent on Hamiltonian structure. 
This section works through complexity limitations of the diabatic Hamiltonian and applicability of query-optimal methods. 
Consideration of these powerful, state-of-the-art methods highlights a need to refine techniques for the class of dense Hamiltonians that define NA-QMD.
A parallelizable, oracle-free circuit structure is motivated, beginning with a thermodynamic perspective of quantum information. 
A detailed quantitative comparison of REQWIEM against three QSP-based constructions and a recent end-to-end fault-tolerant estimate is deferred to Chapter~\ref{ch:comparison}, where $\Tgate$-gate counts, circuit depth, qubit overhead, and Toffoli costs are evaluated on a common benchmark.

\subsection{Complexity Geometry}

Brown and Susskind \cite{brown2018second} conjecture that quantum system evolution with $n$ qubits maps to an auxiliary classical particle's orbit in the $SU(2^n)$ Lie group's manifold. 
Circuit complexity is defined as the geodesic distance from the identity to a target unitary under a metric that penalizes non-local operations.

With a Pauli basis $\{\sigma_I\}$ and Lie algebra $\mathfrak{su}(2^n)$, a Hamiltonian $H$ generating infinitesimal evolution $dU = -iH dtU$ expands as $H = \sum_I J_I \sigma_I,$ where $J_I$ are real coupling coefficients and $I$ ranges over all non-identity Pauli strings.
The complexity metric is defined as a right-invariant Riemannian metric on $SU(2^n)$
\begin{equation}
ds^2 = \sum_{I,J} \mathrm{Tr}\left[ i \, dU \, U^\dagger \sigma_I \right] \mathcal{I}_{IJ} \, \mathrm{Tr}\left[ i \, dU \, U^\dagger \sigma_J \right],
\end{equation}
with penalty factor matrix $\mathcal{I}_{IJ}$.
For $k$-local Hamiltonians (those acting on at most $k$ qubits), penalty factors are set to unity for operators of weight $\leq k$ and exponentially large ($\sim e^{\text{weight}}$) for higher-weight operators
\begin{equation}
\mathcal{I}_{IJ} = \delta_{IJ} \times \begin{cases} 1 & \text{if } |I| \leq k \\ e^{\alpha(|I| - k)} & \text{if } |I| > k \end{cases},
\end{equation}
with weight $|I|$ of Pauli string $\sigma_I$ and penalty strength $\alpha > 0$.

Right-invariance indicates a structural dependence on the implementation of an operator.
Circuit complexity $\mathcal{C}(U)$ is then defined as the geodesic length from the identity
\begin{equation}
\mathcal{C}(U) = \min_{\gamma: I \to U} \int_\gamma ds,
\end{equation}
where the minimum is taken over all paths $\gamma$ connecting the identity $I$ to $U$.

Resource requirements can be stated in terms of two entropies, analogous to common quantities in quantum computing.
Positional entropy $S_{\text{pos}}$ quantifies the computational work required to traverse the unitary manifold, measuring volume of the explored complexity geometry
\begin{equation}
S_{\text{pos}}(t) \propto \langle \mathcal{C}(U(t)) \rangle_{\text{ensemble}},
\end{equation}
For chaotic systems, positional entropy grows linearly with time at rate $\mathcal{K}$ as $S_{\text{pos}}(t) \approx \mathcal{K} t$ until saturation at $S_{\text{pos}}^{\max} \sim e^n$.
This corresponds to time evolution generating computational complexity at a steady rate bounded by the interaction strength.

Kinetic entropy quantifies required information to specify the Hamiltonian, as the Kolmogorov complexity of coupling constants.
\begin{equation}
\mathcal{C}_\kappa(H) = K\left( \{J_I\} \right)
\end{equation}
Here, $K(\cdot)$ denotes the length of a ``shortest program" outputting the set of coupling constants to precision $\delta$.
For a $k$-local Hamiltonian with polynomially many terms, $\mathcal{C}_\kappa(H) = \mathcal{O}(\text{poly}(n))$.
An unstructured Hamiltonian requires specification of $\mathcal{O}(4^n)$ independent coefficients, with kinetic entropy
\begin{equation}
\mathcal{C}_\kappa(H) = \mathcal{O}(4^n \log(1/\delta)).
\end{equation}

Total complexity of a quantum simulation task is bounded
\begin{equation}
\mathcal{C}_{\text{total}} \geq \mathcal{C}_\kappa(H) + \mathcal{C}_{\text{circuit}}(t),
\end{equation}
where $\mathcal{C}_{\text{circuit}}(t)$ is the circuit depth required for time evolution.

Asymptotically optimal simulation algorithms do not circumvent high kinetic entropy cost, with a Hamiltonian description becoming the dominant resource.
These algorithms rely on an ``oracle," being a black-box with an approximate solution for evolving the wavefunction.
Optimality is defined by a minimum number of queries made to the oracle.
If a Hamiltonian decomposes into $\texttt{N}$ terms, the resources required to implement block-encoding and the PREPARE/SELECT oracles scale with $\texttt{N}$, requiring many Toffoli gates and uncomputation (aka ``freeing") of ancilla registers.
With the 1-norm of nonadiabatic systems scaling quadratically with the number of internal states (channels, $\lambda \propto M^2$) and exponentially with the number of degrees of freedom (modes, $\lambda\propto N^d$ for $N$ basis states), QSP and qubitization need an increasingly large oracle, leaving any number of queries impractical for large systems. 
The additional requirement of serialization requires the $N$ terms to be applied without parallelism, prompting the development of oracle-free approaches that distribute necessary resources.

\subsection{Geodesic Curvature and Chaos}

Manifold geometry encodes dynamical properties of the quantum system.
The sectional curvature determines whether nearby geodesics converge or diverge.
For clarity, $R_{\theta_y \theta_x}^{\phantom{\theta_y}\phantom{\theta_x}\theta_x\theta_y}$ denotes this curvature, with
$R_\theta$ later respecified to represent arbitrary phase rotations about a Bloch sphere axis.
For a complexity metric with penalty factor $\mathcal{I}_{zz}$ on non-local directions,
\begin{equation}
R_{\theta_y \theta_x}^{\phantom{\theta_y}\phantom{\theta_x}\theta_x\theta_y} = 4 - 3\mathcal{I}_{zz}.
\end{equation}
When $\mathcal{I}_{zz} > 4/3$, sectional curvature becomes negative and geodesics diverge exponentially.
The Lyapunov exponent $\lambda_L$ governing this divergence satisfies
\begin{equation}
\lambda_L = \sqrt{|R|} \cdot v,
\end{equation}
where $v$ is the velocity of the time-evolution operator along the geodesic with $\lambda_L>0$ indicating a chaotic regime.

This geometric characterization shows that chaotic quantum systems exhibit maximal complexity growth.
These dynamics incur large penalties, resulting in negative sectional curvature and large Lyapunov exponents.
As such, small differences in paths lead to exponentially different final states.
The trajectory through the unitary manifold cannot be shortcut; an algorithm predicting the final state without traversing intermediate steps requires resources comparable to the physical evolution itself.
NA-QMD is intrinsically chaotic near conical intersections and avoided crossings \cite{yamasaki2003chaos, takatsuka2022quantum, takatsuka2024mechanism, markiewicz2001chaos}, necessitating an algorithm with complexity growing linearly in time.

\subsection{Oracle-Based Quantum Simulation: QSP and qubitization}

QSP and qubitization represent the state of the art for Hamiltonian simulation, achieving optimal query complexity in evolution time $t$ and precision $\eps$ \cite{low2017optimal, gilyen2019quantum}.
Once the Hamiltonian is embedded in a larger unitary via block encoding, spectral information is processed and time evolution is implemented. 

Qubitization constructs a quantum walk operator $W$ from the block encoding
\begin{equation}
W = e^{i \arccos(H/\lambda)},
\end{equation}
with one-norm $\lambda$ and eigenvalues $e^{\pm i \arccos(E_k/\lambda)}$ encoding the energy spectrum.
The one-norm follows from a Hamiltonian decomposition into a linear combination of unitaries (LCU)
\begin{equation}
H = \sum_{\ell=0}^{L-1} c_\ell U_\ell,
\end{equation}
where $U_\ell$ are efficiently-implementable unitary operators (typically tensor products of Paulis).
Scalar coefficients $c_\ell > 0$ are used to define $\lambda = \sum_\ell |c_\ell|$.

For time evolution, QSP applies a polynomial transformation to $W$, achieving a bounded error
\begin{equation}
\left\| e^{-iHt} - P_d(W) \right\| \leq \eps,
\end{equation}
where $P_d$ is a degree-$d$ polynomial with $d = \mathcal{O}(\lambda t + \log(1/\eps))$.

The query complexity of qubitization defines the number of calls to the PREPARE and SELECT oracles
\begin{equation}
Q = \mathcal{O}\left( \lambda t + \frac{\log(1/\eps)}{\log\log(1/\eps)} \right),
\end{equation}
and is provably optimal with a quantum lower bound $\Omega(\lambda t)$ for Hamiltonian simulation \cite{berry2015hamiltonian}.

The LCU oracle consists of two components.
PREPARE $P$ loads normalized coefficients into an ancilla register
\begin{equation}
P |0\rangle = \frac{1}{\sqrt{\lambda}} \sum_{\ell} \sqrt{c_\ell} |\ell\rangle,
\end{equation}
SELECT $S$ applies unitaries conditioned on the state of the ancilla
\begin{equation}
S = \sum_{\ell} |\ell\rangle \langle \ell| \otimes U_\ell.
\end{equation}
The block-encoding unitary $U_{\text{BE}}$ is constructed from these two in a PREPARE$^\dagger$-SELECT-PREPARE scheme, as $(P^\dagger \otimes I) S (P \otimes I)$, where
\begin{equation}
\langle 0 | U_{\text{BE}} | 0 \rangle = \frac{H}{\lambda}.
\end{equation}
The normalized Hamiltonian is encoded in the top-left block of $U_{\text{BE}}$.

Optimality is measured here in query complexity, with gate complexity including oracle implementation cost.
For a Hamiltonian decomposed into $L$ terms, PREPARE requires loading $L$ amplitudes into a quantum state, costing $\mathcal{O}(L)$ gates for arbitrary amplitudes, condensing to $\mathcal{O}(\log L)$ for amplitudes with exploitable structure using standard amplitude encoding.
SELECT requires $\mathcal{O}(L)$ controlled operations in the general case.

This gives a total gate complexity
\begin{equation}
G = Q \times G_{\text{oracle}} = \mathcal{O}\left( (\lambda t + \log(1/\eps)) \times L \right),
\end{equation}
for unstructured Hamiltonians.
When $L$ scales polynomially in system size (as for $k$-local Hamiltonians), this remains efficient.
When $L$ scales exponentially, gate complexity becomes intractable.
Chapter~\ref{ch:comparison} makes this gate-level accounting explicit for nonadiabatic molecular dynamics, comparing the total non-Clifford cost of oracle-based and oracle-free architectures on a common benchmark system.

\subsection{First-Quantized Methods}

The work of Georges et al. \cite{georges2025quantum} extends LCU to first-quantized quantum chemistry with arbitrary basis sets, including Gaussian orbitals and dual plane waves.
This approach was presented and popularized by coauthor R\'obert Izs\'ak and offers potential advantages for static properties, with limitations for dynamical simulation.

In first quantization, the electronic Hamiltonian is given in for $\eta$ electrons and $N$ nuclei in $D$ basis functions.
Here, $r_i$ serve as electron position vectors and $R_i$ as nuclear position vectors.

\begin{equation}
H = \sum_{i=1}^{\eta} T_i + \sum_{i<j} V_{ee}(r_i, r_j) + \sum_{i,A} V_{eN}(r_i, R_A).
\end{equation}

Sums run over electron indices rather than orbital occupation numbers.
The LCU decomposition expresses this as Pauli strings acting on registers encoding electron positions.

LCU term scaling acts as $L = \mathcal{O}(D^4)$ for two-body interactions, same as for second quantization.
First-quantized preference emerges in the one-norm
\begin{equation}
\lambda = \sum_{p,q} |\omega'_{pq}| \cdot \eta + \sum_{p,q,r,s} |\omega'_{pqrs}| \cdot \frac{\eta(\eta-1)}{2},
\end{equation}
scaling as $\mathcal{O}(D^3)$ for fixed electron number, compared to $\mathcal{O}(D^4)$ in second quantization.

This protocol's design is especially useful for quantum phase estimation, with critical features of NA-QMD left absent, including unconsidered derivative couplings and nonadiabatic couplings.
For structured basis sets such as dual plane waves with diagonal Coulomb interactions, the number of unique nonzero LCU coefficients reduces to $\mathcal{O}(D^{1.84})$ through sparsity \cite{georges2025quantum}.
This sparsity is a property of the electronic Coulomb operator and does not extend to the addition of arbitrary nonadiabatic coupling terms, being skew-Hermitian ($V_{ij}=-V_{ji}^\dagger$) in the adiabatic representation and Hermitian ($V_{ij}=V_{ji}^\dagger$) in a diabatic representation. 
Excluding spin-orbit interactions, this Hermiticity reduces to a symmetric matrix ($V_{ij}=V_{ji}$).

\subsection{Adiabatic v. Diabatic}
Current oracle-based methods have power in adiabatic regimes and ground state energy estimation.
In the adiabatic representation, the molecular Hamiltonian looks like
\begin{equation}
H_{\text{ad}} = T_{\text{nuc}} + \sum_n |n(R)\rangle V_n(R) \langle n(R)|,
\end{equation}
where $|n(R)\rangle$ are the electronic eigenstates at nuclear configuration $R$ and $V_n(R)$ are the adiabatic potential energy surfaces.
The nuclear kinetic energy operator $T_{\text{nuc}}$ is local and sparse in any reasonable basis.

The potential energy surfaces $V_n(R)$ are smooth, typically well-approximated by polynomial expansions
\begin{equation}
V_n(R) = V_n^{(0)} + \sum_i \omega_i (R_i - R_i^{\text{eq}})^2 + \sum_{ijk} \phi_{ijk} (R_i - R_i^{\text{eq}})(R_j - R_j^{\text{eq}})(R_k - R_k^{\text{eq}}) + \cdots,
\end{equation}
For $d$ vibrational modes, specifying this expansion to order $p$ requires $\mathcal{O}(d^p)$ parameters, polynomial in the number of degrees of freedom.

The kinetic entropy of the adiabatic Hamiltonian is therefore
\begin{equation}
\mathcal{C}_\kappa(H_{\text{ad}}) = \mathcal{O}(\text{poly}(d) \cdot \log(1/\delta))
\end{equation}
This low kinetic entropy enables efficient oracle construction, where PREPARE loads polynomially many structured amplitudes, and SELECT applies polynomially many terms.

In the diabatic representation, the Hamiltonian is
\begin{equation}
H_{\text{dia}} = T_{\text{nuc}} \otimes \mathbf{1}_{\text{el}} + \sum_{m,n} |m\rangle V_{mn}(R) \langle n|,
\end{equation}
where $|m\rangle, |n\rangle$ are fixed diabatic electronic states (independent of $R$) and $V_{mn}(R)$ is the diabatic potential energy matrix (DPEM).
The diagonal elements $V_{mm}(R)$ are diabatic surfaces.
Off-diagonal elements $V_{mn}(R)$ ($m \neq n$) are nonadiabatic couplings responsible for vibronic and rovibronic transitions.

Near conical intersections, the topology of the intersection seam, wavefunction branching, and Berry phase accumulated when encircling the intersection all contribute to a coupling landscape that quickly saturates entropy.

For $M$ coupled electronic states and nuclear coordinates discretized on a grid with $N=2^n$ points per mode across $d$ modes, the total number of independent potential values becomes
\begin{equation}
N_{\text{DPEM}} = M^2 \times N^d
\end{equation}
If these values have no exploitable structure, the kinetic entropy becomes
\begin{equation}
\mathcal{C}_\kappa(H_{\text{dia}}) = \mathcal{O}(M^2 \cdot N^d \cdot \log(1/\delta)),
\end{equation}
which is exponential in the number of vibrational modes $d$.

For the diabatic Hamiltonian, the number of LCU terms scales as
\begin{equation}
L = \mathcal{O}(M^2 \cdot N^d).
\end{equation}
The PREPARE oracle must load $\mathcal{O}(N^d)$ amplitudes for each of the $\mathcal{O}(M^2)$ coupling functions.
Even with quantum amplitude encoding achieving $\mathcal{O}(N^d)$ gates (logarithmic in Hilbert space dimension but exponential in mode number), the construction cost dominates.

For unstructured coupling potentials with typical magnitude $\bar{V}$, the one-norm scales as
\begin{equation}
\lambda \sim M^2 \cdot N^d \cdot \bar{V}.
\end{equation}
The query complexity $\mathcal{O}(\lambda t)$ thus scales exponentially in $d$, independent of any algorithmic optimizations.

The total complexity of qubitization applied to NA-QMD is
\begin{equation}
\mathcal{C}_{\text{total}} = \underbrace{\mathcal{O}(M^2 \cdot N^d)}_{\text{Kinetic Entropy (oracle)}} + \underbrace{\mathcal{O}(\lambda t)}_{\text{Positional Entropy (queries)}}
\end{equation}
Both terms are exponential in $d$.

Ground state-finding, phase estimation, and adiabatic dynamics are important subfields in quantum chemistry where quantum advantage can be realized with sufficiently large, error-corrected hardware.
For these problems, LCU, QSP, and qubitization methods show promise.
Limitations in simulating nonadiabatic, chaotic systems motivate the construction of a new algorithm designed to enact the desired physics on quantum registers to recover dynamic observables. 

\section{Algorithmic Constraints}

Following information theoretic constraints, necessary behaviors and expected complexity constraints are specified.

\subsection{Non-Fast-Forwardability of Nonadiabatic Dynamics}

A Hamiltonian $H$ acting on $n$ qubits is $(T(n), \alpha(n))$-{fast-forwardable} if there exists a quantum circuit of size $\text{poly}(n)$ that simulates evolution for time $t \leq T(n)$ with error at most $\alpha(n)$ \cite{atia2017fast}.
The Atia-Aharonov theorem establishes that if PSPACE $\neq$ BQP, there are physically realizable Hamiltonians that are not $(T(n), 1/4)$-fast-forwardable for any $T(n) = 2^{\omega(\text{poly}(n))}$. 

For sparse Hamiltonians, a tighter lower bound applies \cite{berry2007efficient, childs2009limitations}:
\begin{equation}
Q = \Omega\left( \|H\| \cdot t \right)
\end{equation}
Simulating evolution for time $t$ requires at least $\Omega(t)$ queries to the Hamiltonian oracle, establishing that no algorithmic speedup beyond constant factors is possible.

Taking the vibronic Hamiltonian as an example, kinetic, potential, and coupling terms do not commute 
\begin{equation}
[T_{\text{nuc}}, V_{mn}(R)] \neq 0, \quad [V_{mm}(R), V_{mn}(R)] \neq 0.
\end{equation}
Fast-forwarding techniques for commuting Hamiltonians \cite{atia2017fast} therefore do not apply. 
This lack of commutation is directly considered when calculating the Trotter error in later chapters.

Generic vibronic Hamiltonians with tunable couplings can simulate universal quantum computation.
Conical intersections provide a natural two-level system, and nuclear motion provides continuous control parameters.
By mapping quantum computation to the conical intersection problem, an algorithm that is fast-forwarded would be solving a BQP-complete problem in polynomial time. 

From a locality perspective, nonadiabatic coupling connects all nuclear configurations with appreciable wavefunction amplitude near conical intersections. 
The resulting Hamiltonian would not be sparse nor banded.

The non-fast-forwardability theorem \cite{atia2017fast} implies that a circuit that simulates NA-QMD has a depth that scales at least linearly with physical evolution time
\begin{equation}
D_{\text{circuit}} = \Omega(t / \Delta t_{\min}),
\end{equation}
where $\Delta t_{\min}$ is set by inverse spectral bandwidth $\|H\|^{-1}$.
This limitation is incorporated when calculating the estimated number of timesteps required for simulating ultrafast dynamics.
The statement follows that nonadiabatic dynamics are computationally irreducible, meaning circuit depth must scale linearly with the physical evolution time $t$, prohibiting any further compression of simulation costs without sacrificing accuracy.

\subsubsection{Scrambling Dynamics and Algorithmic Instability}

Beyond non-fast-forwardability, NA-QMD is chaotic and introduces an exponential sensitivity to algorithmic approximations.
Sensitivity can be quantified through Out-of-Time-Ordered Correlators (OTOCs) and the Loschmidt echo.

The Out-of-Time-Ordered Correlator for operators $V$ and $W$ is defined as
\begin{equation}
C_{VW}(t) = \langle [W(t), V]^\dagger [W(t), V] \rangle,
\end{equation}
where $W(t) = e^{iHt} W e^{-iHt}$ is the Heisenberg-evolved operator.
The OTOC measures the degree to which initially commuting operators fail to commute under time evolution.

OTOCs exhibit exponential growth at early times in chaotic systems \cite{maldacena2016bound}
\begin{equation}
C_{VW}(t) \sim \hbar^2 e^{2\lambda_L t},
\end{equation}
where $\lambda_L$ is the quantum Lyapunov exponent, bounded by temperature as $\lambda_L \leq \frac{2\pi k_B T}{\hbar}$.
Systems saturating this bound are termed ``maximally chaotic'' or ``fastest scramblers.''

For molecular systems, Zhang, Wolynes, and Gruebele \cite{zhang2022quantum} demonstrated that vibrational scrambling in polyatomic molecules occurs on femtosecond timescales, with scrambling times $t_*$ scaling with the density of coupled vibrational states.
This would classify inter-mode vibrations as ``slow scramblers" ($2\times\lambda_L\times10\,\text{fs} \approx10^{-2} \text{ at } 1K$).
At conical intersections, the breakdown of the Born-Oppenheimer approximation couples electronic and nuclear degrees of freedom, accelerating scrambling toward the chaos bound with dimensional rescaling finding effective scrambling rates comparable to black holes \cite{zhang2022quantum}.

The Loschmidt echo quantifies sensitivity to perturbations
\begin{equation}
M(t) = \left| \langle \psi_0 | e^{iH_2 t} e^{-iH_1 t} | \psi_0 \rangle \right|^2,
\end{equation}
where $H_1$ is the exact Hamiltonian and $H_2 = H_1 + \delta H$ is a perturbed version.
For chaotic systems, the echo decays exponentially at a rate determined by the Lyapunov exponent \cite{jalabert2001environment}
\begin{equation}
M(t) \sim e^{-\lambda_L t}
\end{equation}
This is independent of perturbation strength in the Lyapunov regime, so errors cannot be made arbitrarily small by improving algorithmic precision.

To relate the information theoretic limits to computation, consider an approximate simulation algorithm that implements $\tilde{U}(t) = e^{-i\tilde{H}t}$ instead of the exact propagator $U(t) = e^{-iHt}$.
The algorithmic error $\delta H = \tilde{H} - H$ may arise from:
\begin{itemize}
\item Trotter-Suzuki decomposition errors in product formula methods
\item Polynomial approximation errors in QSP
\item Finite precision in phase arithmetic
\item Truncation of the LCU expansion
\item Grid discretization (approximating a continuum)
\end{itemize}

For chaotic nonadiabatic dynamics, fidelity between exact and approximate evolutions decays as
\begin{equation}
F(t) = |\langle \psi_{\text{exact}}(t) | \psi_{\text{approx}}(t) \rangle|^2 \sim e^{-\lambda_L t}.
\end{equation}
Maintaining $F(t) > 1 - \eps$ over time, algorithmic error must satisfy
\begin{equation}
\|\delta H\| < \frac{\eps}{\lambda_L t} e^{-\lambda_L t},
\end{equation}
decreasing exponentially with simulation time.

This exponential precision requirement translates directly to resource costs.
In QSP, the polynomial degree $d$ required for precision $\eps$ scales as $d = \mathcal{O}(\log(1/\eps))$ for fixed $t$, but maintaining fixed fidelity over increasing $t$ requires $\eps \sim e^{-\lambda_L t}$, giving
\begin{equation}
d = \mathcal{O}(\lambda_L t).
\end{equation}
Scrambling dynamics thus impose an unavoidable linear-in-$t$ overhead.
For data-loading procedures, overhead appears in the oracle size and cost of querying the oracle, shifting to a per-timestep overhead for product formulas, where it is frequently represented by an $\mathcal{O}(k)$ overhead for $k$ timesteps.

Sensitivity to approximation, compounded by the overhead of scrambling dynamics, creates a fundamental bottleneck for NA-QMD: reliance on a necessarily serial oracle. 
Vibronic couplings in these systems represent information-dense objects that cannot be scalably compressed into a sparse block-encoding. 
Information-theoretic bounds dictate that an oracle representing $N$ unique, non-structured coefficients requires $\mathcal{O}(N)$ operations to load, which translates to greater circuit depth for chaotic molecular dynamics. 
With an unavoidable $\mathcal{O}(t)$ complexity, minimizing circuit depth suggests the distribution of necessary computational overhead.
This dissertation investigates the direct application of the Hamiltonian by applying kinetic and potential energy terms directly to the wavefunction to avoid a serially-applied oracle, achieving parallelism at scale or up to large constant factors. 

If an algorithm like this is constructed, the exponential kinetic entropy is not compressed into oracle depth, but distributed across parallel computational resources.
The time evolution operator is approximated via product formulas, where the kinetic operator $e^{-iT_{\text{nuc}} \Delta t}$ is diagonal in momentum space and can be applied in between quantum Fourier transforms.
The potential operator $e^{-iV(R) \Delta t}$ is diagonal in position space, factorizable into phase rotations in a computational basis
\begin{equation}
e^{-iV(R) \Delta t} = \bigotimes_{R} e^{-i V(R) \Delta t} |R\rangle \langle R|
\end{equation}

This direct approach may accommodate fast scrambling dynamics.
An algorithm following a product formula approach is derived, detailed, and classically emulated in this document.
Product formula errors accumulate linearly rather than being amplified by chaos, since each time step applies the Hamiltonian directly at that instant (up to Trotter error in operator ordering).
The Loschmidt echo for product formulas decays as
\begin{equation}
M(t) \sim 1 - \mathcal{O}\left( \|[T, V]\|^2 \Delta t^2 \cdot t \right),
\end{equation}
for second-order Trotter, which is polynomial in $t$ rather than exponential.

Complexity is generated incrementally through time evolution rather than front-loaded into oracle construction, matching algorithm structure to physical dynamics and enabling efficient simulation even in the chaotic regime.

These results motivate the development of oracle-free simulation methods that distribute kinetic entropy across parallel resources rather than serializing it into circuit depth.
The following sections develop such a method, demonstrating that direct parallel application of the diabatic Hamiltonian achieves efficient quantum simulation of NA-QMD.

The oracle-free, parallelized product formula structure motivated above is not specific to NA-QMD or a common test case in vibronic Hamiltonians. 
Computationally, the algorithm applies a dense, position-dependent operator matrix to a grid-encoded wavefunction across multiple internal channels via binary-weighted rotations and uniformly controlled rotations.
This is a primitive for quantum simulation of coupled-channel dynamics on discretized coordinate spaces, analogous in function to level-2 BLAS routines in classical scientific computing. 
Any problem requiring a smooth potential step in a split-operator decomposition calls this primitive: the vibronic coupling matrices of nonadiabatic photochemistry, the inverse-polynomial and Morse potentials of reactive scattering, the anharmonic double-well surfaces of excited-state intramolecular proton transfer, and the coupled dissociative channels of predissociation and photodissociation all reduce to the same circuit-level operation with different classical preprocessing of rotation angles. 
A quantum analog to this primitive is developed here, tested under qualitatively different conditions to establish an architecture that can be reused for future grid-based quantum simulation. 
The vibronic application \cite{motlagh2025quantum} represents a demanding test case due to the density of the coupling matrix, the multiplicity of electronic channels, and chaotic dynamics near conical intersections.
Scattering and ESIPT applications confirm that the same circuit structures extend without modification to qualitatively different physics.
This architecture provides the real-time evolution subroutine required by quantum Krylov subspace diagonalization protocols, where parallelized application of the propagator directly accelerates the generation of Krylov vectors. 
    Extensions to diffusion problems and coupled nonlinear ODEs through operator splitting are natural and discussed in Chapter~\ref{ch:Application}.

\subsection{Divisions in Quantum Computation}
Now considering how an algorithm should be constructed or framed, the quantum computing landscape finds itself split as emergent limitations of quantum hardware guide implementation. 
Resources, sources of error, and the field's zeitgeist factor into development, discussed in this subsection.

\subsubsection{NISQ v. FTQC}
A term to describe early quantum hardware by John Preskill in 2018, Noisy Intermediate-Scale Quantum (NISQ) computing faces severe scaling limitations.
While NISQ devices point toward ``quantum supremacy" in specific quantum information metrics such as random circuit sampling, broader application falters with uncorrectable noise.
Statistical methods used to mitigate errors on noisy processors incur a sampling overhead that scales exponentially with circuit depth and error rate \cite{eisert2025mind}.
Effective computational volume of NISQ machines hits a ceiling where deep circuits are rendered useless by the accumulation of gate errors. 
Focusing on long-time evolution, error correction becomes the leading topic of discussion \cite{mohammad2025quantum}.

Fault-Tolerant Quantum Computing (FTQC) aims to suppress logical errors exponentially by increasing the code distance of error-correcting codes. 
This transition is marked by the race to demonstrate ``break-even" points where logical qubits outperform physical qubits.
Google Quantum AI's Willow processor recently demonstrated increasing code distances reducing logical error rates \cite{meister2024efficient}.

This dissertation constructs an algorithm that closely resembles those designed for NISQ architectures, but must be weighed on the fault-tolerant scale for any future utility.
NISQ algorithms typically use a native gate set with arbitrary rotations, such as a phase shift $R_z(\theta) = e^{-i \theta Z / 2}$.
Using trapped-ion hardware as an example, amplitude, duration, and pulse shape of an analog microwave or laser pulse achieves a rotation angle $\theta$ on the measurement axis up to some precision \cite{schafer2018fast}. 
NISQ algorithms (like VQE or QAOA) therefore aim to explore continuous parameter spaces easily.
Quantum Error Correction (QEC) protocols by contrast require a discrete set of gates (typically the Clifford group) to identify errors without collapsing the quantum state \cite{bluvstein2025architectural}.
Any gate outside the Clifford group (T, $\sqrt{\text{T}}$) require magic state distillation, where noisy copies distill into high-fidelity ones, dominating resource overhead \cite{wang2025orchestrating}.
QEC codes generally cannot perform non-Clifford gates transversally, with FTQC systems synthesizing arbitrary rotations by consuming magic states.
Restriction is brought by the Eastin-Knill theorem \cite{eastin2009restrictions}, stating that any code capable of correcting local errors gives a set of transversal logical gates forming a finite group. 
As a popular example, the Steane code ([[7,1,3]]) has transveral Clifford gates, but the T gate must be distilled and injected \cite{steane1996multiple}.
Similarly, the Reed-Muller [[15,1,3]] code and some 3D color codes perform the T gate transversally, sacrificing a transversal Hadamard \cite{steane2002quantum}.
Though the circuits presented here apply arbitrary rotations, resource estimates use a conversion to T gates for a given rotation error $\eps_{\theta}$.

\subsubsection{Analog v. Digital}
Simulation of nonadiabatic chemical dynamics, particularly for systems exhibiting conical intersections, represents a primary target for quantum advantage due to the breakdown of the Born-Oppenheimer approximation and the resulting entanglement between electronic and nuclear degrees of freedom.
Analog quantum simulation relies on mapping the Hamiltonian directly onto the quantum device's physical interaction Hamiltonian \cite{georgescu2014quantum}.
Conversely, digital quantum simulation employs a universal set of discrete quantum gates to approximate the time-evolution operator, offering a path to arbitrary precision at the cost of significant gate depth \cite{lloyd1996universal}. 
Analog has demonstrated superior hardware efficiency and coherence resilience on NISQ hardware, while digital largely remains the long-term objective for fault-tolerant implementation with universality and compatibility with quantum error correction (QEC) \cite{preskill2018quantum, postler2022demonstration, campbell2017roads}.
This objective has stood as a cornerstone of quantum computing research, as digital quantum simulation has the potential to emulate any interaction Hamiltonian of arbitrary dimension or scale \cite{feynman1982simulating, lloyd1996universal}.

Efficacy in NA-QMD is determined by an algorithm's ability to encode the diabatic Hamiltonian.
A prominent example of analog methods in the NISQ era is found in recent work by Navickas et al. \cite{navickas2025experimental}, using a mixed qubit-boson trapped-ion simulator. 
By mapping electronic states to internal spin states $H_{el} \propto \sigma_z$ and vibrational modes to motional phonon modes of the trap $H_{vib}\propto a^\dagger a$, they successfully simulated ultrafast dynamics of pyrazine through a conical intersection, naturally handling bosonic statistics. 
Navickas et al. \cite{navickas2025experimental} demonstrated that simulating the Linear Vibronic Coupling (LVC) model for pyrazine required orders of magnitude fewer resources than a comparable digital approach, and suggested the need for Gray code or Jordan-Wigner mappings to represent the vibrational Fock space on qubits.

Analog simulation faces validity challenges regarding scalability and programmability.
$H_{\text{device}}$ must match $H_{\text{target}}$, limiting the simulation to models supported by the native device physics (e.g., harmonic potentials). 
Looking toward the fault-tolerant era, digital algorithms aim to overcome the resource overheads associated with wavefunction encoding. 
A recent proposal by Motlagh et al. \cite{motlagh2025quantum} proposes a digital quantum algorithm for vibronic dynamics based on product formulas (Trotterization). 
Unlike variational approaches, this method focuses on direct time evolution of vibronic Hamiltonians in real space.
The proposal highlights that while digital methods require significant qubit counts (e.g., utilizing many qubits to digitize the continuous vibrational degree of freedom), the use of higher-order Trotter-Suzuki decompositions allow for controlled error bounds.
This contrasts with the difficulty in analog methods of bounding calibration errors, where the mismatch $\delta H = H_{device}-H_{target}$ may be unknown and uncorrectable.

Both paradigms aim to surpass the ``quantum exponential wall" of classical factorization methods. 
A primary goal is to engineer arbitrary system-bath interactions and anharmonic expansions with precise control and universal application.
Analog methods are limited by high noise and decoherence, while cutting edge digital methods have insufficient fidelity, error correction, and coherence times for problems that benefit from a quantum advantage.
Immediate future simulation likely belongs to hybrid analog-digital schemes (like the mixed qubit-boson approach \cite{navickas2025experimental}) that leverage native bosonic modes to compress the Hilbert space.
While a great example that can be reasonably approximated with harmonic potentials and LVC, pyrazine spectra is highly dependent on bilinear terms, and is a convenient ``proof-of-principle" example, used in Chapter~\ref{ch:model_ext} to show nonadiabatic transitions. 
Many molecules however, do not have the eight point group symmetries of pyrazine, and cannot be accurately approximated with a second-order truncation, giving a digital approach broader application for chemical systems. 

\section{Dissertation structure}
This introduction and background helps answer the question ``Where does this thesis fit?"
Reviewing classical methods identifies the ever-rising ceiling of modern computational chemistry.
Quantum information highlights the ability of popular quantum algorithms to simulate sparse Hamiltonians and the strong limitation given by chaotic systems, necessitating a new construction for NA-QMD.
This thesis discusses such a construction, gives numerical validation in simplified chemical models, and calculates scaling for the number of degrees of freedom, number of electronic channels, evolution time, and potential energy structure.
The bulk of the thesis will cover several fundamental topics in chemistry and grid-based simulation.
The circuits presented numerically handle existing problems that scale poorly, extending beyond the trailhead of theoretic justification and harmonic approximation with linear coupling.
Chapter~\ref{ch:wavepacket_prop} will discuss the initialization of a Gaussian wavepacket in a qubit basis, Trotterization in a grid-based simulation, and numerical error accumulation from discretizing a coordinate space. 
Example calculations will be given for relevant problems, along with small-scale quantum hardware results for wavepacket initialization and propagation on a single potential energy surface. 
Chapter~\ref{ch:marcus} introduces the diabatic Hamiltonian and considers the approximation of quantum circuits used to simulate a nonadiabatic transition between coupled harmonic oscillators.
Examples are taken from chemical toy models to real molecules, where variational compression for NISQ-era circuits is performed before larger classical emulation and results recovered from seminal work \cite{ollitrault2020nonadiabatic}.
The Real-time Evolution of Quantum Wavepackets In Explicit Modularity (REQWIEM) algorithm is proposed and proof-of-principle is given through classical emulation in Chapter~\ref{ch:model_ext}.
Two-mode and four-mode pyrazine simulations recover spectral distributions and Berry phase accumulation, and a four-channel Landau-Zener bowtie model is further constructed with convergence analysis for absorption spectra. 
These emulations are compared with multiconfigurational time-dependent Hartree (MCTDH) and classical Hamiltonian diagonalization.
This chapter gives the first numerically validated digital quantum circuits for first-quantized conical intersection dynamics and multi-channel nonadiabatic dynamics, with T-depth improvement across five benchmark systems over current QROM-based proposals \cite{motlagh2025quantum}, calculated in Chapter~\ref{ch:resources}. 

Chapter~\ref{ch:scattering} considers inverse polynomial potentials, deriving circuits to apply inverse polynomials exactly on a binary-encoded grid. 
These circuits are shown to accurately recover phase shifts and scattering amplitudes for a liquid Helium-Helium system. 
This chapter gives the first numerical validation of quantum circuits to accurately evolve scattering systems to reproduce differential cross sections without the classical precomputation of system eigenvalues. 
A direct comparison is made with Gray-encoded circuits~\cite{chang2022improving} to apply these potentials, validating the construction on an operator level. 

Chapter~\ref{ch:smooth_potentials} extends beyond harmonic and inverse polynomials, giving the first derived quantum circuits to exactly apply higher order polynomials, trigonometric functions, exponentials, and smooth analytical functions in a binary qubit basis.
Smooth analytical functions are further extended to nonadiabatic couplings.
Fluorescence dynamics are shown with a system inspired by 2-(2'-hydroxyphenyl)benzothiazole (HBT) fluorescence, using a Gaussian nonadiabatic coupling term to mediate a dissipation/isomerization between cis- and trans- stereoisomers in a keto state (quartic) and an *enol state (harmonic), representing an excited state intramolecular proton transfer (ESIPT) system.
Spectra are reproduced for a NaI$^+$-like system, nonadiabatically coupling a Morse potential to an exponential repulsive potential inducing an avoided crossing. 
This chapter therefore gives the first numerically validated quantum circuits to accurately reproduce spectra for smooth non-polynomial potential, smooth non-polynomial nonadiabatic coupling, and first-quantized nonadiabatic dynamics with a double-well potential, with convergence criteria for register sizes sufficient to capture relevant spectral features.
Chapter~\ref{ch:smooth_potentials} concludes with a derivation for a quantum circuit that can apply an arbitrary 2-dimensional potential energy surface, with binary encoding providing a natural inclusion-exclusion principle that reduces circuits to a minimal number of rotation coefficients in their analytic form.
This circuit is validated with an H\'enon-Heiles potential, recovering recurrence spectra and quantum scars following chaotic behavior.

Chapter~\ref{ch:comparison} provides a five-way quantitative comparison of REQWIEM against three quantum signal processing (QSP) constructions: QSP with a Wigner--Haagerup LCU block encoding, QSP with a kinetic--potential split, and QSP applied in a truncated Fock space. 
These estimates are compared together with the recent end-to-end fault-tolerant estimate of Eklund et al.\ \cite{eklund2026endtoend}.
$\Tgate$-gate counts, circuit depth, Toffoli cost, and qubit overhead are evaluated on the four-mode pyrazine $S_1/S_2$ benchmark, and a Trotter--Dyson duality near conical intersections is established analytically.

Chapter~\ref{ch:Application} considers future quantum hardware application in the fault-tolerant era, future research directions for further optimization, and how the developed architecture extends to quantum mechanical problems placed in a finite basis representation (FBR) / discrete variable representation (DVR) dualism.
REQWIEM presents as an attractive front end to Quantum Krylov Subspace Diagonalization (QKSD) protocols in the parallelized application of real-time evolution. 
Future directions consider customization to handle time-dependent potential surfaces, as are in strong-field and laser excited systems, opening future research directions including bifurcated spectra in strong fields, quantum materials, and astrophysics.
All circuits are derived in Appendix~\ref{app:circuits} and convergence and error calculations are given in Appendix~\ref{app:Error} and Chapter~\ref{ch:wavepacket_prop}.
Additional appendices detail decomposition methods, molecular symmetries, gate synthesis optimization, error calculation, and a recalculation of resource estimates given by previous literature.

\chapter{Wavefunction Propagation}\label{ch:wavepacket_prop}
\section{Wavepacket Initialization}\label{sec:Wavepacket_init}
The first step of wavefunction evolution is to construct an initial state $\psi(x,t=0)=\ket{\psi_0}$.
Two quantum computational methods can be used to perform this initialization, taken from variational quantum eigensolver (VQE) and variational quantum imaginary time evolution (VarQITE) techniques \cite{peruzzo2014variational, motta2020determining}. 
While valuable for intermediate- and large-scale quantum registers ($n\gtrsim12$ qubits), this variational optimization for typical initial shapes can be performed classically for small registers, taking a vector of parameters $\vec{\theta}\in\mathbb{C}^{\mathcal{O}(n^2)}$ and optimizing a parameterized ansatz, dubbed here as a `preparation' circuit $\mathcal{P}(\vec{\theta})$.
A variational optimization follows from the Rayleigh-Ritz quotient \cite{rayleigh1870finding, ritz1909neue}. 
This quotient is bounded by eigenvalues and is minimized variationally to converge on the smallest eigenvalue. 
A typical protocol (e.g. gradient descent) imposes this on a finite Hilbert space by a Fourier decomposition of the Sturm-Liouville equation, whereby parameterizing a circuit to model the quotient allows convergence to an approximate ground state via recursive updates \cite{mcclean2018barren}.
Small registers $n \lesssim12$ is found in this chapter to be sufficient for much of first-quantized wavefunction dynamics, so a classical optimization protocol is adopted to find a good approximation to a Gaussian wavepacket $\ket{\psi_0}\propto e^{-ax^2}$.

\begin{figure}[t]
    \centering  
    \includegraphics[width=0.8\textwidth]{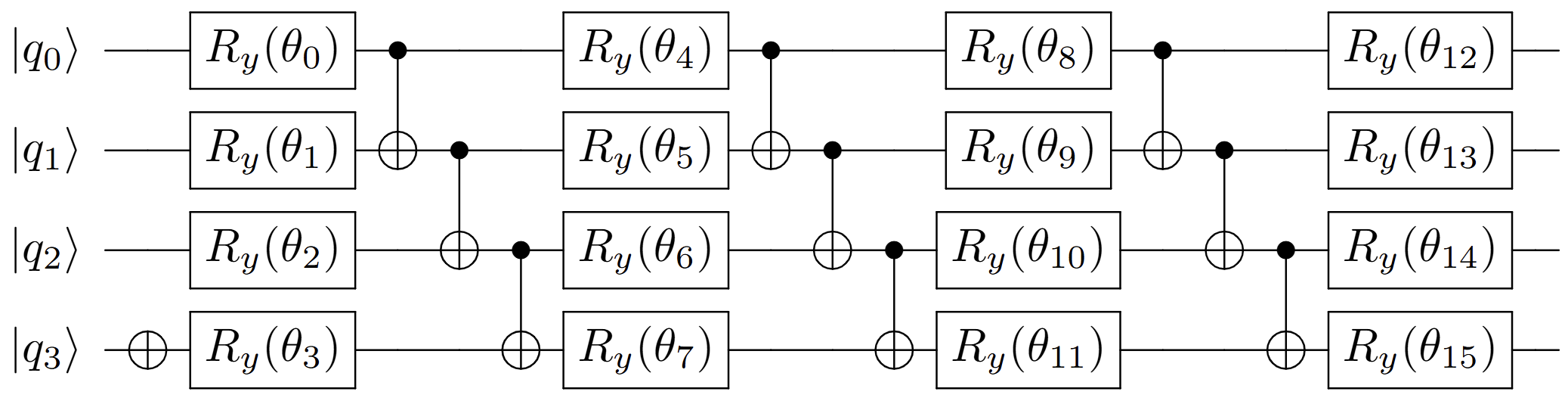}
    \caption{Factorized unitary coupled-cluster ansatz circuit for $n=4$ qubits. Classical emulation can optimize on the ground state of a harmonic oscillator, or maximize state fidelity with a classically-initialized function.}
    \label{fig:UCC-Schematic}
\end{figure}

To follow Ollitrault et al. \cite{ollitrault2020nonadiabatic} and extend the photonic QC implementation of Peruzzo et al. \cite{peruzzo2014variational}, a factorized unitary coupled cluster (UCC) ansatz circuit with $n^2$ variational parameters and $(n-1)^2$ entanglement gates is used (Figure~\ref{fig:UCC-Schematic}). 
The structure of this circuit is Hamiltonian-derived, following the cluster operators from coupled-cluster expansions for singles and doubles \cite{taube2006new}.
The ansatz guarantees that a Gaussian distribution can be constructed up to numerical precision, though it does not guarantee success in finding this global minimum, as parameter space exhibits many local minima and barren plateaus \cite{mcclean2018barren}.

\begin{figure}
    \centering  
    \includegraphics[width=1.0\textwidth]{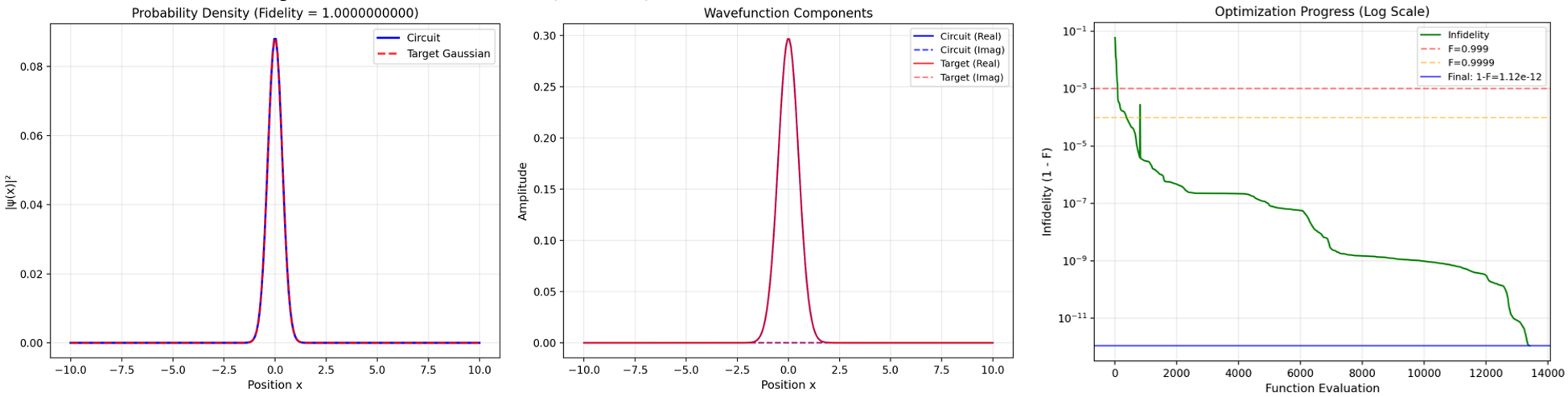}
    \caption{Gaussian wavepacket optimization on $n=8$ qubits. 
    Applied with parallelized gradients, a CPU-based Qiskit backend completed this optimization with a wall time of around three hours.
    Infidelity is defined as $1-F$ with state fidelity $F = |\psi\cdot\tilde{\psi}|^2$ defined between a target wavefunction $\psi$ and approximate wavefunction $\tilde{\psi}$.}
    \label{fig:8q_optimizer_sample}
\end{figure}

For this work, a robust classical optimizer was built to deliver high-fidelity approximations to Gaussian wavefunctions. 
For system sizes $n=3-12$ qubits, the factorized UCC ansatz of Taube and Bartlett \cite{taube2006new} was variationally optimized via classical emulator.
An example optimization is given in Figure~\ref{fig:8q_optimizer_sample}, optimizing for a wavefunction $\psi = Ae^{-x^2/(2\sigma^2)}$, achieving precision of $1-F=10^{-11}$ (eleven 9s) precision for $n=8$ qubits for a box size $L=20$ and Gaussian width of $\sigma=0.5$ (units of distance).
This optimization further shows that no phase is imparted during the optimization, guaranteeing no initial momentum prior to applying the Hamiltonian.

\subsection{Classical Optimization}

This optimization problem is inherently non-convex, exhibiting multiple local minima and susceptibility to barren plateaus where gradients vanish exponentially with system size \cite{mcclean2018barren}.
This exponential behavior makes large systems classically intractable, but for register sizes considered in this work ($n \leq 12$ qubits per dimension), classical emulation of the variational circuit is feasible, enabling high-fidelity state preparation.
This optimization is VQE-compatible \cite{peruzzo2014variational}, though quantum computers are not needed for these small optimizations. 

For variational circuits composed of Pauli rotation gates $R_y(\theta) = e^{-i\theta \sigma_y/2}$, the gradient of the fidelity with respect to each parameter admits an exact analytic expression through the parameter-shift rule \cite{schuld2019evaluating, mitarai2018quantum}:
\begin{equation}\label{eq:parameter_shift}
    \frac{\partial F}{\partial \theta_k} = \frac{1}{2}\left[F\left(\theta_k + \frac{\pi}{2}\right) - F\left(\theta_k - \frac{\pi}{2}\right)\right].
\end{equation}
This formulation requires $2n^2$ circuit evaluations per gradient computation.
While more expensive than finite-difference approximations ($n^2 + 1$ evaluations), the parameter-shift rule provides machine-precision gradients essential for targeted convergence tolerances ($\|\nabla F\| < 10^{-12}$).
After a warm-start initialization, an Adam exploration with warm restarts (cosine annealing) is used to escape shallow minima.
For a threshold fidelity, stochastic basin hopping is employed to escape deeper local minima where Adam stagnates.
Below this threshold, a limited-memory Broyden-Fletcher-Goldfarb-Shanno optimization with bounded variables (L-BFGS-B) with progressively tighter tolerances allows convergence to the bottom of the resulting minima \cite{egidio2021learning}.
This method is a quasi-Newton gradient descent, using first-order gradient information to determine updated direction.
This process does not guarantee a global minimum. 
A ``perfect" optimizer that guarantees such optimization is highly desired in classical computation, though the current protocol typically achieves reliable infidelities $1-F < 10^{-5}$.

\subsubsection{Results and Validation}

\begin{figure}
    \centering
    \includegraphics[width=0.8\textwidth]{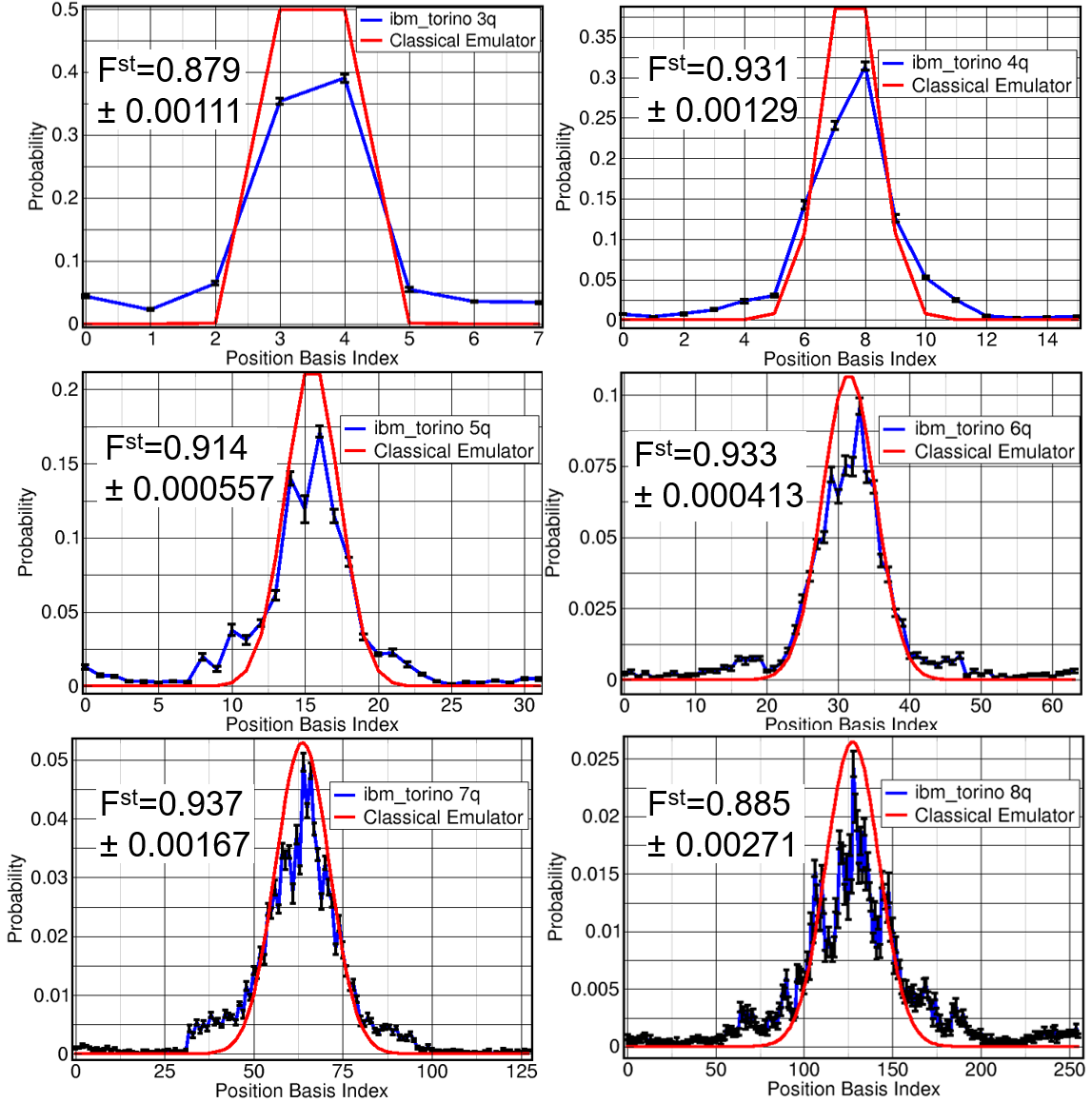}
    \caption{\label{fig:UCC_Hardware_VaryQubits}Quantum circuit compilations of the UCC ansatz applied on $ibm\_torino$.}
    \end{figure}

To test these parameterized ansatzes on quantum hardware, ten independent runs were performed on $n=3-8$ physical qubits, with 1024 weak measurements of the full register per run (2048 for $n=7,8$). 
Error-mitigated results are shown in Figure~\ref{fig:UCC_Hardware_VaryQubits}, with error bars defined as the standard error from the mean for each basis state. 

Gaussian width was set as $\sigma=0.5$ for a box size $L=6.0$. 
As no physical model has yet been imposed, no units are attached to $\sigma$ or box size $L$. 
Trials were performed on $ibm\_torino$, a 133-qubit Heron processor by IBM Quantum with a heavy-hex topology of superconducting transmon qubits.
These are physical qubits, distinguished from the error-corrected logical qubits that will be considered later. 
Gate precisions are discussed later in this chapter, where the triangle inequality gives an upper bound for gate errors and implementation of a quantum circuit $\tilde{U}$ to arbitrary numerical precision gives the operation of a target unitary $U$, such that $||\tilde{U}U^\dagger|| = 0$. 
For a measurement probability $p_i$ and ideal Gaussian probability $q_i$ for basis state $\ket{i}$, a classical state fidelity is defined as
\begin{equation}
    F^{st}=\sum_i \sqrt{p_iq_i}.
\end{equation}
These state fidelities are reported as insets in each respective plot.

\subsubsection{Open-Source software}

\begin{figure}
\centering
\includegraphics[width=0.8\textwidth]{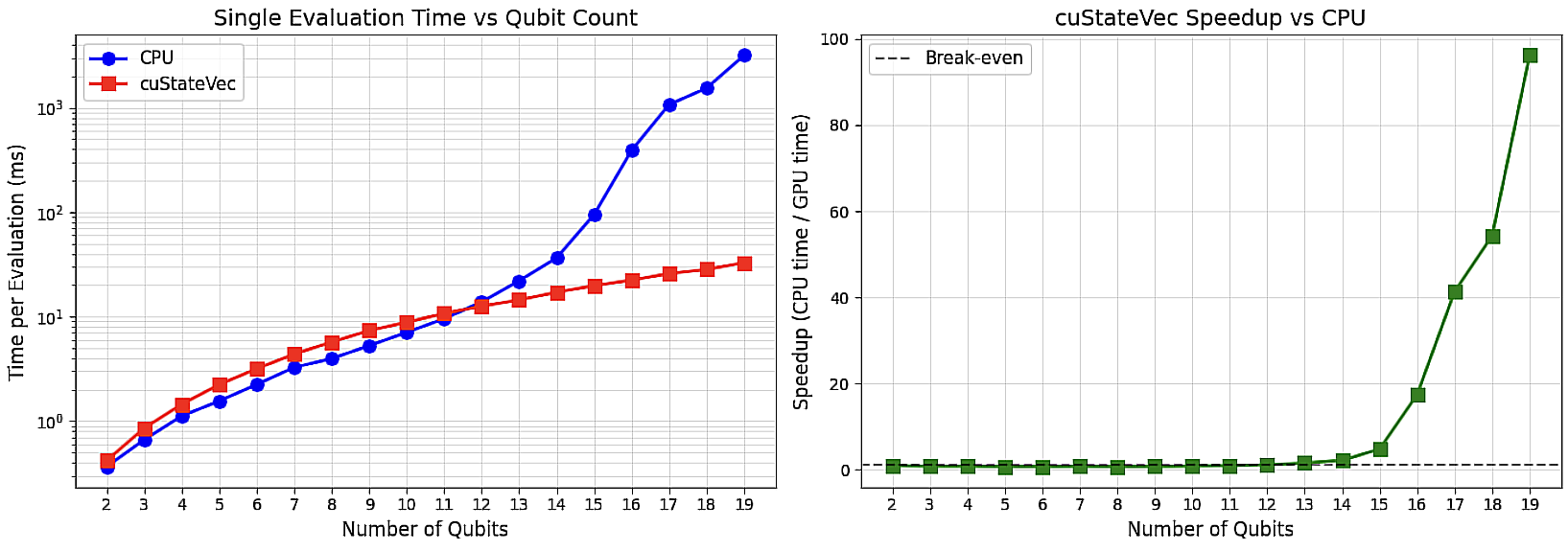}
\caption{\label{fig:GPU_advantage}Evaluation times for various quantum register sizes using the factorized UCC ansatz for Qiskit backend compilation (CPU) and Nvidia cuQuantum (cuStateVec).}
\end{figure}

The use of Nvidia cuQuantum cuStateVector gives significant gradient computation speedup relative to a Qiskit-based CPU optimization, with a projected $>100\times$ gradient calculation speedup for $n\geq20$ (Figure~\ref{fig:GPU_advantage}). 

The optimization protocol was expanded to include Lorentzians and other wavepacket shapes with options for custom ansatzes, and was formalized into WINGS (Wavepacket Initialization on Neighboring Grid States), an open-source tool available via pip install \cite{wings2026}.
While not scientifically novel or building upon scientific theory, WINGS exists as the first open-source, pip installable tool for initializing Gaussian wavefunctions on a grid-like basis. 
The term ``Gaussian optimizer" typically refers to the construction of a Gaussian in continuous phase space intrinsic to the qubit states, categorizing tools like  Pennylane's Strawberry Fields \cite{strawberryfields}, and is avoided for clarity.
Multi-GPU compatibility, warm-starts, and batched optimization allows for future large-scale jobs for classical precomputation.

Once an initial wavefunction is defined and initialized, the grid can be mapped onto a physically meaningful space.
In chemistry, this may be a reaction coordinate, vibrational mode, or degree of freedom.

\section{Simulation Grid}\label{sec:SimGrid}

In simulating dynamics on a grid, Macridin et al. \cite{macridin2018digital} calculates the exponential convergence of observables with increased grid discretization as 
\begin{equation}
    \eps_{\text{disc}}\leq10e^{-(0.51N_x-0.765N_{ph})},
\end{equation}
where $N_x=2^n$ basis states and $N_{ph}$ is the boson cutoff number.
Nonadiabatic dynamics are chaotic in general, following Wigner-Dyson eigenvalue statistics with state transitions \cite{takatsuka2024mechanism}. 
$N_{ph}$ is therefore effectively bounded by thresholds that capture the bulk of the spectrum, incorporated with the Nyquist-Shannon sampling theorem.
A reasonable error of $\eps_{\text{disc}} \approx 1\times10^{-6}$ and $N_{ph}\approx150$ requires around $n=8$ qubits per vibrational mode. 
This error requires $n=7$ qubits if $N_{ph}\leq64$ and $n=6$ qubits for $N_{ph}\leq21$.

This discretization error is readily handled by a closely related, more sensitive requirement for coherent wavefunction dynamics. 
Simulating in a box requires the bulk of the wavefunction to remain away from boundaries. 
Using qubits makes these boundaries periodic.
As a result, the wavefunction will not reflect, but wrap to the other side of the system (alias) if the system is improperly scaled.

The momentum grid is defined as a finite interval $[-\mathcal{N}, \mathcal{N}]$ for a spatial Nyquist limit $\mathcal{N} = \pi/\Delta_x$.
$\mathcal{N}$ must be sufficiently large to capture the bulk of the wavefunction in momentum space without aliasing. 
For a characteristic energy $E_{\text{char}}$, a minimum register size of 
\begin{equation}\label{eq:nmin}
n_{\min} = \left\lceil \log_2\left(\frac{L\sqrt{2\mu E_{\text{char}}}}{\pi}\right) \right\rceil
\end{equation}
qubits are needed to capture coherent wavefunction evolution in momentum space for a particle of mass $\mu$ and a box of size $L$ in position space.

A temporal Nyquist limit is given as the maximum resolvable energy from the time step $\tau$, being $\omega_{\mathcal{N}} = \pi/\tau$, giving an energy resolution of $\Delta E \approx 2\pi/T$.
These are not quantum theoretic principles, instead being fundamental limitations of grid-based simulation and Fourier analysis, where greater grid resolution is required for greater spectral energies, and energy resolution increases with evolution time.
One classical emulation uses a grid-based first-quantized scheme to achieve empirically converged observables \cite{ollitrault2020nonadiabatic}. 
This work kept the bulk of dynamics within this region, though its calculation is not explicitly considered, relying on empirical population transfer results to confirm physical behavior.
Previous proposals for a scalable algorithm simulating the vibronic Hamiltonian, a classic NA-QMD problem, lack numerical validation at the time of writing, asserting $n=4$ qubits per vibrational mode for resource estimations \cite{motlagh2025quantum}. 
These estimates are considered in Chapter~\ref{ch:model_ext} and recalculated in Appendix~\ref{app:Motlagh_Resource}.

Going forward, units are applied to physically meaningful quantities. 
This dissertation will use natural units $c=1, \hbar =1$, with energy in Hartrees ($1.00\,\text{E}_\text{H} \approx 27.2 \,\text{eV}$), distance in bohr radii ($1.00\,\text{a}_0\approx0.0529\,\text{nm}$), mass in electron masses $\text{m}_\text{e}$, and a.u. of time $1.00 \,\hbar/\text{E}_\text{H}\approx0.0242\,\text{fs}$. 
This is frequently reduced to $\text{a.u.}$ (atomic units), acknowledging apparent contexts for distance, momentum, mass, and time.
Alternative units (eV, nm, cm$^{-1}$, fs) are used to compare to previous literature and for absorption spectra, where applicable.

For an example similar to that of pyrazine with a reduced mass $\mu = 2\times10^4 \,\text{m}_\text{e}$ for a vibrational mode, a box of size $L=8.0\, \text{a}_0$, and a characteristic energy $E_{\text{char}}=10.0\,\text{eV} \approx0.367\,\text{E}_\text{H}$, a minimum of $n=9$ qubits are required to capture coherent wavefunction evolution.
If the bulk of $\psi(x)$ remains within $L=4.0\, \text{a}_0$, minimum register size decreases to $n=8$.
This calculation can be performed independently for each mode, optimizing values of $L$ and $n$ according to $\mu$ and $E_{\text{char}}$.
This serves as a necessary heuristic for optimizing quantum computational resources.

Basis states discretize a coordinate grid.
Following a binary encoding, a coordinate $x$ can be mapped in a finite interval $[0,L]$ to qubits $q_j$ as 
\begin{equation}
    x = \Delta_x\sum_{i=0}^{n-1}2^jq_j,
\end{equation}
for a discretization $\Delta_x = L/2^n$ and binary weights $2^j$.
In this way, 
\begin{equation}
    \psi(x) = \ket{\psi} = \ket{q_{n-1}q_{n-2}\cdots q_0} = \ket{q_{n-1}}\otimes \ket{q_{n-2}} \otimes \cdots \otimes \ket{q_0}\,,\quad q=\{0,1\}
\end{equation}

Following minimum requirements for convergent dynamics on a discrete space, temporal limits are highly dependent on Trotterization error for product-formula based algorithms. 
These temporal limits are dependent on the number of grid points, spectral norm, and nonzero commutators between the kinetic operator, the diagonal potential operator, and the off-diagonal coupling operator, detailed in Appendix~\ref{app:Error}. 

\section{Trotterization}

Being a Hamiltonian simulation problem, system evolution follows the time-dependent Schr\"odinger equation (TDSE)
\begin{equation}
    \ket{\psi(t)} = e^{-iHt}\ket{\psi(0)},
\end{equation}
where $e^{-iHt}$ is the time-evolution operator.
Adiabatic evolution propagates a wavefunction $\psi(\mathbf{x})$ on a potential energy surface $V(\mathbf{x})$ with the Hamiltonian $H=T+V$. 
The vector of coordinates $\mathbf{x}$ is currently given as an orthonormal basis of Euclidean coordinates.
The kinetic and potential operators to not commute $[T,V]\neq 0$, so 
\begin{equation}
    e^{-iHt} = e^{-i(T+V)t} \neq e^{-iTt}e^{-iVt}.
\end{equation}
Evolution under the full Hamiltonian cannot be factored into separate, sequential evolutions under $T$ and $V$ without introducing error.
This error is algebraic. 
Following the Baker-Campbell-Hausdorff (BCH) formula,
\begin{equation}\label{eq:BCH}
    \exp{A}\exp{B} = \exp\left\{A + B + \frac{1}{2}[A,B] + \frac{1}{12}\left([A,[A,B]] + [B,[B,A]]\right) + \cdots\right\},
\end{equation}
so the naive product $e^{-iTt}e^{-iVt}$ differs from $e^{-iHt}$ by terms involving nested commutators of $T$ and $V$.
Product formulas handle this by splitting the evolution into $k$ small timesteps of size $\tau = t/k$.
For sufficiently small $\tau$, the commutator corrections become negligible, and the factored evolution converges to the exact propagator.
As a method popularized in physics by classical molecular dynamics, this is frequently referred to as Suzuki-Trotter decomposition, Suzuki-Lie-Trotter decomposition, or simply Trotterization. 
This dissertation will subsequently refer to the product formula as a Trotterization.

\subsection{First- and Second-Order Decompositions}

\begin{figure}[t]
    \centering
    \includegraphics[width=0.8\textwidth]{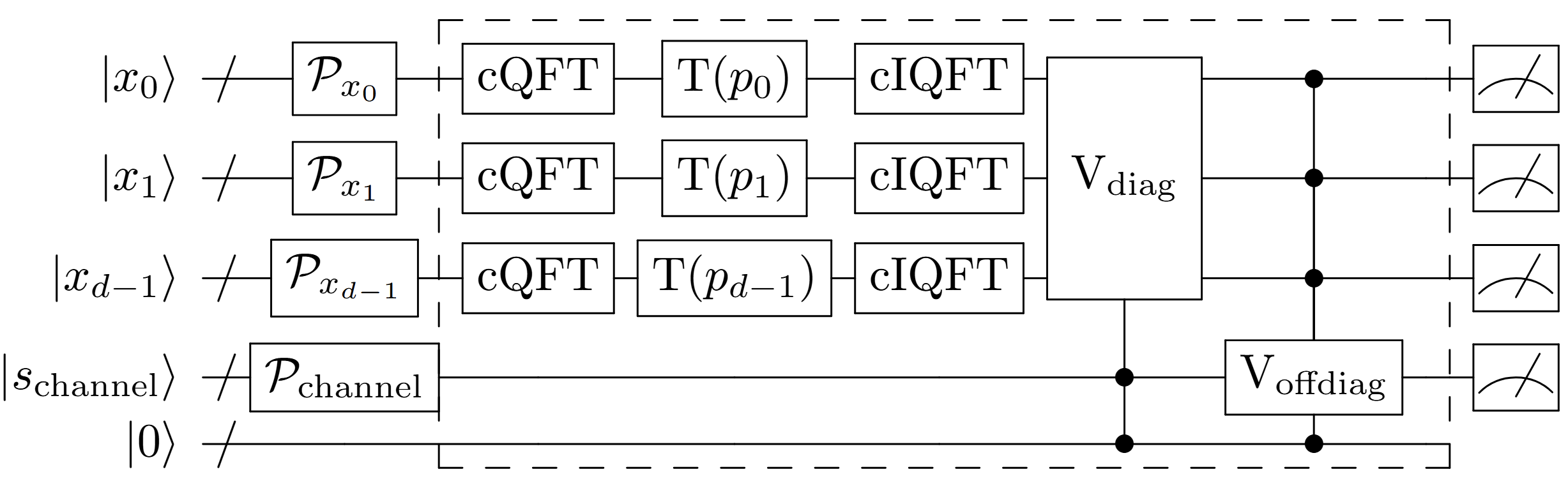}
    \caption{Schematic quantum circuit for first-order Trotterization. 
    In this schematic, control operations depend on potential terms and do not require controls on all qubits. Registers $\ket{x}$ represent coordinates, each with $n$ qubits. $\ket{s_{\text{channel}}}$ indexes electronic channels, with a work ancilla register initialized to $\ket{0\cdots0}=\ket{0}$. Here $p_0$ is the momentum associated with the $0$-th coordinate, not the initial momentum. Operations inside dashed box are iterated for $k$ Trotter steps.}
    \label{fig:Trotter-Schematic}
\end{figure}

The simplest product formula is the first-order (Lie-Trotter) decomposition,
\begin{equation}
    e^{-iH\tau} \approx e^{-iT\tau}e^{-iV\tau},
\end{equation}
with a local error of $\mathcal{O}(\tau^2)$ per step, accumulating to a global error of $\mathcal{O}(\tau)$ over $k = t/\tau$ steps.
For a potential operator $V=V_{\text{diag}} + V_{\text{offdiag}}$, the decomposition expands as 
\begin{equation}
    e^{-iH\tau} \approx e^{-iT\tau}e^{-i{V_\text{diag}}\tau}e^{-i{V_\text{offdiag}}\tau},
\end{equation}
where diagonal and off-diagonal terms do not commute $[V_\text{diag},V_\text{offdiag}]\neq0$. 
A schematic for first-order Trotterization is given in Figure~\ref{fig:Trotter-Schematic}.

A second-order decomposition is brought by Strang splitting, symmetrically decomposing the operator as
\begin{equation}\label{eq:Strang_splitting}
    e^{-iH\tau} = e^{-iT\tau/2}e^{-iV\tau}e^{-iT\tau/2} + \mathcal{O}(\tau^3).
\end{equation}
Symmetrization cancels the leading-order commutator error from Eq.~\ref{eq:BCH}, leaving a local error of $\mathcal{O}(\tau^3)$ per step and a global error of $\mathcal{O}(\tau^2)$.
This is the decomposition used throughout this dissertation.

When composing $k$ timesteps, the trailing half-step kinetic operator of one step and the leading half-step of the next combine:
\begin{equation}
    \left(e^{-iT\tau/2}e^{-iV\tau}e^{-iT\tau/2}\right)^k = e^{-iT\tau/2}\left(e^{-iV\tau}e^{-iT\tau}\right)^{k-1}e^{-iV\tau}e^{-iT\tau/2},
\end{equation}
leaving only two half-step kinetic operators.
Circuit depth is reduced by a (small) constant factor while maintaining the global $\mathcal{O}(\tau^2)$ error of the second-order decomposition \cite{childs2009limitations}.

Operator division further separates the second-order Trotterization to
\begin{equation}
    \left(e^{-iT\tau/2}e^{-iV\tau}e^{-iT\tau/2}\right)^k = \left(e^{-iT\tau/2}e^{-iV_{\text{diag}}\tau/2}e^{-iV_{\text{offdiag}}\tau}e^{-iV_{\text{diag}}\tau/2}e^{-iT\tau/2}\right)^k.
\end{equation}

Higher-order product formulas exist and can further reduce error for a given timestep by canceling additional commutator terms in the BCH expansion.
The fourth-order Yoshida construction is briefly discussed, since its structure illustrates the depth-accuracy tradeoff central to choosing a Trotter order for a given simulation.

\subsection{Higher-Order Product Formulas and Circuit Depth}

The Yoshida construction \cite{yoshida1990construction, suzuki1991general} builds a $2p$-th order integrator by composing three copies of a $(2p-2)$-th order integrator with appropriately chosen substep sizes.
For the fourth-order case, define the second-order symmetric splitting as
\begin{equation}
    \mathcal{S}_2(\tau) = e^{-iT\tau/2}\,e^{-iV_{\text{diag}}\tau/2}\,e^{-iV_{\text{offdiag}}\tau}\,e^{-iV_{\text{diag}}\tau/2}\,e^{-iT\tau/2}.
\end{equation}
The fourth-order integrator is a weighted symmetric splitting of the second-order expansion
\begin{equation}\label{eq:yoshida4}
    \mathcal{S}_4(\tau) = \mathcal{S}_2(s_1\tau)\;\mathcal{S}_2(s_2\tau)\;\mathcal{S}_2(s_1\tau),
\end{equation}
with Yoshida coefficients
\begin{equation}
    s_1 = \frac{1}{2 - 2^{1/3}}\approx 1.3512,\qquad s_2 = -\frac{2^{1/3}}{2 - 2^{1/3}}\approx -1.7024,
\end{equation}
satisfying $2s_1 + s_2 = 1$ (so one full timestep is taken) and $2s_1^3 + s_2^3 = 0$ (canceling the third-order error).

Expanding Eq.~\ref{eq:yoshida4} for the diabatic Hamiltonian, a single fourth-order timestep consists of 15 exponential factors before simplification:
\begin{equation}
\begin{split}
    \mathcal{S}_4(\tau) = \;&e^{-iTs_1\tau/2}\,e^{-iV_{\text{diag}}s_1\tau/2}\,e^{-iV_{\text{offdiag}}s_1\tau}\,e^{-iV_{\text{diag}}s_1\tau/2}\,e^{-iTs_1\tau/2}\\
    &\times e^{-iTs_2\tau/2}\,e^{-iV_{\text{diag}}s_2\tau/2}\,e^{-iV_{\text{offdiag}}s_2\tau}\,e^{-iV_{\text{diag}}s_2\tau/2}\,e^{-iTs_2\tau/2}\\
    &\times e^{-iTs_1\tau/2}\,e^{-iV_{\text{diag}}s_1\tau/2}\,e^{-iV_{\text{offdiag}}s_1\tau}\,e^{-iV_{\text{diag}}s_1\tau/2}\,e^{-iTs_1\tau/2}.
\end{split}
\end{equation}
Adjacent exponentials of the same operator commute and can be merged, as before.
Between consecutive $\mathcal{S}_2$ blocks, trailing kinetic operator half-steps commute and merge, resulting in 11 distinct exponential applications: 3 kinetic, 4 diagonal potential, and 4 off-diagonal coupling, with one pair of kinetic half-steps leading and ending the time evolution.
Iterating over $k$ fourth-order steps reduces this to 2 kinetic term applications.

The per-step circuit depth $D^{(2p)}$ for a $2p$-th order formula scales as
\begin{equation}
    D^{(2p)} = 5^{p-1}\times D^{(2)},
\end{equation}
where $D^{(2)}$ is the depth of a single second-order step and $5^{p-1}$ arises from recursive Suzuki construction \cite{suzuki1991general}, composing five copies of the lower-order integrator at each level. 
The Yoshida form with three copies is a simplification available at the first recursion level.
For $p=2$ (fourth order), this gives $D^{(4)} = 5\times D^{(2)}$ in the Suzuki form, or approximately $3\times D^{(2)}$ in the Yoshida form after boundary merging.

A $2p$-th order formula has global error scaling as $\mathcal{O}(\tau^{2p})$.
Achieving a target error $\eps$ requires
\begin{equation}
    k^{(2p)} = \mathcal{O}\left(\frac{t^{1+1/(2p)}}{\eps^{1/(2p)}}\right),
\end{equation}
timesteps.
The total circuit depth is $k^{(2p)}\times D^{(2p)}$.
For second order, $k^{(2)} = \mathcal{O}(t^{3/2}/\eps^{1/2})$ with per-step depth $D^{(2)}$.
For fourth order, $k^{(4)} = \mathcal{O}(t^{5/4}/\eps^{1/4})$ with per-step depth $\sim 3\text{--}5\times D^{(2)}$.
Whether the reduced step count compensates for the increased per-step cost depends on the ratio $t/\eps$: fourth-order formulas become favorable for long evolution times or low error, as savings from $\eps^{-1/4}$ versus $\eps^{-1/2}$ eventually outweigh the constant-factor depth increase.

For the simulations in this work, second-order Trotterization is the primary method to show proof-of-principle validation of the full algorithm.
Per-step depth $D^{(2)}$ will be dominated by multivariate terms in the on- and off-diagonal coupling circuits, detailed in Chapter~\ref{ch:resources}, and the factor of 3--5$\times$ overhead from fourth-order decomposition would significantly increase already demanding circuit depths.
Timestep requirements are found by error analysis described below and in Appendix~\ref{app:Error} confirming that second-order formulas with moderate $k$ are sufficient for the molecular systems considered.

\subsection{Trotter Error Bounds}

Error introduced by Trotterization is bounded using nested commutators of Hamiltonian terms \cite{childs2021theory}.
For a second-order symmetric decomposition in Eq.~\ref{eq:Strang_splitting}, global error after $k$ timesteps satisfies
\begin{equation}\label{eq:trotter_error_bound}
    \eps_{\text{Trotter}} \leq \frac{t^3}{24k^2}\left(\norm{[V,[V,T]]} + 2\norm{[T,[T,V]]}\right),
\end{equation}
where $\norm{\cdot}$ denotes the spectral norm and $t = k\tau$ is the total evolution time.
Distributing the commutator over its terms allows analysis of term contributions to error bounds.

Equation~\ref{eq:trotter_error_bound} relates to how product formula errors scale with the energy scales of the Hamiltonian.
The spectral norm $\norm{H}$ represents a characteristic energy (aka largest eigenvalue).
For much of grid-based simulation, $\norm{H}$ is dominated by the maximum kinetic energy $T_{\text{max}} = p_{\text{max}}^2/2\mu = \pi^2 N^2/(2\mu L^2)$, growing quadratically with grid resolution.
The nested commutator norms in Eq.~\ref{eq:trotter_error_bound} inherit this scaling, where $\norm{[T,[T,V]]}$ contains factors of $p_{\text{max}}^2 \propto \norm{T}$, so error bounds scale polynomially with $\norm{H}$.

For a Hamiltonian with spectral norm $\Lambda = \norm{H}$, the second-order Trotter error satisfies
\begin{equation}
    \eps_{\text{Trotter}} = \mathcal{O}\left(\frac{t^3\Lambda^3}{k^2}\right),
\end{equation}
requiring $k = \mathcal{O}(t^{3/2}\Lambda^{3/2}/\eps^{1/2})$ steps to achieve error $\eps$.
A $2p$-th order product formula gives $k^{(2p)} = \mathcal{O}((t\Lambda)^{1+1/(2p)}/\eps^{1/(2p)})$ \cite{childs2021theory}.
This must be considered for any error estimation, following grid resolution requirements. 

Post-Trotter methods (QSP, qubitization) have query complexity scaling as $\mathcal{O}(t\Lambda + \log(1/\eps))$ \cite{low2017optimal, low2019hamiltonian}. 
Linear $\Lambda$-dependence becomes significant for large spectral norms, and logarithmic $\eps$-dependence is a strict asymptotic advantage.
Following results from Chapter~\ref{ch:Introduction}, there is exponential overhead in using oracles for dense Hamiltonians, leading to the intractability of QSP and qubitization despite more favorable error scaling. 
Shown in subsequent chapters, the oracle-free structure of product formulas results in lower constant-factor costs, not precluding the existence of a post-trotter method with more favorable overhead and lower error.

\subsection{Evaluating Trotter Error}

A full derivation is deferred to Appendix~\ref{app:Error}, but a procedure is outlined here to apply the bound in Eq.~\ref{eq:trotter_error_bound} to a specific vibronic Hamiltonian.

The Hamiltonian is first decomposed into its constituent operator classes.
For vibronic systems, this typically includes kinetic terms $T_r = p_r^2/2\mu_r$ for each vibrational mode $r$, and potential terms that may be constant, linear ($\propto x_r)$, quadratic ($\propto x_r^2$), or bilinear in pairs of coordinates ($\propto x_rx_s$).
Each potential term may appear in the diagonal or off-diagonal block of the Hamiltonian.
Off-diagonal terms carry electronic coupling operators $\hat{\sigma}_{ij} = \ket{i}\bra{j} + \ket{j}\bra{i}$ with algebraic properties $\hat{\sigma}_{ij}^2 = \hat{p}_i + \hat{p}_j$ factoring through the calculation, where $\hat{p}_i = \ket{i}\bra{i}$.

Nested commutators $[A,[B,T]]$ and $[T,[T,A]]$ are evaluated using standard canonical commutation relations for each pair of operator classes, following identities $[x_r, p_r] = i$, $[x_r^2, p_r^2] = 2i\{p_r, x_r\}$, and the vanishing of commutators between operators acting on different modes, $[x_r, p_s] = 0$ for $r \neq s$.
Electronic operators factor cleanly since $T$ is proportional to the electronic identity, so the nuclear commutator structure is identical for diagonal and off-diagonal potential terms, differing only by an electronic prefactor.

A hierarchy of contributions is revealed based on grid scaling.
$\mathcal{O}(1)$ contributions (grid-independent) include commutators involving only constant and linear potential terms.
The nested commutator $[T,[T,V_{\text{lin}}]]$ vanishes identically since $[p^2, p] = 0$, and $[V_{\text{lin}}, [V_{\text{lin}}, T]] \propto \alpha_r^2/\mu_r$, where $\alpha_r$ is the linear coupling coefficient.
Bilinear terms also produce $\mathcal{O}(1)$ contributions through $[V_{\text{bilin}},[V_{\text{bilin}},T]]$, which yields terms proportional to $\gamma_{rs}^2 L^2/\mu_r$.
Cross-terms between different operator classes (such as diagonal-linear with off-diagonal-bilinear) are similarly bounded by $\mathcal{O}(1)$ expressions.
These terms are subdominant and only contribute a small fraction of total error.
    
 Grid-dependent contributions scale as $\mathcal{O}(N^2)$ and include commutators of the form $[T,[T,V]]$ for quadratic and bilinear potential terms produce bounds proportional to $p_{\text{max}}^2 = \pi^2 N^2/L^2$, reflecting the characteristic energy.
Specifically, dominant contributions arise with the quadratic diagonal term, with $\norm{[T,[T,V_{\text{quad}}]]} \leq \omega_r^2 \pi^2 N^2 / (\mu_r L^2)$, and the bilinear term, with $\norm{[T,[T,V_{\text{bilin}}]]} \leq 2|\gamma_{rs}|\pi^2 N^2 / (\mu_r\mu_s L^2)$.

Collecting all contributions, the error bound takes the form
\begin{equation}\label{eq:trotter_error_summary}
    \eps_{\text{Trotter}} \leq \frac{t^3}{24k^2}\left(C_0 + C_2 N^2\right),
\end{equation}
where $C_0$ aggregates all grid-independent contributions and $C_2$ aggregates the grid-dependent contributions.
$C_2$ depends on molecular parameters, being harmonic frequencies $\omega_r$, bilinear coupling strengths $\gamma_{rs}$ and $\eta_{rs}$, reduced masses $\mu_r$, and the grid domain $L$.
Explicit expressions for $C_0$ and $C_2$ in terms of all Hamiltonian coefficients are given in Appendix~\ref{app:Error}.

\subsection{Implications for Timestep Selection}

Inverting Eq.~\ref{eq:trotter_error_summary} gives a lower bound for number of Trotter steps:
\begin{equation}\label{eq:k_min}
    k \geq \sqrt{\frac{t^3(C_0 + C_2 N^2)}{24\eps}}.
\end{equation}
$N^2$ dependence from quadratic and bilinear terms dominates, giving
\begin{equation}
    k = \mathcal{O}\left(\frac{t^{3/2}N}{\sqrt{\eps}}\right).
\end{equation}
This means the number of Trotter steps grows linearly with the grid resolution $N = 2^n$, so doubling the number of grid points (adding one qubit per mode) doubles the required number of timesteps.
This $\mathcal{O}(N)$ dependence means that choice of register size $n$ directly impacts the total circuit depth through both the per-step gate count and the number of steps required.
Reducing from $n = 10$ to $n = 8$ qubits per mode decreases the required Trotter steps (by a factor of four), at the cost of reduced grid resolution that may compromise accuracy for high-frequency modes.
This tradeoff motivates the per-mode optimization of grid parameters discussed in Section~\ref{sec:SimGrid}, where each degree of freedom can (and likely should) have its own register size, box size, and discretization tailored to its physical requirements for the most resource-efficient simulation.
These customizations preclude some applications of quantum arithmetic, where registers of different scales cannot be added together to leverage structure in parameter weights. 

A final distinction is seen between commutator bounds and empirical Trotter error.
The bounds in Eq.~\ref{eq:trotter_error_bound} are worst-case estimates, with actual error for a given physical system being substantially smaller.
Childs et al. \cite{childs2021theory} notes that the gap between rigorous bounds and error can span orders of magnitude, particularly for structured Hamiltonians where cancellations occur between commutator terms.
For molecular systems considered here, empirical convergence studies (varying $\tau$ until observables stabilize) are used alongside analytical bounds to determine practical timestep requirements.

\section{Kinetic Operator}
The kinetic operator $T = p^2/2\mu$ is tridiagonal in position (configuration) space and diagonal in momentum (dual) space. 
Applying this operator directly on the register without oracles most efficiently performed in momentum space, requiring quantum fourier transforms (QFT), shown in Eq.~\ref{eq:QFT}. 
When applying a QFT directly, a lattice ``site" exists in the center of the Nyquist region. 
To center in the middle of the 1D lattice's Brillouin zone, a NOT operation is attached to the QFT, shifting momentum space and making a centered QFT (cQFT) (Eq.~\ref{eq:cQFT}). 
\begin{equation}\label{eq:QFT}
\begin{split}
    \hat{\text{QFT}} = \left(\sum_{j',k'}\frac{e^{2\pi ik'j' / N}}{\sqrt{N}}\ket{k'}\bra{j'}\right)\text{, }\hat{\text{QFT}}^{-1} & = \left(\sum_{j',k'}\frac{e^{2\pi ik'j' / N}}{\sqrt{N}}\ket{j}\bra{k}\right)
    \end{split}
\end{equation}

\begin{equation}\label{eq:cQFT}
    \hat{\text{cQFT}} = \begin{bmatrix}
        & & & 1 & \cdots & 0 \\
        & \cdots & & & \ddots & \\
        & & & 0 & \cdots & 1 \\
        1 & \cdots & 0 & & & \\
        & \ddots & & & \cdots & \\
        0 & \cdots & 1 & & & \\
    \end{bmatrix}\hat{\text{QFT}},
\end{equation}

As seen in Section~\ref{sec:Wavepacket_init}, a prepared wavefunction is not initialized with an initial momentum $p_0$. 
This can be applied in the kinetic term such that
\begin{equation}
    T = \frac{(p-p_0)^2}{2\mu},
\end{equation}
applying a Lorentz boost to the simulation (box) frame according to initial wavefunction conditions.
Momentum space is also binary encoded, as $p = \Delta_p\sum_j^{n-1}2^jq_j$.
Following Benenti and Strini \cite{benenti2008quantum}, the operator is cast as
\begin{equation}
    \begin{split}
        T & = \frac{1}{2\mu}\left(p^2-2pp_0 + p_0^2\right)\\
        & = \frac{1}{2\mu}\left(\Delta_p^2\sum_{\ell=0}^{n-1}\sum_{j=0}^{n-1}2^jq_j2^\ell q_\ell-2p_0\Delta_p\sum_{j=0}^{n-1}2^jq_j+p_0^2\right).\\
    \end{split}
\end{equation}

Constant $p_0^2 /2\mu=\phi_1$ acts as a global phase, being $\mathcal{O}(1)$ and requiring four operations, as $R_z(\phi_1)$-NOT-$R_z(\phi_1)$-NOT.
Terms linear in $p$ contribute to the movement of the wavepacket, as $n$ $R_z$ gates. 
   $R_z$ gates act as 
\begin{equation}
    R_z(\alpha) \otimes \mathbb{1} = \ket{0}\bra{0}\otimes \mathbb{1} + e^{i\alpha}\ket{1}\bra{1}\otimes \mathbb{1}
\end{equation}
$R_z$ gates therefore commute with other $R_z$ gates
\begin{equation}
    [R_z(\alpha)\otimes\mathbb{1}, R_z(\beta)\otimes\mathbb{1}] = 0.
\end{equation}

Quadratic terms ($\propto p^2$) contribute to wavepacket expansion, as Controlled-$R_z$ ($\mathrm{CR}_z$) gates when $j \neq \ell$ and $R_z$ gates when $j = \ell$. 
A $\mathrm{CR}_z$ gate acts as
\begin{equation}
    \mathrm{CR}_z(\alpha) \otimes\mathbb 1 = \ket{0}\bra{0}\otimes\mathbb{1}\otimes\mathbb{1} + \ket{1}\bra{1}\otimes R_z(\alpha) \otimes \mathbb{1}.
\end{equation}
Two $\mathrm{CR}_z$ gates acting on different tensor spaces commute:
\begin{equation}
    [\mathrm{CR}_z(\alpha) \otimes\mathbb 1, \mathrm{CR}_z(\beta) \otimes\mathbb 1] = 0,
\end{equation}
So, the quadratic term can be reformed as
\begin{equation}
        \frac{p^2}{2\mu} = \frac{1}{2\mu}\Delta_p^2\left(\sum_j^{n-1}2^{2j}q_j+2\sum_{j<\ell}2^{j+\ell}q_jq_\ell\right),
\end{equation}
being $n$ $R_z$ gates and $\binom{n}{2}$ $\mathrm{CR}_z$ gates.
Combining with the other terms, the exponentiated operator $e^{-iT\tau}$ is given as
\begin{equation}
\begin{split}
    & e^{-i\tau \frac{1}{2\mu}\Delta_p^2 \sum_j^{n-1}2^{2j}q_j}\times e^{-i\tau\frac{1}{2\mu}\Delta_p^2\left(2\sum_{j<\ell}2^{j+\ell}q_jq_\ell\right)}\times e^{-i\tau\frac{p_0}{2\mu}\Delta_p\sum_j^{n-1}2^jq_j} \\
    & = e^{-i\tau\frac{1}{2\mu}\Delta_p^2\left(2\sum_{j<\ell}2^{j+\ell}q_jq_\ell\right)}\times e^{-i\tau\frac{1}{2\mu}\Delta_p^2 \sum_j^{n-1}2^{2j}q_j}\times e^{-i\tau\frac{p_0}{2\mu}\Delta_p\sum_j^{n-1}2^jq_j} \\
    & = e^{-i\tau\frac{1}{2\mu}\Delta_p^2\left(2\sum_{j<\ell}2^{j+\ell}q_jq_\ell\right)}\times e^{-i\tau\left(\frac{1}{2\mu}\Delta_p^2 \sum_j^{n-1}2^{2j} + \frac{p_0}{2\mu}\Delta_p\sum_j^{n-1}2^j\right)q_j}.
\end{split}
\end{equation}

\begin{figure}
    \centering
    \includegraphics[width=1.0\textwidth]{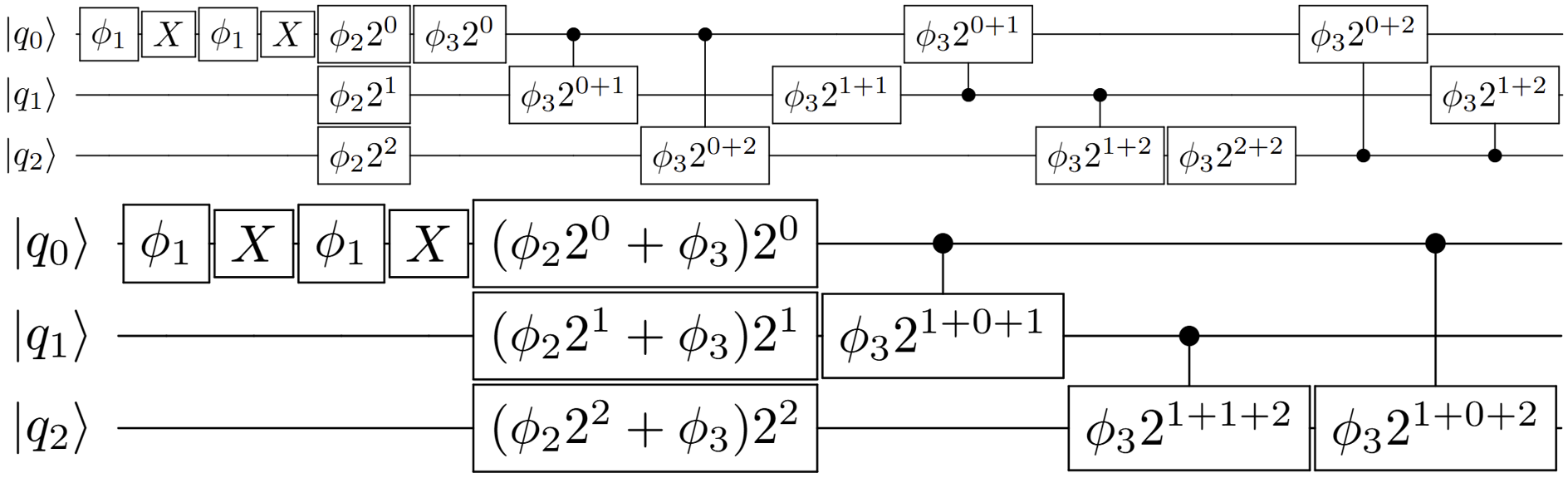}
    \caption[Explicit gate decomposition of quadratic function]{\label{fig:Kinetic_Sample}Explicit gate decomposition of quadratic function~\cite{benenti2008quantum} (above), with equivalent circuit with tensor product commutativity given below. Operations are ordered by global phase ($\phi_1 = -\tau(\eta x_0^2 + \delta)$), first-order terms $\phi_2 =-2\tau\eta x_0 \Delta$, and quadratic terms $\phi_3 = -\tau\eta \Delta^2$, weighted by the binary encoding of each qubit.}
\end{figure}

The total kinetic operator $e^{-iT\tau}$ can therefore be applied in $\binom{n}{2}$ $\mathrm{CR}_z$ gates, $n+2$ $R_z$ gates, and 2 NOT gates.
This operator is placed between two $\mathcal{O}(n^2)$ cQFTs, making the total operation computationally efficient. 
A schematic circuit for $T$ is given in Figure~\ref{fig:Kinetic_Sample}. 

This circuit structure is applied for all quadratic operations. 
For harmonic potential energy surfaces $V(x)\propto x^2$, coefficients are recalculated and the operator is applied in position space.

\begin{figure}[htbp]
    \centering
    \includegraphics[width=0.6\textwidth]{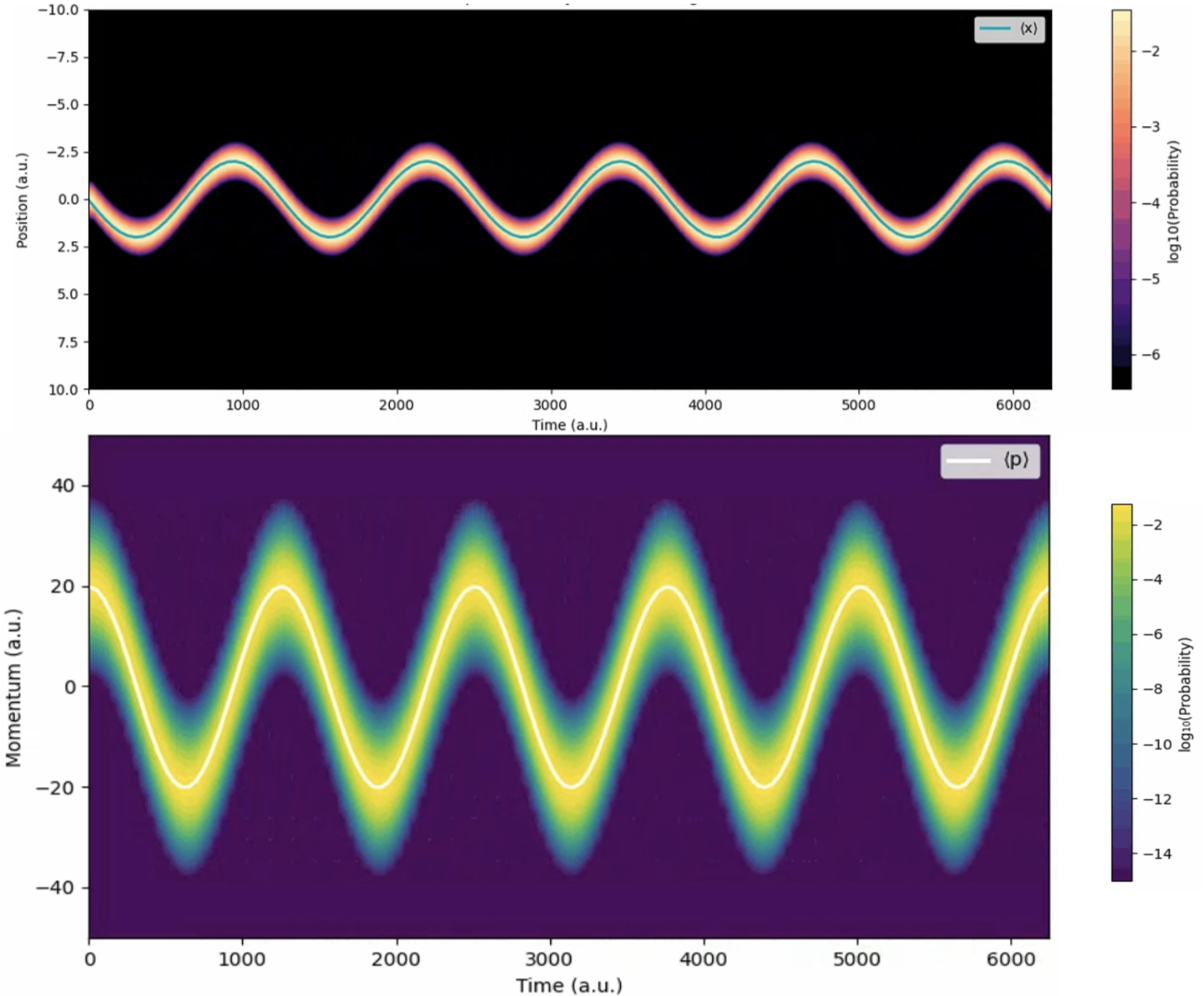}
    \caption{Probability density heatmaps of a wavepacket evolving in a single harmonic potential energy surface.}
    \label{fig:Harmonic-Heatmap}
\end{figure}

A recovery of results by Benenti and Strini \cite{benenti2008quantum} for a wavepacket propagating in a single, 1D harmonic oscillator potential is given in Figure~\ref{fig:Harmonic-Heatmap}. 
The surface $V(x) = \frac{1}{2}\mu \omega x^2$ is used, with $\omega=0.005\,\text{a.u.}^{-1}$ with a wavepacket with mass $\mu=2000\,\text{m}_\text{e}$. 
The wavepacket was given a momentum of $p_0 = 20\,\text{a.u.}$, propagated in a box of size $L=20\,\text{a.u.}$ with $n=10$ qubits and timestep $\tau=5.0\,\text{a.u}$. 
This result serves as a foundation for wavefunction propagation in a first-quantized perspective.

\section{Summary}

This chapter reviews Trotterization and wavepacket initialization, requisite background before considering potentials that are not harmonic, any nonadiabatic dynamics, and first-quantized dynamics on a binary-encoded grid.
Classical emulation recovers seminal results from Benenti and Strini \cite{benenti2008quantum}, while wavepacket initializations show drastic improvement on quantum hardware since 2020 \cite{ollitrault2020nonadiabatic}, with reasonable state fidelity of $0.885\pm0.00271$ for an $n=8$-qubit UCC ansatz on $ibm\_torino$ (Figure~\ref{fig:UCC_Hardware_VaryQubits}). Error-mitigated NISQ architectures handling the UCC ansatz suggests promise, but the application of a necessarily $\mathcal{O}(t)$-depth circuit for nonadiabatic dynamics represents a challenge that cannot be accurately simulated on modern quantum hardware. 

With a foundation for grid-based simulation, product formulas for nonadiabatic systems, and robust optimization for wavepacket initialization, a move toward novel contribution begins in nonadiabatic simulation. 
The next chapter focuses on recovering the results of a linearly-approximated nonadiabatic coupling between harmonic oscillator potentials, and the applicability of approximation to such as system. 
This will be developed and extended to quantum circuits describing nonadiabatic behavior in multiple dimensions (multi-mode), multiple channels (multi-channel), and moving beyond harmonic approximation and linear nonadiabatic coupling to arbitrary smooth finite-difference expressions for both diagonal and off-diagonal potential terms.

\chapter{Marcus Model}\label{ch:marcus}
\section{Introduction}
This chapter begins by defining a structure for quantum registers to perform a simulation of nonadiabatic dynamics. 
Coupled harmonic potentials (Marcus model) are used as a first example, recovering results from previous research with reduced resource costs~\cite{ollitrault2020nonadiabatic}.
Linear approximation is considered for Gaussian nonadiabatic coupling and optimized for system scale, followed by a variational compression to kinetic and diabatic potential terms to assess recovery of rate coefficients with shorter quantum circuits.
This serves as the first explicit scientific contribution of this dissertation.
Once characterization of this toy model is realized, subsequent chapters will develop a complete workflow for real systems.

\section{Quantum Registers}
A complete quantum register is constructed with $d+2$ quantum registers for $d$ coordinates, a register $\ket{s_\text{channel}}$ to store populations of $M$ electronic states, and a register with ``work ancilla" to ease intermediate calculations. 
\begin{equation}
    \begin{split}
        &\ket{x_{n_0}} \otimes \ket{x_{n_1}}\otimes \cdots \otimes \ket{x_{n_d}}\otimes \ket{s_{\text{channel}}}\otimes\ket{a_{\text{work}}} \\
        & = \ket{\psi(\mathbf{x})}\otimes\ket{s_{\text{channel}}}\otimes\ket{a_{\text{work}}},
    \end{split}
\end{equation}

For this dissertation, the Hilbert space generated by the $d$ coordinates will be referred to as ``position space" and its dual as ``momentum space," with the $d$ registers collectively as the ``position register." 
$\ket{s_{\text{channel}}}$ is the ``channel" register and its Hilbert space as ``channel space," storing electronic state populations in its basis states.
The register of work ancilla will be referred to passively, where all operations applied are eventually reversed. The total Hilbert space for the system can be written as 
\begin{equation}
    \mathcal{H}^{tot} = \mathcal{H}^{position}\otimes\mathcal{H}^{channel}\otimes\mathcal{H}^{work}.
\end{equation}

A total of $dn$ qubits are needed in the position register to support $d$ coordinates.
Chapter~\ref{ch:wavepacket_prop} maintains that some coordinates require fewer qubits for a coherent evolution, though $n$ qubits per coordinate is used for clear complexity scaling.
The channel register will require $\lceil\log_2M\rceil$ qubits for $M$ electronic channels.
The number of work ancilla can vary, taken on a case-by-case basis.

Measurement of basis states for each coordinate $i$ in the position register $\ket{x_i}$ gives the probability amplitude $|\psi(x_i)|^2$ for the projection of the wavefunction onto that coordinate. 
The full position register therefore stores $\ket{\psi(\mathbf{x},t)}$ at a given time $t$.
Basis states in the channel register store electronic state populations in a binary encoding. 
In this way, the measurement of $\ket{s_{\text{channel}}}$ yields the population of all $M$ channels up to measurement statistics.

\section{Trotterization Scheme}

A second-order Trotterization scheme (Strang-splitting) is chosen as the standard evolution method.
Following expansions given in Chapter~\ref{ch:wavepacket_prop} for nonadiabatic systems, a second-order Trotter step of size $\tau$ is therefore given by 
\begin{equation}\label{eq:Strang_splitting_with_QFTs}
    e^{-iH\tau}\approx\left(\text{cIQFT}e^{-iT\tau/2}\text{cQFT}e^{-iV_\text{diag}\tau/2}e^{-iV_\text{offdiag}\tau}e^{-iV_\text{diag}\tau/2}\text{cIQFT}e^{-iT\tau/2}\text{cQFT}\right)^k
\end{equation}

Exponentiated operators act on $\ket{\psi}$ from right-to-left, hence operator ordering.
The split kinetic term reforms into a single term $e^{-iT\tau}$ with $\tau = t/k$.
Each effective trotter step therefore only requires one pair of cQFTs, while maintaining the global $\mathcal{O}(\tau^2)$ error of second-order decomposition \cite{childs2021theory}.

The diabatic potential term $V_{\text{diag}}$ is expanded as
\begin{equation}\label{eq:Potential_Term_Expansion}
    e^{-iV_\text{diag}\tau}=e^{-iV_{00}\tau}e^{-iV_{11}\tau}\cdots e^{-iV_{M-1M-1}\tau},
\end{equation}
where $V_{M-1M-1}=V_{M-1M-1}(\mathbf{x})$ is the diabatic potential surface for the $M$-th electronic channel, given as the on-diagonal elements of the potential matrix $V$ (Eq.~\ref{eq:PotentialMatrix}). 
Each diabatic potential surface commutes with other diabatic potential surfaces and are broadly dependent on all coordinates as variables $\mathbf{x} = \ket{x_{n_0}}\otimes\ket{x_{n_1}}\otimes\cdots \otimes\ket{x_{n_d}}$ for $d$ modes.
Each vector element is a coordinate $\ket{x_{n_d}}$, given to be square-integrable in a scaled finite interval $[0,L]$. 
As such, each reaction coordinate can be treated as a Cartesian ``dimension."

$V_{\text{offdiag}} = V_{\text{offdiag}}(\mathbf{x})$ represents all nonadiabatic coupling terms, as the off-diagonal components of Eq.~\ref{eq:PotentialMatrix}. 

\begin{equation}\label{eq:PotentialMatrix}
\hat{V}(\mathbf{x})=
    \begin{pmatrix}
        V_{00}(\mathbf{x}) & V_{01}(\mathbf{x})  & \cdots & V_{0M}(\mathbf{x}) \\
        V_{10}(\mathbf{x}) & V_{11}(\mathbf{x}) & \cdots & \vdots \\
        \vdots & \vdots & \ddots & \vdots \\
        V_{0M}(\mathbf{x}) & \cdots & \cdots & V_{MM}(\mathbf{x})
    \end{pmatrix}
\end{equation}

Each on-diagonal term is controlled by a basis state in channel space. 
Scaling for this feature and an optimization protocol is given in subsequent sections.
Off-diagonal terms act on the channel space as electronic coupling terms, controlled by basis states in position space. 

\begin{figure}[htbp]
    \centering
    \includegraphics[width=1.0\textwidth]{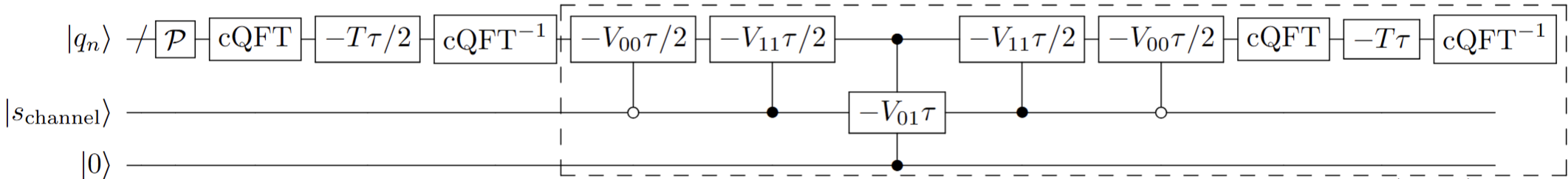}
    \caption{Schematic for a two-channel, one-coordinate Trotterization. 
    Terms in the dashed box are iterated for $k-1$ timesteps.}
    \label{fig:2-channel-schem}
\end{figure}

A harmonic potential is often used as an approximation for the equilibrium geometry of a potential energy surface, efficiently applied on a quantum register in $\mathcal{O}(n^2)$ gates (Chapter~\ref{ch:wavepacket_prop}).
In effect, two diabatic potentials can be applied serially, as in Figure~\ref{fig:2-channel-schem}.
These diabatic potentials are controlled on the state $\ket{s_\text{channel}}$.
$V_{00}$ is controlled on the channel state being $\ket{0}$, while $V_{11}$ is controlled on the channel state being $\ket{1}$, activating according to relative populations for each surface $P_0 = (\langle Z \rangle + 1)/2$, being the expectation value of measuring the qubit on the $z$-axis.
$P_0=0$ is the wavefunction fully in the zeroth channel, $P_0=1$ is the wavefunction fully in the first channel, and $P_0=0.5$ being a coherent superposition between potential surfaces.

\section{Marcus Model}

The Marcus model is a classic functional example for nonadiabatic dynamics governing electron transfer reactions through a reaction coordinate \cite{marcus1956theory, tachiya1993generalization}. 
In the diabatic representation, two electronic states $\ket{0}$ and $\ket{1}$ are associated with harmonic potential energy surfaces $V_0(x)$ and $V_1(x)$ along a nuclear coordinate $x$.

\begin{figure}[t]
\centerline{
\includegraphics[width=100mm]{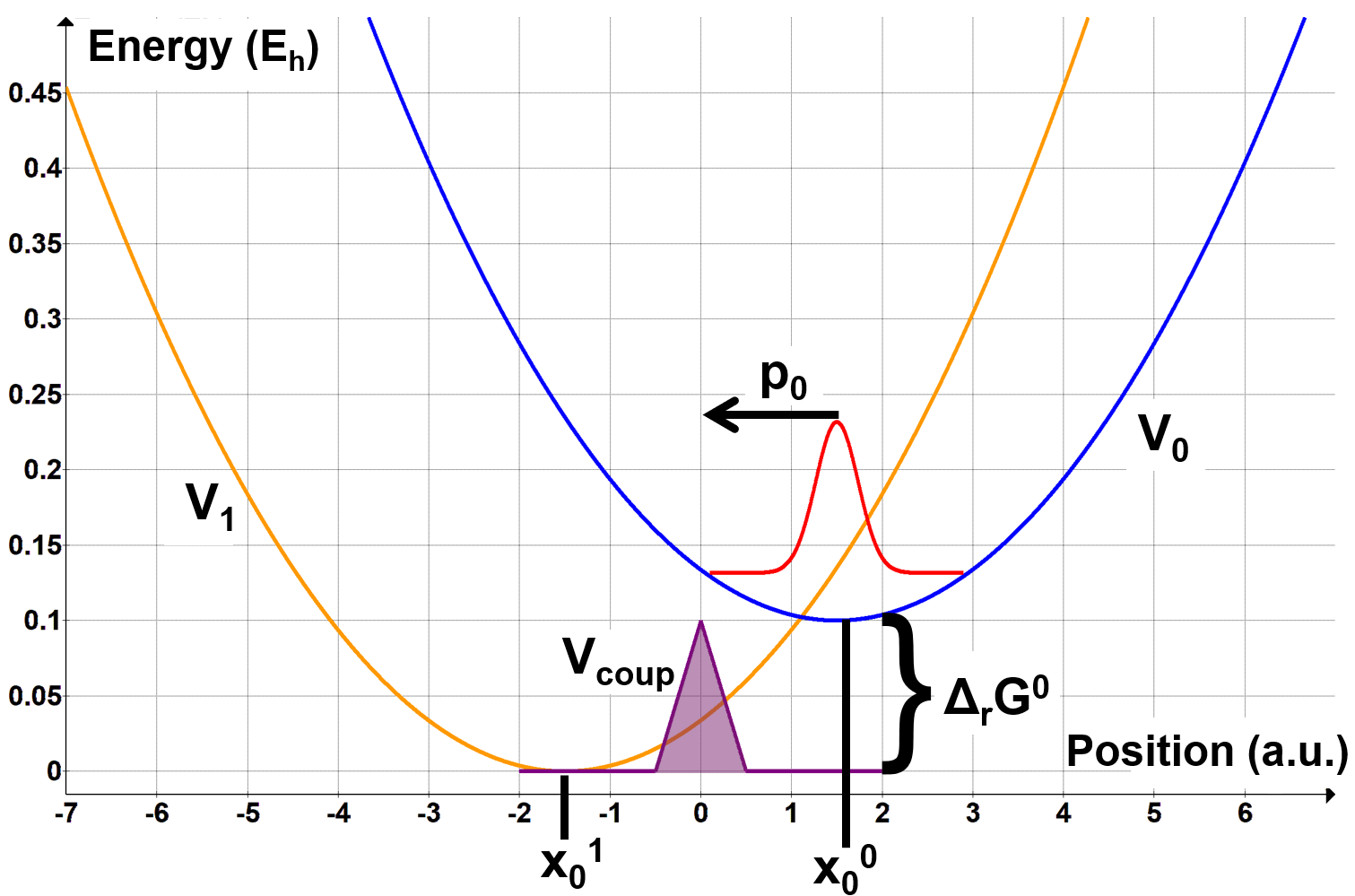}}
\caption{Marcus model for nonadiabatic dynamics with a linearly-approximated Gaussian coupling. 
Two diabatic potential energy surfaces $V_0$ and $V_1$ are offset by $\Delta_r G^{0}$.
\label{fig:marcus_model}}
\end{figure}

The diabatic potentials take the form
\begin{equation}
    V_{ii}(x) = \frac{1}{2} \mu \omega_i^2 (x - x_i^0)^2 + \Delta_r G_i^0,
    \label{eq:diabatic_potentials}
\end{equation}
where $\mu$ is the effective mass, $\omega_i$ is the angular frequency, $x_i^0$ is the equilibrium position, and $\Delta_r G_i^0$ is the energy offset for surface $i$. The electronic coupling $V_{\text{offdiag}}(x)$ is typically localized near a defined reaction center or near the crossing point of the two surfaces (Figure~\ref{fig:marcus_model}).

The full Hamiltonian in the diabatic basis is
\begin{equation}
    H = T \otimes \mathbb{1}_{\text{el}} + V_{00} \otimes \ket{0}\bra{0} + V_{11} \otimes \ket{1}\bra{1} + V_{\text{01}} \otimes \sigma_x,
    \label{eq:marcus_hamiltonian}
\end{equation}
where $K = p^2/2\mu$ is the kinetic energy, $\mathbb{1}_{\text{el}}$ is the identity on the electronic subspace, and $\sigma_x$ couples the diabatic states, with $V_{01}=V_{10}$ for a real nonadiabatic coupling.

Constructing a quantum ciruit for this system proceeds by encoding the nuclear wavepacket on a position register of $n$ qubits, implementing the Trotterized evolution, then measuring the channel qubit to extract population transfer dynamics.
Population of the initial diabatic state as a function of time yields rate coefficients via short-time dynamics \cite{nitzan2024chemical}. 
Previous classical emulations of this model achieved accurate population dynamics using explicit Trotterization on an $n=8$-qubit position register \cite{ollitrault2020nonadiabatic}. 
Probing the behavior of these quantum circuits first requires recovery of results. 
This dissertation confirms recovery in the next subsection before considering the impact of truncated circuits on observables. 

\subsection{Linearly Approximated Coupling}

\begin{figure}[htbp]
    \centering
    \includegraphics[width=0.6\textwidth]{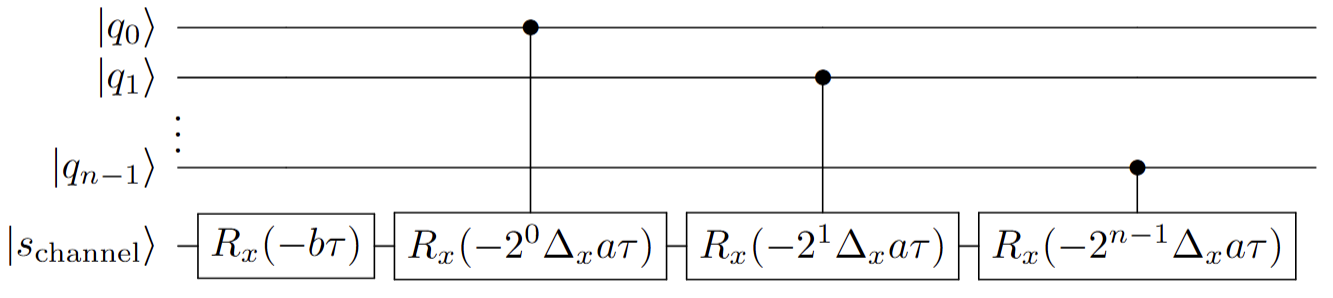}
    \caption{Schematic for a linear nonadiabatic coupling.}
    \label{fig:lin-coup-schem}
\end{figure}

An $M=2$ channel case with linear coupling is considered, quickly transitioning to piecewise linear to recover results from Ollitrault et al. \cite{ollitrault2020nonadiabatic}. 
A linear coupling $V_{01}(x) = ax+b$ acting on $\lceil \log_2 M\rceil =1$ channel qubit can be applied in $\mathcal{O}(n)$ $CR_x$ operations for $a,b\in\mathbb{R}$, given schematically in Figure~\ref{fig:lin-coup-schem}.

A linear approximation can be made by checking wavefunction position.
If part of the wavefunction is beyond a breakpoint $p$, linear coupling can be conditionally applied on that proportion. 
This can be performed by either ripple-carry addition (arithmetic) \cite{cuccaro2004new, ollitrault2020nonadiabatic} or following a Toffoli cascade. 
The Toffoli cascade is found to have a worst-case $\mathcal{O}(n^{1.5})$ Toffoli gate cost, but has less total overhead than the $\mathcal{O}(n\log n)$ quantum addition, being more practical for small $n$ (Appendix~\ref{app:circuits}). 
If breakpoint $p$ is defined as a basis state $\ket{q_{n-1}q_{n-2}\cdots q_1q_0}, \ket{q_j}\in\{0,1\}$, logical comparisons can be made. 
This cascade is an optimization of AND and OR conditional statements, stored into a subset of the work ancilla register, referred to here as ``comparator qubits."

\begin{figure}[htbp]
    \centering
    \includegraphics[width=0.6\textwidth]{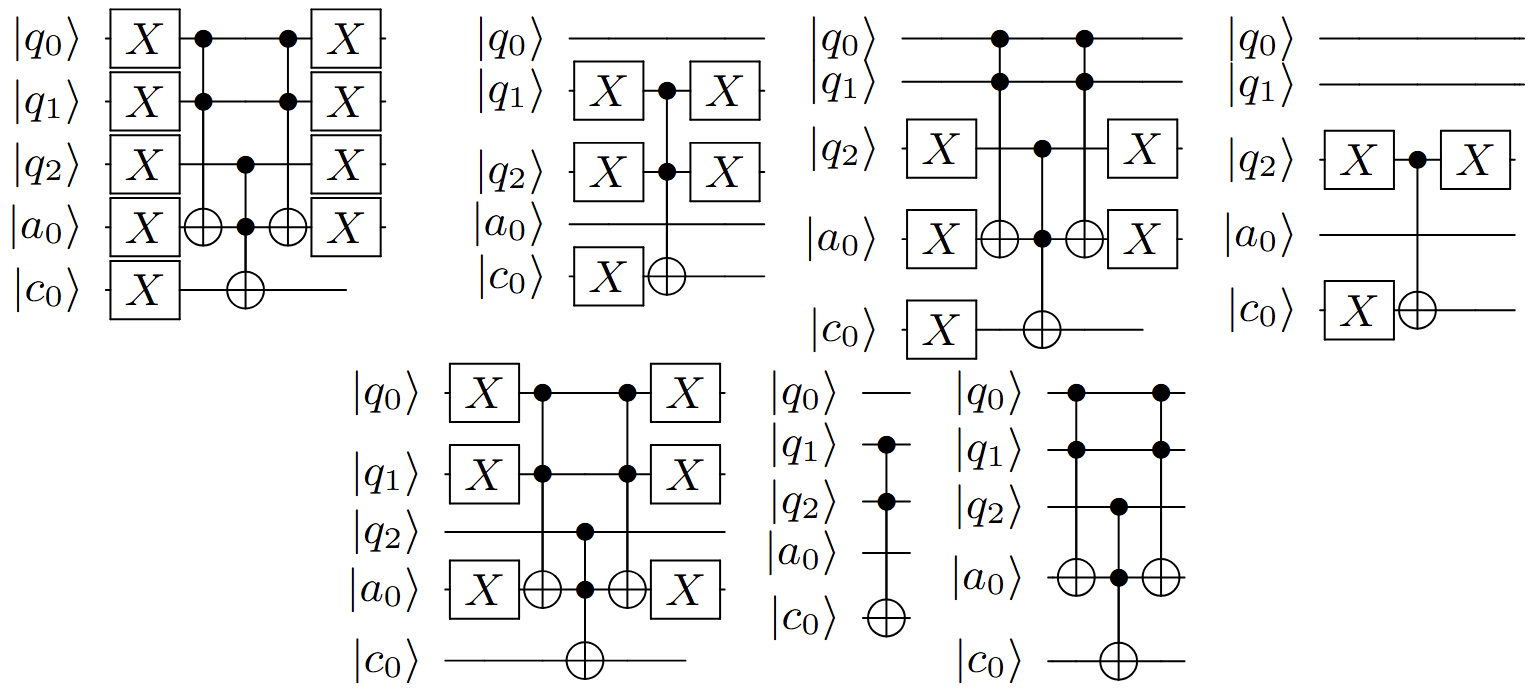}
    \caption{Comparison circuits to test if a 3-qubit state is `greater than' a given integer value from 0 (state $\ket{000}$, top left) to 6 (state $\ket{110}$, bottom right). 
A test for `greater than' state $\ket{111}$ always returns false. 
For a single comparison the minimum number of qubits that need to be checked is always one (see $>3$, top right), occurring only once in a set of $\mathcal{N}$ basis states, and the maximum number to be checked is always $n$, occurring as half of all basis states.
The number of comparisons that need to check $m$ qubits is $2^{m-1}$ for $m \leq n$.}
    \label{fig:3q-comparators}
\end{figure}

Simple examples for each three-qubit `greater-than' comparison is given in Figure~\ref{fig:3q-comparators}

\begin{figure}[htbp]
    \centering
    \includegraphics[width=0.6\textwidth]{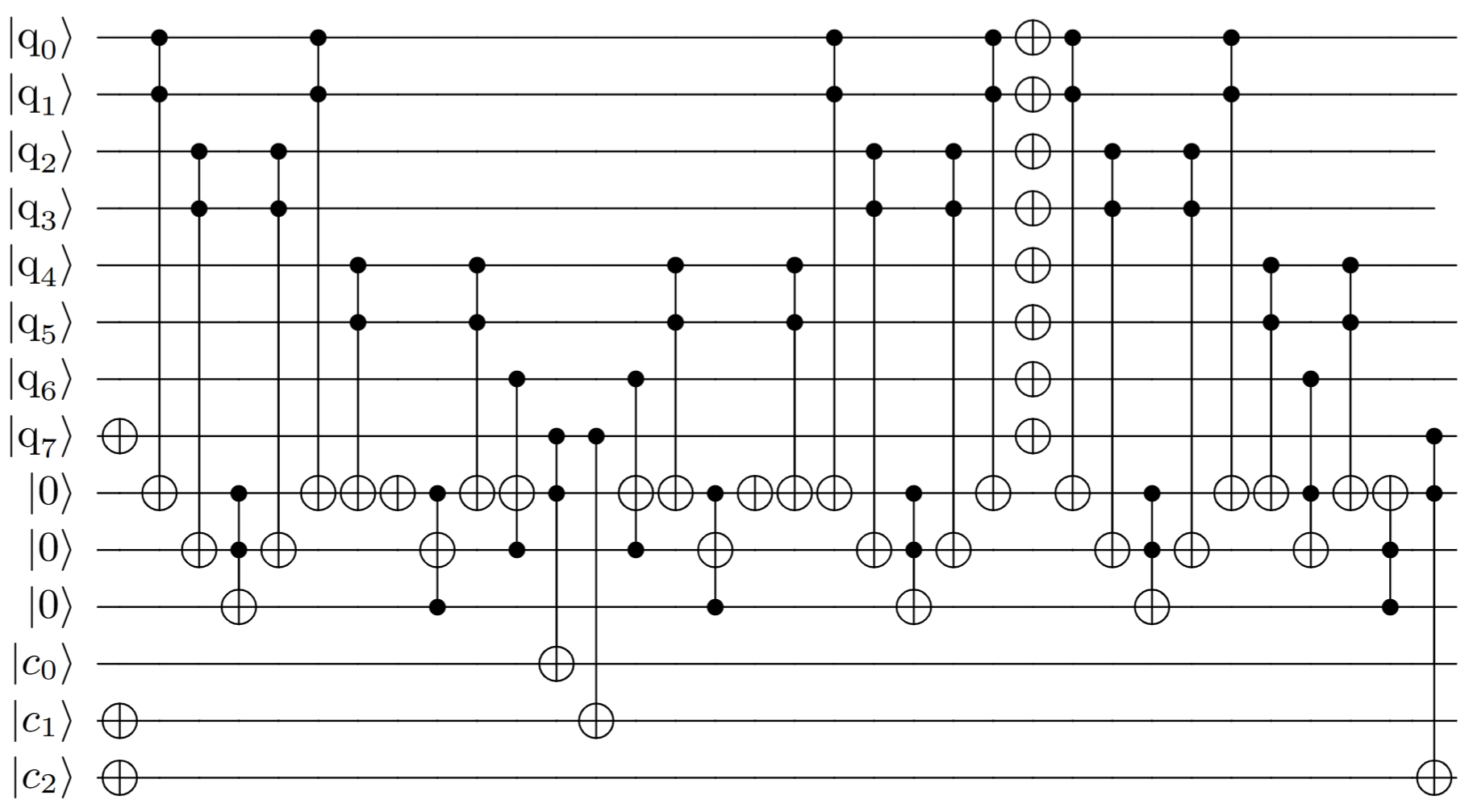}
    \caption{Comparison circuits to test if an 8-qubit state is ``less than'' basis states 127, 128, and 129.}
    \label{fig:Greater-Than}
\end{figure}

Figure~\ref{fig:Greater-Than} presents a circuit used in the current work to determine if the eight-qubit quantum state $\ket{\psi}$ is less than three values, being 127, 128, and 129. `Less than' is chosen since the wavepacket is propagating from right to left in the simulation, defined as the one's complement to each `greater than' operation.
For brevity, only the `write' portion is shown, with `un-write' applied after the coupling has been applied.
The $2^n =256$ discrete positions represent a space of $L = 20\,\text{a.u.}$. 

Characterization of the system is given in the following section. For direct comparison with \cite{ollitrault2020nonadiabatic}, a step function with breakpoints of 127 and 129 was used for nonadiabatic molecular dynamics simulations, given by the same circuit as Figure~\ref{fig:Greater-Than} without the CNOT to qubit $c_1$.

After making each comparison, the channel qubit has rotations about the $x$-axis controlled by the position register and the comparison qubits, given by the desired piecewise linear approximation, as $e^{-i\tau f(x) \sigma_x}$, where
\begin{equation}
    f(x) = \sum_{i=0}^{P-1} \alpha_i x + \beta_i.
\end{equation}

\begin{figure}[htbp]
    \centering
    \includegraphics[width=0.8\textwidth]{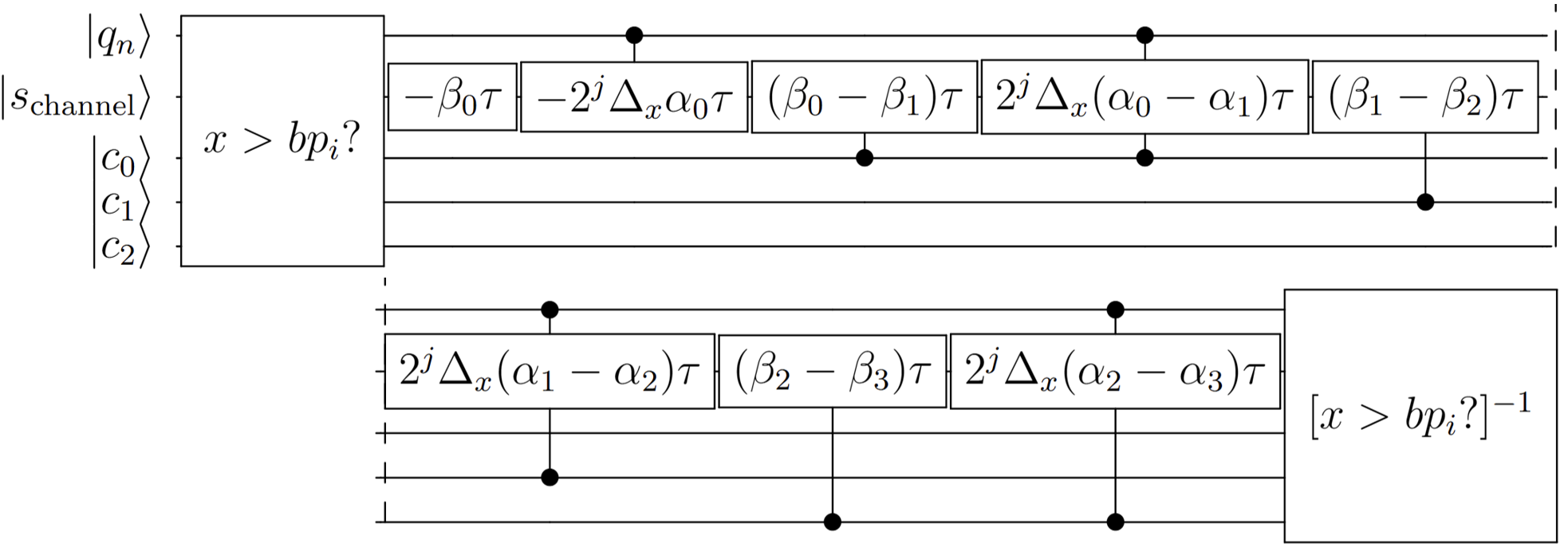}
    \caption{Schematic coupling for each piece of the function, conditioned upon the state of the wavepacket and the state of comparator qubits. 
Each linear term is expanded according to Figure~\ref{fig:lin-coup-schem}, controlled by qubits indexed by $j$.}
    \label{fig:Four-Piece Coupling}
\end{figure}

By the condition of ``if the wavepacket is less than breakpoint $p$,'' a linear coupling can be applied to the entire space of $2^n$ basis states. 
This coupling is ``turned off'' via the truth values of other comparison qubits.
A schematic of this coupling for a $P=4$-piece linear approximation of $f(x)$ is given in Figure~\ref{fig:Four-Piece Coupling}.
Varying coupling parameterizations drastically changes the molecular dynamics (Appendix~\ref{app:greedy}).

Avoiding arithmetic circuits reduces the ``minimal" circuit size for coherent Marcus model dynamics, giving an improvement from the previously stated 18 qubits (Ollitrault et al. \cite{ollitrault2020nonadiabatic}) to $8+\log_2(8) + 3 +\log_2(2) = 15$ qubits, with the final qubit being the channel qubit storing state populations of $M = 2$ potential surfaces. 
Comparison for the $p=3$ breakpoints can be made in $29$ Toffoli gates, compared to $42$ Toffolis necessary for a ripple-carry adder-like comparison \cite{stamatopoulos2020option, cuccaro2004new}.

\begin{figure}[htbp]
    \centering
    \includegraphics[width=1.0\textwidth]{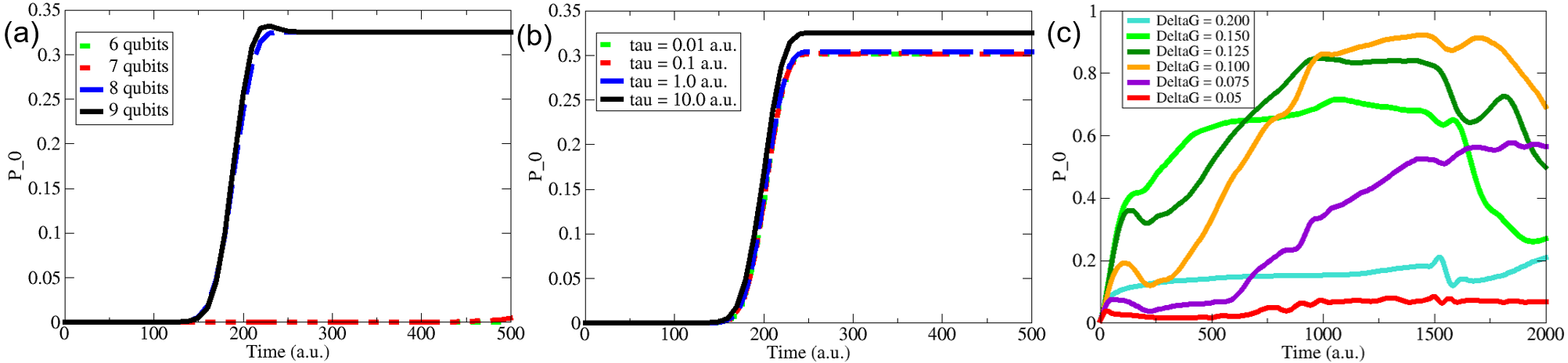}
    \caption{Recovery of results from Ollitrault et al. \cite{ollitrault2020nonadiabatic}. Adiabatic state transition in a Marcus-model type system with $p_0 = -30$, $x_0 = 4$. 
    Parameter variations are given, with position register size ranges from $n=6$ to $n=9$ with $\tau=10.0\,\text{a.u.}$ (left), timestep $\tau$ ranges from $0.01\,\text{a.u.}$ to $10.0\,\text{a.u.}$ with $n=8$ (middle), and energy gap $\Delta G$ ranges from 0.05 to 0.20 with $n=8\,,\tau=10.0\,\text{a.u.}$ 
    The Nyquist limit $\mathcal{N}$ is doubled for every additional qubit in the position register.}
    \label{fig:Ollitrault recovery}
\end{figure}

Convergence is recovered under the same conditions in Figures~\ref{fig:Ollitrault recovery}a and \ref{fig:Ollitrault recovery}b.
A Gaussian wavefunction was propagated at $p_0=-30\,\text{a.u.}$ in a box with size $L=20\,\text{a.u.}$ starting in the right harmonic potential with $\Delta_rG^0=0.0\, \text{E}_{\text{H}}$. 
A 4-piece linear approximation to a Gaussian coupling $V_{01}=0.01e^{-5x^2}$ is applied in a first-order Trotterization, recovering state population transfer over an evolution time of $t=500\,\text{a.u.}$ while varying register size (Figure~\ref{fig:Ollitrault recovery}a) and timestep (Figure~\ref{fig:Ollitrault recovery}b).
Once recovered, the coupling was simplified from a ``triangle" to a step function, being a 3-piece linear approximation.
A greedy search is detailed in Appendix~\ref{app:greedy} to recover results for varying $\Delta_rG^0$. 
This search spanned specifics not detailed in Ollitrault et al. \cite{ollitrault2020nonadiabatic}, and recovers near-identical dynamics when the step function is area-preserving to the Gaussian form and spans three basis states, being $\{127, 128, 129\}$ (Figure~\ref{fig:Ollitrault recovery}c).
Once recovered, these quantum circuits are tuned variationally in an attempt to reduce circuit overhead while preserving accurate rate coefficients.

\section{Variational Compression}

Truncating an analytically constructed circuit can be represented as approximating a unitary $U$ with an ansatz $A$. 
If $\norm{UA^\dagger}=0$, then $U\cong A$, being equivalent up to a global phase. 
In NISQ-centric algorithms, the ansatz is typically seen as being a parameterized quantum circuit of either variational structure or fixed structure \cite{bilkis2021semi, peruzzo2014variational}.
This dissertation will use a fixed structure ansatz, derived from a Schur decomposition.

\subsection{Ansatz Construction}
For unitary, Hermitian matrices, the Schur decomposition is a classic diagonalization $U = WDW^\dagger$,
where $W$ diagonalizes the operator and $D$ is diagonal containing the eigenvalues of $U$. 
This decomposition is unique up to eigenvalue ordering and eigenvector phases, motivating the variational ansatz structure.
By tuning parameters for both the eigenvector matrix $W$ and the diagonal matrix $D$ for a target unitary $U$, a compressed circuit is formed that separates a ``rotation into the eigenbasis'' (via $W$) from ``phase accumulation'' (via $D$). 
Time evolution becomes trivial in the diagonal representation, where evolving for time $N\tau$ rescales the diagonal phases by $N$, enabling fast-forwarding in certain sparse Hamiltonians \cite{atia2017fast}.

Ansatzes for $W$ and $D$ are given using Walsh functions, forming a complete orthonormal basis for functions on binary strings, analogous to the Fourier basis for periodic functions \cite{walsh1923closed}. 
For $n$ binary variables $q_0, q_1, \ldots, q_{n-1} \in \{0,1\}$, the Walsh functions are
\begin{equation}
    W_j(q_0, \ldots, q_{n-1}) = (-1)^{\sum_{k=0}^{n-1} j_k q_k},
    \label{eq:walsh_functions}
\end{equation}
where $j = \sum_{k=0}^{n-1} j_k 2^k$ indexes the function and $j_k \in \{0,1\}$ are its binary digits. 

Any real-valued function can be expanded in the Walsh basis:
\begin{equation}
    f(q_0, \ldots, q_{n-1}) = \sum_{j=0}^{2^n - 1} \hat{f}_j \, W_j(q_0, \ldots, q_{n-1}),
    \label{eq:walsh_expansion}
\end{equation}
with Walsh coefficients $\hat{f}_j = 2^{-n} \sum_q f(q) W_j(q)$.

For quantum circuits, this translates to tensor products of Pauli-$Z$ operators.
Since $Z\ket{q} = (-1)^q \ket{q}$ for $q \in \{0,1\}$, 
\begin{equation}
    W_j(q_0, \ldots, q_{n-1}) = \bra{q_0 \ldots q_{n-1}} \bigotimes_{q=0}^{n-1} Z_q^{j_q} \ket{q_0 \ldots q_{n-1}}.
    \label{eq:walsh_pauli}
\end{equation}
A diagonal unitary $e^{-i\tau f(x)}$ can therefore be written as
\begin{equation}
    e^{-i\tau f(x)} = \prod_{j=0}^{2^n-1} e^{-i\tau \hat{f}_j \bigotimes_{q=0}^{n-1} Z_q^{j_q}}.
    \label{eq:diagonal_walsh}
\end{equation}

The order of a Walsh term is defined as its Hamming weight $|h| = \sum_q h_q$, corresponding to the number of qubits involved in the Pauli-$Z$ tensor product, with order 0 being the identity, order 1 being $R_z$ rotations, and order 2 being parity-based $ZZ$ gates.
Quadratic functions on binary-encoded positions have Walsh expansions truncated at order $2$, giving a scheme for approximating diagonal unitaries.
An ansatz from a Walsh function expansion placed in a Schur decomposition gives a fast-forwardable, variationally optimizable circuit, following Cirstoiu et al. \cite{cirstoiu2020variational}.

\subsection{Variational Fast Forwarding}

Following a Schur decomposition, the VFF ansatz takes the form
\begin{equation}
    A(\vec{\gamma}, \vec{\theta}, \tau) = W(\vec{\gamma}) D(\vec{\theta}, \tau) W^\dagger(\vec{\gamma}),
    \label{eq:vff_ansatz}
\end{equation}
where $W(\vec{\gamma})$ is a parameterized eigenvector unitary and $D(\vec{\theta}, \tau)$ is an independently parameterized diagonal.

The eigenvector unitary $W(\vec{\gamma})$ is constructed using the Baker-Campbell-Hausdorff (BCH) formula:
\begin{equation}
    e^A e^B = e^{A + B + \frac{1}{2}[A,B] + \frac{1}{12}[A,[A,B]] - \frac{1}{12}[B,[A,B]] + \cdots}.
    \label{eq:bch}
\end{equation}
This results in a hardware-efficient ansatz (HEA) structure, alternating layers of single-qubit gates and nearest-neighbor entangling gates to increase expressivity while respecting hardware connectivity constraints \cite{kandala2017hardware}.

The diagonal operator $D(\vec{\theta}, \tau)$ is parameterized as a truncated Walsh expansion. 
Once the ansatz is optimized for timestep $\tau$, evolution for time $N\tau$ can be achieved by simply rescaling the diagonal parameters to $N\vec{\theta}$, while $W(\vec{\gamma})$ remains unchanged. 
This enables simulation of dynamics beyond the coherence time of the quantum hardware, limited only by the accuracy of the initial optimization for compatible Hamiltonians.
To examine the variational optimization of a fast-forwardable Hamiltonian (the kinetic term $T$) and in the approximation of terms in the diabatic Hamiltonian ($T$ and $V$), classical emulations of quantum-assisted quantum compiling (QAQC) are performed in the next subsection.

\subsection{Optimization via Quantum-Assisted Quantum Compiling}

QAQC provides a method to optimize variational circuits using quantum hardware to evaluate the cost function and gradients \cite{khatri2019quantum}. 
This optimization updates parameters in an ansatz $A(\vec{\alpha})$ to approximate a target unitary $U$.

The Local Hilbert-Schmidt Test (LHST) cost function measures average entanglement fidelity across qubit pairs:
\begin{equation}
    C_{\text{LHST}}(U, A) = 1 - \frac{1}{n} \sum_{j=1}^{n} F_e^{(j)},
    \label{eq:lhst_cost}
\end{equation}
where $F_e^{(j)}$ is the entanglement fidelity for the $j$-th qubit pair. 
The LHST cost satisfies $C_{\text{LHST}} = 0$ if and only if $A = e^{i\phi} U$ for some global phase $\phi$, making it a faithful measure of unitary equivalence \cite{khatri2019quantum}.

\begin{figure}[t]
\centerline{
\includegraphics[width=65mm]{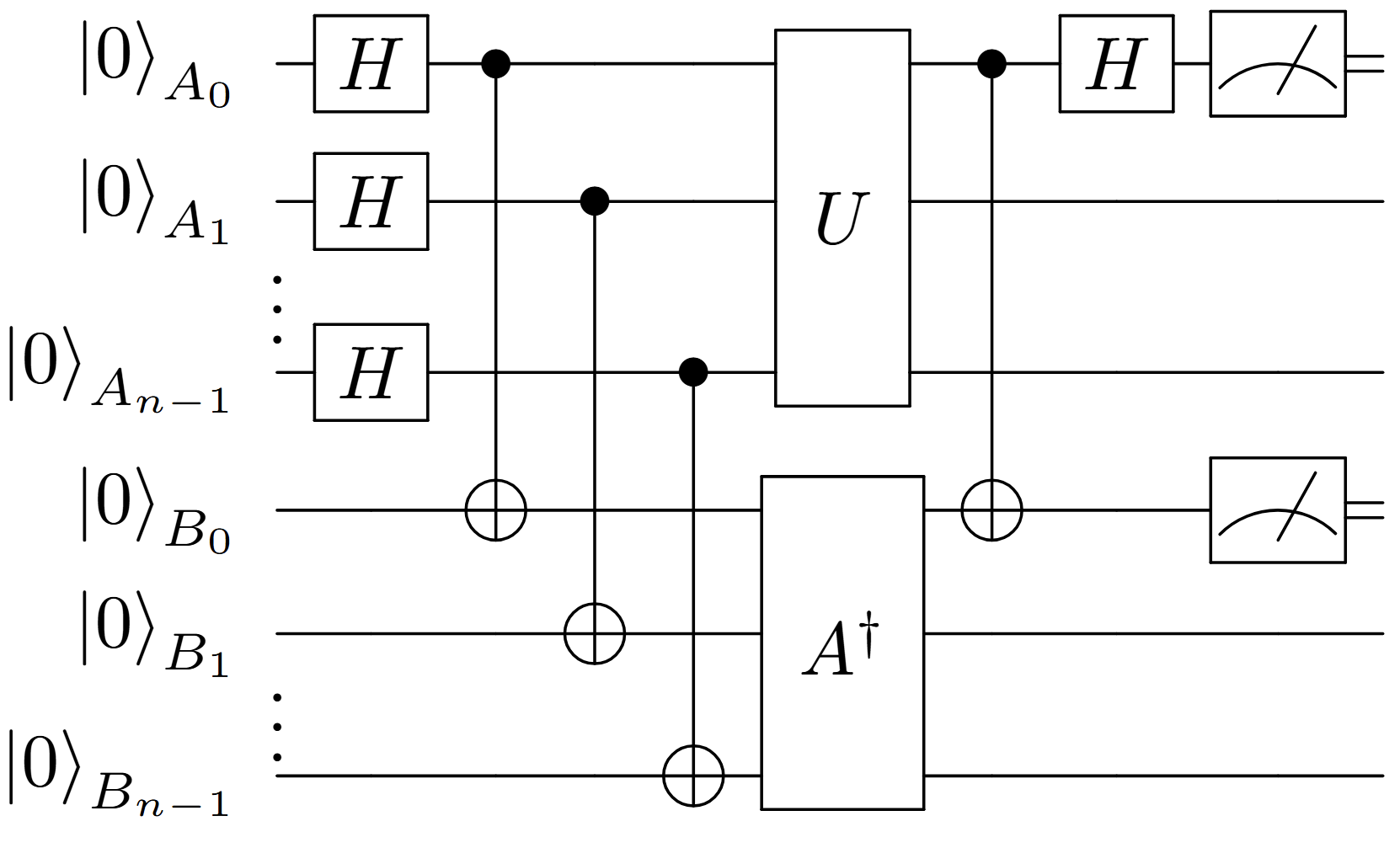}}
\caption{\label{fig:qaqc_schematic} 
Quantum-assisted quantum compiling (QAQC) circuit for measuring entanglement fidelity. 
Bell pairs are prepared between corresponding qubits in registers $A$ and $B$. 
Bell basis measurement extracts entanglement fidelity $F_e^{(j)}$ for each qubit pair, requiring measurement of all $n$ pairs of qubits.}
\end{figure}

Entanglement fidelity is measured by preparing Bell pairs between corresponding qubits in two $n$-qubit registers, applying the target $U$ to register $A$ and the ansatz inverse $A^\dagger$ to register $B$, and measuring the probability of returning to the initial Bell state (Figure~\ref{fig:qaqc_schematic}). 
This requires $2n$ qubits total and provides a noise-resilient cost function: hardware noise affects both $U$ and $A$ similarly, partially canceling in the fidelity measurement \cite{sharma2020noise}.

Gradients for optimization are computed using the parameter-shift rule (Eq.~\ref{eq:parameter_shift}),
enabling gradient-based optimization using only cost function measurements.

\begin{figure}[t]
\includegraphics[width=\columnwidth]{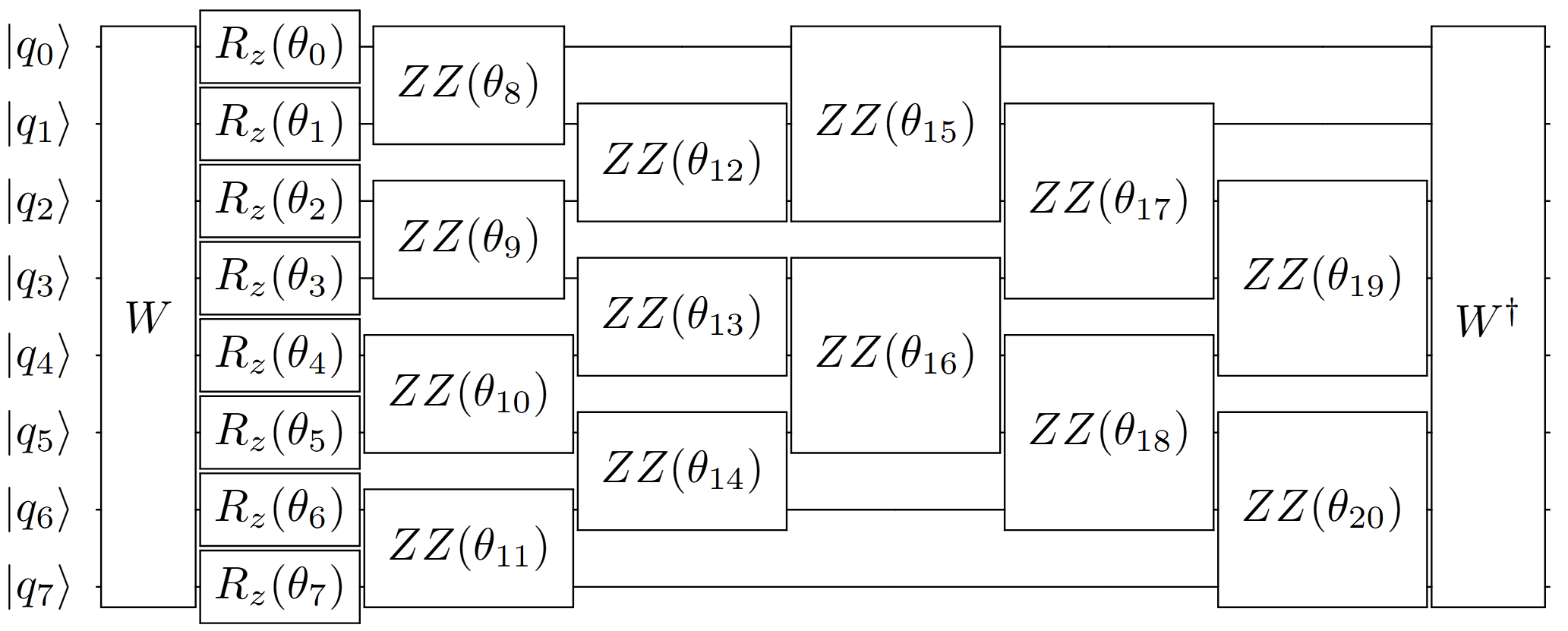}
\caption{\label{fig:l-local-ansatz} 
Ansatz structure for $l=3$-local truncation on $n=8$ qubits. 
Gates are organized by locality: $l=1$ (single-qubit $R_z$), $l=2$ (nearest-neighbor $ZZ$), and $l=3$ (next-nearest-neighbor $ZZ$).
The structure $W$ follows the first three sublayers of $D$, truncated at $l=2$-local $ZZ$ terms.}
\end{figure}

Beginning with the ansatz structure, $ZZ$ operations have a `locality' defined by the distance between its control and target qubits on a linear qubit array. 
A single-qubit $R_z$ gate is 1-local, a $ZZ$ gate between nearest-neighbor qubits is 2-local, and so forth.
Following Eq.~\ref{eq:diagonal_walsh} and organizing by increasing locality,
\begin{equation}
    D_l(\vec{\theta}, \tau) = \prod_{m=1}^{l} \left( \prod_{\substack{j \in S_m \\ \text{locality}(j) \leq l}} e^{i\theta_j \bigotimes_{q=1}^{n}(Z_q)^{j_q}} \right).
    \label{eq:l_local}
\end{equation}
An example of an $l=3$-local ansatz cirucuit for $D_l$ is given in Figure~\ref{fig:l-local-ansatz}.

This truncation reduces circuit depth by eliminating long-range $ZZ$ operations that would require SWAP gates on hardware with limited connectivity (e.g. superconducting transmons). 
Approximation is tunable when these long-range operations are removed as a primary result of this chapter.

When truncating to $l < n$, the remaining variational parameters ``compensate'' during optimization. 
For a quadratic target, first-order terms ($R_z$ sublayer) converge to a single effective parameter $\alpha_0$, and the second-order terms ($ZZ$ sublayers) converge to a second effective parameter $\alpha_1$, weighted by respective binary weights. Both relate to the original quadratic coefficients through
\begin{equation}
    \alpha_0 \propto \phi_2 = -2\tau \eta x_0 \Delta, \quad \alpha_1 \propto \phi_3 = -\tau \eta \Delta^2,
    \label{eq:parameter_relations}
\end{equation}
where $\eta = m\omega^2/2$ for a harmonic potential. 
This reduces the effective parameter space from $\mathcal{O}(n^2)$ to two (2) for a quadratic approximation.
Presented optimizations do not advantage this relationship. 
Recovery of these parameters through an uninformed gradient descent validates the method, reaching the global minimum.

Ansatz parameters are optimized using the Local Hilbert-Schmidt Test (LHST) cost function (Eq.~\ref{eq:lhst_cost})~\cite{khatri2019quantum}:
 The entanglement fidelity is measured by initializing Bell pairs between corresponding qubits in two registers, applying $U$ to one register and $A^\dagger$ to the other, and measuring overlap with the initial Bell state:
\begin{equation}
    F_e^{(j)} = \text{Tr}\left[ \ket{\Phi^+}\bra{\Phi^+}_{A_j B_j} \left(\mathcal{E}_j \otimes I_{B_j}\right)\left(\ket{\Phi^+}\bra{\Phi^+}_{A_j B_j}\right) \right],
    \label{eq:entanglement_fidelity}
\end{equation}
where $\mathcal{E}_j$ is the quantum channel acting on qubit $A_j$ after tracing out all other qubits in register $A$.

Gradients are computed using the parameter-shift rule. 
For eigenvector parameters $\theta_k$, Eq.~\ref{eq:parameter_shift} expands as 
\begin{equation}
\begin{split}
    \frac{\partial C_{\text{LHST}}}{\partial \gamma_k} = \frac{1}{2} \Big[ & C_{\text{LHST}}(U, W_+^k D W^\dagger) - C_{\text{LHST}}(U, W_-^k D W^\dagger) \\
    & + C_{\text{LHST}}(U, W D (W_+^k)^\dagger) - C_{\text{LHST}}(U, W D (W_-^k)^\dagger) \Big],
\end{split}
\label{eq:grad_theta}
\end{equation}
where $W_\pm^k$ denotes $W$ with parameter $\theta_k$ shifted by $\pm\pi/2$. For diagonal parameters $\theta_\ell$:
\begin{equation}
    \frac{\partial C_{\text{LHST}}}{\partial \theta_\ell} = \frac{1}{2} \Big[ C_{\text{LHST}}(U, W D_+^\ell W^\dagger) - C_{\text{LHST}}(U, W D_-^\ell W^\dagger) \Big].
    \label{eq:grad_gamma}
\end{equation}

Adam optimization is employed with these gradients, using the adaptive learning rate to escape local minima and navigate the $\mathcal{O}(n^2) \mod 2\pi$ parameter landscape. 
Classical emulation of the full $2n$-qubit QAQC protocol is performed for $n \leq 5$.
For $n > 5$, the Hilbert-Schmidt distance $\| UA^\dagger\|_{\text{HS}}^2$ is directly minimized and rescaled to match $C_{\text{LHST}}$.
In all cases, optimizations converge to global minima by comparing the optimized diagonal parameters to analytical expressions derived from the explicit quadratic expansion (Figure~\ref{fig:optimization_results}).
Once truncated, new operators $T',V_0', V_1'$ replace Trotter terms $T,V_0,V_1$ to approximate Marcus model dynamics.

Each compressed term uses the $l$-local truncated ansatz from Eq.~\ref{eq:l_local}, optimized to achieve $C_{\text{LHST}} \leq 0.01$. 
For an 8-qubit position register, the kinetic term is approximated as being $l=4$-local, and the potential terms $V_0'$,$V_1'$ are truncated to be $l=6$-local for linear qubit connectivity and $l=4$-local for a ring topology.

Rate coefficients from short-time population dynamics can be calculated using the approximated Trotter evolution, comparing results for step sizes $\tau = 1$ a.u. and $\tau = 10$ a.u., given in Section~\ref{sec:Results}.
Gate precision and circuit depth prevent current hardware from applying meaningful quantum simulation, so $n=15$ ideal logical qubits are classically emulated using the Qiskit SDK, with evolution proceeding as a noiseless statevector simulation.

\section{The Global Minimum}
Following Chapter~\ref{ch:wavepacket_prop}, the quadratic operator is cast as
\begin{equation}
\begin{split}
    & e^{-i\tau\phi_3 \sum_j^{n-1}2^{2j}q_j}\times e^{-i\tau\phi_3\left(2\sum_{j<\ell}2^{j+\ell}q_jq_\ell\right)}\times e^{-i\tau\phi_2\sum_j^{n-1}2^jq_j} \\
    & = e^{-i\tau\phi_3\left(2\sum_{j<\ell}2^{j+\ell}q_jq_\ell\right)}\times e^{-i\tau\phi_3 \sum_j^{n-1}2^{2j}q_j}\times e^{-i\tau\phi_2\sum_j^{n-1}2^jq_j} \\
    & = e^{-i\tau\phi_3\left(2\sum_{j<\ell}2^{j+\ell}q_jq_\ell\right)}\times e^{-i\tau(\phi_3 \sum_j^{n-1}2^{2j}q_j + \phi_2\sum_j^{n-1}2^jq_j)},
\end{split}
\end{equation}

For clarity, $\ell$ is not directly related to $l-$locality and is merely an index.
This expression shows that the compressed ansatz with $l=n$-local terms in the second-order Walsh expansion is equivalent to the explicit ordered operations from \cite{benenti2008quantum}, confirmed by the translation between phase-based gates ($\mathrm{CR}_z$) and parity-based gates ($ZZ$).  
For NISQ-era computation, it is advantageous to optimize the variational circuit using the native gate set on the desired hardware, even with equivalent Hamiltonian expressibility, given that transpilation does not guarantee the most efficient conversion.

An $l-$local truncation method is well-suited for superconducting quantum computers, minimizing the number of long-range controlled operations. 
Other truncation methods may be preferable for trapped-ion systems, such as the removal of gates whose analytical solution result in rotation angles $\theta\,\text{mod}\,2\pi$ being small or the replacement of multi-controlled gates with fewer control qubits.
Although trapped-ion systems are not explicitly considered, increasing entanglement fidelity in a diagonal space requires additional gates to increase the circuit's expressibility i.e., an additional gate cannot be exactly represented using the tensor product spaces of previously added gates \cite{sim2019expressibility}.
After a transformation (e.g. QFT) or application of a gate that is not unitary in the position register's Hilbert space (e.g. nonadiabatic coupling), this guideline does not hold.

\section{\label{sec:Results} Results}
\begin{figure}[t]
\includegraphics[width=\columnwidth]{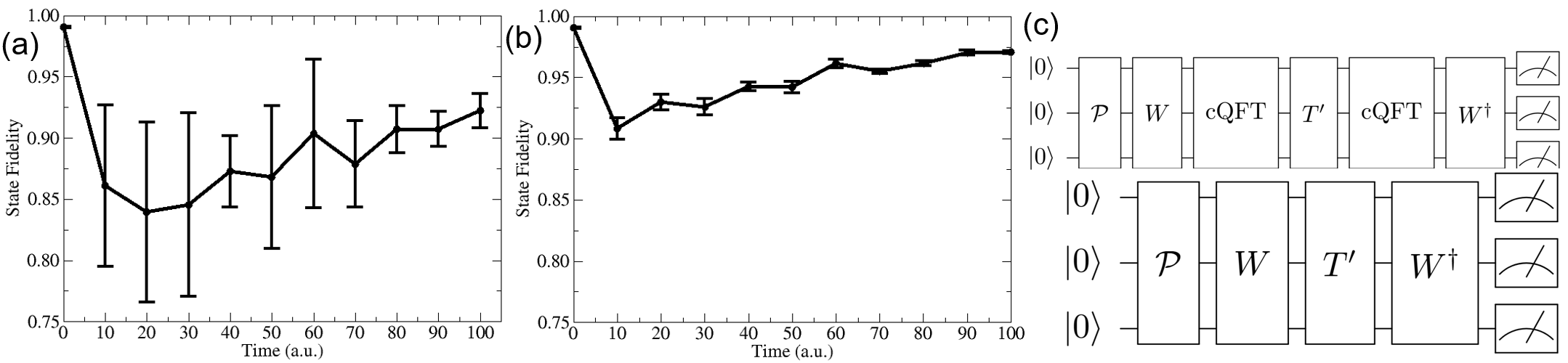}
\caption{\label{fig:fast_forwarding} 
Fast-forwarding demonstration on $ibm\_brisbane$. (a) $C_{LHST}$ for the $l=1$-local isolated kinetic term is $1.031\times10^{-6}$. (b) $C_{LHST}$ for the full kinetic operator with $l=1$-local diagonal is $4.494\times10^{-3}$. (c) Circuit schematics for each quantum simulation, consisting of three physical qubits each.}
\end{figure}

Fast-forwarding of the compressed kinetic operator is demonstrated on $ibm\_brisbane$, a 127-qubit superconducting transmon chip by IBM Quantum.
Starting from the initialized wavepacket, $N = 0$ to $10$ timesteps of the $l=1$-local compressed kinetic operator are applied with $\tau = 10$ a.u., sandwiched between cQFTs (Figure~\ref{fig:fast_forwarding}).

Highlighting the use of $W$, the hardware-efficient $ZZ$ scheme was altered to variationally compress the cQFTs. 
$ZZ$ operations can only apply operations that skew the measurement axis, meaning the effect of the QFT (moving to momentum space) cannot be sufficiently captured. 
Using the same block structure, each $ZZ$ block was replaced with a general $U(4)$ gate, parameterized by 15 rotation parameters each \cite{vatan2004optimal}.
The full application of the kinetic term $\text{cQFT}-T-\text{cIQFT}$ was then variationally optimized via classical emulation of the QAQC algorithm.
The resulting circuit has no $l=3$-local operations or SWAP gates.
Results show improved state fidelity under the same initial conditions, emphasizing a tradeoff between hardware error and approximation error.
Circuit expressibility can largely be accounted for in nearest-neighbor operations for small circuit sizes, with an exponentially suppressed impact for increased $n$ \cite{sim2019expressibility}. 
The results in Figure~\ref{fig:fast_forwarding} therefore may not easily extend to larger systems, with additional recognition of variational parameters per $U(4)$ gate compared to one variational parameters for a $ZZ$ gate.

The state fidelity remained above 0.80 through $N = 10$ timesteps, demonstrating that the compressed ansatz structure is compatible with near-term hardware noise levels.

\begin{figure}[t]
\includegraphics[width=\columnwidth]{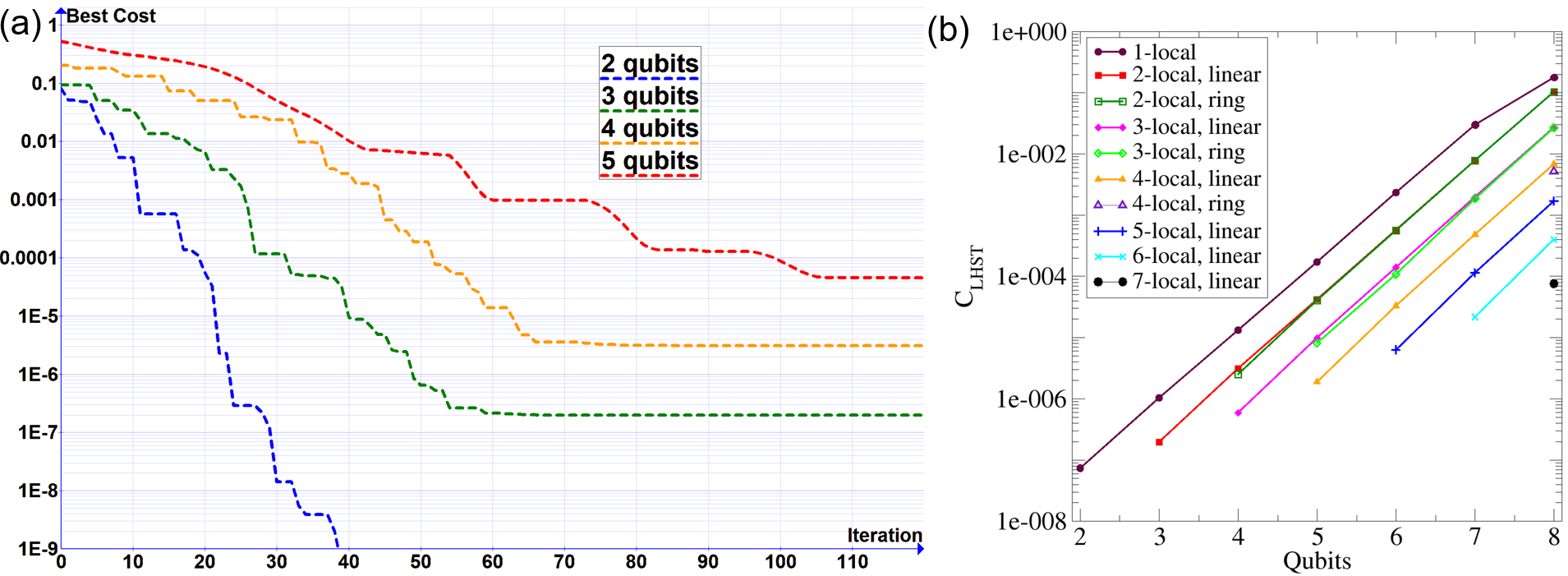}
\caption{\label{fig:optimization_results} 
LHST cost versus locality parameter $l$ for the kinetic operator. 
(a) Sample optimizations of kinetic operator with 2-local ansatz with linear topology. 
Costs were verified as global minima by relation of variational parameters to analytically-derived parameterizations of quadratic operators.
(b) Global minima for linear, ring circuit topologies for varying register sizes, found by QAQC (2-5 qubits) and Hilbert-Schmidt distance minimization (6-8 qubits).}
\end{figure}

Kinetic and potential operators are optimized to $l$-local ansatzes across register sizes $n = 3$ to $n = 8$ qubits, considering both linear and ring circuit topologies. Figure~\ref{fig:optimization_results} shows the LHST cost as a function of locality $l$ for each configuration.

For the kinetic operator, $l = 4$-local truncation achieves $C_{\text{LHST}} < 0.01$ for all tested register sizes with linear topology. 
The ring topology achieves the same threshold at $l = 3$ for $n \leq 6$ qubits. 
Potential operators need longer-range gates for the same fidelity, requiring $l = 6$ for linear, $l = 4$ for ring topology. 
This is likely due to their spatial centering near the middle of the position basis, where binary encoding concentrates high-order bit contributions.

For small times, the population $\mathcal{P}_0(t)$ decays according to Fermi's golden rule:
\begin{equation}\label{eq:short_time}
    \mathcal{P}_0(t) \approx 1 - k t + \mathcal{O}(t^2),
\end{equation}
where $k$ is the rate coefficient. 
$\mathcal{P}_0(t)$ is computed by measuring the expectation value of the channel qubit, $\mathcal{P}_0 = (\langle Z_{\text{obj}} \rangle + 1)/2$, after $N$ Trotter steps of duration $\tau$.

To extract $k$, linear regression on $\mathcal{P}_0(t)$ is performed for $t \leq 100$ a.u., where the short-time approximation remains valid. The slope of the linear fit yields $-k$. 
Repeating this procedure for driving forces $\Delta G^0 \in [0, 0.3]$ a.u.\ maps out the rate coefficient as a function of $\Delta G^0$.

\begin{figure}[t]
\includegraphics[width=\columnwidth]{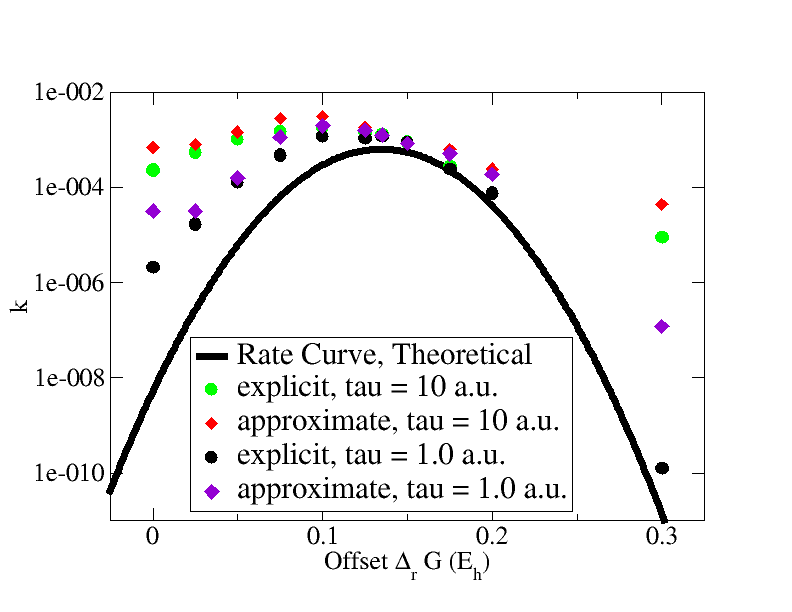}
\caption{\label{fig:rate_coefficients} 
Marcus model rate coefficients extracted from short-time population dynamics. 
Solid line: theoretical Marcus rate (Eq.~\ref{eq:marcus_rate}). 
Symbols: quantum simulation results using explicit circuits (green, black) and compressed $l$-local ansatzes (red, purple) at different Trotter step sizes.}
\end{figure}

Figure~\ref{fig:rate_coefficients} compares rate coefficients computed using explicit Trotter circuits to those obtained with compressed $l$-local ansatzes. The theoretical Marcus rate is given by~\cite{marcus1956theory, nitzan2024chemical}:
\begin{equation}
    k_{\text{Marcus}} = \frac{2\pi}{\hbar} |V_{\text{offdiag}}|^2 \frac{1}{\sqrt{4\pi \lambda k_B T}} \exp\left(-\frac{(\Delta G^0 + \lambda)^2}{4\lambda k_B T}\right),
    \label{eq:marcus_rate}
\end{equation}
where $|V_{\text{offdiag}}|$ is the electronic coupling matrix element, $\lambda$ is the reorganization energy, and $T$ is the temperature. For a zero-temperature wavepacket simulations, the rate is determined by the Franck-Condon overlap between the initial wavepacket position and the crossing region. 
Extending from Fermi's golden rule for short time dynamics, projection of the wavefunction from state 2 onto the initial wavefunction (cross-correlation) is required for higher-order expansions.

The compressed circuits ($l=4$ for kinetic, $l=6$ for potential terms) recover rate coefficients that retain similar structure to theoretical predictions and explicit circuit results, albeit with noticeable deviation. 
This shows that truncation error introduced by $l$-local compression retains behavior of physically relevant observables, but significantly impacts accuracy.
The theoretical peak rate is associated to where the bottom of $V_0$ intersects the right side of $V_1$, where the most accurate results are found for the Trotterization incorporating substituted $T',V_0',V_1'$ terms.

\section{\label{sec:Discussion} Discussion}

\subsection{Resource Scaling}
Total qubit count for simulating nonadiabatic dynamics comprises of registers with distinct functions.
A $d$-dimensional wavepacket requires $d \log_2(\mathcal{N})$ qubits to represent $\mathcal{N}$ grid points per dimension; for the one-dimensional case studied here ($d=1$), an $n$-qubit position register encodes $2^n$ discrete positions.
Dynamics on $M$ diabatic surfaces requires a channel register of size $\lceil \log_2(M) \rceil$; for $M = 2$ surfaces, a single channel qubit suffices.

The coupling operator introduces additional qubit overhead.
A piecewise linear coupling function with $P$ pieces requires $P-1$ comparator qubits to store threshold truth values, plus ancilla qubits for intermediate calculations. 
Using the bespoke comparison circuits (Fig.~\ref{fig:Greater-Than}), the ancilla register scales as $\lceil \log_2 n \rceil$ for $n > 3$ qubits.
The total qubit count for a one-dimensional, two-surface system with $P$-piece coupling is therefore
\begin{equation}
    n_{\text{total}} = n + 1 + (P-1) + \lceil \log_2 n \rceil = n + P + \lceil \log_2 n \rceil.
    ,\label{eq:total_qubits}
\end{equation}
For an 8-qubit position register with 3-piece coupling ($P=3$) and 1 channel qubit, this yields $n_{\text{total}} = 8 + 3 + 3  + 1= 15$ qubits, compared to 18 qubits reported in Ref.~\cite{ollitrault2020nonadiabatic} using a different comparator construction.

Gate count is most notably reduced in tensor product commutativity, reducing $n(n-1)$ $\mathrm{CR}_z$ gates by a factor of two per quadratic term.
A conversion to more common $ZZ$ gates therefore requires $2n(n-1)$ CNOT gates and $n(n-1)$ $R_z$ gates.
Analytical compression reduces gate count by a factor of two $n(n-1)$ CNOT gates, prior to SWAP gates.

Table~\ref{tab:gate_counts} summarizes the explicit and compressed gate counts for the kinetic and potential operators across register sizes $n = 4, 6, 8$, broken down by gate type.
For $n \leq l$, the compressed ansatz includes all Walsh terms and the $W/W^\dagger$ layers add overhead; compression savings appear only when $l < n$ eliminates long-range interactions requiring SWAP routing.

\begin{table}[t]
\caption{\label{tab:gate_counts} Logical gate counts for explicit (Ex.) versus $l$-local compressed (Comp.) operators on a linear qubit topology. $ZZ$ counts represent parity-rotation gates ($ZZ = \text{CNOT}\cdot R_z \cdot \text{CNOT}$). For controlled operators $V_0, V_1$, each $ZZ$ becomes a $CZZ$ requiring $2$ Toffoli gates; $W$ and $W^\dagger$ remain unconditional. The compressed ansatz uses one brick-wall layer for $W$. SWAP routing costs are architecture-dependent and not included; the maximum locality column indicates the longest-range interaction, which determines SWAP overhead on linear hardware.}
\begin{tabular}{l|l|c|c|c|c|c|c|c|c|c}
 & & & \multicolumn{2}{c}{$ZZ$/$CZZ$} & \multicolumn{2}{c}{$R_z$} & \multicolumn{2}{c}{Toffoli} & \multicolumn{2}{c}{Max $l$} \\
$n$ & Op. & $l$ & Ex. & Comp. & Ex. & Comp. & Ex. & Comp. & Ex. & Comp. \\
\hline
4 & $T$ & 4 & 12 & 6 & 16 & 10 & --- & --- & 4 & 4 \\
4 & $V_0$ & 4 & 12 & 6 & 16 & 10 & 24 & 12 & 4 & 4 \\
4 & $V_1$ & 4 & 12 & 6 & 16 & 10 & 24 & 12 & 4 & 4 \\
\hline
6 & $T$ & 4 & 24 & 12 & 30 & 18 & --- & --- & 6 & 4 \\
6 & $V_0$ & 6 & 30 & 15 & 36 & 21 & 60 & 30 & 6 & 6 \\
6 & $V_1$ & 6 & 30 & 15 & 36 & 21 & 60 & 30 & 6 & 6 \\
\hline
8 & $T$ & 4 & 36 & 18 & 44 & 26 & --- & --- & 8 & 4 \\
8 & $V_0$ & 6 & 50 & 25 & 58 & 63 & 100 & 50 & 8 & 6 \\
8 & $V_1$ & 6 & 50 & 25 & 58 & 63 & 100 & 50 & 8 & 6 \\
\end{tabular}
\end{table}

Variational compression as it was performed most notably reduces SWAP gate cost.
If performed on a superconducting transmon chip, long-range interactions require $2d-3$ SWAP gates to move qubits that are a distance $d$ apart. 
For an $n=8$-qubit circuit, the $l=4$-local truncation of the kinetic term removes $4$ $l=5$-local $ZZ$ gates, $3$ $l=6$-local $ZZ$ gates, $2$ $l=7$-local $ZZ$ gates, and $1$ $l=8$-local $ZZ$ gate ($10$ gates total), saving $20$ CNOT gates and $10$ $R_z$ gates from the $ZZ$ decompositions alone, plus all SWAP routing overhead for those long-range interactions, while retaining an entanglement fidelity of $0.99$.
Controlled operations $V_0,V_1$ replace CNOT cost with Toffolis ($ZZ \rightarrow CZZ$), being far more resource-expensive.  
The $l=6$-local truncation therefore only saves $1$ $l=7$-local $CZZ$ gate, though this one operation alone requires $132$ CNOT gates and $14$ T gates for a full operation, resetting to initial positions. 

\subsection{Limitations and Sources of Error}

Several factors limit the accuracy of this compressed simulation.
The $l$-local truncation introduces systematic error by construction, controlled by some threshold $C_{\text{LHST}} < 0.01$. 
This error propagates through Trotter steps, accumulating coherently as $\mathcal{O}(N \cdot C_{\text{LHST}})$ for $N$ steps.
For the rate coefficient calculations presented here, even small truncation error leads to a large deviation in observables, leaving simulation highly sensitive to approximation.

Quantum hardware simulations suffer from gate errors, decoherence, and readout errors, degrading state fidelity beyond that which is considered in this chapter.
The fast-forwarding demonstration on $ibm\_brisbane$ showed quick fidelity decay from 0.99 at initialization (Figure~\ref{fig:fast_forwarding}), but leveling off over longer times, suggesting a possible implementation in adiabatic systems or prior to subjecting the wavefunction to nonadiabatic influences.
Error mitigation techniques such as zero-noise extrapolation could extend useful simulation depth \cite{temme2017error, kandala2019error}, but significant progress needs to be made before meaningful nonadiabatic dynamics simulations can be applied, even to the scale of a single dimension with harmonic potentials.
The QAQC cost function may contain local minima, particularly for large $n$ and $l$. 
Convergence to global minima is verified by comparing optimized parameters to analytical expressions for all tested configurations, giving variational parameters of $A^\dagger$ no prior knowledge of $U$. 
For production use, having an unknown global minimum poses practical difficulty, eased by informed initial guesses based on analytical structure.

\section{\label{sec:Conclusions} Conclusion}

Trotterized quadratic operators can be variationally compressed for quantum simulation of nonadiabatic molecular dynamics. 
The method employs a VFF ansatz with $l$-local truncation of the Walsh series expansion, optimized using the QAQC protocol. 
The compressed ansatz exactly represents any quadratic operator when $l=n$ for linear qubit topology or $l = \lceil (n+1)/2 \rceil$ for ring topology, achieving $C_{\text{LHST}} = 0$ up to a scalar phase rotation.
Reducing $l$ below these limits introduces bounded approximation error.
Remaining variational parameters compensate during optimization, converging to effective values determined by the polynomial coefficients, reducing parameter space from $\mathcal{O}(n^2)$ to $2$ for quadratic polynomials.
Marcus model rate coefficients computed using truncated ansatzes ($l=4$ for kinetic, $l=6$ for potential terms) retain behavior but deviate from theoretical predictions, demonstrating preservation of physically relevant observables, as well as a high sensitivity to controlled circuit approximation.
Recent wavepacket initializations on $ibm\_torino$ achieved fidelities of $0.933 \pm 0.000413$ for $n = 6$ qubits (Figure~\ref{fig:UCC_Hardware_VaryQubits}), showing improved error mitigation for larger register operations, but with statistically lower fidelities for $n=3$ (Chapter~\ref{ch:wavepacket_prop}).
Fast-forwarded dynamics on $ibm\_brisbane$ maintained fidelity above 0.80 through 10 timesteps with directly applied QFTs and above 0.90 with QFTs absorbed in the diagonalization (Figure~\ref{fig:fast_forwarding}), validating potential utility of the compressed ansatz structure for diagonalizable Hamiltonians.
An $l$-local compression reduces circuit depth from $\mathcal{O}(n^2)$ to $\mathcal{O}(l\cdot n)$, with the locality parameter $l$ determined by the desired approximation accuracy.

Several extensions of this work merit investigation, but are not considered in this dissertation.
The Walsh expansion extends to degree $k > 2$ by including multi-qubit terms ($ZZZ$, $ZZZZ$, etc.). 
This would enable compression of anharmonic potentials relevant to molecular dynamics beyond the harmonic approximation, including Morse potentials and polynomial fits to ab initio potential energy surfaces \cite{braams2009permutationally, jiang2016potential}.
Extension to dense operators such as position-dependent kinetic terms or non-local couplings would significantly increase parameter space, but may potentially enable compression of entire Trotter steps.
As quantum hardware transitions toward fault tolerance, comparing variational compression of adiabatic terms to QSP-based methods at equivalent logical error rates will clarify a practical approach for different problem classes~\cite{yoder2025tour, shaw2025lowering}. 
In Chapter~\ref{ch:comparison}, a bounded-error quantum resource estimation is performed across benchmark systems to identify the bottleneck in first- and second-quantized QSP circuits in simulating nonadiabatic systems.
Error correction suggests a need to optimize while minimizing the number of non-transversal gates for fault-tolerant architectures, opening new areas of optimization for quantum operations applied in parallel.

The next route considered in this dissertation is scaling the analytical model. 
While multi-coordinate simulations are briefly mentioned by Ollitrault et al. \cite{ollitrault2020nonadiabatic}, a literature review currently finds no numerically validated schemes that extend to conical intersections. 
The previously proposed multi-channel extension for parameterized quantum circuits~\cite{ollitrault2020nonadiabatic} does not scale optimally (Appendix~\ref{app:basic_coupling}), though is intuitive for small systems. 
As natural extensions for a scalable and efficient algorithm for NA-QMD, these open problems are handled and validated in the next chapter.

\chapter{Model Extension}\label{ch:model_ext}

As mentioned in Chapter~\ref{ch:Introduction}, a quantum algorithm for NA-QMD with practical utility should efficiently handle strong coupling in the multidimensional and multi-channel extensions. 
Protocols for both extensions are described in this chapter, scaling the parameterized product formula motif of Ollitrault et al.~\cite{ollitrault2020nonadiabatic} to be efficient and parallel.
The resulting algorithm, dubbed Real-Time Evolution of Quantum Wavepackets In Explicit Modularity (REQWIEM), has improved depth scaling and extends beyond the limitations of phase gradient approaches~\cite{motlagh2025quantum}.
Literature for comparison is sparse, with large-scale quantum algorithms for dynamics still being relatively new. Motlagh et al.~\cite{motlagh2025quantum} being the only proposal found claiming to efficiently handle multi-mode, multi-channel vibronic systems.
This chapter focuses on detailing the circuit structure and associated costs, while the next chapter (Chapter~\ref{ch:resources}) is dedicated to resource estimation and comparison of both the parameterized quantum circuits of REQWIEM and the QROM loading procedure. 
This resource comparison focuses on fault-tolerant application, including rotation synthesis estimations into total overhead cost. 
Advantages and limitations to both methods are presented, with extensivity of REQWIEM considered in later chapters in the handling of hard core, anharmonic, and finite-difference potential energy surfaces and couplings.

A background of characteristic features of vibronic systems is given, beginning with conical intersections (Section~\ref{sec:conical_intersections}), being a topological feature in multidimensional dynamical systems. 
These features are foundational in NA-QMD and are critical for accurate physics.
The vibronic Hamiltonian, energy redistribution, and population recurrence are discussed in the remainder of Section~\ref{sec:background}, transitioning to a full decomposition of the REQWIEM algorithm (Appendix~\ref{app:reqwiem-algorithm}). 

\section{Background}\label{sec:background}
\subsection{Conical Intersections}\label{sec:conical_intersections}

The molecular wavefunction in the adiabatic representation expands in terms of electronic eigenstates $\phi_i(\mathbf{r}; \mathbf{R})$ and nuclear amplitudes $\chi_i(\mathbf{R})$, given as
\begin{equation}\label{eq:adiabatic_expansion}
    \psi(\mathbf{r}, \mathbf{R}) = \sum_i \chi_i(\mathbf{R})\phi_i(\mathbf{r}; \mathbf{R}),
\end{equation}
where $\phi_i(\mathbf{r}; \mathbf{R})$ satisfy $\hat{H}_e(\mathbf{r};\mathbf{R})\phi_i = E_i(\mathbf{R})\phi_i$ at fixed nuclear configuration $\mathbf{R}$~\cite{born1927quantenmechanik}.
Substituting into the full time independent Schr\"{o}dinger equation and projecting onto electronic state $\phi_j$ yields the coupled nuclear equations
\begin{equation}\label{eq:coupled_nuclear}
    \left[-\frac{1}{2\mu}\nabla_R^2 + E_j(\mathbf{R})\right]\chi_j(\mathbf{R}) - \frac{1}{2\mu}\sum_i\left[2\mathbf{F}_{ji}\cdot\nabla_R + G_{ji}\right]\chi_i(\mathbf{R}) = \mathcal{E}\chi_j(\mathbf{R}),
\end{equation}
with first-order (derivative) and second-order nonadiabatic coupling matrix elements are
\begin{equation}\label{eq:nacme}
    \mathbf{F}_{ji}(\mathbf{R}) = \langle \phi_j | \nabla_R \phi_i \rangle, \quad G_{ji}(\mathbf{R}) = \langle \phi_j | \nabla_R^2 \phi_i \rangle.
\end{equation}
The Born-Oppenheimer approximation neglects $\mathbf{F}_{ji}$ and $G_{ji}$ entirely, reducing the problem to uncoupled motion on each surface $E_j(\mathbf{R})$.
This approximation is valid when adiabatic surfaces are well-separated in energy, since first-order coupling can be written via the Hellmann-Feynman theorem
\begin{equation}\label{eq:hellmann_feynman_nacme}
    \mathbf{F}_{ji}(\mathbf{R}) = \frac{\langle \phi_j | \nabla_R \hat{H}_e | \phi_i \rangle}{E_i(\mathbf{R}) - E_j(\mathbf{R})}, \quad i \neq j,
\end{equation}
diverging as $E_i \to E_j$~\cite{baer2006beyond}.
Where these adiabatic surfaces are degenerate defines a conical intersection for dimensions $d\geq2$ (an avoided crossing for $d=1$), with multi-state nuclear dynamics~\cite{teller1937crossing}.

Conical behavior stems from the degeneracy manifold, following from the electronic Hamiltonian. 
For a real symmetric $M \times M$ electronic Hamiltonian parameterized by $d$ nuclear coordinates, exact degeneracy between two states $i$ and $j$ requires simultaneously satisfying two constraints on $d$ parameters~\cite{wigner1929behaviour}:
\begin{equation}\label{eq:degeneracy_conditions}
    H_{ii}(\mathbf{R}) - H_{jj}(\mathbf{R}) = 0, \quad H_{ij}(\mathbf{R}) = 0,
\end{equation}
The intersection seam is formed as a submanifold of dimension $d - 2$.
For diatoms ($d = 1$), two conditions cannot be uniquely satisfied with one free parameter, recovering the `avoided crossing' rule.
Polyatomic molecules ($d \geq 2$) can have a seam with codimension 2 in nuclear configuration space.
Conical topology appears in the branching space, as the two-dimensional orthogonal complement.

\begin{figure}[t]
\centering
\includegraphics[width=0.6\columnwidth]{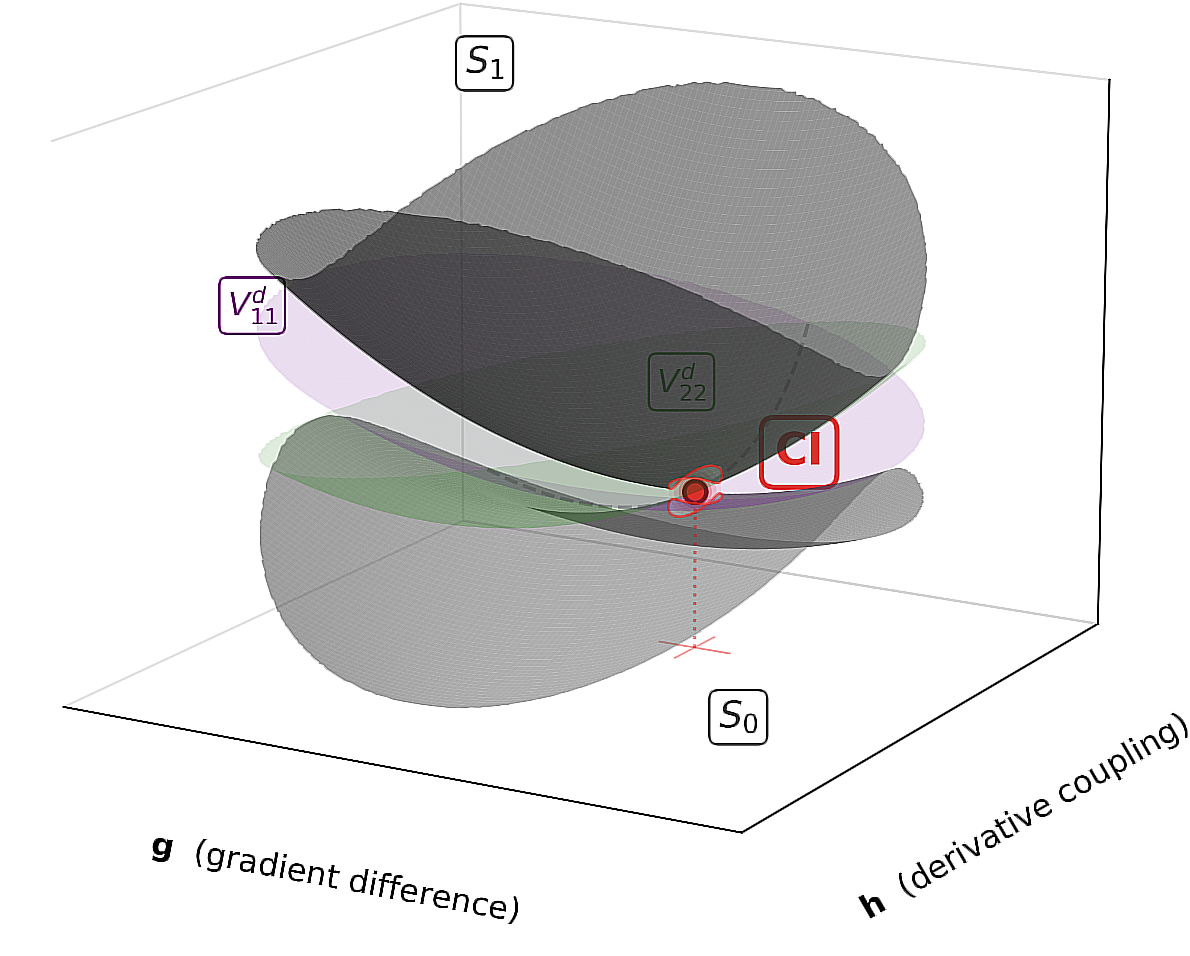}
\caption{\label{fig:Conical_schem}
Schematic for a conical intersection formed by two quadratic potentials. 
In the diabatic representation, the two potentials $V_{00}$ and $V_{11}$ cross and form an intersection seam.
The adiabatic representation $S_0$, $S_1$ has a conical intersection (red dot) in the branching space ($g,h$).}
\end{figure}

To characterize local structure, consider a two-state electronic Hamiltonian in the diabatic representation near a reference point $\mathbf{R}_0$ on the seam.
Expanding to first order in displacement $\delta\mathbf{R} = \mathbf{R} - \mathbf{R}_0$, the $2\times 2$ potential matrix takes the form
\begin{equation}\label{eq:diabatic_expansion}
    \mathbf{V}(\mathbf{R}) = \bar{E}(\mathbf{R})\mathbb{1} + \begin{pmatrix}
        -s & h \\ h & s
    \end{pmatrix},
\end{equation}
with mean energy $\bar{E}(\mathbf{R}) = \frac{1}{2}\text{Tr}[\mathbf{V}]$, $s = \frac{1}{2}(V_{11} - V_{00}) = \mathbf{g}\cdot\delta\mathbf{R}$, and $h = V_{01} = \mathbf{h}\cdot\delta\mathbf{R}$.
The vectors $\mathbf{g} = \frac{1}{2}\nabla_R(E_2 - E_1)$ (gradient difference) and $\mathbf{h} = \nabla_R H_{01}$ (derivative coupling direction) span the branching plane~\cite{yarkony1996diabolical, atchity1991potential}.
Diagonalizing Eq.~\ref{eq:diabatic_expansion} yields adiabatic energies
\begin{equation}\label{eq:cone_energies}
    E_\pm(\mathbf{R}) = \bar{E}(\mathbf{R}) \pm \sqrt{s^2 + h^2} = \bar{E}(\mathbf{R}) \pm \sqrt{(\mathbf{g}\cdot\delta\mathbf{R})^2 + (\mathbf{h}\cdot\delta\mathbf{R})^2}.
\end{equation}
Energy splitting becomes $\Delta E = 2\rho\sqrt{g^2\cos^2\theta + h^2\sin^2\theta}$ in polar coordinates, vanishing linearly in $\rho$ at the apex and giving an elliptical double-cone topology with eccentricity $e = |g^2 - h^2|/(g^2 + h^2)$ (circular when $g=h$)  (Figure~\ref{fig:Conical_schem}).

The adiabatic-to-diabatic transformation connects the two representations by a unitary rotation
\begin{equation}\label{eq:adt}
    \begin{pmatrix} \phi_1^{\text{ad}} \\ \phi_2^{\text{ad}} \end{pmatrix} = \begin{pmatrix} \cos\alpha & -\sin\alpha \\ \sin\alpha & \cos\alpha \end{pmatrix}\begin{pmatrix} \phi_1^{\text{d}} \\ \phi_2^{\text{d}} \end{pmatrix},
\end{equation}
where the mixing angle $\alpha(\mathbf{R}) = \frac{1}{2}\arctan(2h/s) = \frac{1}{2}\arctan(h\sin\theta/(g\cos\theta))$ satisfies $\nabla_R \alpha = F_{12}$, following Eq.~\ref{eq:nacme})~\cite{baer1975adiabatic, mead1982conditions}.
Derivative coupling is replaced by smooth off-diagonal potential elements $V_{ij}(\mathbf{R})$, being finite at the conical intersection.
The mixing angle can take multiple values, since traversing a closed loop $\mathcal{C}$ encircling the seam in the branching plane results in $\theta$ advancing by $2\pi$ while $\alpha$ advances by only $\pi$.
Adiabatic wavefunctions acquire a sign change
\begin{equation}\label{eq:sign_change}
    \phi_i^{\text{ad}}(\mathbf{R}) \xrightarrow{\mathcal{C}} -\phi_i^{\text{ad}}(\mathbf{R}),
\end{equation}
first recognized by Herzberg and Longuet-Higgins~\cite{herzberg1963electronic, longuet1975intersection}, accumulating a `Berry' phase $\gamma = \oint_\mathcal{C} F_{12}\cdot d\mathbf{R} = \pi$ for any loop enclosing a single conical intersection~\cite{berry1984quantal, mead1979determination}.
Quantum numbers are subsequently shifted by half-integers, with altered selection rules near the intersection~\cite{mead1980molecular, juanes2005theoretical}.
In the diabatic representation, geometric phase is automatically incorporated through smooth coupling $V_{ij}(\mathbf{R})$~\cite{baer2006beyond, domcke2004conical}.
For systems with $d \geq 3$ internal coordinates, the intersection seam becomes a $(d-2)$-dimensional manifold, with seams between multiple electronic states constraining accessible relaxation pathways~\cite{yarkony2001nuclear, matsika2011nonadiabatic}.

Conical intersections control ultrafast non-radiative transitions in polyatomic molecules.
A photoexcited nuclear wavepacket propagates toward the conical intersection, where divergent nonadiabatic coupling induces population transfer to the lower surface on timescales of tens to hundreds of femtoseconds~\cite{kasha1950characterization, domcke2004conical, domcke2012role}.
Branching ratios between product channels are determined by conical intersection topography (peaked vs. sloped), with seam dimension relative to a space constructed by vibrational modes~\cite{atchity1991potential, yarkony2005statistical}.
In photochemistry, conical intersections govern common interactions, such as retinal isomerization in rhodopsin~\cite{polli2010conical, garavelli1997relaxation}, photostability of DNA nucleobases~\cite{matsika2004radiationless, perun2005ab}, and branching in ring-opening reactions~\cite{robb2000computational, levine2007isomerization}, making Berry phase accumulation a valuable heuristic during numerical validation of a quantum algorithm. 

For systems described by a vibronic Hamiltonian with $d$ nuclear degrees of freedom and $M$ coupled electronic states, conical intersection seam structure and the functional form of the off-diagonal diabatic couplings $V_{ij}(\mathbf{x})$ dictate computational complexity of faithful simulation.
The second-order vibronic Hamiltonian of K\"{o}ppel, Domcke, and Cederbaum~\cite{kouppel1984multimode, cederbaum1981multimode} captures simple conical topology through linear and quadratic coupling terms, discussed in the following section.
Multivariate and mode-mixing terms are essential for physically correct relaxation dynamics consistent with Kasha's rule and are dominant contribution to quantum gate counts~\cite{kasha1950characterization}.
It is these terms that make tensor decomposition/hypercontraction infeasible in the limit of strong correlation and low symmetry (Chapter~\ref{ch:wavepacket_prop}).
A quantum algorithm aiming to efficiently simulate NA-QMD must be made efficient in the physically complete extension that can handle real chemistry. 
Here, a greater focus is made to develop an accurate, flexible protocol for simulation, leaving the majority of circuit approximation to future work.

\subsection{Vibronic Dynamics}

Photochemistry provides a wealth of example systems with ultrafast dynamics mediated by conical intersections, directing a first focus to vibronic systems, where electronic and vibrational degrees of freedom are strongly coupled.
K\"oppel, Domcke, and Cederbaum give an expansion of the vibronic Hamiltonian for nonadiabatic dynamics~\cite{kouppel1984multimode, domcke2004conical, cederbaum1981multimode}.
Expanding around the equilibrium geometry, a second-order truncation is typically used (Eq.~\ref{eq:vibronic_ham_full}), with quartic force fields more recently incorporated, improving accuracy with increased available computational resources~\cite{fortenberry2025quantum}.

In diabatic form, the vibronic Hamiltonian for $M$ electronic states and $d$ vibrational modes is written as an expansion around the ground-state equilibrium geometry $(x_i - x_i^{\text{eq}})\mapsto x_i$:
\begin{equation}\label{eq:vibronic_ham_full}
\begin{split}
    \hat{H} &= \sum_i^d \frac{1}{2\mu_i}\left(-\frac{\partial^2}{\partial x_i^2} + \mu_i^2\omega_i^2x_i^2\right)\mathbf{1}_M + \mathbf{E}^{(0)} \\
    &+ \sum_{i \in \mathcal{G}_1}\boldsymbol{a}_i\, x_i + \sum_{(i,j) \in \mathcal{G}_2}\boldsymbol{b}_{ij}\, x_i x_j \\
    &+ \sum_{i \in \mathcal{G}_3}\boldsymbol{c}_i\, x_i + \sum_{(i,j) \in \mathcal{G}_4}\boldsymbol{d}_{ij}\, x_i x_j,
\end{split}
\end{equation}
where $\mathbf{1}_M$ is the $M\times M$ identity in electronic space and $\mathbf{E}^{(0)} = \text{diag}(E_1^{(0)}, \ldots, E_M^{(0)})$ is the matrix of vertical state energies at the reference geometry. 
Index sets $\{\mathcal{G}_{1,2,3,4}\}$ define symmetry-allowed terms, associated to the irreducible representations (irreps) of the molecule being studied. 
These terms are critical for resource estimation and operator parallelization and are a significant focus in this chapter and the following resource estimation chapter.

Ultrafast dynamics like Kasha's rule appear since the rate of internal conversion between upper excited states ($S_n \to S_{n-1}$) is orders of magnitude faster ($10^{-11}$\,\text{s}--$10^{-14}\,\text{s}$) than radiative decay ($10^{-9}\,\text{s}$)~\cite{kasha1950characterization, englman1970energy}.
When a molecule passes through a conical intersection, electronic energy is dumped into specific ``tuning" and ``coupling" vibrational modes. 
If that energy stays in those modes, the wavepacket can coherently oscillate back up the potential surface (recurrence), slowing down the effective relaxation.
To be irreversible and fast (Kasha-compliant), that energy must be rapidly spread into the bath of other vibrational modes.
Bilinear terms are such an algebraic route, representing mode mixing~\cite{yarkony1996diabolical, domcke2012role}. 
As Duschinsky rotations in a basis transformation context, bilinear terms create non-separability between different vibrational degrees of freedom, allowing stretches in mode $x_a$ to modulate the effective potential of mode $x_b$~\cite{duschinsky1937importance, small1971herzberg}.
Energy transfers from active coupling modes to dark bath modes, requiring efficient handling for quantum algorithms that try to have a meaningful advantage over classical methods. 

\subsection{Intramolecular Vibrational Energy Redistribution \\ and Mode Coupling}\label{sec:ivr}

To understand the vibronic Hamiltonian, the topic of intramolecular vibrational energy redistribution (IVR) is imperative.
The vibronic Hamiltonian truncated at linear order in mode coordinates (the linear vibronic coupling aka LVC model) treats each vibrational mode as independent displaced harmonic oscillators on each diabatic surface.
In this approximation, the potential matrix for $M$ electronic states and $d$ vibrational modes takes the separable form
\begin{equation}\label{eq:lvc_hamiltonian}
    V_{ij}^{\text{LVC}}(\mathbf{x}) = \delta_{ij}\left(E_i^{(0)} + \sum_r^d a_r^{(i)} x_r\right) + (1 - \delta_{ij})\sum_r^d c_r^{(ij)} x_r,
\end{equation}
where $a_r^{(i)}$ are on-diagonal linear coupling constants and $c_r^{(ij)}$ are off-diagonal vibronic coupling constants mediating population transfer between states $i$ and $j$.
Separability of Equation ~\ref{eq:lvc_hamiltonian} in the mode index $r$ leads to an irreversible energy deposition into a given vibrational mode, given by passage through the conical intersection.
Without inter-mode coupling, the nuclear wavepacket oscillates coherently along tuning and coupling coordinates, producing persistent recurrences in the autocorrelation function and periodic repopulation of the upper diabatic surface, being physically incorrect for molecules with more than a few active modes~\cite{nesbitt1996vibrational, gruebele2004quantum}.

IVR arises from anharmonic and inter-mode coupling terms that break Hamiltonian separability.
Leading contributions for vibronic Hamiltonians come from second-order (quadratic and bilinear) terms in the Taylor expansion of the diabatic potential matrix around the reference geometry $\mathbf{x}_0$:
\begin{equation}\label{eq:second_order_potential}
    V_{ij}(\mathbf{x}) = V_{ij}^{\text{LVC}}(\mathbf{x}) + \frac{1}{2}\sum_{r,s}^d V_{rs}^{(ij)} x_r x_s,
\end{equation}
where $V_{rs}^{(ij)} = (\partial^2 V_{ij}/\partial x_r \partial x_s) |_{\mathbf{x}_0}$ are the second-order coupling constants.
On the diagonal of the electronic matrix ($i = j$), $V_{rs}^{(ii)}$ reduces to the on-diagonal coefficients $a_{rs}^{(i)}$ of Eq.~\ref{eq:vibronic_ham_full}; on the off-diagonal ($i \neq j$), it reduces to the bilinear vibronic couplings $c_{rs}^{(ij)}$.
State-dependent frequency shifts $\omega_r^{(i)} \neq \omega_r^{(j)}$ occur when $r = s$, modifying curvature of each diabatic surface along mode $r$, while distinct vibrational degrees of freedom couple for $r \neq s$, enabling energy flow between modes and driving IVR.
Bilinear terms $a_{rs}^{(i)} x_r x_s$ on the electronic matrix diagonal produce state-dependent mode mixing (Duschinsky rotation) on each surface; terms $c_{rs}^{(ij)} x_r x_s$ (off-diagonal) create position-dependent vibronic coupling strengths depending on two mode coordinates.

A chemical perspective highlights the importance of these terms in an chemical dynamics algorithm, with bilinear couplings understood through Duschinsky rotations~\cite{duschinsky1937importance}.
The normal modes $\{x_r\}$ of a molecule are defined at a specific nuclear geometry, typically the ground-state equilibrium.
Equilibrium geometry shifts upon electronic excitation, changing the Hessian ($\partial^2 E/\partial R_i \partial R_j$) and rotating and displacing the normal modes of excited state $\{x_r'\}$
\begin{equation}\label{eq:duschinsky}
    x_r' = \sum_s J_{rs} x_s + \Delta_r,
\end{equation}
with Duschinsky rotation matrix $\mathbf{J}$ and displacement $\boldsymbol{\Delta}$~\cite{duschinsky1937importance, small1971herzberg}.
The matrix $\mathbf{J}$ is orthogonal but generally not diagonal, with off-diagonal elements quantifying projection of the normal mode $x_s$ of the ground state onto mode $x_r'$ of the excited state.
Expressing excited-state potentials in terms of ground-state normal modes generates bilinear cross terms:
\begin{equation}\label{eq:duschinsky_potential}
    V'(\mathbf{x}) = \frac{1}{2}\sum_r \omega_r'^2\left(\sum_s J_{rs}x_s + \Delta_r\right)^2 = \frac{1}{2}\sum_{r,s,t}\omega_r'^2 J_{rs}J_{rt}x_sx_t + \cdots
\end{equation}
The terms $x_sx_t$ with $s \neq t$ appear whenever $\mathbf{J}$ has nonzero off-diagonal elements, i.e., whenever the excited-state normal modes are rotated with respect to the ground state.
Molecules near a conical intersection have significant Duschinsky mixing, as intersection geometry is displaced from both equilibrium structures and potential curvature changes rapidly in the branching space~\cite{yarkony1996diabolical, ichino2009quasidiabatic}.
A diabatic representation absorbs the mode rotation in a second-order coupling constant $c_{rs}^{(ij)}$.

Vibrational excitations initially localized in given mode are redistributed among modes connected by $J_{rs}\neq0$ elements.
Consider as an example a wavepacket prepared by vertical excitation onto diabatic surface $\ket{s_2}$, localized in modes $x_t$ (tuning) and $x_c$ (coupling).
In the LVC model, the subsequent evolution factorizes, with $x_{tune}$ oscillating with period $2\pi/\omega_{tune}$, $x_{coup}$ mediates periodic population transfer at the intersection, and all other modes $x_{bath}$ remain in their ground vibrational state.
Including bilinear terms $a_{\text{tune,bath}}^{(i)}x_{\text{tune}}x_{\text{bath}}$ introduces a coupling with an effective strength proportional to the instantaneous displacement $\langle x_{\text{tune}}(t)\rangle$.
Bilinear terms act as time-dependent driving forces on each bath mode with amplitudes set by the displacement of active modes.
The tuning mode excites coupled bath modes, exciting further modes through their own bilinear couplings, continuing until energy is distributed statistically among all accessible modes~\cite{nesbitt1996vibrational, gruebele2004quantum}.
Rate of transfer depends on the density of bath modes near resonance with active modes and on the magnitude of the coupling constants, both determined by the Duschinsky matrix and mode frequencies.

For accurate nonadiabatic dynamics beyond $\sim$50~fs, inclusion of bilinear terms is essential~\cite{raab1999molecular, worth2004beyond}.
Raab et al.\cite{raab1999molecular} demonstrated that the four-mode LVC model of pyrazine, while reproducing the short-time population transfer through the $S_2/S_1$ conical intersection, fails to capture the irreversible decay of the $S_2$ population observed in the full 24-mode calculation and in experiment.
Adding bilinear couplings between active modes ($\nu_{6a}$, $\nu_{10a}$) and bath modes ($\nu_1$, $\nu_{9a}$, etc.) quantitatively recovers the correct relaxation timescale and suppresses the artificial recurrences in reduced-dimensional models.
This supports the notion that a simulation aspiring to recover correct physics (Kasha-compliant relaxation, correct branching ratios) must include these terms, as they provide the only algebraic route for energy dissipation from the conical-intersection's active subspace into the full vibrational manifold.

\subsection{Recurrence in Nonadiabatic Wavepacket Dynamics}\label{sec:recurrence}

The propagated wavefunction $\ket{\psi(t)} = e^{-iHt}\ket{\psi(0)}$ can be expanded in the eigenbasis $\{\ket{E_\alpha}\}$ of $H$. 
With $c_\alpha = \braket{E_\alpha}{\psi(0)}$, the autocorrelation function takes the form
\begin{equation}\label{eq:autocorr_expansion}
    C(t) = \braket{\psi(0)}{\psi(t)} = \sum_\alpha |c_\alpha|^2 e^{-iE_\alpha t}.
\end{equation}
This is a quasiperiodic function whose Fourier transform yields the spectral density $I(E) = \sum_\alpha |c_\alpha|^2 \delta(E - E_\alpha)$, recovering the absorption spectrum in the limit of complete time evolution.
Recurrence manifests as a revival in $|C(t)|^2$ when the relative phases $e^{-i(E_\alpha - E_\beta)t}$ between significantly populated eigenstates rephase constructively.
For a wavepacket dominated by a set of nearly equally spaced levels with mean spacing $\Delta E$, partial revivals occur at multiples of $t_{\text{rec}} \approx 2\pi/\Delta E$, producing a series of peaks in $|C(t)|$ that decay in amplitude due to anharmonicity and spectral broadening~\cite{robinett2004quantum, heller1981semiclassical}.
Strong recurrence in the frequency domain corresponds to a well-resolved comb of spectral peaks, while rapid decoherence from coupling to a dense bath of modes produces broad, featureless absorption bands.
An exponential damping envelope $|C(t)| \sim e^{-\Gamma t}$ corresponds to a Lorentzian linewidth $\Gamma$, providing a direct connection between the temporal decay of coherence and spectral broadening observed in ultrafast experiments~\cite{heller1981semiclassical, tannor2007introduction}.

Diabatic populations of surface $\ket{s_j}$ are defined as
\begin{equation}\label{eq:diabatic_population}
    P_j(t) = \int |\braket{s_j}{\psi(\mathbf{x},t)}|^2 \, d\mathbf{x} = \text{Tr}_{\text{nuc}}\left[\braket{s_j}{\hat{\rho}(t)}{s_j}\right],
\end{equation}
tracing over all nuclear degrees of freedom, with $\hat{\rho}(t) = \ket{\psi(t)}\bra{\psi(t)}$ being the full density operator.
Population transfer through a conical intersection produces characteristic oscillation in $P_j(t)$, where the wavepacket bifurcates after an initial passage through the conical intersection~\cite{domcke2004conical, worth2004beyond}.
Recurrence in population dynamics occurs when the wavepacket on either surface returns to the conical intersection with sufficient momentum to undergo a second nonadiabatic transition.
With $d$ active modes, recurrence time is governed by the tuning coordinate's vibrational period, $t_{\text{rec}} \approx 2\pi/\omega_{\text{tune}}$, acting as damped oscillations in $P_j(t)$ superimposed on a decay toward thermal equilibrium~\cite{kouppel1984multimode, stock1995resonance}.
Each recurrence event contributes a secondary peak in the autocorrelation whose amplitude reflects the overlap between the returning wavepacket and the initial state.
Phase accumulated along the closed trajectory determines whether successive recurrences interfere constructively or destructively~\cite{berry1984quantal, ryabinkin2017geometric}.
Irreversible population transfer occurs with rapid IVR~\cite{domcke2012role, worth2004beyond}, with timescales for the initial ballistic motion through the conical intersection ($\sim 10$--$50$~fs) and IVR-driven thermalization ($\sim 100$~fs--$1$~ps)  observable in both $|C(t)|$ and $P_j(t)$~\cite{raab1999molecular, worth2004beyond}.
Spectra and populations are sensitive to recurrence, and are used as the primary observables for numerical validations in this dissertation, extractable by repeated measurement of one qubit for $C(t)$ and the channel register for $M$ populations.
With features of the vibronic Hamiltonian and vibronic dynamics specified, the algorithmic contribution central to this dissertation can now be constructed and validated for vibronic systems. 
Later chapters will extend beyond the harmonic approximation, though vibronic systems are the only systems for which a reasonable resource comparison can be made~\cite{motlagh2025quantum}.

\section{Hamiltonian Decomposition}\label{sec:algorithm}
This section details the full decomposition used in REQWIEM. 
The Hamiltonian is decomposed into kinetic, on-diagonal, and off-diagonal potential terms $H=T+V_\text{d}+V_{\text{od}}$, then further decomposed based on gate type and implementation.
A full depth analysis is found in Chapter~\ref{ch:resources}, as the use of a large-scale parameterized quantum circuit requires rotation synthesis.

\subsection{Multi-Mode Systems}\label{sec:multimode}

Following Eq.~\ref{eq:vibronic_ham_full} for the vibronic Hamiltonian, the size of the channel register is given by $m=\lceil \log_2 M\rceil$, storing electronic channel populations as basis states.

Though frequently given in mass-weighted dimensionless coordinates, reduced masses $\mu_i$ for each vibrational mode are included in the statevector simulations performed in this dissertation.
The coefficient matrices are diagonal or off-diagonal in the electronic index:
\begin{equation}
    (\boldsymbol{a}_i)_{ss'} = a_i^{(s)}\delta_{ss'}, \quad 
    (\boldsymbol{a}_{ij})_{ss'} = a_{ij}^{(s)}\delta_{ss'}, \quad
    (\boldsymbol{c}_i)_{ss'} = c_i^{(ss')}(1-\delta_{ss'}), \quad
    (\boldsymbol{c}_{ij})_{ss'} = c_{ij}^{(ss')}(1-\delta_{ss'}).
\end{equation}

Index sets $\mathcal{G}_1$--$\mathcal{G}_4$ are determined by the point group $\mathcal{G}$ of the molecule and the irreducible representations (irreps) $\Gamma_s$ of the electronic states.
A term contributes to the Hamiltonian only if the direct product of all symmetry labels contains the totally symmetric representation $\Gamma_{\text{sym}}$; terms violating this condition vanish identically by the Wigner--Eckart theorem.
The four sets are:
\begin{align}
    \mathcal{G}_1 &= \left\{i \;\middle|\; \Gamma(x_i) = \Gamma_{\text{sym}}\right\}, \label{eq:G1}\\
    \mathcal{G}_2 &= \left\{(i,j) \;\middle|\; \Gamma(x_i) \otimes \Gamma(x_j) = \Gamma_{\text{sym}},\; i \leq j\right\}, \label{eq:G2}\\
    \mathcal{G}_3 &= \left\{i \;\middle|\; \exists\, s\neq s': \Gamma_s \otimes \Gamma(x_i) \otimes \Gamma_{s'} \supset \Gamma_{\text{sym}}\right\}, \label{eq:G3}\\
    \mathcal{G}_4 &= \left\{(i,j) \;\middle|\; \exists\, s\neq s': \Gamma_s \otimes \Gamma(x_i) \otimes \Gamma(x_j) \otimes \Gamma_{s'} \supset \Gamma_{\text{sym}},\; i < j\right\}. \label{eq:G4}
\end{align}

The intra-mode quadratic subset is distinguished as $\mathcal{G}_2^{=} = \{(r,r)\} $ (always $d$ elements, contributing
  $x_r^2$ terms) from the bilinear subset $\mathcal{G}_2^{<} = \{(r,s) \in \mathcal{G}_2 : r < s\}$ (contributing $x_rx_s$ cross-mode terms).
  Gate count formulas reference $|\mathcal{G}_2^{<}|$ for the bilinear pairs, equaling $\binom{d}{2}=d(d-1)/2$ in the lowest symmetry ($C_1$).
  
$\mathcal{G}_2$ includes the diagonal quadratic terms $i=j$, where $\Gamma(x_i^2) = \Gamma_{\text{sym}}$ holds unconditionally in non-degenerate point groups (Section~\ref{sec:symmetry}), as well as the bilinear on-diagonal terms $i \neq j$ restricted to same-symmetry mode pairs.
The set $\mathcal{G}_1$ contains only totally symmetric modes, acting as tuning modes displacing the equilibrium along each diabatic surface.
$\mathcal{G}_3$ contains modes whose symmetry matches the electronic product $\Gamma_s \otimes \Gamma_{s'}$ for at least one pair of states, acting as coupling modes that mediate population transfer at conical intersections.
The set $\mathcal{G}_4$ contains mode pairs whose combined symmetry matches the electronic product for at least one off-diagonal block, acting as bilinear vibronic couplings (BVC).

Pyrazine is a prototypical example. 
This molecule ($D_{2h}$, $S_1 \sim B_{3u}$, $S_2 \sim B_{2u}$) has a particularly sparse form, and consequently is well characterized by MCTDH simulation~\cite{raab1999molecular}.
$\mathcal{G}_1$ contains the 5 modes of $A_g$ symmetry.
$\mathcal{G}_2$ contains all 24 diagonal quadratic terms plus the bilinear pairs with $\Gamma(x_i)\otimes\Gamma(x_j) = A_g$.
$\mathcal{G}_3$ contains modes satisfying $\Gamma(x_i) = B_{3u}\otimes B_{2u} = B_{1g}$; this is the single mode $\nu_{10a}$ of $B_{1g}$ symmetry, making $|\mathcal{G}_3|=1$~\cite{innes1988electronic}.
$\mathcal{G}_4$ contains mode pairs satisfying $\Gamma(x_i)\otimes\Gamma(x_j) = B_{1g}$, selecting complementary-symmetry pairs; $|\mathcal{G}_4| = 29$ for the 24-mode model.
The complete quantitative analysis of symmetry-allowed terms for all systems used as benchmarks is given in Appendix~\ref{app:symmetry_appendix}.

\subsection{Kinetic Energy}\label{sec:kinetic}

The kinetic operator $T = \sum_{r=0}^{d-1}p_r^2 / 2\mu_r$ is diagonal in momentum space.
Each mode's contribution is implemented as in Chapter~\ref{ch:wavepacket_prop}, where
\begin{enumerate}
    \item cQFT on the $n_r$-qubit position register $\ket{x_r}$, transforming to momentum space.
    \item Binary-decomposed phase kicks for $p_r^2/2\mu_r$: $n_r$ single-qubit $R_z$ gates (from $q_\ell^2 = q_\ell$) plus $\binom{n_r}{2}$ two-qubit $\mathrm{CR}_z$ gates (cross terms $q_\ell q_{\ell'}$), with angles $\theta_{\ell} = \tau\Delta_{p_r}^2 \cdot 2^{2\ell}/2\mu_r$ and $\theta_{\ell \ell'} = \tau\Delta_{p_r}^2\cdot 2^{\ell+\ell'+1}/2\mu_r$.
    \item cIQFT to return to position space.
\end{enumerate}

The kinetic term involves no channel register interaction, being identical on all electronic surfaces (the $\mathbf{1}_M$ prefactor in Eq.~\ref{eq:vibronic_ham_full}).
All $d$ modes execute in parallel, giving kinetic depth $\mathcal{O}(n^2)$ independent of $d$.
In the second-order Trotter decomposition (Chapter~\ref{ch:wavepacket_prop}), the first kinetic half-step's QFT can be absorbed into the initial state preparation shift operator, saving $2d$ QFT applications over the full simulation (Chapter~\ref{ch:wavepacket_prop}).

  Per kinetic evaluation (all $d$ modes in parallel):
  \begin{align}
      R_z^{(\text{kin})} &= \frac{dn(n+1)}{2}, &
      C_{\text{CNOT}}^{(\text{kin})} &= dn(n-1), &
      T^{(\text{kin})} &= 0.
  \end{align}
  The associated cQFT/cIQFT pair adds $dn(n{-}1)/2$ rotations and
  $d[n(n{-}1) + 3\lfloor n/2 \rfloor]$ CNOTs per invocation.
  Over $k$ Trotter steps with boundary merging, there are $k{+}1$
  kinetic evaluations and $k{+}1$ cQFT/cIQFT pairs.

\subsection{On-Diagonal Potential: UCR-Multiplexed Protocol}\label{sec:mux_alg}

The on-diagonal potential $V_\text{d} = \sum_{s=0}^{M-1}|s\rangle\langle s|\otimes V_{ss}(\mathbf{x})$ applies $M$ distinct surface operators conditioned on the electronic state.
The protocol assigns each class of potential term its own decomposition, using quantum multiplexing for single-qubit terms (constant, linear, and diagonal-quadratic contributions) and uniformly controlled rotations for multi-qubit terms (intra-mode quadratic cross-terms and inter-mode bilinear terms).
The channel register serves exclusively as a read-only control, enabling fan-out (Section~\ref{sec:fanout}) to $d$ parallel copies so that all vibrational modes execute simultaneously.

The reference surface's multi-qubit contributions fold directly into the UCR decomposition of Tiers~2 and~3 (see below).
Since a Gray-code uniformly controlled rotation (UCR) operation UCR$_Z^{(m)}$ produces exactly $M$ rotations and $M - 1$ CNOTs regardless of the angle vector, encoding the full surface-dependent angle $\theta_j$ (including $j = 0$) costs no more than encoding the correction angle $\Delta\theta_j = \theta_j - \theta_0$ with $\theta_0 = 0$.
This eliminates the need for a separate unconditional application of the reference surface's cross-terms, saving $d\binom{n}{2}$ rotations and $dn(n{-}1)$ CNOTs per half-step from intra-mode terms alone, with analogous savings for bilinear terms.

\subsubsection{Protocol Overview}\label{sec:mux_protocol}

The on-diagonal block proceeds in seven phases, summarized here.

\begin{enumerate}
    \item \textbf{Reference surface (single-qubit terms).}
    Apply the single-qubit contributions of the reference surface $V_{00}(\mathbf{x})$ unconditionally on the vibrational register: the constant energy $E_0^{(0)}$, linear gradients, and diagonal part of the quadratic curvature.
    No channel register interaction is required; all $d$ modes execute in parallel.
    Multi-qubit cross-terms from $V_{00}$ are {not} applied here; they are absorbed into the Tier~2 and Tier~3 UCR circuits.

    \item \textbf{Truth-value computation.}
    For each correction surface $s=1,\ldots,M-1$ with $\mathrm{popcount}(s)\geq 2$, compute a joint truth value $b_s = \bigwedge_{k\in S_s}s_k$ into a dedicated work ancilla via a Toffoli cascade.
    Surfaces with $\mathrm{popcount}(s)=1$ require no truth value; the corresponding channel qubit $s_k$ serves directly.

    \item \textbf{Fan-out.}
    Fan out the $m$ channel register qubits and the $(M-1-m)$ truth-value ancillas to $d$ copies via binary CNOT trees, producing a total of $(M-1)(d-1)$ entangled copies.
    Each vibrational mode receives a local set of $M-1$ control qubits.

    \item \textbf{Tier~1: single-qubit corrections} (quantum multiplexing, parallel across modes).
    For each correction surface $s = 1,\ldots,M-1$, apply single-qubit angle corrections as $\mathrm{CR}_z$ gates controlled on the local copy of the truth value or channel qubit. 
    This covers constant energy offsets $\Delta E_s$, linear gradient differences $\Delta\alpha_s^{(r)}$, and diagonal-quadratic curvature differences 

    \item \textbf{Tier~2: intra-mode quadratic cross-terms} ($\mathrm{UCR}_Z^{(m)}$).
    For each qubit pair $(\ell,\ell')$ within mode $r$, apply a CNOT sandwich enclosing a uniformly controlled rotation $\mathrm{UCR}_Z^{(m)}$ with the $m$ channel register qubits as controls, encoding the {full} surface-dependent curvature (shared harmonic plus surface correction) for all $M$ surfaces at that site.

    \item \textbf{Tier~3: inter-mode bilinear terms} ($\mathrm{UCR}_Z^{(m)}$, chromatic scheduling).
    For each qubit pair $(q_\ell^{(r)}, q_p^{(s)})$ across modes $r\neq s$, apply the same CNOT--UCR--CNOT structure using the channel register, encoding the {full} bilinear coupling $a_j^{(rs)}$ for all $M$ surfaces.
    Mode pairs with disjoint register sets execute in parallel; the schedule is determined by the chromatic number $\chi(\mathcal{G}_2^<)$ of the mode-pair conflict graph (Section~\ref{sec:chromatic_index}).

    \item \textbf{Fan-in and uncomputation.}
    Reverse the fan-out CNOT trees (Phase~3), then reverse the Toffoli cascades (Phase~2) to restore all ancillas to $|0\rangle$.
\end{enumerate}

\subsubsection{Phase~1: Reference Surface}\label{sec:mux_ref}

The reference surface $V_{00}(\mathbf{x})$ contributes single-qubit terms that are applied unconditionally.
After binary expansion of each position operator $x_r = \Delta_r\sum_{\ell=0}^{n-1}2^\ell q_\ell^{(r)}$, the constant energy $E_0^{(0)}$, linear gradient $\alpha_0^{(r)}x_r$, and diagonal part of the quadratic curvature $(\tfrac{1}{2}\mu_r\omega_r^2 + a_0^{(r)})\cdot 2^{2\ell}\Delta_r^2\,q_\ell^{(r)}$ (using $q_\ell^2 = q_\ell$) combine into a single $R_z$ gate per qubit with angle
\begin{equation}
    \varphi_{0,\ell}^{(r)} = \tau\bigl[\alpha_0^{(r)}\,2^\ell\,\Delta_r + \bigl(\tfrac{1}{2}\mu_r\omega_r^2 + a_0^{(r)}\bigr)\,2^{2\ell}\,\Delta_r^2\bigr],
\end{equation}
plus one global-phase $R_z$ per mode absorbing the constant $E_0^{(0)}/d$.

No multi-qubit gates appear in Phase~1; the intra-mode cross-terms and bilinear terms of the reference surface are absorbed into the Tier~2 and Tier~3 UCR circuits (Sections~\ref{sec:mux_tier2}--\ref{sec:mux_tier3}), where the $j = 0$ UCR angle carries the full reference curvature rather than zero.
The Phase~1 gate count is therefore
\begin{equation}\label{eq:V00_gates}
    \text{Phase~1:}\quad
    d(1 + n)\;\text{rotations},\quad
    0\;\text{CNOTs},\quad
    0\;\text{Toffoli}.
\end{equation}

\subsubsection{Phases~2--3: Truth-Value Computation and Fan-Out}\label{sec:mux_fanout}

$M-1$ correction surfaces are indexed by nonzero $m$-bit strings $s$.
Of these, $\binom{m}{1}=m$ have $\mathrm{popcount}(s)=1$ and use the corresponding channel qubit $s_k$ directly as a single-bit control.
The remaining
\begin{equation}
    N_{\mathrm{tv}} = \sum_{\alpha=2}^{m}\binom{m}{\alpha} = M - 1 - m,
\end{equation}
surfaces require a Toffoli cascade to compute $b_s = \bigwedge_{k\in S_s}s_k$ into a dedicated work ancilla.
A surface with $\mathrm{popcount}(s) = \alpha$ requires $\alpha - 1$ Toffoli gates in a sequential cascade.
Cascades for distinct surfaces use disjoint ancilla targets and can overlap where their channel-qubit inputs are compatible.
The total Toffoli count for computing and later uncomputing all truth values is
\begin{equation}\label{eq:mux_toffoli_overhead}
    T_{\mathrm{truth}} = 2\sum_{\alpha=2}^{m}\binom{m}{\alpha}(\alpha - 1) = M(m-2) + 2 = \mathcal{O}(M\log M),
\end{equation}
with maximum cascade depth $m-1$.

After truth-value computation, $m$ channel register qubits and $N_{\mathrm{tv}} = M-1-m$ truth-value ancillas are fanned out to $d$ copies via binary CNOT trees.
Each of the $M-1$ source qubits is copied to $d-1$ fresh ancillas, requiring one CNOT per copy qubit per tree level.
Since source qubits are disjoint, all $M-1$ trees execute in parallel.

\begin{equation}\label{eq:fanout_gates}
    \text{Phases~2--3:}\quad
    T_{\mathrm{truth}} \text{ Toffoli},\quad
    (M-1)(d-1) \text{ CNOTs}.
\end{equation}

Fan-out produces entangled GHZ-like copies, not independent clones.
Each vibrational mode~$r$ now possesses a local control set $\mathcal{C}^{(r)} = \{s_0^{(r)},\ldots,s_{m-1}^{(r)},b_{s_1}^{(r)},\ldots\}$ of $M-1$ qubits, enabling fully parallel execution across modes.

The ancilla overhead for the on-diagonal block is $(M-1-m)$ truth-value qubits plus $(M-1)(d-1)$ fan-out copies, totaling $(M-1)d - m$ ancillas.
These are reused from the off-diagonal fan-out pool, which requires at most $m(d-1)$ copies; since $M-1 \geq m$, the on-diagonal allocation subsumes the off-diagonal requirement.

\subsubsection{Tier~1: Constant and Linear Corrections}\label{sec:mux_tier1}

The single-qubit correction for each correction surface $s = 1,\ldots,M-1$ is
\begin{equation}
    \Delta V_s^{(\mathrm{sq})}(\mathbf{x}) = \Delta E_s + \sum_{r=1}^{d}\sum_{\ell=0}^{n-1}\Delta\varphi_{s,\ell}^{(r)}\;q_\ell^{(r)},
\end{equation}
where $\Delta E_s = E_s^{(0)} - E_0^{(0)}$ is the vertical energy offset and
\begin{equation}
    \Delta\varphi_{s,\ell}^{(r)} = \tau\bigl[(\alpha_s^{(r)} - \alpha_0^{(r)})\,2^\ell\,\Delta_r + (a_s^{(r)} - a_0^{(r)})\,2^{2\ell}\,\Delta_r^2\bigr],
\end{equation}
encodes combined linear and diagonal-quadratic difference for qubit $\ell$ in mode~$r$.
This absorbs the diagonal part of any surface-dependent curvature correction into the same gate as the linear gradient, avoiding a separate quadratic tier for single-qubit terms.

Each correction is controlled on a single local qubit: the channel copy $s_k^{(r)}$ if $\mathrm{popcount}(s) = 1$, or the truth-value copy $b_s^{(r)}$ otherwise.
All corrections are applied as controlled-$R_z$ gates, each decomposed via the parity identity:
\begin{equation}
    \mathrm{CR}_z(\theta)_{c\to t} = \mathrm{CNOT}_{c\to t}\;\cdot\;R_z(\theta)_t\;\cdot\;\mathrm{CNOT}_{c\to t},
\end{equation}
requiring 1 rotation and 2 CNOTs per gate.
The constant offset $\Delta E_s$ is implemented as a $\mathrm{CR}_z(\tau\,\Delta E_s/d)$ with the correction distributed equally across $d$ modes, absorbed into the same gate site as the per-qubit linear correction by targeting a designated qubit $\ell_0$ per mode.

Each correction uses a different control qubit but acts on the same vibrational targets, leading corrections within a mode to serialize across the $M-1$ surfaces.
However, all $d$ modes execute their respective correction chains in parallel using local control copies.

Per correction surface, per mode: $(1 + n)$ rotations and $2(1 + n)$ CNOTs.
\begin{equation}\label{eq:tier1_gates}
    \text{Tier~1:}\quad
    d(M-1)(1 + n)\;\text{rotations},\quad
    2d(M-1)(1+n)\;\text{CNOTs},\quad
    0\;\text{Toffoli}.
\end{equation}

\subsubsection{Tier~2: Intra-Mode Quadratic Cross-Terms}\label{sec:mux_tier2}

Surface-dependent quadratic curvature includes the shared harmonic contribution and generates $\binom{n}{2}$ intra-mode cross-term sites $q_\ell^{(r)}\,q_{\ell'}^{(r)}$ with $\ell < \ell'$ per mode.
Each site requires a rotation angle varying across $M$ electronic surfaces.
Angles encode the full curvature by folding the reference surface's contribution into UCR.
\begin{equation}
    \theta_j^{(r,\ell,\ell')} = \tau\,\kappa_j^{(r)}\cdot 2^{\ell+\ell'+1}\,\Delta_r^2,
    \qquad \kappa_j^{(r)} = \tfrac{1}{2}\mu_r\omega_r^2 + a_j^{(r)},
\end{equation}
where $\kappa_j^{(r)}$ is the total curvature on surface $j$, combining the shared harmonic contribution $\tfrac{1}{2}\mu_r\omega_r^2$ with the surface-dependent correction $a_j^{(r)}$.
No entry of the angle vector is constrained to be zero, and the gate count is independent of the angle values.

Each site is implemented by a CNOT sandwich enclosing a $\mathrm{UCR}_Z^{(m)}$ whose $m$ control qubits are the original channel register:
\begin{equation}\label{eq:RZZ_UCR_ondiag}
    \mathrm{CNOT}_{\ell'\to\ell}\;\longrightarrow\;\mathrm{UCR}_Z^{(m)}\bigl(\{\theta_j\};\;\mathcal{C}_{\mathrm{channel}}\to q_\ell^{(r)}\bigr)\;\longrightarrow\;\mathrm{CNOT}_{\ell'\to\ell}.
\end{equation}
The Gray-code decomposition of $\mathrm{UCR}_Z^{(m)}$ produces $M = 2^m$ rotations interleaved with $M - 1$ CNOTs per site, with circuit angles $\{\phi_g\}$ obtained from the target angles $\{\theta_j\}$ by classical Walsh--Hadamard preprocessing (Appendix~\ref{app:binary_decomp},~\ref{app:offdiag}).
Including the two outer CNOTs from the $ZZ$ sandwich:

Per site: $M$ rotations and $(M+1)$ CNOTs.
\begin{equation}\label{eq:tier2_gates}
    \text{Tier~2:}\quad
    d\,M\,\frac{n(n-1)}{2}\;\text{rotations},\quad
    d\,(M+1)\,\frac{n(n-1)}{2}\;\text{CNOTs},\quad
    0\;\text{Toffoli}.
\end{equation}
This is the same rotation count as a correction-only formulation, but eliminates the $d\binom{n}{2}$ rotations and $dn(n{-}1)$ CNOTs that would otherwise be required for an unconditional reference-surface application.

\subsubsection{Tier~3: Inter-Mode Bilinear Terms}\label{sec:mux_tier3}

On-diagonal bilinear terms $a_{rs}^{(j)}\,x_r\,x_s$ for mode pairs $(r,s) \in \mathcal{G}_2^{<}$ couple qubits across different mode registers.
As with Tier~2, the reference surface's bilinear coupling is absorbed into UCR, encoding the full angle
\begin{equation}
    \theta_j^{(r,s,\ell,p)} = \tau\,a_j^{(rs)}\cdot 2^{\ell+p}\,\Delta_r\,\Delta_s,
\end{equation}
for all $M$ surfaces, with $j = 0$ carrying the reference coupling $a_0^{(rs)}$ (potentially nonzero).
Each of the $n_r\times n_s$ qubit-pair sites $(q_\ell^{(r)}, q_p^{(s)})$ is implemented identically to Tier~2, with the CNOT sandwich bridging qubits on separate mode registers:
\begin{equation}
    \mathrm{CNOT}_{q_p^{(s)}\to q_\ell^{(r)}}\;\longrightarrow\;\mathrm{UCR}_Z^{(m)}\bigl(\{\theta_j\};\;\mathcal{C}_{\mathrm{channel}}\to q_\ell^{(r)}\bigr)\;\longrightarrow\;\mathrm{CNOT}_{q_p^{(s)}\to q_\ell^{(r)}}.
\end{equation}

Per site: $M$ rotations and $(M+1)$ CNOTs, identical to Tier~2.
\begin{equation}\label{eq:tier3_gates}
    \text{Tier~3:}\quad
    |\mathcal{G}_2^{<}|\,M\,n^2\;\text{rotations},\quad
    |\mathcal{G}_2^{<}|\,(M+1)\,n^2\;\text{CNOTs},\quad
    0\;\text{Toffoli}.
\end{equation}

Mode pairs $(r,s)$ and $(r',s')$ with $\{r,s\}\cap\{r',s'\} = \emptyset$ act on disjoint vibrational registers and use disjoint channel register qubits, so they execute in parallel.
The mode-pair conflict graph $\mathcal{G}_2$ has edges between pairs sharing a mode; its chromatic index $\chi'(\mathcal{G}_2)$ determines the number of sequential color classes.
By Vizing's theorem applied to the edge-coloring of $K_d$, $\chi'(\mathcal{G}_2) \leq d-1$ for even~$d$ and $\chi'(\mathcal{G}_2) \leq d$ for odd~$d$.

Because Tier~3 operates on cross-mode qubit pairs while Tiers~1--2 are intra-mode, the two groups cannot overlap: the vibrational qubits of a mode participating in a Tier~3 bilinear must be free from Tier~1--2 operations.
Tiers~1 and~2 therefore complete before Tier~3 begins, though the fan-out remains active throughout all three tiers.

\subsubsection{Phase~4: Fan-In and Uncomputation}\label{sec:mux_unfanout}

After all three tiers, fan-out CNOT trees are reversed (fan-in), restoring the entangled copies to $|0\rangle$ and disentangling the channel register from the ancilla pool.
Toffoli cascades of Phase~2 are then reversed, returning the truth-value ancillas to $|0\rangle$.
\begin{equation}\label{eq:fanin_gates}
    \text{Phase~4:}\quad
    (M-1)(d-1)\;\text{CNOTs (fan-in)},\quad
    T_{\mathrm{truth}}\;\text{Toffoli (uncompute)}.
\end{equation}

\subsubsection{Combined Gate Counts}\label{sec:mux_total}

Table~\ref{tab:ondiag_gates} collects the gate counts from all phases for one on-diagonal half-step $e^{-iV_\mathrm{d}\tau/2}$.
The per-Trotter-step cost is twice these values.
Toffoli gates appear only for truth-value computation and uncomputation; Tiers~1 and 3 are Toffoli-free.

\begin{table}[htbp]
\centering
\caption{On-diagonal potential gate counts per half-step.
$M = 2^m$ electronic channels, $d$ vibrational modes, $n$ qubits per mode, $|\mathcal{G}_2^{<}|$ symmetry-allowed bilinear mode pairs.
The Gray-code UCR$_Z^{(m)}$ contributes $M$ rotations and $M - 1$ CNOTs per site (acyclic traversal); the additional CNOTs arise from outer $ZZ$ sandwiches.
Per-Trotter-step costs are $2\times$ these values.}\label{tab:ondiag_gates}
\begin{tabular}{lccc}
\toprule
\textbf{Component} & $\boldsymbol{R_z}$ & \textbf{CNOT} & \textbf{Toffoli} \\
\midrule
Phase~1: $V_{00}$ (single-qubit) & $d(1 + n)$ & $0$ & $0$ \\[6pt]
Phase~2: truth values & --- & --- & $T_{\mathrm{truth}}/2$ \\[4pt]
Phase~3: fan-out & --- & $(M-1)(d-1)$ & --- \\[4pt]
Tier~1: corrections & $d(M-1)(1+n)$ & $2d(M-1)(1+n)$ & $0$ \\[4pt]
Tier~2: quadratic UCR & $\dfrac{dMn(n-1)}{2}$ & $\dfrac{d(M+1)n(n-1)}{2}$ & $0$ \\[6pt]
Tier~3: bilinear UCR & $|\mathcal{G}_2^{<}|\,Mn^2$ & $|\mathcal{G}_2^{<}|\,(M+1)\,n^2$ & $0$ \\[6pt]
Phase~4: fan-in + uncompute & --- & $(M-1)(d-1)$ & $T_{\mathrm{truth}}/2$ \\[4pt]
\midrule
\textbf{Total (half-step)} & $N_{R_z}^{(\mathrm{d})}$ & $N_{\mathrm{CX}}^{(\mathrm{d})}$ & $T_{\mathrm{truth}}$ \\
\bottomrule
\end{tabular}
\end{table}

The total rotation count per half-step is
\begin{equation}\label{eq:ondiag_Rz_total}
    N_{R_z}^{(\mathrm{d})} = dM(1+n) + \frac{dMn(n-1)}{2} + |\mathcal{G}_2^{<}|\,Mn^2
    = dM\!\left(1 + \frac{n(n+1)}{2}\right) + |\mathcal{G}_2^{<}|\,Mn^2,
\end{equation}
with asymptotic complexity $\mathcal{O}(Md^2n^2)$ in $C_1$ symmetry where $|\mathcal{G}_2^{<}| = \binom{d}{2}$.
The total CNOT count per half-step is
\begin{equation}\label{eq:ondiag_CX_total}
    N_{\mathrm{CX}}^{(\mathrm{d})} = 2(M-1)(d-1) + 2d(M-1)(1+n) + \frac{d(M+1)n(n-1)}{2} + |\mathcal{G}_2^{<}|\,(M+1)\,n^2,
\end{equation}
with the same $\mathcal{O}(Md^2n^2)$ complexity.
The total Toffoli count per half-step is
\begin{equation}\label{eq:ondiag_Tf_total}
    N_{\mathrm{Tf}}^{(\mathrm{d})} = T_{\mathrm{truth}} = M(m-2)+2 = \mathcal{O}(M\log M),
\end{equation}
and per Trotter step $2T_{\mathrm{truth}} = 2[M(m-2)+2]$.

The folded formulation eliminates the unconditional cross-term application that would appear in a separated Phase~1.
A separated approach would require, per half-step:
\begin{itemize}
    \item $d\binom{n}{2}$ additional rotations (one per intra-mode cross-term site on the reference surface),
    \item $dn(n-1)$ additional CNOTs (two per unconditional $ZZ$ gate),
    \item $|\mathcal{G}_2^{<,\mathrm{ref}}|\,n^2$ additional rotations and $2|\mathcal{G}_2^{<,\mathrm{ref}}|\,n^2$ additional CNOTs for reference bilinear terms,
\end{itemize}
where $|\mathcal{G}_2^{<,\mathrm{ref}}|$ is the number of bilinear pairs with nonzero reference coupling.
These gates are absorbed into Tier~2 and Tier~3 UCR circuits, costing the same whether $\theta_0 = 0$ or $\theta_0 \neq 0$.
The total savings for the lowest symmetry ($C_1$) per half-step are $d\binom{n}{2} + \binom{d}{2}n^2$ rotations and $dn(n{-}1) + d(d{-}1)n^2$ CNOTs.

\subsection{Off-Diagonal Potential: XOR-Class Fragmentation}\label{sec:offdiag_alg}

The off-diagonal vibronic coupling
\begin{equation}\label{eq:Vod_def}
    V_{\text{od}} = \sum_{\substack{i,j=0 \\ i < j}}^{M-1} c_{ij}(\mathbf{x})\,\bigl(|i\rangle\langle j| + |j\rangle\langle i|\bigr),
\end{equation}
drives population transfer between diabatic electronic states, governing nonadiabatic dynamics in the algorithm.
Unlike the on-diagonal block, the off-diagonal block contains Pauli-$X$ operators on the channel register, storing electronic channel populations~\cite{ollitrault2020nonadiabatic}.
Though schemes can be defined to apply a coupling in this basis, scalable depth advantage is found by bringing $V_{\text{od}}$ to a diagonal form before $R_z$-based circuits can be applied.

For $M = 2$ the reduction is straightforward: a single Hadamard gate on the channel qubit sends $X \mapsto Z$, and the transformed operator is diagonal.
For $M > 2$, the $\binom{M}{2}$ coupling pairs carry distinct coupling functions $c_{ij}(\mathbf{x})$.
A global Hadamard $H^{\otimes m}$ fails to diagonalize $V_{\text{od}}$, producing residual off-diagonal Pauli-$X$ terms on spectator qubits (Appendix~\ref{app:offdiag}).
An alternative treatment, following Motlagh et al.~\cite{motlagh2025quantum}, partitions $V_{\text{od}}$ into XOR-class fragments and diagonalizes each fragment independently with a lightweight Clifford circuit.

\subsubsection{XOR-Class Partition}\label{sec:xor_partition}

Each pair $(i,j)$ with $i < j$ is assigned to a XOR class $C_\zeta$ labelled by $\zeta = i \oplus j$, where $\oplus$ denotes bitwise XOR on $m$-bit channel indices.
$\binom{M}{2}$ pairs partition into $M - 1$ disjoint classes containing $M/2 = 2^{m-1}$ pairs.
The off-diagonal potential decomposes as
\begin{equation}\label{eq:Vod_fragment_sum}
    V_{\text{od}} = \sum_{\zeta=1}^{M-1} H_\zeta,
    \qquad
    H_\zeta = \sum_{(i,j)\in C_\zeta} c_{ij}(\mathbf{x})\,G_{ij},
\end{equation}
where each fragment $H_\zeta$ collects coupling pairs whose channel indices differ in bit positions specified by the support $S(\zeta) = \{q : \zeta_q = 1\}$.
Hamming weight $h(\zeta) = |S(\zeta)|$ records how many channel qubits participate in the swap, while the remaining $\kappa(\zeta) = m - h(\zeta)$ qubits are spectators whose values label distinct pairs within the class.

\subsubsection{Per-Fragment Clifford Diagonalization}\label{sec:clifford_diag}

Each fragment is diagonalized by a fragment-specific circuit $U_\zeta$ that acts only on the channel register.

Within the support $S(\zeta) = \{s_1, s_2, \ldots, s_h\}$, one qubit $s_1$ is designated the `target.'
A cascade of $h(\zeta) - 1$ CNOT gates, each with $s_1$ as `control' and $s_k$ ($k \geq 2$) as `target,'
\begin{equation}\label{eq:cnot_focusing}
    \prod_{k=2}^{h(\zeta)} \mathrm{CNOT}_{s_1 \to s_k},
\end{equation}
concentrates the multi-qubit flip $X_S = \bigotimes_{q \in S} X_q$ into a single-qubit operator $X_{s_1}$.
After the cascade, non-target support qubits $s_2, \ldots, s_h$ acquire identical bit values on both sides of each swap pair and become additional spectators, enlarging the spectator set from $\kappa(\zeta)$ to $m - 1$ qubits regardless of $h(\zeta)$.

Applying $H_{s_1}$ converts $X_{s_1} \mapsto Z_{s_1}$, producing a fully diagonal operator:
\begin{equation}\label{eq:block_diag_main}
    U_\zeta\,H_\zeta\,U_\zeta^\dagger
    = \left[\sum_{(i,j)\in C_\zeta} c_{ij}(\mathbf{x})\;|\tilde\sigma(i)\rangle\langle\tilde\sigma(i)|_{\tilde{S^c}}\right] \otimes Z_{s_1},
\end{equation}
where $\tilde{S^c} = \{0,\ldots,m-1\}\setminus\{s_1\}$ is the $(m-1)$-qubit extended spectator set and $\tilde\sigma(i)$ is the extended spectator configuration that uniquely addresses pair $(i,j)$.
The eigenvalue on computational basis state $|k\rangle$ of the channel register is $(-1)^{k_{s_1}} c_{\tilde\sigma(k)}(\mathbf{x})$, where $m-1$ non-target bits select which pair's coupling function appears, and the target bit provides a sign $\pm 1$.

The focusing unitary $U_\zeta$ requires $h(\zeta) - 1$ CNOTs and one Hadamard gate, with circuit depth $h(\zeta)$.
Its adjoint $U_\zeta^\dagger$ has the same cost, applied in reverse order after the UCR block.

\subsubsection{UCR Implementation Within Each Fragment}\label{sec:ucr_offdiag}

Once $H_\zeta$ has been diagonalized, its time evolution $\exp(-iU_\zeta H_\zeta U_\zeta^\dagger\,\tau)$ is implemented by the same $R_z$-based machinery used for on-diagonal blocks with one structural difference: the UCR$_Z$ gates have $m - 1$ control qubits (the extended spectator set $\tilde{S^c}$) rather than the (up to) $m$ controls used in the on-diagonal UCR$_Z^{(m)}$.
This follows from the block-diagonal structure established in Eq.~\ref{eq:block_diag_main}: the $m - 1$ non-target bits select between $M/2$ coupling functions, and the target qubit $s_1$ contributes through a CNOT sandwich rather than as a UCR control.

Each rotation site on the vibrational register is implemented as
\begin{equation}\label{eq:ucr_offdiag_site}
    \mathrm{CNOT}_{s_1\to\ell}\;\;\mathrm{UCR}_Z^{(m-1)}(\ell;\;\tilde{S^c})\;\;\mathrm{CNOT}_{s_1\to\ell},
\end{equation}
where the outer CNOT pair encodes the $Z_{s_1}$ sign flip and the inner UCR$_Z^{(m-1)}$ selects among the $2^{m-1}$ coupling-function values.
The Gray-code decomposition of each UCR$_Z^{(m-1)}$ produces $2^{m-1}$ rotations interleaved with $2^{m-1} - 1$ CNOTs per site.
For a cross-term (two-qubit) site the total cost is $2^{m-1}$ rotations and $2^{m-1} + 3$ CNOTs; for a single-qubit site, $2^{m-1}$ rotations and $2^{m-1} + 1$ CNOTs.
These costs are independent of the Hamming weight $h(\zeta)$, since the focusing Clifford uniformly reduces the control count to $m - 1$.

\subsubsection{Symmetric Trotter Integration}\label{sec:trotter_offdiag}

Since distinct XOR-class fragments $H_\zeta$ and $H_{\zeta'}$ do not generally commute, a first-order product $\prod_\zeta e^{-iH_\zeta\tau}$ would introduce $\mathcal{O}(\tau^2)$ error that would dominate the $\mathcal{O}(\tau^3)$ accuracy of the global second-order Trotter step.
Fragments are symmetrized into a second-order Strang split product, following a second-order Trotterization:
\begin{equation}\label{eq:symmetric_product_main}
    \mathcal{V}_\text{od}(\tau) = \prod_{\zeta=1}^{M-1} e^{-iH_\zeta\tau/2} \;\cdot\; \prod_{\zeta=M-1}^{1} e^{-iH_\zeta\tau/2}
    = e^{-iV_{\text{od}}\tau} + \mathcal{O}(\tau^3).
\end{equation}
A forward sweep applies each fragment at half-angle in ascending $\zeta$ order; the backward sweep repeats in descending order.
Asymmetric commutator error from the forward sweep is canceled by the backward sweep, preserving $\mathcal{O}(\tau^3)$ global error.

$2(M-1)$ fragment applications consists of the sequence: focusing Clifford $U_\zeta$, channel register fan-out, mode-parallel UCR block, fan-in, unfocusing $U_\zeta^\dagger$.
After every $U_\zeta^\dagger$, the channel register returns to the diabatic basis, so inter-fragment interfacing is clean and no frame propagation is required.

The full second-order Trotter step for the complete Hamiltonian $H = T + V_{\text{d}} + V_{\text{od}}$ takes the form
\begin{equation}\label{eq:full_trotter_main}
    \mathcal{U}(\tau) = e^{-iT\tau/2}\,e^{-iV_{\text{d}}\tau/2}\,\mathcal{V}_\text{od}(\tau)\,e^{-iV_{\text{d}}\tau/2}\,e^{-iT\tau/2} + \mathcal{O}(\tau^3).
\end{equation}

\subsubsection{Interferometric Population Transfer: Mach-Zehnder Analogy}\label{sec:mach_zehnder}

The per-fragment protocol implements population transfer analogously to a Mach-Zehnder interferometer, providing physical intuition for how the XOR-class circuit produces correct nonadiabatic dynamics.

Within each XOR class $C_\zeta$, every pair $(i,j)$ with $i \oplus j = \zeta$ evolves as an independent two-level system.
The fragment Hamiltonian restricted to the subspace $\{|i\rangle, |j\rangle\}$ is $c_{ij}(\mathbf{x})\,G_{ij}$, where $G_{ij} = |i\rangle\langle j| + |j\rangle\langle i|$ acts as a position-dependent beamsplitter.
Time evolution under this restriction gives the transition amplitude
\begin{equation}\label{eq:rabi_transition}
    \langle b|\,e^{-iH_\zeta\tau}\,|a\rangle_{\text{channel}} =
    \begin{cases}
    \cos\bigl(c_{a,a\oplus\zeta}(\mathbf{x})\,\tau\bigr)\,\delta_{ab} - i\sin\bigl(c_{a,a\oplus\zeta}(\mathbf{x})\,\tau\bigr)\,\delta_{b,a\oplus\zeta}, & a\oplus b \in \{0, \zeta\}, \\[4pt]
    \delta_{ab}, & \text{otherwise},
    \end{cases}
\end{equation}
as a standard Rabi oscillation between states $|a\rangle$ and $|a\oplus\zeta\rangle$ at a position-dependent frequency $c_{a,a\oplus\zeta}(\mathbf{x})$.
Population transfer probability for each pair is
\begin{equation}\label{eq:beamsplitter_reflectivity}
    \mathcal{T}_{a\to b}^{(\zeta)}(\mathbf{x},\tau) = \sin^2\!\bigl(c_{ab}(\mathbf{x})\,\tau\bigr),
\end{equation}
where the vibronic coupling function $c_{ab}(\mathbf{x})$ controls the beamsplitter reflectivity, just as Rabi frequency governs population inversion in a two-level atom driven by a laser field.
States not connected by the XOR label $\zeta$ are spectators, passing through the fragment unaffected, analogous to an optical mode that bypasses the beamsplitter entirely.

For a four-state system ($M=4$, $m=2$), the three XOR classes $\zeta \in \{01, 10, 11\}$ each implement $M/2 = 2$ independent beamsplitters.
Class $\zeta = 01$ couples $(|00\rangle,|01\rangle)$ and $(|10\rangle,|11\rangle)$; class $\zeta = 10$ couples $(|00\rangle,|10\rangle)$ and $(|01\rangle,|11\rangle)$; class $\zeta = 11$ couples $(|00\rangle,|11\rangle)$ and $(|01\rangle,|10\rangle)$.
The sequential application of all three classes in the symmetric Trotter product connects every pair of states through at least one beamsplitter, recovering the full off-diagonal coupling $V_{\text{od}}$.

The spectral mapping theorem guarantees
\begin{equation}\label{eq:spectral_map}
    U_\zeta^\dagger\;\exp\!\bigl(-i\,U_\zeta H_\zeta U_\zeta^\dagger\,\tau\bigr)\;U_\zeta = \exp(-iH_\zeta\tau),
\end{equation}
so the circuit produces the same unitary as direct exponentiation of $H_\zeta$.
Physically, the focusing Clifford rotates the channel register into a frame where the multi-qubit swap $X_S = \bigotimes_{q\in S}X_q$ is concentrated onto a single target qubit $s_1$ (Section~\ref{sec:clifford_diag}), which the Hadamard then converts to a $Z_{s_1}$ phase.
In this frame, the beamsplitter action is encoded as a position-dependent $Z$-rotation.
Unfocusing Cliffords restore the diabatic basis, returning the channel register to its physical meaning before subsequent fragment action.
Since $U_\zeta$ acts only on the $m$-qubit channel register while populations are defined by tracing over the vibrational register (Eq.~\ref{eq:diabatic_population}), Clifford conjugation and partial trace commute and measuring the channel register after unfocusing yields the same populations as if the beamsplitter had been applied directly.

This finds analogy in the frame invariance of a physical interferometer, where inserting matched polarization rotators before and after a beamsplitter does not alter the fringe pattern, provided the rotators are inverses of each other.
The Clifford pair $U_\zeta$, $U_\zeta^\dagger$ plays precisely this role, rotating into a convenient computational frame and back without introducing any physical effect on population dynamics.

The off-diagonal step applies $M-1$ non-commuting beamsplitter layers in a symmetric order (Eq.~\ref{eq:symmetric_product_main}).
The palindromic structure cancels the leading commutator error between layers.
Expanding through second order,
\begin{equation}\label{eq:palindrome_cancellation}
    \mathcal{V}_{\text{od}}(\tau) = I - i\tau V_{\text{od}} - \frac{\tau^2}{2}V_{\text{od}}^2 + \mathcal{O}(\tau^3),
\end{equation}
the cancellation symmetrically combines $H_\zeta H_{\zeta'} + H_{\zeta'}H_\zeta = \{H_\zeta, H_{\zeta'}\}$ to eliminate the antisymmetric commutator $[H_\zeta, H_{\zeta'}]$ that would appear in a single-sweep product.
The $-i$ phase is required by unitarity, producing the correct interference pattern when fragments compose, vanishing from observable populations. 
This corresponds to a Mach-Zehnder interferometer measuring intensity $\propto|E|^2$, not field amplitudes $E$.
Since $\langle a|V_{\text{od}}|a\rangle = 0$ (the off-diagonal potential has no diagonal matrix elements in the diabatic basis), first-order population changes vanish identically.
Leading population update is second order:
\begin{equation}\label{eq:pop_update_MZ}
    P_a(\tau) = P_a(0) - \tau^2\!\sum_{b \neq a} \int |c_{ab}(\mathbf{x})|^2\bigl(|\psi_a(\mathbf{x})|^2 - |\psi_b(\mathbf{x})|^2\bigr)\,d\mathbf{x} + \mathcal{O}(\tau^3),
\end{equation}
where the sum runs over all diabatic states $b$ coupled to $a$, and the integrand depends on local population imbalance weighted by the squared coupling strength.
Following a short-time expansion of the exact propagator $e^{-iV_{\text{od}}\tau}$, this serves as confirmation that the layered beamsplitter protocol faithfully reproduces nonadiabatic population transfer to Trotter accuracy.

The per-fragment approach is verified by tracing the diabatic population dynamics through the Clifford sandwich.
Within each fragment $\zeta$, the operator $H_\zeta$ couples states pairwise: each pair $(i,j) \in C_\zeta$ evolves as an independent $2 \times 2$ Rabi oscillation with frequency $c_{ij}(\mathbf{x})$, while states not connected by $\zeta$ are unaffected.
Clifford conjugation preserves population transfer, where after the complete symmetric product $\mathcal{V}_2(\tau)$, the diabatic populations satisfy the second-order update:
\begin{equation}\label{eq:pop_update_main}
    P_a(\tau) = P_a(0) - \tau^2\!\sum_{b \neq a} |c_{ab}|^2\bigl(P_a(0) - P_b(0)\bigr) + \mathcal{O}(\tau^3).
\end{equation}

\section{Channel Register Fan-Out}\label{sec:fanout}

Both the on-diagonal potential (Section~\ref{sec:mux_alg}) and the per-fragment off-diagonal potential (Section~\ref{sec:offdiag_alg}) use information derived from the channel register as a read-only control for position-dependent rotations on the vibrational register.
This read-only property enables a fan-out protocol that copies relevant control bits to $d$ ancilla registers so that channel-controlled operations on different modes execute in parallel.
The construction is closely related to the preparation of a GHZ state: each ancilla copy is entangled with the original, not independently duplicated.

The two potential blocks differ in what is fanned out:
\begin{itemize}
    \item \textbf{On-diagonal:} $M - 1$ truth-value ancillas computed by Toffoli cascade (Section~\ref{sec:mux_protocol}) are fanned out.
    Each mode requires a local copy of every truth-value bit so that state-dependent correction rotations of Tier~1 can execute in parallel.
    $m$ channel register qubits are fanned out alongside $M-1-m$ truth-value ancillas (totaling $M-1$ source qubits), so that Tiers 1-3 all execute in parallel across modes using local copies.
    \item \textbf{Off-diagonal:} $m$ channel qubits are fanned out {within each fragment application}, after the focusing Clifford $U_\zeta$ has rendered the channel register read-only.
    Because the symmetric Trotter product applies $2(M-1)$ fragments sequentially, fan-out and fan-in are executed once per fragment, for a total of $2(M-1)$ invocations per Trotter step.
\end{itemize}

\subsubsection{Binary-Tree Protocol}\label{sec:fanout_tree}

The fan-out of $w$ control bits to $d$ copies uses a binary tree of CNOTs:
\begin{enumerate}
    \item \textbf{Layer~1:} Each of the $w$ original control bits entangled to its first ancilla copy ($w$ CNOTs, depth~1).
    \item \textbf{Layer~$\ell$} ($\ell = 2, \ldots, \lceil\log_2 d\rceil$): Each existing copy fans out to one new copy ($w \cdot \min(2^{\ell-1},\,d - 2^{\ell-1})$ CNOTs, depth~1).
    \item After $\lceil\log_2 d\rceil$ layers, $d$ copies of the $w$-bit control string are available, one per mode.
\end{enumerate}
The total CNOT count per invocation is $w(d - 1)$, and the circuit depth is $\lceil\log_2 d\rceil$.
Fan-in (uncomputation) mirrors the fan-out at the same cost: $w(d - 1)$ CNOTs, depth $\lceil\log_2 d\rceil$.

\subsubsection{Validity Condition}\label{sec:fanout_validity}

Fan-out preserves correctness iff the qubits remain read-only during the parallel block.
Both potential blocks satisfy this condition by construction:
\begin{itemize}
    \item \textbf{On-diagonal (multiplexing):} Truth-value ancillas are computed once before fan-out and serve only as single-qubit controls for the Tier~1 correction rotations; they are never the target of a rotation or a CNOT within the parallel block.
    \item \textbf{Off-diagonal (per-fragment):} After the focusing Clifford $U_\zeta$, the transformed fragment $U_\zeta H_\zeta U_\zeta^\dagger$ is diagonal in the computational basis of the channel register (Section~\ref{sec:clifford_diag}).
    The channel qubits therefore serve exclusively as UCR controls and are never modified by the vibrational-register gates.
    The fan-in restores the original channel register before the unfocusing Clifford $U_\zeta^\dagger$ is applied.
\end{itemize}

\section{Molecular Symmetries}\label{sec:symmetry}

\subsection{Selection Rules for Vibronic Coupling Terms}

Consider a vibronic Hamiltonian matrix element $V_{ij}$ between diabatic electronic states $\ket{\psi_i}$ and $\ket{\psi_j}$ transforming as irreps $\Gamma_i$ and $\Gamma_j$ of the molecular point group $\mathcal{G}$.
A coupling operator $\hat{O}(\mathbf{x})$ contributes to $V_{ij}$ only if the matrix element
\begin{equation}\label{eq:symmetry_integral}
    \langle \psi_i | \hat{O}(\mathbf{x}) | \psi_j \rangle \neq 0,
\end{equation}
requiring that the direct product contains the totally symmetric representation $\Gamma_{\text{sym}}$ of $\mathcal{G}$ (Wigner-Eckart Theorem):
\begin{equation}\label{eq:selection_rule}
    \Gamma_i \otimes \Gamma(\hat{O}) \otimes \Gamma_j \supset \Gamma_{\text{sym}},
\end{equation}
For non-degenerate point groups (only one-dimensional irreps, including $D_{2h}$, $C_{2v}$, and $C_s$), every normal mode $x_\alpha$ transforms as a single irrep $\Gamma_\alpha$ with character $\chi_\alpha(\hat{R}) = \pm 1$ under each symmetry operation $\hat{R}$.
Representations of operators appearing in the vibronic expansion are determined by direct product rules.

Operator symmetries follow for each order in the vibronic expansion. 
The linear operator $x_\alpha$ transforms as $\Gamma_\alpha$ itself.
The diagonal quadratic operator $x_\alpha^2$ transforms as
\begin{equation}\label{eq:q_squared_symmetry}
    \Gamma(x_\alpha^2) = \Gamma_\alpha \otimes \Gamma_\alpha = \Gamma_{\text{sym}},
\end{equation}
since for one-dimensional representations, $\chi_\alpha(\hat{R})^2 = (\pm 1)^2 = +1$ for all $\hat{R} \in \mathcal{G}$, regardless of whether $x_\alpha$ is symmetric or antisymmetric.
This is specific to non-degenerate point groups, where for groups admitting multidimensional representations (e.g., $C_{3v}$, $D_{3h}$), the symmetric square $[\Gamma_\alpha \otimes \Gamma_\alpha]_S$ may contain non-totally-symmetric components.
Bilinear operators $x_\alpha x_\beta$ with $\alpha \neq \beta$ transform as $\Gamma_\alpha \otimes \Gamma_\beta$, and are not generally symmetric.

Substituting into Eq.~\ref{eq:selection_rule}, selection rules for each coupling type are:

\begin{center}
\renewcommand{\arraystretch}{1.3}
\begin{tabular}{lll}
\hline
Intrastate linear ($a_\alpha^{(i)} x_\alpha$) & $\Gamma_\alpha$ & $\Gamma_\alpha = \Gamma_{\text{sym}}$ \\
Intrastate quadratic ($a_{\alpha\alpha}^{(i)} x_\alpha^2$) & $\Gamma_{\text{sym}}$ & Always allowed \\
Intrastate bilinear ($a_{\alpha\beta}^{(i)} x_\alpha x_\beta$) & $\Gamma_\alpha \otimes \Gamma_\beta$ & $\Gamma_\alpha = \Gamma_\beta$ \\
Interstate linear ($c_\alpha^{(ij)} x_\alpha$) & $\Gamma_\alpha$ & $\Gamma_i \otimes \Gamma_\alpha \otimes \Gamma_j = \Gamma_{\text{sym}}$ \\
Interstate quadratic ($c_{\alpha\alpha}^{(ij)} x_\alpha^2$) & $\Gamma_{\text{sym}}$ & $\Gamma_i = \Gamma_j$ \\
Interstate bilinear ($c_{\alpha\beta}^{(ij)} x_\alpha x_\beta$) & $\Gamma_\alpha \otimes \Gamma_\beta$ & $\Gamma_i \otimes \Gamma_\alpha \otimes \Gamma_\beta \otimes \Gamma_j = \Gamma_{\text{sym}}$ \\
\hline
\end{tabular}
\end{center}

\subsection{Counting Symmetry-Allowed Terms}
Intrastate quadratic terms $x_\alpha^2$ are always symmetry-allowed on every diabatic surface, reflecting the physical requirement that each surface possesses a well-defined curvature along every mode at the reference geometry.
Interstate quadratic terms $c_{\alpha\alpha}^{(ij)}x_\alpha^2$ vanish identically whenever $\Gamma_i \neq \Gamma_j$, since the totally symmetric operator $x_\alpha^2$ cannot bridge electronic states of different symmetry.
This eliminates an entire class of coupling terms from the vibronic Hamiltonian without approximation, directly reducing the number of off-diagonal circuit operations.

For intrastate bilinear terms (Duschinsky mixing), the condition $\Gamma_\alpha = \Gamma_\beta$ restricts the allowed mode pairs to those sharing the same irrep.
Distributing $d$ modes as $\{n_\Gamma\}$ across representations of $\mathcal{G}$ results in $\sum_\Gamma \binom{n_\Gamma}{2}$ bilinear terms per surface, compared to $\binom{d}{2}$ in the generic case.
For interstate bilinear terms, the condition $\Gamma_\alpha \otimes \Gamma_\beta = \Gamma_i \otimes \Gamma_j$ selects mode pairs from complementary representations determined by electronic product symmetry, with $\sum_{\Gamma_\alpha \otimes \Gamma_\beta = \Gamma_i \otimes \Gamma_j} n_{\Gamma_\alpha} n_{\Gamma_\beta}$ terms.
Since bilinear terms dominate gate complexity at large $d$ for both REQWIEM and phase gradient methods, these reductions have the most significant impact on total resource requirements.
To identify scaling features, a graph theoretic representation is chosen to quickly identify parallelization advantages.

\subsection{Mode-Pair Conflict Graph and Circuit Depth Reduction}\label{sec:chromatic_index}

Symmetry-allowed bilinear pairs determine both gate count and circuit depth.
While gate count scales with the total number of allowed pairs $|\mathcal{G}_2|$ or $|\mathcal{G}_4|$, circuit depth is governed by the degree of parallel execution.
Bilinear terms coupling registers $(r,s)$ and $(r',s')$ can execute simultaneously if and only if $\{r,s\}\cap\{r',s'\} = \emptyset$, acting on disjoint vibrational registers.
The parallelization problem is equivalent to edge coloring a mode-pair conflict graph, whose chromatic index $\chi$ directly enters the depth formulas of Chapter~\ref{ch:resources}.

\subsubsection{Graph Construction}

For a given index set of symmetry-allowed bilinear pairs ($\mathcal{G}_2$ (on-diagonal) or $\mathcal{G}_4$ (off-diagonal)), a mode-pair conflict graph $G = (V, E)$ is defined as follows.
The vertex set $V = \{1,2,\ldots,d\}$ consists of $d$ vibrational modes.
The edge set $E$ contains an edge $(r,s)$ for each symmetry-allowed bilinear pair.
Two bilinear operations can execute in parallel if and only if their corresponding edges share no vertex; a maximum set of simultaneously executable operations is a matching in $G$.
The minimum number of sequential layers needed to execute all bilinear terms is the chromatic index (edge chromatic number) $\chi'(G)$, being the minimum number of colors such that when each edge of $G$ is assigned a color, no two edges incident on the same vertex share a color.
Each color class constitutes one parallel execution layer.
Throughout, $\chi$ denotes $\chi'(G)$ for the relevant conflict graph.
Distinct graphs arise for on-diagonal ($\mathcal{G}_2^<$) and off-diagonal ($\mathcal{G}_4$) bilinear sets, denoted $\chi(\mathcal{G}_2^<)$ and $\chi(\mathcal{G}_4)$ respectively when relevant.

\subsubsection{Scaling in the Absence of Symmetry ($C_1$)}

For $C_1$ symmetry (lowest symmetry group), all $\binom{d}{2}$ bilinear mode pairs are allowed, so $G = K_d$ (the complete graph on $d$ vertices).
By Vizing's theorem~\cite{vizing1964estimate}, the chromatic index of any simple graph satisfies $\Delta(G) \leq \chi'(G) \leq \Delta(G) + 1$, where $\Delta(G)$ is the maximum vertex degree.
For $K_d$, every vertex has degree $d - 1$, and the chromatic index is known exactly~\cite{vizing1964estimate}:
\begin{equation}\label{eq:chi_Kd}
    \chi'(K_d) = \begin{cases}
        d - 1, & d \text{ even},\\
        d, & d \text{ odd}.
    \end{cases}
\end{equation}
In both cases $\chi = \Theta(d)$.
For even $d$, Vizing's bound is tight, with a round-robin tournament schedule~\cite{gross2018graph} providing explicit edge coloring with exactly $d - 1$ colors, each color class being a perfect matching of $d/2$ edges.
For odd $d$, the extra color arises because no perfect matching exists; the optimal coloring consists of $d$ near-perfect matchings of $(d-1)/2$ edges each.

Circuit depth for diagonal and off-diagonal terms therefore depend on chromatic index $\chi$.
Combined fan-out and coloring reduction is derived in Chapter~\ref{ch:resources} and applied to both REQWIEM and the approach of Motlagh et al.~\cite{motlagh2025quantum}, resulting in a $\sim d/2$-fold depth reduction in the former, beyond fan-out parallelism of intra-mode terms.

\subsubsection{Scaling with Molecular Symmetry}

For molecules with non-trivial point group symmetry, selection rules of Section~\ref{sec:symmetry} restrict the edge set $E$ to a proper subgraph of $K_d$.
Conflict graphs decompose according to the representation structure of the point group.
For on-diagonal bilinear terms, the condition $\Gamma_\alpha = \Gamma_\beta$ partitions edges into cliques: within each irrep $\Gamma$, the $n_\Gamma$ modes of that symmetry form a complete subgraph $K_{n_\Gamma}$, and no edges connect modes of different symmetry.
The conflict graph is therefore a disjoint union of complete graphs,
\begin{equation}\label{eq:ondiag_graph}
    G_{\text{d}} = \bigsqcup_\Gamma K_{n_\Gamma},
\end{equation}
with chromatic index
\begin{equation}\label{eq:chi_ondiag}
    \chi'(G_{\text{d}}) = \max_\Gamma \chi'(K_{n_\Gamma}) = \max_\Gamma
    \begin{cases}
        n_\Gamma - 1, & n_\Gamma \text{ even},\\
        n_\Gamma, & n_\Gamma \text{ odd},
    \end{cases}
\end{equation}
Since $n_\Gamma \leq d$ for every $\Gamma$ and typically $n_\Gamma \ll d$ when modes are distributed across many representations, this yields substantial depth reductions, scaling as $\mathcal{O}(d)$.\
A conflict graph comes from a line graph of connectivity, promoting each edge to a vertex.
Wherever these vertices share an endpoint in $K_5$, they share an edge in the conflict graph. 
When they conflict, they cannot be applied simultaneously, involving a common mode coordinate.

The number of terms that can be applied simultaneously is visually apparent for the Petersen graph, as the complement to the conflict graph by connecting two term-vertices if they do not share a mode.
An independent set in the conflict graph is a clique in the Petersen graph, and vice versa. 
Finding the maximum number of terms that can be applied simultaneously is equivalent to finding the maximum clique in the Petersen graph, being two for $A_g$ modes for pyrazine. 
A proper coloring of $L(K_5)$ corresponds to a clique cover of the Petersen graph, ordering the terms in the most efficient way~\cite{gross2018graph}.

\begin{figure}
    \centerline{
    \includegraphics[width=180mm,scale=0.5]{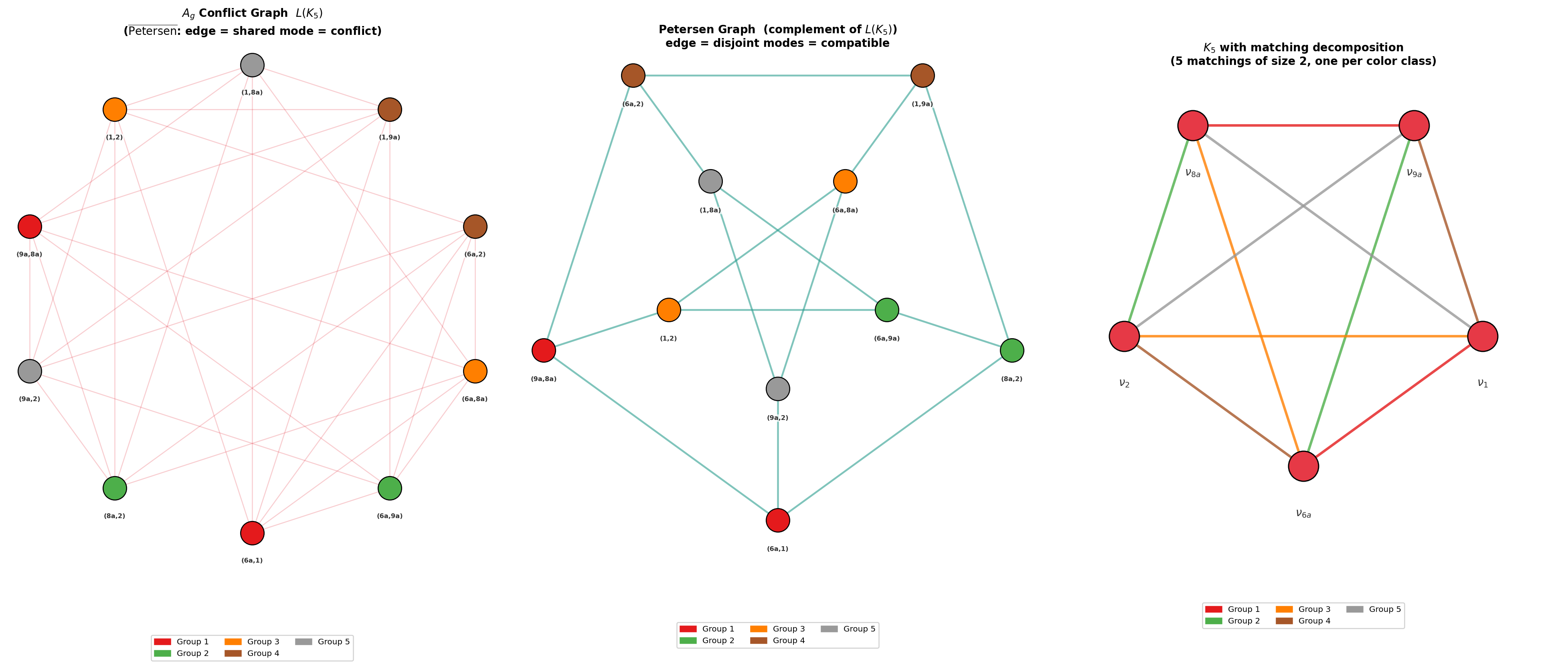}
    }
    \caption{\label{fig:ag_conflict} (Left) Conflict graph for $A_g$ symmetry modes. 
    A single vibrational mode cannot act as a control or target from two different bilinear terms, so depth advantage from parallelism is defined by how many bilinear terms can be applied independently.
    (Middle) Petersen graph for $A_g$ symmetry modes, delineating which bilinear terms can be applied simultaneously.
    (Right) Connectivity graph for $A_g$ symmetry modes, being all-to-all, forming a complete $K_5$ graph for the five $A_g$ modes.}
\end{figure}

An example conflict graph is given for the $A_g$ symmetry group's on-diagonal bilinear terms of a $d=24$-mode pyrazine model \cite{raab1999molecular}, alongside its complementary Petersen graph and connectivity graph in Figure~\ref{fig:ag_conflict}. 

Emphasis is made in the graph theoretic representation of bilinear terms in quantum circuits, as further discussion in Chapter~\ref{ch:smooth_potentials} extends applicable problems to nonlinear, multivariate terms.
Conflict graphs and complementary Petersen graphs serve as natural ways to maximize circuit parallelism, according to system symmetries and hardware connectivity.
The full conflict graph for 24-mode pyrazine can be found in Appendix~\ref{app:symmetry_appendix}.

\subsubsection{Benchmark Examples}

The scaling of $\chi$ across the symmetry regimes encountered in $d=24$-mode pyrazine is summarized in Table~\ref{tab:chromatic_scaling}.

\begin{table}[ht]
\centering
\renewcommand{\arraystretch}{1.25}
\caption{Chromatic index $\chi$ of the bilinear mode-pair conflict graph for representative symmetry regimes.
$|\mathcal{G}_2|$ and $|\mathcal{G}_4|$ denote the number of symmetry-allowed on-diagonal and off-diagonal bilinear pairs, respectively.
$\binom{d}{2}$ is the $C_1$ (no symmetry) pair count for comparison.
The depth reduction factor is $\binom{d}{2}/\chi$, representing the parallelism gain from graph coloring relative to fully sequential execution.
}\label{tab:chromatic_scaling}
\begin{tabular}{l c c c c c c}
\hline\hline
Regime & $d$ & $\binom{d}{2}$ & Allowed pairs & $\Delta(G)$ & $\chi$ & Depth reduction \\
\hline
$C_1$ (no symmetry) & $d$ & $\binom{d}{2}$ & $\binom{d}{2}$ & $d{-}1$ & $d{-}1$ or $d$ & $\sim d/2$ \\[4pt]
Pyrazine ($D_{2h}$, on-diag.) & 24 & 276 & 31 & 5 & 5 & $55\times$ \\
Pyrazine ($D_{2h}$, off-diag.) & 24 & 276 & 29 & 5 & 5 & $55\times$ \\[4pt]
LVC (no bilinear) & $d$ & $\binom{d}{2}$ & 0 & 0 & 0 & $\infty$ \\
\hline\hline
\end{tabular}
\end{table}

For pyrazine ($D_{2h}$, $d = 24$), the on-diagonal conflict graph has 24 modes distributed across eight irreps of $D_{2h}$.
The chromatic index $\chi(\mathcal{G}_2^<) = 5$ reflects the largest same-symmetry clique ($A_g$ modes), compared to $d - 1 = 23$ in $C_1$, giving a $4.6\times$ reduction in the number of sequential color layers and a $55\times$ reduction relative to fully sequential execution of all 276 possible pairs.
The off-diagonal also exhibit reduced cost, with only 29 pairs satisfying $B_{1g}$ product symmetry, giving $\chi(\mathcal{G}_4) = 5$ and a $55\times$ depth reduction relative to sequential execution.

For a linear vibronic coupling (LVC) model, bilinear terms are absent entirely ($|\mathcal{G}_2^{<}| = |\mathcal{G}_4| = 0$), so $\chi = 0$ and the bilinear depth contribution vanishes.
The circuit depth is then determined entirely by the intra-mode terms (parallelized via fan-out) and scales as $\mathcal{O}(kM^2 n^2)$, independent of $d$.

Depth reduction from graph coloring complements fan-out parallelism of Section~\ref{sec:fanout}: fan-out eliminates the factor of $d$ from intra-mode terms (each mode executes independently on its own copy of the channel register), while graph coloring reduces the bilinear depth from $\binom{d}{2}$ to $\chi \leq d$.
Together, these two parallelism mechanisms reduce the off-diagonal depth from $\mathcal{O}(kM^2 d^2 n^2)$ (fully sequential, no fan-out, no coloring) to $\mathcal{O}(kM^2 \chi n^2)$, giving a worst-case combined reduction factor of $\sim d^2/\chi \approx d$, or substantially more when molecular symmetry restricts the allowed pairs.
This section therefore shows explicitly that circuit parallelism for nonadiabatic dynamics is directly related to a conflict graph coloring representation of terms allowed by molecular symmetry. 

\section{Numerical Validation}

\subsection{Two-mode Pyrazine Simulation}\label{sec:Pyrazine}
A grid-based real-time evolution of a multi-mode quantum dynamical system is given in this section.
Using the REQWIEM algorithm, gate-based quantum circuits were built using the Qiskit SDK and classically compiled in a statevector simulation. 
A two-mode simulation is given as a proof-of-principle for bilinear terms. 
The Nyquist range limit requires $n=9$ qubits per mode results in a simulation totaling $9+9+1=19$ logical, noiseless (ideal) qubits (Chapter~\ref{ch:wavepacket_prop})

There are several prototypical examples of simple conical intersection problems used in quantum chemistry. 
These are long-standing since the late 1970s to test numerous classical computational methods, including ML-MCTDH and DMRG (Chapter~\ref{ch:Introduction}).
As suggested by previous discussion, pyrazine (a diazine) is chosen as a prime example, with many diazenes having many advantageous qualities for simulation and analysis including ease of synthesis, easily accessible vibrational frequencies (low-UV to near-IR), and high symmetry. 
Following Raab et al.~\cite{raab1999molecular}, the $S_2\rightarrow S_1$ absorption spectrum is considered as a wavepacket propagation problem to confirm the simulation of conical intersections, comparing spectra to MCTDH results~\cite{raab1999molecular}. 

The non-zero terms in the Hamiltonian for only the $v6a$ and $v10a$ modes ($x$ and $y$, respectively) are therefore
\begin{equation}\label{eq:PyrazineHamiltonian}
\begin{split}
    \hat{H} & = \frac{p_x^2}{2\mu_x}+\frac{p_y^2}{2\mu_y}+\begin{pmatrix}
        \frac{1}{2}\mu_x\omega_x^2 & 0 \\ 0 & \frac{1}{2}\mu_x\omega_x^2
    \end{pmatrix}x^2 +\begin{pmatrix}
        \frac{1}{2}\mu_y\omega_y^2 & 0 \\ 0 & \frac{1}{2}\mu_y\omega_y^2
    \end{pmatrix}y^2  \\
    &+\begin{pmatrix}
        -\Delta E & 0 \\ 0 & \Delta E
    \end{pmatrix}
    + \begin{pmatrix}
        a_1 & 0 \\ 0 & b_1
    \end{pmatrix}x + \begin{pmatrix}
        a_2 & 0 \\ 0 & b_2
    \end{pmatrix}x^2\\&+\begin{pmatrix}
        a_3 & 0 \\ 0 & b_3
    \end{pmatrix}y^2+\begin{pmatrix}
        0 & c_1 \\ c_1 & 0
    \end{pmatrix}y + \begin{pmatrix}
        0 & c_2 \\ c_2 & 0
    \end{pmatrix}xy,
    \end{split}
\end{equation}

For the present example, bath modes are not included.
Parameters were chosen to match Raab et al.~\cite{raab1999molecular, woywod1994characterization, stock1995resonance}, and are given in Appendix~\ref{app:pyrazine_params}.
$x$ and $y$ are substituted variables for vibrational mode coordinates $x_i$, representing the $v6a$ and $v10a$ vibrational modes of pyrazine.
Quantum registers are defined for each coordinate, with $n_x=n_y=9$ qubits defining a $2^{9}\times2^{9}$ grid in a box of size $L_x=L_y=8\,\text{a.u}$. 
A vertical excitation is applied from $S_1$ to $S_2$ by rotating the $\lceil \log_2(2)\rceil=1$ channel qubit with a single $R_x(\theta)$ gate, where $\theta = 2\arccos(\sqrt{P_0})$. 
The initial state superposition was set as $P_0=0.75,P_1=0.25$.
The total number of ideal qubits in this classical emulation is therefore $9+9+1=19$.

The mappings $x \mapsto (x-L_x/2)$ and $y \mapsto (y - L_y/2)$ centers the origin of the qubit grid, giving additional shifts to coefficients. 
This does not increase gate count but must be noted for correct dynamics, with on-diagonal terms $V_{00}(x,y)$ and $V_{11}(x,y)$ being a sum of linear and quadratic terms, extending from the harmonic approximation of the potential wells. 

\begin{figure}
    \centerline{
    \includegraphics[width=80mm,scale=0.5]{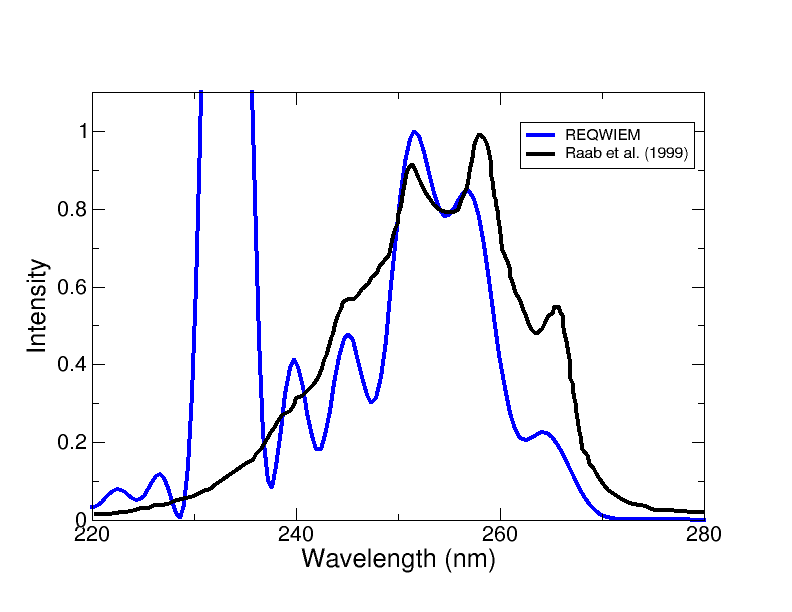}
    \includegraphics[width=70mm,scale=0.5]{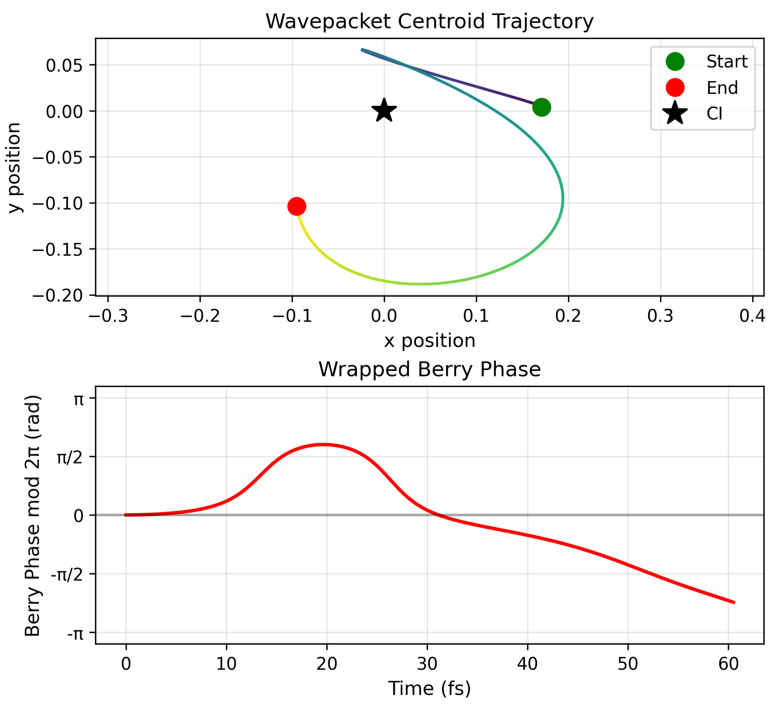}
    }
    \caption{\label{fig:Pyrazine_Compare} (left) Spectra comparison for a two-mode pyrazine system simulated as a real-time evolution using REQWIEM (blue) and a four-mode MCTDH simulation conducted by Raab et al.~\cite{raab1999molecular}. Data were extracted through digitization~\cite{rohatgi2026webplotdigitizer}. REQWIEM results are renormalized around the peak at 251.5 nm. (right) Centroid trajectory of wavepacket initialized at $x=0.175,y=0$ and extracted Berry phase. Spectral intensities are strongly dependent on wavefunction trajectory.
}
\end{figure}

The emulation proceed with a total evolution time $t=2500\,\text{a.u.}\approx 60.47\,\text{fs}$ with a timestep of $t=1.0\,\text{a.u.}\approx 2.419\times10^{-2} \,\text{fs}$. 
The following steps are performed classically for numerical validation of the propagator.
At each timestep, the cross-correlation function is taken as
\begin{equation}
    C(t) = \braket{\psi_1(0)}{\psi_2(t)}.
\end{equation}
This acts as a time-dependent overlap between the initial state and the time-evolved state and can be applied with an electronically-controlled SWAP test, requiring an additional $2n$ qubits. 
Two damping factors are added to the correlation function, the first being an exponential damping $f(t) = e^{-t/\tau_{damp}}$, giving a phenomenological lifetime broadening as a Lorentzian in the frequency domain. 
The second is a Blackman window function to suppress the sides of the autocorrelation function and is typical in grid-based signal processing to dampen Gibbs' ringing.
A fast Fourier transform is then applied to recover the absorption spectrum, shown in Figure~\ref{fig:Pyrazine_Compare}. 
Results can be compared to seminal work by Raab et al.~\cite{raab1999molecular} by imposing the spectra onto a digitized extraction of their four-mode MCTDH results, performed using WebPlotDigitizer~\cite{rohatgi2026webplotdigitizer}.
The additional two modes ($v_1\,, v_{9a}$) from their plot act as bath modes and broaden the spectra closer to what would be seen experimentally.
An additional peak appears around $233$nm, corresponding to the energy at the Franck-Condon excitation point.
This is an artifact of initializing away from equilibrium with a two dimensional Gaussian, with no IVR to dissipate energy and prevent its recurrence.
The five primary peaks from Raab et al.~\cite{raab1999molecular} are faithfully recovered, supporting the proof-of-principle for REQWIEM as an effective method for real-time quantum dynamics in the multi-mode extension.
The Berry phase is calculated classically by assuming coherence, first calculating the centroid of the localized wavepacket $\braket{x_i}(t) = \int x_i|\psi(r,t)|^2dr$.
The cumulative Berry phase $\gamma(t)$ for a semiclassical trajectory is given as half the winding angle $d\theta(t) = (xdy-ydx)/r^2$ as $\gamma(t) = \gamma(t-\tau)+0.5d\theta$.

\begin{figure}
    \centerline{
    \includegraphics[width=150mm,scale=0.5]{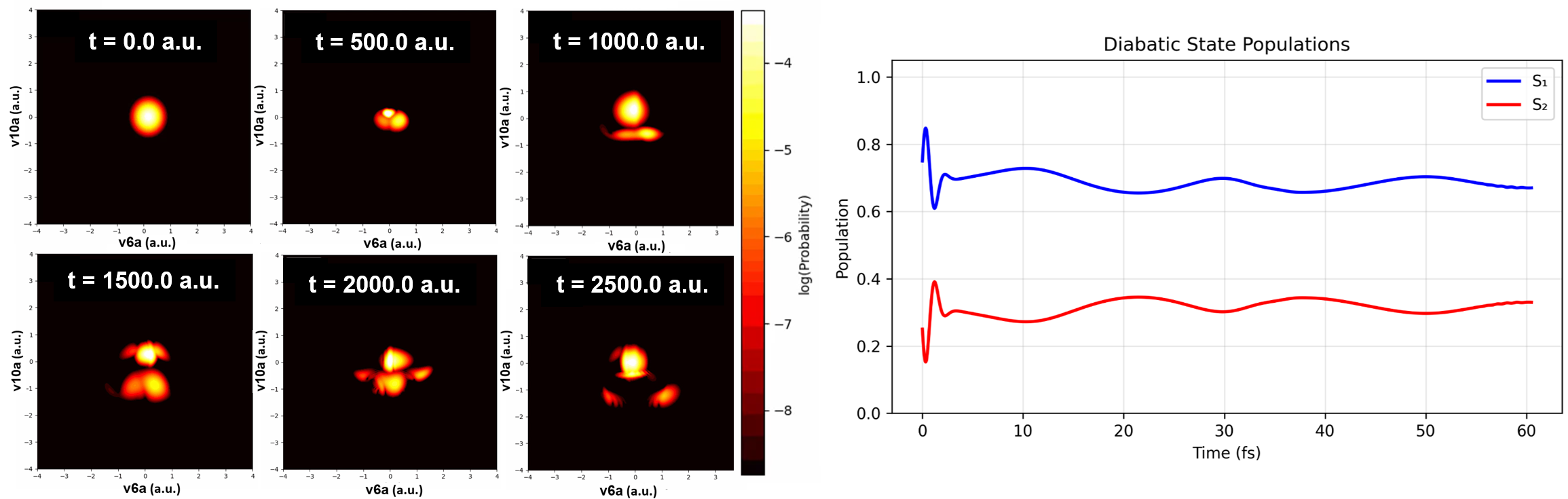}
    }
    \caption{\label{fig:population-and-shape} (left) Snapshots of probability amplitude in position space over time evolution. Visualization shows log(Probability) for clarity.
    (right) Population dynamics over evolution. 
    A sharp population transfer at the beginning of the evolution followed by recurrence every 20 fs shows correct qualitative behavior of a two-mode simulation.
}
\end{figure}

Population dynamics and snapshots of probability amplitudes are shown in Figure~\ref{fig:population-and-shape}.
A fast population transfer is characteristic of initializing mostly in the $S_2$ state.
A recurrence period of around 20 fs with slowly decreasing transfer peaks confers the wavepacket spreading out between the two modes without any bath modes or dark states.

Agreement between REQWIEM and converged MCTDH results for the five primary spectral peaks confirms that the exact gate-level implementation of all Hamiltonian terms produces physically correct dynamics.

\subsection{Four-Mode Pyrazine}\label{sec:4Mode_Pyr}
This section works to scale the classical emulation in Section~\ref{sec:Pyrazine}.
Two bath modes ($v1,v9a$) are added alongside ($v6a,v10a$) to follow the reduced dimensional model considered in Raab et al. \cite{raab1999molecular}.
The box size was reduced from $L=8\,\text{a.u.}$ to $L=4\,\text{a.u.}$, in order to reduce the number of qubits per register from $n=9$ to $n=8$ while retaining the same Nyquist range and grid discretization as the two-mode model, introducing a risk of aliased dynamics from periodic boundary conditions for wavefunction tails.
The total register size for the four-mode model is therefore $8+8+8+8+1=33$ qubits.
The total evolution time was $t=2500\,\text{a.u.}$ with Trotter step size $\tau=1.0\,\text{a.u}$.

The Hamiltonian includes additional terms, including on-diagonal bilinear terms between $v6a\,,v1\,,$ and $v9a$ and off-diagonal bilinear terms between $v1\,,v9a$ and $v10a$. 
Full parameterization can be found in Appendix~\ref{app:pyrazine_params}, using parameters from Raab et al. \cite{raab1999molecular}.

A statevector simulation with $33$ qubits requires special handling, as one full statevector is represented by $2^{33}\approx8.6\times10^9$ amplitudes, converting to $\sim137.4\,\text{GB}$. 
This exceeds tractability for the CPU-based Qiskit SDK, so Nvidia cuQuantum was used, as a Qiskit-compatible GPU-based software kit.
As NA-QMD is characterized by strong correlations, this problem is not suitable for tensor network decomposition to give a computational speedup. 
CuStateVec is used for circuit compilation instead of CuTensorNet as a result.
The program was compiled and ran on the Georgia Advanced Computing Resource Center's Sapelo2 cluster, using four Nvidia A100-SXM4-80GB GPUs for the statevector simulation.
This is a safe number of GPUs required for this computation, requiring additional space for intermediate computations.
The wavefunction was not tracked to reduce I/O overhead.
Total wall time for the full $33$-qubit emulation is approximately 150 hours per run (600 GPU hours per run), with memory-saving operations for intermediate computations. 

Wall clock time can be optimized in several ways for future work. 
The applied circuit is compiled using a static series of primitive gates.
Upon multi-GPU compilation, qubits must be repeatedly re-indexed to perform local multiplications on single GPUs.
In other words, rows and columns of the density matrix are repeatedly swapped upon classical circuit compilation.
Clustering commuting terms in each operator (e.g. $CR_z$ operations in $V_\text{diag}$) according to locally-defined GPU indexes can save large constant factors, especially when compounding between Trotter steps and operations. 
Such an optimization is beneficial in future application on quantum hardware, significantly reducing routing costs by minimizing the path traversed by all qubits.

\begin{figure}
    \centerline{
    \includegraphics[width=\textwidth]{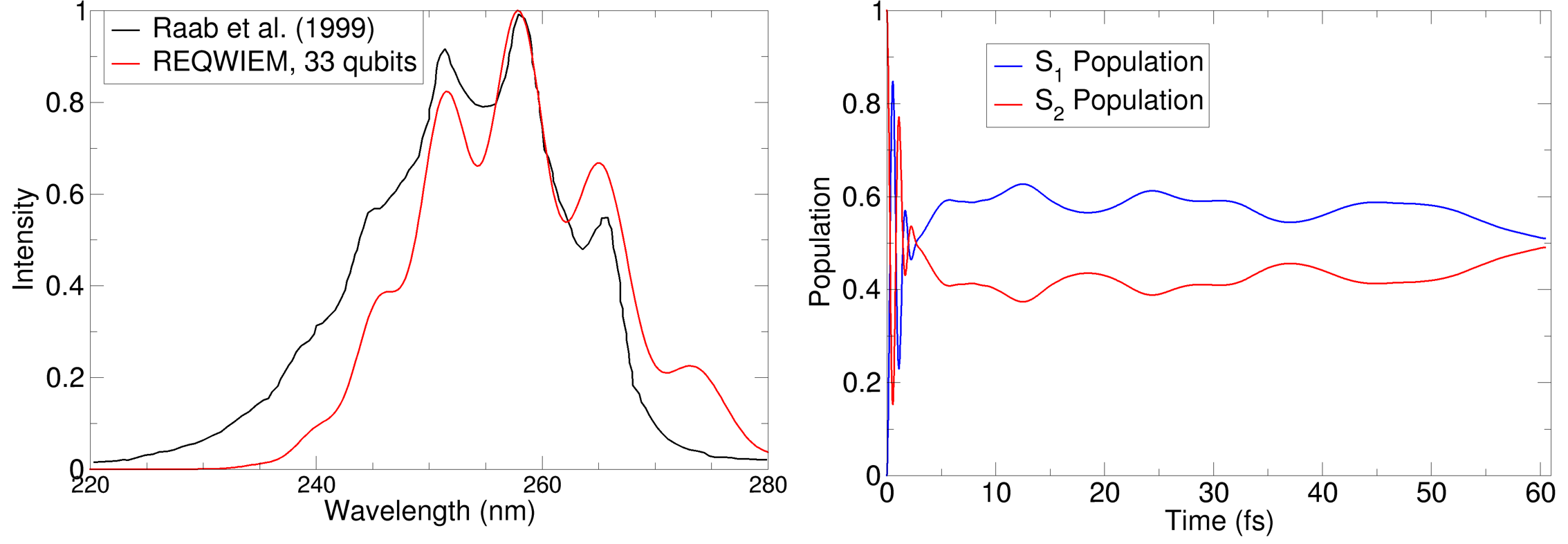}
    }
    \caption{\label{fig:4-Mode_Pyrazine} (Left) Spectra comparison for four-mode pyrazine using REQWIEM (red), compared to Raab et al. \cite{raab1999molecular} (black). 
    Data from \cite{raab1999molecular} was digitized using WebPlotDigitizer \cite{rohatgi2026webplotdigitizer}.
    (Right) Populations of $S_2$ and $S_1$ over time.
}
\end{figure}

For a full statevector simulation, computational resources double for each additional qubit.
This is a well-known phenomenon in the emulation of quantum circuits, highlighting an advantage of quantum computation. 
A consequence remains where emulating with $n=9$ qubits per mode (33 to 37 qubit emulation) requires 32-64 mutually compatible GPUs with exponential computation time compounded with MPI overhead.
For reference, the Frontier supercomputer at Oak Ridge National Laboratory in Oak Ridge, Tennessee, USA, has $37,888$ AMD Instinct M1250X GPUs. 
This is sufficient to simulate a $46-47$-qubit statevector for chaotic systems where negligible efficiency is brought by tensor decomposition), prior to computation time considerations.
This is the bottleneck of classical compiling quantum circuits simulating dense Hamiltonians.
The resulting spectra and populations are shown in Figure~\ref{fig:4-Mode_Pyrazine}.

\subsection{Multi-channel Validation Beyond Two Electronic States}\label{sec:LZ}
A simple four-channel extension considers a Landau-Zener bowtie model, consisting of four linear potential surfaces and four constant couplings. 
As shown in other examples, the surfaces need not be linear, nor the couplings constant. 
In the diabatic basis, the Hamiltonian takes the form
\begin{equation}
    \hat{H}(x) = -\frac{\partial^2}{\partial x^2}\frac{1}{2\mu}\mathbb{1}^4 + \begin{bmatrix}
        V_{00}(x) & \lambda_{01} & 0 & \lambda_{03} \\
        \lambda_{10} & V_{11}(x) & \lambda_{12} & 0 \\
        0 & \lambda_{21} & V_{22}(x) & \lambda_{23} \\
        \lambda_{30} & 0 & \lambda_{32} & V_{33}(x)
    \end{bmatrix},
\end{equation}
where
\begin{equation}
    \begin{split}
        V_{00}(x) & =\beta_1x + E+ V_0 \\
        V_{11}(x) & = -\beta_1x +V_0 \\
        V_{22}(x) & = \beta_2x +V_0 \\
        V_{33}(x) & = -\beta_2x - E + V_0.
    \end{split}
\end{equation}
Diabatic state slopes are given as $\beta_{1,2}$. 
$E$ represents an energy bias as a vertical offset, and $V_0$ is a reference energy.
The Landau-Zener bowtie model is exceptionally well-suited to exhibit the multi-channel extension, with classical numerical solutions easy to construct and solve without significant compute time.
The four-channel example exercises the XOR-class fragmentation with $M = 4$ ($m = 2$), meaning three XOR classes and six fragment applications per Trotter step. 
The focusing Clifford varies across fragments (different Hamming weights) (Algorithm~\ref{alg:single_frag}).
A Gaussian wavefunction with width $\sigma = 0.5$ represents a particle with mass $\mu = 2000.0~\text{a.u.}$ on a grid with $n=12$ qubits in the $V_{00}$ channel.
The box size is $L=80~\text{a.u.}$ with periodic boundary conditions. 
Electronic channels are stored in $\lceil\log_2(4)\rceil=2$ channel qubits, totaling $12+2=14$ qubits for this emulation.
Initial position is set to $x=40~\text{a.u.}$ (middle of the box), with $p_0 = 0~\text{a.u.}$, a total evolution time of $5000~\text{a.u.}$ with a timestep of $\tau = 0.1~\text{a.u.}$

Quantum circuit compilation is validated against a classical diagonalization on the same finite grid with $2^{12} = 4096$ grid points.
The classical reference constructs the full $4N\times 4N$ coupled Hamiltonian matrix, where the kinetic energy operator is represented as an FFT-based circulant matrix that remains exact on the periodic finite grid, matching the QFT-based kinetic energy used by the quantum circuit.
The coupled Hamiltonian is then diagonalized directly (via \texttt{scipy.linalg.eigh}), yielding all eigenvalues $\{E_k\}$ and eigenvectors $\{|\phi_k\rangle\}$.
From the decomposition, the autocorrelation function is computed as
\begin{equation}
    C(t) = \langle\psi_0|e^{-i\hat{H}t}|\psi_0\rangle = \sum_k |c_k|^2\, e^{-iE_k t},
\end{equation}
where $c_k = \langle\phi_k|\psi_0\rangle$ are expansion coefficients of the initial state.
Fourier transforming with Gaussian damping ($\tau_d = 1000~\text{a.u.}$) and Blackman window identical to the time evolution yields the autocorrelation spectrum.
Since many eigenstates carry nonzero overlap with the initial wavepacket across four coupled potential surfaces, a comb-like eigenvalue structure results; the windowed Fourier transform broadens these into resolved spectral peaks.

To quantify spectral convergence across the parameter space, the Wasserstein-1 distance (earth mover's distance) is computed between the classical reference spectrum and each quantum simulation trial spectrum~\cite{villani2009optimal}. 
Both spectra are restricted to a common energy window and $L^1$-normalized to define probability distributions, and the one-dimensional Wasserstein-1 distance
\begin{equation}
    W_1(P_{\text{ref}}, P_{\text{trial}}) = \inf_{\gamma \in \Gamma(P_{\text{ref}}, P_{\text{trial}})} \int |E - E'|\, d\gamma(E, E'),
\end{equation}
measures the minimum cost of transporting one spectral distribution into the other.
$\gamma(E,E')$ acts as a a joint probability distribution on pairs $(E,E')$ with imposed marginals $P_{\text{ref}}$ and $P_{\text{trial}}$. 
`$\inf$' denotes that this integral is unbounded on $\gamma$.
This metric is well-suited to spectral comparison, being sensitive to both peak position shifts and intensity redistribution, unlike pointwise metrics requiring spectra to be evaluated on identical grids.
A parameter sweep over grid resolution ($n = 7$--$12$ qubits), Trotter step size ($\tau = 0.1$, $0.5$, $1.0~\text{a.u.}$), and box size ($L = 10$, $20$, $40$, $80~\text{a.u.}$) maps the convergence landscape, with $W_1 \to 0$ indicating agreement with the exact result. 
This convergence analysis is presented in Appendix~\ref{app:pyrazine_params}.

\begin{figure}
    \centerline{
    \includegraphics[width=\textwidth]{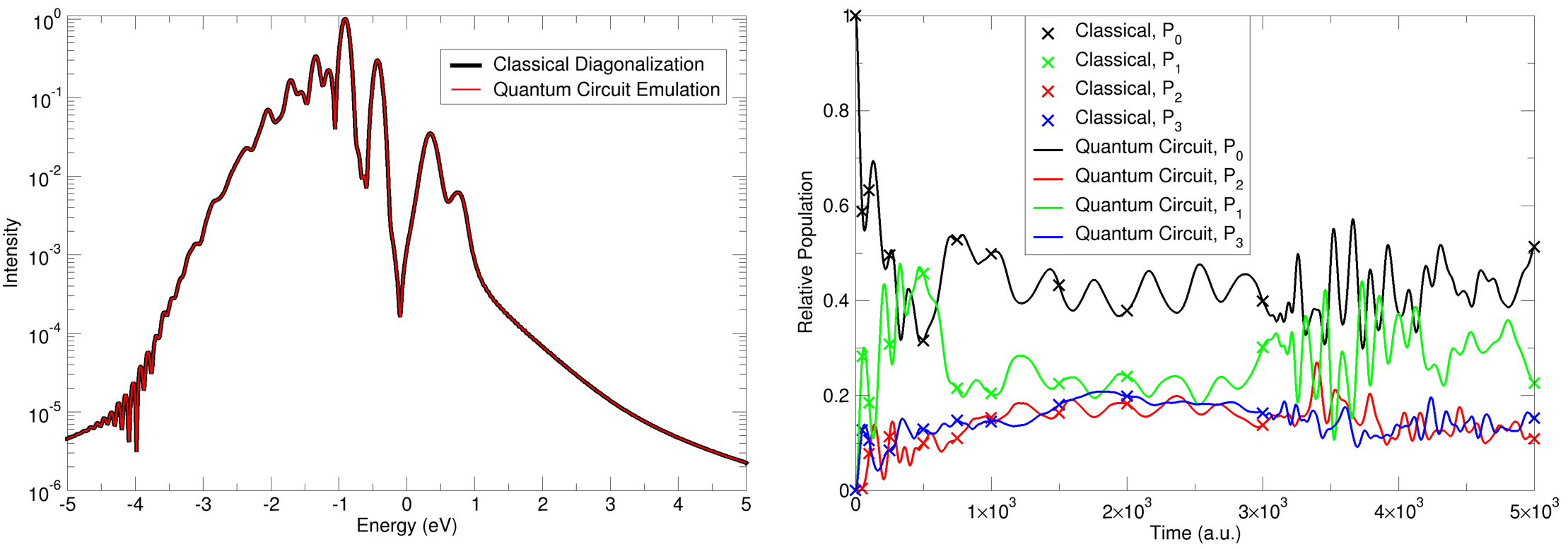}
    }
    \caption{\label{fig:LZ_Bowtie_comparison} (Left) Comparison between spectra obtained from Trotterized quantum dynamics (REQWIEM, red) and classical exact diagonalization (black). The classical reference constructs and diagonalizes the full $4N\times 4N$ Hamiltonian on the same finite grid, with spectral convergence quantified by the Wasserstein-1 distance. With matching parameters, this distance is $W_1=6.86\times10^{-7}\,\text{eV}$ with a Trotter step $\tau=2.42\times10^{-3}\,\text{fs}$.
    (Right) Population dynamics for each diabatic surface. Xs denote classically derived populations at select times, with a $<0.5\%$ difference from quantum circuit compilation.}
\end{figure}

Classical quantum circuit emulation results are compared to the exact diagonalization in Figure~\ref{fig:LZ_Bowtie_comparison}. 
The classical diagonalization follows the same initial conditions ($N = 4096$ grid points, $L=80~\text{a.u.}$, $\mu = 2000~\text{a.u.}$) and the same Hamiltonian.
The real-time evolution algorithm converges on the same spectral structure as the classical exact diagonalization with sufficient system metrics ($n$, $t$, $\tau$, $L$), as confirmed by monotonically decreasing Wasserstein-1 distance with increasing grid resolution and decreasing timestep, supporting REQWIEM as a viable time-evolution algorithm in the multi-channel case.
This convergence analysis is given in Appendix~\ref{app:pyrazine_params}
This emulation also serves as a numerical proof-of-principle for a quantum algorithm modeling multi-channel quantum dynamics.

\section{Summary}

This chapter specifies the REQWIEM algorithm, detailing the application of each term in the Hamiltonian as parameterized quantum circuits. 
The multi-mode and multi-channel extensions are numerically validated against classical methods (MCTDH, Lanczos) for two-mode, four-mode, and four-channel systems spanning 14-33 ideal qubits, using classical compilation of quantum circuits (Figures~\ref{fig:LZ_Bowtie_comparison}, \ref{fig:4-Mode_Pyrazine}, and~\ref{fig:Pyrazine_Compare}).
The fan-out protocol was not implemented in these validations, requiring additional qubits for intermediate operations. 
Converged spectra, dampened recurrence in the four-mode pyrazine model, and Berry phase accumulation around the conical intersection numerically validate REQWIEM as an algorithm capable of handling NA-QMD (Figures~\ref{fig:Pyrazine_Compare} and~\ref{fig:population-and-shape}). 

Gate count scales with the number of variables, as $\mathcal{O}(kM^2d^2n^2)$ with $d$ modes, $n$ qubits per mode, and $M$ electronic channels, with $k\propto t$.
REQWIEM contains additive terms that are subdominant to $\mathcal{O}(kM^2d^2n^2)$, including $M\log m$ Toffoli gates for truth value ancilla computation, and $\mathcal{O}(\log d)$ layers of CNOTs for channel register fan-out.
Both are given to increase algorithmic parallelism.
Even though proof is not given that REQWIEM has optimal gate {count}, optimal gate scaling is clear for a grid-based representation from the information theoretic bound.
A detailed resource analysis is given in the next chapter, directly comparing REQWIEM to the phase gradient approach of Motlagh et al. \cite{motlagh2025quantum}, following several adjustments for consistent system complexity and parameterizations.
Symmetry constraints and channel register fan-out scalably reduce circuit depth to $\mathcal{O}(kM\chi n^2)$, where $\chi \propto d$, with explicit depth calculations performed in the forthcoming chapter
. 

\chapter{Resource Analysis}\label{ch:resources}
Multidimensional, multi-channel quantum algorithms for NA-QMD have just begun to be considered.
Ollitrault et al. \cite{ollitrault2020nonadiabatic} suggests extensivity of their rotation-based, first-quantized approach, but is restricted to one mode with linearly approximated coupling, while Motlagh et al. \cite{motlagh2025quantum} relies on phase gradients loading data from a QROM oracle, with published resource estimations omitting Duschinsky mixing and bilinear vibronic coupling terms. 
Both restrict themselves to a second-order expansion of the vibronic Hamiltonian. 
REQWIEM directly extends Ollitrault et al. \cite{ollitrault2020nonadiabatic} and incorporates improved application of multi-controlled gate operations. 
This chapter gives a quantitative resource comparison between REQWIEM and Motlagh et al. \cite{motlagh2025quantum}, assuming validity of the latter sans numerical validation. 
Resources are recalculated from the existing literature \cite{motlagh2025quantum}, following discussion on sufficient model complexity and system size requirements (Section~\ref{sec:direct_comparison}).
Detailed methodology for this recalculation can be found in Appendix~\ref{app:Motlagh_Resource}.
Once a comparison is made, an estimated quantum advantage is found for REQWIEM when compared against compute times for ML-MCTDH.

To review notation, $d$ represents the number of vibrational modes, $M$ is the number of electronic channels, $m=\lceil \log_2 M \rceil$ is the number of channel qubits in $\ket{s_{\text{channel}}}$, and $n$ is the number of qubits per mode register.
All $d$ registers are taken as having $n$ qubits for ease of calculation, though optimization can result in large constant factors of depth reduction for both methods.
Five benchmark systems are defined, four taken from Motlagh et al. \cite{motlagh2025quantum}, being Anthracene ($d=21$, $M=6$, $n=10$) \cite{pradhan2022design}, a $(NO)_4-$Anthracene dimer ($d=19$, $M=5$, $n=10$) \cite{pradhan2023triplet}, an Anthracene-Fullerene dyad ($d=246$, $M=4$, $n=10$) \cite{xie2015full}, and a reduced dimensional Anthracene-Fullerene dyad ($d=11, M=4, n=10$) \cite{xie2015full}.
The fifth benchmark is pyrazine, as given in Raab et al. \cite{raab1999molecular} with $d=24, M=2,n=10$.
A position register size $n=10$ is kept constant between modes despite potential for optimization to clarify scaling behaviors. 
As discussed in Chapter~\ref{ch:wavepacket_prop}, the size of each register should be chosen according to the characteristic energy, optimizing grid resolution and Trotter error. 
Resources given here are used as a standard between REQWIEM and Motlagh et al. \cite{motlagh2025quantum} for an apples-to-apples comparison between QROM-loading and rotation synthesis.

Resource estimates primarily rely on Non-Clifford depth and $T$-gate count, extended to $T$-depth and full algorithmic depth.
Focus is directed toward the exact gate counts of REQWIEM, with additional information on resource estimates in Appendix~\ref{app:Motlagh_Resource}.
A broader comparison against quantum signal processing and qubitization methods, which offer asymptotically optimal query complexity but incur oracle-construction overhead, is given in Chapter~\ref{ch:comparison}.

\section{REQWIEM Gate Counts and Circuit Depth}\label{sec:reqwiem_costs}

Exact gate counts for rotations, CNOTs, and Toffolis are derived for each component of the REQWIEM Trotter step, alongside per-step cost formulas in terms of the symmetry index sets $\mathcal{G}_1$, $\mathcal{G}_2$ (on-diagonal) and $\mathcal{G}_3$, $\mathcal{G}_4$ (off-diagonal).
Throughout, a mode indexed by $r$ is taken to have $n_r$ qubits (grid points $N_r = 2^{n_r}$, spacing $\Delta_r = L_r/N_r$).

On-diagonal and off-diagonal potentials are polynomials in the position operators $x_r$.
In the computational basis, $|x_r\rangle = |\Delta_r(x_r - N_r/2)\rangle$, with the integer $x_r = \sum_{\ell=0}^{n_r-1} 2^\ell q_\ell^{(r)}$ encoded in the qubit register.
Any polynomial function of $x_r$ decomposes into a multilinear polynomial in the individual qubits $\{q_\ell^{(r)}\}$, since $q_\ell^2 = q_\ell$ for binary variables.
Linear, diagonal-quadratic, and bilinear decompositions yield $n_r$ single-qubit terms, $\binom{n_r}{2}$ intra-mode cross terms, and $n_r n_s$ inter-mode cross terms respectively.
Full derivations for these operations are given in Appendix~\ref{app:binary_decomp}.

The term `depth' initially neglects all rotation synthesis costs for fault-tolerant architectures.
This is later converted into $T$-depth under rotation synthesis heuristics, where a full depth is later calculated.

\subsection{Kinetic energy}\label{sec:kin_cost}

The kinetic operator $T = \sum_r p_r^2/2\mu_r$ is diagonal in the momentum basis and involves no electronic-state dependence.
Each mode $r$ requires a forward QFT, phase kicks for $p_r^2$ using the binary decomposition of $p_r^2$, and an inverse QFT.
Since there is no electronic-state multiplexing, each single-qubit contribution is a plain $R_z$, and each intra-mode cross pair uses a CNOT--$R_z$--CNOT sandwich.

Gate counts per mode:
\begin{equation}\label{eq:kin_per_mode}
    R_{\text{kin}}^{(r)} = n_r + \binom{n_r}{2}, \qquad
    C_{\text{kin}}^{(r)} = n_r(n_r - 1), \qquad
    T_{\text{kin}}^{(r)} = 0.
\end{equation}
Cross terms $q_\ell q_{\ell'}$ within each mode register require only a CNOT--$R_z$--CNOT sequence ($2$ CNOTs and $1$ rotation per pair), with no ancilla or Toffoli gate.
The QFT contributes additional CNOTs and single-qubit gates:
$C_{\text{QFT}}^{(r)} = 3n_r(n_r - 1) + 6\lfloor n_r/2\rfloor$.

Totals (summed over all modes, counted once per kinetic evaluation):
\begin{equation}\label{eq:kin_total}
    R_{\text{kin}} = \sum_r R_{\text{kin}}^{(r)}, \qquad
    T_{\text{kin}} = 0.
\end{equation}
All $d$ modes are independent, so the kinetic depth equals the cost of a single mode:
\begin{equation}\label{eq:kin_depth}
    D_{\text{kin}} = 2\,D_{\text{cQFT}}(n_{\max}) + n_{\max} + \binom{n_{\max}}{2},
\end{equation}
where $n_{\max} = \max_r n_r$ and $D_{\text{cQFT}}(n)$ is the depth of cQFT on $n$ qubits.

\subsection{On-diagonal potential}\label{sec:ondiag_cost}

The on-diagonal potential uses the three-tier hybrid decomposition of Chapter~\ref{ch:model_ext}.
Costs are derived for one half-step $e^{-iV_{\text{d}}\tau/2}$; the per-Trotter-step cost is twice these values.

\paragraph{Tier~1: Reference surface (uncontrolled).}
The full linear potential of the reference surface $s = 0$ is applied unconditionally.
This contributes $d$ constant-phase rotations and $dn$ single-qubit $R_z$ gates (from the linear and diagonal-quadratic terms, using $q_\ell^2 = q_\ell$), with no CNOT or Toffoli overhead:
\begin{equation}\label{eq:ondiag_ref}
    R_z^{(\text{ref})} = d(1 + n), \qquad
    C^{(\text{ref})} = 0, \qquad
    T^{(\text{ref})} = 0.
\end{equation}

\paragraph{Tier~1: Correction surfaces (truth-value--controlled).}
For each of the $M - 1$ correction surfaces $s \neq 0$, the single-qubit correction potential
\begin{equation}
    \Delta V_s^{(\mathrm{sq})}(\mathbf{x}) = \Delta E_s + \sum_{r=1}^{d}\sum_{\ell=0}^{n-1}\Delta\varphi_{s,\ell}^{(r)}\;q_\ell^{(r)},
\end{equation}
where $\Delta E_s = E_s^{(0)} - E_0^{(0)}$ and
\begin{equation}
    \Delta\varphi_{s,\ell}^{(r)} = \tau\bigl[(\alpha_s^{(r)} - \alpha_0^{(r)})\,2^\ell\,\Delta_r + (a_s^{(r)} - a_0^{(r)})\,2^{2\ell}\,\Delta_r^2\bigr],
\end{equation}
encodes the combined linear gradient and diagonal-quadratic curvature difference for qubit~$\ell$ in mode~$r$.
Because intrastate quadratic terms $x_r^2$ are always symmetry-allowed, curvature corrections $a_s^{(r)} - a_0^{(r)}$ can be nonzero on all $d$ modes, not only the $|\mathcal{G}_1|$ modes with nonzero linear coupling.
The constant offset $\Delta E_s$ distributes equally across $d$ modes, absorbed into a designated qubit $\ell_0$ per mode as $\mathrm{CR}_z(\tau\,\Delta E_s/d)$.
Each correction is controlled on a single local qubit: the channel copy $s_k^{(r)}$ if $\mathrm{popcount}(s) = 1$, or the truth-value copy $b_s^{(r)}$ otherwise, with each $\mathrm{CR}_z$ decomposed as $\mathrm{CNOT}_{c\to t}\cdot R_z(\theta)_t \cdot \mathrm{CNOT}_{c\to t}$ (1~rotation, 2~CNOTs per gate).

Per correction surface $s$, per mode $r$: $(1 + n)$ rotations and $2(1 + n)$ CNOTs.
Summing over all $d$ modes and $M - 1$ correction surfaces:
\begin{equation}\label{eq:ondiag_corr}
    R_z^{(\text{corr})} = d(M{-}1)(1 + n), \qquad
    C^{(\text{corr})} = 2\,d(M{-}1)(1 + n), \qquad
    T^{(\text{corr})} = 0.
\end{equation}

Total Toffoli overhead for computing and uncomputing truth-value ancillas across all surfaces is:
\begin{equation}\label{eq:mux_toffoli}
    T_{\text{truth}} = 2\!\sum_{\alpha=2}^{m}\binom{m}{\alpha}(\alpha - 1) = m \cdot 2^m - 2^{m+1} + 2.
\end{equation}

\paragraph{Tier~2: Intra-mode quadratic cross terms (UCR$_Z^{(m)}$).}
State-dependent curvatures $a_{rr}^{(s)} x_r^2$ produce $\binom{n}{2}$ intra-mode cross sites per mode after binary expansion.
Each site is implemented via a CNOT--UCR$_Z^{(m)}$--CNOT sandwich, requiring $M$ rotations and $(M + 1)$ CNOTs per site.
Tier~2 totals are:
\begin{equation}\label{eq:ondiag_tier2}
    R^{(\text{T2})} = M\,d\binom{n}{2}, \qquad
    C^{(\text{T2})} = (M{+}1)\,d\binom{n}{2}, \qquad
    T^{(\text{T2})} = 0,
\end{equation}
Since Phase~3 fans out the $m$ channel register qubits alongside truth-value ancillas, each mode possesses a local copy of the full $m$-qubit control set.
The Gray-code CNOT sequence within each UCR$_Z^{(m)}$ depends only on $m$, so all $d$ copies undergo identical bit-flip schedules and remain synchronized throughout the traversal; mode-dependent data enters only through $R_z$ angles, acting on independent vibrational targets.
All $d$ modes execute their Tier~2 circuits in parallel, with $\binom{n}{2}$ sites serialized within each mode.

\paragraph{Tier~3: Bilinear cross terms (UCR$_Z^{(m)}$).}
The on-diagonal bilinear terms $a_{rs}^{(s)} x_r x_s$ for mode pairs $(r,s) \in \mathcal{G}_2^{<}$ couple qubits across different mode registers.
Per site: $M$ rotations, $(M + 1)$ CNOTs, $0$ Toffoli.
The Tier~3 totals are:
\begin{equation}\label{eq:ondiag_tier3}
    R^{(\text{T3})} = M\,|\mathcal{G}_2^{<}|\,n^2, \qquad
    C^{(\text{T3})} = (M{+}1)\,|\mathcal{G}_2^{<}|\,n^2, \qquad
    T^{(\text{T3})} = 0,
\end{equation}
As with Tier~2, each mode pair uses fanned-out channel copies for its UCR$_Z^{(m)}$ controls.
Mode pairs $(r,s)$ and $(r',s')$ with $\{r,s\}\cap\{r',s'\} = \emptyset$ act on disjoint vibrational registers and disjoint channel copies, executing in parallel.
Mode-pair conflict graphs determine chromatic scheduling; Tier~3 is serialized over $\chi(\mathcal{G}_2^{<})$ color classes, with all pairs in each class executing simultaneously.
Tier~3 operates on cross-mode qubit pairs while Tiers~1--2 are purely intra-mode, the vibrational qubits of a mode participating in a Tier~3 bilinear term must be free from Tier~1--2 operations; Tiers~1 and~2 therefore complete before Tier~3 begins.

\paragraph{Half-step totals.}
Summing the reference, $M - 1$ corrections, and Tier~2/3 contributions:
\begin{equation}\label{eq:Rd_half}
    R_{\text{d}}^{(\text{half})} = \underbrace{d\,M\,(1{+}n)}_{\text{Tier~1}} + \underbrace{M\,d\binom{n}{2}}_{\text{Tier~2}} + \underbrace{M\,|\mathcal{G}_2^{<}|\,n^2}_{\text{Tier~3}}
    = M\!\left[d\!\left(1 + n + \binom{n}{2}\right) + |\mathcal{G}_2^{<}|\,n^2\right],
\end{equation}
where the Tier~1 reference $d(1{+}n)$ and corrections $d(M{-}1)(1{+}n)$ combine into $dM(1{+}n)$.
The total CNOT count per half-step (excluding fan-out and truth-value Toffoli) is:
\begin{equation}\label{eq:Cd_half}
    C_{\text{d}}^{(\text{half})} = 2\,d(M{-}1)(1{+}n) + (M{+}1)\!\left[d\binom{n}{2} + |\mathcal{G}_2^{<}|\,n^2\right].
\end{equation}
The only Toffoli gates are those for truth-value computation:
\begin{equation}\label{eq:Td_half}
    T_{\text{d}}^{(\text{half})} = T_{\text{truth}} = m \cdot 2^m - 2^{m+1} + 2.
\end{equation}

\paragraph{Depth per half-step (with fan-out).}
Phase~3 fans out both the $m$ channel register qubits and the $M - 1 - m$ truth-value ancillas to $d$ local copies, enabling mode-parallel execution across all tiers.
Fan-out encompasses Tiers~1--3; fan-in occurs after Tier~3 completes:
\begin{equation}\label{eq:Dd_half}
\begin{split}
    D_{\text{d}}^{(\text{half})} &= \underbrace{2D_{\text{truth}}}_{\text{compute/uncompute}} + \underbrace{2\lceil\log_2 d\rceil}_{\text{fan-out/in}} + \underbrace{(n{+}1) + 3(M{-}1)}_{\text{Tier~1 (mode-parallel)}} \\
    &\quad + \underbrace{\binom{n}{2}(2M{+}1)}_{\text{Tier~2 (mode-parallel)}} + \underbrace{\chi(\mathcal{G}_2^{<})\,n^2\,(2M{+}1)}_{\text{Tier~3 (color-serial)}},
\end{split}
\end{equation}
where $D_{\text{truth}}$ is the depth of the Toffoli cascade for truth-value computation.
The Tier~1 reference contributes depth $n + 1$ (all modes parallel, uncontrolled).
For each correction surface~$s$, the $(1{+}n)$ $\mathrm{CR}_z$ gates per mode share a read-only control but target distinct vibrational qubits, executing in parallel at depth~$3$; after fan-out, all $d$ modes execute simultaneously, giving per-surface correction depth $3$ independent of both $d$ and $n$.
The $M - 1$ surfaces serialize (same vibrational targets, different controls), giving total Tier~1 correction depth $3(M{-}1)$.
Tier~2 is mode-parallel: each mode's $\binom{n}{2}$ intra-mode UCR$_Z^{(m)}$ circuits run on independent vibrational registers and synchronized local channel copies, serialized only over the $\binom{n}{2}$ sites within each mode.
Tier~3 serializes over $\chi(\mathcal{G}_2^{<})$ chromatic layers; within each layer, with all scheduled mode pairs executing simultaneously on disjoint registers and disjoint channel copies.
The dominant depth contribution is Tier~3 for systems with many bilinear pairs, being the terms that drive IVR.

\subsection{Off-diagonal potential}\label{sec:offdiag_cost}

Off-diagonal coupling uses XOR-class fragmentation with per-fragment Clifford diagonalization and UCR$_Z^{(m-1)}$.
The symmetric Trotter product $\mathcal{V}_\text{od}(\tau)$ applies each of $M - 1$ fragments twice (forward and backward), for a total of $2(M - 1)$ fragment applications per step.
The costs below are for the full off-diagonal step.

Each fragment $\zeta$ contributes a rotation count that is independent of $\zeta$, since the focusing Clifford uniformly reduces the UCR control count to $m - 1$.
The $M/2 = 2^{m-1}$ rotations per site arise from the Gray-code decomposition of UCR$_Z^{(m-1)}$.

Prior to group theoretic constraints, the rotation sites per fragment decompose into four classes:
\begin{center}
\renewcommand{\arraystretch}{1.15}
\begin{tabular}{l c c c}
\hline\hline
Site type & Count & $R_z$/site & CNOT/site \\
\hline
Constant-phase & $d$ & $M/2$ & $M/2 + 1$ \\
Single-qubit (linear) & $dn$ & $M/2$ & $M/2 + 1$ \\
Intra-mode cross ($r^2$) & $d\binom{n}{2}$ & $M/2$ & $M/2 + 3$ \\
Bilinear cross ($r s$) & $|\mathcal{G}_4|\,n^2$ & $M/2$ & $M/2 + 3$ \\
\hline\hline
\end{tabular}
\end{center}
Constant-phase and single-qubit sites each consist of a $\mathrm{CNOT}_{s_1\to\ell}$--UCR$_Z^{(m-1)}$--$\mathrm{CNOT}_{s_1\to\ell}$ sandwich, where the Gray-code UCR contributes $M/2$ rotations interleaved with $M/2 - 1$ CNOTs and the outer $s_1$ pair adds $2$ CNOTs, totaling $M/2 + 1$ per site.
Intra-mode cross sites and bilinear cross sites include an outer CNOT pair for the $ZZ$ encoding ($\mathrm{CNOT}_{\ell'\to\ell}$ before and after the $s_1$ sandwich), adding $2$ more CNOTs for a total of $M/2 + 3$ per site.

The total rotation count per fragment is
\begin{equation}\label{eq:Rzeta_2}
    R_\zeta = \frac{M}{2}\,S_{\text{sites}}, \qquad
    S_{\text{sites}} = d(1{+}n) + d\binom{n}{2} + |\mathcal{G}_4|\,n^2,
\end{equation}
where $S_{\text{sites}}$ is the total number of rotation sites per
fragment.
The aggregate rotation count across all $2(M - 1)$ fragment
applications is:
\begin{equation}\label{eq:Rod}
   R_{\text{od}} = 2(M{-}1) \cdot \frac{M}{2}\,S_{\text{sites}} = M(M{-}1)\,S_{\text{sites}}.
\end{equation}

Off-diagonal CNOTs decompose into three sources: focusing Cliffords,
fan-out/fan-in, and UCR circuits.

The per-fragment UCR CNOT count (uniform across $\zeta$) is:
\begin{equation}\label{eq:Nucr_frag}
\begin{split}
    N^{(\text{UCR})} &= d\!\left[\bigl(\tfrac{M}{2}{+}3\bigr)\binom{n}{2}
      + \bigl(\tfrac{M}{2}{+}1\bigr)n
      + \bigl(\tfrac{M}{2}{+}1\bigr)\right]
      + |\mathcal{G}_4|\,n^2\bigl(\tfrac{M}{2}{+}3\bigr) \\
    &= d\bigl(\tfrac{M}{2}{+}1\bigr)(1{+}n)
      + d\bigl(\tfrac{M}{2}{+}3\bigr)\binom{n}{2}
      + |\mathcal{G}_4|\,n^2\bigl(\tfrac{M}{2}{+}3\bigr),
\end{split}
\end{equation}
where the first line groups by site type and the second line collects constant-phase and single-qubit terms (both with coefficient $M/2 + 1$).

The total off-diagonal CNOT count per Trotter step is:
\begin{equation}\label{eq:Cod}
    C_{\text{od}} = \underbrace{4\!\left[m\cdot 2^{m-1} - M + 1\right]}_{\text{Clifford}}
      + \underbrace{4(M{-}1)\,m(d{-}1)}_{\text{fan-out/in}}
      + \underbrace{2(M{-}1)\,N^{(\text{UCR})}}_{\text{UCR circuits}}.
\end{equation}

The Toffoli count $T_{\text{od}}$ is zero.
No Toffoli gates appear in the off-diagonal block: all interactions are encoded through the CNOT--UCR$_Z^{(m-1)}$--CNOT sandwich structure.

Each fragment application has depth:

\begin{equation}\label{eq:Dfrag}
    D_\zeta = \underbrace{2h(\zeta)}_{\text{Clifford pair}}
      + \underbrace{2\lceil\log_2 d\rceil}_{\text{fan-out/in}}
      + \underbrace{D^{(\text{UCR})}}_{\text{UCR block}},
\end{equation}

where the UCR block depth includes mode-parallel intra-mode contributions and color-serialized bilinear contributions:

\begin{equation}\label{eq:Ducr_perfrag}
    D^{(\text{UCR})} = \underbrace{(M{+}3)\binom{n}{2}
      + (M{+}1)\,n + (M{+}1)}_{\text{intra-mode (mode-parallel)}}
      + \underbrace{\chi(\mathcal{G}_4)\,n^2\,(M{+}3)}_{\text{bilinear (color-serial)}}.
\end{equation}

The depth per site type is $M + 1$ for constant-phase and single-qubit sites ($s_1$ sandwich $+$ Gray-code UCR), and $M + 3$
for cross-term and bilinear sites (additional outer CNOT pair for $ZZ$ encoding).
The intra-mode terms (constant-phase, single-qubit, and intra-mode cross) are parallelized across all $d$ modes via fan-out of the $m$ channel qubits. 
Bilinear terms serialize over $\chi(\mathcal{G}_4)$ coloring layers of the off-diagonal mode-pair conflict graph.
The UCR depth is independent of $\zeta$ because the focusing Clifford uniformly produces $m - 1$ UCR controls regardless of $h(\zeta)$ (using $M = 2^m$ throughout: $M + 3 = 2^m + 3$, etc.).

The symmetric product applies each fragment twice, giving
\begin{equation}
\begin{split}
    D_{\text{od}} &= 2\!\sum_{\zeta=1}^{M-1} D_\zeta
      = 2(M{-}1)\bigl[D^{(\text{UCR})} + 2\lceil\log_2 d\rceil\bigr]
        + 4\!\sum_{\zeta=1}^{M-1}h(\zeta) \\
    &= 2(M{-}1)\bigl[D^{(\text{UCR})} + 2\lceil\log_2 d\rceil\bigr] + 2mM,
\end{split}
\end{equation}
where the identity $\sum_{\zeta=1}^{M-1}h(\zeta) = m \cdot 2^{m-1}$ gives aggregate Clifford depth.
The dominant term is the UCR contribution $2(M{-}1)\,D^{(\text{UCR})} = \mathcal{O}(M^2 n^2 \chi)$, where $\chi = \chi(\mathcal{G}_4)$ is the chromatic index of the off-diagonal mode-pair conflict graph.

\subsection{Per-Trotter-Step Fan-Out Costs}\label{sec:fanout_costs}

CNOT costs of the fan-out and fan-in operations, aggregated over one full Trotter step, are collected below.

Each of the two half-step applications $e^{-iV_{\text{d}}\tau/2}$ requires one fan-out and one fan-in of the $m$ channel register qubits and $M - 1 - m$ truth-value ancillas (totaling $M - 1$ source qubits.
The per-invocation cost is $(M-1)(d-1)$ CNOTs, giving a total of
\begin{equation}\label{eq:fanout_ondiag}
    C_{\text{fan}}^{(\text{d})} = 4(M-1)(d-1)
\end{equation}
CNOTs at depth $4\lceil\log_2 d\rceil$.
Channel register fan-out enables mode-parallel execution of Tiers~2 and~3 (via local UCR$_Z^{(m)}$ copies), while fan-out of the truth-value ancillas enables mode-parallel execution of Tier~1 corrections.

Each of the $2(M-1)$ fragment applications fans out and fans in the $m$ channel qubits (after the focusing Clifford $U_\zeta$).
The per-invocation cost is $m(d-1)$ CNOTs, giving a total of
\begin{equation}\label{eq:fanout_offdiag}
    C_{\text{fan}}^{(\text{od})} = 4(M-1)\,m(d-1)
\end{equation}
CNOTs at depth $4(M-1)\lceil\log_2 d\rceil$.

Reduction from $\binom{d}{2}$ sequential bilinear layers (without graph coloring) to $\chi \leq d$ parallel layers yields a depth improvement factor of
\begin{equation}\label{eq:coloring_speedup}
    \frac{\binom{d}{2}}{\chi'(K_d)} = \frac{d}{2} \quad (d \text{ even}),
    \qquad \frac{d-1}{2} \quad (d \text{ odd}),
\end{equation}
representing an additional $\sim d/2$ reduction beyond mode-parallel fan-out.

The condition $\Gamma_\alpha \otimes \Gamma_\beta = \Gamma_i \otimes \Gamma_j$ selects edges between modes of complementary representations for off-diagonal bilinear terms.
The resulting graph is generally a union of complete bipartite subgraphs $K_{n_{\Gamma_\alpha}, n_{\Gamma_\beta}}$, one for each pair $(\Gamma_\alpha, \Gamma_\beta)$ satisfying the selection rule.
By K\"{o}nig's theorem~\cite{gross2018graph, konig1931grafok}, the chromatic index of a bipartite graph equals its maximum degree:
\begin{equation}\label{eq:chi_bipartite}
    \chi'(K_{n_{\Gamma_\alpha}, n_{\Gamma_\beta}})
      = \max(n_{\Gamma_\alpha}, n_{\Gamma_\beta}),
\end{equation}
For the full off-diagonal conflict graph $G_{\text{od}}$, the chromatic index satisfies
\begin{equation}\label{eq:chi_offdiag}
    \chi'(G_{\text{od}}) \leq \Delta(G_{\text{od}}) + 1,
\end{equation}
where $\Delta(G_{\text{od}})$ is the maximum number of symmetry-allowed off-diagonal bilinear partners of any single mode.
When the component bipartite subgraphs share no vertices (i.e., each mode participates in at most one complementary-symmetry pairing), the bound tightens to $\chi'(G_{\text{od}}) = \Delta(G_{\text{od}})$.

The aggregate fan-out CNOT cost per Trotter step is
\begin{equation}\label{eq:fanout_total}
    C_{\text{fan}}^{(\text{step})}
      = C_{\text{fan}}^{(\text{d})} + C_{\text{fan}}^{(\text{od})}
      = 4(M-1)(d-1)(1 + m),
\end{equation}
with total fan-out depth $4M\lceil\log_2 d\rceil$.

The depth reduction factor per invocation (on-diagonal or off-diagonal) is
\begin{equation}\label{eq:fanout_reduction}
    \frac{D_{\text{sequential}}}{D_{\text{fan-out}}}
      = \frac{d\,D_{\text{mode}}}{D_{\text{mode}} + 2\lceil\log_2 d\rceil}
      \;\xrightarrow{D_{\text{mode}} \gg \log d}\; d,
\end{equation}
where $D_{\text{mode}} = \mathcal{O}(M\,n^2)$ from the Gray-code UCR decomposition, which dominates $\log_2 d$ for all benchmark systems considered.
Effective depth reduction is therefore $d$-fold for intra-mode terms within each invocation.

The off-diagonal block benefits from this reduction within each of the $2(M-1)$ fragment applications: mode-parallel intra-mode UCR depth per fragment is $(M{+}3)\binom{n}{2} + (M{+}1)(n{+}1)$, independent of $d$.
Without fan-out, the same UCR circuits would require $d$ sequential passes, increasing off-diagonal depth by a factor of $d$ and making it the dominant bottleneck for systems with many vibrational modes. 
The on-diagonal block benefits identically: after the Tier~2 fix (Section~\ref{sec:ondiag_cost}), the intra-mode contribution $\binom{n}{2}(2M{+}1)$ is mode-parallel, and only the bilinear Tier~3 is serialized over the chromatic schedule.
This is the most significant depth reduction in REQWIEM, being an absolute factor of $d$ reduction relative to gate count, unavailable to phase-gradient architectures, which are restricted to serial execution.

\subsection{Ancilla Requirements and Reuse}\label{sec:fanout_ancilla}

On-diagonal fan-out uses $(M-1)(d)-m$ ancilla qubits to store both channel register copies and truth value ancilla copies.
While not required, these truth value ancilla reduce the number of CNOT gates per UCR ladder from $m+2$ to $1+2$ for degree-$2$ polynomial terms controlled on the $m$-qubit channel register.
Off-diagonal fan-out requires $m(d-1)$ ancilla qubits for $d - 1$ copies of the $m$ channel qubits.
Since $M - 1 \geq m$ for all $M \geq 2$ (with equality at $M = 2$), on-diagonal allocation is the larger of the two.
Off-diagonal fragments reuse a subset of the on-diagonal ancilla pool, so total fan-out ancilla budget is
\begin{equation}\label{eq:fanout_ancilla}
    A_{\text{fan}} = (M-1)(d-1).
\end{equation}
These ancillas are allocated once and shared across all on-diagonal and off-diagonal invocations; no additional qubits are introduced by the fan-out protocol beyond this pool.
After each on-diagonal half-step, truth values are uncomputed by reversing the Toffoli cascade, freeing those work qubits for reuse as scratch space during the off-diagonal block if needed.

\subsection{Assembled per-step cost}\label{sec:step_assembly}

The second-order Trotter step consists of $2$ kinetic half-steps, $2$ on-diagonal half-steps, and $1$ off-diagonal full step.

Per step rotation count:
\begin{equation}\label{eq:Rstep}
    R_{\text{step}} = 2\,R_{\text{kin}}
      + 2\,R_{\text{d}}^{(\text{half})} + R_{\text{od}}.
\end{equation}
The off-diagonal contribution $R_{\text{od}} = M(M{-}1)\,S_{\text{sites}}$ scales as $\mathcal{O}(M^2\,d\,n^2)$ and dominates for $M \geq 4$, exceeding the on-diagonal contribution $2\,R_{\text{d}}^{(\text{half})} = \mathcal{O}(M\,d\,n^2)$ by a factor
of $\mathcal{O}(M)$.
For $M = 2$, off-diagonal and on-diagonal rotation counts are comparable.

Per step Toffoli count:
\begin{equation}\label{eq:Tstep}
    T_{\text{step}} = 2\,T_{\text{d}}^{(\text{half})}
      = 2(m \cdot 2^m - 2^{m+1} + 2).
\end{equation}
The only Toffoli gates in the circuit are those required for truth-value ancilla computation in the on-diagonal half-steps.
They are not ``necessary" for computation, but act to reduce overall depth.
Neither the kinetic nor off-diagonal block contributes any Toffoli gates.
For all benchmark systems ($m \leq 3$), $T_{\text{step}} \leq 20$ per step,  making the Toffoli contribution to the fault-tolerant $T$-gate budget negligible.

Per step CNOT count:
\begin{equation}\label{eq:Cstep}
    C_{\text{step}} = 2\,C_{\text{kin}}
      + 2\bigl[C_{\text{d}}^{(\text{half})}
      + C_{\text{fan}}^{(\text{d},\text{half})}\bigr] + C_{\text{od}},
\end{equation}
where
$C_{\text{fan}}^{(\text{d},\text{half})} = 2(M{-}1)(d{-}1)$ is the per-half-step fan-out/fan-in cost for the on-diagonal block.
The off-diagonal CNOTs $C_{\text{od}}$ (Eq.~\ref{eq:Cod}) include the Clifford, fan-out, and UCR contributions.

Per step depth:
\begin{equation}\label{eq:Dstep}
    D_{\text{step}} = 2\,D_{\text{kin}} + 2\,D_{\text{d}}^{(\text{half})} + D_{\text{od}}.
\end{equation}
Off-diagonal depth $D_{\text{od}} = \mathcal{O}(M^2 n^2 \chi(\mathcal{G}_4))$ arises from
$2(M{-}1)$ sequential fragment applications, each contributing $\mathcal{O}(Mn^2 \chi(\mathcal{G}_4))$ after intra-fragment fan-out.
The on-diagonal depth $2D_{\text{d}}^{(\text{half})}$ is dominated by color-serial Tier~3 bilinear terms, contributing $\mathcal{O}(\chi(\mathcal{G}_2^{<})\,M\,n^2)$; Tier~2 is mode-parallel after fan-out, contributing a subdominant term scaling as $\mathcal{O}(M\,n^2)$.
For systems with $M \geq 4$, the off-diagonal depth dominates.
For high numbers of vibrational modes $d$ with $M = 2$ electronic channels and many symmetry-allowed bilinear pairs, the on-diagonal Tier~3 may dominate via $\chi(\mathcal{G}_2^{<})$.
Overall depth complexity per simulation is
\begin{equation}
    D_{\text{sim}} = \mathcal{O}\!\left(k\bigl[M^2 n^2 \chi(\mathcal{G}_4)
      + \chi(\mathcal{G}_2^{<})\,M\,n^2\bigr]\right),
\end{equation}
where $k$ is the Trotter step count.

\subsection{Explicit cost expressions (uniform $n$, $C_1$ symmetry)}\label{sec:explicit_costs}

Asserting uniform grid size $n_r = n$ for all modes and $C_1$ symmetry, where $|\mathcal{G}_1| = d$, $|\mathcal{G}_2^{<}| = |\mathcal{G}_4| = \binom{d}{2}$, and $\chi(\mathcal{G}_2^{<}) = \chi(\mathcal{G}_4) \leq d - 1$, gate count is summarized below:

On-diagonal rotations:
\begin{equation}\label{eq:Rd_uniform}
    R_{\text{d}}^{(\text{half})}
      = Md\!\left[1 + n + \binom{n}{2} + \frac{(d{-}1)\,n^2}{2}\right].
\end{equation}

Off-diagonal rotations:
\begin{equation}\label{eq:Rod_uniform}
    R_{\text{od}} = M(M{-}1)\!\left[d(1{+}n) + d\binom{n}{2}
      + \binom{d}{2}n^2\right] = M(M{-}1)\,S_{\text{sites}}.
\end{equation}

Kinetic rotations:
\begin{equation}\label{eq:Rkin_uniform}
    R_{\text{kin}} = d\!\left[n + \binom{n}{2}\right].
\end{equation}

Per-step rotation count:
\begin{equation}\label{eq:Rstep_uniform}
    R_{\text{step}} = 2\,R_{\text{kin}}
      + 2\,R_{\text{d}}^{(\text{half})} + R_{\text{od}}.
\end{equation}
The off-diagonal contribution dominates for $M \geq 4$: the ratio
\begin{equation}\label{eq:od_ondiag_ratio}
    \frac{R_{\text{od}}}{2\,R_{\text{d}}^{(\text{half})}}
      \approx \frac{M - 1}{2}
\end{equation}
grows linearly with the number of electronic states. 
The approximation holds when bilinear terms dominate both $R_{\text{od}}$ and $R_\text{d}^{(\text{half})}$.
For $M = 2$, the ratio is $\approx 1/2$: off-diagonal is cheaper than on-diagonal because each fragment's UCR$_Z^{(0)}$ reduces to a bare $R_z$ ($M/2 = 1$ rotation per site) while the on-diagonal UCR$_Z^{(1)}$ contributes $M = 2$ rotations per site.
For $M \geq 4$, the ratio exceeds unity and off-diagonal blocks become the dominant rotation cost.

Toffoli count:
\begin{equation}\label{eq:Tstep_uniform}
    T_{\text{step}} = 2\,T_{\text{truth}}
      = 2(m \cdot 2^m - 2^{m+1} + 2).
\end{equation}

An alternative AND-ancilla implementation of the bilinear interactions replaces the UCR CNOT cascade with a Toffoli-based iteration that reduces the CNOT count at the cost of $2n$ Toffoli gates per bilinear pair per correction channel.
This trades CNOT depth for Toffoli gates and may be preferable in architectures permitting CZZ gates, such as recent progress in neutral atom arrays \cite{huang2024zap}.
This trade-off is not discussed in present resource estimations, but may be significant with further technological progress.

On-diagonal depth per half-step:
\begin{equation}\label{eq:Dd_uniform}
\begin{split}
    D_{\text{d}}^{(\text{half})} &= 2D_{\text{truth}}
      + 2\lceil\log_2 d\rceil + (n{+}1) + 3(M{-}1) \\
    &\quad + \binom{n}{2}(2M{+}1)
      + (d{-}1)\,n^2(2M{+}1).
\end{split}
\end{equation}
This collects truth-value Toffoli cascades, fan-out/fan-in, and the mode-parallel Tier~1: the reference surface contributes depth $n + 1$ and the $M - 1$ correction surfaces each contribute depth~$3$ (the $(1{+}n)$ $\mathrm{CR}_z$ gates per mode share a read-only control but target distinct qubits, executing in parallel).
The second line collects mode-parallel Tier~2 ($\binom{n}{2}$ intra-mode sites $\times$ depth $2M + 1$ per site, independent of $d$ after fan-out) and color-serialized Tier~3 ($\chi = d - 1$ color layers $\times$ $n^2$ sites $\times$ depth $2M + 1$).
Tier~3 dominates for systems with many bilinear pairs.

Off-diagonal depth:
\begin{equation}\label{eq:Dod_uniform}
    D_{\text{od}} = 2(M{-}1)\!\left[D^{(\text{UCR})}
      + 2\lceil\log_2 d\rceil\right] + 2mM,
\end{equation}
with
\begin{equation}\label{eq:Ducr_uniform}
    D^{(\text{UCR})} = (M{+}3)\binom{n}{2} + (M{+}1)\,n
      + (M{+}1) + (d{-}1)\,n^2(M{+}3).
\end{equation}

Per-step depth:
\begin{equation}\label{eq:Dstep_uniform}
    D_{\text{step}} = 2\,D_{\text{kin}}
      + 2\,D_{\text{d}}^{(\text{half})} + D_{\text{od}}.
\end{equation}
Off-diagonal depth scales as $\mathcal{O}(M^2 n^2 d)$ (from $2(M{-}1)$ fragments each contributing $\mathcal{O}(Mn^2 d)$ via the
color-serialized bilinear terms), while the on-diagonal depth scales as $\mathcal{O}(dMn^2)$ from the color-serial Tier~3 (Tier~2 is mode-parallel and contributes only $\mathcal{O}(Mn^2)$). 

On-diagonal depth per half-step is dominated by the color-serial Tier~3 bilinear terms, scaling as $\chi(\mathcal{G}_2^{<})\,n^2\,(2M{+}1) = (d{-}1)\,n^2\,(2M{+}1)$ in $C_1$ symmetry; mode-parallel Tier~2 contributes only $\binom{n}{2}(2M{+}1)$, subdominant for $d \gg 1$. 
Off-diagonal depth is dominated by $2(M{-}1)$ sequential fragment applications, each contributing
$(d{-}1)\,n^2\,(M{+}3)$ from the bilinear UCR block, giving $2(M{-}1)(d{-}1)\,n^2\,(M{+}3)$ overall.
The depth ratio (comparing bilinear contributions, dominant for large~$d$) is approximately:
\begin{equation}\label{eq:depth_ratio}
    \frac{D_{\text{od}}}{2\,D_{\text{d}}^{(\text{half})}}
      \approx \frac{2(M{-}1)(d{-}1)\,n^2\,(M{+}3)}
               {2\,(d{-}1)\,n^2\,(2M{+}1)}
      = \frac{(M{-}1)(M{+}3)}{2M{+}1}
      \;\xrightarrow{\text{large M}}\; \approx\frac{M}{2},
\end{equation}
For $M = 2$, this ratio is $5/5 = 1.0$: the off-diagonal and on-diagonal bilinear depths are comparable, since both serialize over
the same $d - 1$ color layers at the same per-site depth ($M + 3 = 2M + 1 = 5$).
For $M = 8$, the ratio is $\approx 4.5$, and off-diagonal depth clearly dominates.

Overall depth complexity is $\mathcal{O}(k\,M^2\,n^2\,\chi)$, where $\chi = \chi(\mathcal{G}_4) = d - 1$ in $C_1$ symmetry and $k$ is the Trotter step count.
The on-diagonal contribution $\mathcal{O}(k\,\chi(\mathcal{G}_2^{<})\,M\,n^2)$ is sub-leading by a factor of $M$.

\subsection{Role of symmetry}\label{sec:cost_symmetry}

The symmetry index sets $\mathcal{G}_1$--$\mathcal{G}_4$ directly control the number of nonzero terms:

\begin{itemize}
    \item \textbf{On-diagonal linear} ($\mathcal{G}_1$): only totally symmetric
    modes contribute to tuning displacements.
    For $D_{2h}$, $|\mathcal{G}_1| = d_{A_g}$.
    Under the hybrid decomposition, Tier~1 multiplexing reduces the correction rotation count from $(M{-}1)dn$ (UCR) to    $(M{-}1)|\mathcal{G}_1|n$, a saving of $(M{-}1)(d-|\mathcal{G}_1|)n$ per half-step.

    \item \textbf{On-diagonal quadratic}: Always allowed.

    \item \textbf{On-diagonal bilinear} ($\mathcal{G}_2^<$): only same-irrep pairs.
    For pyrazine ($d = 24$, $D_{2h}$): $|\mathcal{G}_2^<| = 31$ vs.\ $\binom{24}{2} = 276$ without symmetry.

    \item \textbf{Off-diagonal linear} ($\mathcal{G}_3$): modes with $\Gamma(x_r) = \Gamma_s \otimes \Gamma_{s'}$.

    \item \textbf{On-diagonal quadratic}: Always forbidden.

    \item \textbf{Off-diagonal bilinear} ($\mathcal{G}_4$): mode pairs with $\Gamma_r \otimes \Gamma_s = \Gamma_s \otimes \Gamma_{s'}$.
    For pyrazine (24-mode): $|\mathcal{G}_4| = 29$.
\end{itemize}

For $C_1$ symmetry, $|\mathcal{G}_1| = d$ and $|\mathcal{G}_2^<| = \binom{d}{2}$, so the Tier~1 linear savings vanish and the hybrid reduces to the pure-UCR formulae with a negligible constant-phase correction of $(M{-}1)(d{-}1)$ per half-step.
Tables~\ref{tab:ondiag_structure} and~\ref{tab:offdiag_structure} collect the on- and off-diagonal cost structures.

\subsection{Cost tables}\label{sec:cost_tables}

\begin{table}[ht]
\centering
\caption{On-diagonal cost structure per half-step (three-tier hybrid).}\label{tab:ondiag_structure}
\renewcommand{\arraystretch}{1.25}
\begin{tabular}{l l l c c}
\hline\hline
Tier & Term & Source & Rotations & Toffoli \\
\hline
1 & Single-qubit & Constant $+$ linear $x_r$ $+$ diag-quad $x_r^2$
  & $d\,M\,(1{+}n)$
  & $0$ \\
2 & Intra-mode cross & $x_r^2$ cross
  & $M\!\cdot\!d\,\binom{n}{2}$
  & $0$ \\
3 & Bilinear cross & $x_rx_s$,\; $(r,s)\!\in\!\mathcal{G}_2^{<}$
  & $M\!\cdot\!|\mathcal{G}_2^{<}|\,n^2$
  & $0$ \\
\hline
\multicolumn{2}{l}{Total} &
  & Eq.~\ref{eq:Rd_half}
  & $0$ \\
\hline\hline
\end{tabular}
\end{table}

\begin{table}[ht]
\centering
\caption{Off-diagonal cost structure per full step.}\label{tab:offdiag_structure}
\renewcommand{\arraystretch}{1.25}
\begin{tabular}{l c c c}
\hline\hline
 & Comp.\ basis & Hadamard basis (per class) & Toffoli ($\mathcal{H}$) \\
\hline
LVC per mode $r$, class $\zeta$ & $n_r \cdot 2^{m-h+1}$ $R_x$ & $n_r \cdot M$ $R_z$ & $0$ \\
BVC per pair, class $\zeta$ & $n_rn_s \cdot 2^{m-h+2}$ $R_x$ & $n_rn_s \cdot M$ $R_z$ & $0$ \\
\hline
Sum over classes & $\sum_\zeta (\cdots)$ & $(M{-}1) \times (\cdots)$ & -- \\
Fan-out eligible & No & Yes & -- \\
\hline\hline
\end{tabular}
\end{table}

\subsection{Binary Arithmetic Encoding of Rotation Angles}\label{sec:binary_encoding}

Each vibrational coordinate $x_r$ is encoded in an $n_r$-qubit register as $x_r = \Delta_r \sum_{\ell=0}^{n_r-1} 2^{\ell-n_r} q_\ell + x_r^{(\min)}$, where $q_\ell \in \{0,1\}$ is the $\ell$-th qubit.
Substituting this binary expansion into polynomial diabatic potentials produces rotation angles that factorize exactly as a state-dependent coupling constant times a binary weight factor:
\begin{equation}\label{eq:angle_factorisation}
    \theta_{\boldsymbol\ell}^{(s)} = c^{(s)} \times 2^{f(\boldsymbol\ell,\,\mathbf{n})},
\end{equation}
where $c^{(s)}$ is the potential energy surface parameter on surface $s$ and $f(\boldsymbol\ell, \mathbf{n})$ is a linear function of qubit indices and register widths.
Linear terms yield $n_r$ single-qubit rotations per mode, quadratic terms yield $\binom{n_r}{2}$ intra-mode cross-pair rotations, and bilinear terms yield $n_r n_s$ inter-mode cross-pair rotations (see Appendix~\ref{app:binary_decomp}).

Number of rotation sites per mode is fixed by polynomial degree and grid size, independent of functional form.
Gate counts in Sections~\ref{sec:ondiag_cost}--\ref{sec:offdiag_cost} are tight for the polynomial model, not approximations subject to smoothness or truncation conditions.

Binary weight factors pass through the channel-register Walsh--Hadamard transform as a site-independent multiplicative constant.
For on-diagonal UCR$_Z^{(m)}$, the $M$-point Walsh transform acts exclusively on the $M$-dimensional vector of coupling constants $(c^{(0)}, \ldots, c^{(M-1)})$; for off-diagonal UCR$_Z^{(m-1)}$ within each XOR-class fragment, the $(M/2)$-point transform acts on $M/2$ coupling values addressed by spectators.
Since $M$ is small and fixed (typically $2$--$8$), all Walsh coefficients are retained exactly and no spectral truncation is performed.
This contrasts with grid-based Walsh encodings~\cite{chang2022improving}, where Walsh transforms act on $2^n$ grid-sampled potential values and the resulting coefficient decay depends on the smoothness of $V$.
This is discussed further with regard to scattering potentials in Chapter~\ref{ch:scattering}.

\subsection{Circuit Depth Summary}\label{sec:parallelism}

Beyond gate counts, circuit depth determines execution time and error accumulation.
REQWIEM exploits several layers of parallelism inherent in the vibronic Hamiltonian structure, though the XOR-class fragmentation of the off-diagonal block introduces sequential fragment applications that were absent in the two-state case.

The kinetic operator acts independently on each vibrational register.
All $d$ modes execute simultaneously, reducing kinetic depth from $\mathcal{O}(dn^2)$ to $\mathcal{O}(n^2)$.
For potential terms, channel-register fan-out enables parallel execution across modes:
\begin{itemize}
    \item \textbf{On-diagonal (all three tiers):}
    Phase~3 fans out both $m$ channel register qubits and $M - 1 - m$ truth-value ancillas to $d$ local copies.
    Tier~1 corrections execute in parallel across modes using the truth-value copies.
    Tiers~2 and~3 execute their UCR$_Z^{(m)}$ circuits on local copies of the channel register; the Gray-code CNOT schedule depends only on $m$ (not on mode-specific coupling constants), so all $d$ copies remain synchronized throughout the traversal.
    Tier~2 is therefore fully mode-parallel, and Tier~3 is mode-parallel and color-serial.

    \item \textbf{Off-diagonal (per-fragment):}
    Within each fragment application, the $m$ channel qubits are fanned out after the focusing Clifford $U_\zeta$, enabling mode-parallel UCR$_Z^{(m-1)}$ execution.
    This reduces the per-fragment intra-mode UCR depth from $\mathcal{O}(d^2M^2n^2)$ to $\mathcal{O}(dMn^2)$.
\end{itemize}

Bilinear cross-mode terms $x_r x_s$ and $x_{r'} x_{s'}$ with $\{r,s\} \cap \{r',s'\} = \emptyset$ act on disjoint registers and
execute simultaneously.
The $|\mathcal{G}_2^{<}|$ on-diagonal and $|\mathcal{G}_4|$ off-diagonal mode pairs reduce to $\chi(\mathcal{G}_2^{<})$ and $\chi(\mathcal{G}_4)$ sequential layers respectively, where $\chi$ is the chromatic index of the mode-pair conflict graph.
For $C_1$ symmetry, $\chi \leq d - 1$, giving a further reduction from $\binom{d}{2}$ to $(d - 1)$ layers.

For on-diagonal Tier~1, quantum multiplexing applies the base surface $V_{00}$ with zero channel overhead and reduces correction depth from $\mathcal{O}(M)$ sequential controlled blocks to $\mathcal{O}(m)$ levels in the binary hierarchy, with disjoint control sets executing in parallel on fanned-out copies.
For off-diagonal potential terms, each fragment's UCR$_Z^{(m-1)}$ applies $M/2$ rotations per site via Gray-code decomposition, serialized at the single-qubit level within each UCR gate but parallelized across modes via fan-out.

The $M - 1$ XOR-class fragments do not commute with one another and must be applied sequentially.
The symmetric Trotter product $\mathcal{V}_\text{od}(\tau)$ applies each fragment twice (forward and backward) for a total of $2(M - 1)$ sequential fragment applications per Trotter step.
This sequential composition is the principal depth overhead introduced by the XOR-class fragmentation relative to the (invalid for $M > 2$) single-pass Hadamard approach.

Per-step depth with all parallelism enabled is
\begin{equation}\label{eq:depth_assembled}
    D_{\text{step}} = \underbrace{2D_{\text{kin}}}_{\text{kinetic half-steps}}
      + \underbrace{2D_{\text{d}}}_{\text{on-diag half-steps}}
      + \underbrace{D_{\text{od}}}_{\text{off-diag full step}},
\end{equation}
with contributions:
\begin{align}
    D_{\text{kin}} &= \mathcal{O}(n^2),
    \label{eq:Dkin}\\[4pt]
    D_{\text{d}} &= \underbrace{2\lceil\log_2 d\rceil
      + (n{+}1) + 3(M{-}1)}_{\text{Tier~1 (mode-parallel)}}
      + \underbrace{(2M{+}1)\binom{n}{2}}_{\text{Tier~2 (mode-parallel)}}
      + \underbrace{\chi(\mathcal{G}_2^{<})\,(2M{+}1)\,n^2}_{\text{Tier~3 (color-serial)}},
    \label{eq:Dd}\\[4pt]
    D_{\text{od}} &= 2(M{-}1)\bigl[D^{(\mathrm{UCR})}
      + 2\lceil\log_2 d\rceil\bigr] + 2mM,
    \label{eq:Dod}
\end{align}
Per-fragment intra-mode UCR depth is
\begin{equation}\label{eq:Ducr_frag}
    D^{(\mathrm{UCR})} = (2^m{+}3)\binom{n}{2} + (2^m{+}1)\,n
      + (2^m{+}1)
      + \chi(\mathcal{G}_4)\,(2^m{+}3)\,n^2.
\end{equation}

Off-diagonal depth dominates for $M \geq 4$ and scales as $\mathcal{O}(M^2 n^2 \chi(\mathcal{G}_4))$, since each of the $2(M-1)$ fragments contributes $\mathcal{O}(Mn^2\chi(\mathcal{G}_4))$.
For the on-diagonal block, the color-serial Tier~3 contributes $\mathcal{O}(\chi(\mathcal{G}_2^{<})\,M\,n^2)$; Tier~2 is
mode-parallel via fan-out and contributes only $\mathcal{O}(Mn^2)$, which is subdominant for systems with bilinear coupling.
The overall depth complexity per simulation is
\begin{equation}\label{eq:depth_overall}
    D_{\text{sim}} = \mathcal{O}\!\left(k\bigl[M^2 n^2
      \chi(\mathcal{G}_4)
      + \chi(\mathcal{G}_2^{<})\,M\,n^2\bigr]\right),
\end{equation}
where $k$ is the Trotter step count. The off-diagonal fan-out within each fragment provides a depth reduction of approximately $d$ relative to a mode-serial implementation of the same fragmented scheme, preserving utility of fan-out architecture despite the $2(M-1)$-fold fragment serialization.

In accumulator-based implementations such as the QROM-loaded phase gradient register of Motlagh et al.~\cite{motlagh2025quantum}, every coupling term within a Trotter step writes to a shared phase register sequentially, imposing depth $\Omega(\text{\#\,terms} \times D_{\text{QROM+add}})$ with no possibility of parallelization.
For the off-diagonal block, this gives accumulator depth $\mathcal{O}(M^2 d n \cdot D_{\text{QROM}})$, incurring a factor of $\mathcal{O}(d)$ not present in REQWIEM.
Depth advantage of the UCR-based approach over the accumulator architecture therefore persists under XOR-class correction, since fan-out parallelism operates within each fragment independently of how many fragments are needed.

\begin{table}[ht]
\centering
\small
\caption{Gate scaling terms in REQWIEM Trotterization.
SQC refers to `single-qubit Clifford' operators, as
$\{$NOT, Y, Z, H, S$\}$.
Depth complexities assume channel-register fan-out is enabled; $\chi$ denotes the chromatic index of the symmetry-allowed mode-pair conflict graph, satisfying $\chi = \Theta(d)$ for $C_1$ symmetry and $\chi \leq \Delta(G) + 1$ for non-degenerate point groups, where $\Delta(G)$ is the maximum vertex degree.
$\chi_{\textrm{d}}$ and $\chi_{\textrm{od}}$ refer to the chromatic indices of the on-diagonal and off-diagonal conflict graphs, respectively; in general $\chi_{\textrm{d}} \leq \chi_{\textrm{od}}$.}
\label{tab:gate_scaling}
\begin{tabular}{llll}
\hline
Term & Gate & Complexity & Depth Complexity \\
\hline
Preparation & $R_z$    & $\mathcal{O}(dn^2)$     & $\mathcal{O}(n^2)$ \\
            & SQC      & $\mathcal{O}(dn)$        & \\
            & CNOT     & $\mathcal{O}(dn^2)$      & \\
            & Toffoli  & $\mathcal{O}(0)$          & \\
Kinetic     & $R_z$    & $\mathcal{O}(kdn^2)$     & $\mathcal{O}(kn^2)$ \\
            & SQC      & $\mathcal{O}(kdn)$        & \\
            & CNOT     & $\mathcal{O}(kdn^2)$      & \\
            & Toffoli  & $\mathcal{O}(0)$           & \\
On-Diagonal & $R_z$    & $\mathcal{O}(kMd^2n^2)$  & $\mathcal{O}(kM\chi_{\textrm{d}} n^2)$ \\
Potential   & SQC      & $\mathcal{O}(kM\log M)$   & \\
            & CNOT     & $\mathcal{O}(kMd^2n^2)$   & \\
            & Toffoli  & $\mathcal{O}(kM\log M)$   & \\
Off-Diagonal & $R_z$   & $\mathcal{O}(kM^2 d^2 n^2)$ & $\mathcal{O}(kM^2\chi_{\textrm{od}} n^2)$ \\
Potential    & SQC     & $\mathcal{O}(kM)$            & \\
             & CNOT    & $\mathcal{O}(kM^2 d^2 n^2)$  & \\
             & Toffoli & $\mathcal{O}(0)$              & \\
\hline
\end{tabular}
\end{table}

\subsection{Gate-Count and Depth Comparison}\label{sec:optimality}

XOR-class fragmentation required for $M > 2$ electronic states changes the resource landscape of the off-diagonal block relative to the two-state case.
This section considers these gate-count and depth properties and compares them with the accumulator-based
approach of Ref.~\cite{motlagh2025quantum}, which shares the same XOR-class fragmentation and focusing Clifford structure but differs in how the diagonalized fragment Hamiltonians are exponentiated.
Results are summarized in Table~\ref{tab:gate_scaling}.

\subsubsection{Rotation count and the information-theoretic bound}
\label{sec:rotation_bound}

$\binom{M}{2}$ independent coupling functions $\{c_{ij}(\mathbf{x})\}_{i<j}$ each require at least one independently
parameterized rotation per grid site to encode their values, giving an information-theoretic lower bound of
\begin{equation}\label{eq:info_lower_bound}
    R_{\text{od}}^{(\min)} = \binom{M}{2}\,S_{\text{sites}}
      = \frac{M(M-1)}{2}\,S_{\text{sites}},
\end{equation}
parameterized gates, where $S_{\text{sites}} = d(1{+}n) + d\binom{n}{2} + |\mathcal{G}_4|\,n^2$ is the total number of rotation sites per coupling pair ($|\mathcal{G}_4| = \binom{d}{2}$ in $C_1$ symmetry).

REQWIEM's off-diagonal rotation count is
\begin{equation}\label{eq:Rod_reqwiem}
    R_{\text{od}} = M(M{-}1)\,S_{\text{sites}}
      = 2\,R_{\text{od}}^{(\min)}.
\end{equation}
The factor of~$2$ arises entirely from the symmetric Trotter product $\mathcal{V}_{\text{od}}(\tau)$, which applies each of the $M - 1$ fragments twice (forward and backward) to maintain $\mathcal{O}(\tau^3)$ accuracy.
Within a single sweep direction, each fragment contributes $(M/2)\,S_{\text{sites}}$ rotations, and $M - 1$ fragments sum to
$M(M{-}1)/2 \cdot S_{\text{sites}} = R_{\text{od}}^{(\min)}$, exactly saturating the lower bound.
The factor-of-$2$ overhead remains as an unavoidable cost of second-order Trotter accuracy, shared equally by any symmetric product-formula approach.

For the on-diagonal block, the three-tier hybrid decomposition yields a rotation count of
$R_{\text{d}}^{(\text{half})} = M[d(1{+}n{+}\binom{n}{2}) + |\mathcal{G}_2^{<}|n^2]$ per half-step (Eq.~\ref{eq:Rd_half}).
The Tier~1 reference and correction contributions combine into $dM(1{+}n)$, absorbing both linear and diagonal-quadratic single-qubit terms (Section~\ref{sec:ondiag_cost}).
Tiers~2 and~3 contribute $Md\binom{n}{2}$ and $M|\mathcal{G}_2^{<}|n^2$ respectively, each at $M$ rotations per site from the Gray-code UCR$_Z^{(m)}$ decomposition.
The total on-diagonal rotation count per step is $2R_{\text{d}}^{(\text{half})}$, scaling as $\mathcal{O}(Md^2n^2)$.

\subsubsection{Focusing Clifford overhead}\label{sec:focusing_overhead}

The per-fragment focusing Clifford $U_\zeta$ consists of $h(\zeta) - 1$ CNOTs and one Hadamard gate, where its adjoint $U_\zeta^\dagger$ has the same cost.
The total focusing CNOT count across all $2(M - 1)$ fragment applications is
\begin{equation}\label{eq:focusing_total}
    C_{\text{focus}} = 4\!\left[m \cdot 2^{m-1} - M + 1\right]
      = \mathcal{O}(mM),
\end{equation}
and the total focusing depth is $2mM$.
These are non-parameterized gates that encode no coupling-function information, reducing a multi-qubit Pauli string
$X_S$ to a single-qubit $X_{s_1}$ amenable to Hadamard diagonalization.

Focusing overhead is asymptotically negligible, where $C_{\text{focus}} / R_{\text{od}} = \mathcal{O}(m / (M\,S_{\text{sites}}))$ vanishes as the grid size or mode count grows.
For benchmark vibronic systems considered in here, focusing CNOTs contribute fewer than
$0.1\%$ of the total gate count.
Nevertheless, focusing overhead represents a structural distinction from on-diagonal blocks, where the channel register is never
modified and no non-parameterized channel-register gates are needed.

\subsubsection{Architectural comparison: UCR versus phase gradient
accumulator}\label{sec:ucr_vs_accumulator}

Both REQWIEM and the approach of Motlagh et al.~\cite{motlagh2025quantum} employ the same XOR-class fragmentation and focusing Clifford to block-diagonalize off-diagonal fragments.
For the on-diagonal potential, REQWIEM uses the three-tier hybrid decomposition, while the phase gradient method uses a QROM--arithmetic--accumulator pipeline as for the off-diagonal terms.
The two methods therefore diverge in how diagonal operators are exponentiated:
\begin{itemize}
    \item \textbf{REQWIEM (UCR):} Each diagonal block (whether on-diagonal or a focused off-diagonal fragment) is decomposed into
    classically precomputed rotation angles via Gray-code UCR$_Z^{(p)}$ ($p = m$ for on-diagonal, $p = m{-}1$ for off-diagonal fragments).
    Phase~3 of the on-diagonal protocol fans out both the $m$ channel register qubits and the $M - 1 - m$ truth-value ancillas; within each off-diagonal fragment, $m$ channel qubits are fanned out after the focusing
    Clifford.
    In both cases, all $d$ modes execute their UCR circuits in parallel on independent copies of the control qubits.
    The channel register (or its fan-out copies) is read-only throughout the UCR block.

    \item \textbf{Phase gradient accumulator (Ref.~\cite{motlagh2025quantum}):} Each diagonal eigenvalue is loaded via QROM into a shared phase gradient register, and phases are accumulated through controlled semi-classical additions.
    The accumulator is a shared write target: every coupling term, whether on-diagonal or off-diagonal, intra-mode or
    bilinear, must write to it sequentially, precluding mode-parallel execution regardless of available ancillas.
\end{itemize}

\noindent
Consequences of this architectural distinction are summarized in Table~\ref{tab:structural_comparison}.

\begin{table}[t]
  \centering
  \caption{Architectural comparison of diagonal-block exponentiation strategies within the shared XOR-class fragmentation framework.}
  \label{tab:structural_comparison}
  \begin{tabular}{lcc}
    \toprule
    \textbf{Property}
    & \textbf{PG accumulator}
    & \textbf{REQWIEM} \\
    \midrule
    Channel register
    & Read-only (QROM address)
    & Read-only (UCR control) \\
    Phase register
    & Shared write target
    & N/A (no phase register) \\
    Focusing Cliffords
    & $C_{\text{focus}}$ (identical)
    & $C_{\text{focus}}$ (identical) \\
    On-diag mode parallelism
    & None (serializes)
    & Yes (all three tiers) \\
    Off-diag mode parallelism
    & None (serializes)
    & Yes (per-fragment fan-out) \\
    Per-fragment off-diag depth
    & $\Omega(d \cdot D_{\text{QROM+add}})$
    & $D^{(\text{UCR})} + 2\lceil\log_2 d\rceil$ \\
    On-diag depth per half-step
    & $\Omega(d \cdot D_{\text{terms}})$
    & $D_{\text{d}}^{(\text{half})}$
      (Eq.~\ref{eq:Dd_half}) \\
    Fan-out depth reduction
    & None
    & $\approx d$-fold (intra-mode) \\
    Ancilla overhead
    & $b$ bits $+$ QROM
    & $(M{-}1)(d{-}1)$ fan-out copies \\
    \bottomrule
  \end{tabular}
\end{table}

Distinction lies in circuit depth rather than gate count.
Both approaches incur the same $\mathcal{O}(M^2\,S_{\text{sites}})$ rotation count (up to constant factors determined by the QROM precision model), and both share the $\mathcal{O}(mM)$ focusing Clifford overhead.
UCR gains its advantage by treating the channel register as a passive control bus: since UCR targets only vibrational qubits (never the channel register or a shared accumulator), all $d$ modes can be targeted concurrently through fan-out.

A per-fragment depth reduction of $\sim d$ results for intra-mode terms, yielding an overall off-diagonal depth of $\mathcal{O}(M^2 n^2 \chi(\mathcal{G}_4))$.
Bilinear cross-mode terms require qubits from two modes simultaneously and cannot be fully mode-parallelized; they are instead serialized over $\chi(\mathcal{G}_4)$ chromatic layers of the mode-pair conflict graph, with all disjoint pairs in each layer executing simultaneously on independent fan-out copies.

For the on-diagonal block, full channel register fan-out enables mode-parallel execution of all three tiers (Section~\ref{sec:ondiag_cost}).
Tier~1 corrections execute at depth $3(M{-}1)$ (read-only truth-value controls with distinct vibrational targets).
Tier~2 intra-mode UCR circuits execute at depth $\binom{n}{2}(2M{+}1)$, independent of $d$, since each mode
operates on a local copy of the $m$-qubit control set. 
Tier~3 bilinear terms are color-serialized at depth $\chi(\mathcal{G}_2^{<})\,n^2\,(2M{+}1)$.
On-diagonal depth is therefore $\mathcal{O}(\chi(\mathcal{G}_2^{<})\,M\,n^2)$, dominated by Tier~3.
In the accumulator architecture, all on-diagonal terms including intra-mode sites that REQWIEM parallelizes must write serially to the shared phase register, giving on-diagonal accumulator depth $\mathcal{O}(d \cdot D_{\text{QROM+add}})$ per evaluation.

\section{Direct Comparison with Phase Gradient Methods}
\label{sec:direct_comparison}

REQWIEM is compared to the phase gradient approach of Motlagh et al.~\cite{motlagh2025quantum} on five benchmark vibronic
systems spanning a range of electronic-state counts ($m = 1$--$3$) and vibrational mode counts ($d = 11$--$246$).
As detailed in the cited work (Pradhan and Zeng, \cite{pradhan2023triplet}), the $(NO)_4$-Anthracene dimer is more accurately modeled with quartic diabatic surfaces and Morse-like couplings ($\propto(1-e^{-x})^2$). 
Resource estimations for this system therefore do not reflect potential simulability, only serving as an additional pedagogical comparison between REQWIEM and phase gradient methods.

Once again, $d$ denotes the number of vibrational modes, $M = 2^m$ the number of padded electronic states
($m = \lceil\log_2 M\rceil$, with $M$ the physical state count), $n = 10$ the grid qubits per mode, and $\chi = \chi'(K_d)$ the chromatic index of the complete mode-pair graph ($d - 1$ for even $d$, $d$ for odd $d$.
All estimates use a precision-adjusted phase accumulator register size $b = 45$ (Appendix~\ref{app:Motlagh_Resource}).

For a comparison between the two methods, architectural analysis first compares both methods in units of non-Clifford layers, as Toffoli layers for the phase gradient method and rotation ($R_z$) layers for REQWIEM, isolating the structural advantages of each circuit architecture independently of fault-tolerant compilation details.
Second, the fault-tolerant analysis decomposes both circuits into a universal gate set, providing an apples-to-apples comparison in $T$-depth and total decomposed depth, accounting for different per-operation costs of Toffoli gates and synthesized rotations.

\subsection{Phase Gradient Method: Corrected Cost Formulas}
\label{sec:pg_summary}

The phase gradient algorithm implements each diagonal-block exponentiation (whether on-diagonal or a focused off-diagonal fragment) through a four-stage pipeline \cite{motlagh2025quantum}: (i)~a QROM lookup loads a classically precomputed coefficient into a quantum register;
(ii)~arithmetic circuits (out-of-place squarers or multipliers) compute the grid-point--dependent factors $x_r^2$ or $x_r x_s$;
(iii)~controlled semi-classical additions accumulate the resulting phase into a shared $b$-bit phase gradient register; and
(iv)~the QROM and arithmetic are uncomputed.
Every coupling term in every fragment traverses this pipeline sequentially, since the shared phase gradient register is a write
target that serializes all terms.

\subsubsection{Corrected parameters}
Resource estimates published by Motlagh et al.~\cite{motlagh2025quantum} use $n = 4$ qubits per vibrational mode ($N = 16$ grid points) and a $b = 20$-bit phase gradient register.
Four corrections are applied to bring these estimates to physically
justified values (full derivations in
Appendix~\ref{app:Motlagh_Resource}):
\begin{enumerate}
    \item \textbf{Grid resolution}
    (Chapter~\ref{ch:wavepacket_prop}): $n = 4 \to n = 10$
    ($N = 16 \to 1024$).
    Converged MCTDH calculations require $32$--$64$ primitive basis functions per mode~\cite{raab1999molecular, worth2004beyond};
    $n = 4$ violates the Nyquist criterion for nonadiabatic dynamics $n_{\max} \sim 8$--$10$ (up to $15-20$ for extremely energetic systems, Chapter~\ref{ch:wavepacket_prop}).
    \item \textbf{Phase precision}: $b = 20 \to b = 45$. 
    At $n = 10$, bilinear term one-norms grow by a factor of $64$ relative to $n = 4$; $b = 20$ ($\textrm{precision} \approx
    6 \times 10^{-6}$) becomes insufficient, while $b = 45$ ($\textrm{precision} \approx 1.8 \times 10^{-13}$) provides a
    comfortable margin for accumulation across $\mathcal{O}(M^2)$ terms per step.
    This margin safely handles coherent accumulation of errors and additional Trotter overhead.
    \item \textbf{Bilinear terms included:} Published estimates omit all bilinear cross-mode terms $x_r x_s$, which account for
    $\binom{d}{2}$ on-diagonal and $\binom{d}{2}$ off-diagonal mode pairs in $C_1$ symmetry (See Motlagh et al. \cite{motlagh2025quantum} Supplemental Material).
    These terms dominate gate count for most systems and are essential for correct physics (Chapter~\ref{ch:model_ext}).
    \item \textbf{Trotter error recalibrated}: The combined effect of a grid-induced norm increase ($\beta_{\text{grid}} = 64$) and the bilinear commutator contributions ($\beta_{\text{BVC}} = 1 + \eta^2 N_{\text{BVC}}^{(\text{tot})}/M$) amplifies the commutator sum by a factor of $\beta = \beta_{\text{grid}} \times \beta_{\text{BVC}}$, requiring $\sqrt{\beta}\times$ more Trotter steps to maintain a given target error.
\end{enumerate}

\subsubsection{Per-term Toffoli costs}
At the corrected parameters ($n = 10$, $b = 45$), out-of-place squaring costs $C_{\text{sq}} = k(k{-}1)/2 = 45$ Toffoli, the out-of-place multiplier costs $C_{\text{mult}} = n^2 = 100$ Toffoli, and the controlled semi-adder into the $b$-bit phase register costs $C_{\text{semi}}(n_b) = n_b b - n_b(n_b{-}1)/2$, giving $C_{\text{semi}}(10) = 405$ and $C_{\text{semi}}(20) = 710$ Toffoli~\cite{gidney2018halving, su2021fault}.
Each term additionally incurs two QROM lookups (forward $+$ uncompute) at cost $2C_{\text{QROM}}^{(\zeta)}$, where $C_{\text{QROM}}^{(\zeta)} = 2^{m - h(\zeta)}$ for XOR class $\zeta$ with Hamming weight $h(\zeta)$.
A linear term costs $421$ Toffoli, a diagonal quadratic $816$, and a bilinear cross-mode term $926$ (all for the diagonal fragment $\zeta = 0$ at $M = 6$, $C_{\text{QROM}}^{(0)} = 8$).

\subsubsection{Per-step Toffoli count}
Second-order Strang splitting with kinetic merging gives a per-step cost of
\begin{equation}\label{eq:step_cost_summary}
    \Omega_{\text{step}} = C_{\text{kin}}
      + 2\,C_{V_{\text{d}}} + C_{V_{\text{od}}},
\end{equation}
where $C_{\text{kin}}$, $C_{V_{\text{d}}}$, and $C_{V_{\text{od}}}$ are the kinetic, on-diagonal, and off-diagonal fragment costs.
The off-diagonal cost $C_{V_{\text{od}}}$ sums over all $M-1$ XOR classes, each with QROM cost reduced by the focusing Clifford.
In $C_1$ symmetry, the bilinear terms dominate: for Anth/C$_{60}$ ($d = 246$, $M = 4$), the bilinear contribution accounts for
$>99\%$ of $\Omega_{\text{step}}$.

\subsubsection{Serial circuit depth}
The phase gradient architecture is inherently serial: within each term, the QROM $\to$ arithmetic $\to$ accumulate $\to$ uncompute
pipeline forms a single data-dependency chain; across terms within a fragment, the shared phase gradient register enforces sequential execution; and across fragments, Trotter splitting is sequential.
The non-Clifford (Toffoli) depth therefore equals the Toffoli count:
\begin{equation}\label{eq:DM_serial}
    D_{\text{PG}} = \Omega_{\text{step}}.
\end{equation}

The combined effect of the four corrections increases the per-step Toffoli count by $5$--$123\times$ (depending on the system) and the Trotter step count by $8$--$60\times$, yielding total Toffoli counts $361$--$17{,}100\times$ larger than the published values, dependent on the benchmark system's number of modes and electronic channels.

\section{Architectural analysis: non-Clifford layer comparison}
\label{sec:arch_analysis}

The two algorithms employ different non-Clifford primitives: every non-Clifford operation in the phase gradient circuit
is a Toffoli gate, while every non-Clifford operation in REQWIEM is a parameterized $R_z$ rotation.
Isolating circuit-architectural advantages of each approach begins with a comparison in units of non-Clifford layers, counting
Toffoli layers for the phase gradient method and rotation layers for REQWIEM.
This comparison reveals the depth separation mechanism but overestimates fault-tolerant advantage, because one rotation layer and one Toffoli layer have different costs when decomposed into $T$-gates ($C_T^{(\textrm{rot})} = 23$ versus
$d_T^{(\textrm{Toff})} = 4$, a $5.75\times$ asymmetry that favors the Toffoli-based method) (Appendix~\ref{app:tgate_pipeline}).

Again deviating from published estimates \cite{motlagh2025quantum}, all estimates in this section use the physical molecular point group symmetry of each benchmark system.
Symmetry enters in two ways by restricting the number of nonzero bilinear coupling pairs ($|\mathcal{G}_2^{<}|$ on-diagonal,
$|\mathcal{G}_4|$ off-diagonal), reducing the rotation count in both methods, and by determining the chromatic index of the mode-pair conflict graph ($\chi'(\mathcal{G}_2^{<})$, $\chi'(\mathcal{G}_4)$) governing the depth of color-serialized bilinear layers in REQWIEM.
The first effect benefits both methods equally; the second benefits REQWIEM exclusively, being able to parallelize.

\subsection{Non-Clifford Depth}
Clifford gates (CNOTs, Hadamards, $S$ gates) are excluded from the depth count; in a fault-tolerant architecture, the assumption is made that these operations can be implemented transversally at negligible cost relative to non-Clifford operations synthesized from $T$-gates.
Precise calculation of $H\,,S$ gates is determined on native hardware gate sets, where the entanglement gate CNOT is decomposed into transversal gates and native entanglement operations (echo cross-resonance, M\o lmer-S\o rensen, CZ).

The phase gradient architecture is purely serial: within each Trotter step, the QROM--arithmetic--accumulate pipeline writes to a shared phase register, and all terms execute sequentially.
Its non-Clifford depth is therefore $D_{\textrm{PG}} = \Omega_{\textrm{step}}$ (one Toffoli layer per Toffoli gate).

Non-Clifford depth counts only $R_z$ layers in REQWIEM.
A Gray-code UCR$_Z^{(p)}$ with $P = 2^p$ angles contributes $P$ rotation layers (the interleaved CNOT transitions are Clifford and do not contribute to non-Clifford depth).
A CNOT--UCR$_Z^{(p)}$--CNOT sandwich for cross-pair terms has the same $P$ rotation layers; the outer CNOTs add only Clifford depth.
REQWIEM non-Clifford depth with full fan-out is
\begin{equation}\label{eq:D_reqwiem_comparison}
    D_{\textrm{R}} = 2\,D_{\textrm{kin}}^{(\textrm{rot})}
      + 2\,D_{\textrm{d}}^{(\textrm{rot})}
      + D_{\textrm{od}}^{(\textrm{rot})},
\end{equation}
with contributions:

\begin{equation}\label{eq:Dkin_rot}
        D_{\textrm{kin}}^{(\textrm{rot})}
      = n + \tbinom{n}{2} \qquad \text{(modes parallel, QFT is Clifford)},
\end{equation}
\begin{equation}\label{eq:Dd_rot}
    D_{\textrm{d}}^{(\textrm{rot})}
       = \underbrace{M}_{\text{Tier~1}}
       + \underbrace{\tbinom{n}{2}\,M}_{\text{Tier~2}}
       + \underbrace{\chi'(\mathcal{G}_2^{<})\, n^2\, M}_{\text{Tier~3}}
       \;=\; M\!\left[1 + \tbinom{n}{2}
         + \chi'(\mathcal{G}_2^{<})\,n^2\right],
\end{equation}
\begin{equation}\label{eq:Dod_rot}
            D_{\textrm{od}}^{(\textrm{rot})}
      = 2(M{-}1)\!\left[\left(1{+}n{+}\tbinom{n}{2}\right)\!\frac{M}{2}
        + \chi'(\mathcal{G}_4)\,n^2\,\frac{M}{2}\right], \qquad \text{per step}
\end{equation}

To isolate the two mechanisms that drive non-Clifford layer advantage, a depth fraction is defined
\begin{equation}\label{eq:depth_decomp}
    \frac{D_{\textrm{PG}}}{D_{\textrm{R}}}
      = \underbrace{\frac{D_{\textrm{PG}}}{D_{\textrm{R}}^{(\textrm{ser})}}
        }_{\displaystyle \alpha\;\text{(per-term overhead)}}
        \;\times\;
        \underbrace{\frac{D_{\textrm{R}}^{(\textrm{ser})}}{D_{\textrm{R}}^{(\textrm{fan})}}
        }_{\displaystyle \phi\;\text{(fan-out factor)}},
\end{equation}
where $D_{\textrm{R}}^{(\textrm{ser})}$ is the REQWIEM non-Clifford depth without fan-out (all modes and mode pairs serial, channel register shared between successive UCR operations).
In fully serial execution, every rotation is sequential, so $D_{\textrm{R}}^{(\textrm{ser})} \approx R_{\textrm{step}}$
(the total rotation count) and consequently $\alpha \approx \Omega_{\textrm{step}}/R_{\textrm{step}} = \Omega/R$ (recall $\Omega$ as circuit depth).
This identity confirms that $\alpha$ captures the intrinsic per-term cost ratio between the QROM--adder pipeline (Toffoli gates per term) and the UCR scheme (rotations per site) when both operate serially.

\subsection{Parallelism}
The fan-out factor $\phi$ captures mode-level parallelism exclusive to REQWIEM.
Since the phase gradient architecture writes all terms to a shared accumulator, it has no analogous parallelization opportunity;
$\phi$ therefore represents a structural advantage that cannot be closed by improvements to the phase gradient pipeline.

Table~\ref{tab:depth_decomp} presents the REQWIEM rotation count $R_{\textrm{step}}$ and non-Clifford depth $D_{\textrm{R}}^{(\textrm{fan})}$ for each benchmark system under physical molecular symmetry.
The fan-out factor $\phi = D_{\textrm{R}}^{(\textrm{ser})}/D_{\textrm{R}}^{(\textrm{fan})}$ is determined entirely by REQWIEM's circuit architecture and can be computed independently of the phase gradient cost.

Three sources of parallelism contribute to $\phi$:
\begin{itemize}
    \item \textbf{Mode fan-out} (intra-mode sites in both blocks):
    Fan-out of the channel register and truth-value ancillas parallelizes across $d$ modes, giving $\phi_{\textrm{intra}} = d$ for intra-mode terms in Tiers~1--2 (on-diagonal) and all intra-mode terms within each off-diagonal
    fragment.
    \item \textbf{Site parallelism} (on-diagonal Tier~1 only):
    Within each correction surface, $(1{+}n)$ rotations on distinct target qubits execute in a single non-Clifford layer, collapsing $(1{+}n)$ sequential rotation layers into $1$.
    Truth-value control ancillas are read-only, but the same argument does not apply to UCR sites in Tiers~2--3 or the off-diagonal block, where the Gray-code traversal serializes rotations within each UCR.
    \item \textbf{Graph coloring} (bilinear sites in both blocks): $|\mathcal{G}_2^{<}|$ (on-diagonal) and $|\mathcal{G}_4|$ (off-diagonal) mode pairs are parallelized into $\chi'(\mathcal{G}_2^{<})$ and $\chi'(\mathcal{G}_4)$ color layers respectively, giving $\phi_{\textrm{bilin}} = |\mathcal{G}|/\chi'$.
\end{itemize}
The composite $\phi$ is a weighted average of contributions,
dominated by bilinear cross-mode terms (which account for ${>}95\%$ of per-fragment rotation count for $d \geq 11$).
For systems with $C_1$ symmetry (Anth/C$_{60}$), the bilinear contribution gives $\phi_{\textrm{bilin}} = \binom{d}{2}/\chi'(K_d)
\approx d/2$, as confirmed by $\phi = 123.3 \approx 246/2$ for the full system.
For systems with nontrivial point group symmetry, the restricted mode-pair graph has a lower $|\mathcal{G}|/\chi'$ ratio than $K_d$
(the disjoint union structure admits fewer pairs per color class), yielding a smaller $\phi$.
Pyrazine ($D_{2h}$) achieves $\phi = 8.4$, compared to the hypothetical $C_1$ value of $\phi \approx 12$, reflecting the
reduction from $|\mathcal{G}|/\chi' \approx 6$ (physical) versus $d/2 = 12$ ($C_1$) in the bilinear sector.
Absolute depth, however, decreases much more steeply than $\phi$: pyrazine's fan-out depth $D_{\textrm{R}}^{(\textrm{fan})} = 3{,}406$ is $4.2\times$ smaller under $D_{2h}$ than under $C_1$, because symmetry eliminates the majority of bilinear terms from both the numerator and denominator of $\phi$ while shrinking $D_{\textrm{R}}^{(\textrm{fan})}$.

\subsection{Per-term overhead $\alpha$}
The per-term overhead $\alpha = \Omega_{\textrm{step}} / R_{\textrm{step}}$ requires a symmetry-corrected Toffoli count $\Omega_{\textrm{step}}$ from the phase gradient method.
Since $\alpha$ captures a ratio of per-term costs between the two pipelines, its value is expected to be approximately preserved
under symmetry reduction, where the per-term cost of a bilinear QROM lookup$+$multiplier$+$adder ($\sim 900$ Toffoli at $n = 10$, $b = 45$) versus a bilinear UCR ($M$ or $M/2$ rotations) is intrinsic to each architecture. 
Symmetry-corrected $\Omega_{\textrm{step}}$ values for each benchmark system are derived in (Appendix) and will complete the entries in Table~\ref{tab:depth_decomp}.
For the present analysis, $\alpha =1.16-5.13$, being approximately preserved under symmetry reduction $\pm 7 \%$, confirming that the capture of an intrinsic per-term cost ratio.
The range of $\alpha$ holds for all systems (since per-term Toffoli cost exceeds the per-term rotation count for any $m$), and
the total non-Clifford layer advantage $\alpha \phi$ is bounded below by $\phi$ alone.

\subsection{Limitation of the non-Clifford layer metric}
Non-Clifford layer advantages reported in Table~\ref{tab:depth_decomp} overstate the fault-tolerant execution time advantage due to a false equivalence between one rotation layer and one Toffoli layer, made intentionally to highlight structural properties.
In a fault-tolerant architecture, each Toffoli decomposes into $d_T^{(\textrm{Toff})} = 4$ sequential $T$-gate layers, whereas each $R_z$ rotation synthesized via Ross--Selinger-PyZX-TODD pipeline decomposes into $C_T^{(\textrm{rot})}$ sequential $T$-gate layers (baseline $C_T = 23$) (Appendix~\ref{app:tgate_pipeline}).
One rotation layer is therefore $C_T/d_T^{(\textrm{Toff})} = 23/4 = 5.75\times$ more expensive in $T$-depth than one Toffoli layer.
To obtain a more accurate metric for fault-tolerant advantage, the non-Clifford layer ratio must be corrected by this factor.

\begin{table}[ht]
\centering
\small
\caption{REQWIEM non-Clifford depth components and fan-out factor
under physical molecular symmetry.
$\chi'_{\textrm{d}} \equiv \chi'(\mathcal{G}_2^{<})$ and
$\chi'_{\textrm{od}} \equiv \chi'(\mathcal{G}_4)$ are the
on-diagonal and off-diagonal chromatic indices.
$D_{\text{PG}}$: total non-Clifford depth from the phase gradient method.
$D_{\textrm{R}}^{(\textrm{ser})} \approx R_{\textrm{step}}$:
rotation layers without fan-out (all modes serial).
$D_{\textrm{R}}^{(\textrm{fan})}$: rotation layers with fan-out
(Eqs.~\ref{eq:Dkin_rot}--\ref{eq:Dod_rot}), decomposed into
kinetic ($2D_{\textrm{kin}}$), on-diagonal ($2D_{\textrm{d}}$),
and off-diagonal ($D_{\textrm{od}}$) contributions.
$\alpha = D_{\text{PG}} / D_{\textrm{R}}^{(\textrm{ser})}$: per-term overhead
from the phase gradient method.
$\phi = D_{\textrm{R}}^{(\textrm{ser})}/D_{\textrm{R}}^{(\textrm{fan})}$:
fan-out parallelism factor.
$\alpha\phi$: complete non-Clifford layer ratio.
}\label{tab:depth_decomp}
\renewcommand{\arraystretch}{1.2}
\begin{tabular}{l c c c r r r c c c}
\hline\hline
System & $d$ & $\chi'_{\textrm{d}}$ & $\chi'_{\textrm{od}}$
  & $D_{\text{PG}}$
  & $D_{\textrm{R}}^{(\textrm{ser})}$
  & $D_{\textrm{R}}^{(\textrm{fan})}$
  & $\alpha$
  & $\phi$
  & $\alpha\phi$ \\
\hline
Pyrazine
    & 24 & 5 & 5
    & $1.47\times10^5$
    & $2.87\times10^4$
    & $3.41\times10^3$
    & $5.13$ & $8.4$ & $43$ \\
(NO)$_4$-Anth
    & 19 & 7 & 19
    & $1.29\times 10^6$
    & $1.11\times10^6$
    & $1.22\times10^5$
    & $1.16$ & $9.1$ & $11$ \\
(NO)$_4$-Anth Dimer
    & 21 & 9 & 15
    & $9.87\times10^5$
    & $8.42\times10^5$
    & $1.02\times10^5$
    & $1.17$ & $8.2$ & $10$ \\
Anth/C$_{60}$
    & 11 & 11 & 11
    & $3.02\times10^5$
    & $1.24\times10^5$
    & $2.32\times10^4$
    & $2.45$ & $5.3$ & $13$ \\
Anth/C$_{60}$ (full)
    & 246 & 245 & 245
    & $1.39\times10^8$
    & $6.06\times10^7$
    & $4.91\times10^5$
    & $2.30$ & $123.3$ & $283$ \\
\hline\hline
\end{tabular}
\end{table}

\section{Error Budget}\label{sec:error}

\subsection{Error sources and budget allocation}

Total simulation error in REQWIEM arises from three independent sources: Trotter decomposition error from the split-operator scheme, discretization error from the finite grid representation, and rotation synthesis error from the fault-tolerant implementation of parameterized gates.

The triangle inequality bounds the total error as
\begin{equation}\label{eq:error_total}
\eps_{\text{total}}
  \leq \eps_{\text{Trotter}}
     + \eps_{\text{disc}}
     + \eps_{\text{synth}},
\end{equation}
where each term represents the operator norm distance between the ideal
and implemented evolution. For chemical applications, a typical target is
$\eps_{\text{total}} \leq 1.6 \times 10^{-3}$~E$_\text{H}$
(1~kcal/mol, ``chemical accuracy'').
An error budget is allocated as
$\eps_{\text{Trotter}} \leq 8 \times 10^{-4}$,
$\eps_{\text{disc}} \leq 5 \times 10^{-4}$, and
$\eps_{\text{synth}} \leq 3 \times 10^{-4}$.
This allocation reflects relative controllability of each error source: Trotter error dominates for short simulations but is reduced by increasing step count $k$; discretization error is set by the physical problem and qubit budget; synthesis error is determined by $T$-gate resources.

\subsection{Trotter error}

The second-order Suzuki--Trotter decomposition implements
\begin{equation}\label{eq:trotter_approx}
\tilde{U}(\tau) = e^{-iT\tau/2}\, e^{-iV\tau}\, e^{-iT\tau/2},
\end{equation}
as an approximation to $U(\tau) = e^{-iH\tau}$ for $H = T + V$ and
time step $\tau = t/k$.

The second-order Trotter error is bounded by nested
commutators~\cite{childs2021theory, suzuki1991general}:
\begin{equation}
\eps_{\text{Trotter}}
  \leq \frac{t^3}{24k^2}
       \bigl(\|[V,[V,T]]\| + 2\|[T,[T,V]]\|\bigr).
\end{equation}

For kinetic operator $T = P^2/2\mu$ on a grid with $N = 2^n$ points and $p_{\max} = \pi N/L$, direct calculation gives
(Appendix~\ref{app:Error}): $[T,[T,V]]$ vanishes for linear terms but scales as $O(N^2)$ for quadratic and bilinear terms, with $[p^2, \{p,x\}] = -4ip^2$, introducing an additional factor of $p_{\max}^2 \propto N^2$.

For a vibronic Hamiltonian with bilinear couplings, the Trotter error bound becomes:
\begin{equation}\label{eq:trotter_bilinear}
\eps_{\text{Trotter}}
  \leq \frac{t^3}{24k^2}\left(C_0 + C_2 N^2\right),
\end{equation}
where $C_0 = O(\alpha^2/\mu)$ collects linear contributions and $C_2 = O(\gamma/\mu^2)$ collects bilinear contributions.

To maintain target error $\eps$, the required step count scales as:
\begin{equation}
k_{\text{bilinear}}
  = O\!\left(\frac{t^{3/2} N}{\sqrt{\eps}}\right).
\end{equation}

Step count increases for a given target error as the number of grid points increases. 
The system's characteristic energy (Nyquist limit) doubles with every additional qubit, commensurately increasing spectral norms present in the timestep calculation. 
For large-scale simulation, it therefore becomes necessary to optimize each mode's register size to a characteristic energy scale to minimize Trotter error. 
For the two-mode pyrazine validation simulation ($N = 512$, $n = 9$), the $k = 2500$ steps used provides substantial margin for Trotter accuracy, chosen primarily for sufficient spectral resolution ($\delta\omega = 2\pi/(k\,\tau)$). The full derivation of the commutator norm bounds is given in Appendix~\ref{app:Error}. 
A safe margin of $n=10$ provides a margin for similar systems with greater energetic properties, particularly for systems whose diabatic states have large energy gaps.
In many cases, $n=10$ may unnecessarily increase spectral norm, but it used as a benchmark between methods to ensure converged observables.

\subsection{Rotation synthesis error}\label{sec:synth_error}

On fault-tolerant quantum computers, arbitrary single-qubit rotations $R_z(\theta)$ are synthesized from the discrete Clifford$+T$ gate set.
The Ross--Selinger algorithm achieves optimal scaling for the per-gate approximation error $\varepsilon_\theta$, requiring
$N_T(\varepsilon_\theta) = 3\log_2(1/\varepsilon_\theta) + O(\log\log(1/\varepsilon_\theta))$ $T$-gates~\cite{dawson2005solovay, ross2016optimal}.
With additional improvements from mixed fallback synthesis, PyZX inter-block optimization, and T-Optimization by Diagonal Decomposition (TODD) optimization, a conservative
estimate yields a $T$-gate reduction factor of $\sim\!6.8\times$ relative to bare Ross--Selinger~\cite{ross2016optimal}.

For a circuit with $N_R$ parameterized rotations, each synthesized to precision $\varepsilon_\theta$, the total synthesis error accumulates coherently as $\eps_{\text{synth}} \leq N_R \cdot \varepsilon_\theta$, giving the per-gate precision requirement
$\varepsilon_\theta \leq \eps_{\text{synth}} / N_R$.

The total number of parameterized rotations per Trotter step is $R_{\textrm{step}}$; the total rotation count over the full simulation is $N_R = k \cdot R_{\textrm{step}}$.
For the two-mode pyrazine validation model ($d = 2$, $M = 2$, $n = 9$), $R_{\textrm{step}} \approx 1.2 \times 10^3$, yielding
$N_R \approx 3.0 \times 10^6$ at $k = 2500$ Trotter steps. 
For the stated threshold of $\eps_{\text{synth}} \leq 3 \times 10^{-4}$~E$_\text{H}$, the per-gate precision is $\varepsilon_\theta \approx 1.0 \times 10^{-10}$, which is comfortably achievable with $C_T \approx 100$ $T$-gates per rotation via Ross--Selinger, or $C_T \approx 15$ after the $6.8\times$ optimization factor.

For the benchmark systems of Section~\ref{sec:Fault-Tolerant_Estimates},
the rotation counts are substantially larger (Table~\ref{tab:comprehensive}), and the per-gate precision tightens accordingly.
The most demanding case is Anth/C$_{60}$ ($d = 246$), with $N_R \approx 2.0 \times 10^{11}$ rotations at $k = 3274$ steps,
requiring $\varepsilon_\theta \approx 1.5 \times 10^{-15}$ and $C_T \approx 148$ via bare Ross--Selinger ($C_T \approx 22$ after
optimization).
This is within the range where the baseline compilation parameter $C_T = 23$ used throughout Section~\ref{sec:Fault-Tolerant_Estimates} to be self-consistent.
The classical emulations presented in Chapter~\ref{ch:model_ext} use double-precision floating-point arithmetic for all rotation angles ($\varepsilon_\theta \approx 10^{-16}$), and $\varepsilon_\theta = 10^{-14}$ is used for all $T$-gate estimations given here, providing conservative margin over the precision required by the error budget.

\section{Cost metrics}\label{sec:cost_metrics}

The phase gradient method incurs $\Omega_{\textrm{step}}$ Toffoli gates per Trotter step.
REQWIEM incurs $R_{\textrm{step}}$ parameterized rotations (Eq.~\ref{eq:Rstep}) and a negligible number of Toffoli gates ($T_{\textrm{step}} \leq 20$ for all benchmark systems; Eq.~\ref{eq:Tstep}).
The $T$-gate budget is
\begin{equation}\label{eq:T_budgets}
    T^{(\textrm{PG})} = C_T^{(\textrm{Toff})}\,\Omega_{\textrm{step}}, \qquad
    T^{(\textrm{R})} = C_T^{(\textrm{rot})}\,R_{\textrm{step}} + C_T^{(\textrm{Toff})}\,T_{\textrm{step}},
\end{equation}
where $C_T^{(\textrm{Toff})} = 4$ and $C_T^{(\textrm{rot})}$ is the rotation synthesis cost (baseline $C_T^{(\textrm{rot})} = 23$ via empirical Ross--Selinger fit~\cite{ross2016optimal}).
A crossover
\begin{equation}\label{eq:crossover}
    C_T^{(\textrm{rot},\star)}
      = \frac{C_T^{(\textrm{Toff})}\,\Omega_{\textrm{step}}}{R_{\textrm{step}}}
      = C_T^{(\textrm{Toff})} \cdot \frac{\Omega}{R},
\end{equation}
is the synthesis cost at which both methods have equal $T$-gate count.
Below this value, REQWIEM holds the $T$-gate advantage.
Under physical symmetry, the ratio $\Omega/R$ ranges from $1.16$ ((NO)$_4$-Anth, $m = 3$) to $5.13$ (pyrazine, $m = 1$), yielding crossover values $C_T^{\star} = 4.7$--$20.5$ (Table~\ref{tab:CT_sensitivity}).

\section{Fault-tolerant analysis: T-depth comparison} \label{sec:ft_analysis}

To compare execution time, both circuit design approaches are decomposed into the Clifford + $T$ gate set.
Two depth metrics are reported:
\begin{enumerate}
    \item \textbf{$T$-depth}: number of sequential $T$-gate layers
    (Clifford layers contribute zero $T$-depth).
    \item \textbf{Total decomposed depth}: the number of sequential
    layers of any type ($T$, CNOT, or single-qubit Clifford).
\end{enumerate}
This decomposition converts both algorithms to a common unit of cost, eliminating per-operation asymmetry between Toffoli gates and synthesized rotations that inflates the non-Clifford layer comparison of Section~\ref{sec:arch_analysis}.

\subsection{Decomposition rules}
Each Toffoli gate decomposes into a circuit on three qubits with $T$-count $C_T^{(\textrm{Toff})} = 4$ (via the temporary logical AND~\cite{gidney2018halving}), $T$-depth $d_T^{(\textrm{Toff})} = 4$, and total decomposed depth $d_{\textrm{tot}}^{(\textrm{Toff})} = 12$ (including interleaved CNOTs and Hadamards).
Each $R_z(\theta)$ rotation is given to precision $\varepsilon$ using $C_T^{(\textrm{rot})}$ $T$-gates via the Ross--Selinger protocol~\cite{ross2016optimal}, producing an alternating sequence of $T$ and Clifford gates on a single qubit with $T$-depth $d_T^{(\textrm{rot})} = C_T^{(\textrm{rot})}$ and total depth $d_{\textrm{tot}}^{(\textrm{rot})} = 2C_T^{(\textrm{rot})} - 1$.

\subsection{QROM (phase gradient)}
The architecture is purely serial and composed entirely of Toffoli
gates; the decomposed depths are therefore:
\begin{equation}\label{eq:DM_decomp}
    D_{\textrm{PG}}^{(T)}
      = d_T^{(\textrm{Toff})} \times \Omega_{\textrm{step}}
      = 4\,\Omega_{\textrm{step}}, \qquad
    D_{\textrm{PG}}^{(\textrm{tot})}
      = d_{\textrm{tot}}^{(\textrm{Toff})} \times \Omega_{\textrm{step}}
      = 12\,\Omega_{\textrm{step}}.
\end{equation}

\subsection{REQWIEM}
This circuit method interleaves rotation layers and CNOT layers.
Let $D_{\textrm{R}}^{(\textrm{rot})}$ denote the number of rotation layers in a fan-out-enabled circuit (Eqs.~\ref{eq:Dkin_rot}--\ref{eq:Dod_rot}) and $D_{\textrm{R}}^{(\textrm{CNOT})}$ the number of CNOT layers (Gray-code transitions, fan-out, focusing Cliffords).
After decomposition, each rotation layer expands into $C_T^{(\textrm{rot})}$ $T$-layers and $C_T^{(\textrm{rot})} - 1$ interleaved Clifford layers, while each CNOT layer remains unchanged:
\begin{align}
    D_{\textrm{R}}^{(T)} &= C_T^{(\textrm{rot})}
      \times D_{\textrm{R}}^{(\textrm{rot})},
    \label{eq:DR_T}\\
    D_{\textrm{R}}^{(\textrm{tot})} &= (2C_T^{(\textrm{rot})} {-} 1)
      \times D_{\textrm{R}}^{(\textrm{rot})}
      + D_{\textrm{R}}^{(\textrm{CNOT})}.
    \label{eq:DR_tot}
\end{align}
The per-step rotation-layer count $D_{\textrm{R}}^{(\textrm{rot})} = D_{\textrm{R}}^{(\textrm{fan})}$ is identical to the fan-out-enabled non-Clifford depth of Table~\ref{tab:depth_decomp}; negligible Toffoli contributions from truth-value computation ($T_{\textrm{step}} \leq 20$) is omitted.

\subsection{Connection to the architectural analysis}
$T$-depth ratio connects directly to the non-Clifford layer
decomposition of Section~\ref{sec:arch_analysis}:
\begin{equation}\label{eq:Tdepth_from_arch}
    \frac{D_{\textrm{PG}}^{(T)}}{D_{\textrm{R}}^{(T)}}
    = \frac{d_T^{(\textrm{Toff})}}{C_T^{(\textrm{rot})}}
      \times \underbrace{\frac{D_{\textrm{PG}}}{D_{\textrm{R}}^{(\textrm{fan})}}
      }_{\alpha\,\phi}
    = \frac{4}{C_T^{(\textrm{rot})}} \times \alpha\,\phi.
\end{equation}
The factor $4/C_T^{(\textrm{rot})} = 4/23 \approx 0.17$ at baseline is the ``tax'' REQWIEM pays for using synthesized rotations instead of Toffoli gates: each non-Clifford layer in REQWIEM costs $5.75\times$ more $T$-depth than a Toffoli layer.
The architectural advantage $\alpha\phi$ must exceed $C_T^{(\textrm{rot})}/d_T^{(\textrm{Toff})} = 5.75$ for REQWIEM to
hold a $T$-depth advantage.

For all benchmark systems, $\alpha\phi$ ranges from $10\times$ (dimer) to $283\times$ (Anth/C$_{60}$ full), with pyrazine at $43\times$ (Table~\ref{tab:depth_decomp}), comfortably exceeding this threshold.

\subsection{$T$-depth results}
Table~\ref{tab:Tdepth_comparison} presents the $T$-depth comparison.
REQWIEM achieves a $T$-depth advantage across all five systems, ranging from $1.7\times$ ((NO)$_4$-Anth Dimer, $d = 21$, $m = 3$) to $49\times$ (Anth/C$_{60}$ full, $d = 246$, $m = 2$), with pyrazine at $7.5\times$.
These estimates account for the full per-operation cost asymmetry between the two non-Clifford primitives.

The $T$-depth advantage improves as the rotation synthesis cost $C_T^{(\textrm{rot})}$ decreases (Table~\ref{tab:Tdepth_CT}), because the per-rotation $T$-depth shrinks while the per-Toffoli $T$-depth ($d_T^{(\textrm{Toff})} = 4$) is fixed.
At the Hamming weight phasing regime ($C_T = 10$), the $T$-depth advantage improves from $1.3\times$ to $3.9\times$ for the dimer and from $49\times$ to $113\times$ for the full Anth/C$_{60}$ system.
In the limit $C_T^{(\textrm{rot})} \to 4$ (equal to the Toffoli $T$-count), the $T$-depth ratio converges to the non-Clifford layer
advantage $\alpha\phi$: from $10\times$ to $283\times$.
Future advances in rotation synthesis will therefore benefit REQWIEM in both $T$-gate count and $T$-depth simultaneously.

\subsection{Total decomposed depth results}
Table~\ref{tab:decomposed} presents the total decomposed depth (including CNOT and single-qubit Clifford layers) after Toffoli and rotation decomposition.
Total-depth advantages are uniformly larger than the $T$-depth advantages: $2.5$--$72\times$ versus $1.7$--$49\times$.
This amplification arises from an asymmetry in depth inflation factors and improved depth scaling in REQWIEM.
Every layer in the phase gradient circuit is a Toffoli gate, so depth inflates uniformly by $d_{\textrm{tot}}^{(\textrm{Toff})}/d_T^{(\textrm{Toff})} = 12/4 = 3\times$ going from $T$-depth to total depth.
In contrast, REQWIEM's circuit is composed of both rotation layers ($\sim\!29$--$38\%$ of pre-decomposition depth) and CNOT layers ($\sim\!62$--$71\%$); only rotation layers inflate (by $(2C_T - 1)/C_T = 45/23 \approx 2\times$), while the CNOT layers
contribute no $T$-depth and add only modestly to total depth. 
The effective inflation factor for REQWIEM ($\sim\!14$--$18\times$; Table~\ref{tab:inflation}) is comparable to a phase gradient approach's $12\times$, but the phase-gradient method's pre-decomposition depth is already $\alpha\phi$ times larger; decomposition preserves and slightly amplifies this gap.

\begin{table}[ht]
\centering
\small
\caption{$T$-depth comparison after Clifford$+T$ decomposition.
$D_{\textrm{PG}}^{(T)} = 4\,\Omega_{\textrm{step}}$
(serial Toffoli, $T$-depth $4$ each).
$D_{\textrm{R}}^{(T)} = C_T^{(\textrm{rot})} \times
D_{\textrm{R}}^{(\textrm{rot})}$ (synthesized rotations,
$T$-depth $C_T$ each; baseline $C_T = 23$).
The $T$-depth ratio satisfies
Eq.~\ref{eq:Tdepth_from_arch}:
$(4/C_T) \times \alpha \times \phi$.
All estimates use physical molecular symmetry.
}\label{tab:Tdepth_comparison}
\renewcommand{\arraystretch}{1.2}
\begin{tabular}{l c r r c c c c}
\hline\hline
System & $d$
  & $D_{\textrm{PG}}^{(T)}$
  & $D_{\textrm{R}}^{(T)}$
  & $\alpha$ & $\phi$
  & $\alpha\phi$
  & \shortstack{$T$-depth \\ ratio} \\
\hline
Pyrazine
    & 24 & $5.89\times 10^5$ & $7.83 \times 10^4$
    & $5.13$ & $8.4$ & $43\times$ & $7.5\times$ \\
(NO)$_4$-Anth
    & 19 & $5.15 \times 10^6$ & $2.80\times 10^6$
    & $1.16$ & $9.1$ & $11\times$ & $1.8\times$ \\
(NO)$_4$-Anth Dimer
    & 21 & $3.95 \times 10^6$ & $2.35 \times 10^6$
    & $1.17$ & $8.2$ & $10\times$ & $1.7\times$ \\
Anth/C$_{60}$ ($d\!=\!11$)
    & 11 & $1.21\times 10^6$ & $5.32 \times 10^5$
    & $2.45$ & $5.3$ & $13\times$ & $2.3\times$ \\
Anth/C$_{60}$ ($d\!=\!246$)
    & 246 & $5.56 \times 10^8$ & $1.13 \times 10^7$
    & $2.30$ & $123.3$ & $283\times$ & $49.2\times$ \\
\hline\hline
\end{tabular}
\end{table}

\begin{table}[ht]
\centering
\small
\caption{Total decomposed depth after decomposition into
$\{T,\,\textrm{CNOT},\,H,\,S\}$.
$D_{\textrm{PG}}^{(\textrm{tot})} = 12\,\Omega_{\textrm{step}}$
(total depth $12$ per Toffoli).
$D_{\textrm{R}}^{(\textrm{tot})} =
(2C_T - 1)\,D_{\textrm{R}}^{(\textrm{rot})}
+ D_{\textrm{R}}^{(\textrm{CNOT})}$ (rotation synthesis plus
unchanged CNOT layers).}\label{tab:decomposed}
\renewcommand{\arraystretch}{1.2}
\begin{tabular}{l r r c}
\hline\hline
System
  & $D_{\textrm{PG}}^{(\textrm{tot})}$
  & $D_{\textrm{R}}^{(\textrm{tot})}$
  & \shortstack{Total depth \\ ratio} \\
\hline
Pyrazine
  & $1.77 \times 10^6$ & $1.62 \times 10^5$ & $10.9\times$ \\
(NO)$_4$-Anth
  & $1.54\times10^6$ & $5.68\times10^5$ & $2.7\times$ \\
(NO)$_4$-Anth Dimer
  & $1.18 \times 10^7$ & $4.78 \times 10^6$ & $2.5\times$ \\
Anth/C$_{60}$ ($d\!=\!11$)
  & $3.63 \times 10^6$ & $1.09\times 10^6$ & $3.3\times$ \\
Anth/C$_{60}$ ($d\!=\!246$)
  & $1.67 \times 10^9$ & $2.31 \times 10^7$ & $72.3\times$ \\
\hline\hline
\end{tabular}
\end{table}

\begin{table}[ht]
\centering
\small
\caption{Depth inflation from Clifford$+T$ decomposition.}\label{tab:inflation}
\renewcommand{\arraystretch}{1.2}
\begin{tabular}{l r r c c}
\hline\hline
System
  & \shortstack{Pre-decomp.\ depth\\(PG)}
  & \shortstack{Pre-decomp.\ depth\\(REQWIEM)}
  & \shortstack{Phase Gradient\\inflation}
  & \shortstack{REQWIEM\\inflation} \\
\hline
Pyrazine           & $147{,}373$       & $11{,}692$      & $12.0\times$ & $13.8\times$ \\
(NO)$_4$-Anth      & $1{,}287{,}067$   & $327{,}180$     & $12.0\times$ & $17.4\times$ \\
Dimer              & $986{,}767$       & $272{,}380$     & $12.0\times$ & $17.5\times$ \\
Anth/C$_{60}$ ($d\!=\!11$)  & $302{,}137$       & $69{,}808$      & $12.0\times$ & $15.6\times$ \\
Anth/C$_{60}$ ($d\!=\!246$) & $139{,}026{,}162$ & $1{,}473{,}872$ & $12.0\times$ & $15.7\times$ \\
\hline\hline
\end{tabular}
\end{table}

\begin{table}[ht]
\centering
\small
\caption{$T$-depth advantage
$D_{\textrm{PG}}^{(T)}/D_{\textrm{R}}^{(T)}$ as a function of rotation synthesis cost $C_T^{(\textrm{rot})}$.
}\label{tab:Tdepth_CT}
\renewcommand{\arraystretch}{1.2}
\begin{tabular}{l c c c c}
\hline\hline
\multirow{2}{*}{System}
  & \multicolumn{4}{c}{$T$-depth advantage} \\
\cline{2-5}
  & $C_T = 30$ & $C_T = 23$ & $C_T = 10$ & $C_T = 4$ \\
\hline
Pyrazine
    & $5.8\times$ & $7.5\times$ & $17.3\times$ & $43.3\times$ \\
(NO)$_4$-Anth
    & $1.4\times$ & $1.8\times$ & $4.2\times$ & $10.6\times$ \\
(NO)$_4$-Anth Dimer
    & $1.3\times$ & $1.7\times$ & $3.9\times$ & $9.6\times$ \\
Anth/C$_{60}$ ($d\!=\!11$)
    & $1.7\times$ & $2.3\times$ & $5.2\times$ & $13.1\times$ \\
Anth/C$_{60}$ ($d\!=\!246$)
    & $37.7\times$ & $49.2\times$ & $113.2\times$ & $283.1\times$ \\
\hline\hline
\end{tabular}
\end{table}

\subsection{$T$-gate count comparison}\label{sec:Tgate_comparison}

The $T$-gate comparison is strongly $m$-dependent
(Tables\ref{tab:Tdepth_CT}-~\ref{tab:comparison}).
At the baseline synthesis cost ($C_T^{(\textrm{rot})} = 23$,
$C_T^{(\textrm{Toff})} = 4$), the Toffoli-to-rotation ratio
$\Omega/R$ ranges from $5.13$
(pyrazine, $m = 1$) down to $1.16$ ((NO)$_4$-Anth, $m = 3$).
The resulting $T$-gate ratio $T^{(\textrm{R})}/T^{(\textrm{PG})}$ spans
$1.12\times$ (pyrazine, nearly neutral) to $4.94\times$
((NO)$_4$-Anth, substantial REQWIEM penalty).

The physical origin is as follows.
In the off-diagonal block, each UCR$_Z^{(m-1)}$ contributes
$M/2 = 2^{m-1}$ rotations per grid site, and the $2(M-1)$ fragment
applications yield a total of $M(M-1)\,S_{\textrm{sites}}$ rotations.
For $m = 1$ ($M = 2$), each fragment's UCR$_Z^{(0)}$ reduces to a
single $R_z$ per site, giving a modest rotation count that is
competitive with the QROM--arithmetic cost.
For $m = 3$ ($M = 8$), the UCR$_Z^{(2)}$ contributes $4$ rotations per
site across $14$ fragment applications, and the $56\,S_{\textrm{sites}}$
aggregate rotation count is only marginally smaller than the phase gradient
Toffoli count $\Omega_{\textrm{step}}$. 
At $23/4 = 5.75\times$ higher $T$-gate cost per operation, the penalty compounds.

\begin{table}[ht]
\centering
\small
\caption{Sensitivity of the $T$-gate ratio
$T^{(\textrm{R})}/T^{(\textrm{PG})}$ to the rotation synthesis cost
$C_T^{(\textrm{rot})}$.
REQWIEM's budget scales as
$C_T^{(\textrm{rot})} \cdot R_{\textrm{step}}$
while the phase gradient budget
$4\,\Omega_{\textrm{step}}$ is independent of
$C_T^{(\textrm{rot})}$, so the ratio scales linearly.
``Crossover'' $C_T^{(\textrm{rot},\star)} =
4\,\Omega_{\textrm{step}}/R_{\textrm{step}}$
is the synthesis cost at which the two methods achieve equal $T$-gate
cost.
Values below $1.0$ (\textbf{bold}) indicate REQWIEM advantage.
}\label{tab:CT_sensitivity}
\renewcommand{\arraystretch}{1.2}
\begin{tabular}{l c c c c c}
\hline\hline
\multirow{2}{*}{System}
  & crossover
  & \multicolumn{4}{c}{$T^{(\textrm{R})}/T^{(\textrm{PG})}$} \\
\cline{3-6}
  & $C_T^{(\textrm{rot},\star)}$
  & $C_T = 30$ & $C_T = 23$ & $C_T = 10$ & $C_T = 4$ \\
\hline
Pyrazine
  & $20.5$ & $1.46$ & $1.12$
  & $\mathbf{0.49}$ & $\mathbf{0.19}$ \\
(NO)$_4$-Anth
  & $4.7$ & $6.45$ & $4.94$
  & $2.15$ & $\mathbf{0.86}$ \\
(NO)$_4$-Anth Dimer
  & $4.7$ & $6.40$ & $4.91$
  & $2.13$ & $\mathbf{0.85}$ \\
Anth/C$_{60}$ ($d\!=\!11$)
  & $9.8$ & $3.07$ & $2.35$
  & $1.02$ & $\mathbf{0.41}$ \\
Anth/C$_{60}$ ($d\!=\!246$)
  & $9.2$ & $3.27$ & $2.51$
  & $1.09$ & $\mathbf{0.44}$ \\
\hline\hline
\end{tabular}
\end{table}

The crossover values $C_T^{(\textrm{rot},\star)}$
(Table~\ref{tab:CT_sensitivity}) reflect this $m$-dependence:
\begin{itemize}
    \item $m = 1$ (pyrazine): $C_T^{\star} = 20.5$,
    comfortably below baseline.
    REQWIEM achieves a $T$-gate advantage at any
    $C_T^{(\textrm{rot})} \leq 20$, including the Hamming weight
    phasing regime ($C_T \approx 10$), where it holds a
    $2.0\times$ $T$-gate advantage.
    \item $m = 2$ (Anth/C$_{60}$): $C_T^{\star} \approx 9$--$10$,
    within the Hamming weight phasing regime.
    At $C_T = 10$, the ratio is $1.02$--$1.09\times$ (nearly neutral).
    \item $m = 3$ ((NO)$_4$-Anth): $C_T^{\star} = 4.7$,
    unreachable with current synthesis techniques.
    The $T$-gate penalty persists at $\sim\!2\times$ even in the
    Hamming weight regime ($C_T = 10$).
\end{itemize}

\begin{table}[ht]
\centering
\small
\caption{Per-step $T$-gate comparison at baseline compilation
($C_T^{(\textrm{rot})} = 23$, $C_T^{(\textrm{Toff})} = 4$).
$\Omega_{\textrm{step}}$: Toffoli count for the phase gradient method.
$R_{\textrm{step}}$: rotation count for REQWIEM.
$\Omega/R$ governs the crossover condition~\ref{eq:crossover}: REQWIEM achieves a $T$-gate advantage
when $C_T^{(\textrm{rot})} < C_T^{(\textrm{Toff})}(\Omega/R)$.
}\label{tab:comparison}
\renewcommand{\arraystretch}{1.2}
\begin{tabular}{l r r c r r c}
\hline\hline
System
  & $\Omega_{\textrm{step}}$
  & $R_{\textrm{step}}$
  & $\Omega/R$
  & $T^{(\textrm{PG})}$
  & $T^{(\textrm{R})}$
  & $T^{(\textrm{R})}\!/T^{(\textrm{PG})}$ \\
\hline
Pyrazine
  & $1.47\times10^5$ & $2.87\times10^5$ & $5.13$
  & $5.89\times10^5$ & $6.60\times10^5$ & $1.12$ \\
(NO)$_4$-Anth
  & $1.29\times 10^6 $ & $1.11\times 10^6$ & $1.16$
  & $5.15\times 10^6$ & $2.55\times 10^7$ & $4.94$ \\
(NO)$_4$-Anth Dimer
  & $9.87\times 10^6$ & $8.42\times 10^6$ & $1.17$
  & $3.95\times 10^6$ & $1.94\times 10^7$ & $4.91$ \\
Anth/C$_{60}$ ($d\!=\!11$)
  & $3.02\times 10^6$ & $1.24\times 10^6$ & $2.45$
  & $1.21\times 10^6$ & $2.84\times 10^6$ & $2.35$ \\
Anth/C$_{60}$ ($d\!=\!246$)
  & $1.39\times 10^8$& $6.06\times 10^7$ & $2.30$
  & $5.56\times 10^8$ & $1.39\times 10^9$ & $2.51$ \\
\hline\hline
\end{tabular}
\end{table}

\section{Interpretation and regime map}\label{sec:interpretation}

The phase gradient algorithm resides in the Toffoli cost sector, where every improvement must come from better QROM constructions, arithmetic circuits, or Toffoli factory throughput.
REQWIEM costs improve for better rotation synthesis protocols.

\subsection{Depth advantage (structural, universal)}
The UCR Gray-code cascade is intrinsically less deep per coupling site than the QROM--arithmetic pipeline ($\alpha \geq 1$, i.e., fewer non-Clifford layers per term even in serial execution).
Further, intra-fragment fan-out parallelizes across modes ($\phi \approx d/2$), exploiting a structural feature of the read-only channel register.
The combined non-Clifford layer advantage $\alpha\phi$ ranges from $10\times$ to $283\times$ (Table~\ref{tab:depth_decomp}).
After accounting for a $5.75\times$ per-operation cost asymmetry between rotations and Toffoli gates, fault-tolerant $T$-depth advantage remains with REQWIEM, with $1.7$--$49\times$ improvement at baseline synthesis (Table~\ref{tab:Tdepth_comparison}), growing to $4$--$283\times$ in the Hamming weight regime ($C_T = 10$).
The total decomposed depth advantage ($2.5$--$72\times$; Table~\ref{tab:decomposed}) is uniformly larger than $T$-depth advantage due to an asymmetric inflation of Toffoli versus rotation layers, both being robust across all five systems and all compilation parameters examined.

\subsection{Regime map}
Results trace out largely independent forms of advantage. as compiled in Table~\ref{tab:combined_summary}.. 
The fan-out factor scales as $\phi \approx d/2$ for large $d$ (bilinear graph-coloring reduction dominates), yielding $T$-depth advantages that grow from $1.7$--$2.3\times$ at $d = 11$--$21$ to $49\times$ at $d = 246$ (baseline $C_T = 23$).
Serial overhead $\alpha \geq 1$ for all systems ensures that fan-out amplifies an existing advantage rather than compensating
for a deficit.
Off-diagonal rotation count scales as $M(M-1)\,S_{\textrm{sites}}$, producing $\Omega/R$ ratios that decrease from $5.1$ ($m = 1$) to $1.2$ ($m = 3$).
At baseline synthesis costs, crossover shifts from $C_T^{\star} = 20.5$ (REQWIEM-favourable) down to $C_T^{\star} = 4.7$ (phase-gradient-favourable).
Reasonable future reduction in rotation synthesis cost alleviate the penalty for $m \leq 2$ but not for $m = 3$.

\begin{table}[ht]
\centering
\small
\caption{Summary of per-step comparison at baseline compilation ($C_T^{(\textrm{rot})} = 23$, $C_T^{(\textrm{Toff})} = 4$).
Non-Clifford layer advantages reflect the architectural parallelism of REQWIEM but overestimate the fault-tolerant advantage by a factor of $C_T/d_T^{(\textrm{Toff})} = 5.75$.
$T$-depth and total decomposed depth are apples-to-apples fault-tolerant metrics; both show REQWIEM advantage across all systems.
$T$-gate comparison is $m$-dependent: systems with $m = 1$ are nearly neutral, while $m = 3$ systems incur a substantial
rotation-count overhead. All estimates use physical molecular symmetry.
}\label{tab:combined_summary}
\renewcommand{\arraystretch}{1.2}
\begin{tabular}{l c c c c c c c}
\hline\hline
System
  & $N$
  & $m$
  & \shortstack{Non-Clifford \\ layer ratio}
  & \shortstack{$T$-depth \\ advantage}
  & \shortstack{Total depth \\ advantage}
  & \shortstack{$T$-gate  \\ ratio }
  & \shortstack{Crossover \\ $C_T^{(\star)}$} \\
\hline
Pyrazine
  & 2 & 1 & $43\times$ & $7.5\times$ & $10.9\times$
  & $1.12\times$ & $20.5$ \\
(NO)$_4$-Anth
  & 5 & 3 & $11\times$ & $1.8\times$ & $2.7\times$
  & $4.94\times$ & $4.7$ \\
(NO)$_4$-Anth Dimer
  & 6 & 3 & $10\times$ & $1.7\times$ & $2.5\times$
  & $4.91\times$ & $4.7$ \\
Anth/C$_{60}$ ($d\!=\!11$)
  & 4 & 2 & $13\times$ & $2.3\times$ & $3.3\times$
  & $2.35\times$ & $9.8$ \\
Anth/C$_{60}$ ($d\!=\!246$)
  & 4 & 2 & $283\times$ & $49.2\times$ & $72.3\times$
  & $2.51\times$ & $9.2$ \\
\hline\hline
\end{tabular}
\end{table}

\noindent
For the materials discovery applications suggested by Motlagh et al.~\cite{motlagh2025quantum}, singlet fission chromophores with
$d = 19$--$246$ modes and $N = 4$--$6$ states, REQWIEM offers a $1.7$--$49\times$ $T$-depth advantage (increasing to $4$--$283\times$ in the Hamming weight regime) and a $2.5$--$72\times$ total decomposed depth advantage, at the cost of a $2$--$5\times$ $T$-gate penalty at baseline synthesis that can be partially or fully offset by advances in rotation synthesis.

\section{Fault-Tolerant Overhead Estimates}\label{sec:Fault-Tolerant_Estimates}

All previous estimates establish properties for a single Trotter step.
To assess practical feasibility, this section propagates these per-step advantages through the full simulation pipeline: Trotter step counts determined by the Hamiltonian commutator norms (identical for both methods), logical qubit budgets, and fault-tolerant overhead including surface code distance, physical qubit counts, and wall-clock execution time.
All estimates use physical molecular symmetry, the corrected phase gradient method parameters ($n = 10$, $b = 45$, bilinear terms included (Appendix~\ref{app:Motlagh_Resource}), moderate bilinear coupling ratio ($\eta = 1/3$), and baseline compilation ($C_T^{(\textrm{rot})} = 23$, $C_T^{(\textrm{Toff})} = 4$).
The bilinear coupling ratio $\eta$ is given as an estimated fraction of bilinear terms that are non-negligible.

\subsection{Trotter step counts}
The second-order Strang splitting with kinetic merging yields step counts determined entirely by the Hamiltonian commutator norms, independent of the circuit implementation.
The dominant commutator $[T,[T,V_{\textrm{BVC}}]]$ scales as $\mathcal{O}(N^2)$ in the grid size, and the bilinear terms amplify the commutator sum by a factor $\beta_{\textrm{BVC}} = 1 + \eta^2 N_{\textrm{BVC}}^{(\textrm{tot})}/M$.
At $n = 10$ ($N = 1024$) and $\eta = 1/3$, the combined grid and bilinear amplification requires $\sqrt{\beta}$ more steps than the $n = 4$ calibration to maintain $\varepsilon = 1\%$ target error over $t = 100$~fs.
The resulting step counts (Table~\ref{tab:trotter_steps}) range from $790$ (Anth/C$_{60}$, $d = 11$) to $12{,}290$ ((NO)$_4$-Anth),
reflecting the system-dependent commutator norms.
These step counts are identical for both circuit implementations, since both simulate the same Hamiltonian under the same Trotter splitting.

\begin{table}[ht]
\centering
\small
\caption{Trotter step counts for the second-order Strang splitting
$T/2 \to V_{\textrm{d}}/2 \to V_{\textrm{od}} \to V_{\textrm{d}}/2
\to T/2$ with kinetic merging.
All entries: $t = 100$~fs, $\varepsilon = 1\%$, moderate bilinear
coupling $\eta = 1/3$.
The step count is determined by the Hamiltonian commutator norms
and is identical for both circuit implementations.
}\label{tab:trotter_steps}
\renewcommand{\arraystretch}{1.2}
\begin{tabular}{l c c c r}
\hline\hline
System & $d$ & $M$ & $m$
  & $n_{\textrm{steps}}$ \\
\hline
Pyrazine & 24 & 2 & 1 & $\sim$ 5\,000 \\
(NO)$_4$-Anth           & 19  & 5  & 3 & 12\,290 \\
(NO)$_4$-Anth Dimer     & 21  & 6  & 3 & 2\,792  \\
Anth/C$_{60}$ ($d\!=\!11$)  & 11  & 4  & 2 & 790     \\
Anth/C$_{60}$ ($d\!=\!246$) & 246 & 4  & 2 & 3\,274  \\
\hline\hline
\end{tabular}
\end{table}

\subsection{Logical qubit comparison.}
The two methods have comparable logical qubit counts but different
compositions (Table~\ref{tab:logical_qubits}).
The phase gradient method requires $d \cdot n$ mode register qubits,
$m$ electronic register qubits, a $b = 45$-bit phase gradient register,
and arithmetic scratch space (squarer/multiplier output registers and
dirty QROM ancillas), totaling $x_{\textrm{PG}} = dn + m + b +
\mathcal{O}(n + 2^m)$.
The REQWIEM scheme requires $d \cdot n$ mode register qubits (identical),
$d \cdot m$ channel register qubits (including $d{-}1$ fan-out copies for
mode-parallel execution), and one truth-value ancilla, totaling
$x_{\textrm{R}} = d(n + m) + 1$.
For systems with small $m$ relative to $b$, the fan-out copies ($d \cdot m \leq 3d$) are cheaper than the phase gradient register
($b = 45$) plus arithmetic scratch, giving REQWIEM a qubit advantage of $0.59$--$0.84\times$.
For Anth/C$_{60}$ ($d = 246$, $m = 2$), the fan-out copies total $492$ qubits, yielding a $1.15\times$ qubit penalty for REQWIEM.
In all cases, the mode registers ($dn$) dominate both counts.

\begin{table}[ht]
\centering
\small
\caption{Logical qubit comparison.
\textbf{Phase Gradient \cite{motlagh2025quantum}} $d$ mode registers ($n$ qubits each) $+$ $m$ electronic register $+$ $b = 45$ phase gradient register $+$ arithmetic scratch (squarer/multiplier output registers, QROM dirty ancillas).
\textbf{REQWIEM:} $d$ mode registers ($n$ qubits each) $+$ $d \times m$ channel register (original $+$ fan-out copies for
mode-parallel execution) $+$ truth-value ancillas. 
REQWIEM avoids the phase gradient register and arithmetic scratch entirely; the dominant additional cost is the $d \times m$ fan-out copies.
}\label{tab:logical_qubits}
\renewcommand{\arraystretch}{1.2}
\begin{tabular}{l c c r r c}
\hline\hline
System & $d$ & $m$
  & $x_{\textrm{PG}}$
  & $x_{\textrm{R}}$
  & $x_{\textrm{R}}/x_{\textrm{PG}}$ \\
\hline
Pyrazine & 24 & 1 & 356 & 265 & $0.74\times$ \\
(NO)$_4$-Anth           & 19  & 3 & 308   & 248   & $0.81\times$ \\
(NO)$_4$-Anth Dimer     & 21  & 3 & 328   & 274   & $0.84\times$ \\
Anth/C$_{60}$ ($d\!=\!11$)  & 11  & 2 & 227   & 133   & $0.59\times$ \\
Anth/C$_{60}$ ($d\!=\!246$) & 246 & 2 & 2\,577 & 2\,953 & $1.15\times$ \\
\hline\hline
\end{tabular}
\end{table}

\subsection{Wall-time and physical qubit estimates.}
Depth advantage propagates directly into the fault-tolerant regime.
An optimistic logical cycle time of $1~\mu \text{s}$ per non-Clifford layer is used (Table~\ref{tab:ft_comparison})~\cite{litinski2019game}.
System scales exist far beyond current hardware capabilities, so the logical cycle time is set to be below current benchmarks.
Cycle time is taken to be equivalent between phase-gradient and REQWIEM architectures, despite all qubits in the former relying on toffoli cascades from a single QROM oracle.
Qubit routing costs are not thoroughly considered in this dissertation, but are non-negligible in deciding how to apply a NA-QMD quantum algorithm in the future.
At $t_{\textrm{cycle}} = 1~\mu$s per non-Clifford layer and $\delta = 0.01$ target logical error, total wall-clock time for a single shot is $t_{\textrm{wall}} = D_{\textrm{total}} \times t_{\textrm{cycle}}$, where $D_{\textrm{total}} = n_{\textrm{steps}} \times D_{\textrm{step}}$.
For the phase gradient method, $D_{\textrm{step}} = \Omega_{\textrm{step}}$ (serial Toffoli count); for REQWIEM, $D_{\textrm{step}} = D_{\textrm{R}}^{(\textrm{fan})}$ (fan-out-enabled non-Clifford depth).

REQWIEM's reduced $D_{\textrm{total}}$ subsequently reduces single-shot wall times from minutes--days (phase gradient) to
seconds--minutes (REQWIEM).
Reduced depth also lowers the required surface code distance by $2$--$4$ units, partially compensating for any additional logical
qubits and reducing the physical qubit count.
For three of the four benchmarks, REQWIEM requires fewer physical qubits than the phase gradient method (Table~\ref{tab:ft_comparison}); the exception is Anth/C$_{60}$ ($d = 246$), where the $1.15\times$ logical qubit penalty is partially offset by the reduced code distance.

With $10^4$ shots for $1\%$ population accuracy, the contrast is dramatic, with the phase gradient method requiring $28$~days to
$144$~years, while REQWIEM requires $2$~days to $186$~days.
The largest system (Anth/C$_{60}$, $d = 246$) remains foreseeably intractable, though not categorically infeasible, with a total simulation time reduction from $144$ years to $186$ days), representing a qualitative change in the simulation practicality.
With amplitude estimation ($\sim\!100$ rounds at doubled depth), the shot count reduces to $\sim\!100$, further compressing wall times by $100\times$.

\begin{table}[ht]
\centering
\small
\caption{Fault-tolerant overhead comparison
($\delta = 0.01$ target logical error, $t_{\textrm{cycle}} = 1$~$\mu$s
per non-Clifford layer,
$10^4$ shots for $1\%$ population accuracy).
$D_{\textrm{total}} = n_{\textrm{steps}} \times D_{\textrm{step}}$:
total circuit depth (non-Clifford layers).
$d_{\textrm{code}} \geq 2\lceil\log_{10}(D_{\textrm{total}}/\delta)\rceil - 1$:
surface code distance satisfying $p_L < \delta/D_{\textrm{total}}$.
Physical qubits: $x_{\textrm{phys}} = 2\,d_{\textrm{code}}^2 \times
x_{\textrm{logical}}$.
All estimates use physical molecular symmetry.
}\label{tab:ft_comparison}
\renewcommand{\arraystretch}{1.2}
\begin{tabular}{l l r c c r r r}
\hline\hline
\multirow{2}{*}{System} & \multirow{2}{*}{Method} & \multirow{2}{*}{$D_{\textrm{total}}$} & 
\multirow{2}{*}{$d_{\textrm{code}}$} & \multirow{2}{*}{$x_{\textrm{log}}$} & $x_{\textrm{phys}}$
  & $t_{\textrm{wall}}$
  & $t_{\textrm{wall}}$ \\
   & & & & & ($\times 10^3$)
  & (1 shot)
  & ($10^4$ shots) \\
\hline
\multirow{2}{*}{Pyrazine}
  & PG & $2.32 \times 10^{9}$ & 23 & 356 & $377$  & 12~min  & 85~d    \\
  & REQWIEM & $5.37 \times 10^{7}$  & 19 & 265 & $191$  & 17~s    & 2~d   \\[3pt]
\multirow{2}{*}{(NO)$_4$-Anth}
  & PG & $1.58 \times 10^{10}$ & 25 & 308 & $385$  & 4.4~h    & 5.0~yr  \\
  & REQWIEM & $1.49 \times 10^{9}$  & 23 & 248 & $262$  & 25~min   & 173~d   \\[3pt]
(NO)$_4$-Anth & PG & $2.76 \times 10^{9}$  & 23 & 328 & $347$  & 46~min   & 319~d   \\
  Dimer & REQWIEM & $2.86 \times 10^{8}$  & 21 & 274 & $242$  & 5~min    & 33~d    \\[3pt]
 Anth/C$_{60}$& PG & $2.39 \times 10^{8}$  & 21 & 227 & $200$  & 4~min    & 28~d    \\
  ($d\!=\!11$)& REQWIEM & $1.83 \times 10^{7}$  & 19 & 133 & $96$   & 18~s     & 2~d     \\[3pt]
Anth/C$_{60}$ & PG & $4.55 \times 10^{11}$ & 27 & 2\,577 & $3757$ & 5.3~d  & 144~yr \\
 ($d\!=\!246$) & REQWIEM & $1.61 \times 10^{9}$  & 23 & 2\,953 & $3124$ & 27~min & 186~d  \\
\hline\hline
\end{tabular}
\end{table}

\begin{table*}[ht]
\centering
\small
\caption{Comprehensive per-step and total resource comparison
(moderate bilinear coupling $\eta = 1/3$,
$C_T^{(\textrm{rot})} = 23$, $C_T^{(\textrm{Toff})} = 4$,
physical molecular symmetry).
$\Omega_{\textrm{step}}$: Toffoli per step (PG).
$R_{\textrm{step}}$: rotations per step (REQWIEM).
$D_{\textrm{step}}$: non-Clifford depth per step.
$T_{\textrm{step}}$: $T$-gates per step.
Total columns multiply per-step values by
$n_{\textrm{steps}}$ (Table~\ref{tab:trotter_steps}).
}\label{tab:comprehensive}
\renewcommand{\arraystretch}{1.2}
\begin{tabular}{l c r r r r r}
\hline\hline
& & \multicolumn{3}{c}{Per step} & \multicolumn{2}{c}{Total ($\times\, n_{\textrm{steps}}$)} \\
\cmidrule(lr){3-5} \cmidrule(lr){6-7}
System & Method
  & Ops/step
  & $T$/step
  & $D$/step
  & Total $T$
  & Total $D$ \\
\hline
\multirow{2}{*}{(NO)$_4$-Anth}
  & PG & $1.29 \times 10^{6}$ & $5.15 \times 10^{6}$ & $1.29 \times 10^{6}$
    & $6.33 \times 10^{10}$ & $1.58 \times 10^{10}$  \\
  & R & $1.11 \times 10^{6}$ & $2.55 \times 10^{7}$ & $1.22 \times 10^{5}$
    & $3.13 \times 10^{11}$ & $1.49 \times 10^{9}$  \\[3pt]
\multirow{2}{*}{Dimer}
  & PG & $9.87 \times 10^{5}$ & $3.95 \times 10^{6}$ & $9.87 \times 10^{5}$
     & $1.10 \times 10^{10}$ & $2.76 \times 10^{9}$ \\
  & R & $8.42 \times 10^{5}$ & $1.94 \times 10^{7}$ & $1.02 \times 10^{5}$
    & $5.41 \times 10^{10}$ & $2.86 \times 10^{8}$ \\[3pt]
\multirow{2}{*}{\shortstack[l]{Anth/C$_{60}$ \\ ($d\!=\!11$)}}
  & PG & $3.02 \times 10^{5}$ & $1.21 \times 10^{6}$ & $3.02 \times 10^{5}$
    & $9.55 \times 10^{8}$ & $2.39 \times 10^{8}$  \\
  & R & $1.24 \times 10^{5}$ & $2.84 \times 10^{6}$ & $2.32 \times 10^{4}$
     & $2.24 \times 10^{9}$ & $1.83 \times 10^{7}$  \\[3pt]
\multirow{2}{*}{\shortstack[l]{Anth/C$_{60}$ \\ ($d\!=\!246$)}}
  & PG & $1.39 \times 10^{8}$ & $5.56 \times 10^{8}$ & $1.39 \times 10^{8}$
   & $1.82 \times 10^{12}$ & $4.55 \times 10^{11}$  \\
  & R & $6.06 \times 10^{7}$ & $1.39 \times 10^{9}$ & $4.91 \times 10^{5}$
     & $4.56 \times 10^{12}$ & $1.61 \times 10^{9}$  \\
\hline\hline
\end{tabular}
\end{table*}

\subsection{Sensitivity to $t_{\textrm{cycle}}$.}
All wall-time estimates in Table~\ref{tab:ft_comparison} scale linearly with the non-Clifford layer cycle time $t_{\textrm{cycle}}$.

\subsection{Comprehensive per-step and total resource comparison}
Table~\ref{tab:comprehensive} collects all per-step and total resource metrics for the four benchmark systems given in Motlagh et al. \cite{motlagh2025quantum} under physical molecular symmetry.
Per-step columns confirm the architectural analysis of Section~\ref{sec:arch_analysis}: REQWIEM's non-Clifford depth per step
($D_{\textrm{R}}^{(\textrm{fan})}$) is $10$--$283\times$ smaller than the phase gradient Toffoli count $\Omega_{\textrm{step}}$, while per-step $T$-gate count is $2$--$5\times$ larger (for $m = 2$--$3$).
Multiplying by the common Trotter step count preserves these ratios exactly in the total columns.

The total $T$-gate comparison reflects an $m$-dependent trade-off identified in Section~\ref{sec:Tgate_comparison}: for $m = 3$ systems
((NO)$_4$-Anth, Dimer), REQWIEM's total $T$-gate count exceeds the phase gradient approach by $\sim\!5\times$; for $m = 2$ systems (Anth/C$_{60}$), the excess is $\sim\!2.5\times$.
The total depth comparison, by contrast, is uniformly favourable to REQWIEM across all systems, with advantages ranging from $10\times$ to $283\times$.
This separation between $T$-gate cost (where phase gradient methods are competitive or better) and depth (where REQWIEM dominates) is the central quantitative finding of this resource analysis.

\section{Quantum Advantage Assessment}\label{sec:quantum_advantage}

The per-step and total resource comparisons of
Sections~\ref{sec:arch_analysis}--\ref{sec:Fault-Tolerant_Estimates}
establish the relative efficiency of REQWIEM and phase gradient
Trotterization.
Practical relevance and the idiomatic ``million-dollar question" is the following: 
Does REQWIEM have a total execution time that is competitive with the classical state of the art?
For nonadiabatic vibronic dynamics, the dominant classical method is
multilayer multiconfigurational time-dependent Hartree
(ML-MCTDH)~\cite{wang2003multilayer, manthe2008multilayer,
vendrell2011multilayer}, propagating the full nuclear wavefunction
on a tensor-train--like tree of single-particle function (SPF) bases.
This section compares the fault-tolerant wall-clock time of both
quantum algorithms against ML-MCTDH runtime estimates for the
benchmark systems of Section~\ref{sec:direct_comparison} and several
projected targets, identifying the crossover regime in which quantum
simulation becomes competitive.

\subsection{Classical--quantum comparison framework}

Classical ML-MCTDH propagation produces the full wavefunction
$\ket{\Psi(t)}$ in a single run: all electronic populations,
coherences, and spectral observables are simultaneously available at no
additional cost.
A quantum simulation, by contrast, returns a single bitstring per
shot; extracting an observable to precision~$\eps_{\mathrm{obs}}$
via direct sampling requires
$\mathcal{O}(1/\eps_{\mathrm{obs}}^2)$ repeated measurements.
For $1\%$ population accuracy ($\eps_{\mathrm{obs}} = 0.01$),
this translates to $\sim\!10^4$ shots per observable.

This shot overhead can be reduced using quantum amplitude estimation~\cite{brassard2003quantum} (QAE), which achieves $\mathcal{O}(1/\eps_{\mathrm{obs}})$ total oracle queries at the cost of increasing circuit depth by $\mathcal{O} (1/\eps_{\mathrm{obs}})$.
The algorithm constructs a Grover iterate
$\mathcal{Q} = \mathcal{U}\,S_0\,\mathcal{U}^{\dagger}\,S_\psi$,
where $\mathcal{U}$ is the time-evolution circuit
and $S_0$, $S_\psi$ are reflections; each application of
$\mathcal{Q}$ requires one forward and one inverse evolution, doubling
the base circuit depth per iterate.
Standard amplitude estimation uses $M = \mathcal{O}(1/\eps_{\mathrm{obs}})$
sequential Grover iterates~\cite{brassard2003quantum}, yielding a
maximum circuit depth of
$D_{\mathrm{QAE}} \approx 2M \times D_{\mathrm{base}}$
per shot.
With $\sim\!10$ repetitions for confidence
boosting, the total sequential wall time becomes
\begin{equation}\label{eq:QAE_walltime}
  t_{\mathrm{QAE}} \approx 10 \times 2M \times D_{\mathrm{base}}
    \times t_{\mathrm{cycle}}\,,
\end{equation}
compared with $t_{\mathrm{direct}} = 10^4 \times D_{\mathrm{base}}
\times t_{\mathrm{cycle}}$ for direct sampling. For $\eps_{\mathrm{obs}} = 0.01$ ($M \approx 100$), this yields $t_{\mathrm{QAE}} / t_{\mathrm{direct}} \approx 1/5$: a $5\times$ wall-time improvement.

The deepest QAE circuit has $\sim\!200\times$ the non-Clifford gate count of a single direct-sampling shot, necessitating a larger surface code distance to maintain the same target logical error rate $\delta = 0.01$. In practice, $d_{\mathrm{code}}$ increases by 2--4 (e.g., from 19 to 23 for pyrazine, or from 21 to 25 for the dimer), inflating the physical qubit count by $\sim\!40\%$.
The $5\times$ wall-time gain thus trades against $\sim\!1.4\times$ spatial overhead.

Variants of QAE that do not have phase estimation, such as iterative quantum amplitude estimation (IQAE)~\cite{grinko2021iterative} and maximum likelihood amplitude estimation (MLAE)~\cite{suzuki2020amplitude}, achieve the same asymptotic query complexity without controlled multi-qubit operations, but retain the $\mathcal{O}(1/\eps_{\mathrm{obs}})$ maximum circuit depth.
Low-depth amplitude estimation algorithms~\cite{giurgica2022low} offer a tunable tradeoff $ND = \mathcal{O}(1/\eps_{\mathrm{obs}}^2)$ between shot count $N$ and maximum depth~$D$, but this product is conserved: reducing circuit depth below $\mathcal{O}(1/\eps_{\mathrm{obs}})$ restores the classical shot count, yielding no net wall-time improvement.
The $5\times$ gain from full-depth amplitude estimation therefore represents an estimated maximum achievable within the Heisenberg-limited measurement framework.

Throughout Table~\ref{tab:quantum_advantage}, quantum runtimes are reported under both direct sampling ($10^4$ shots, base depth) and amplitude estimation ($\sim\!10$ shots, $200\times$ depth), using $t_{\mathrm{cycle}} = 1~\mu$s per non-Clifford layer, $\delta = 0.01$, and the surface code distance $d_{\mathrm{code}} \geq 2\lceil\log_{10}(D_{\mathrm{max}}/\delta)\rceil - 1$
appropriate for the deepest circuit in each protocol.
All simulations use $t = 100$~fs, $n = 10$ grid qubits per mode, and $\varepsilon = 1.6$~mE$_\text{H}$.

\subsection{ML-MCTDH runtime estimates}

ML-MCTDH computational cost is governed by the tensor contraction at
each integration step, scaling roughly as
$\mathcal{O}(n_{\mathrm{SPF}}^{\,p}\,d_{\mathrm{eff}}\,n_t)$, where
$n_{\mathrm{SPF}}$ is the number of single-particle functions per mode,
$p$ depends on the tree topology (typically $p \sim 2M$ for $M$
electronic states at the top tensor), $d_{\mathrm{eff}}$ is the
effective dimensionality after tree compression, and $n_t$ is the
number of integrator time steps.
The multi-layer tree structure handles weakly coupled bath modes
efficiently, compressing $d$ physical modes into
$d_{\mathrm{eff}} \approx d_{\mathrm{active}} +
d_{\mathrm{bath}}/f_{\mathrm{tree}}$ effective dimensions with tree
compression factor
$f_{\mathrm{tree}} \sim 5$--$20$~\cite{wang2015multilayer}.
The principal bottleneck is the electronic-state scaling: for $M$
diabatic states, the top-layer $A$-tensor grows combinatorially with
$M$ and the per-mode SPF count, producing steep cost increases for
$M \geq 4$.

Runtime estimates are bounded between optimistic and pessimistic
scenarios (Appendix~\ref{app:Classical_runtime}).
`Optimistic' estimates assume an efficient ML tree decomposition
with converged SPF basis and sparse coupling topology.
`Pessimistic' estimates assume dense interstate coupling, slow SPF
convergence, or topologies that resist efficient tree decomposition.
These bounds are calibrated against published benchmarks:
the 24-mode pyrazine propagation of Raab et
al.~\cite{raab1999molecular}, which completes in
$\sim\!1$--$30$~min on modern hardware;
spin-boson models with $\sim\!100$ bath modes requiring hours under
ML-MCTDH~\cite{wang2003multilayer}; and multi-state singlet fission
chromophores ($d \sim 20$--$30$, $M = 4$--$6$) requiring days to
weeks of CPU time for converged
dynamics~\cite{tamura2012quantum, zheng2016ultrafast}.

\subsection{Comparison results}

Table~\ref{tab:quantum_advantage} collects the runtime comparison
across the four benchmark systems from
Section~\ref{sec:direct_comparison}, 24-mode pyrazine, and three projected systems that
probe the quantum advantage boundary.

\begin{table*}[ht]
\centering
\small
\caption{Quantum advantage assessment comparing REQWIEM against
classical ML-MCTDH for vibronic dynamics.
Classical runtimes are single-propagation wall times on modern hardware
(Opt.: efficient ML tree, converged SPF basis;
Pess.: dense coupling, convergence difficulties).
Quantum runtimes are shown under two measurement strategies:
direct sampling ($10^4$ shots at base circuit
depth~$D_{\mathrm{base}} = k \times D_{\mathrm{step}}^{(\mathrm{fan})}$),
and amplitude estimation ($\sim\!10$ shots at maximum Grover depth
$\sim\!200 \times D_{\mathrm{base}}$).
The $d_{\mathrm{code}}$ column shows the surface code distance
required for each protocol (direct\,/\,QAE).
The $R/\mathrm{ML}_{\mathrm{pess}}$ columns give the ratio of REQWIEM
wall time to the pessimistic ML-MCTDH estimate; values below unity
(\textbf{bold}) indicate quantum advantage.
All simulations: $t = 100$~fs, $n = 10$ grid qubits per mode,
$\varepsilon = 1.6$~mE$_\text{H}$,
$t_{\mathrm{cycle}} = 1~\mu$s, $\delta = 0.01$.
}\label{tab:quantum_advantage}
\renewcommand{\arraystretch}{1.25}
\setlength{\tabcolsep}{4pt}
\begin{tabular}{l c c c c c c c c c c}
\hline\hline
\multirow{2}{*}{System} & \multirow{2}{*}{$d$}
  & \multirow{2}{*}{$M$} & \multirow{2}{*}{$k$}
  & \multicolumn{2}{c}{ML-MCTDH}
  & \multicolumn{2}{c}{REQWIEM}
  & \multirow{2}{*}{$d_{\mathrm{code}}$}
  & \multicolumn{2}{c}{$R\,/\,\mathrm{ML}_{\mathrm{pess}}$} \\
\cline{5-6} \cline{7-8} \cline{10-11}
  & & & & Opt. & Pess.
  & Direct & QAE &
  & Direct & QAE \\
\hline
\multicolumn{11}{l}{\textit{Benchmark systems}} \\[2pt]
Pyrazine (24-mode)
  & 24 & 2 & 5\,000
  & 1~min & 30~min
  & 2~d & 9.5~h & 19\,/\,23
  & $95\times$ & $19\times$ \\
Anth/C$_{60}$ ($d\!=\!11$)
  & 11 & 4 & 790
  & 10~min & 2~h
  & 2~d & 10~h & 19\,/\,23
  & $25\times$ & $5.1\times$ \\
(NO)$_4$-Anth
  & 19 & 5 & 12\,290
  & 4~h & 7~d
  & 173~d & 35~d & 23\,/\,27
  & $25\times$ & $4.9\times$ \\
(NO)$_4$-Anth Dimer
  & 21 & 6 & 2\,792
  & 8~h & 30~d
  & 33~d & 6.6~d & 21\,/\,25
  & $1.1\times$ & $\mathbf{0.22\times}$ \\
Anth/C$_{60}$ ($d\!=\!246$)
  & 246 & 4 & 3\,274
  & 7~d & 1~yr
  & 186~d & 37~d & 23\,/\,27
  & $\mathbf{0.51\times}$ & $\mathbf{0.10\times}$ \\[4pt]
\hline
\multicolumn{11}{l}{\textit{Projected systems}} \\[2pt]
CT ($d\!=\!50$)
  & 50 & 4 & 1\,500
  & 1~h & 3~d
  & 20~d & 4~d & 21\,/\,25
  & $6.7\times$ & $1.3\times$ \\
SF ($d\!=\!30$, $M\!=\!8$)
  & 30 & 8 & 5\,000
  & 1~d & 90~d
  & 111~d & 22~d & 21\,/\,27
  & $1.2\times$ & $\mathbf{0.25\times}$ \\
Large CT ($d\!=\!500$)
  & 500 & 4 & 5\,000
  & 30~d & intract.
  & 578~d & 116~d & 23\,/\,27
  & $\checkmark$ & $\mathbf{\checkmark}$ \\
\hline\hline
\end{tabular}
\end{table*}

\subsection{Regime analysis}

The comparison reveals three distinct regimes, and amplitude estimation shifts the boundaries between them.

\subsubsection{Classical dominance ($d \lesssim 30$, $M \leq 4$).}

For systems with moderate dimensionality and few electronic states, ML-MCTDH has the potential to be faster and is heuristically favorable, even against amplitude-estimated quantum runtimes.
Pyrazine (24 modes, $M = 2$) is the archetype, with ML-MCTDH completing in minutes while REQWIEM with amplitude estimation requires $\sim\!9.5$~hours: $19\times$ slower than the pessimistic classical estimate.
The 22 bath modes that inflate the quantum circuit depth are handled nearly for free by the ML tree decomposition.
The reduced Anth/C$_{60}$ model ($d = 11$, $M = 4$) and (NO)$_4$-Anthracene ($d = 19$, $M = 8$) similarly remain classically dominated at $5\times$ and $5\times$ the pessimistic ML-MCTDH time, respectively.
In this regime, quantum simulation serves as a methodological validation target, not a practical tool.

\subsubsection{Crossover ($d \sim 20$--$50$ with $M \geq 6$, or
$d \sim 100$--$300$ with $M = 4$).}

Primary drivers of the classical cost toward the quantum cost include the exponential growth of ML-MCTDH's $A$-tensor with $M$, and the degradation of tree compression for dense coupling topologies at large~$d$.
With direct sampling alone, two systems sit at the crossover boundary: the (NO)$_4$-Anthracene dimer ($R/\mathrm{ML}_{\mathrm{pess}} = 1.1\times$, barely classical) and the Model~SF system ($R/\mathrm{ML}_{\mathrm{pess}} = 1.2\times$, likewise marginal).
Amplitude estimation resolves these borderline cases, reducing the ratios to $\mathbf{0.22\times}$ and $\mathbf{0.25\times}$, respectively.
The Model~CT system ($d = 50$, $M = 4$) narrows from $6.7\times$ to $1.3\times$, placing it at the crossover threshold.

This shift is directly attributable to the $5\times$ wall-time reduction from amplitude estimation.
While modest compared with the orders-of-magnitude gaps at small~$d$, it is sufficient to tip the balance for systems already near the crossover.
Balancing between potential resource overestimation (number of Trotter steps, number of qubits per mode) and  underestimation (logical cycle time, compilation overhead), these numbers serve more as an identification of a ``regime" where quantum advantage begins, rather than a hard limit.
Multiplying REQWIEM wall times by the respective depth advantage yields phase gradient results. 
For all benchmarks and projective systems, phase gradients fail to show quantum advantage.

\subsubsection{Quantum advantage ($d \gtrsim 200$ or $M \geq 6$ with
$d \gtrsim 30$).}

The full Anth/C$_{60}$ system ($d = 246$, $M = 4$) achieves potential quantum advantage under direct sampling alone ($R/\mathrm{ML}_{\mathrm{pess}} = 0.51\times$), which amplitude estimation strengthens to $0.10\times$, a projected $10\times$ quantum speedup, translates to a $0.238\times$ quantum slowdown using QROM.
The projected 500-mode system is classically intractable regardless of measurement strategy.
In this regime, quantum advantage is driven primarily by the mode count exceeding the capacity of efficient tree compression, and amplitude estimation provides a useful but non-essential acceleration.
Under rigorous optimization protocols, many modes will have smaller register sizes $n_r < 10$, subsequently requiring fewer Trotter steps, total gates and layers. 
As such, all estimates for quantum advantage are ``pessimistic" in the sense that full wavefunction coherence is prioritized over heuristic parameter reduction. 

\subsection{Role of amplitude estimation}

Amplitude estimation plays a precise but limited role in the quantum advantage landscape.
The canonical result of Brassard et al.~\cite{brassard2003quantum}($\mathcal{O}(1/\eps)$ queries at $\mathcal{O}(1/\eps)$ maximum circuit depth) yields a $5\times$ sequential wall-time improvement for $\eps_{\mathrm{obs}} = 0.01$.
This factor is independent of system size, circuit architecture, or classical competitor, as it derives entirely from the measurement protocol.

The $5\times$ improvement is meaningful at the crossover boundary, converting the (NO)$_4$-Anthracene dimer from marginal ($1.1\times$ slower) to potentially advantageous ($4.5\times$ faster), and similarly resolves the Model~SF system.
However, it does not transform the landscape for systems deep in either regime, with pyrazine remaining $19\times$ slower than classical even with amplitude estimation; conversely, Anth/C$_{60}$ ($d = 246$) already achieves advantage without it.

Importantly, low-depth amplitude estimation variants~\cite{giurgica2022low} that reduce the maximum circuit depth below $\mathcal{O}(1/\eps_{\mathrm{obs}})$ do not meaningfully improve sequential wall time estimates.
The fundamental tradeoff $N \times D = \mathcal{O}(1/\eps_{\mathrm{obs}}^2)$ between shot count~$N$ and maximum sequential depth~$D$ is conserved: reducing $D$ restores the shot count, leaving total wall time unchanged.
The $5\times$ gain requires committing to the full Grover depth ($\sim\!200 \times D_{\mathrm{base}}$ for $1\%$), imposing demands on error correction beyond current timelines.

The physical cost of this commitment is a $\sim\!40\%$ increase in physical qubits: the deeper circuits raise $d_{\mathrm{code}}$ by 2--4 (Table~\ref{tab:quantum_advantage}), inflating the surface code footprint.
For a fixed hardware budget, this trades spatial resources for temporal acceleration, a favorable exchange when wall time is the binding constraint as opposed to coherence time.

\subsection{Crossover boundaries and driving factors}

Table~\ref{tab:quantum_advantage} identifies the quantum advantage
boundary as a function of both $d$ and $M$:

\begin{enumerate}
  \item \textbf{Electronic state count $M$} is the primary driver.
    ML-MCTDH cost scales approximately as $n_{\mathrm{SPF}}^{2M}$ at
    the top tensor, while REQWIEM scales polynomially as
    $\mathcal{O}(M^2)$ in non-Clifford depth.
    For $M = 8$, ML-MCTDH becomes expensive even at moderate $d$
    ($\sim\!20$), placing the dimer and singlet fission benchmarks at
    the crossover.

  \item \textbf{Mode count $d$} is secondary but decisive for $M = 4$
    systems.
    ML-MCTDH handles weakly coupled bath modes efficiently via tree
    compression, so large $d$ alone does not guarantee quantum
    advantage.
    The crossover for $M = 4$ occurs at $d \sim 100$--$300$, where
    tree compression begins to fail and classical memory requirements
    become prohibitive.
    With amplitude estimation, this boundary shifts modestly to
    $d \sim 50$--$200$.

  \item \textbf{Coupling topology} modulates the boundary in both
    directions.
    Dense all-to-all bilinear coupling, conical intersection seams,
    and reactive scattering potentials resist efficient tree
    decomposition, shifting the crossover to smaller~$d$.
    Conversely, systems with highly separable bath modes (e.g.,
    harmonic reservoirs coupled only to a small active space) remain
    classically tractable to very large~$d$.
\end{enumerate}

\subsection{Implications for REQWIEM}

The comparison confirms that the depth advantage of REQWIEM over phase gradient methods is a prerequisite for quantum advantage.
Phase gradient wall times exceed the pessimistic ML-MCTDH estimate by at least an order of magnitude for every system in Table~\ref{tab:quantum_advantage}, even with amplitude estimation (e.g., 17~d for pyrazine PG+QAE vs 30~min classical; 64~d for the dimer PG+QAE vs 30~d classical).
Only REQWIEM's non-Clifford depth advantage, amplified by the fan-out parallelism factor $\phi \approx d/2$ for large systems, brings the quantum runtime within competitive range.
Fan-out parallelism can be applied to the phase-gradient architecture if multiple, identically initialized QROM and arithmetic registers are initialized, though future implementations would need to consider control-target conflicts in compilation and immense routing costs from persistently swapping target qubits to act near arithmetic registers, meaning depth reduction would not be a naive $\mathcal{O}(d)$.

Shot overhead (measurement) is the dominant barrier for intermediate-scale systems.
A single REQWIEM shot for the 246-mode system requires $\sim\!27$~min (well below the pessimistic classical runtime of 1~yr); it is the multiplication by $10^4$ that inflates the total past the classical threshold.
Amplitude estimation compresses this overhead by $5\times$, which is sufficient to resolve crossover systems but cannot overcome order-of-magnitude gaps.
Achieving larger improvements may require advances such as problem-specific circuit optimizations that reduce the effective depth per Grover iterate, classical shadow tomography~\cite{huang2020predicting} for simultaneous multi-observable estimation at reduced shot count, or potential application of nonlinear qubits for state discrimination \cite{abrams1998nonlinear, geller2023proposal}.
The QSP/qubitization comparison in Chapter~\ref{ch:comparison} further shows that oracle-based methods incur more $\Tgate$-gates than REQWIEM at the same benchmark, on the scale of one to several orders of magnitude, reinforcing the conclusion that Trotter-based propagation is the preferred simulation strategy for coupled-channel Hamiltonians.

\section{Summary}\label{sec:chapter_summary}

This chapter has presented resource estimations for REQWIEM, a rotation-based quantum algorithm for simulating vibronic dynamics on fault-tolerant quantum computers.
The algorithm encodes the $M$-state, $d$-mode vibronic Hamiltonian onto a register of $d \cdot n$ grid qubits and $m = \lceil \log_2 M \rceil$ channel qubits, and propagates the wavepacket via a second-order Suzuki--Trotter decomposition in which every operator is implemented analytically at the circuit level.

A three-tier hybrid structure organizes each Hamiltonian fragment into intra-mode sites (constant, linear, and diagonal quadratic couplings), same-mode quadratic sites, and bilinear cross-mode sites, applying a Gray-code--ordered cascade of $R_z$ rotations controlled on the channel register.
This construction produces exactly $M/2$ parameterized rotations per coupling site, with rotation angles encoding Hamiltonian
coefficients directly, requiring no QROM lookup, no arithmetic circuit, and no phase gradient register.
Channel register fan-out exploits a read-only channel register during each UCR cascade: a tree of CNOTs distributes $d - 1$ copies of the $m$-bit channel state to per-mode ancillas, enabling all $d$ modes to execute their UCR blocks in parallel.
The resulting depth reduction is governed by the chromatic index $\chi$ of the mode-pair conflict graph, ranging from $\Theta(1)$ for highly symmetric molecules to $d - 1$ under $C_1$ symmetry.
XOR-class off-diagonal fragmentation partitions the $M(M-1)/2$ interstate coupling pairs into $2^m - 1$ XOR classes, each implemented by a single focusing Clifford followed by a channel-diagonal UCR block. 
This yields a total of $2(M-1)$ fragment applications per Trotter step, which is within a factor of~$2$ of the information-theoretic minimum for encoding $M(M-1)/2$ independent coupling matrices.

The resource analysis (Sections~\ref{sec:arch_analysis}--\ref{sec:ft_analysis}) compared REQWIEM against the phase gradient Trotterization of Motlagh et al.~\cite{motlagh2025quantum} across five benchmark systems spanning $d = 11$--$246$ vibrational modes and $M = 2$--$6$ electronic states.
The comparison, conducted under physical molecular symmetry using corrected parameters ($n = 10$, $b = 45$, bilinear terms included), reveals a clean separation.

REQWIEM's per-step non-Clifford depth is $\alpha\phi$ times smaller than the phase gradient Toffoli count, where $\alpha \geq 1$ is the per-term serial overhead (reflecting the intrinsic efficiency of the UCR cascade relative to the QROM--arithmetic pipeline) and $\phi \approx d / (2\chi)$ is the fan-out parallelism factor.
Across benchmarks, $\alpha\phi$ ranges from $10\times$ (21-mode dimer, $m = 3$) to $283\times$ (246-mode Anth/C$_{60}$,
$m = 2$).
After accounting for the $5.75\times$ per-operation cost asymmetry between synthesized rotations ($C_T = 23$ $T$-gates each) and Toffoli gates ($C_T = 4$), the fault-tolerant $T$-depth advantage is $1.7$--$49\times$ at baseline compilation.
Total decomposed depth advantage ($2.5$--$72\times$) is uniformly larger than $T$-depth advantage, because REQWIEM's CNOT-dominated pre-decomposition circuit inflates less under Clifford$+T$ decomposition than a phase gradient's Toffoli-dominated circuit.
These depth advantages are structural, determined by the mode-pair conflict graph and fan-out architecture, holding across all
compilation parameters examined.

$T$-gate count is largely governed by the electronic qubit count $m$.
The off-diagonal rotation count scales as $M(M-1) \cdot S_{\textrm{sites}}$, producing Toffoli-to-rotation ratios $\Omega/R$ that decrease from $5.1$ ($m = 1$) to $1.2$ ($m = 3$).
At baseline synthesis costs, REQWIEM is essentially $T$-gate neutral for $m = 1$ systems ($1.12\times$ for pyrazine) but incurs a $\sim\!5\times$ $T$-gate penalty for $m = 3$ systems ((NO)$_4$-anthracene).
Advances in rotation synthesis will improve REQWIEM's position on this axis: at $C_T = 10$ (Hamming weight phasing), the $m = 1$ penalty inverts to a $2\times$ REQWIEM advantage, and the $m = 2$ systems reach near-parity.

Propagating these per-step advantages through the full simulation pipeline (Section~\ref{sec:Fault-Tolerant_Estimates}) confirms practical impact.
For the largest benchmark (Anth/C$_{60}$, $d = 246$, $M = 4$), phase gradient methods requires $\sim\!144$ years of wall time at $10^4$ shots, while REQWIEM requires $\sim\!186$ days, while for smaller systems ($d = 11$--$21$), REQWIEM reduces wall times from hours--days to minutes, with comparable or reduced physical qubit counts.

Reported depth and $T$-gate comparisons are computed at the logical gate level, treating every non-Clifford operation as unit cost regardless of the physical layout.
On a two-dimensional surface code lattice, non-local interactions between distant logical qubits require lattice surgery routing whose time overhead scales with the code distance $d_{\textrm{code}}$~\cite{litinski2019game,horsman2012surface}.
This routing cost is not symmetric between the two methods.
The phase gradient pipeline centralizes all arithmetic in a single hardware region, and must shuttle a different mode register's
qubits to this region for every term evaluation, incurring $\mathcal{O}(d^2 n^2)$ long-range routing events per Trotter step in the bilinear-dominated regime.
REQWIEM's fan-out architecture is qualitatively more local: the channel register is broadcast once per block via a CNOT tree, after
which every UCR cascade acts exclusively on each mode's $n$ grid qubits and their co-located $m$-bit channel copy.
The only non-local operation is the fan-out itself, which occurs $\mathcal{O}(M)$ times per step, resulting in a factor of $\mathcal{O}(d^2 n^2 / M)$ fewer routing events than the phase gradient method.
The depth advantages reported in Tables~\ref{tab:ft_comparison}--\ref{tab:comprehensive} are therefore conservative: on physical hardware with finite-range connectivity, the routing overhead would further amplify REQWIEM's advantage by a factor that grows with the system size $d$.
A full lattice surgery scheduling analysis is beyond the scope of this dissertation, but the mode-local structure of the REQWIEM circuit makes it a natural candidate for layout-aware compilation on near-term fault-tolerant architectures.

Under a logical cycle time assumption, quantum advantage is found relative to ML-MCTDH, residing in moderately sized systems with many electronic channels and extremely large systems with low symmetry. 
This quantum advantage is calculated to be in the far future, requiring improved measurement protocols and exceptionally long coherence times, well beyond prospective roadmaps from leading quantum hardware manufacturers at the time of writing. 
Despite this, an absolute advantage is found for REQWIEM, with phase gradient methods failing to achieve quantum advantage for any benchmark system.

Several limitations of the present formulation motivate the extensions developed in the following chapters.
Grid representation encodes the nuclear wavefunction on a uniform mesh of $N = 2^n$ points per mode, with the kinetic energy implemented via QFT.
This representation handles polynomial potentials through the UCR site structure (each monomial $x_r^{a_1} x_s^{a_2}$ maps to a set of grid sites with analytically computable rotation angles), but it does not naturally accommodate potentials with singularities, hard walls, or multi-valued character.
Reactive scattering problems, in which the nuclear wavepacket encounters a repulsive wall or passes through a conical intersection
seam, require potential energy surfaces that diverge or become discontinuous in certain regions of configuration space.
Smooth molecular potentials fitted to ab initio data, such as Morse oscillators, Lennard-Jones interactions, or spline interpolants, involve transcendental functions that cannot be represented as finite-degree polynomials on the grid.
Higher-order anharmonic couplings (cubic, quartic, and beyond the quadratic--bilinear truncation used here) introduce additional cross-mode terms that expand the site structure and modify the conflict graph coloring, but do not change the algorithmic framework.
Each of these extensions preserves the core REQWIEM architecture (UCR cascades, channel-register fan-out, XOR-class fragmentation) while adapting the site structure and rotation angle computation to the specific potential form.
The next chapters, Chapter~\ref{ch:scattering} and Chapter~\ref{ch:smooth_potentials} develops these extensions systematically.

\chapter{Quantum Scattering}\label{ch:scattering}

Scattering cross sections are a fundamental observable in atomic, molecular, and materials physics, encoding collision outcomes between particles, governing reaction rates, transport coefficients, spectral line shapes, and a predictor for the thermodynamic properties of liquids and gases \cite{courtney2021comprehensive, stancil1999ab}.
A small-scale view of scattering begins to probe quantum materials like ultracold gasses and crystal structures, where scattering observables correspond in interaction models to that which is seen in experiment .

Scattering simulation in many-body systems and systems with many internal degrees of freedom once again confronts a ``curse of dimensionality" (Chapter~\ref{ch:Introduction}), following from the same principles as in vibronic models. 
Coordinates/modes are typically defined internally, found in terms of the center of mass and interatomic angles and distances \cite{joachain1975quantum}.
Classical approaches mitigate this scaling through time-independent close-coupling methods, semiclassical approximations, or subspace techniques that isolate partial eigenspectra \cite{geller2015universal, ito1992semi, krivoruchenko2007semiclassical}.
Classical models continue to develop for collisional problems, being highly effective in restricted regimes but ultimately face the exponential wall when applied to multichannel or many-body scattering \cite{yang2025full, wang2025multilayer, mihalik2025accurate}.

Extending beyond vibronics, quantum computing may handle this problem more easily.
Recent work has demonstrated eigenstate preparation via rodeo algorithms \cite{stancil2016towards, choi2021rodeo, qian2024demonstration}, scattering simulations in condensed matter and quantum chromodynamics \cite{preston2025nonadiabatic, chawdhry2025quantum}, and quantum algorithms for phase-shift extraction that propagate classically precomputed eigenstates \cite{turro2024evaluation, gustafson2021real, sharma2024scattering}.
These techniques are crucial in their own subfields and in reducing measurement overhead of phase shift extraction, though chemical scattering has been understudied, with classical precomputation of eigenstates remaining exponentially complex and the Schr\"odinger equation solution left open.

In this chapter, the on-diagonal potential energy surfaces introduced in previous chapters are now considered as inverse polynomial functions, with quantum circuit emulations representing an atomic scattering event without eigenstate precomputation.
A Gaussian wavepacket is evolved under a Trotterized Hamiltonian whose potential energy surface (PES) is a single Lennard--Jones interaction~\cite{lennard1924determination}.
The chosen example describes a collision of two neutral, ground-state Helium atoms with a closed-form analytical gate composition on a binary-encoded qubit basis.
The potential energy surface requires $\mathcal{O}(2^n)$ quantum gates per trotter step, with no oracle overhead and no polynomial approximation, following information-theoretic bounds on loading data, as a linearly-scaling coordinate requires exponentially many gates to load such a function without quantum arithmetic~\cite{shende2005synthesis}.
A direct comparison with quantum arithmetic finds that repeated multipliers and squarers require higher overhead than implementing the operation directly for most reasonable circuit sizes.
In the derivation of these surfaces, two exponentially scaling resources are identified, the first being the number of quantum gates to construct the circuit, the other a one-time classical precomputation of rotation angles. 
These angles are calculated ``one-time" for a register with a given number of qubits and a given polynomial power, to be cached and reweighted for any future simulation involving a PES of this shape. 

Phase shifts are extracted from the time-evolved wavefunction and used to recover the deflection function, scattering amplitude, and differential and integral cross sections.
Results are validated against time-independent solutions of the radial Schr\"odinger equation using the Numerov method, a fourth-order numerical solver outputting phase shifts.
Operator fidelity is validated by calculating the Hilbert-Schmidt distance between the binary circuit and an independently derived binary-reflected Gray code (BRGC) construction \cite{chang2022improving}. 
The Hilbert-Schmidt distance is found to vanish to numerical precision, providing an operator-level validation free of classical approximation.

A single surface can be considered adiabatically, making post-Trotter techniques (QSP, qubitization) favorable for long times~\cite{low2017optimal, gilyen2019quantum}. 
Methods and results shown in this chapter therefore do not show a standalone advantage relative to existing work, rather serving as a module in the REQWIEM toolkit for swapping diagonal diabatic terms.
As the circuit is analytically derived and closed-form, it establishes a ground-truth benchmark for variational and approximate methods, enabling error quantification via operator-level metrics (e.g. Hilbert-Schmidt distance). 
The general multilinear expansion underlying the circuit construction and its corresponding UCR composition are presented in Chapter~\ref{ch:smooth_potentials}.

The remainder of this chapter is organized as follows.
Section~\ref{sec:method} develops the inverse polynomial circuit and a repeated-squaring algorithm for parameter computation.
Section~\ref{sec:simulation} details a validation procedure for the He--He system, while Section~\ref{sec:results} presents the extracted scattering observables and their validation.
Section~\ref{sec:quantum_arithmetic} calculates a threshold where quantum arithmetic gives a reduced gate count for Lennard-Jones potential systems.
Section~\ref{sec:discussion} analyzes the complexity-theoretic implications, periodic boundary effects, and benchmarking infrastructure enabled by the exact circuit, and Section~\ref{sec:scattering_summary} summarizes the contribution with transition into finite-difference potential surfaces.

\section{Method}
\label{sec:method}

\subsection{Problem Setup}
\label{sec:problem_setup}

Scattering off a hard-core potential typically solves the radial time-independent Schr\"odinger equation. 
REQWIEM handles this by means of a real-time evolution, making the problem time-dependent,
\begin{equation}
    \label{eq:radial_TISE}
    [T + V_{\text{eff}}]\psi(r, t) = E\psi(r, t),
\end{equation}
where as before $T = (p-p_0^2)/2\mu$ is the radial kinetic energy operator, and $V_{\text{eff}}$ is an effective potential containing both the interatomic interaction and a centrifugal barrier for a given partial wave $\ell$.
Reduced mass of a two-body collision system is $\mu = m_1 m_2/(m_1 + m_2)$, and $p_0$ denotes the initial relative momentum applied as a Lorentz boost, since the variational UCC ansatz does not encode momentum (Chapter~\ref{ch:wavepacket_prop}).

Eigenstates of Eq.~\ref{eq:radial_TISE} are taken to be unknown, as a necessary precondition for scalable quantum advantage: an algorithm that begins from classically precomputed eigenstates does not overcome the exponential cost of eigendecomposition itself.
The initialized Gaussian wavepacket represents a superposition of standing waves,
\begin{equation}
    \label{eq:gaussian_wp}
    \psi(r, t_0) = \left(\frac{1}{2\pi\delta^2}\right)^{1/4} \exp\!\left[-\frac{(r - r_0)^2}{4\delta^2}\right],
\end{equation}
where $\delta$ is the spatial width and $r_0$ is the initial mean position of the wavepacket.
Initialization follows procedures detailed in Chapter~\ref{ch:wavepacket_prop}, with propagation following similarly with the kinetic term $T$ being applied between two cQFTs as $\mathcal{O}(n^2)$ gates.

Instead of a nuclear coordinate, a typical collision between for an ion/atom or two noble gas atoms uses a spatial coordinate, or internally defined coordinates (e.g. Jacobi coordinates).
In the present context, a single coordinate $r$ is encoded in a binary position basis on a register of $n$ qubits, just as before with vibrational mode coordinates
\begin{equation}
    \label{eq:binary_position}
    r = \Delta \sum_{j=0}^{n-1} 2^j\, q_j,
\end{equation}
where $q_j \in \{0, 1\}$ are the individual qubit states.

\subsection{Constructing the Inverse Polynomial Circuit}
\label{sec:constructing_pes}

The central challenge addressed in this chapter is representing $f(r) = 1/r^\alpha$ for a degree $\alpha$ as a quantum circuit acting on the binary-encoded position states.
Unlike the kinetic operator, which is polynomial in its conjugate variable, the potential $V(r) \propto 1/r^\alpha\,,\alpha\in\mathbb{Z}^+$ is a non-polynomial function of the position variable encoded in the qubit register.
Classical evaluation of $1/r$ proceeds by separately handling numerator and denominator for each basis state, reduced to a trivial operation.
In a quantum context, the potential must act as a global function on the entire register.

\subsubsection{Multilinear Expansion}

The function $f(r) = 1/r$ is asserted to be representable exactly on a discrete grid defined by binary variables $q_j \in \{0,1\}$ via a multilinear expansion,
\begin{equation}
    \label{eq:multilinear}
    \frac{1}{r} = \sum_{S \subseteq \{0, 1, \ldots, n-1\}} a_S \prod_{j \in S} q_j.
\end{equation}
This follows from the fact that any real-valued function defined on the $2^n$ vertices of the Boolean hypercube $\{0,1\}^n$ admits a unique multilinear representation, as a multilinear extension over the power set of qubit indices \cite{o2014analysis}.
Each subset $S$ labels a monomial $\prod_{j \in S} q_j$, and the expansion contains $2^n$ terms (including the empty product $S = \emptyset$, which gives a constant offset).

\subsubsection{M\"obius Inversion for Coefficients}

Coefficients $a_S$ are obtained by M\"obius inversion on the Boolean lattice.
Substituting Eq.~\ref{eq:binary_position} into $f(r) = 1/r$ and evaluating at all $2^n$ vertices yields a system of linear equations.
Its solution is given by
\begin{equation}
    \label{eq:mobius_coulomb}
    a_S = \sum_{T \subseteq S} (-1)^{|S| - |T|} \cdot \frac{1}{\Delta \sum_{j \in T} 2^j},
\end{equation}
where $|S|$ and $|T|$ denote the cardinalities of sets $S$ and $T$, respectively, adopting the convention $\sum_{j \in \emptyset} 2^j = 0$.
The sum over all subsets $T$ of $S$ contains $2^{|S|}$ terms for each $S$, and the full set of $2^n$ coefficients can be computed exactly in $\mathcal{O}(3^n)$ classical operations ($\sum_{S} 2^{|S|} = 3^n$ by the binomial theorem applied to the power set).
Denominators $\Delta\sum_{j \in T} 2^j$ are sums of distinct powers of two scaled by $\Delta$, so coefficients admit exact rational arithmetic.
All scaling factors can be factored out of the recursive coefficient calculation, added back for problem-specific application.

Sums of fixed-size integers (32-bit, 64-bit) can be computed in $\mathcal{O}(1)$ time per classical operation, though additional overhead for higher precision typically requires $\mathcal{O}(N)$ time or $\mathcal{O}(\log N)$ time for carry-lookahead adders. 
In the numerical validation of this work, the $\texttt{get\_prec}()$ tool, native to C, is used to obtain 50 decimal places of precision.
This degree of precision is not strictly necessary, but is used to make the classical proxy of ``synthesis error" negligible relative to discretization and Trotter error.
A thorough derivation of the multilinear expansion and M\"obius inversion for $f(r) = 1/r$ is given in Appendix~\ref{app:circuits}.

\subsubsection{Exponentiation as Phase Rotations}

The potential energy operator is diagonal in the position basis and acts as a phase,
\begin{equation}
    U = e^{-iV(r)\tau} = e^{-ic f(r)\tau},
\end{equation}
where $c$ collects all physical prefactors with Trotter step $\tau$.
Substituting Eq.~\ref{eq:multilinear},
\begin{equation}
    U = \exp\!\left[-ic\tau \sum_{S} a_S \prod_{j \in S} q_j\right]
      = \prod_{S \subseteq \{0,\ldots,n-1\}} \exp\!\left[-ic\tau\, a_S \prod_{j \in S} q_j\right].
\end{equation}
All terms commute, and factorization into a product over subsets $S$ $\prod_{j \in S} q_j$ is diagonal in the computational basis.
Each factor is written here as a $|S|$-fold controlled $R_z$ rotation,
\begin{equation}
    \label{eq:controlled_rz_product}
    U = \prod_{S \subseteq \{0, 1, \ldots, n-1\}} C^{|S|}\!R_z(2\pi c\, a_S),
\end{equation}
where $C^{|S|}\!R_z(\theta)$ denotes an $R_z(\theta)$ gate conditioned on all qubits in $S$ being in the $\ket{1}$ state.
This product contains $2^n - 1$ nontrivial terms (excluding $S = \emptyset$, which contributes only a global phase).

\subsubsection{Explicit Coulomb Potential}

Considering immediate application, the Coulomb potential $V(r) = \alpha/r$ is represented on the binary qubit basis as
\begin{equation}
    \label{eq:coulomb_circuit}
    V_{\text{Coulomb}}(r) = \alpha \sum_{S \subseteq \{0,1,\ldots,n-1\}}
      \left(\sum_{T \subseteq S} (-1)^{|S|-|T|} \cdot \frac{1}{\Delta\sum_{j \in T} 2^j}\right)
      \prod_{j \in S} q_j.
\end{equation}
This is implemented by the circuit of Eq.~\ref{eq:controlled_rz_product} with $\mathcal{O}(2^n - 1)$ multi-controlled $R_z$ gates prior to UCR decomposition.

This constitutes the first closed-form circuit implementation for inverse polynomial potentials on a binary-encoded qubit basis that we are aware of.
No oracle is required, no block-encoding overhead is introduced, and no polynomial approximation of the target function is made.

\subsubsection{Coulomb Potential Demonstration}
\label{sec:coulomb_demo}

The M\"obius inversion function for circuit parameters can be tested through the classical compilation of a quantum circuit modeling a proton moving in a Coulomb potential $V(r) = 1/r$ on a register of $n = 9$ qubits.
The particle is initialized with mass $m = 1836\,\text{a.u.}$ at position $x_0 = 10\,\text{a.u.}$ with initial momentum $p_0 = -15\,\text{a.u.}$ and wavepacket width $\delta = 1.0\,\text{a.u.}$, in a box of size $L = 40\,\text{a.u.}$
The symmetric Trotter step of Eq.~\ref{eq:trotter2} is applied for $k = 800$ timesteps.

\begin{figure}[t]
    \centering
    \includegraphics[width=\textwidth]{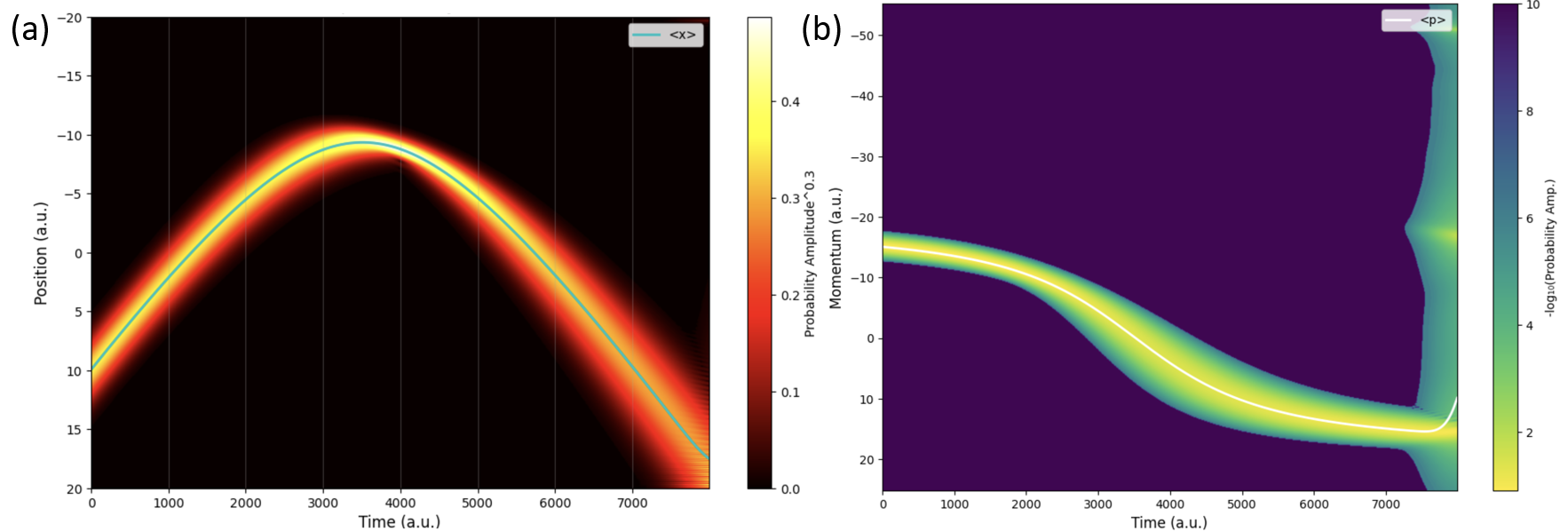}
    \caption{\label{fig:Coulomb}
    Position and momentum heatmaps for a proton in a Coulomb potential. 
    Data represents weighted probability densities from a classical emulation with $n=9$ ideal qubits.}
\end{figure}

Position- and momentum-space heatmaps of the time-evolved probability density are shown in Figure~\ref{fig:Coulomb}.
The position heatmap displays probability amplitudes scaled by a power of $0.3$ for visualization; the momentum heatmap plots $-\log(|\tilde{\psi}|^2)$, where $\tilde{\psi}$ is obtained by measuring the qubit register after a centered QFT.
Near $t \approx 7500\,\text{a.u.}$, the wavepacket begins to interact with the ``back side" of the Coulomb potential due to the periodic boundary conditions intrinsic to the binary position basis, illustrating the aliasing effect discussed in Section~\ref{sec:discussion}.
This demonstration confirms that the inverse polynomial circuit of Eq.~\ref{eq:coulomb_circuit} correctly encodes a $1/r$ potential and produces physically expected dynamics, including reflection from the Coulomb singularity and subsequent free propagation prior to boundary contamination.

\subsection{Repeated Squaring Algorithm}
\label{sec:repeated_squaring}

The effective potential for scattering contains terms proportional to $1/r^\gamma$ for integer $\gamma > 1$ (e.g., the centrifugal barrier $\propto 1/r^2$ and the Lennard--Jones terms $\propto 1/r^6, 1/r^{12}$).
Rather than independently deriving the multilinear expansion for each power, higher inverse powers are found by algebraically squaring and combining the base expansion of $1/r$.

\subsubsection{Construction of $1/r^2$}

Squaring the multilinear expansion of Eq.~\ref{eq:multilinear} gives
\begin{equation}
    \frac{1}{x^2}
    = \left(\sum_{S_1} a_{S_1}\prod_{j \in S_1} q_j\right)
      \left(\sum_{S_2} a_{S_2}\prod_{j \in S_2} q_j\right)
    = \sum_{S_1}\sum_{S_2} a_{S_1}\,a_{S_2}\prod_{j \in S_1} q_j \prod_{j \in S_2} q_j,
\end{equation}
Since $q_j^2 = q_j$ (Boolean idempotency), collapsing the product $\prod_{j \in S_1} q_j \cdot \prod_{j \in S_2} q_j = \prod_{j \in S_1 \cup S_2} q_j$, reducing to a single sum over the power set, as
\begin{equation}
    \label{eq:one_over_x2}
    \frac{1}{x^2} = \sum_{T \subseteq \{0,1,\ldots,n-1\}} b_T^{(2)} \prod_{j \in T} q_j,
\end{equation}
with coefficients
\begin{equation}
    \label{eq:b_coeffs_2}
    b_T^{(2)} = \sum_{\substack{S_1, S_2 \subseteq \{0,1,\ldots,n-1\} \\ S_1 \cup S_2 = T}} a_{S_1} \cdot a_{S_2},
\end{equation}
Since the union constraint $S_1 \cup S_2 = T$ means $S_1$ and $S_2$ are subsets of $T$, the number of $(S_1, S_2)$ pairs contributing to each $b_T^{(2)}$ is $(2^{|T|})^2 = 4^{|T|}$.

\subsubsection{General $1/r^\gamma$ via Union--Convolution}

Induction extends the construction to arbitrary integer powers $\gamma$.
For any $\gamma \geq 1$,
\begin{equation}
    \label{eq:one_over_xm}
    \frac{1}{x^\gamma} = \sum_{T \subseteq \{0,1,\ldots,n-1\}} b_T^{(\gamma)} \prod_{j \in T} q_j,
\end{equation}
where coefficients are given by the union--convolution
\begin{equation}
    \label{eq:b_coeffs_m}
    b_T^{(\gamma)} = \sum_{\substack{S_1, S_2, \ldots, S_\gamma \subseteq \{0,1,\ldots,n-1\} \\ S_1 \cup S_2 \cup \cdots \cup S_\gamma = T}} a_{S_1} \cdot a_{S_2} \cdots a_{S_\gamma}.
\end{equation}

\subsubsection{Gate Count Invariance}

Gate count is invariant under the power $\gamma$, beginning at the maximum possible value for a diagonal operator.
By tensor product space commutativity, coefficients $b_T^{(\gamma)}$ are absorbed into the same $|T|$-fold controlled $R_z$ gate as the base $a_T$ coefficient.
Implementing $1/r^\gamma$ requires exactly the same circuit as $1/r$ with only the rotation angles changed.
By extension, any potential of the form
\begin{equation}
    V(r) = \sum_i \gamma_i\, x^{-m_i}, \qquad m_i \in \mathbb{Z}^+,
\end{equation}
can be implemented with no more gates than a single $1/r$ term.
Coefficients from each power are computed independently and then summed into a single set of $2^n - 1$ rotation angles, one per subset $S$.

\subsubsection{Centrifugal Potential}

As an immediate application, the centrifugal barrier for partial wave $\ell$ with reduced mass $\mu$ is
\begin{equation}
    \label{eq:centrifugal}
    V_{\text{cf}}(r) = \frac{\ell(\ell+1)}{2\mu r^2}
    = \frac{\ell(\ell+1)}{2\mu} \sum_{T \subseteq \{0,1,\ldots,n-1\}} b_T^{(2)} \prod_{j \in T} q_j,
\end{equation}
where the $b_T^{(2)}$ coefficients are those of Eq.~\ref{eq:b_coeffs_2} and the physical prefactor $\ell(\ell+1)/2\mu$ multiplies each rotation angle.

\subsubsection{Classical Precomputation of Coefficients}

Base coefficients $\{a_S\}$ depend only on register size $n$ and grid spacing $\Delta$; union--convolution coefficients $\{b_T^{(\gamma)}\}$ depend additionally on the power $\gamma$.
Physical prefactors (coupling constants, partial-wave quantum numbers, energy scales) enter as multiplicative factors on rotation angles, so coefficients need only be computed once for a given $n$ and can be reused across all simulations with different physical parameters.

Rotation angles for larger registers involve coefficients whose magnitudes span many orders of magnitude, often exceeding classical double-precision floating-point limits (Section~\ref{sec:hs_validation}).
However, exact rational arithmetic is available for all $1/r^\gamma$ coefficients, since the denominators are sums of distinct powers of two.

\subsection{Effective Potential Assembly}
\label{sec:effective_potential}

Establishing that a sum of inverse polynomial terms shares the same circuit, a complete effective potential for the scattering problem can be constructed.
The Lennard--Jones effective potential~\cite{lennard1924determination} for partial wave $\ell$ is
\begin{equation}
    \label{eq:Veff}
    V_{\text{eff}}(r) = V_{\text{LJ}}(r) + V_{\text{cf}}(r)
    = 4\eps\left[\left(\frac{\sigma_\text{LJ}}{r}\right)^{12} - \left(\frac{\sigma_\text{LJ}}{r}\right)^{6}\right]
    + \frac{\ell(\ell+1)}{2\mu r^2},
\end{equation}
where $\eps$ and $\sigma_\text{LJ}$ are the Lennard--Jones well depth and length parameters, respectively.

Each term in Eq.~\ref{eq:Veff} expands independently on the qubit basis using the repeated squaring algorithm of Section~\ref{sec:repeated_squaring}:
\begin{itemize}
    \item The repulsive wall $(\sigma_\text{LJ}/r)^{12} \propto r^{-12}$ uses coefficients $b_T^{(12)}$ with prefactor $4\eps\,\sigma_\text{LJ}^{12}$.
    \item The attractive well $-(\sigma_\text{LJ}/r)^{6} \propto r^{-6}$ uses coefficients $b_T^{(6)}$ with prefactor $-4\eps\,\sigma^{6}$.
    \item The centrifugal barrier $\propto r^{-2}$ uses coefficients $b_T^{(2)}$ with prefactor $\ell(\ell+1)/(2\mu)$.
\end{itemize}
For each subset $T$, values are summed into a single effective rotation angle,
\begin{equation}
    \label{eq:effective_angle}
    \theta_T = 2\pi\tau\left[
        4\eps\,\sigma_\text{LJ}^{12}\, b_T^{(12)}
      - 4\eps\,\sigma_\text{LJ}^{6}\, b_T^{(6)}
      + \frac{\ell(\ell+1)}{2\mu}\, b_T^{(2)}
    \right],
\end{equation}
The full effective potential can be written as
\begin{equation}
    \label{eq:Veff_circuit}
    e^{-iV_{\text{eff}}\tau}= \prod_{T \subseteq \{0,1,\ldots,n-1\}} C^{|T|}\!R_z(\theta_T),
\end{equation}
using $|T|$-controlled $R_z$ rotations, decomposed by UCR into CNOT ladders and $\mathcal{O}(2^n-1)$ uncontrolled $R_z$ rotations (Chapter~\ref{ch:model_ext}, Appendix~\ref{app:binary_decomp}).

\section{Simulation}
\label{sec:simulation}

\subsection{Trotterization Procedure}
\label{sec:trotterization}

The time-evolution operator $U = e^{-iH t}$ is approximated by a second-order (symmetric) Trotter--Suzuki decomposition~\cite{suzuki1991general}.
Writing the Hamiltonian as $H = T + V_{\text{eff}}$, the propagator for a single timestep $\tau$ is
\begin{equation}
    \label{eq:trotter2}
    U(\tau) = e^{-iH\tau} \approx e^{-iT\tau /2}\, e^{-iV_{\text{eff}}\tau}\,e^{-iT\tau /2} + \mathcal{O}(\tau^3),
\end{equation}
The full evolution over a time $t = k\tau$ is obtained by iterating this symmetric step,
After state preparation, the symmetric Trotter step is iterated over $k$ timesteps, with the full statevector being recorded for each timestep.

\subsection{Simulation Parameters}
\label{sec:sim_params}

The system under study is a low-energy scattering of two neutral ground-state helium atoms, modeled by a 12--6 Lennard--Jones interaction.
Potential parameters are chosen to approximate fluid helium at a temperature of $2.044\,\text{K}$ \cite{tchouar2003thermodynamic}:
\begin{equation}
    \epsilon = 3.236 \times 10^{-5}\,\text{E}_h,\, \sigma_\text{LJ} = 4.831\,\text{a.u.},\,m_{\text{He}^4} \approx 7298\,\text{a.u.}\,p_0\approx-0.2172\,\text{a.u.}
\end{equation}
For identical interacting particles, the reduced mass is $\mu = m_{\text{He}}/2 \approx 3649\,m_\text{e}$~\cite{CODATA2022_online}.

The Gaussian wavepacket of Eq.~\ref{eq:gaussian_wp} is initialized with width $\delta = 10.0\,\text{a.u.}$ and mean position $r_0 = 100\,\text{a.u.}$.
The convention chosen increases $r$ from left to right on the grid and negative $p_0$ corresponds to leftward propagation toward $r = 0$.
The simulation box has size $L = 200\,\text{a.u.}$, chosen to provide a sufficiently large asymptotic region for phase-shift extraction after the collision without interference from the periodic boundary conditions intrinsic to the binary basis.
A quantum register of $n = 13$ qubits discretizes the box into $N = 2^{13} = 8192$ grid points with resolution $\Delta = L/N \approx 2.44 \times 10^{-2}\,\text{a.u.}$

The wavepacket is propagated for a total evolution time $t = 5 \times 10^6\,\text{a.u.}$ using a Trotter timestep $\tau = 1\,\text{a.u}$.
This long evolution time ensures that the wavepacket traverses the interaction region and develops an asymptotic tail extending well beyond the range of the effective potential, from which converged phase shifts can be extracted.

\begin{figure}[t]
    \centering
    \includegraphics[width=0.8\textwidth]{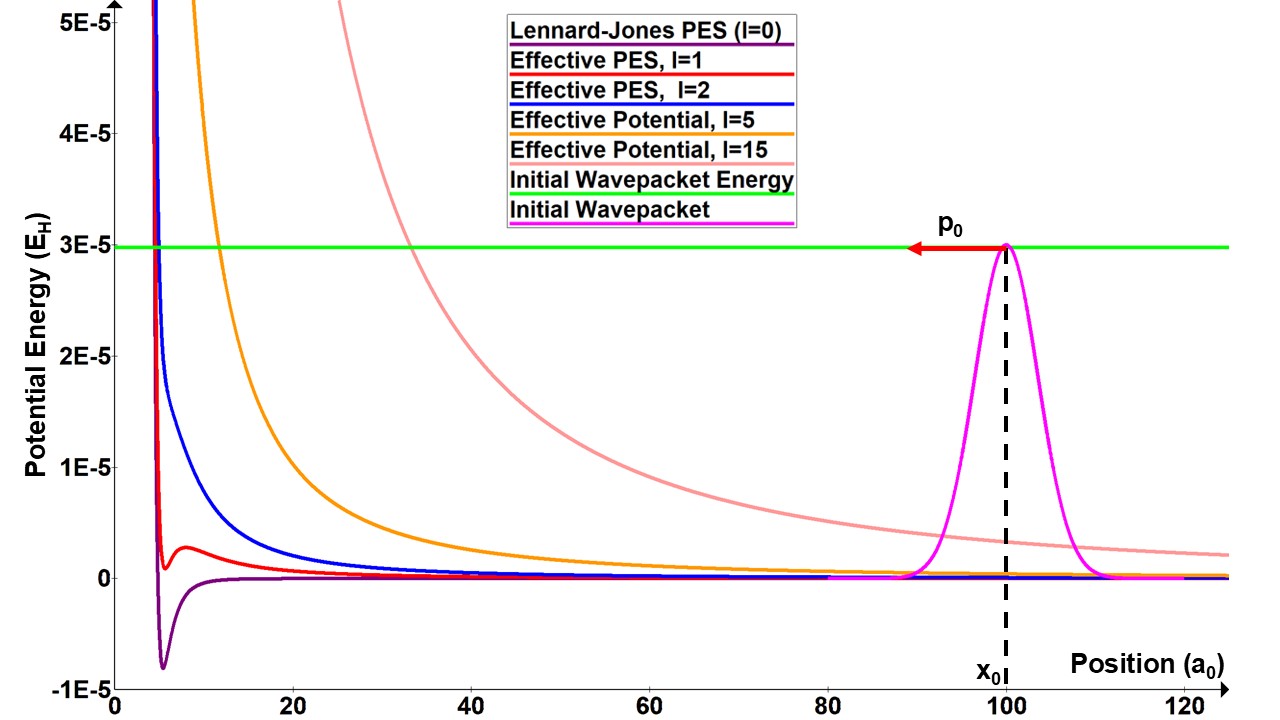}
    \caption{\label{fig:Scattering_Schematic}
    Schematic for the quantum scattering simulation performed via
    error-free statevector simulation.
    A Gaussian wavepacket is initialized at $r_0 = 100\,\text{a.u}$.
    The effective potential $V_{\text{eff}} = V_{\text{LJ}}(r) +
    V_{\text{cf}}(r)$ is shown, with individual components indicated.
    Wavepacket not to scale.}
\end{figure}

Two sets of 19 partial-wave simulations were performed, spanning $\ell = [0,18]$.
The first set uses the full effective potential $V_{\text{eff}}$, while the second set is used as a reference simulation without the Lennard--Jones potential, using only $V_{\text{cf}}$. 
The position basis was scaled to the simulation box size and Nyquist range to capture the majority of relevant eigenstates with sufficient grid resolution and box size, defining a reliable asymptotic region for the majority of partial waves.
A schematic of the simulation geometry is shown in Figure~\ref{fig:Scattering_Schematic}.

\subsection{Phase Shift Comparison}
\label{sec:sim_types}

To validate the accuracy of the quantum algorithm's representation of the closed-form PES, the radial Schr\"odinger equation is solved using the Numerov method~\cite{johnson1977new, johnson1978renormalized} on a classical computer, providing a time-independent benchmark by numerically propagating the scattering wavefunction outward to large interatomic distances $r$.
The solution is matched to the known asymptotic form (regular and irregular free-particle functions) to extract the phase shift directly.
By comparing to a time-independent approach, error from Trotterization, circuit representation, and grid discretization is cumulated.

\section{Results}
\label{sec:results}

\subsection{Phase Shift Extraction}
\label{sec:phase_shift_extraction}
Phase shifts are extracted from the time-evolved wavepacket using a spatial Fourier transform (momentum-space decomposition) of full wavefunction snapshots recorded after scattering~\cite{tannor2007introduction}.
At any time $t$ after the wavepacket has interacted with the potential, the spatial FFT of $\psi(r, t)$ decomposes the wavefunction into momentum components, with $k > 0$ corresponding to outgoing (scattered) waves and $k < 0$ to incoming waves.

Because the Trotter propagation is performed in a simulation frame with Hamiltonian $H = (p - p_0)^2/(2\mu) + V_\text{eff}$, the wavepacket must first be boosted to the laboratory frame.
This is readily handled by multiplying the recorded real-space wavefunction by a plane-wave phase $e^{ik_0 x}$, where $k_0 = p_0/\hbar < 0$ is the initial wavenumber, so that the incoming peak appears at $k_\text{lab} = k_0 < 0$ and the outgoing peak at $k_\text{lab} = |k_0| > 0$.
The lab-frame momentum representation is then obtained via a discrete Fourier transform of the boosted wavefunction,
\begin{equation}
    \label{eq:spatial_fft}
    \tilde{\psi}(k, t) = \mathrm{FFT}\!\left[\psi(r, t)\, e^{ik_0 r}\right].
\end{equation}
Only the outgoing components $\tilde{\psi}(k > 0, t)$ are retained for the $S$-matrix extraction.

The $\ell$-dependent $S$-matrix element $S_{\mathrm{LJ}}(k)$ is extracted as a ratio of outgoing amplitudes, with the reference denominator constructed differently for $\ell = 0$ and $\ell > 0$.

For partial waves with $\ell > 0$, the centrifugal barrier reflects a portion of the incoming wavepacket in both the full ($\epsilon_\mathrm{LJ} \neq 0$) and reference ($\epsilon_\mathrm{LJ} = 0$) simulations, producing well-defined outgoing amplitudes in each case.
The interaction $S$-matrix element is obtained as
\begin{equation}
    \label{eq:smatrix_ratio}
    S_{\mathrm{LJ}}(k) = \frac{\tilde{\psi}_\mathrm{full}(k > 0,\, t)}{\tilde{\psi}_\mathrm{ref}(k > 0,\, t)} = e^{2i\delta_{\mathrm{LJ}}(\ell,\, E)},
\end{equation}
where $E = \hbar^2 k^2 / (2\mu)$.

For $s$-wave ($\ell=0$) scattering, the reference simulation with $V_\mathrm{eff} = 0$ has no reflecting barrier and therefore produces no outgoing wave, precluding a direct ratio.
Instead, the outgoing amplitude is related to the known incoming amplitude through the scattering relation
\begin{equation}
    \label{eq:swave_scattering}
    \tilde{\psi}_\mathrm{out}(k > 0,\, t) = S_0(k)\, A_\mathrm{in}(-k)\, e^{-i(k - k_0)^2 t/(2\mu\hbar)},
\end{equation}
where $A_\mathrm{in}(-k)$ is the initial wavepacket amplitude at the mirror momentum $-k$ in the lab frame.
Following a similar procedure to $\ell>0$, the lab-frame boost and FFT from Eq.~\ref{eq:spatial_fft} are applied to $\psi_0(r)$, where by evaluating at mirror indices corresponding to $-k$,
\begin{equation}
    \label{eq:mirror_amp}
    A_\mathrm{in}(-k) = \mathrm{FFT}\!\left[\psi_0(r)\, e^{ik_0 r}\right]\bigg|_{k' = -k}.
\end{equation}
Free propagation phase is computed in the simulation frame as $e^{-i(k - k_0)^2 t/(2\mu\hbar)}$, and the $s$-wave $S$-matrix follows from
\begin{equation}
    \label{eq:swave_smatrix}
    S_0(k) = \frac{\tilde{\psi}_\mathrm{out}(k,\, t)}{A_\mathrm{in}(-k)\, e^{-i(k - k_0)^2 t/(2\mu\hbar)}}.
\end{equation}

\begin{figure}[t]
    \centering
    \includegraphics[width=\textwidth]{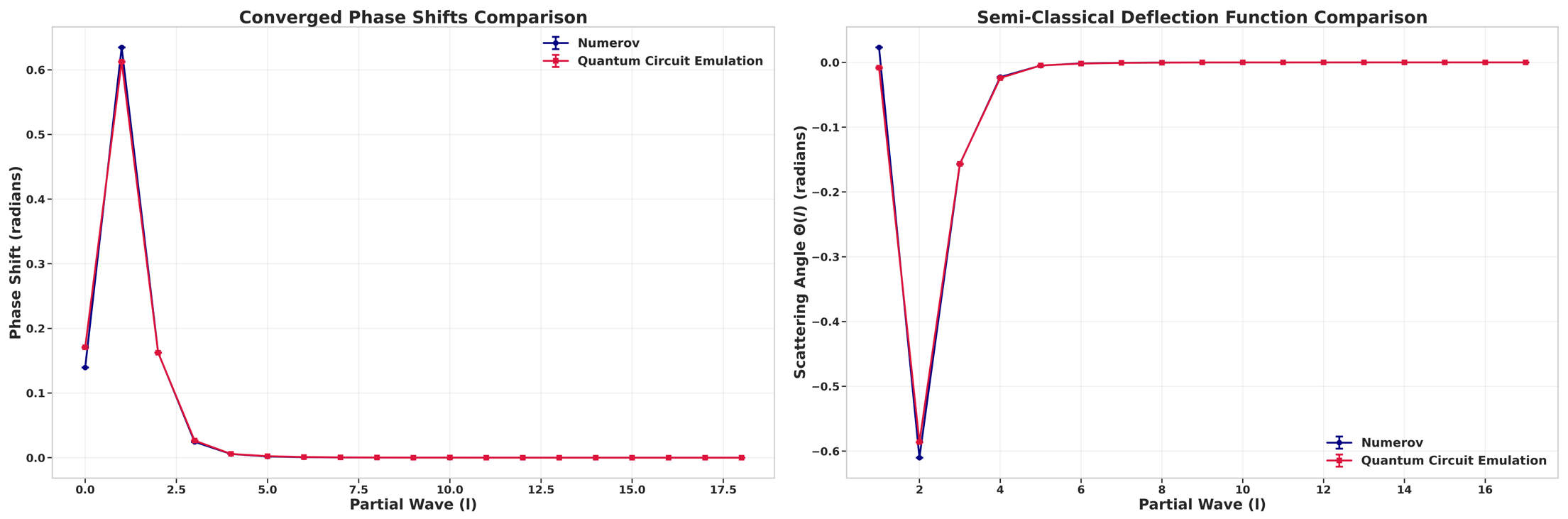}
    \caption{\label{fig:Shfits+Deflection function}
    (Left) Phase Shifts calculated by spatial FFT applied to tracked wavefunction via time-evolving wavepacket with quantum circuits (red), compared against Numerov phase shift calculation.
    (Right) Deflection function comparison, tracking the difference between $\delta_{\ell}$ and $\delta_{\ell+1}$.}
\end{figure}

Final values for partial waves are calculated by averaging $500$ snapshots once the phase shifts have converged. 
These values are shown in Figure~\ref{fig:Shfits+Deflection function} compared against the time-independent result, alongside the deflection function $\Theta(\ell) = 2(\delta_\ell - \delta_{\ell+1})$~\cite{joachain1975quantum}.

\subsection{Scattering Observables}
\label{sec:scattering_observables}

\begin{figure}[t]
    \centering
    \includegraphics[width=\textwidth]{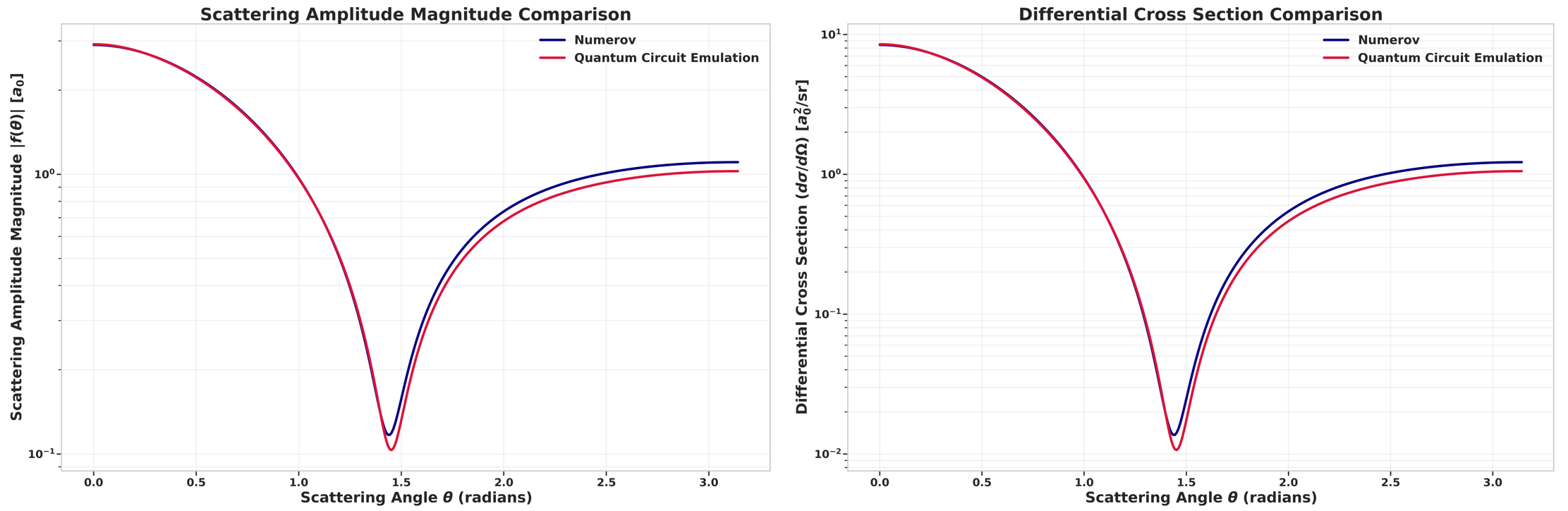}
    \caption{\label{fig:Scattering amplitude + Diff xsec}
    (Left) Scattering amplitude comparison between quantum circuit emulation and Numerov.
    (Right) Derived differential cross section, given as $|f(\theta)|^2$}
\end{figure}

From the phase shifts $\{\delta_{\text{LJ}}(\ell, E)\}$, the scattering amplitude is derived as~\cite{joachain1975quantum}
\begin{equation}
    \label{eq:scattering_amplitude}
    f(\theta, E) = \frac{1}{2ik}\sum_{\ell=0}^{\ell_{\max}}(2\ell + 1)\left(e^{2i\delta_\ell} - 1\right)P_\ell(\cos\theta),
\end{equation}
from which the differential and total cross sections follow:
\begin{equation}
    \frac{d\sigma}{d\Omega} = |f(\theta, E)|^2, \qquad
    \sigma_{\text{tot}}(E) = \frac{4\pi}{k^2}\sum_{\ell=0}^{\ell_{\max}}(2\ell + 1)\sin^2\delta_\ell.
\end{equation}
All observables are compared directly with the classical Numerov simulation (time-independent) described in Section~\ref{sec:sim_types}, with results presented in Figure~\ref{fig:Scattering amplitude + Diff xsec}.
Following similarity in phase shifts, differential cross sections largely agree between each method, validating quantum circuits for quantum scattering without the classical precomputation of eigenvalues.
Integral cross sections were found to be $15.20\,\text{a}_0^2$ and $14.53\,\text{a}_0^2 \pm 0.01\,\text{a}_0^2$ for the Numerov and quantum circuit emulations, respectively.

\subsection{Operator Validation}
\label{sec:hs_validation}

\begin{figure}[t]
    \centering
    \includegraphics[width=\textwidth]{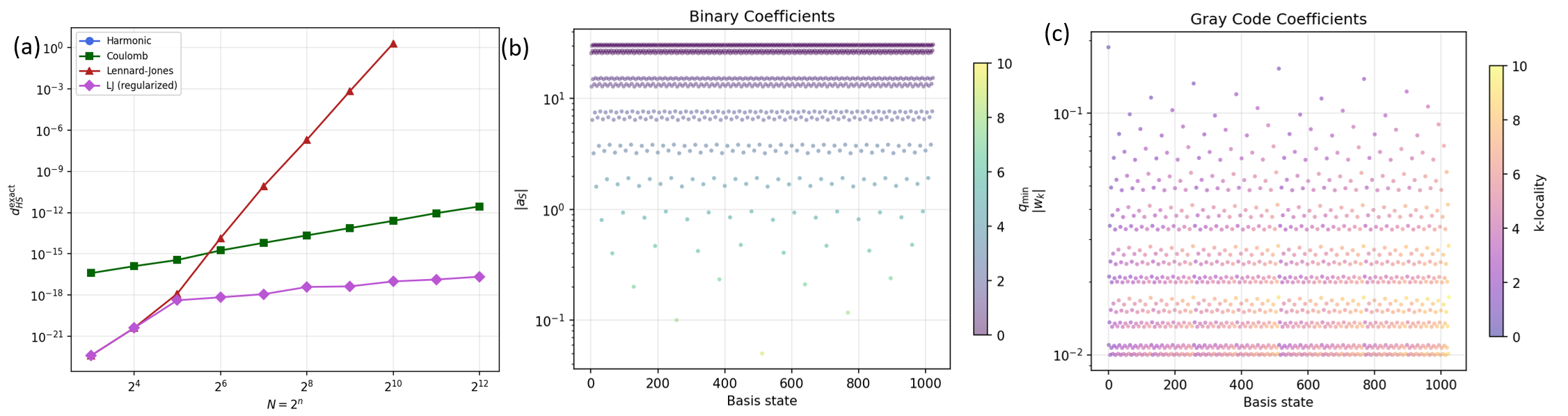}
    \caption{\label{fig:BRGC-compare}
    a) Hilbert-Schmidt distance for varying register sizes between binary-encoded circuit and BRGC circuit.
    Harmonic data (blue) directly overlap with the regularized Lennard-Jones data (magenta), upon equating the spectral norm of each potential term.
    b) Weighted binary coefficients for Coulomb potential on $n=10$ qubits. 
    Color gradient denotes the number of control qubits prior to UCR decomposition.
    c) Weighted Walsh coefficients for BRGC potential on $n=10$ qubits.
    Color gradient denotes the $k$-locality of the associated operation, being the Manhattan distance between two operative qubits.}
\end{figure}

The Hilbert--Schmidt distance is calculated for $r^2$, $1/r$, and the Lennard Jones potential, between the binary-encoded potential circuit and the independently derived binary-reflected gray code (BRGC) construction of Chang et al.\ (2022) \cite{chang2022improving}.
The two unitary matrices are found to agree to numerical precision, with the Hilbert--Schmidt distance vanishing to the level of double-precision arithmetic ($\sim 10^{-15}$).
For a full Lennard--Jones potential, coefficients scale $\propto (2^n)^{12-1}$ according to Eq.~\ref{eq:b_coeffs_m} with $\gamma_{S_\gamma} \propto 2^n$. 
When computed to 64-bit precision, repeated squaring coherently accumulates machine precision error, increasing the distance between each circuit construction, as well as between circuits and target potential shape (Figure~\ref{fig:BRGC-compare}a). This leads to extreme differences for an unregularized Lennard--Jones potential, with unweighted parameters increasing as $(2^n)^{12-1}$ accumulating significant error, impacting operator accuracy near the left boundary.
This effect is suppressed under regularization, where the Lennard--Jones potential is trimmed to the characteristic energy of the incoming particle.
This ``trimming" sets a threshold maximum energy for $V_{\text{LJ}}$, then uses a finite-difference multilinear expansion to calculate coefficients. 
This circuit is also closed-form on the interval $[0,L]$ and is detailed in Chapter~\ref{ch:smooth_potentials}.
Coulomb and Harmonic potentials follow a consistent error accumulation, with exponential scaling in $n$ following from the exponential number of double-precision arithmetic operations with coherent error accumulation. 
Parameter magnitudes of the Coulomb potential on $n=10$ qubits for the binary and BRGC methods are given in Figures~\ref{fig:BRGC-compare}b and \ref{fig:BRGC-compare}c, where magnitude distributions become inverted and rescaled.
This sensitivity to parameter magnitude suggests that, prior to the $\mod 2\pi$ operation to convert the parameter to a rotation angle, parameters need additional precision in classical precomputation.

Binary multilinear decomposition and the BRGC Walsh--Hadamard decomposition produce operators identical up to numerical precision, validating the circuit construction without reliance on any classical approximation or scattering-specific observable.
As mentioned by Chang et al. \cite{chang2022improving} and confirmed numerically, BRGC coefficients blow up near the hard-core singularity, a trait shared with the binary encoding. 
While these reduce to values between $0$ and $2\pi$ in practice, large precise values need special handling in classical precomputation to store far more decimal places than double-precision values, as discussed before.

\section{Considering Quantum Arithmetic}
\label{sec:quantum_arithmetic}

An alternative to the multilinear expansion of Section~\ref{sec:constructing_pes} and BRGC protocol of Chang et al. \cite{chang2022improving} is to compute $1/r^\gamma$ via quantum arithmetic circuits acting on ancilla registers, reducing gate count from exponential to polynomial in $n$ at the cost of additional qubits and Toffoli gates.
This section analyzes the overhead of such an approach for the Lennard--Jones $1/r^{12}$ potential used in the He--He scattering validation, comparing against the UCR-decomposed multilinear scheme.

The quantum arithmetic strategy transforms the binary-encoded position register $|r\rangle$ into a fixed-point representation of $1/r^{12}$ via a sequence of reversible arithmetic operations, applying diagonal phase rotations on the result register, then uncomputing the arithmetic to restore all ancilla qubits to $|0\rangle$.
The pipeline consists of five forward steps:
\begin{enumerate}
    \item \textbf{Reciprocal:} Compute $|r\rangle|0^{n_a}\rangle \to |r\rangle|s_1 = 1/r\rangle$ into an $n_a$-bit ancilla register.
    \item \textbf{Square (1):} Compute $|s_1\rangle|0^{n_a}\rangle \to |s_1\rangle|s_2 = 1/r^2\rangle$.
    \item \textbf{Square (2):} Compute $|s_2\rangle|0^{n_a}\rangle \to |s_2\rangle|s_3 = 1/r^4\rangle$.
    \item \textbf{Square (3):} Compute $|s_3\rangle|0^{n_a}\rangle \to |s_3\rangle|s_4 = 1/r^8\rangle$.
    \item \textbf{Multiply:} Compute $|s_3\rangle|s_4\rangle|0^{n_a}\rangle \to |s_3\rangle|s_4\rangle|s_5 = 1/r^{12}\rangle$.
\end{enumerate}
At this point, the phase $e^{-i\alpha\tau\cdot s_5}$ is applied as $n_a$ single-qubit $R_z$ rotations on the output register $|s_5\rangle$, exploiting the binary encoding of $s_5$:
\begin{equation}
    \label{eq:phase_from_binary_arith}
    e^{-i\alpha\tau\, s_5}
    = \prod_{\ell=0}^{n_a-1} R_z\!\left(2\alpha\tau\,\Delta_a\, 2^\ell\right),
\end{equation}
where $\Delta_a$ is the least-significant-bit weight of the fixed-point representation.
Reversal through Steps~5 through~1 uncomputes all intermediate registers.
Because the kinetic operator is diagonal in the momentum basis, the full arithmetic pipeline must be computed and uncomputed for every application of the potential operator within the Trotter step: twice per second-order step (Eq.~\ref{eq:trotter2}), reduced to once if there are no off-diagonal terms preventing $V_\text{diag}$ self-commutation.

The Lennard--Jones attractive term $1/r^6$ is obtained analogously, branching after Step~3: one multiplies $|s_2\rangle$ and $|s_3\rangle$ to obtain $1/r^{2} \times 1/r^{4} = 1/r^6$.
The centrifugal term $1/r^2$ is available directly from Step~2.
In all cases, the same pipeline structure applies, and the total resources are dominated by the $1/r^{12}$ path.

Arithmetic precision $n_a$ of the ancilla registers is set by the dynamic range of the target function.
On the binary grid of Eq.~\ref{eq:binary_position}, the smallest nonzero position is $r_{\min} = \Delta = L/2^n$, and the maximum value of the inverse twelfth power is $(1/\Delta)^{12} = (2^n/L)^{12}$.
The phase angle applied by Eq.~\ref{eq:phase_from_binary_arith} reaches a maximum of
\begin{equation}
    \label{eq:theta_max}
    \theta_{\max} = |\alpha\tau| \cdot (2^n/L)^{12},
\end{equation}
where $\alpha = 4\epsilon\sigma_\text{LJ}^{12}$ for the repulsive Lennard--Jones term.
Because $R_z(\theta)$ is $2\pi$-periodic, only $\theta \bmod 2\pi$ affects the quantum state.
Computing the residue to precision $\varepsilon_\theta$ requires knowing $\theta$ to absolute precision $\varepsilon_\theta$, leaving the number of bits to be
\begin{equation}
    \label{eq:na_requirement}
    n_a = \left\lceil \log_2\!\left(\frac{\theta_{\max}}{\varepsilon_\theta}\right)\right\rceil,
\end{equation}

For the He--He system of Section~\ref{sec:sim_params} with $n = 13$, $L = 200\,\text{a.u.}$, $\epsilon = 3.236 \times 10^{-5}\,\text{E}_h$, $\sigma_\text{LJ} = 4.831\,\text{a.u.}$, and $\tau = 1\,\text{a.u.}$:
\begin{equation}
    \alpha\tau = 4\epsilon\sigma_\text{LJ}^{12}\tau \approx 2.09 \times 10^{4}\,\text{a.u.},
    \qquad
    (2^{13}/200)^{12} \approx 2.23 \times 10^{19},
\end{equation}
giving $\theta_{\max} \approx 4.67 \times 10^{23}$.
With the per-gate synthesis precision $\varepsilon_\theta = 10^{-10}$ consistent with the error budget given in (Chapter~\ref{ch:resources}).
\begin{equation}
    \label{eq:na_hehe}
    n_a = \left\lceil \log_2(4.67 \times 10^{33})\right\rceil = 113.
\end{equation}
Arithmetic register sizes exceed position register size by nearly an order of magnitude ($n_a/n \approx 8.7$), a consequence $r^{-12}$ scaling amplifying the dynamic range of the grid.
With regularization (shifting the singularity left of the box), this number falls to $n_a \approx \log_2(4.67\times ||V_\text{reg}||\times10^{-10})$.

\subsection{Toffoli gate counts}

Building blocks of the quantum arithmetic circuit are multipliers, squarers, and a reciprocal circuit.
Resource estimates below use the schoolbook quantum multiplier~\cite{haner2018optimizing, thapliyal2019quantum}, being more space-efficient than QFT-based multipliers for moderate bit-widths.

\begin{center}
\renewcommand{\arraystretch}{1.25}
\begin{tabular}{l c c c}
\hline\hline
Operation & Toffoli & Output qubits & Work qubits \\
\hline
Multiplier ($n_a \times n_a$) & $2n_a^2$ & $2n_a$ & $\mathcal{O}(n_a)$ \\
Squarer ($n_a \times n_a$) & $n_a(n_a - 1)/2 + n_a$ & $2n_a$ & $\mathcal{O}(n_a)$ \\
Reciprocal (Newton, $p$ iter.) & $\leq 4n_a^2$ & $n_a$ & $\mathcal{O}(n_a)$ \\
\hline\hline
\end{tabular}
\end{center}
The multiplier Toffoli count comprises $n_a^2$ partial-product AND gates and $n_a^2$ Toffoli gates for carry propagation in the $n_a$ ripple-carry additions, accumulating partial products into a $2n_a$-bit output register~\cite{cuccaro2004new}.
The squarer has $\binom{n_a}{2}$ distinct pairs plus $n_a$ diagonal terms, yielding $n_a(n_a+1)/2$ Toffoli gates.
The reciprocal circuit uses Newton--Raphson iteration $y_{k+1} = y_k(2 - r\, y_k)$, converging quadratically from an initial approximation to $n_a$-bit precision in $p = \lceil\log_2 n_a\rceil$ iterations~\cite{munoz2018quantum, haner2018optimizing}.
Each iteration requires two multiplications and one subtraction at the current working precision; the geometric sum over iteration precisions $2^k$ ($k = 0, \ldots, p-1$) gives a total dominated by the final-precision iteration, bounded by $4n_a^2$ Toffoli gates.

Squaring and multiplication produce $2n_a$-bit results.
Only the leading $n_a$ bits are retained, truncated via Bennett's uncomputation trick~\cite{bennett1973logical}: the full $2n_a$-bit product is computed, the leading $n_a$ bits are copied to a clean register via $n_a$ CNOT gates, and the full product is uncomputed, returning the $2n_a$-bit work register to $|0\rangle$.
This doubles Toffoli cost for each truncated operation but uses only $n_a$ persistent output qubits.

Denoting the truncated Toffoli costs as $\tilde{T}_{\text{sq}} = n_a(n_a + 1)$ (compute + uncompute) and $\tilde{T}_{\text{mult}} = 4n_a^2$, the forward-pass Toffoli count for the $1/r^{12}$ pipeline is
\begin{equation}
    \label{eq:toff_forward}
    T_{\text{fwd}} = \underbrace{4n_a^2}_{\text{REC}}
                   + \underbrace{3\,n_a(n_a + 1)}_{\text{3 squarings}}
                   + \underbrace{4n_a^2}_{\text{MULT}}
                   = 11n_a^2 + 3n_a.
\end{equation}
Including the reverse pass (uncomputation of all five operations),
\begin{equation}
    \label{eq:toff_total}
    T_{\text{arith}} = 2\,T_{\text{fwd}} = 22\,n_a^2 + 6\,n_a.
\end{equation}

\subsection{Ancilla overhead}

At the point of phase application (between Steps~5 and the uncompute of Step~5), all five intermediate registers must coexist:
\begin{equation}
    \label{eq:ancilla_arith}
    A_{\text{arith}} = 5\,n_a + A_{\text{work}},
\end{equation}
where $A_{\text{work}} = \mathcal{O}(n_a)$ reusable work qubits support carry propagation within each arithmetic operation.
Taking $A_{\text{work}} \approx n_a$, the total ancilla count is $A_{\text{arith}} \approx 6\,n_a$.

\subsection{Rotation costs}
After producing $|s_5\rangle = |1/r^{12}\rangle$ in $n_a$-bit fixed-point precision, the phase $e^{-i\alpha\tau \cdot s_5}$ is applied via Eq.~\ref{eq:phase_from_binary_arith} as $n_a$ single-qubit $R_z$ rotations.
No multi-controlled rotations are required: the value of $1/r^{12}$ is stored in the ancilla register, and each qubit $q_\ell^{(\text{anc})}$ directly controls a known rotation angle $2\alpha\tau\Delta_a 2^\ell$.

Following Chapter~\ref{ch:resources}, each Toffoli gate decomposes into $C_T^{(\text{Toff})} = 4$ $T$-gates at $T$-depth $d_T^{(\text{Toff})} = 4$ and total decomposed depth $d_{\text{tot}}^{(\text{Toff})} = 12$~\cite{gidney2018halving}.
Each $R_z$ rotation is synthesized at baseline cost $C_T^{(\text{rot})} = 23$ $T$-gates per rotation, with $T$-depth $d_T^{(\text{rot})} = 23$ and total decomposed depth $d_{\text{tot}}^{(\text{rot})} = 45$~\cite{ross2016optimal}.

The per-application $T$-gate count for the arithmetic scheme is
\begin{equation}
    \label{eq:T_arith}
    T^{(\text{arith})} = C_T^{(\text{Toff})} \cdot T_{\text{arith}} + C_T^{(\text{rot})} \cdot n_a
    = 4(22\,n_a^2 + 6\,n_a) + 23\,n_a
    = 88\,n_a^2 + 47\,n_a,
\end{equation}

For the UCR-decomposed multilinear scheme, the per-application $T$-gate count is
\begin{equation}
    \label{eq:T_ucr}
    T^{(\text{UCR})} = C_T^{(\text{rot})} \cdot (2^n - 1) = 23\,(2^n - 1),
\end{equation}
since the circuit contains $2^n - 1$ unconditional $R_z$ rotations (Section~\ref{sec:complexity}) and zero Toffoli gates.

The five arithmetic operations execute sequentially (each depending on the output of the previous step).
Using schoolbook multiplication with ripple-carry adders, the Toffoli depth sums to 

\begin{equation}
    N^{(\text{Toff})} \approx \underbrace{ 4n_a \cdot p}_{\text{Reciprocal}} + \underbrace{2n_a}_{\text{Squaring}} +  \underbrace{4n_a}_{\text{Multiplying}}
\end{equation}

The multiplier depth of $2n_a$ per computation (with ripple-carry) reflects $n_a$ sequential partial-product additions, each of depth $\mathcal{O}(1)$ for the AND gate and $\mathcal{O}(n_a)$ for carry propagation, applied across rows to yield an overall depth of $2n_a$.
Truncation doubles the depth (compute + uncompute).
Each Newton iteration has depth $\sim 4n_a$ (two multiplications at depth $2n_a$ each, plus an $\mathcal{O}(n_a)$-depth subtraction); however, only the final iteration operates at full $n_a$-bit precision, with earlier iterations at geometrically smaller bit-widths.
Total Newton depth is dominated by the last iteration: $\leq 4n_a \cdot p$ with $p = \lceil\log_2 n_a\rceil$.

The forward-pass Toffoli depth is
\begin{equation}
    \label{eq:depth_forward}
    D_{\text{fwd}}^{(\text{Toff})} = 4n_a p + 3 \times 2n_a + 4n_a = 4n_a(p + \tfrac{5}{2}),
\end{equation}
and the full compute--phase--uncompute depth is
\begin{equation}
    \label{eq:depth_total_arith}
    D_{\text{arith}}^{(\text{Toff})} = 2D_{\text{fwd}}^{(\text{Toff})} = 8n_a(p + \tfrac{5}{2}).
\end{equation}
The $n_a$ phase rotations contribute additional depth $n_a$ in $R_z$ layers, applied in a single layer.
Conservatively serializing the phase rotations, the full $T$-depth is
\begin{equation}
    \label{eq:Tdepth_arith}
    D_{\text{arith}}^{(T)} = d_T^{(\text{Toff})} \cdot D_{\text{arith}}^{(\text{Toff})} + C_T^{(\text{rot})} \cdot n_a
    = 32\,n_a(p + \tfrac{5}{2}) + 23\,n_a,
\end{equation}

For the UCR scheme, the $2^n - 1$ rotation layers are fully serial (each UCR rotation acts on a potentially overlapping set of qubits), giving $T$-depth
\begin{equation}
    \label{eq:Tdepth_ucr}
    D_{\text{UCR}}^{(T)} = C_T^{(\text{rot})} \cdot (2^n - 1) = 23\,(2^n - 1),
\end{equation}
and total decomposed depth
\begin{equation}
    \label{eq:Dtot_ucr}
    D_{\text{UCR}}^{(\text{tot})} = (2C_T^{(\text{rot})} - 1)(2^n - 1) + (2^n - 2) = 45\,(2^n - 1) + (2^n - 2),
\end{equation}
where the second term accounts for the $2^n - 2$ interleaved CNOT layers of the UCR cascade.

\subsection{Comparison for He--He}

Table~\ref{tab:arith_vs_ucr} collects the per-application resource comparison for the He--He scattering system, under the decomposition parameters $C_T^{(\text{rot})} = 23$ and $C_T^{(\text{Toff})} = 4$.

\begin{table}[ht]
\centering
\small
\caption{Per-application resource comparison between the UCR-decomposed multilinear circuit (Section~\ref{sec:constructing_pes}) and the quantum arithmetic pipeline for $1/r^{12}$ on $n = 13$ position qubits.
Arithmetic precision $n_a = 113$ bits (Eq.~\ref{eq:na_hehe}).
Newton iteration count $p = \lceil\log_2 113\rceil = 7$.
All counts are per half-Trotter-step application of the potential; the full symmetric Trotter step applies the potential twice.
}\label{tab:arith_vs_ucr}
\renewcommand{\arraystretch}{1.25}
\begin{tabular}{l r r c}
\hline\hline
Metric & UCR scheme & Arith.\ scheme & Ratio (Arith/UCR) \\
\hline
Position qubits & $13$ & $13$ & $1.0\times$ \\
Ancilla qubits & $0$ & $678$ & --- \\[3pt]
Toffoli gates & $0$ & $2.81\times10^5$ & --- \\
$R_z$ rotations & $8.19\times10^3$ & $113$ & $0.014\times$ \\
CNOT gates & $8.19\times10^3$ & $\mathcal{O}(n_a^2)$ & --- \\[3pt]
$T$-gate count & $1.88\times10^5$ & $1.13\times10^6$ & $6.0\times$ \\
$T$-depth & $1.88\times10^5$ & $2.76\times10^5$ & $1.5\times$ \\
Total decomposed depth & $3.77\times10^5$ & $8.37\times10^5$ & $2.2\times$ \\
\hline\hline
\end{tabular}
\end{table}

This relationship depends on the number of Newton iterations through the reciprocal; using carry-lookahead adders~\cite{draper2004logarithmic} reduces the depth of each multiplication from $\mathcal{O}(n_a)$ to $\mathcal{O}(\log n_a)$, yielding $D_{\text{arith}}^{(T)} \approx 32\,n_a\,\log_2 n_a\,(p + 5/2)/n_a = 32\,\log_2 n_a\,(p + 5/2)$ and a more favorable depth comparison at the expense of $\mathcal{O}(n_a)$ additional ancilla for the lookahead circuit.

\subsection{Crossover analysis}

A $T$-gate crossover occurs when $T^{(\text{arith})} = T^{(\text{UCR})}$:
\begin{equation}
    \label{eq:crossover_arith}
    88\,n_a^2 + 47\,n_a = 23\,(2^n - 1),
\end{equation}
For the Lennard--Jones potential with parameters comparable to the He--He system, the arithmetic precision scales as $n_a \approx 12n + c_0$, where $12n$ arises from $(1/\Delta)^{12} = (2^n/L)^{12}$ and $c_0 = \lceil\log_2(|\alpha\tau|/\varepsilon_\theta)\rceil$ is a system-dependent constant.
Taking $n_a \approx 9n$ as representative (including physical prefactors),
\begin{equation}
    88(9n)^2 \approx 7{,}128\,n^2,
\end{equation}
and the crossover $7{,}128\,n^2 = 23 \times 2^n$ gives $n^{\star} \approx 15$.
Hundreds of arithmetic qubits is highly restrictive, exacerbated in the multi-channel, multi-mode extension.
Future research may consider the application of multipliers for similar potential surfaces, though classical emulation is restricted in this regime.
In a fault-tolerant era where logical qubits are cheap, this optimization may resurface to trade circuit depth for register size.

\section{Discussion}
\label{sec:discussion}

Error-free statevector simulation produces a clean Trotterized evolution: phase shifts converge with simulation time, and the extracted deflection function, scattering amplitude, and differential and integral cross sections agree with the time-independent classical benchmark (Figures~\ref{fig:Shfits+Deflection function} and~\ref{fig:Scattering amplitude + Diff xsec}).
By considering the implications of these analytically-derived circuit, practical limitations become clear, positioning this construction as a building block within REQWIEM for nonadiabatic dynamics.

\subsection{Complexity Analysis}
\label{sec:complexity}

The inverse polynomial circuit (Eq.~\ref{eq:controlled_rz_product}) requires $\mathcal{O}(2^n - 1)$ multi-controlled $R_z$ gates prior to decomposition into native two-qubit gates, reducing to $\mathcal{O}(2^{n-1})$ layers after CNOT cascade decomposition.
The function $f(r) =1/r$ evaluated on a linearly spaced binary grid of $N = 2^n$ points has no known symmetry in the multilinear basis, leaving all $2^n$ coefficients $a_S$ of Eq.~\ref{eq:mobius} to be nonzero and distinct, requiring $\mathcal{O}(2^n)$ independent rotation angles, specifying $\mathcal{O}(2^n)$ independent function values.
Whether ancilla-mediated non-arithmetic constructions can circumvent this bound for non-diagonal intermediate operations remains open \cite{o2014analysis}.

Exponential scaling is in simulation box size, not in physical dimensionality of the problem.
Since the function $V(r) = 1/r^{\alpha}$ would replace a harmonic function in the REQWIEM algorithm, multidimensional and multi-channel extensions retain polynomial complexity. 
Following Chapter~\ref{ch:resources}, the full gate complexity of a scattering problem becomes $\mathcal{O}(kM^2d^22^n)$ and depth complexity becomes $\mathcal{O}(kM^2d2^n)$, where error scaling is absorbed in the Trotter step $k$.

\subsection{Comparison to Oracle-Based Methods}

Quantum signal processing (QSP), qubitization, and the Harrow--Hassidim--Lloyd (HHL) algorithm still have $\mathcal{O}(2^n)$ gate complexity, while retaining an optimal query complexity for a single-channel system, achieved by block-encoding the target Hamiltonian through an oracle and approximating matrix functions via polynomial transformations \cite{harrow2009quantum, low2017optimal, gilyen2019quantum}.

For a single effective potential, the only limitation is that a block-encoding decomposition requires $\mathcal{O}(2^n)$ terms, and the Hamiltonian offers no shortcut through structured access patterns.
Once encoded, the simulation proceeds adiabatically, and query optimality gives absolute advantage for long times over product formulas.
Shifting the potential to center at an estimated penetration depth $\propto 1/\sqrt{E}$ (for collision energy $E$) reduces the spectral range of $V_{\text{eff}}$, allowing high-accuracy approximation.

Aside from these techniques, this adiabatic simulation is efficiently handled in the interaction picture \cite{georges2025quantum}. 
The purpose of this chapter is not to gain a depth advantage in the single-potential case, but to build a diagonal potential to include in multi-mode, nonadiabatic problems, where oracle cost grows exponentially in the number of modes (Chapter~\ref{ch:Introduction}).
In this limit, the overhead from the interaction picture scales poorly with the number of orbitals, and without efficient handling of nonadiabatic coupling, the problem remains open for architectures like REQWIEM.
A full gate-count comparison between REQWIEM and three QSP constructions, including the spectral-range penalty that drives $\Tgate$-gate costs in coupled-channel settings, is given in Chapter~\ref{ch:comparison}.

\subsection{Periodic Boundary Effects}
\label{sec:periodic_boundary}

Periodicity is a property of the binary-encoded position basis shared by all grid-based quantum simulations on qubit registers.
Absorbing boundary conditions, which are standard in classical grid-based propagation, are not straightforwardly implementable in unitary quantum evolution without introducing ancilla-mediated dissipation or post-selection~\cite{sunnergren1998radiative, jin2023quantum}.
This direction remains valuable for future work, likely in a post-trotter implementation of non-hermitian Hamiltonians or in the Schr\"odingerization of the complex part by dilating coordinate space~\cite{jin2023quantum}.
Regardless, a simulation must be designed such that observables can be extracted before boundary effects contaminate the wavefunction.

\section{Summary}
\label{sec:scattering_summary}

This chapter presents the first real-time evolution of an atomic scattering event without the classical precomputation of eigenstates using quantum circuits.
An initialized Gaussian wavepacket evolving in a hard-core potential describes the collision of two neutral ground-state helium atoms, propagated under a second-order Trotterized Hamiltonian, with the effective potential implemented as a closed-form analytical gate composition on a binary-encoded qubit basis.

Exact quantum circuits for potentials of the form $V(r) = \sum_i \alpha_i r^{-\gamma_i}$ with $\gamma_i \in \mathbb{Z}^+$ can faithfully and accurately reproduce dynamics, requiring $\mathcal{O}(2^n - 1)$ multi-controlled $R_z$ gates whose count does not depend on polynomial power $\gamma$ and under summation of multiple terms.
Phase shifts extracted via spatial FFT with reference subtraction recover the deflection function, scattering amplitude, and differential and integral cross sections in agreement with independent classical simulations (Figures~\ref{fig:Shfits+Deflection function} and~\ref{fig:Scattering amplitude + Diff xsec}).
The Hilbert--Schmidt distance between the binary circuit and the BRGC construction of Chang et al.\ \cite{chang2022improving} vanishes to numerical precision, providing operator-level validation free of classical approximation (Figure~\ref{fig:BRGC-compare}).

Complexity analysis establishes $\mathcal{O}(2^n)$ as an upper bound for exact implementation, with exponential scaling in simulation box size rather than dimensionality.
The exact circuit is oracle-free and, for the low-energy He--He system studied here, serves primarily as a ground-truth benchmark for variational compression, $T$-gate budgeting, and machine-learning-guided circuit optimization if direct implementation proves impractical in the long term.

The single-surface scattering problem corresponds to the $M=1$ limit of the REQWIEM on-diagonal block.
For $M > 1$ electronic channels, each surface $V_{ii}(r)$ carries its own interaction parameters and partial-wave quantum numbers; the multilinear coefficients and repeated-squaring algorithm generate rotation angles for each surface independently, assembled into UCR angle vectors encoding all $M$ surfaces into uniformly controlled rotations with $m = \lceil\log_2 M\rceil$ channel register qubits as controls.
Gate cost per rotation site increases from one unconditional $R_z$ to $M$ rotations interleaved with $M-1$ CNOTs, but the number of rotation sites and the gate count invariance under polynomial power $m$ remain unchanged, and precomputed coefficient libraries can be reused across channels by rescaling with surface-dependent prefactors.
The total on-diagonal depth is $\mathcal{O}(M \cdot 2^n)$ per Trotter step, reducible to $\mathcal{O}(Mn^2)$ through efficient approximation of each surface; efficient approximability is guaranteed by syllogism from the Chang et al.~\cite{chang2022improving} proof for the BRGC encoding, though impact on observables requires future work.
Off-diagonal depth is $\mathcal{O}(M^2 P)$ for piecewise-linear couplings with $P$ pieces, rising to $\mathcal{O}(M^2 \cdot 2^n)$ for arbitrary smooth coupling functions.
Extension to multi-mode scattering ($d > 1$) places each spatial or internal mode on its own $n$-qubit register, with the inverse polynomial circuit providing the intra-mode potential, inter-mode bilinear terms handled by Tier~3 of the on-diagonal protocol, and fan-out protocols adding $(M-1)(d-1)$ ancilla qubits for parallel execution.

The inverse polynomial circuit addresses a specific class of potentials, but molecular dynamics extends well beyond $1/r^\gamma$.
Anharmonic corrections, Morse-like repulsive walls, excited-state potential energy surfaces with arbitrary curvature, and inseparable multivariable interactions all require smooth but non-polynomial functions of the position coordinate.
The next chapter develops a general multilinear expansion for arbitrary smooth functions $f(r)$ on the binary qubit basis, exhibiting UCR angle vector composition to allow multiple potential terms to be combined into a single circuit at zero gate overhead.
In this way, any on-diagonal potential surface and any off-diagonal coupling expressible as functions defined by finite-difference, whether polynomial, inverse polynomial, exponential, or tabulated from ab initio calculation, can be encoded as analytic, closed-form circuits, composed with multi-channel and multi-mode infrastructure.

\chapter{Smooth Potentials}\label{ch:smooth_potentials}

\section{Introduction}\label{sec:smooth_intro}

As constructed in Chapter~\ref{ch:model_ext}, the REQWIEM architecture decomposes the diabatic Hamiltonian into kinetic, on-diagonal potential, and off-diagonal coupling terms, implementing each as parameterized quantum circuits composed of uniformly controlled rotations (UCRs), CNOT cascades, and Clifford focusing unitaries.
Numerical validation in Chapters~\ref{ch:model_ext} and~\ref{ch:scattering} demonstrates reproduction of nonadiabatic population transfer, Berry phase accumulation, absorption spectra, and elastic scattering phase shifts for systems whose diabatic potential energy surfaces are polynomial or inverse polynomial functions of the nuclear coordinates.
These functional forms are popular, covering the standard second-order vibronic Hamiltonian of K\"{o}ppel, Domcke, and Cederbaum~\cite{kouppel1984multimode, cederbaum1981multimode} and a small set of hard-core interactions, but are highly restrictive, being approximations themselves of electronic structures.

Molecular potential energy surfaces routinely require functional forms beyond finite-degree polynomials.
Bond-dissociative modes are often fit to Morse oscillators $V(x) = D[1 - e^{-A(x-x_0)}]^2$, whose exponential character encodes finite dissociation energy and anharmonic level spacing~\cite{morse1929diatomic, lang2026quantum}.
Short-range repulsive interactions follow exponential or inverse-power laws; molecule--surface coupling profiles take $\tanh$-type or Gaussian forms; and proton transfer coordinates in excited-state intramolecular proton transfer (ESIPT) systems are governed by quartic or higher-order double-well potentials reflecting tautomer bistability~\cite{zhao2012excited, sedgwick2018excited}.
Ab initio electronic structure calculations produce pointwise potential energy data that are fitted to these analytic forms for computational convenience and often result in a potential energy surface that has no closed polynomial form. 
As an example, expanding to second order around an equilibrium geometry removes the possibility of dissociation, with greater distortion of higher energy states by altering barrier heights and shifting transition-state geometries.

This chapter extends the REQWIEM circuit architecture from the polynomial regime to arbitrary smooth potentials in an orthogonal set of coordinates following the condition that the kinetic operator $T(\mathbf{p}) = \sum_r p_r^2/2\mu_r$ depends only on momentum, admitting alternation between position and momentum space via quantum Fourier transforms.
Internal coordinates, Jacobi coordinates, and coordinate substitutions are set aside, to be considered in Chapter~\ref{ch:Application}.

The extension rests on a multilinear expansion that represents an arbitrary real-valued function $f(x)$ on the $2^n$-point computational grid.
Any function of binary-encoded position admits a unique decomposition into $2^n$ multilinear monomials $\prod_{j \in S} q_j$, with coefficients determined by an inclusion--exclusion formula evaluated at the grid points.
This expansion produces a set of commuting diagonal phase operators that are structurally identical to the binary-decomposed polynomial terms of Chapter~\ref{ch:model_ext}.
The inverse polynomial case $f(x) = 1/x^k$ was derived and demonstrated for cold atomic scattering in Chapter~\ref{ch:scattering}; the present chapter generalizes to arbitrary $f$.

This chapter establishes that the multilinear expansion can implement an arbitrary $f(x)$ as $2^n$ diagonal phase operators, implemented as a CNOT cascade with parity-based $R_z$ (or UCR$_Z^{(m)}$) rotations. 
The UCR composition absorbs surface-dependent coefficients $a_S^{(s)}$ into the angle vector, providing a factor-of-$M$ composition advantage over QROM methods for non-separable analytical potentials.
The depth advantage from conflict graph-based parallelism (Chapter~\ref{ch:resources}) holds for all functional forms and amplifies with the introduction of analytical potentials. 
This advantage is preserved under the additive decomposition of Hamiltonians containing both polynomial and analytical terms.

Section~\ref{sec:grid_resolution} establishes grid resolution requirements for nonlinear potential energy surfaces.
Section~\ref{sec:multilinear} derives the multilinear M\"{o}bius expansion and its Gray-code UCR implementation.
Section~\ref{sec:ucr_composition} shows the composition with UCR multiplexing and derives generalized gate counts and depths.
Section~\ref{sec:numerical_validation} provides numerical validation on three physically motivated systems.
Section~\ref{sec:smooth_summary} summarizes the scope, limitations, and conclusions.

\section{Grid Resolution Requirements for Nonlinear Potential Energy Surfaces}
\label{sec:grid_resolution}

Chapter~\ref{ch:wavepacket_prop} establishes that $n = 8$--$9$ qubits per vibrational mode typically suffices for converged nonadiabatic dynamics with quantum harmonic oscillators.
Resource estimates in Chapter~\ref{ch:resources} use a conservative $n=10$ to handle more energetic systems as a conservative estimate.
Nonlinear, anharmonic potential surfaces impose more stringent requirements, dictated by spectral and spatial structures.

\subsection{Morse Potential Convergence}
\label{sec:morse_resolution}

The Morse potential,
\begin{equation}
    V_{\mathrm{Morse}}(x) = D_e \left(1 - e^{-\alpha(x - x_0)}\right)^2,
    \label{eq:morse}
\end{equation}
supports a finite number of bound states $n_{\mathrm{max}} = \lfloor 2D_e/\omega_e - 1/2 \rfloor$, where $\omega_e = \alpha\sqrt{2D_e/\mu}$, with exact eigenvalues
\begin{equation}
    E_\nu = \omega_e\!\left(\nu + \tfrac{1}{2}\right)
    - \frac{\omega_e^2}{4D_e}\!\left(\nu + \tfrac{1}{2}\right)^2,
    \qquad \nu = 0, 1, \ldots, n_{\mathrm{max}}.
    \label{eq:morse_exact_eigenvalues}
\end{equation}
These exact eigenvalues provide a benchmark for grid convergence.
Classically calculating these eigenvalues on a spatial grid numerically (Appendix~\ref{app:grid_convergence}) shows that lower and mid-range bound states reach spectroscopic accuracy ($< 1$~cm$^{-1}$) by $n = 6$--$8$ qubits ($64--256$ grid points) for systems representative of H--metal, N--O, and C--O interactions.
The highest bound states exhibit a persistent error that plateaus with increasing $n$ at fixed box size $L$, reflecting boundary truncation rather than grid-spacing limitation, and is resolved by increasing $L$ and commensurately adding more qubits. 

An aliasing diagnostic $\|V\|_\infty / T_{\max}$ gives the ratio of the maximum potential value on the grid to the maximum representable kinetic energy.
If total control over $V$ is assumed, $\|V\|_\infty$ can be truncated to a characteristic energy as the spectral norm of $V$, and $T_{\max}$ is the kinetic energy associated to the spatial Nyquist limit (Chapter~\ref{ch:wavepacket_prop}). 
When this ratio exceeds unity, the potential generates local momenta beyond the grid's Nyquist limit, providing a heuristic for potential energy surface regularization.
For H--metal and H$_2$--surface systems, this ratio drops below unity only at $n = 12$; for stiffer C--O and N--O stretches, $n \geq 14$--$15$ is required.
Unlike inverse polynomial functions, quantum arithmetic cannot be advantaged for these larger register sizes for some real-valued $V(x) =f(x)$.

A scattering simulation requires $L \geq 200\,\text{a.u.}$ to accommodate the asymptotic wavepacket while maintaining grid spacing $\Delta x \lesssim 0.02\,\text{a.u.}$ to resolve the potential well, giving
\begin{equation}\label{eq:n_scattering}
    n_{\mathrm{scat}} = \left\lceil \log_2\!\left(\frac{L_{\mathrm{scat}}}{\Delta x_{\mathrm{req}}}\right) \right\rceil,
\end{equation}
which yields $n_{\mathrm{scat}} = 11$--$13$ depending on potential stiffness (Appendix~\ref{app:grid_convergence}).
During scattering, the Trotter step applies $e^{-i\tau V(x)}$ at every grid point; aliased phase kicks at the repulsive wall produce irreversible corruption compounding at each step, disturbing spectral, phase, or population observables.
In nonadiabatic scattering, population transfer distributes amplitude across vibrational levels up to the dissociation threshold, and undersampled upper levels produce incorrect branching ratios.
Full spectral completeness reinforces $n \geq 12$ as a lower bound, used as the benchmark value in this chapter's computational overhead comparisons.

\subsection{Register Size Summary}
\label{sec:register_summary}

Table~\ref{tab:echar_summary} summarizes the minimum register size $n_{\min}$ across four progressively demanding regimes.
Characteristic energy analysis underlying these estimates is developed in Chapter~\ref{ch:wavepacket_prop}; detailed convergence tables and the resolution tier analysis are given in Appendix~\ref{app:grid_convergence}.

\begin{table}[t]
\centering
\caption{Minimum register sizes $n_{\min}$ per vibrational mode for four regimes of nonlinear potential energy surfaces.
All values assume representative reactive-mode parameters; system-specific parameterizations may shift entries by $\pm 1$ qubit.}
\label{tab:echar_summary}
\begin{tabular}{l l c}
\hline\hline
Regime & Primary $E_{\text{char}}$ driver & $n_{\min}$ \\
\hline
Harmonic (reference) & Highest populated level & 8--9 \\[3pt]
Morse bound state & Repulsive wall at $Q_{\min}$ & 10--11 \\[3pt]
Coupled bound state & $\|\mathbf{V}\|_{\text{op}} + \Delta E_{\text{rep.}}$ & 11--12 \\[3pt]
Morse scattering & $\|V\|_\infty + E_{\text{kin}}$; extended $L$ & 12--13 \\[3pt]
Coupled multi-surface scattering & All of above & 12--14 \\
\hline\hline
\end{tabular}
\end{table}

The $n = 12$ threshold for reactive modes is the working value adopted throughout this chapter.
Convergence analysis shows that systems with a higher characteristic energy (strongly excited, deeper potential wells) may require up to $n = 15$ for spectroscopic accuracy (Appendix~\ref{app:grid_convergence}).
These requirements apply equally to all grid-based first-quantized architectures encoding the potential on a $2^n$-point computational grid.

\section{Multilinear Expansion of Arbitrary Diagonal Functions}
\label{sec:multilinear}

This section establishes the extension of the REQWIEM circuit architecture to arbitrary real-valued functions on the computational grid.
A full diagonal unitary $e^{-i\tau f(x)}$ is compiled via a Gray-code uniformly controlled rotation (UCR) into exactly $2^n$ rotations and $2^n - 1$ CNOT gates, achieving the information-theoretic optimum in the chosen representation~\cite{mottonen12006decompositions}.

\subsection{Multilinear Expansion}\label{sec:multilinear_theorem}

Let $f\colon \{0, 1, \ldots, 2^n - 1\} \to \mathbb{R}$ be an arbitrary real-valued function defined on the $n$-qubit computational basis states of a single vibrational mode register.
The integer argument $x = \sum_{\ell=0}^{n-1} 2^\ell q_\ell$ is encoded in the qubit register $\{q_\ell\}_{\ell=0}^{n-1}$ with $q_\ell \in \{0,1\}$.
Any function of $x$ admits a unique multilinear expansion
\begin{equation}\label{eq:multilinear_expansion}
    f(x) = \sum_{S \subseteq \{0, 1, \ldots, n-1\}} a_S \prod_{j \in S} q_j\,,
\end{equation}
where the sum runs over all $2^n$ subsets of the qubit index set, the empty product is unity ($\prod_{j \in \emptyset} q_j = 1$), and the $a_S \in \mathbb{R}$ are the multilinear coefficients.

The set of multilinear monomials $\{\prod_{j \in S} q_j\}_{S \subseteq \{0,\ldots,n-1\}}$ forms a basis for the vector space of all real-valued functions on $\{0,1\}^n$~\cite{o2014analysis}.
Binary encoding bijects $2^n$ grid points and vertices of the Boolean hypercube $\{0,1\}^n$, giving a unique, closed-form representation on the finite interval. 
The functions considered here are smooth, but the expansion~\ref{eq:multilinear_expansion} does not require smoothness. 

\subsection{Coefficients via Inclusion--Exclusion}
\label{sec:mobius_coefficients}

Coefficients $a_S$ are computed by M\"{o}bius inversion on the Boolean lattice~\cite{o2014analysis}.
Defining the finite difference operator in qubit $j$ as $\Delta_j f(x) = f(x + 2^j) - f(x)$, and the multi-index finite difference for subset $S$ as $\Delta_S f(x) = \prod_{j \in S} \Delta_j f(x)$, where the operators commute ($\Delta_j \Delta_k = \Delta_k \Delta_j$), coefficients are given by the inclusion--exclusion formula:
\begin{equation}\label{eq:mobius_coefficients}
    {a_S = (\Delta_S f)(0) = \sum_{T \subseteq S} (-1)^{|S| - |T|}\, f\!\left(\sum_{j \in T} 2^j\right)}
\end{equation}
This formula evaluates $f$ at $2^{|S|}$ grid points whose binary representations are vertices of the $|S|$-dimensional subcube indexed by $S$.
Full derivations of the finite difference operators and their multi-variable extensions are given in Appendix~\ref{app:circuits}.

For small subsets, the formula reduces to:
\begin{align}
    a_\emptyset &= f(0)\,, \label{eq:a_empty}\\[4pt]
    a_{\{j\}} &= f(2^j) - f(0)\,, \label{eq:a_single}\\[4pt]
    a_{\{j,k\}} &= f(2^j + 2^k) - f(2^j) - f(2^k) + f(0)\,, \label{eq:a_pair}\\[4pt]
    a_{\{j,k,\ell\}} &= f(2^j+2^k+2^\ell) - f(2^j+2^k) - f(2^j+2^\ell) - f(2^k+2^\ell) \notag\\
    &\quad + f(2^j) + f(2^k) + f(2^\ell) - f(0)\,. \label{eq:a_triple}
\end{align}
The empty-set coefficient $a_\emptyset = f(0)$ is the function value at the origin, single-element coefficients are finite differences, and the pairwise coefficient $a_{\{j,k\}}$ are discrete analogs of the second mixed partial derivative.

For a linear function $f(x) = c_0 + c_1 x$, all $a_S$ with $|S| \geq 2$ vanish.
For a quadratic polynomial $f(x) = c_0 + c_1 x + c_2 x^2$, all $a_S$ with $|S| \geq 3$ vanish, recovering the Tier~1--2 structure of Chapter~\ref{ch:model_ext}.
For a general smooth function, coefficients at all orders $|S| = 0, 1, \ldots, n$ are generically nonzero, requiring the full $2^n$-term expansion.

\subsection{Diagonal Operator Form}
\label{sec:diagonal_operator}

As a diagonal operator on the vibrational register, the expansion~\ref{eq:multilinear_expansion} takes the form
\begin{equation}\label{eq:f_diagonal}
    {f}(x) = \sum_{S} a_S \prod_{j \in S} \frac{I - {\sigma}_j^{(z)}}{2}
    = \sum_{S} \frac{a_S}{2^{|S|}} \prod_{j \in S} (I - {\sigma}_j^{(z)})\,,
\end{equation}
where each product is a projector onto the subspace with $q_j = 1$ for all $j \in S$.
Every term is diagonal in the computational basis and commutes with every other term, with diagonal entries $\langle x | {f} | x \rangle = f(x)$ for $x = 0, \ldots, 2^n - 1$.
The time-evolution operator $e^{-i\tau {f}(x)}$ is therefore a diagonal unitary with $2^n$ independent phase entries, establishing $2^n$ as the information-theoretic lower bound on the number of rotation gates required for exact implementation.

\subsection{Gray-Code UCR Implementation}
\label{sec:graycode_ucr}

Implementing each multilinear term independently incurs $\mathcal{O}(n \cdot 2^n)$ CNOT gates (Appendix~\ref{app:binary_decomp}). 
Instead, the full diagonal unitary $e^{-i\tau {f}(x)}$ is compiled as a single uniformly controlled $Z$-rotation UCR$_Z^{(n)}$ on the $n$-qubit vibrational register, achieving the information-theoretic optimum of $\mathcal{O}(2^n)$ gates.

\subsubsection{Walsh--Hadamard Transform}
\label{subsubsec:walsh_compilation}

Each of the $2^n$ diagonal phases $\{\tau\, f(x)\}_{x=0}^{2^n-1}$ are converted to $2^n$ circuit rotation angles $\{\phi_k\}_{k=0}^{2^n-1}$ by the Walsh--Hadamard transform:
\begin{equation}\label{eq:walsh_transform_position}
    \phi_k = \frac{1}{2^n} \sum_{x=0}^{2^n-1} (-1)^{\langle k, x \rangle}\, \tau\, f(x)\,,
\end{equation}
where $\langle k, x \rangle = \sum_{\ell=0}^{n-1} k_\ell\, x_\ell$ is the bitwise inner product, and the $1/2^n$ prefactor follows a M\"ott\"onen decomposition normalization~\cite{mottonen12006decompositions}.
This transform is the same Walsh--Hadamard transform used for the channel-register UCR (Appendix~\ref{app:binary_decomp}), applied here to the position register.

\subsubsection{Gray-Code Circuit Structure}
\label{subsubsec:graycode_circuit}

The UCR$_Z^{(n)}$ is implemented by traversing the $2^n$ computational basis states in Gray-code order $\{g_0, g_1, \ldots, g_{2^n-1}\}$, where consecutive Gray-code words differ in exactly one bit position.
The circuit consists of $2^n$ rotation gates interleaved with $2^n - 1$ CNOT gates:
\begin{enumerate}
    \item \textbf{Initialize.} Begin with all qubits in their input state.
    \item \textbf{For $k = 0, 1, \ldots, 2^n - 1$:}
    \begin{enumerate}
        \item If $k > 0$: apply $\mathrm{CNOT}_{c(k) \to t(k)}$, where the bit that flips at step $k$ is $\ell(k) = \nu_2(k)$ (the 2-adic valuation of $k$).
        \item Apply $R_z(\phi_{\pi(k)})$ on the target qubit, where $\pi$ maps circuit step $k$ to the Gray-code ordered Walsh-coefficient index.
    \end{enumerate}
\end{enumerate}
Consecutive Gray-code words differ in exactly one bit, so a single CNOT suffices to advance the parity register between basis states, avoiding redundant build-and-teardown of CNOT cascades.

\subsubsection{Gate Count}
\label{subsubsec:graycode_gate_count}

The Gray-code UCR$_Z^{(n)}$ circuit requires exactly:
\begin{equation}\label{eq:graycode_gates}
    N_{\mathrm{CNOT}} = 2^n - 1\,, \qquad N_{R_z} = 2^n\,.
\end{equation}
The total gate count is $2^{n+1} - 1$, and the circuit depth is $2^{n+1} - 1$ (each gate is sequential in the single-mode case).
Both counts match the lower bound of $2^n$ parameters for an arbitrary diagonal unitary on $n$ qubits, up to an additive constant~\cite{shende2005synthesis, mottonen2004quantum}.

\subsubsection{Channel-Register Connection and Tier Recovery}
\label{subsubsec:channel_and_tiers}

The Gray-code compilation of the position-register diagonal is structurally identical to the M\"ott\"onen decomposition used for the channel-register UCR$_Z^{(m)}$ in the multi-surface setting (Appendix~\ref{app:binary_decomp}).
The two UCR levels compose hierarchically when both registers are present, replacing the single $R_z(\phi_k)$ with a channel-register UCR$_Z^{(m)}$ at each position-register site with $M$ surface-dependent angles.

The multilinear expansion~\ref{eq:multilinear_expansion} together with the Gray-code UCR extends the tier structure of Chapter~\ref{ch:model_ext} to arbitrary order:

\begin{center}
\renewcommand{\arraystretch}{1.3}
\begin{tabular}{l l l}
\hline\hline
Subset order $|S|$ & Chapter~\ref{ch:model_ext} & Present chapter \\
\hline
$|S| = 0$ & Global phase (constant shift) & $a_\emptyset = f(0)$ \\
$|S| = 1$ & Tier~1: single-qubit $R_z$ corrections & Linear finite diff. \\
$|S| = 2$ & Tier~2: intra-mode UCR$_Z^{(m)}$ & Quadratic finite diff. \\
$|S| \geq 3$ & (Not present for quadratic potentials) & Higher-order multilinear \\
\hline\hline
\end{tabular}
\end{center}

For a quadratic polynomial, all coefficients with $|S| \geq 3$ vanish, and the multilinear expansion reduces exactly to Tiers~1--2.
For functions requiring a full expansion, Gray-code UCR is strictly superior to the per-term cascade approach.
A hybrid strategy is natural: use the Tier~1--2 structure for polynomial modes, and the Gray-code UCR$_Z^{(n)}$ for modes carrying analytical or tabulated functions.

\subsection{Classical Preprocessing}
\label{sec:classical_preprocessing}

Coefficients~\ref{eq:mobius_coefficients} are precomputed classically via the fast M\"{o}bius transform in $\mathcal{O}(n \cdot 2^n)$ operations, or $\mathcal{O}(2^{|S|})$ per individual coefficient if computed selectively.
For inverse polynomials $f(x) = 1/x^\gamma$ and rational functions with integer or rational coefficients, grid values are rational numbers and all coefficients can be computed in exact rational arithmetic, eliminating floating-point rounding entirely (Chapter~\ref{ch:scattering}) and significantly reducing rotation calculation overhead.

Transcendental functions (Morse, exponential, Gaussian, $\tanh$) can use double-precision arithmetic ($\varepsilon_{\mathrm{mach}} \approx 10^{-16}$).
The inclusion--exclusion sum~\ref{eq:mobius_coefficients} involves $2^{|S|}$ terms with alternating signs; the worst-case coefficient error is bounded by
\begin{equation}\label{eq:coeff_error}
    \varepsilon_{a_S} \lesssim 2^{|S|} \times \varepsilon_{\mathrm{mach}} \times \max_{T \subseteq S} |f(\textstyle\sum_{j \in T} 2^j)|\,,
\end{equation}
giving $\varepsilon_{a_S} \lesssim 2^n \times 10^{-16}$ in the worst case ($|S| = n$).
For $n = 12$ this is $\sim 4 \times 10^{-13}$; for $n = 15$, $\sim 3 \times 10^{-12}$.
Where this approaches the rotation synthesis precision $\varepsilon_\theta \sim 10^{-10}$--$10^{-14}$, extended-precision arithmetic is the fallback (e.g., Python \texttt{Decimal} or C \texttt{get$\_$prec()}).
These functions are treated no differently than an ab-initio surface, with each grid point evaluated individually, instantiating a finite difference.

\section{Composition with UCR Multiplexing and Generalized Gate Counts}
\label{sec:ucr_composition}

The multilinear expansion composes with the channel-register multiplexing protocols of Chapter~\ref{ch:model_ext}, preserving the full on-diagonal and off-diagonal circuit architecture without modification.
This section re-presents the composition mechanism and derives generalized site counts, rotation counts, CNOT counts, and circuit depths for a Hamiltonian containing both polynomial and analytical potential terms.
Neglecting bilinear terms, depth of the generalized circuit is independent of the number of modes $d$, preserving the fan-out parallelism established in the polynomial regime.

\subsection{On-Diagonal Multi-Surface Composition}
\label{sec:ondiag_composition}

Consider an analytical potential $f_\kappa^{(s)}({x}_\kappa)$ in the diagonal electronic Hamiltonian on surface $s$ for vibrational mode $\kappa$.
Surface dependence enters through the potential parameters: different equilibrium positions $r_\kappa^{(s)}$, well depths $D_\kappa^{(s)}$, and stiffness constants $A_\kappa^{(s)}$ on each electronic state.
The multilinear expansion on surface $s$ is
\begin{equation}\label{eq:surface_dependent_expansion}
    f_\kappa^{(s)}(x) = \sum_{S \subseteq \{0, \ldots, n-1\}} a_S^{(s)}\, \prod_{j \in S} q_j^{(\kappa)}\,,
\end{equation}
where $a_S^{(s)}$ is computed from $f_\kappa^{(s)}$ via Eq.~\ref{eq:mobius_coefficients}.
Different surfaces produce different coefficient vectors $\mathbf{a}_S = (a_S^{(0)}, \ldots, a_S^{(M-1)})$ at each subset $S$, but the set of subsets is identical across surfaces.

The on-diagonal potential for mode $\kappa$ across all $M$ surfaces is
\begin{equation}\label{eq:ondiag_analytical}
    {V}_\kappa^{(\mathrm{an})} = \sum_{s=0}^{M-1} \ket{s}\!\bra{s} \otimes {f}_\kappa^{(s)}
    = \sum_{S} \left(\sum_{s=0}^{M-1} a_S^{(s)}\, \ket{s}\!\bra{s}\right) \otimes \prod_{j \in S} \frac{I - {\sigma}_j^{(z)}}{2}\,.
\end{equation}
At each subset site $S$ in the Gray-code walk, surface-dependent information enters exclusively through the $M$-dimensional angle vector
\begin{equation}\label{eq:angle_vector}
    \boldsymbol{\theta}_S = \bigl(\tau\, a_S^{(0)},\; \tau\, a_S^{(1)},\; \ldots,\; \tau\, a_S^{(M-1)}\bigr)\,,
\end{equation}
processed by the Walsh--Hadamard transform $\boldsymbol{\phi} = H_M \, \boldsymbol{\theta}_S$ to produce circuit angles $\{\phi_g\}_{g=0}^{M-1}$.
The transform is norm-preserving ($\|\boldsymbol{\phi}\|_2 = \|\boldsymbol{\theta}_S\|_2$). 
Resulting $M$ rotations are interleaved with $M - 1$ Gray-code CNOT transitions on the channel register, independent of the vibrational-register structure at the site.

Consequently, the UCR gate at each site $S$ is structurally identical for every functional form.
The quantum circuit is agnostic to the functional form, which enters through the classical computation of $a_S^{(s)}$

To make this concrete, consider a single site $S$ for a Morse potential on two surfaces ($M = 2$, $m = 1$).
The Walsh--Hadamard transform gives circuit angles
\begin{equation}\label{eq:M2_example}
    \phi_0 = \frac{\tau}{2}\bigl(a_S^{(0)} + a_S^{(1)}\bigr)\,, \qquad
    \phi_1 = \frac{\tau}{2}\bigl(a_S^{(0)} - a_S^{(1)}\bigr)\,.
\end{equation}
The circuit at this site consists of a CNOT cascade (depth $|S|-1$) $\to$ $R_z(\phi_0)$ $\to$ CNOT on channel qubit $\to$ $R_z(\phi_1)$ $\to$ CNOT on channel qubit $\to$ reverse cascade.
If two surfaces have identical parameters, then $\phi_1 = 0$ and the UCR reduces to a single $R_z$; if the parameters differ, the full UCR is required, but the gate count is unchanged.

\subsection{Off-Diagonal Composition}
\label{sec:offdiag_composition}

Off-diagonal coupling functions $c_{ij}^{(\mathrm{an})}(x_\kappa)$ such as Gaussian or $\tanh$-type couplings admit the multilinear expansion~\ref{eq:multilinear_expansion}.
Within each XOR-class fragment $\zeta$, the focusing Clifford $U_\zeta$ of Chapter~\ref{ch:model_ext} reduces the off-diagonal interaction to a diagonal form, and the diagonalized coupling function is expanded as
\begin{equation}\label{eq:offdiag_analytical}
    {c}_\kappa^{(\zeta)}(x) = \sum_{S} a_S^{(\zeta)}\, \prod_{j \in S} q_j^{(\kappa)}\,,
\end{equation}
with a UCR$_Z^{(m-1)}$ applying $M/2$ rotations at each site $S$, controlled on the $(m-1)$-qubit reduced channel register.
XOR-class fragmentation, focusing Clifford, and symmetric Trotter product $\mathcal{V}_{\mathrm{od}}(\tau)$ are all agnostic to the functional form of the coupling: they depend only on the electronic-state index structure, not on the nuclear-coordinate dependence.
The fragmentation produces $2(M-1)$ fragments for $M$ electronic states regardless of whether the coupling function is linear, Gaussian, exponential, or tabulated (Appendix~\ref{app:offdiag}).

\subsection{Tier Structure and Site Counts}
\label{sec:gen_sites}

The tier structure preserves the existing tiers of Chapter~\ref{ch:model_ext}, adding a new tier (Tier~2$'$) for higher-order multilinear terms.

\subsubsection{Tier~1: reference potential and single-qubit corrections (unchanged)}
The polynomial constant ($|S| = 0$) and linear ($|S| = 1$) terms are unchanged from Chapter~\ref{ch:model_ext}.
Analytical functions contribute at these orders, but their contributions are absorbed into existing Tier~1 rotations by angle addition:
\begin{align}
    \theta_{\emptyset}^{(s)} &\leftarrow \theta_{\emptyset}^{(s)} + \tau\, a_\emptyset^{(s)}\,, \label{eq:tier1_absorption_const}\\
    \theta_{\{j\}}^{(s)} &\leftarrow \theta_{\{j\}}^{(s)} + \tau\, a_{\{j\}}^{(s)}\,, \qquad j = 0, \ldots, n-1\,. \label{eq:tier1_absorption_linear}
\end{align}
This absorption incurs zero additional gates.

\subsubsection{Tier~2: intra-mode quadratic cross-terms (shared sites).}
The $\binom{n_\kappa}{2}$ intra-mode cross-pair sites from the polynomial decomposition remain.
When an analytical function also has nonzero $a_{\{j,k\}}$ coefficients at the same two-qubit subsets, its contributions are absorbed by angle addition at the shared sites:
\begin{equation}\label{eq:tier2_sites}
    S_\kappa^{(2)} = \binom{n_\kappa}{2}\,.
\end{equation}

\subsubsection{Tier~2$'$: higher-order multilinear terms (new sites)}
Each analytical function contributes sites at $|S| \geq 3$ that are not present in the polynomial decomposition. 
For higher-degree polynomials, these terms also exist, with gate scaling as $\mathcal{O}(n^\alpha)$ for a polynomial of degree $\alpha$.
Explicit derivations for quartic $\propto x^4$, sextic $\propto x^6$, and octic $\propto x^8$ polynomials are given in Appendix~\ref{app:circuits}. 
If $N_f^{(\kappa)}$ distinct analytical basis functions appear in mode $\kappa$ across all electronic surfaces, the Tier~2$'$ site count is:
\begin{equation}\label{eq:tier2prime_sites}
    S_\kappa^{(2')} = N_f^{(\kappa)} \sum_{k=3}^{n_\kappa} \binom{n_\kappa}{k} = N_f^{(\kappa)} \left(2^{n_\kappa} - 1 - n_\kappa - \binom{n_\kappa}{2}\right).
\end{equation}
Each site is implemented as a CNOT-cascade--UCR$_Z^{(m)}$--cascade sandwich, with cascade depth $|S| - 1$ and UCR depth $2M - 1$.

\subsubsection{Tier~3: inter-mode bilinear terms (unchanged)}
The inter-mode bilinear coupling terms and their chromatic scheduling are identical to Chapter~\ref{ch:model_ext}, producing $n_r \cdot n_{r'}$ sites per coupled mode pair and a depth contribution $D_{\mathrm{Tier\,3}} = \chi \cdot n^2 \cdot (2M + 1)$, where $\chi$ is the chromatic index of the mode-pair conflict graph.
These exist apart from the univariate multilinear function. 
When considering multivariate multilinear functions, these bilinear terms will be absorbed in the total circuit.

\subsubsection{Total site count}
The total number of rotation sites per mode (with no bilinear terms) is
\begin{equation}\label{eq:gen_sites_full}
    S_\kappa^{(\mathrm{gen})} = \underbrace{(1 + n_\kappa)}_{\text{Tier~1}} + \underbrace{\binom{n_\kappa}{2}}_{\text{Tier~2}} + \underbrace{N_f^{(\kappa)}\!\left(2^{n_\kappa} - 1 - n_\kappa - \binom{n_\kappa}{2}\right)}_{\text{Tier~2$'$ ($|S| \geq 3$)}},
\end{equation}
For modes with no analytical functions ($N_f^{(\kappa)} = 0$), this reduces to the polynomial expression of Chapter~\ref{ch:resources}.
For modes with a single analytical function ($N_f^{(\kappa)} = 1$) at uniform register size $n$, the total simplifies to $S_\kappa^{(\mathrm{gen})} = 2^n$.

Table~\ref{tab:site_counts} gives numerical values for representative register sizes.

\begin{table}[t]
\centering
\caption{Site counts per mode for uniform register size $n$, with $N_f = 1$ analytical function.
Tier~1 and Tier~2 sites are shared between polynomial and analytical contributions.
Tier~2$'$ sites ($|S| \geq 3$) arise from the multilinear expansion.
The total equals $2^n$ for $N_f = 1$.}
\label{tab:site_counts}
\begin{tabular}{c r r r r r}
\hline\hline
$n$ & Tier~1 & Tier~2 & Tier~2$'$ & Total & $2^n$ \\
    & $(1+n)$ & $\binom{n}{2}$ & $\sum_{k \geq 3}\binom{n}{k}$ & & \\
\hline
7  & 8   & 21   & 99      & 128   & 128 \\
8  & 9   & 28   & 219     & 256   & 256 \\
9  & 10  & 36   & 466     & 512   & 512 \\
10 & 11  & 45   & 968     & 1024  & 1024 \\
11 & 12  & 55   & 1981    & 2048  & 2048 \\
12 & 13  & 66   & 4017    & 4096  & 4096 \\
13 & 14  & 78   & 8100    & 8192  & 8192 \\
14 & 15  & 91   & 16278   & 16384 & 16384 \\
\hline\hline
\end{tabular}
\end{table}

Tier~2$'$ dominates the site count for all register sizes $n \geq 7$.
At $n = 12$, the 4017 Tier~2$'$ sites constitute 98\% of the total, with the 79 polynomial sites (Tiers~1--2) contributing only 2\%.
The hybrid strategy follows: modes governed by polynomial potentials use the Tier~1--2 structure at $\mathcal{O}(n^2)$ cost, while reactive or dissociative modes carrying analytical functions use the Gray-code UCR at $\mathcal{O}(2^n)$ cost.

\subsection{Generalized Gate Counts}
\label{sec:gen_gate_counts}

Each rotation site produces $M$ rotations (on-diagonal, via UCR$_Z^{(m)}$) or $M/2$ rotations (off-diagonal, via UCR$_Z^{(m-1)}$), exactly as in the polynomial regime.

\subsubsection{On-diagonal rotations (per half-step)}
\begin{equation}\label{eq:Rd_gen}
    R_{\mathrm{d}}^{(\mathrm{half,gen})} = M\!\left[\sum_{\kappa=1}^{d} S_\kappa^{(\mathrm{gen})} + |\mathcal{G}_2^{<}|\, n^2\right],
\end{equation}
where the first term sums intra-mode sites (Tiers~1, 2, 2$'$) across all $d$ modes and the second counts inter-mode bilinear sites (Tier~3) across on-diagonal coupled mode pairs $\mathcal{G}_2^{<}$.
With uniform register size and $N_f = 1$ on every mode: $R_{\mathrm{d}}^{(\mathrm{half,gen})} = M[d \cdot 2^n + |\mathcal{G}_2^{<}|\, n^2]$.

\subsubsection{Off-diagonal rotations (per step)}
\begin{equation}\label{eq:Rod_gen}
    R_{\mathrm{od}}^{(\mathrm{gen})} = M(M-1)\!\left[\sum_{\kappa=1}^{d} S_\kappa^{(\mathrm{gen})} + |\mathcal{G}_4|\, n^2\right],
\end{equation}
where $|\mathcal{G}_4|$ counts off-diagonal coupled mode pairs and the $M(M-1)$ prefactor accounts for $M/2$ rotations per UCR$_Z^{(m-1)}$ across $2(M-1)$ fragments.

\subsubsection{Total rotations per Trotter step}
\begin{equation}\label{eq:Rtot_gen}
    R_{\mathrm{tot}}^{(\mathrm{gen})} = 2\, R_{\mathrm{d}}^{(\mathrm{half,gen})} + R_{\mathrm{od}}^{(\mathrm{gen})}\,.
\end{equation}
Both expressions reduce to the polynomial rotation counts of Chapter~\ref{ch:resources} when $N_f^{(\kappa)} = 0$ for all modes.

\subsubsection{CNOT counts}
Each Tier~2$'$ site with $|S| = k$ contributes $2(k-1)$ cascade CNOTs (forward and reverse).
The total cascade CNOT count for the analytical sites in a single mode is
\begin{equation}\label{eq:cascade_cnots}
    C_{\mathrm{cascade}}^{(\kappa)} = N_f^{(\kappa)} \cdot 2\sum_{k=3}^{n} (k-1)\binom{n}{k} = N_f^{(\kappa)}\!\left[n \cdot 2^n - 2^{n+1} + 2n + 2 - 2\binom{n}{2}\right],
\end{equation}
For $N_f^{(\kappa)} = 1$ and $n = 12$: $C_{\mathrm{cascade}}^{(\kappa)} = 40{,}854$.
Each on-diagonal site additionally contributes $M - 1$ Gray-code CNOTs on the channel register (or its fanned-out copies), and each off-diagonal site contributes $M/2 - 1$.
The total on-diagonal CNOT count for a single analytical function on mode $\kappa$ is
\begin{equation}\label{eq:gen_total_cnot_ondiag}
    N_{\mathrm{CNOT}}^{(\mathrm{on\text{-}diag},f)} = (2^{n_\kappa} - 1) + 2^{n_\kappa}(M - 1) \approx M \cdot 2^{n_\kappa}\,.
\end{equation}
The fan-out tree distributing the $m$ channel-register qubits to $d$ local copies requires $m \cdot (d - 1)$ CNOTs per fan-out operation, unchanged from Chapter~\ref{ch:model_ext}.

\subsection{Depth Independence from $d$}
\label{sec:depth_summary}

Depth independence from $d$ immediately follows from additive functions $f(\mathbf{x}) = f_0(x_0) + f_1(x_1)+ \text{etc.}$ 
The complete on-diagonal depth per half-step decomposes as
\begin{equation}\label{eq:Dd_gen_full}
    D_{\mathrm{d}}^{(\mathrm{half,gen})} = D_{\mathrm{fan}} + D_{\mathrm{Tier\,1}} + D_{\mathrm{Tier\,2}} + D_{\mathrm{Tier\,2'}} + D_{\mathrm{Tier\,3}} + D_{\mathrm{unfan}}\,,
\end{equation}
with contributions:
\begin{align}
    D_{\mathrm{fan}} = D_{\mathrm{unfan}} &= \lceil \log_2 d \rceil\,, \label{eq:depth_fan}\\[4pt]
    D_{\mathrm{Tier\,1}} &= (1 + n)(2M - 1)\,, \label{eq:depth_tier1}\\[4pt]
    D_{\mathrm{Tier\,2}} &= \binom{n}{2}(2M + 1)\,, \label{eq:depth_tier2}\\[4pt]
    D_{\mathrm{Tier\,3}} &= \chi \cdot n^2 \cdot (2M + 1)\,. \label{eq:depth_tier3}
\end{align}
The Tier~2$'$ depth, serialized over all $|S| \geq 3$ sites within the bottleneck mode, is
\begin{equation}\label{eq:depth_tier2prime}
    D_{\mathrm{Tier\,2'}} = N_f^{(\max)} \sum_{k=3}^{n} \binom{n}{k}\bigl[2(k-1) + (2M-1)\bigr]\,,
\end{equation}
where $N_f^{(\max)} = \max_\kappa N_f^{(\kappa)}$.
For large $n$ and $N_f^{(\max)} = 1$, this simplifies to
\begin{equation}\label{eq:depth_tier2prime_approx}
    D_{\mathrm{Tier\,2'}} \approx (2M + n - 1) \cdot 2^n\,,
\end{equation}
using the average cascade depth $\bar{s} \approx n/2$ at large $n$.

Complete on-diagonal depth per half-step is
\begin{equation}\label{eq:Dd_gen_summary}
\begin{split}
    D_{\mathrm{d}}^{(\mathrm{half,gen})} &= 2\lceil\log_2 d\rceil + (1+n)(2M-1) + \tbinom{n}{2}(2M+1) \\
    &\quad + N_f^{(\max)}\!\left(2^n - 1 - n - \tbinom{n}{2}\right)\!(2M + 2\bar{s} - 1) + \chi\, n^2(2M+1)\,.
\end{split}
\end{equation}
The $d$-dependence enters only through the logarithmic fan-out term and the chromatic index $\chi$.
Tiers~1, 2, and 2$'$ are all executed in parallel across modes on local channel-register copies and are therefore independent of $d$.

The depth ratio of the smooth-potential extension to the polynomial regime quantifies the per-mode overhead:
\begin{equation}\label{eq:tier_ratio}
    \frac{D_{\mathrm{Tier\,2'}}}{D_{\mathrm{Tier\,2}}} \approx \frac{(2M + n - 1) \cdot 2^n}{\binom{n}{2}(2M+1)} \approx \frac{2^{n+1}}{n^2}\,,
\end{equation}
where the last approximation holds for $n \gg M$.
At $n = 12$: $D_{\mathrm{Tier\,2'}}/D_{\mathrm{Tier\,2}} \approx 57$.
This ratio is the same for both REQWIEM and QROM architectures, as both process $2^n$ grid points per analytical function.
The off-diagonal depth per step follows the same structure: within each of the $2(M-1)$ XOR-class fragments, the intra-mode analytical sites are mode-parallel, and the inter-mode sites follow the chromatic schedule.
The total Trotter step depth is $D_{\mathrm{step}}^{(\mathrm{gen})} = 2\, D_{\mathrm{d}}^{(\mathrm{half,gen})} + D_{\mathrm{od}}^{(\mathrm{gen})}$, independent of $d$ up to the logarithmic fan-out and the chromatic Tier~3 terms.
This structural independence is the foundation of the depth advantage over QROM methods, where the serial accumulator forces a baseline $D \propto d$.

\section{Numerical Validation}
\label{sec:numerical_validation}

This section focuses on three concrete systems that exercise distinct classes of analytical potentials: an ESIPT model with quartic on-diagonal and Gaussian off-diagonal potentials (Section~\ref{sec:esipt_validation}), a Morse-repulsive scattering model with exponential-class potentials (Section~\ref{sec:morse_validation}), and a modified H\'{e}non-Heiles system testing inseparable multidimensional dynamics and automatic circuit sparsity (Section~\ref{sec:henon_heiles_validation}).
The last system extends the univariate multilinear function to a multivariate multilinear potential, where any coefficients found to be zero can be removed from the circuit, reducing circuit depth to the efficient ($\mathcal{O}(\text{poly}(n))$ case.

All classical emulations are performed by classically compiling quantum circuits, with coefficients computed by the M\"{o}bius transform at double precision (transcendental functions).

\subsection{ESIPT System: HBT Model}
\label{sec:esipt_validation}

\subsubsection{Model Specification}
\label{subsubsec:esipt_model}

The ESIPT model is based on 2-(2$'$-hydroxyphenyl)-benzothiazole (HBT), a prototypical proton-transfer chromophore whose excited-state intramolecular proton transfer (ESIPT) dynamics have been characterized by ultrafast spectroscopy~\cite{lochbrunner2000ultrafast, rini2003femtosecond} and multireference electronic structure calculations~\cite{barbatti2009ultrafast, aquino2005excited, pijeau2017excited}.
A simplified model comprises one nuclear degree of freedom ($d=1$), representing the proton-transfer coordinate along the intramolecular O--H$\cdots$N hydrogen bond, and two electronic states ($M=2$): an excited bright state (S$_1$) whose on-diagonal potential is an asymmetric quartic double well representing keto--enol tautomerism, and a dark ground state (S$_0$) with a harmonic potential.
The Hamiltonian is
\begin{equation}
\label{eq:esipt_H}
  {H}
  = \frac{{p}^{2}}{2m_p}\,\mathbb{1}_{2}
  + \begin{pmatrix}
      V_{\text{harm}}(x) & V_{01}(x) \\[4pt]
      V_{01}(x)         & V_{\text{quart}}(x)
    \end{pmatrix},
\end{equation}
with $m_p = 1836\;\text{a.u.}$ and
\begin{subequations}
\label{eq:esipt_potentials}
\begin{align}
  V_{\text{harm}}(x)
    &= \tfrac{1}{2}\,k\,x^{2},
    \qquad k = m_p\,\omega_{\text{OH}}^{2},
    \label{eq:V_harm}
  \\[4pt]
  V_{\text{quart}}(x)
    &= c_{0} + c_{1}\,x + c_{2}\,x^{2} + c_{4}\,x^{4},
    \label{eq:V_quart}
  \\[4pt]
  V_{01}(x)
    &= A\,\exp\!\Bigl[-\frac{x^{2}}{2\sigma^{2}}\Bigr].
    \label{eq:V_coupling}
\end{align}
\end{subequations}

The parameters, collected in Table~\ref{tab:esipt_params}, are calibrated against experimental and computational data for the HBT S$_1$/S$_0$ surface structure.
The O--H stretching frequency $\omega_{\text{OH}} = 0.014\;E_h/\hbar$ ($\approx 3070\;\text{cm}^{-1}$) is consistent with the intramolecularly hydrogen-bonded phenol stretch in the enol tautomer~\cite{elsaesser1991initial, rini2003femtosecond}.
The vertical excitation energy $E_{\text{vert}} = 0.136\;E_h$ ($\approx 3.70\;\text{eV}$, $335\;\text{nm}$) reproduces the experimental S$_0 \to$ S$_1$ absorption maximum of HBT in nonpolar solvents~\cite{elsaesser1989transient, lochbrunner2000ultrafast}, consistent with RI-CC2 and CASPT2 vertical excitation energies of $3.7$--$3.9\;\text{eV}$~\cite{aquino2005excited, sobolewski1999ab}.
The proton migration distance $q_{0} = 8/7\;a_{0} \approx 1.14\;a_{0}$ ($0.60\;\text{\AA}$) matches computed displacements of the proton along the O--N vector during ESIPT, where time dependent DFT and RI-CC2 optimizations yield O--H bond elongations of $1.1$--$1.5\;a_0$ between the enol and keto minima~\cite{barbatti2009ultrafast, kungwan2012effect}.

The S$_1$ surface is tilted, where the keto tautomer is preferentially stabilized by excited-state electronic redistribution, producing a gradient from the Franck--Condon region toward the keto minimum~\cite{barbatti2009ultrafast, sobolewski1999ab, pijeau2017excited}.
At the RI-CC2 level, no enol minimum exists on S$_1$; the surface is monotonically downhill toward the keto form~\cite{aquino2005excited}.
This asymmetry is parameterized by a linear tilt coefficient $c_{1} = -\Delta E_{\text{asym}}/(2q_{0})$, with $\Delta E_{\text{asym}} = 0.10\;\text{eV}$, stabilizing the keto minimum ($x = +q_0$) by $100\;\text{meV}$ relative to the enol minimum ($x = -q_0$).

\begin{figure}[t]
     \centering
     \includegraphics[width=0.8\textwidth]{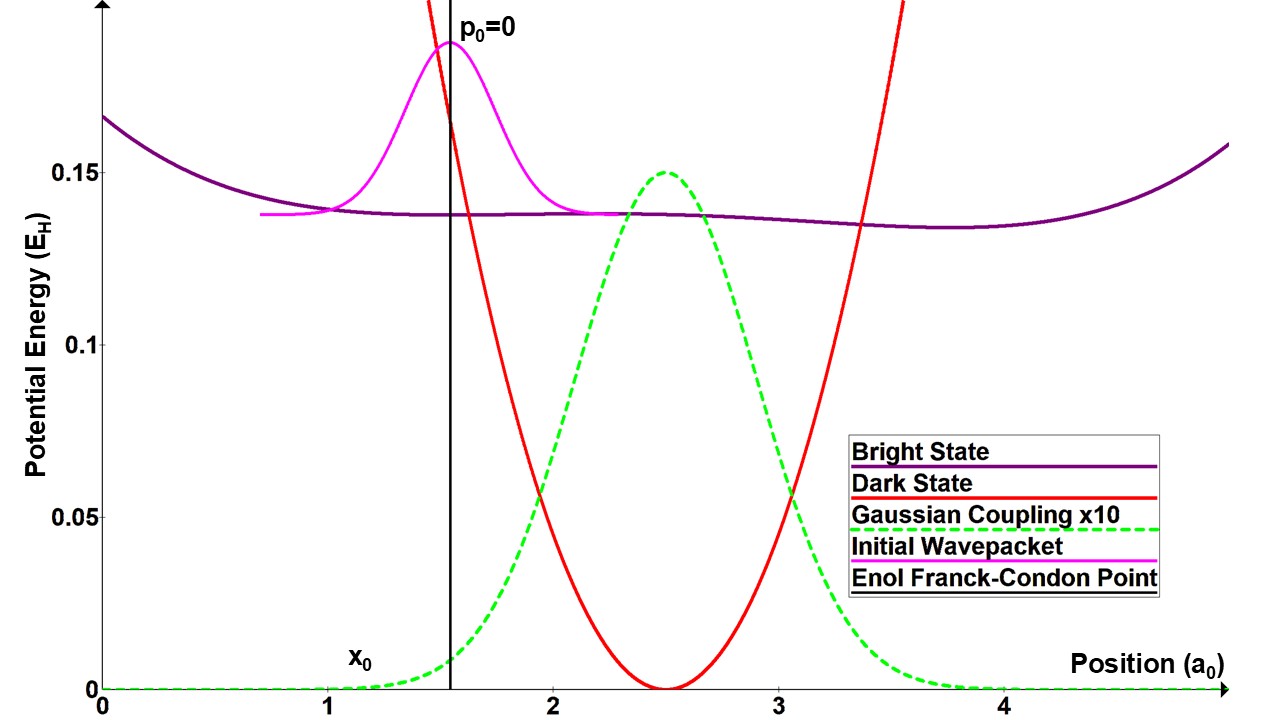}
     \caption{
     Potential energy surface schematic showing the asymmetric quartic
     (purple), harmonic (red), and Gaussian coupling $V_{01}$
     ($\times 10$, dashed green).  The Franck--Condon point at
     $x = -q_0$ sits on a slope directed toward the keto minimum.}
     \label{fig:esipt_schematic}
\end{figure}

The Gaussian coupling (Figure~\ref{fig:esipt_schematic}) has amplitude $A = 0.015\;E_h$ ($408\;\text{meV}$) and spatial width $\sigma = 0.40\;a_0$, centered at $x = 0$.
The diabatic surfaces cross at $x \approx \pm 0.87\;a_0$, where the coupling evaluates to $V_{01} = 38\;\text{meV}$.
At the quartic minima ($x = \pm q_0$) the coupling is exponentially suppressed to $V_{01}(q_0) \approx 7\;\text{meV}$, ensuring that the adiabatic surfaces are well separated at the keto and enol geometries while mixing occurs primarily in the barrier region, consistent with the strongly adiabatic regime expected for intramolecular proton transfer~\cite{hammes2015proton}.

\begin{table}[t]
\centering
\caption{Model parameters for the HBT ESIPT system in atomic units.
  Values in parentheses give approximate conversions to conventional units.}
\label{tab:esipt_params}
\begin{tabular}{@{}l l l l@{}}
\hline\hline
Parameter & Symbol & Value & \\
\hline
Proton mass             & $m$                     & 1836    & a.u.\ (CODATA 2022)    \\
Box half-width          & $d = L/2$               & 2.5        & $a_{0}$                \\
O--H frequency          & $\omega_{\text{OH}}$    & 0.014      & $E_h/\hbar$ (3070 cm$^{-1}$) \\
Harmonic force const.\  & $k = m\omega^{2}$       & 0.3599     & $E_h/a_{0}^{2}$        \\
Vertical excitation     & $E_{\text{vert}}$       & 0.136      & $E_h$ (3.70 eV)        \\
Migration distance      & $q_{0}$                 & $8/7$      & $a_{0}$ (0.60 \AA)     \\
Symmetric barrier       & $V_{\text{b}}$          & $1.84 \times 10^{-3}$ & $E_h$ (50 meV)   \\
Asymmetry               & $\Delta E_{\text{asym}}$ & $3.68 \times 10^{-3}$ & $E_h$ (100 meV)  \\
Quartic coefficient     & $c_{4}$                 & $1.078 \times 10^{-3}$ & $E_h/a_{0}^{4}$  \\
Quadratic coefficient   & $c_{2}$                 & $-2.815 \times 10^{-3}$ & $E_h/a_{0}^{2}$ \\
Linear (tilt) coeff.    & $c_{1}$                 & $-1.608 \times 10^{-3}$ & $E_h/a_{0}$     \\
Constant offset         & $c_{0}$                 & 0.1378     & $E_h$ (3.75 eV)        \\
Coupling amplitude      & $A$                     & 0.015      & $E_h$ (408 meV)        \\
Coupling width          & $\sigma$                & 0.40       & $a_{0}$                \\
Wavepacket width        & $\delta$                & 0.20       & $a_{0}$                \\
\hline\hline
\end{tabular}
\end{table}

\subsubsection{Simulation Protocol}
\label{subsubsec:esipt_protocol}

The initial state is a Gaussian wavepacket of width $\delta = 0.20\;a_0$ centered at the Franck--Condon point ($x_0 = -q_0$) on the quartic surface with zero initial momentum, representing a vertical excitation.
The width matches the harmonic zero-point amplitude $\sigma_{\text{ZPE}} = \sqrt{\hbar/(m\omega_{\text{OH}})} = 0.197\;a_0$ to within $2\%$.

The $(n+1)$-qubit register encodes four observable sectors defined by two qubits: the channel qubit selects the electronic surface ($|0\rangle_s \to V_{\text{quart}}$, S$_1$; $|1\rangle_s \to V_{\text{harm}}$, S$_0$), while the most significant qubit (the ``high bit") of the position register selects the spatial half ($|0\rangle_{high} \to x < 0$, enol well; $|1\rangle_{high} \to x > 0$, keto well).
At $t = 0$ the full population resides in the enol/S$_1$ sector.
ESIPT corresponds to adiabatic population flow from enol/S$_1$ to keto/S$_1$, while electronic relaxation is the nonadiabatically-mediated property.

The simulation uses $n = 10$ position qubits ($N = 1024$ grid points, $\Delta x = 4.88 \times 10^{-3}\;a_0$) plus one ancilla qubit encoding the electronic surface.
Time propagation employs a second-order (Strang) Trotter splitting,
\begin{equation}
\label{eq:strang}
  e^{-iH\tau/\hbar}
  \approx
  e^{-iT\tau/(2\hbar)}\,
  e^{-iV_\text{d}\tau/(2\hbar)}\,
  \mathcal{V}_2(\tau)
    e^{-iV_\text{d}\tau/(2\hbar)}\,
    e^{-iT\tau/(2\hbar)}\,,
\end{equation}
where the off-diagonal $\mathcal{V}_2(\tau)$ is an exponentiated Gaussian.
This equation takes a similar form to the XOR-fragmented term in Chapter~\ref{ch:model_ext}, but a two-channel system only has one XOR class, reducing this term to $\mathcal{V}_2(\tau) = e^{-iV_\text{od}\tau}$
Step size is set to $\tau = 0.1\;\text{a.u.}$ ($2.4\;\text{as}$) for a total propagation time of $T = 10{,}000\;\text{a.u.}$ ($242\;\text{fs}$).

\subsubsection{Exact Diagonalization Benchmark}
\label{subsubsec:esipt_exact}

A Trotter-error-free reference is obtained by instantiating the two-surface Hamiltonian~\ref{eq:esipt_H} on a real-space grid of $N = 2048$ points spanning $x \in [-2.5,\,2.5]\;a_0$ ($\Delta x \approx 2.4 \times 10^{-3}\;a_0$).
The kinetic energy is discretized with a three-point finite-difference stencil, yielding a real symmetric matrix of dimension $2N \times 2N = 4096 \times 4096$ with block structure
\begin{equation}
\label{eq:H_block}
  \mathbf{H}
  = \begin{pmatrix}
      T + V_{\text{quart}}(x)
        & V_{01}(x) \\[4pt]
      V_{01}(x)
        & T + V_{\text{harm}}(x)
    \end{pmatrix}.
\end{equation}
Full diagonalization returns the eigenspectrum $\{E_k, |\chi_k\rangle\}$, from which the autocorrelation function is obtained without time-stepping error as $C(t) = \sum_k |c_k|^2 e^{-iE_k t/\hbar}$, where $c_k = \langle\chi_k|\Psi(0)\rangle$.
The spectrum is computed using the same Blackman window, exponential damping ($\tau_d = 5000\;\text{a.u.}$), and fourfold zero-padded FFT as the quantum circuit simulation, so that any differences between the two spectra arise solely from the Trotter decomposition.

\subsubsection{Trotter Error Analysis}
\label{subsubsec:trotter_error}

Second-order Strang splitting~\ref{eq:strang} introduces a per-step error governed by the nested commutators of the kinetic and potential operators~\cite{childs2021theory}.
For a wavepacket with root-mean-square momentum $p_{\text{rms}} = \hbar/(2\delta) \approx 2.5\;a_0^{-1}$ (kinetic energy $T_{\text{phys}} = p_{\text{rms}}^2/(2m) = 46\;\text{meV}$) propagating on a potential of depth $V_{\text{max}} \approx 3.75\;\text{eV}$, the per-step error magnitudes are
\begin{align}
  \varepsilon_2
  &\sim \frac{\tau^3}{12}\,T_{\text{phys}}\,V_{\text{max}}
    \bigl(V_{\text{max}} + T_{\text{phys}}\bigr)
  \approx 5 \times 10^{-4}\;\text{meV},
\end{align}
The physically relevant phase accumulation per step, $\tau\,p_{\text{rms}}^2/(2m\hbar) = 1.7 \times 10^{-3}\;\text{rad}$, remains well within the stable regime.
Over $N_{\text{steps}} = 10^5$ steps, the symmetric structure of the Strang splitting ensures partial cancellation of odd-order error contributions, suppressing cumulative phase error relative to the first-order decomposition.

\subsubsection{Proton Transfer Dynamics}
\label{subsubsec:esipt_dynamics}

\begin{figure}[t]
     \centering
     \includegraphics[width=\textwidth]{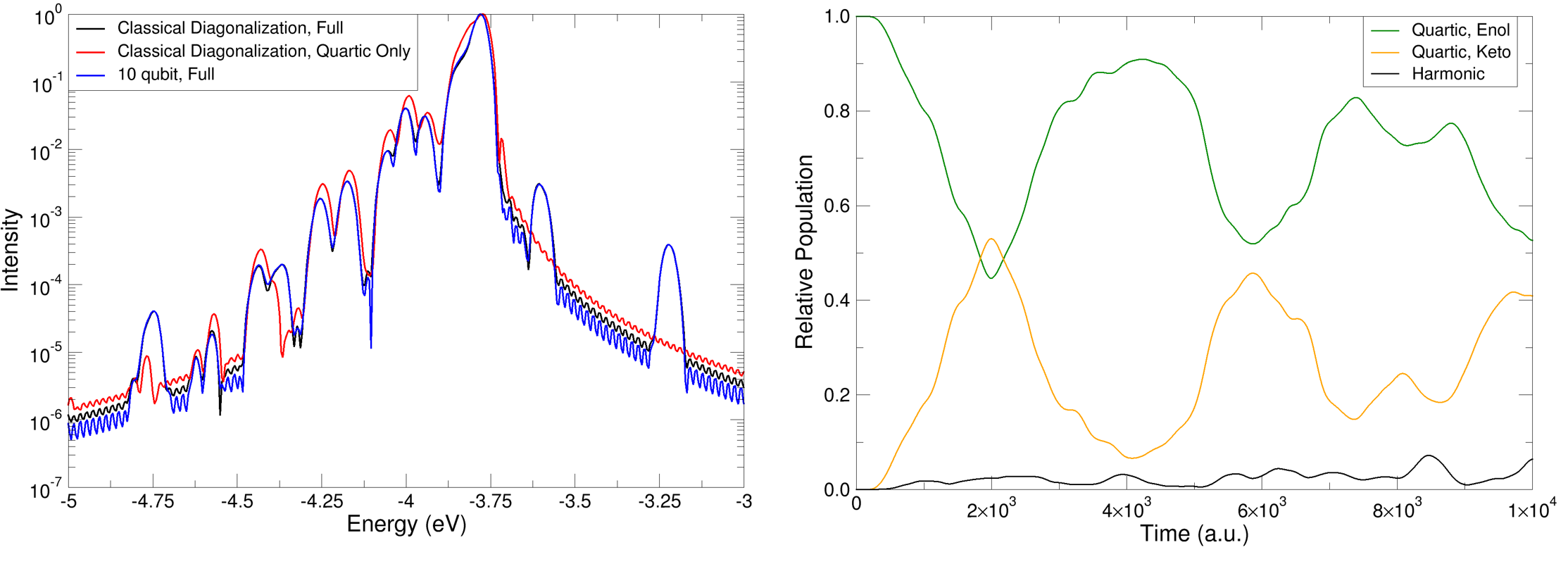}
     \caption{Observable benchmarks for the HBT ESIPT model.
     (left)~Autocorrelation spectrum from $|C(t)|$: black, exact coupled
     diagonalization; red, isolated quartic; blue, second-order
     Trotterized quantum circuit ($\tau = 0.1\;\text{a.u.}$,
     10~position qubits $+$ 1~ancilla qubit).  
     The quartic progression is shifted by ${\sim}5$--$15\;\text{meV}$ by the Gaussian coupling;
     the discrepancy near $-4.3$ to $-4.5\;\text{eV}$ arises from
     avoided crossings between quartic states and harmonic $m = 11$.
          (right)~Three-sector population dynamics from the Trotterized quantum circuit simulation: the keto/S$_1$ population rises to ${\sim}0.53$ at $48\;\text{fs}$, continuing to ${\sim}0.91$ by $97\;\text{fs}$,
     while the total S$_0$ population remains below $5\%$. 
     Coherent recurrences modulate the proton transfer with periods consistent
     with the dominant quartic eigenstate beat frequencies.}
     \label{fig:esipt_observables}
\end{figure}

Figure~\ref{fig:esipt_observables}(b) presents the time-resolved population dynamics in the three principal register sectors: enol/S$_1$, keto/S$_1$, and total S$_0$.

The proton transfer proceeds ballistically under the asymmetric quartic gradient.
The keto/S$_1$ population rises monotonically from zero, reaching approximately $0.53$ at $t \approx 2000\;\text{a.u.}$ ($48\;\text{fs}$), at which point the enol and keto populations on S$_1$ are approximately equal. 
This population transfer is modulated by the coupling strength, with equal populations representing an a point in time where the molecular configuration is in a superposition between tautomers.
This timescale matches the experimental keto$^{*}$ stimulated emission rise of $30$--$50\;\text{fs}$ measured by femtosecond fluorescence up-conversion~\cite{lochbrunner2000ultrafast} and mid-infrared transient absorption~\cite{rini2003femtosecond}, confirming that the asymmetric quartic potential produces the correct ballistic driving force.

Following the initial transfer, the wavepacket exhibits coherent oscillations between the two tautomers.
The enol/S$_1$ population continues rising to a maximum of approximately $0.91$ near $t \approx 4000\;\text{a.u.}$ ($97\;\text{fs}$), then drops sharply to ${\sim}0.5$ near $t \approx 5000\;\text{a.u.}$ ($121\;\text{fs}$) as the wavepacket partially returns to the keto well.
A second recurrence carries the enol population back to ${\sim}0.83$ by $t \approx 7500\;\text{a.u.}$ ($181\;\text{fs}$), followed by a third partial return around $t \approx 9000\;\text{a.u.}$ ($218\;\text{fs}$).
The oscillation amplitude is damped by dephasing among the multiply populated quartic eigenstates: the initial Franck--Condon wavepacket projects predominantly onto $n = 2$ ($|c_2|^2 = 0.46$), with strong secondary contributions from $n = 3$ ($|c_3|^2 = 0.32$) and $n = 4$ ($|c_4|^2 = 0.13$), spanning a spectral width of $\Delta E = 46\;\text{meV}$.
Dominant beat frequencies between these levels predict recurrence periods of $T_{2\leftrightarrow 3} = 2\pi\hbar/\Delta E_{23} = 128\;\text{fs}$ and $T_{2\leftrightarrow 4} = 57\;\text{fs}$, consistent with the observed oscillation timescales and within the $60$--$140\;\text{fs}$ range of coherent vibrational dynamics reported in time-resolved fluorescence of HBT~\cite{lochbrunner2000ultrafast}.

The total S$_0$ population (harmonic surface) remains below $10\%$ throughout the $242\;\text{fs}$ simulation, oscillating without secular growth.
This suppression is consistent with the off-resonant Rabi picture.
At the Franck--Condon point, the coupling-to-gap ratio is $V_{01}(-q_0)/\Delta E_{\text{gap}} = 6.9\;\text{meV}/2.65\;\text{eV} = 0.003$, predicting negligible transfer probability.
The few-percent leakage arises when the wavepacket transiently passes through $x \approx 0$ during proton migration, where the coupling peaks at $408\;\text{meV}$ against a gap of $3.75\;\text{eV}$, giving $(V/\Delta)^2 \approx 0.012$.
The S$_0$ population distributes approximately equally between the enol and keto spatial halves, indicating that electronic leakage occurs primarily in the barrier region where $V_{01}$ is maximal, and the leaked amplitude then spreads symmetrically on the harmonic surface.
This confirms that proton transfer proceeds largely adiabatically on S$_1$ with only small, though non-negligible nonadiabatic leakage, consistent with the strongly coupled regime of intramolecular proton transfer~\cite{hammes2015proton}.
ESIPT systems that have significant nonadiabatic character will have a stronger coupling constant, tunable as a single value in classical preprocessing.

\subsubsection{Spectral Benchmarks}
\label{subsubsec:esipt_spectrum}

Figure~\ref{fig:esipt_observables}(a) compares autocorrelation spectra from three calculations: exact diagonalization of the full coupled two-surface Hamiltonian (black), exact diagonalization of the isolated quartic potential (red), and the second-order Trotterized quantum circuit simulation of the full Hamiltonian (green).
In the main quartic progression between $-3.7$ and $-4.2\;\text{eV}$, the three traces are in close agreement, with peak positions matching to within the spectral resolution.
The quartic-only spectrum (red) is displaced by ${\sim}2$--$15\;\text{meV}$ from the coupled result (black), reflecting perturbative level shifts induced by the Gaussian coupling; the Trotterized circuit (blue) tracks the coupled reference with comparable fidelity.
The most significant discrepancy between the traces appears in the noise floor.
Isolated features attributable to harmonic-surface eigenstates appear in the coupled diagonalization (black) near $-3.2\;\text{eV}$, where they are absent from the quartic-only trace, confirming the expected redistribution of spectral weight by the interchromophore coupling. 
The quantum circuit Trotterization is not initialized with a momentum ($p_0=0$), making these states largely inaccessible and not present in the spectrum.

\subsection{Morse-Repulsive System: NaI Model}
\label{sec:morse_validation}

\subsubsection{Model Specification}
\label{subsubsec:morse_model}

The Morse-repulsive model is parameterized after sodium iodide (NaI), with femtosecond photodissociation dynamics first resolved by Zewail et al.~\cite{zewail1993femtochemistry,mokhtari1990direct} and
is a standard benchmark for nonadiabatic wavepacket methods~\cite{engel1989quantum,braun1996reflection,batista1997nonadiabatic}.
The model has one coordinate ($d=1$), being the Na--I internuclear separation $R$, and two diabatic electronic states ($M=2$): an ionic ground state (Na$^{+}$I$^{-}$, $X\,{}^{1}\Sigma^{+}$) modeled by a Morse potential, and a covalent excited state (Na$^{*}$\,+\,I) modeled by a repulsive exponential.
The diabatic Hamiltonian is
\begin{equation}
\label{eq:morse_H}
  {H}
  = \frac{{p}^{2}}{2\mu}\,\mathbb{1}_{2}
  + \begin{pmatrix}
      V_{\text{ionic}}(R) & V_{01}(R) \\[4pt]
      V_{01}(R)           & V_{\text{cov}}(R)
    \end{pmatrix},
\end{equation}
with $\mu = 35{,}480\;\text{a.u.}$ ($19.464\;\text{amu}$, the Na--I reduced mass) and
\begin{subequations}
\label{eq:morse_potentials}
\begin{align}
  V_{\text{ionic}}(R)
    &= D_{e}\bigl[1 - e^{-\beta(R - R_{e})}\bigr]^{2}
       - D_{e} + \Delta E_{\infty},
    \label{eq:V_morse}
  \\[4pt]
  V_{\text{cov}}(R)
    &= A\,e^{-\alpha R},
    \label{eq:V_rep}
  \\[4pt]
  V_{01}(R)
    &= W_{0}\,\exp\!\bigl[-\gamma\,(R - R_{c})^{2}\bigr].
    \label{eq:V_coupling_morse}
\end{align}
\end{subequations}
The energy zero is the covalent asymptote Na($3s\;{}^{2}\!S_{1/2}$)\,+\,I(${}^{2}\!P_{3/2}$).
The Morse dissociation limit is therefore the ionic asymptote
$\Delta E_{\infty} = \mathrm{IP}(\text{Na}) - \mathrm{EA}(\text{I})
= 2.080\;\text{eV}$~\cite{NIST_ASD}, and the Morse minimum lies at
$\Delta E_{\infty} - D_{e} = -1.12\;\text{eV}$ below the covalent zero.
All parameters are collected in Table~\ref{tab:morse_params}.

Equilibrium bond length $R_{e} = 5.122\;a_{0}$ is taken from
microwave spectroscopy~\cite{rusk1962millimeter}, with well depth
$D_{e} = 3.20\;\text{eV}$ ($0.1176\;E_{h}$) from photofragment
translational spectroscopy~\cite{berry1957interaction} corrected for
zero-point energy.
The Morse exponent
$\beta = \omega_{e}\sqrt{\mu/(2D_{e})} = 0.459\;a_{0}^{-1}$
reproduces the harmonic frequency
$\omega_{e} = 259\;\text{cm}^{-1}$~\cite{NIST_Diatomic,rusk1962millimeter}.

The repulsive amplitude $A = 0.1236\;E_{h}$ ($3.363\;\text{eV}$) and
decay constant $\alpha = 0.04288\;a_{0}^{-1}$ are jointly constrained
the vertical excitation energy at $R_{e}$,
\begin{equation}
\label{eq:morse_vert_exc}
  \Delta E_{\text{abs}}
  = V_{\text{cov}}(R_{e}) - V_{\text{ionic}}(R_{e})
  = A\,e^{-\alpha R_{e}} - (\Delta E_{\infty} - D_{e})
  = 0.1404\;E_{h}
  \approx 3.82\;\text{eV}\;(325\;\text{nm}),
\end{equation}
matching the experimental absorption maximum~\cite{davidovits1967ultraviolet}. 
They are also constrained by the diabatic crossing at the
Coulomb radius $R_{c} = 1/\Delta E_{\infty} = 13.1\;a_{0}$, giving
$V_{\text{ionic}}(R_{c}) = V_{\text{cov}}(R_{c}) = 1.918\;\text{eV}$.
The decay constant $\alpha = 0.04288\;a_{0}^{-1}$ reflects Na$^{*}$\,+\,I covalent repulsion, whose range
is set by the Na($3s$) valence orbital radius.

Gaussian coupling~\eqref{eq:V_coupling_morse} is centered at the
crossing radius $R_{c} = 13.1\;a_{0}$.
The amplitude $W_{0} = 3.85 \times 10^{-4}\;E_{h}$ ($10.5\;\text{meV}$)
is chosen to reproduce the experimentally observed per-crossing escape
fraction of ${\sim}\,11\%$~\cite{zewail1993femtochemistry,mokhtari1990direct},
as verified by the Landau--Zener calculation in
Section~\ref{subsubsec:morse_stokes}.
The adiabatic gap at the crossing is $2W_{0} = 21\;\text{meV}$.
The width $\gamma = 0.01\;a_{0}^{-2}$ confines the nonadiabatic
interaction to within
$\Delta R \approx 1/\sqrt{2\gamma} \approx 7\;a_{0}$ of the crossing.

The initial state is a coherent equal-amplitude superposition of both
diabatic surfaces, testing propagation fidelity on both surfaces
and benchmarking nonadiabatic population transfer at the crossing.

The simulation box spans $R \in [0, L]$ with $L = 100\;a_{0}$
and $N = 2^{12} = 4096$ grid points
($\Delta R = 0.0244\;a_{0}$), using $n = 12$ position qubits plus
one channel qubit.
The propagation time $t = 10^{5}\;\text{a.u.}$ ($2.42\;\text{ps}$)
with Trotter step $\tau = 1.0\;\text{a.u.}$ ($24.2\;\text{as}$)
captures at least two nonadiabatic crossing events.

\begin{table}[t]
\centering
\caption{Model parameters for the NaI Morse-repulsive system in atomic units.
Parenthetical values give approximate conversions.
Energy reference is the covalent asymptote
Na($3s$)\,+\,I(${}^{2}\!P_{3/2}$) $= 0$.}
\label{tab:morse_params}
\begin{tabular}{@{}l l l l@{}}
\hline\hline
Parameter & Symbol & Value & \\
\hline
Reduced mass             & $\mu$               & $35{,}480$                    & a.u.\ ($19.464$ amu)                \\
Box length               & $L$                 & 100                           & $a_{0}$ ($52.9$ \AA)                \\
Grid qubits              & $n$                 & 12                            & ($N = 4096$ points)                 \\
Trotter step             & $\tau$              & 1.0                           & a.u.\ ($24.2$ as)                   \\
Propagation time         & $T$                 & $100{,}000$                   & a.u.\ ($2.42$ ps)                   \\[4pt]
\multicolumn{4}{@{}l}{\emph{Ionic Morse surface}} \\
Well depth               & $D_{e}$             & 0.1176                        & $E_{h}$ ($3.20$ eV)                 \\
Equilibrium distance     & $R_{e}$             & 5.122                         & $a_{0}$ ($2.711$ \AA)               \\
Morse exponent           & $\beta$             & 0.459                         & $a_{0}^{-1}$                        \\
Ionic asymptote          & $\Delta E_{\infty}$ & 0.07644                       & $E_{h}$ ($2.080$ eV)                \\
Harmonic frequency       & $\omega_{e}$        & $1.182 \times 10^{-3}$        & $E_{h}/\hbar$ ($259$ cm$^{-1}$)     \\[4pt]
\multicolumn{4}{@{}l}{\emph{Covalent repulsive surface}} \\
Repulsive amplitude      & $A$                 & 0.1236                        & $E_{h}$ ($3.363$ eV)                \\
Decay constant           & $\alpha$            & 0.04288                       & $a_{0}^{-1}$                        \\
Asymptote                & $V_{\infty}$        & 0                             & $E_{h}$                             \\[4pt]
\multicolumn{4}{@{}l}{\emph{Gaussian diabatic coupling}} \\
Coupling amplitude       & $W_{0}$             & $3.85 \times 10^{-4}$         & $E_{h}$ ($10.5$ meV)                \\
Crossing radius          & $R_{c}$             & 13.1                          & $a_{0}$ ($6.93$ \AA)                \\
Gaussian width           & $\gamma$            & 0.01                          & $a_{0}^{-2}$                        \\[4pt]
\multicolumn{4}{@{}l}{\emph{Initial wavepacket}} \\
Center                   & $R_{0}$             & 5.122                         & $a_{0}$ ($= R_{e}$)                 \\
Width                    & $\delta$            & 0.15                          & $a_{0}$                             \\
Mixing angle             & $\theta$            & $\pi/4$                       & ($P_{\text{ion}} = P_{\text{cov}} = 0.5$)  \\
Initial momentum         & $p_{0}$             & 0                             & a.u.                                \\
\hline\hline
\end{tabular}
\end{table}

\subsubsection{Predicted Dynamics and Spectral Signatures}
\label{subsubsec:morse_stokes}

The covalent component accelerates outward on the repulsive surface.
Classical integration of the equations of motion gives an arrival time
at $R_{c}$ of $t_{\text{arr}} \approx 11{,}900\;\text{a.u.}$
($288\;\text{fs}$) with radial velocity
\begin{equation}
\label{eq:v_crossing}
  v_{c}
  = \sqrt{\frac{2}{\mu}\bigl[V_{\text{cov}}(R_{e}) - V_{\text{cov}}(R_{c})\bigr]}
  = 1.27 \times 10^{-3}\;\text{a.u.}
\end{equation}
Difference between the diabatic surfaces at $R_{c}$ is
$|\Delta F| = 5.72 \times 10^{-3}\;E_{h}/a_{0}$,
giving a Landau--Zener escape probability~\cite{zener1932non}
\begin{equation}
\label{eq:LZ_prob}
  P_{\text{esc}} = \exp\!\Bigl(
    -\frac{2\pi\,W_{0}^{2}}{\hbar\,v_{c}\,|\Delta F|}
  \Bigr)
  = e^{-0.128}
  = 0.88,
\end{equation}
corresponding to a per-crossing trapping probability of
$P_{\text{trap}} = 1 - P_{\text{esc}} \approx 12\%$, consistent with
the experimental value of
${\sim}\,11\%$~\cite{zewail1993femtochemistry,mokhtari1990direct,engel1989quantum}.

The Gaussian coupling region spans $\Delta R \approx 7\;a_{0}$.
With a mean velocity of $v_{c} = 1.27 \times 10^{-3}\;\text{a.u.}$, the wavepacket requires $\Delta t_{\text{trav}} \approx 5{,}500\;\text{a.u.}$ ($133\;\text{fs}$)
to traverse it.
Free-particle spreading time $t_{\text{sp}} = 2\mu\delta^{2} \approx 1{,}600\;\text{a.u.}$ is much shorter than the transit time, giving a spatial width at $R_{c}$ of
$\sigma(t_{\text{arr}}) \approx 1.1\;a_{0}$.
The crossing event therefore extends over $\sim$5{,}500--6{,}000~a.u.,
producing a temporally resolved population-transfer ramp.

The ${\sim}\,12\%$ of the nonadiabatically
transferred covalent state at $R_c$ is deposited into
high-lying Morse vibrational states.
The crossing energy
$V_{\text{ionic}}(R_{c}) = 1.918\;\text{eV}$
lies $3.04\;\text{eV}$ above the Morse minimum.
The outer turning point of $v = 154$ lies at $R_{\text{out}} = 13.1\;a_{0} \approx R_{c}$: the trapped
wavepacket's classical orbit grazes the crossing radius at every
oscillation, enabling Zewail-type recurrent population transfer.

Following Zewail (1993)~\cite{zewail1993femtochemistry}, a ``staircase" population transfer
requires a dephasing time not less than the mean oscillation period. 
At $v = 154$, $T_{\text{Morse}} \approx 23{,}700\;\text{a.u.}$
($573\;\text{fs}$), being a factor of $4.4\times$ longer than the harmonic
period $2\pi/\omega_{e} = 5{,}300\;\text{a.u.}$
The Gaussian coupling deposits population over a range of $v$-states
spanning the energy width
$\Delta E \approx |dV_{\text{ionic}}/dR|_{R_{c}} \times \Delta R
\approx 520\;\text{meV}$ (${\pm}\,70$ vibrational quanta). 
The classical period varies significantly
($T \approx 9{,}000\;\text{a.u.}$ at $v = 80$ to
$T \approx 700{,}000\;\text{a.u.}$ at $v = 195$).
The resulting dephasing time,
$t_{\text{deph}} = \pi/\Delta\omega \approx 4{,}600\;\text{a.u.}$
($111\;\text{fs}$), is $5\times$ shorter than the mean oscillation
period.
The trapped wavepacket loses spatial coherence before a single classical orbit~\cite{du1988effect}, disallowing a clean staircase structure.
Individual $v$-components still follow outer turning
points and traverse the coupling region, but arrive at staggered
times, producing a continuous, sub-percent population drift rather than
resolved steps.

The main covalent fraction (${\sim}\,88\%$ of the initial covalent
population) that escapes past $R_{c}$ continues outward with
asymptotic velocity
$v_{\infty} = \sqrt{2\,V_{\text{cov}}(R_{e})/\mu} = 2.37 \times 10^{-3}\;\text{a.u.}$
This fragment reflects from the ``hard wall" $R = L$, formed by the back of the diabatic surfaces, and returns through $R_{c}$ after a round-trip time
\begin{equation}
\label{eq:t_roundtrip}
  T_{\text{rt}}
  \approx \frac{2(L - R_{c})}{v_{\infty}}
  \approx 73{,}500\;\text{a.u.}\quad(L = 100\;a_{0}),
\end{equation}
producing a second population-transfer step near
$t \approx t_{\text{arr}} + T_{\text{rt}} \approx 85{,}000\;\text{a.u.}$
A larger box would push this recurrence beyond the
simulation window
($T_{\text{rt}} \approx 116{,}000\;\text{a.u.} > T$),
recovering single-crossing behavior.

\subsubsection{Classical Diagonalization Benchmark}
\label{subsubsec:morse_exact}

The full two-surface Hamiltonian~\eqref{eq:morse_H} is diagonalized on a real-space grid of $N$ points spanning $R \in [0, L]$.
The potential term takes a block structure
\begin{equation}
\label{eq:morse_H_block}
  \mathbf{V}
  = \begin{pmatrix}
      V_{\text{ionic}}(R_{i})
        & V_{01}(R_{i}) \\[4pt]
      V_{01}(R_{i})
        & V_{\text{cov}}(R_{i})
    \end{pmatrix},
\end{equation}
with total dimension $2N \times 2N$.

The quantum circuit applies $T(p) = p^{2}/(2\mu)$ in momentum space, where the QFT maps the
position register to the conjugate momentum basis with exact quadratic dispersion (circulant).

\begin{equation}
\label{eq:T_FFT}
  \mathbf{T} = \mathbf{F}^{\dagger}\,
  \operatorname{diag}\!\Bigl[\frac{\hbar^{2}k_{j}^{2}}{2\mu}\Bigr]\,
  \mathbf{F},
  \qquad
  k_{j} = \frac{2\pi j}{N\,\Delta R},
  \quad j = 0,\dots, N-1,
\end{equation}
where $\mathbf{F}$ is the discrete Fourier transform matrix.

Full diagonalization via \textsc{lapack} (\texttt{dsyevd}) returns the complete eigenspectrum $\{E_{k}, |\chi_{k}\rangle\}$.
Each eigenstate is projected onto the surface blocks to determine its ionic character
$P_{k}^{\text{ionic}} = \sum_{i}|\chi_{k}^{(1)}(R_{i})|^{2}$.
The Trotter-free autocorrelation is then
$C(t) = \sum_{k}|c_{k}|^{2}\,e^{-iE_{k}t/\hbar}$, with
$c_{k} = \langle\chi_{k}|\Psi(0)\rangle$.
The spectrum is computed using the same processing pipeline as the
circuit simulation: exponential damping ($\tau_{\text{damp}} = 30{,}000\;\text{a.u.}$), Blackman window, and $4\times$ zero-padded FFT.

\subsubsection{Observable Benchmarks}
\label{subsubsec:morse_observables}

\begin{figure}[t]
    \centering
    \includegraphics[width=\columnwidth]{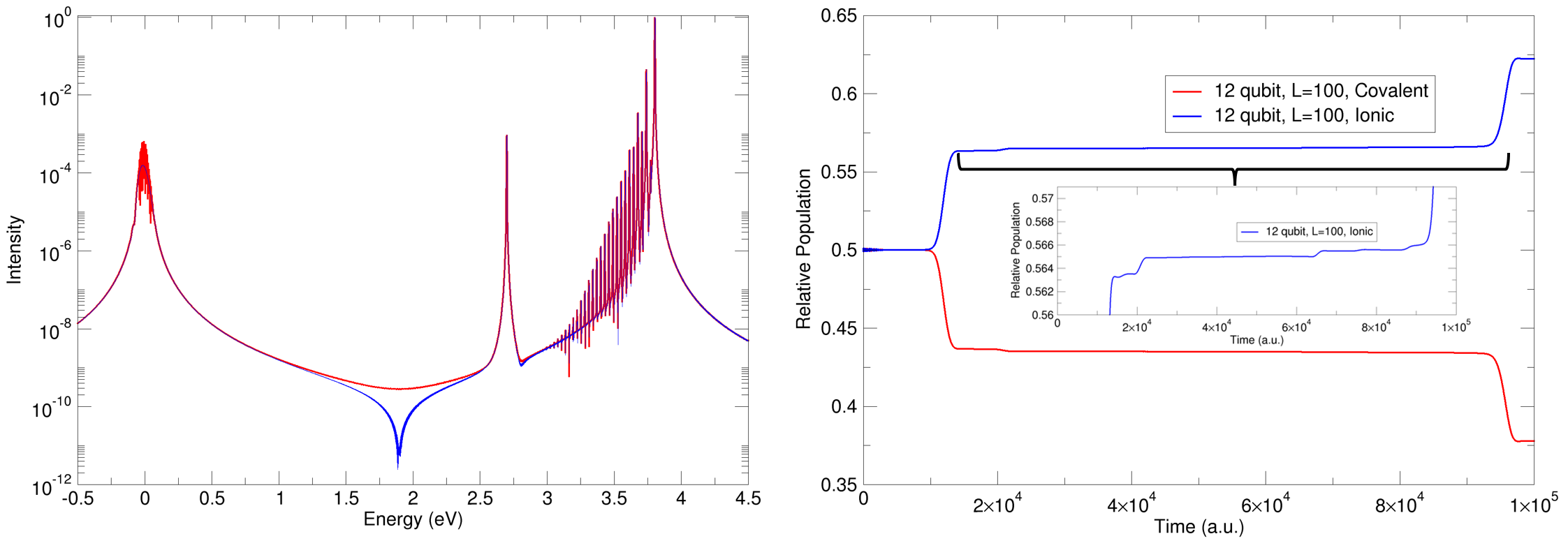}
    \caption{Observable benchmarks for the NaI model.
    (left)~Absorption spectrum from the Fourier transform of the
    autocorrelation function.
    Classical diagonalization (red) and quantum circuit emulation (blue)
    agree on features above the Trotter noise floor (${\sim}\,10^{-5}$).
    The Morse vibrational progression ($3.0$--$3.85\;\text{eV}$) and
    covalent continuum (${\sim}\,0\;\text{eV}$) are separated by a
    gap with Trotter-limited depth ($\propto \tau^{2}$).
    (right)~Electronic population dynamics showing two Landau--Zener
    transfer events, each trapping ${\approx}\,12\%$ of the remaining
    covalent population.
    The first step ($t \approx 12{,}000\;\text{a.u.}$) is the physical
    nonadiabatic crossing; the second ($t \approx 90{,}000\;\text{a.u.}$)
    is a finite-box recurrence.
    Inset: fine structure of the ionic population on the plateau,
    showing temporally resolved crossing and slow
    drift from anharmonically dephased trapped-state dynamics.}
    \label{fig:morse_observables}
\end{figure}

Figure~\ref{fig:morse_observables}(right) shows diabatic
populations from the circuit emulation.
Both surfaces begin at $P = 0.50$ by construction.
Two discrete population-transfer events are visible:

\begin{enumerate}
\item \emph{First crossing} ($t \approx 12{,}000\;\text{a.u.}$,
  $290\;\text{fs}$). The covalent wavepacket reaches $R_{c}$ and
  transfers $\Delta P \approx 0.065$ to the ionic surface, raising
  $P_{\text{ionic}}$ from $0.500$ to $0.565$.
  The effective trapping fraction
  $\Delta P / P_{\text{cov}} = 0.065/0.50 = 13\%$ agrees with the
  Landau--Zener prediction of $12\%$.

\item \emph{Box-reflection recurrence} ($t \approx 90{,}000\;\text{a.u.}$,
  $2.18\;\text{ps}$). The covalent fragment returns through $R_{c}$ and transfers an additional
  $\Delta P \approx 0.059$.
  The ratio of successive step heights
  $\Delta P_{2}/\Delta P_{1} = 0.059/0.065 = 0.91$ agrees with
  the expected geometric decay $P_{\text{esc}} = 0.88$:
  each crossing removes the same fraction ($12\%$) of the
  \emph{remaining} covalent population.
  The step separation of ${\sim}\,78{,}000\;\text{a.u.}$ is consistent
  with the round-trip estimate~\eqref{eq:t_roundtrip} of
  $73{,}500\;\text{a.u.}$
\end{enumerate}

The inset in Figure~\ref{fig:morse_observables}(right) reveals the first Landau--Zener event.
Rather than a sharp step, the ionic population rises over a ramp
spanning ${\sim}\,6{,}000\;\text{a.u.}$ ($145\;\text{fs}$), consistent
with the time for the dispersed wavepacket
($\sigma \approx 1.1\;a_{0}$ at $R_{c}$) to traverse the
$7\;a_{0}$-wide coupling region at $v_{c} = 1.27 \times 10^{-3}\;\text{a.u.}$
A gradual upward drift of ${\sim}\,0.001$ in $P_{\text{ionic}}$ over
${\sim}\,50{,}000\;\text{a.u.}$ is clearly visible, coming from the ${\sim}\,6\%$ of population trapped onto
high-lying ionic Morse states ($v \approx 80$--$180$) at the first
crossing.
Trapped $v$-components oscillate with their own classical period and individually traverse
$R_{c}$ at staggered times, indicating anharmonic dephasing.

Figure~\ref{fig:morse_observables}(left) shows the absorption spectrum
obtained from the Fourier transform of the autocorrelation function.
Under the FFT convention
($\tilde{C}(f) = \sum_{n} C(t_{n})\,e^{-2\pi i f t_{n}}$),
spectral peaks appear at displayed energies
$E_{\text{disp}} = -E_{k} \times E_{h} + E_{\text{offset}}$,
where $E_{\text{offset}} = 2.70\;\text{eV}$ is the vertical
excitation gap.

\begin{enumerate}
\item \emph{Morse vibrational progression (3.0--3.85 eV).}
  A comb of narrow peaks corresponds to the ionic Morse vibrational states $v = 0, 1, 2, \ldots$
  The $v = 0$ peak at $3.80\;\text{eV}$ carries the largest weight
  ($|c_{0}|^{2} \approx 0.48$).
  Peak spacing of ${\sim}\,32\;\text{meV}$ ($259\;\text{cm}^{-1}$)
  reflects $\omega_{e}$, with the Franck--Condon envelope decaying
  steeply: $v = 1$ is suppressed by ${\sim}\,20\times$ relative to $v = 0$,
  and states beyond $v \approx 5$ fall below $10^{-6}$ relative intensity.

\item \emph{(1.7--2.5 eV).}
  A deep minimum separates the Morse progression from the covalent
  continuum.
  The minimum at ${\sim}\,1.86\;\text{eV}$ maps to Morse
  $v \approx 50$, which carries negligible Franck--Condon weight from
  either surface.
  The exact benchmark reaches ${\sim}\,10^{-11}$ in this region.
  The circuit emulation shows a smooth floor at
  ${\sim}\,10^{-6}$ to $10^{-8}$, reflecting Trotter error raising the noise floor.

\item \emph{Covalent continuum (${\sim}\,0\;\text{eV}$).}
  The broad feature centered near $E_{\text{disp}} \approx 0$
  arises from the covalent half of the initial state.
  $V_{\text{cov}}(R_{e}) = 2.70\;\text{eV}$ maps to
  $E_{\text{disp}} \approx 0$ under the sign inversion, 
  populating a range of classical scattering states in the continuum.
\end{enumerate}

A peak at $E_{\text{disp}} = 2.70\;\text{eV}$ corresponds
to the $f = 0$ (DC) bin of the FFT.
This feature shifts or vanishes under alternative DC-removal strategies
and carries no physical content.

The circuit emulation's spectrum agrees with the exact benchmark for
all features above ${\sim}\,10^{-5}$ relative intensity.
Below this, the circuit trace exhibits a smooth noise floor, arising from spectral weight leakage.

\subsection{Modified H\'{e}non-Heiles Potential}
\label{sec:henon_heiles_validation}

\subsubsection{Model Specification}
\label{subsubsec:hh_model}

The modified H\'{e}non-Heiles system tests the multilinear expansion on an inseparable multidimensional potential surface.
Any function $f$ on the grid $\{0,\dots,2^n{-}1\}\times\{0,\dots,2^m{-}1\}$ admits the closed-form expansion
\begin{equation}\label{eq:2d-expansion}
    f(x,y) = \sum_{\substack{S\subseteq\{0,\dots,n-1\}\\[2pt]T\subseteq\{0,\dots,m-1\}}} c_{S,T}\;\prod_{j\in S}q_j\;\prod_{k\in T}r_k,
\end{equation}
with $c_{S,T}$ given by  
\begin{equation}\label{eq:2d-coeff}
    c_{S,T} \;\equiv\; \bigl(\Delta_S^{(x)}\Delta_T^{(y)} f\bigr)(0,0) \;=\; \sum_{A\subseteq S}\sum_{B\subseteq T}(-1)^{|S\setminus A|+|T\setminus B|}\; f\!\left(\sum_{j\in A}2^j,\;\sum_{k\in B}2^k\right).
\end{equation}
A detailed derivation is given in Appendix~\ref{app:circuits}.

The set of monomials $\{\prod_{j\in S}q_j\prod_{k\in T}r_k\}_{S,T}$ forms a basis for all functions on $\{0,1\}^{n+m}$ (the multilinear polynomials over GF(2) extended to $\mathbb{R}$). 
Extending from the univariate case, since the binary encoding maps the grid bijectively to $\{0,1\}^{n+m}$, any function on the grid has a unique expansion in this basis.
The coefficients are determined by the requirement that the expansion reproduce $f$ at all grid points, equivalent to Eq.~\ref{eq:2d-coeff} by the M\"{o}bius inversion formula on the Boolean lattice.

Extending to $d$-variate potential energy surfaces, they take the form 
\begin{equation}\label{eq:d-expansion}
    f(x_1,\dots,x_d) = \sum_{S_1,\dots,S_d} c_{S_1,\dots,S_d}\;\prod_{i=1}^d\prod_{j\in S_i}q_j^{(i)}\,,
\end{equation}
with 
\begin{equation}
    c_{S_1,\dots,S_d} = \sum_{A_1\subseteq S_1}\cdots\sum_{A_d\subseteq S_d}(-1)^{\sum_{i=1}^d |S_i\setminus A_i|}\;f\!\left(\sum_{j\in A_1}2^j,\;\dots,\;\sum_{j\in A_d}2^j\right)\,,
\end{equation}
for subsets $S_i\subseteq\{0,\dots,n_i{-}1\}$, $i=1,\dots,d$.
The circuit consists of at most $2^{n_1+\cdots+n_d}$ multi-controlled $R_z$ gates, each with controls drawn from up to $d$ registers simultaneously.

The H\'{e}non-Heiles system, when decomposed into gates using the multivariate expansion, introduces rotation angles that are zero. 
Counting the nonzero terms shows that a multivariate multilinear expansion, if reducible, will output a minimum number of rotation angles to construct the circuit, corresponding to the minimum number of controlled operations. 

The model comprises two coupled nuclear modes ($d = 2$) with $n = 12$ qubits per mode (24 total qubits, $4096 \times 4096$ grid), simulated by GPU-accelerated state-vector emulation on ideal, noiseless qubits via cuStateVec~\cite{nvidia2023cuquantum}.
The potential is
\begin{equation}\label{eq:hh_potential}
    V(x, y) = \tfrac{1}{2} k (x^2 + y^2) + \lambda(x^2 y - y^3/3) + \varepsilon\, (x^2 + y^2)^3\,,
\end{equation}
where the standard H\'{e}non-Heiles cubic coupling $\lambda(x^2 y - y^3/3)$ is the inseparable interaction, and the sextic regularization $\varepsilon\,(x^2+y^2)^3$ maintains bounded classical dynamics at energies near and above the classical escape energy $E_{\mathrm{esc}} = k^3/(6\lambda^2)$.
Parameters are collected in Table~\ref{tab:hh_params}.

\begin{table}[t]
\centering
\caption{Parameters for the modified H\'{e}non-Heiles model.}
\label{tab:hh_params}
\begin{tabular}{l l l}
\hline\hline
Parameter & Symbol & Value \\
\hline
Qubits per mode & $n$ & 12 (4096 grid points per axis) \\
Total qubits & $n_{\mathrm{tot}}$ & 24 \\
Mass & $\mu$ & $2000\,\text{a.u.}$ \\
Harmonic constant & $k$ & $0.2\,\text{a.u.}$ \\
Anharmonic coupling & $\lambda$ & 0.09--0.10 a.u. \\
Sextic regularization & $\varepsilon$ & $10^{-4}$--$10^{-5}$ a.u. \\
Box length & $L_x = L_y$ & 16--30 a.u. \\
Wavepacket width & $\sigma$ & 0.2--0.3 a.u. \\
Time step & $\tau$ & 0.5 a.u. \\
Propagation time & $T$ & 500--10,000 a.u. \\
\hline\hline
\end{tabular}
\end{table}

Time evolution employs symmetric second-order Trotter splitting (Strang splitting), alternating half-step potential phases $e^{-i\tau V/2}$ with a full-step kinetic propagator $e^{-i\tau T}$ applied in momentum space via the centered QFT.
The kinetic energy uses the gate-based polynomial decomposition (Tier~1--2 of Chapter~\ref{ch:model_ext}), while the potential energy is applied as a diagonal phase operator on the position register.
This is the Trotter structure of the REQWIEM architecture with $M=0$ channels, now applied to a two-register inseparable potential.

\begin{figure}[h]
     \centering
     \includegraphics[width=\textwidth]{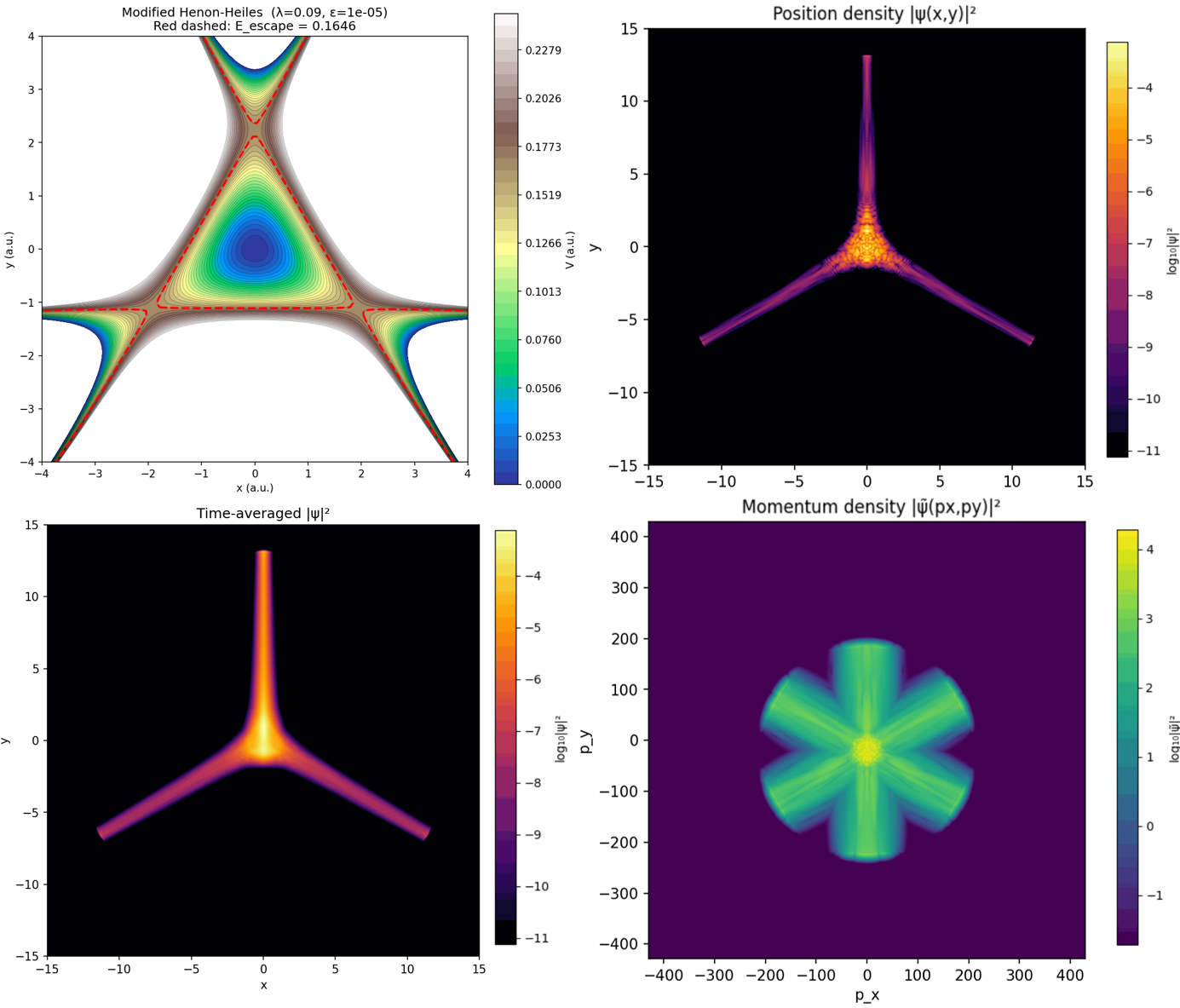}
     \caption{Overview of the modified H\'{e}non-Heiles system at $n = 12$ qubits per mode (24 total qubits).
     Upper left: potential energy surface $V(x,y)$ with $\lambda = 0.09$, $\varepsilon = 10^{-5}$; the red dashed contour marks the classical escape energy $E_{\mathrm{esc}} = 0.1646\;\text{a.u.}$
     Upper right: instantaneous position-space probability density $|\psi(x,y)|^2$ (log scale) at $t=10^4\,\text{a.u}$.
     Lower left: time-averaged density $\bar{\rho}(x,y)$ showing probability concentration along classical periodic orbit channels and the triangular confinement geometry.
     Lower right: momentum-space density $|\tilde{\psi}(p_x,p_y)|^2$ (log scale) at $t=10^4\,\text{a.u}$.}
     \label{fig:hh_snapshot}
\end{figure}

Figure~\ref{fig:hh_snapshot} presents an overview of the system.
The modified H\'{e}non-Heiles potential (upper right) exhibits C$_3$ symmetry with three saddle points at radius $r_{\mathrm{saddle}} = k/\lambda$ from the origin, through which the wavepacket can escape when its energy exceeds $E_{\mathrm{esc}}$.
The red dashed contour marks the classical escape energy.
An instantaneous snapshot of the position-space probability density (upper left) captures the wavepacket partially escaping through all three exit channels, confirming that the inseparable cubic coupling correctly routes amplitude along the symmetry-equivalent valleys.
The corresponding momentum-space density (lower left), obtained by applying the QFT to both registers, displays the conjugate six-fold structure reflecting the three preferred directions of outward flow and three of inward return.
The time-averaged density (lower right) accumulates probability over the full propagation interval, revealing the triangular confinement geometry and enhanced density along the classical periodic orbits connecting the potential center to each saddle point.

\subsubsection{Recurrence Structure and Classical Correspondence}
\label{subsubsec:hh_dynamics}

The inseparability of the H\'{e}non-Heiles potential prevents factorization into single-coordinate functions $g(x) \cdot h(y)$, making the system available as a benchmark for multidimensional potentials with nontrivial classical phase-space structure~\cite{henon1964applicability}.
This section validates the circuit-generated dynamics by comparing quantum recurrence timescales to the classical periodic orbit of the $y$-axis, quantifying the expected deviations from dimensional effects and eigenstate dephasing.

The $y$-axis ($x = 0$) is an invariant subspace of the H\'{e}non-Heiles dynamics: since $\partial V/\partial x|_{x=0} = 2\lambda \cdot 0 \cdot y = 0$ for all $y$, a trajectory initialized with $x = 0$, $p_x = 0$ remains on the $y$-axis for all time.
The motion reduces to a one-dimensional oscillation in the effective potential $V(0,y) = \tfrac{1}{2}ky^2 - \tfrac{1}{3}\lambda y^3 + \varepsilon y^6$, whose period is uniquely determined by the energy.
At $E = 0.10\;\text{a.u.}$, the asymmetric turning points are $y_{\min} = -1.00$ and $y_{\max} = 1.61\;\text{a.u.}$, reflecting the cubic term that softens the potential for $y > 0$ and steepens it for $y < 0$.
The classical period, obtained by symplectic integration from the upper turning point, is $T_{\mathrm{cl}} = 712.0\;\text{a.u}$.

Full variational integration of the 20-dimensional orbit-plus-monodromy system yields the Gutzwiller invariants~\cite{gutzwiller1971periodic}: classical action $S = 69.2$, Maslov index $\mu = 4$ (from four zero crossings of $\det(\Phi_{qq})$ per period), stability type ``stable'' with Greene's residue $R = 0.030$.
Symplecticity is preserved to machine precision ($\det(\Phi) = 1.000$), and the trivial eigenvalue pair deviation $|\lambda - 1| = 1.3 \times 10^{-3}$ confirms closure of the periodic orbit.
The anharmonic period dispersion curve $T(E)$ (Figure~\ref{fig:hh_dispersion}, upper panel), computed at 300 energies by symplectic integration, increases monotonically ($dT/dE = 1413\;\text{a.u.}$), a negative anharmonicity inherited from the cubic term.

\begin{figure}[h]
     \centering
     \includegraphics[width=\textwidth]{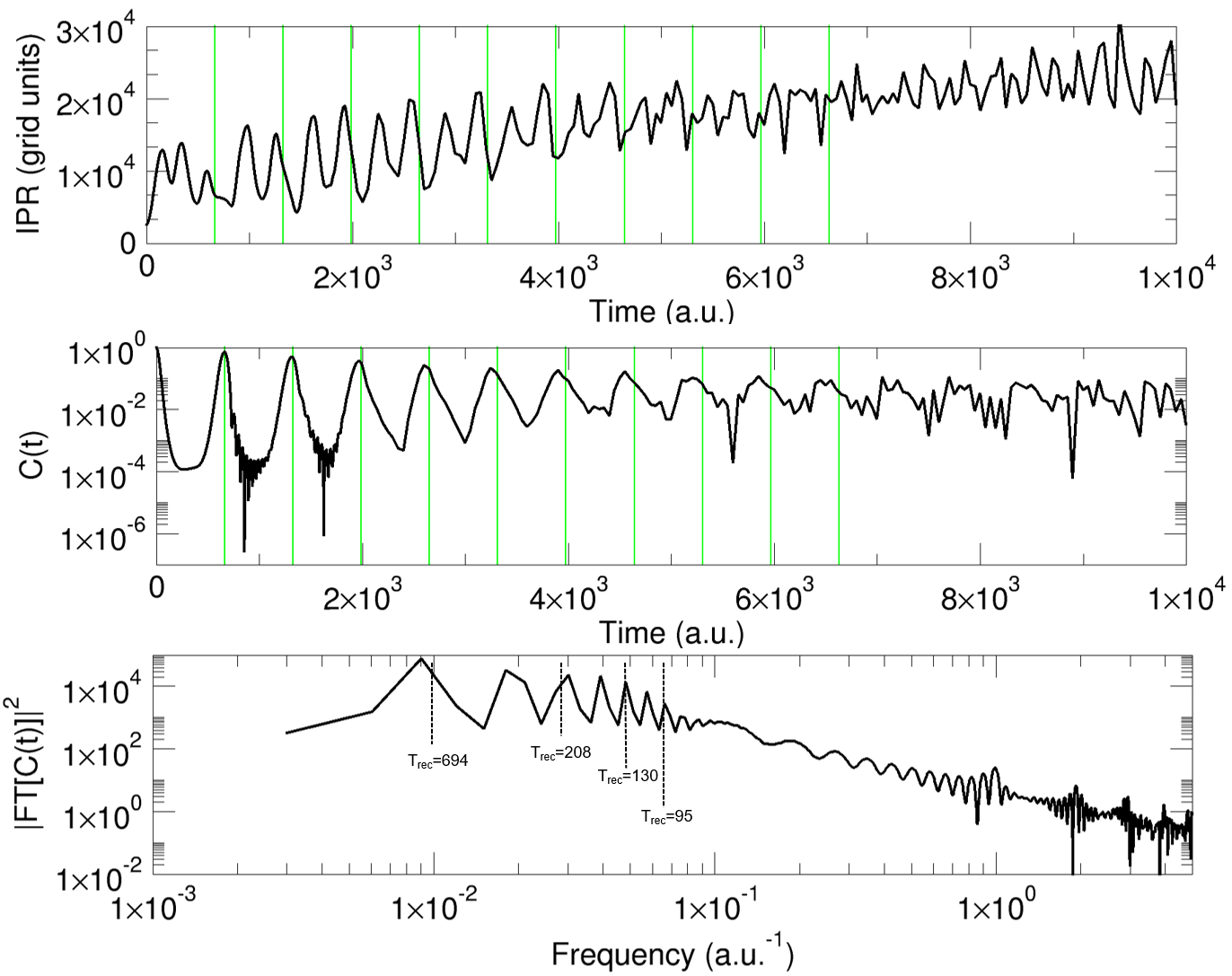}
     \caption{Autocorrelation diagnostics for the modified H\'{e}non-Heiles model at 24 qubits with initial energy $E = 0.10\;\text{a.u.}$ ($E/E_{\mathrm{esc}} = 0.61$).
     Upper: inverse participation ratio ($|\psi|^4$) showing regular periodicity on half-period intervals before decohering over time.
     Middle: autocorrelation $C(t) = |\langle\psi(0)|\psi(t)\rangle|^2$ showing a strong initial recurrence near $t \approx 663\;\text{a.u.}$ (scar strength $11.4\times$ ergodic baseline), with subsequent recurrences at contracting intervals.
     Lower: power spectrum $|\mathrm{FT}[C(t)]|^2$ with labeled recurrence periods $T_{\mathrm{rec}}$.
     Blue and cyan dashed lines mark the harmonic frequency $\omega_0 = \sqrt{k/\mu}$ and its first overtone.}
     \label{fig:hh_recurrence}
\end{figure}

\begin{figure}[h]
     \centering
     \includegraphics[width=\textwidth]{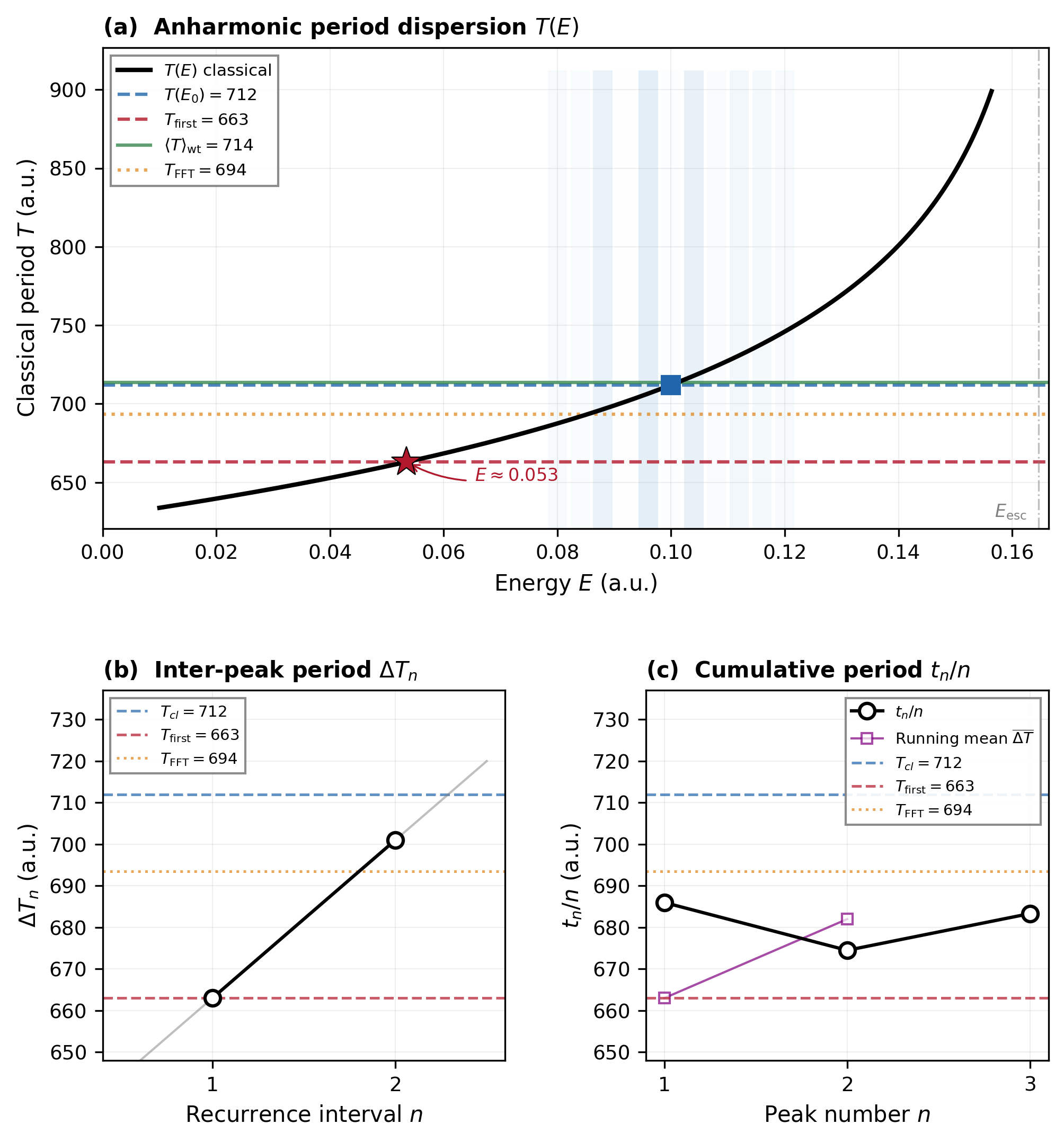}
     \caption{Classical orbit analysis and quantum-classical correspondence at $E = 0.10\;\text{a.u.}$
     Upper: anharmonic period dispersion $T(E)$ of the $y$-axis orbit (black curve), with the classical period at $E_0$ (blue dashed), the eigenstate-weighted average $\langle T \rangle$ (green), and the first quantum recurrence period (red dashed).
     The quantum recurrence at $T_{\mathrm{first}} = 663$ falls below $T(E_0) = 712$ due to transverse stiffening by the cubic coupling.
     The red star marks the energy at which $T(E)$ equals the first recurrence period, corresponding to a substantially lower effective energy than $E_0$.
     Lower left: inter-peak recurrence period $\Delta T_n$ showing contraction as dephasing isolates the dominant eigenstate spacing.
     Lower right: convergence of the cumulative average recurrence period $t_n/n$.}
     \label{fig:hh_dispersion}
\end{figure}

Figure~\ref{fig:hh_recurrence} presents the autocorrelation function $C(t) = |\langle\psi(0)|\psi(t)\rangle|^2$ and its power spectrum for an initial Gaussian wavepacket at $E = 0.10\;\text{a.u.}$ ($E/E_{\mathrm{esc}} = 0.61$).
The autocorrelation exhibits a strong initial recurrence near $t \approx 663\;\text{a.u.}$, with a scar strength (maximum recurrence amplitude divided by the ergodic baseline $\bar{C} \approx 0.022$) of $11.4\times$.
Subsequent recurrence peaks arrive at times slightly earlier than integer multiples of this fundamental period, a contraction effect analyzed below.
The power spectrum identifies a dominant peak at $T_{\mathrm{FFT}} = 694\;\text{a.u.}$, which represents a time-averaged frequency over the full propagation window and differs from the first-recurrence period due to dephasing dynamics.
Secondary spectral peaks appear at $T_{\mathrm{rec}} = 347$, $231$, $173$, $130$, $109$, $95$, $77$, and $63\;\text{a.u}$.

The first quantum recurrence occurs at $T_{\mathrm{first}} \approx 663\;\text{a.u.}$, establishing a clear offset from the classical $y$-axis orbit period:
\begin{equation}\label{eq:period_hierarchy}
    T_{\mathrm{first}} \approx 663 < T_{\mathrm{FFT}} = 694 < T_{\mathrm{cl}} = 712\;\text{a.u.}
\end{equation}
This $6.9\%$ shift to shorter quantum periods rules out the standard anharmonic dispersion mechanism, which would predict $T_{\mathrm{quantum}} > T_{\mathrm{cl}}$ for a potential with $dT/dE > 0$, and instead reflects the transverse stiffening of the effective potential by the cubic coupling.

The classical orbit is confined to the one-dimensional invariant subspace $x = 0$, where it experiences only $V(0,y)$.
The quantum wavepacket, by contrast, has finite width $\sigma = 0.15\;\text{a.u.}$ in both coordinates and samples the full two-dimensional potential from the first timestep.
The cross-term $\lambda x^2 y$ generates a transverse curvature $\partial^2 V/\partial x^2 = k + 2\lambda y$, which adds a correction $2\lambda y$ to the harmonic curvature $k = 0.2$.
At the time-averaged displacement $\langle y \rangle \approx 0.8\;\text{a.u.}$ along the orbit, this correction is $2\lambda \langle y \rangle \approx 0.14$, a $72\%$ enhancement of the transverse confinement.
Since the wavepacket spends the majority of each oscillation cycle in the $y > 0$ region (the upper turning point $y_{\max} = 1.61$ is $60\%$ farther from the origin than $|y_{\min}| = 1.00$), the time-averaged transverse stiffening is net positive and substantial.
The stiffer effective potential increases the eigenstate spacings $\Delta E_n$ of the full two-dimensional Hamiltonian relative to those of the one-dimensional invariant-subspace Hamiltonian, shortening the beat periods $2\pi/\Delta E_n$ and producing quantum recurrences at $T < T_{\mathrm{cl}}$.

For comparison, the sextic regularization contributes less than $0.2\%$ of the potential energy at the turning points ($\varepsilon y^6 \approx 1.7 \times 10^{-4}\;\text{a.u.}$ versus $V_{\mathrm{total}} = 0.10\;\text{a.u.}$) and a curvature correction $30\varepsilon y^4 \approx 2 \times 10^{-3}$ that is two orders of magnitude smaller than the cubic transverse stiffening.
The sextic term serves a structural role (ensuring a fully discrete spectrum at all energies) but does not contribute to the observed period shift at $E/E_{\mathrm{esc}} = 0.61$.

Peak-by-peak analysis of the autocorrelation function (Figure~\ref{fig:hh_dispersion}, lower panels) reveals a contracting recurrence structure: successive peaks arrive at times slightly earlier than integer multiples of $T_{\mathrm{first}}$, indicating that the effective inter-peak period decreases with each subsequent recurrence.
This contraction reflects the progressive isolation of the dominant eigenstate spacing by dephasing.

The initial wavepacket projects onto multiple eigenstates whose pairwise beat frequencies span a range $\delta\omega$.
The first recurrence reflects the population-weighted average of all contributing eigenstate spacings, producing a period $T_{\mathrm{first}} \approx 663\;\text{a.u.}$ that represents the full coherent superposition.
At later times, eigenstate pairs whose beat frequencies are incommensurate with the dominant spacing destructively interfere, and the surviving constructive interference is increasingly dominated by the single most strongly populated pair.
If this dominant pair has a slightly larger spacing $\Delta E$ (shorter period) than the population-weighted mean, the effective recurrence period contracts toward the spacing of the dominant pair.

The time-averaged FFT period $T_{\mathrm{FFT}} = 694\;\text{a.u.}$ does not correspond to a single physical timescale but represents an intermediate value biased by the window function and the time-varying coherence.
When the C$_3$-symmetric green reference lines are placed at multiples of $T_{\mathrm{first}} \approx 663\;\text{a.u.}$ rather than at $T_{\mathrm{FFT}}$, the first recurrence peak aligns to within the spectral resolution, while later peaks progressively precede the reference, consistent with the contracting inter-peak period.

The close agreement between the first quantum recurrence and the classical orbit period (within $7\%$), combined with the correct sign of the transverse stiffening correction, the expected contraction dynamics from multistate dephasing, and the reproduction of the correct C$_3$-symmetric escape topology, provides strong evidence that the two-register Trotterized propagation faithfully reproduces the inseparable multidimensional dynamics of the H\'{e}non-Heiles system.
For comparison, a Trotter error large enough to produce a $7\%$ period shift would require a step size $\tau \gtrsim 5\;\text{a.u.}$, an order of magnitude larger than the simulation value $\tau = 0.5\;\text{a.u.}$

\subsubsection{Quantum Scarring}
\label{subsubsec:hh_scarring}

The time-averaged position density $\bar{\rho}(x,y)$ accumulated over the propagation interval is compared to the microcanonical ergodic density $\rho_{\mathrm{mc}} \propto \sqrt{E - V(x,y)}$ to identify quantum scarring signatures~\cite{heller1984bound}.
The scar excess ratio $\log_{10}(\bar{\rho}/\rho_{\mathrm{mc}})$ localizes regions of enhanced (red) or depleted (blue) probability density relative to the ergodic baseline.

\begin{figure}[t]
     \centering
     \includegraphics[width=\textwidth]{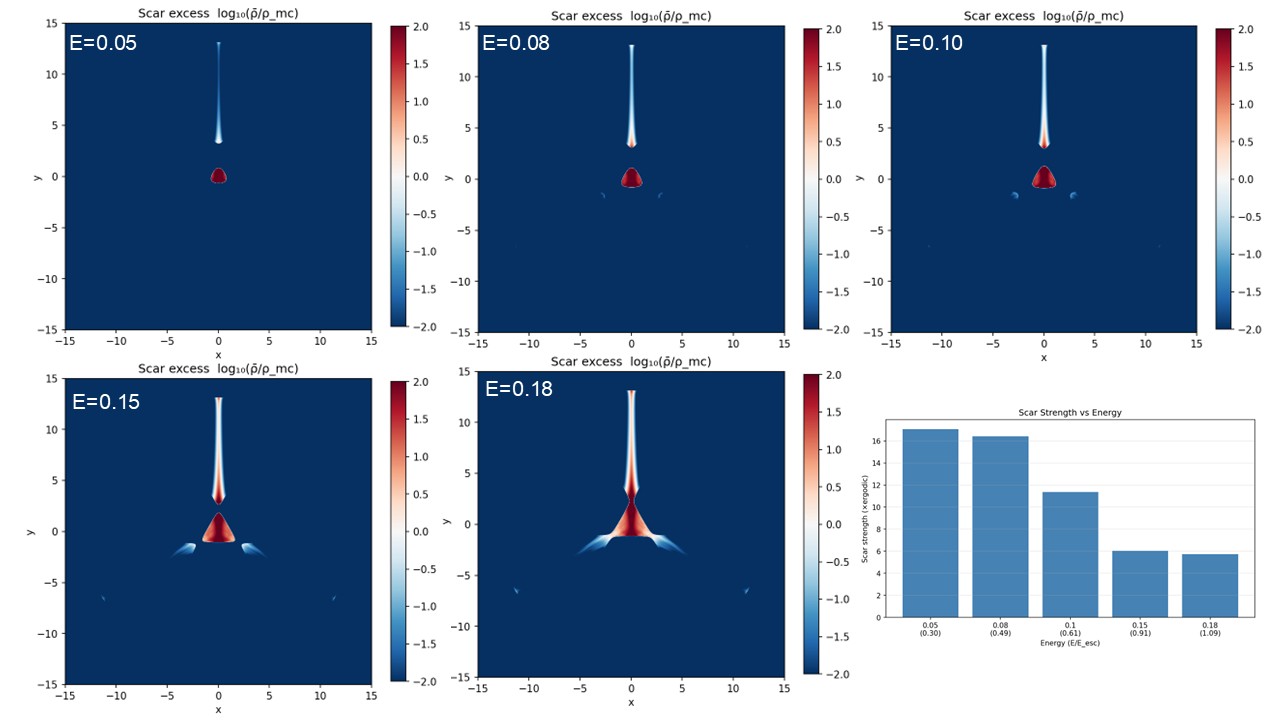}
     \caption{Quantum scarring in the modified H\'{e}non-Heiles system.
     Panels show the scar excess ratio $\log_{10}(\bar{\rho}/\rho_{\mathrm{mc}})$ for initial energies $E = 0.05$, $0.08$, $0.10$, $0.15$, and $0.18\;\text{a.u.}$ (corresponding to $E/E_{\mathrm{esc}} = 0.30$, $0.49$, $0.61$, $0.91$, and $1.09$).
     Red indicates density enhancement above the microcanonical baseline; blue indicates depletion.
     The dominant scar along the $y$-axis linear orbit weakens with increasing energy as the classical dynamics transitions from regular to chaotic.
     Lower right: scar strength (kurtosis excess) versus energy, showing the expected monotonic decrease.}
     \label{fig:hh_scarring}
\end{figure}

Figure~\ref{fig:hh_scarring} presents the scar excess for five initial energies spanning the range $E/E_{\mathrm{esc}} = 0.30$ to $1.09$.
At low energies ($E = 0.05\;\text{a.u.}$), the scar excess is strongly concentrated along the $y$-axis linear orbit, with the wavepacket density exceeding the ergodic prediction by up to two orders of magnitude along this trajectory.
As the initial energy increases, the classically accessible region expands and the scar pattern evolves: at $E = 0.10\;\text{a.u.}$ ($E/E_{\mathrm{esc}} = 0.61$), secondary scar features associated with the triangular orbit family become visible at the lower vertices of the potential.
Near the escape threshold ($E = 0.15$ and $0.18\;\text{a.u.}$), the scar excess broadens as the wavepacket begins to access the exit channels, and the triangular scarring intensifies as the classical dynamics approaches the homoclinic tangle of the saddle points.

The scar strength, quantified by the kurtosis excess of $\bar{\rho}/\rho_{\mathrm{mc}}$ (lower right panel of Figure~\ref{fig:hh_scarring}), decreases monotonically with increasing energy~\cite{heller1984bound}.
This trend is consistent with the expected transition from regular to chaotic classical dynamics in the H\'{e}non-Heiles system: at low energies, phase space is dominated by stable islands surrounding the periodic orbits, producing strong localization; as $E \to E_{\mathrm{esc}}$, the chaotic sea expands and the wavefunction distributes more uniformly, reducing the scar contrast.
The circuit-generated scar maps reproduce the known energy dependence of quantum scarring in this system, confirming that the Trotterized two-register propagation correctly captures the semiclassical localization structure.

\subsubsection{Circuit Sparsity and Gate Counts}
\label{subsubsec:hh_sparsity}

The H\'{e}non-Heiles potential (Eq.~\ref{eq:hh_potential}), despite being inseparable, is a finite-degree polynomial.
The harmonic term is degree 2 (Tier~1--2 in the REQWIEM decomposition), the cubic coupling is degree 3, and the sextic regularization is degree 6.
When the two-register multilinear M\"{o}bius expansion (Eq.~\ref{eq:2d-expansion}) is applied to this potential, the expansion coefficients $c_{S,T}$ are exactly zero for all subsets with $|S| + |T| > 6$ (or $|S| + |T| > 3$ if the sextic term is omitted).
The full $2^{n_x + n_y} = 2^{24} \approx 1.7 \times 10^7$ coefficient expansion collapses to $\sum_{k=0}^{a} \binom{n}{k}$ nonzero terms, where $a$ is the polynomial degree per coordinate.

\begin{figure}
    \centering
    \includegraphics[width=\textwidth]{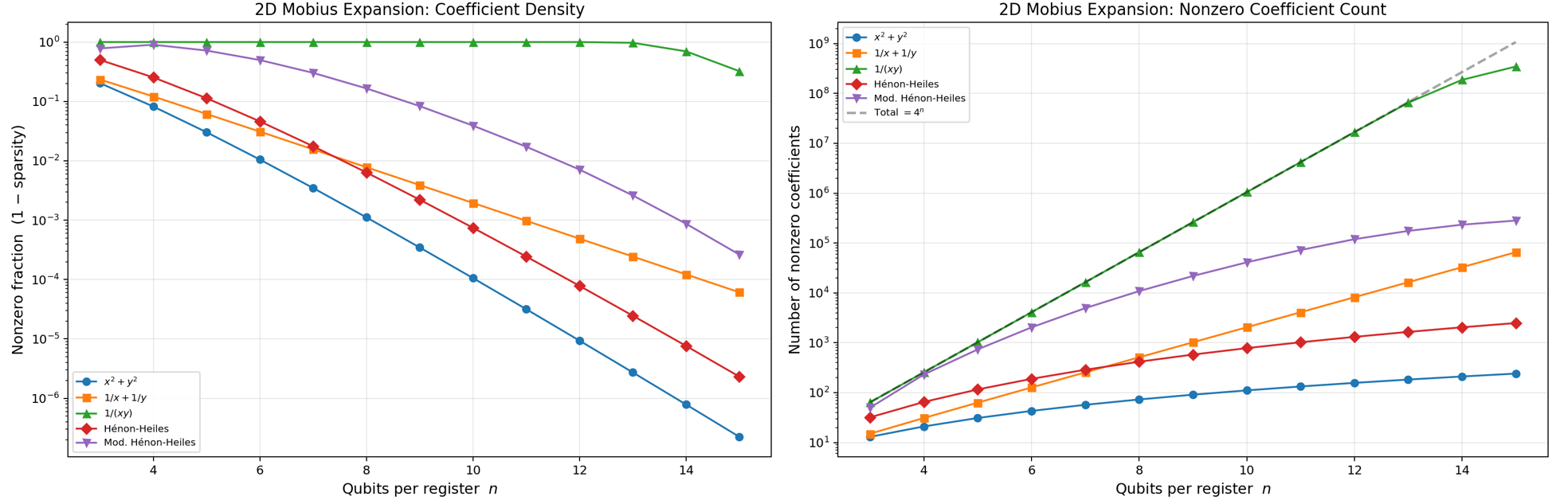}
    \caption{Circuit sparsity and gate count reduction for the modified H\'{e}non-Heiles model.
    Left: Fraction of nonzero coefficients. $f(x,y)=1/(xy)$ is observed as having all nonzero coefficients, where divergence from the maximum for larger registers is from coefficient magnitudes decreasing below of $10^{-16}$, a threshold set to prevent numerical error propagation in other functions.
    Right: rotation gate count versus register size $n$ (qubits per mode), comparing the full $2^{2n}$-term multilinear expansion (dashed) with the polynomial-reduced count after sparsity elimination (solid).
    At $n = 12$, the reduction exceeds 99.6\%, from ${\sim}1.7 \times 10^7$ to ${\sim}71{,}000$ gates.}
    \label{fig:hh_sparsity}
\end{figure}

Figure~\ref{fig:hh_sparsity} quantifies this reduction.
This calculation uses commuting tensor products to reduce the number of gates where possible, extending to long-double precision ($80$-bit with $64$-bit mantissa) for higher accuracy. 
A threshold value of $10^{-16}$ was set for numerical stability, with only extremely small values contributing to the asymptotically flat behavior of $f(x,y)=1/(xy)$ being truncated.
The left panel displays the coefficient magnitude distribution from the M\"{o}bius transform, organized by subset order $|S| + |T|$.
The left panel shows the nonzero fraction of coefficients, corresponding to the fraction of gates that need be applied for the 2D potential. 
Separable quadratic functions retain $\mathcal{O}(n^2)$ gate scaling, and separable reciprocal functions retain $\mathcal{O}(2^n)$ scaling. 
The unmodified H\'enon-Heiles potential has a cubic inter-mode term $\propto x^2y$ that significantly increases gate count, but retains polynomial complexity as $\mathcal{O}(n^3)$, while the modified variant scales as $\mathcal{O}(n^6)$.
A simple example of a maximally structured inseparable potential is $f(x,y)=1/(xy)$, requiring all $4^n$ gates have nonzero coefficients. 
The right panel shows the corresponding rotation gate count as a function of register size $n$, comparing the full $2^{2n}$ expansion (dashed) against the polynomial-reduced count (solid), confirming the collapse from exponential to polynomial scaling.

This validates the central sparsity claim of the hybrid strategy (Section~\ref{sec:gen_sites}): the general $\mathcal{O}(2^n)$ multilinear expansion, when applied to a degree-$a$ polynomial potential, automatically recovers the $\mathcal{O}(n^a)$ scaling of the dedicated polynomial decomposition.
Table~\ref{tab:hh_sparsity} gives numerical values for representative register sizes.

\begin{table}[t]
\centering
\caption{Nonzero coefficients for the modified H\'{e}non-Heiles model at various $n$.
The count matches the polynomial site prediction exactly, confirming automatic sparsity recovery.
The cubic-only column gives the count for $\varepsilon = 0$ (degree 3); the full column includes the sextic regularization (degree 6).}
\label{tab:hh_sparsity}
\begin{tabular}{c r r r r}
\hline\hline
$n$ & $2^{2n}$ (full expansion) & Cubic only ($a=3$) & Full ($a=6$) & Sparsity \\
\hline
7  & 16,384     & 238 & 1,869 & 88.6$\%$ \\
8  & 65,536     & 352 & 4,636 & 92.9$\%$\\
10 & 1,048,576  & 680 & 21,044 & 98.0$\%$\\
12 & 16,777,216 & 1,168 & 71,150& 99.6$\%$ \\
\hline\hline
\end{tabular}
\end{table}

The H\'{e}non-Heiles system also clarifies the interaction between single-register and cross-register terms.
The harmonic part $\frac{1}{2}k(x^2 + y^2)$ is separable and produces only $|S_x| \leq 2$, $|S_y| = 0$ and $|S_x| = 0$, $|S_y| \leq 2$ terms.
The cubic coupling $\lambda(x^2 y - y^3/3)$ produces cross-register terms with $|S_x| + |S_y| \leq 3$, including the genuinely inseparable sites $|S_x| \geq 1, |S_y| \geq 1$.
The sextic regularization $\varepsilon(x^2+y^2)^3$ produces terms up to $|S_x| + |S_y| = 6$, partially separable under algebraic expansion.
Tier~2$'$ dominates the total site count at $n = 12$, but the exponential suppression of higher-order coefficients for polynomial potentials ensures that the circuit depth remains $\mathcal{O}(\mathrm{poly}(n))$ rather than $\mathcal{O}(2^n)$.

\section{Summary}
\label{sec:smooth_summary}
\subsection{Scope of Application}
\label{sec:scope}

The REQWIEM architecture, as extended in this chapter, can simulate any molecular Hamiltonian of the form
\begin{equation}\label{eq:general_hamiltonian}
    {H} = {T} + \sum_{s} \ket{s}\!\bra{s} \otimes {V}^{(s)}(\mathbf{x}) + \sum_{i < j} \left(\ket{i}\!\bra{j} + \ket{j}\!\bra{i}\right) \otimes {G}_{ij}(\mathbf{x})\,,
\end{equation}
where ${T}$ is the kinetic energy operator (separable in Cartesian coordinates), ${V}^{(s)}(\mathbf{x})$ is the on-diagonal potential on a diabatic surface $s$, and ${G}_{ij}(\mathbf{x})$ are off-diagonal coupling functions.
The only structural requirement is that the coordinates are encoded in position registers via the standard binary encoding, such that the kinetic operator $T$ remains dependent on momentum $p$, not position $x$. 
Chapter~\ref{ch:Application} discusses coordinate transformations and variable placement to maintain this condition, as commonly exhibited in many-body systems.

On-diagonal and off-diagonal potentials can be arbitrary combinations of polynomials of any degree, handled by the binary decomposition for degree $\leq 2$ and by the multilinear (or custom) expansion for degree $> 2$; transcendental functions of single coordinates; inverse polynomials and rational functions, treated via analytical, rational arithmetic as established in Chapter~\ref{ch:smooth_potentials}; and tabulated ab initio data with no analytical form required (one is guaranteed on a discretized finite interval).
Different functional forms can coexist within the same Trotter step: harmonic modes use the $\mathcal{O}(n^2)$ polynomial decomposition, dissociative or reactive modes use the $\mathcal{O}(2^n)$ multilinear expansion, and inverse polynomial modes use exact rational arithmetic, multiplexed through the same UCR protocol and fan-out infrastructure.

The span of physical problems accessible to REQWIEM therefore currently includes nonadiabatic photochemistry (internal conversion, intersystem crossing, conical intersection dynamics), molecule--surface scattering with Morse/exponential potentials and analytical coupling functions, gas-phase reactive scattering on multiple Born--Oppenheimer surfaces, Coulombic and long-range charge-transfer dynamics, and model Hamiltonians (spin--boson, Holstein, Peierls) where bath spectral densities determine potential shapes that need not be polynomial.

\subsection{Comparison with Existing Methods}
\label{sec:method_comparison}

The smooth-potential extension positions REQWIEM relative to QROM methods, including a recent proposal for Morse potentials~\cite{lang2026quantum, motlagh2025quantum}.
The full architectural comparison for vibronic dynamics, including the per-term non-Clifford overhead $\alpha_{\mathrm{smooth}} = (2+b)/M$, the fan-out parallelism $\phi \approx d/2$, and the $T$-gate crossover analysis, is given in Chapter~\ref{ch:resources}.
The combined advantage $(2+b)\,d/(2M)$ grows linearly with both the phase register precision and the mode count, and holds for any functional form.

QROM caching optimization of Ref.~\cite{lang2026quantum} fails for non-separable surface-dependent potentials, when potentials motivate a smooth-potential extension.
For polynomial terms where the caching optimization succeeds, the two architectures have comparable per-term overhead, and the REQWIEM advantage reduces to the fan-out factor alone.
A broader comparison against QSP with Wigner--Haagerup LCU block encoding, QSP with kinetic--potential splitting, QSP in a truncated Fock space, and the end-to-end estimate of Eklund et al.\ \cite{eklund2026endtoend}, is given in Chapter~\ref{ch:comparison}.
That analysis finds $\Tgate$-gate ratios ranging from $1.5\times$ (QSP Fock) to $\sim 3.1\times 10^9\times$ (Eklund et al.) on the pyrazine benchmark, with the gap amplifying at relaxed synthesis precision due to REQWIEM's optional-Toffoli architecture.

\subsection{Limitations}
\label{sec:smooth_limitations}

The $\mathcal{O}(2^n)$ site count per analytical function per mode is the dominant cost of the smooth-potential extension.
This exponential scaling is information-theoretic and shared with the QROM approach, being a well-known bottleneck of first-quantized, grid-based approaches~\cite{kassal2008polynomial, babbush2018encoding}.
As a chosen representation, merely placing the system on a grid defined by qubit basis states does not change this bottleneck, though often branded as $\mathcal{O}(N)$ scaling, where $N=2^n$.
The limitation is mitigated given an observation in molecular modeling, where most molecular systems require analytical functions on only a small number of reactive modes ($N_f^{(\kappa)} = 1$ for $\sim$1--6 modes), with remaining modes handled at polynomial cost.

The kinetic energy operator separability enabling QFT-based kinetic propagator requires Cartesian coordinates.
Curvilinear coordinates introduce coordinate-dependent kinetic energy terms that are not diagonal in any simple basis (e.g. Ammonia,~\cite{hause2008vibrationally}). 
Extending REQWIEM to naturally consider curvilinear coordinates is nontrivial and is not addressed in the present work.

REQWIEM is intrinsically a Trotterized time evolution method. 
Second-order symmetric Trotter splitting incurs error $\mathcal{O}(\tau^3)$ per step, requiring $N_{\mathrm{Trotter}} = \mathcal{O}(t/\tau)$ steps for total propagation time $t$.
If advances in qubitization, quantum signal processing, or lower-error split-operator methods are integrated with nonadiabatic coupling, fan-out, and coherent evolution, a more practical NA-QMD algorithm may result for future quantum computers~\cite{low2017optimal, gilyen2019quantum}.

Perhaps most importantly, gate counts and depths assume ideal gate execution.
The fault-tolerant $T$-depth analysis of Chapter~\ref{ch:resources} accounts for rotation synthesis overhead but does not include surface code, error-correcting code costs, or routing overhead.
The relative comparison between REQWIEM and QROM is preserved under any uniform compilation strategy, with an aforementioned note on the excess routing costs brought by QROM (Chapter~\ref{ch:resources}).

\subsection{Conclusion}
\label{sec:concluding}

Multilinear M\"{o}bius expansion allows any real-valued function of the nuclear coordinates to be encoded as a diagonal phase operator on the vibrational register, multiplexed across electronic surfaces via the UCR angle vector, and executed in parallel across vibrational modes via channel-register fan-out.
The harmonic regime of Chapters~\ref{ch:model_ext} and ~\ref{ch:resources} is recovered as a special case $a_S = 0$ for $|S| \geq 3$; the inverse polynomial regime of Chapter~\ref{ch:scattering} is recovered as the special case where exact rational arithmetic eliminates floating-point error; and the general smooth-potential regime of this chapter subsumes both.

The three claims stated in Section~\ref{sec:smooth_intro} are established. 
A multilinear expansion provides a closed-form, truncation-free decomposition of arbitrary $f(x)$ into $2^n$ diagonal phase operators, implementable as CNOT cascades with a single $R_z$ (or UCR$_Z^{(m)}$) rotation per term.
UCR composition absorbs surface-dependent coefficients $a_S^{(s)}$ into the angle vector at zero gate overhead, providing a factor-of-$M$ composition advantage over QROM methods for non-separable analytical potentials.
Depth advantage is universal, holding for all functional forms, amplifies with the introduction of analytical potentials ($\alpha_{\mathrm{smooth}} \geq \alpha_{\mathrm{poly}}$), and is preserved under additive decomposition of Hamiltonians containing both polynomial and analytical terms.

The smooth-potential extension transforms REQWIEM into a general-purpose architecture for quantum simulation of nuclear dynamics on arbitrary potential energy surfaces.
The architecture is now limited not by the functional form of the potential but by the grid resolution ($n \leq 14$--$15$), the Cartesian coordinate restriction of the kinetic operator, and the Trotter step count.
Prior to Chapter~\ref{ch:Application}, which considers the complete REQWIEM framework holistically, Chapter~\ref{ch:comparison} provides a quantitativecomparison of the Trotter-based architecture against QSP-based alternatives, establishing that the depth and $\Tgate$-gate advantages demonstrated for QROM methods extend when measured against QSP and qubitization.
Chapter~\ref{ch:Application} then considers future quantum hardware application in the fault-tolerant era, future research directions for further optimization, and how the developed architecture extends to quantum mechanical problems placed in a finite basis representation (FBR) / discrete variable representation (DVR) dualism.

\chapter{Comparison with Polynomial Quantum Simulation Methods}\label{ch:comparison}

The preceding chapters developed REQWIEM as an oracle-free, Trotter-based propagator for nonadiabatic quantum molecular dynamics on diabatic potential energy surfaces.
Gate counts were derived in Chapter~\ref{ch:resources}, circuit depth was analyzed in Section~\ref{sec:reqwiem_costs}, and the extension to arbitrary smooth potentials was completed in Chapter~\ref{ch:smooth_potentials}.
A natural question resides in the ``oracle-free" nature itself, and whether powerful algorithms for Hamiltonian simulation like quantum signal processing (QSP) could achieve lower cost for the same physical problem~\cite{low2017optimal, motlagh2024generalized}.

QSP achieves \emph{optimal query complexity} for Hamiltonian simulation: the number of calls to a block-encoding oracle scales as $\mathcal{O}(\alpha t + \log(1/\eps))$, where $\alpha$ is the block-encoding normalization and $t$ the evolution time~\cite{low2017optimal, gilyen2019quantum}.
Translation from query complexity to gate complexity, then calculating T-gates and T-depth requires specifying the oracle, where block-encoding costs for a diabatic Hamiltonian on a position-space grid become the dominant expense.
As Childs et al.\ note, product formulas can outperform asymptotically optimal methods precisely when commutator structure makes the Hamiltonian ``easier than its norm suggests''~\cite{childs2021theory}.
This chapter decomposes this note in model systems.

This chapter provides a quantitative, five-way comparison between:
\begin{enumerate}[label=(\roman*)]
    \item REQWIEM (Chapters~\ref{ch:wavepacket_prop}--\ref{ch:smooth_potentials}),
    \item QSP with Walsh--Hadamard LCU block encoding (WH-LCU),
    \item QSP with a two-term $T{+}V$ split block encoding,
    \item QSP in the Fock (harmonic oscillator) basis,
    \item the full-Coulomb end-to-end algorithm of Eklund et al.~\cite{eklund2026endtoend}.
\end{enumerate}
All comparisons use the pyrazine $S_1/S_2$ four-mode QVC model of Raab et al.~\cite{raab1999molecular} at converged parameters ($d=4$ modes, $M=2$ electronic states, $n=10$ qubits per mode, $t=4000$~a.u., $\eps = 10^{-3}$), with physical parameters given in Appendix~\ref{app:pyrazine_params}.
The chapter concludes with a consideration of the non-perturbative convergence barrier near conical intersections and its implications for perturbative and polynomial simulation methods.

\section{QSP on Grid-Discretized Diabatic Surfaces}\label{sec:qsp_grid}

\subsection{Walsh--Hadamard LCU block encoding}\label{sec:wh_lcu}

The vibronic Hamiltonian (Eq.~\ref{eq:vibronic_ham_full}) can be decomposed as a linear combination of unitaries (LCU) by expanding the diagonal potential surfaces in the Walsh--Hadamard basis~\cite{childs2012hamiltonian}.
On a grid of $N = 2^n$ points per mode in $d$ dimensions, the diagonal surface $V_{ss}(Q_1,\ldots,Q_d)$ has up to $N^d = 2^{nd}$ distinct values.
The Walsh--Hadamard expansion requires up to $2^{nd}$ terms, giving a block-encoding normalization
\begin{equation}\label{eq:alpha_wh}
    \alpha_{\mathrm{WH}} = \sum_\ell |\alpha_\ell| = \|V\|_1^{(\mathrm{WH})},
\end{equation}
which grows exponentially with $nd$.
Explicitly, for a QVC Hamiltonian with maximum bilinear coefficient $|a_{rs}|_{\max}$ and grid spacing $\Delta$,
\begin{equation}\label{eq:alpha_wh_scaling}
    \alpha_{\mathrm{WH}} \sim \binom{d}{2} |a_{rs}|_{\max}\, \Delta^2 \cdot 2^{2(n-1)} = \mathcal{O}(d^2 \cdot 4^n).
\end{equation}
At $n=10$, $\alpha_{\mathrm{WH}} \approx 3.45 \times 10^4$; at $n=12$, $\alpha_{\mathrm{WH}} \approx 5.52 \times 10^5$, confirming a factor of $\sim\!16\times$ per two additional qubits.

The QSP degree for Hamiltonian simulation is~\cite{low2017optimal}
\begin{equation}\label{eq:dqsp}
    d_{\mathrm{QSP}} = \alpha\, t + \log_2(1/\eps),
\end{equation}
and each query to the WH-LCU oracle requires $\sim\!32$ Toffoli gates for the PREPARE/SELECT network.
At the pyrazine benchmark parameters, the WH-LCU encoding produces $d_{\mathrm{QSP}} \approx 1.38 \times 10^8$ queries and $\sim\!1.77 \times 10^{10}$ total T-gates, being $138.5\times$ more expensive than REQWIEM.

\subsection{Two-term $T{+}V$ split block encoding}\label{sec:tv_split}

A more efficient strategy block-encodes $H = T + V$ as a two-term LCU, with separate encodings for the kinetic and potential operators.
$T$ is diagonal in the momentum basis (accessed via QFT) and $V$ is diagonal in the position basis, each can be block-encoded at normalization cost equal to its spectral radius:
\begin{equation}\label{eq:alpha_tv}
    \alpha_{T{+}V} = \|T\|_{\max} + \|V\|_{\max}.
\end{equation}
The kinetic cutoff follows from the grid's maximum representable momentum (cf.\ Eq.~\ref{eq:nmin}):
\begin{equation}\label{eq:tmax}
    T_{\max} = \sum_r \frac{\pi^2}{2\mu_r \Delta_r^2} \propto \frac{4^n}{L^2},
\end{equation}
where $\Delta_r = L_r/2^n$ is the grid spacing for mode $r$.
This characteristic energy (Chapter~\ref{ch:wavepacket_prop}) acts as a \emph{spectral range penalty}: $\alpha_{T{+}V}$ is dominated by $T_{\max}$ at high resolution, even though physically populated kinetic energy $\langle T \rangle \sim 0.004$~a.u.\ is many orders of magnitude smaller.

\begin{table}[H]
\centering
\caption{Spectral range penalty: $\alpha_{T{+}V}$ scaling with grid resolution ($d=4$, $M=2$, pyrazine).}
\label{tab:spectral_range}
\begin{tabular}{ccccc}
\toprule
$n$ & $\Delta x$ (a.u.) & $T_{\max}$/mode (lightest) & $\alpha_{T{+}V}$ & $d_{\mathrm{QSP}}$ \\
\midrule
6   & 0.3125  & 0.03     & 3.1     & $\sim 1.2 \times 10^4$  \\
8   & 0.0781  & 0.44     & 14.7    & $\sim 5.9 \times 10^4$  \\
10  & 0.0195  & 7.05     & 232.2   & $6.93 \times 10^4$      \\
12  & 0.00488 & 112.73   & $\sim 3{,}700$ & $\sim 1.5 \times 10^7$  \\
\bottomrule
\end{tabular}
\end{table}

Each QSP query to the $T{+}V$ block encoding requires $\sim\!1{,}800$ Toffoli gates (for arithmetic circuits and QFT rotations within the SELECT oracle).
At $n = 10$, the total T-gate cost, including both Toffoli decomposition ($C_T^{(\mathrm{Toff})} = 4$ per Toffoli) and QFT rotation synthesis ($C_T^{(\mathrm{rot})} = 47$ per $R_z$ at $\epsilon_{\text{synth}} = 10^{-10}$ via Ross--Selinger~\cite{ross2016optimal}), is $\sim\!1.68 \times 10^9$: a ratio of $13.2\times$ over REQWIEM.

\begin{table}[H]
\centering
\caption{QSP block-encoding comparison at the pyrazine benchmark ($n=10$, $d=4$, $M=2$).}
\label{tab:qsp_encodings}
\begin{tabular}{lcccc}
\toprule
Encoding & $\alpha$ & $d_{\mathrm{QSP}}$ & Toffoli/query & Total T-gates \\
\midrule
WH-LCU      & $3.45\times 10^4$ & $1.38\times 10^8$ & 32     & $1.77\times 10^{10}$ \\
$T{+}V$ split & 232.2           & $6.93\times 10^4$ & 1{,}800 & $1.68\times 10^{9}$  \\
\bottomrule
\end{tabular}
\end{table}

\subsection{T-gate accounting}\label{sec:tgate_accounting}

Both Toffoli gates and $R_z$ rotations consume magic states on a fault-tolerant architecture.
The Toffoli gate decomposes into $C_T^{(\mathrm{Toff})} = 4$ T-gates~\cite{gidney2018halving}, and each $R_z(\theta)$ rotation is synthesized via the Ross--Selinger protocol~\cite{ross2016optimal}, reduced by the mixed-fallback and spider fusion (Appendix~\ref{app:tgate_pipeline} to $23$--$30$ T-gates per rotation~\cite{ross2016optimal}.
Both sources of T-gates consume one magic state each~\cite{bravyi2005universal}, making the total T-gate count is cost metric applied throughout this chapter.

\section{QSP in the Fock Basis}\label{sec:qsp_fock}

An alternative to grid discretization represents nuclear degrees of freedom in the harmonic oscillator (Fock) basis, truncated at $N_{\max}$ quanta per mode.
Each mode requires $n_{\mathrm{fock}} = \ceil{\log_2 N_{\max}}$ qubits.
The QVC Hamiltonian in this basis takes the form
\begin{equation}\label{eq:H_fock}
    H = \sum_k \omega_k \bigl(a_k^\dagger a_k + \tfrac{1}{2}\bigr) \otimes I_M + \sum_{ij,k} \kappa_k^{(ij)} (a_k + a_k^\dagger) \ket{i}\!\bra{j} + \cdots
\end{equation}
where creation/annihilation operators have bandwidth (sparsity) $s = 2d^2 + 1$ nonzero entries per row in the Fock representation.

Block encoding via sparse-access qubitization~\cite{babbush2018encoding} gives normalization $\alpha_{\mathrm{2Q}} = d^2 N_{\max}\, \max(|\kappa|, |\gamma|, \ldots)$.
For the four-mode pyrazine model at the MCTDH-converged Fock cutoff $N_{\max} = 30$~\cite{raab1999molecular, worth2004beyond}: $s = 33$, $\alpha_{\mathrm{2Q}} = 22.0$, $d_{\mathrm{QSP}} = 88{,}164$, and total logical qubits $= 70$~\ref{app:qsp_costs}.

\begin{table}[H]
\centering
\caption{QSP(Fock) scaling with Fock cutoff ($d=4$, $n=10$ REQWIEM baseline $= 1.28\times 10^8$ T-gates).}
\label{tab:fock_nmax}
\begin{tabular}{ccccccc}
\toprule
$N_{\max}$ & $n_{\mathrm{fock}}$ & $s$ & $\alpha_{\mathrm{2Q}}$ & $d_{\mathrm{QSP}}$ & T-gates & Ratio \\
\midrule
$10^\dagger$  & 4 & 33 & 11.4 & 45{,}765   & $7.87\times 10^7$ & 0.6$\times$ \\
20            & 5 & 33 & 16.7 & 66{,}964   & $1.43\times 10^8$ & 1.1$\times$ \\
30            & 5 & 33 & 22.0 & 88{,}164   & $1.88\times 10^8$ & 1.5$\times$ \\
40            & 6 & 33 & 27.3 & 109{,}363  & $3.03\times 10^8$ & 2.4$\times$ \\
100           & 7 & 33 & 59.1 & 236{,}558  & $9.08\times 10^8$ & 7.1$\times$ \\
\bottomrule
\multicolumn{7}{l}{\footnotesize $^\dagger$Sub-converged for pyrazine; the 0.6$\times$ ratio is not physically meaningful.}
\end{tabular}
\end{table}

At the converged cutoff $N_{\max} = 30$, QSP(Fock) produces $1.88 \times 10^8$ T-gates against REQWIEM's $1.28 \times 10^8$: a ratio of only $1.5\times$.
This is the most competitive QSP encoding for harmonic or near-harmonic systems.
The advantage grows rapidly with both $N_{\max}$ (anharmonicity) and $d$ (dimensionality): at $d=24$ the ratio increases to $38.1\times$ in the harmonic limit.

The sparsity formula $s = 2d^2 + 1$ counts only on-diagonal electronic block connections; including off-diagonal coupling increases $s$ by $\mathcal{O}(d)$ per state pair, making the reported QSP(Fock) costs conservative lower bounds.

\section{The Anharmonic Wall}\label{sec:anharmonic_wall}

The QSP(Fock) near-parity at $N_{\max} = 30$ is specific to the \emph{harmonic} limit of the QVC model.
For arbitrary smooth potential energy surfaces $V(Q_1,\ldots,Q_d)$ (Ch.~\ref{ch:smooth_potentials}), the Fock-basis oracle must encode $V$ via QROM lookup of a $d$-dimensional tensor.
The effective sparsity then scales as
\begin{equation}\label{eq:seff_qrom}
    s_{\mathrm{eff}} = \mathcal{O}(N_{\max}^d),
\end{equation}
because every grid point in the $d$-dimensional Fock space must be individually addressed.

REQWIEM avoids this entirely, whereby the multilinear expansion of $V$ (Eq.~\ref{eq:multilinear_expansion}) composes with the UCR multiplexing protocol of Chapter~\ref{ch:smooth_potentials}, optimizing Toffoli-free application through the Gray-code construction (Eq.~\ref{eq:graycode_gates}).

\begin{table}[H]
\centering
\caption{Anharmonic regime: QSP(Fock) vs.\ REQWIEM ($d=4$, $M=2$, $n=10$, REQWIEM baseline $= 1.28\times 10^8$).}
\label{tab:anharmonic}
\begin{tabular}{lccccc}
\toprule
PES type & $N_{\max}$ & $s$ & $\alpha$ & QSP T-gates & Ratio \\
\midrule
Harmonic             & 30  & 33       & 22.0       & $1.88\times 10^{8}$  & 1.5$\times$    \\
Anharmonic ($\lambda\!=\!0.1$) & 36  & 49 & 37.4 & $4.28\times 10^{8}$  & 3.4$\times$    \\
Anharmonic ($\lambda\!=\!0.5$) & 60  & 49 & 56.3 & $6.44\times 10^{8}$  & 5.0$\times$    \\
Morse ($D_e\!=\!5.0$~eV) & 122 & 196  & 420.5      & $7.27\times 10^{9}$  & 57.0$\times$   \\
Double-well          & 60  & 65       & 74.7       & $8.70\times 10^{8}$  & 6.8$\times$    \\
Conical intersection & 80  & 33       & 48.5       & $7.46\times 10^{8}$  & 5.8$\times$    \\
\textbf{Arbitrary smooth} & \textbf{80} & \textbf{194{,}481} & \textbf{286{,}049} & $\mathbf{9.02\times 10^{12}}$ & $\mathbf{70{,}706\times}$ \\
\bottomrule
\end{tabular}
\end{table}

The QROM scaling with dimensionality is shown in Table~\ref{tab:qrom_scaling}.
For the physically general problem (arbitrary smooth PES, $d > 4$), QSP(Fock) suffers the same exponential blowup as QSP(WH-LCU).
The effective sparsity $s_{\mathrm{eff}}$ is an optimistic estimate; the true QROM size is $\mathcal{O}(N_{\max}^d)$, again making reported ratios conservative lower bounds for non-polynomial potentials.

\begin{table}[H]
\centering
\caption{Arbitrary smooth PES: QROM scaling with dimensionality ($n=10$, $N_{\max}=80$).}
\label{tab:qrom_scaling}
\begin{tabular}{ccccc}
\toprule
$d$ & $s_{\mathrm{eff}}$ & QSP T-gates & REQWIEM T-gates & Ratio \\
\midrule
2  & 441                 & $4.78\times 10^{9}$    & $3.53\times 10^{7}$ & 135$\times$             \\
4  & 194{,}481           & $9.02\times 10^{12}$   & $1.28\times 10^{8}$ & $70{,}706\times$        \\
6  & $8.58\times 10^{7}$ & $5.33\times 10^{16}$   & $2.77\times 10^{8}$ & $1.93\times 10^{8}\times$ \\
8  & $3.78\times 10^{10}$& $6.53\times 10^{20}$   & $4.82\times 10^{8}$ & $1.35\times 10^{12}\times$ \\
\bottomrule
\end{tabular}
\end{table}

\section{Full-Coulomb Baseline: Eklund et al.}\label{sec:eklund}

Eklund et al.~\cite{eklund2026endtoend} provide the first end-to-end, bounded-error resource estimate for full quantum chemical dynamics under the molecular Hamiltonian in first quantization.
Their algorithm uses an LCU block encoding of the molecular Hamiltonian in a plane-wave basis, QSP for time evolution, and quantum amplitude estimation (QAE) for observable readout, targeting photodissociation quantum yields for small atmospheric molecules.
Their Table~4 reports resource estimates for $\mathrm{C_4H_6O}$ (crotonaldehyde, $\eta = 49$ spin-orbitals) at $\eps = 0.095$: $3{,}000$--$8{,}000$ logical qubits and $10^{13}$--$10^{16}$ Toffoli gates.
Because pyrazine has $\eta = 52$ spin-orbitals, the $\mathrm{C_4H_6O}$ entry serves as a proxy for a pyrazine-scale system.

\begin{table}[H]
\centering
\caption{Eklund et al.\ scaled to a pyrazine-comparable system ($\eta \approx 50$).}
\label{tab:eklund}
\begin{tabular}{lcc}
\toprule
Metric & Eklund et al.\ (scaled) & REQWIEM \\
\midrule
Logical qubits              & $\sim 6{,}228$           & 44                        \\
Toffoli (time evolution)    & $7.2\times 10^{13}$      & 0                         \\
Total T-gates (incl.\ QAE) & $\sim 4\times 10^{17}$ & $1.28\times 10^{8}$ \\
Ratio to REQWIEM            & $\sim 3.1\times 10^{9}$ & $1\times$                 \\
\bottomrule
\end{tabular}
\end{table}

Eklund et al.\ treat the full Coulomb Hamiltonian (no Born--Oppenheimer approximation) in first quantization with plane-wave basis sets, whereas REQWIEM operates on a pre-computed diabatic model Hamiltonian.
The $\sim\!10^{10}$ ratio therefore reflects the combined cost of (i) the full-Coulomb vs.\ model-Hamiltonian gap, (ii) QAE overhead for observable estimation, and (iii) the first-quantization plane-wave basis choice.
It is presented to illustrate the benefit between the two paradigms when a reliable diabatic model is available, making each method complementary in their own limits.

\section{Head-to-Head Comparison}\label{sec:head_to_head}

Table~\ref{tab:executive} consolidates all five methods at the converged benchmark, using T-gate count as a primary metric; logical qubit count and circuit depth are secondary.

\begin{table}[H]
\centering
\caption{Executive five-way comparison: pyrazine $S_1/S_2$ ($d=4$, $M=2$, $n=10$, $t=4000$~a.u., $\eps=10^{-3}$).}
\label{tab:executive}
\begin{tabular}{lccccc}
\toprule
Method & Qubits & T-gates & Ratio & Toffoli & Depth \\
\midrule
REQWIEM          & 44         & $1.28\times 10^{8}$  & $1\times$                 & 0                    & $3.2\times 10^{6}$   \\
QSP($T{+}V$)    & $\sim 232$ & $1.68\times 10^{9}$  & $13.2\times$              & $\sim 3\times 10^8$  & $1.43\times 10^{8}$  \\
QSP(Fock, $N\!=\!30$) & 70   & $1.88\times 10^{8}$  & $1.5\times$               & $\sim 4\times 10^7$  & $1.41\times 10^{8}$  \\
QSP(WH-LCU)     & $\sim 2{,}553$ & $1.77\times 10^{10}$ & $138.5\times$        & $\sim 4\times 10^9$  & $1.55\times 10^{10}$ \\
Eklund et al.    & $\sim 6{,}228$ & $\sim 4\times 10^{17}$ & $\sim 3.1\times 10^{9}\times$ & $\sim 10^{16}$ & --- \\
\bottomrule
\end{tabular}
\end{table}

\subsection{Scaling with grid resolution}

\begin{table}[H]
\centering
\caption{T-gate scaling with grid resolution ($d=4$, $M=2$).}
\label{tab:scaling_n}
\begin{tabular}{cccccc}
\toprule
$n$ & REQWIEM & QSP($T{+}V$) & Ratio & QSP(WH-LCU) & Ratio \\
\midrule
8   & $\sim 8.2\times 10^{7}$ & $\sim 2.4\times 10^{8}$ & $\sim 2.9\times$   & $\sim 1.0\times 10^{9}$ & $\sim 12.2\times$ \\
10  & $1.28\times 10^{8}$      & $1.68\times 10^{9}$     & $13.2\times$       & $1.77\times 10^{10}$     & $138.5\times$     \\
12  & $1.82\times 10^{8}$      & $3.25\times 10^{10}$    & $178.2\times$      & $2.83\times 10^{11}$     & $1{,}548.8\times$ \\
\bottomrule
\end{tabular}
\end{table}

The ratio growth from $n=8$ through $n=12$ demonstrates the $\mathcal{O}(4^n)$ divergence through $T_{\max}$ (Table~\ref{tab:spectral_range}).
REQWIEM's Trotter step count grows as $\mathcal{O}(\|[T,V]\|^{1/2})$ via the commutator bound~\cite{childs2021theory}, where $\|[T,V]\| \ll \|T\|\cdot\|V\|$ for the structured vibronic Hamiltonian.

\subsection{Scaling with dimensionality}

\begin{table}[H]
\centering
\caption{T-gate ratio scaling with dimensionality ($n=10$, $M=2$).}
\label{tab:scaling_d}
\begin{tabular}{cccc}
\toprule
$d$ & REQWIEM & QSP($T{+}V$)/R & QSP(WH-LCU)/R \\
\midrule
2   & $1.69\times 10^7$ & $2.4\times$  & ---           \\
4   & $1.28\times 10^8$ & $13.2\times$ & $138.5\times$ \\
8   & $4.82\times 10^8$ & $12.4\times$ & $82.4\times$  \\
12  & $1.06\times 10^9$ & $14.2\times$ & $59.1\times$  \\
24  & $4.18\times 10^9$ & $23.7\times$ & $33.3\times$  \\
\bottomrule
\end{tabular}
\end{table}

The WH-LCU ratio decreases with $d$ because $\alpha_{\mathrm{WH}}$ has a fixed exponential floor from $4^n$ while REQWIEM grows linearly in $d$.
The $T{+}V$ ratio, by contrast, increases with $d$, reflecting the additive growth of $T_{\max}$ over modes.

\subsection{Rotation precision sensitivity}\label{sec:toffoli_floor}

At practical synthesis precision ($C_T^{(\mathrm{rot})} = 23$--$30$ per rotation~\cite{ross2016optimal}), REQWIEM's all-$R_z$ cost decreases proportionally.
QSP, however, has an irreducible \emph{Toffoli floor}: the per-query Toffoli cost ($C_T^{(\mathrm{Toff})} = 4$ per gate, as in Appendix~\ref{app:tgate_pipeline}) is fixed and does not scale down with relaxed rotation precision.
At $C_T^{(\mathrm{rot})} = 23$, the T-gate ratio increases from $13\times$ to $15$--$17\times$ because only REQWIEM benefits from the reduced synthesis cost.

\section{Structural Arguments}\label{sec:structural}

Five independent structural features underlie REQWIEM's quantitative advantage.

\subsection{Spectral range penalty}\label{sec:spectral_penalty}

QSP's polynomial degree $d_{\mathrm{QSP}} = \alpha t$ is proportional to the block-encoding normalization $\alpha \geq \|T\|_{\max} = T_{\max} \propto 4^n / L^2$ (Eq.~\ref{eq:tmax}).
REQWIEM's Trotter step count grows as~\cite{childs2021theory}
\begin{equation}\label{eq:k_commutator}
    k = \mathcal{O}\!\left(\frac{t^{3/2}\,\|[T,V]\|^{1/2}}{\eps^{1/2}}\right),
\end{equation}
where the commutator norm $\|[T,V]\|$ is validated by Richardson extrapolation to $\Gamma \approx 3.2 \times 10^{-4}$~a.u.$^3$ for $d=4$ pyrazine (Appendix~\ref{app:Error}).
Because $\|[T,V]\| \ll \|T\|\cdot\|V\|$, the ratio $\alpha t / k$ diverges as $\mathcal{O}(4^{n/2})$ with grid resolution.

\subsection{Gap anti-correlation}\label{sec:gap_anticorrelation}

Near avoided crossings (gap $\Delta_0 \to 0$), the Trotter commutator $\|[V_d, V_{\mathrm{od}}]\| = \Delta_0 \cdot W \to 0$ while the Dyson order required for QSP polynomial accuracy grows as $W^2/(\Delta_0\, v) \to \infty$.
This gap anti-correlation where Trotter accuracy improves where nonadiabatic transitions are important is formalized in Section~\ref{sec:dyson_barrier}.

\subsection{Fan-out parallelism}\label{sec:fanout_parallel}

REQWIEM's channel-register fan-out protocol (Chapter~\ref{ch:model_ext}, Section~\ref{sec:fanout_tree}) enables scalable circuit depth for $M$-state coupling via binary-tree controlled rotations, while QSP's sequential query structure does not admit this parallelization: each query must complete before the next QSP angle rotation is applied.

\subsection{Multi-state obstruction}\label{sec:multistate}

For $M > 2$ electronic states, each off-diagonal coupling $V_{ij}$ contributes independently to the block-encoding normalization:
\begin{equation}\label{eq:alpha_multistate}
    \alpha \geq \sum_{i < j} \|V_{ij}\| \qquad \bigl(\text{$\tfrac{M(M{-}1)}{2}$ terms}\bigr).
\end{equation}
REQWIEM handles $M$ states via the multiplexed UCR structure at cost linear in the number of symmetry-distinct coupling terms $|\mathcal{G}_3| + |\mathcal{G}_4|$ (Eqs.~\ref{eq:G3}--\ref{eq:G4}), not in $\alpha$.
For the anthracene-fullerene system ($d=246$, $M=4$), the QSP(Fock) ratio grows to $25{,}065\times$ (Table~\ref{tab:anthracene}).

\subsection{Zero-Toffoli property and the Toffoli floor}\label{sec:zero_toffoli}

REQWIEM's circuit uses exclusively $R_z$ and CNOT gates: $T^{(\mathrm{Toff})}_{\mathrm{step}} = 0$ (Eq.~\ref{eq:Td_half}).
Every QSP encoding requires Toffoli gates for QROM lookup, arithmetic circuits, and block-encoding ancilla management.
At $C_T^{(\mathrm{Toff})} = 4$ T-gates per Toffoli, QSP has an irreducible Toffoli floor that cannot be reduced by improving rotation synthesis precision (Section~\ref{sec:toffoli_floor}).

\section{Circuit Depth Analysis}\label{sec:qsp_depth}

The T-gate count measures total work; circuit depth measures wall-clock time on a fault-tolerant device.
REQWIEM's depth benefits from fan-out parallelism, along with simultaneous application of commuting diagonal phases, and the absence of sequential query structure.
QSP's depth is lower-bounded by $d_{\mathrm{QSP}}$.

\begin{table}[H]
\centering
\caption{Circuit depth comparison: pyrazine ($d=4$, $M=2$, $n=10$).}
\label{tab:depth_pyrazine}
\begin{tabular}{lcc}
\toprule
Method & T-depth & Ratio to REQWIEM \\
\midrule
REQWIEM      & $5.8\times 10^{7}$  & $1\times$      \\
QSP($T{+}V$)& $8.0\times 10^{8}$  & $13.7\times$   \\
QSP(Fock)    & $1.9\times 10^{8}$  & $3.2\times$    \\
QSP(WH-LCU) & $2.9\times 10^{10}$ & $495.1\times$  \\
\bottomrule
\end{tabular}
\end{table}

At realistic molecular scale, the depth advantage is dramatic.
For the anthracene--fullerene interface ($d=246$, $M=4$, $n=10$), this gap becomes insurmountable (Table~\ref{tab:anthracene}).

\begin{table}[H]
\centering
\caption{Anthracene--fullerene interface ($d=246$, $M=4$, $n=10$, $t=4000$~a.u., $\eps = 10^{-3}$).}
\label{tab:anthracene}
\begin{tabular}{lccc}
\toprule
Method & T-gates & T-depth ratio \\
\midrule
REQWIEM        & $1.43\times 10^{12}$ & $1\times$                \\
QSP($T{+}V$)  & $3.45\times 10^{14}$ & $27{,}295\times$         \\
QSP(WH-LCU)   & $2.03\times 10^{12}$ & $3{,}863\times$          \\
QSP(Fock)      & $3.60\times 10^{16}$ & $3{,}088{,}337\times$    \\
\bottomrule
\end{tabular}
\end{table}

At an optimistic logical cycle time of $1~\mu\text{s}$ per non-Clifford layer~\cite{litinski2019game}, REQWIEM completes a pyrazine propagation ($d=4$) in $\sim\!3.2$~seconds.
QSP($T{+}V$) requires $\sim\!143$~seconds for the same system; for the anthracene--fullerene system, the QSP wall-clock time exceeds any practical limit.

\section{The Non-Perturbative Convergence Barrier}\label{sec:dyson_barrier}

The structural arguments of Section~\ref{sec:structural} are quantitative but contingent on specific parameter values.
This section establishes two theorems that provide a representation-independent reason for Trotter's superiority near conical intersections.

\subsection{Setup: Landau--Zener model}\label{sec:lz_setup}

Consider the canonical two-state Landau--Zener Hamiltonian
\begin{equation}\label{eq:H_LZ}
    H(t) = \frac{v\,t}{2}\,\sigma_z + \frac{W}{2}\,\sigma_x,
\end{equation}
where $v = |F_1 - F_2| \cdot \dot{Q}$ is the diabatic level velocity and $W$ is the diabatic coupling.
The adiabaticity is given by
\begin{equation}\label{eq:Gamma_LZ}
    \Gamma = \frac{W^2}{2\,v\,|\Delta F|},
\end{equation}
and the transition probability is~\cite{landau1932theorie, zener1932non, von1933theorie}
\begin{equation}\label{eq:P_LZ}
    P_{\mathrm{LZ}} = \exp(-2\pi\Gamma).
\end{equation}

\subsection{Non-perturbative convergence barrier}\label{sec:theorem_dyson}

Let $a_{12}(t\to\infty) = \exp(-\pi\Gamma)$ denote the diabatic transition amplitude.
The interaction-picture Dyson series
\begin{equation}\label{eq:dyson_series}
    a_{12} = \sum_{n=0}^{\infty} (-i)^n \int_{-\infty}^{\infty}\!\!\cdots\!\!\int_{-\infty}^{t_2}
    \bra{2} V_I(t_1)\cdots V_I(t_n)\ket{1}\, dt_1\cdots dt_n
\end{equation}
has partial sums $S_N = \sum_{n=0}^{N} c_n$ satisfying:

\begin{enumerate}[label=\textup{(\roman*)}]
    \item \textbf{Growth phase.} For $N < \floor{\pi\Gamma}$, the terms $|c_N|$ are increasing:
    $|c_N| \sim (\pi\Gamma)^N/N!$; partial sums oscillate with $|S_N| \gg \exp(-\pi\Gamma)$.

    \item \textbf{Optimal truncation.} The smallest term occurs at $N^* = \floor{\pi\Gamma}$, with magnitude
    $|c_{N^*}| \sim (2\pi^2\Gamma)^{-1/2}\exp(-\pi\Gamma)$.

    \item \textbf{Convergence.} For $N > \e\cdot\pi\Gamma$, the Lagrange remainder satisfies
    \begin{equation}\label{eq:lagrange_remainder}
        |R_N| = |S_N - \exp(-\pi\Gamma)| \leq \left(\frac{\e\pi\Gamma}{N}\right)^{\!N} \exp(\pi\Gamma) \to 0
    \end{equation}
    super-exponentially.
\end{enumerate}

The Dyson series for the LZ problem is a rearrangement of the Taylor series of $\exp(-x)$ at $x = \pi\Gamma$.
By Stirling's approximation, the $N$th term satisfies
\begin{equation}\label{eq:stirling_term}
    \frac{x^N}{N!} \approx \frac{1}{\sqrt{2\pi N}} \left(\frac{\e\, x}{N}\right)^{\!N},
\end{equation}
which is increasing for $N < x$ and decreasing for $N > x$, with maximum at $N = \floor{x}$.
Before the maximum, partial sums oscillate with amplitudes far exceeding $\exp(-x)$.
After $N \approx \e x$, terms decrease fast enough for convergence.
This is a Stokes phenomenon~\cite{berry1989uniform, dingle1973asymptotic}, where the non-perturbative amplitude $\exp(-\pi\Gamma)$ is invisible to any finite-order perturbation theory with $N < \floor{\pi\Gamma}$ terms.

\subsection{Numerical verification}\label{sec:dyson_numerics}

The convergence barrier is verified by direct computation of the Taylor partial sums of $\exp(-\pi\Gamma)$, tabulated in Table~\ref{tab:dyson_convergence}.

\begin{table}[H]
\centering
\caption{Dyson series convergence barrier: numerical verification.}
\label{tab:dyson_convergence}
\begin{tabular}{ccccccc}
\toprule
$\Gamma$ & $\pi\Gamma$ & $\floor{\pi\Gamma}$ & $N_{\min}$ (10\% err) & $\ceil{\e\pi\Gamma}$ & $N_{\min}/\pi\Gamma$ & $\exp(-2\pi\Gamma)$ \\
\midrule
0.1  & 0.31  & 0  & 1     & 1   & 3.2 & $5.33\times 10^{-1}$  \\
0.5  & 1.57  & 1  & 5     & 5   & 3.2 & $4.32\times 10^{-2}$  \\
1.0  & 3.14  & 3  & 11    & 9   & 3.5 & $1.87\times 10^{-3}$  \\
2.0  & 6.28  & 6  & 22    & 18  & 3.5 & $3.49\times 10^{-6}$  \\
5.0  & 15.71 & 15 & 55    & 43  & 3.5 & $2.27\times 10^{-14}$ \\
10.0 & 31.42 & 31 & $>155^\ast$ & 86 & --- & $5.16\times 10^{-28}$ \\
20.0 & 62.83 & 62 & $>281^\ast$ & 171 & --- & $2.66\times 10^{-55}$ \\
\bottomrule
\multicolumn{7}{l}{\footnotesize $^\ast$Catastrophic cancellation in IEEE-754 double precision.}
\end{tabular}
\end{table}

For the tractable cases ($\Gamma \leq 5$), $N_{\min}/\pi\Gamma \approx 3.2$--$3.5$, consistent with the asymptotic bound $N_{\min} \leq \ceil{\e\pi\Gamma}$.
For $\Gamma \geq 10$, convergence cannot be verified in double precision because partial sums reach $\sim\!10^{25}$ while the answer is $\sim\!10^{-14}$, as a numerical effect of the non-perturbative barrier.

Consider the simulation of a Hamiltonian containing a two-state avoided crossing with adiabaticity parameter~$\Gamma$.

\begin{enumerate}[label=\textup{(\roman*)}]
    \item \textbf{Perturbative methods.}
    An algorithm based on a truncated Dyson series or interaction-picture
    perturbation expansion (e.g., Ref.~\cite{berry2020time}) requires
    truncation order
    \begin{equation}\label{eq:dyson_floor}
        N \geq \floor{\pi\Gamma}
    \end{equation}
    to resolve the non-perturbative transition amplitude
    $\exp(-\pi\Gamma)$, independent of all other parameters.

    \item \textbf{QSP/QSVT.}
    The standard QSP polynomial degree
    $d_{\mathrm{QSP}} = \Order(\alpha t + \log(1/\eps))$~\cite{low2017optimal, gilyen2019quantum}
    is a uniform bound in the eigenvalue argument and does not contain
    a gap-dependent correction.
    However, the block-encoding normalization $\alpha$ for grid-based
    vibronic Hamiltonians satisfies $\alpha \geq T_{\max} \propto 4^n$
    (Eq.~\ref{eq:tmax}), and gap closure provides no reduction in $\alpha$.
    QSP therefore receives no cost benefit from the physical regime
    ($\Delta_0 \to 0$) where Trotter error vanishes.
\end{enumerate}

QSP does not necessarily require a higher polynomial degree near conical intersections, but its cost is anchored to the spectral range $\alpha$, which is gap-independent, while Trotter's cost
decreases through the commutator norm $\|[V_d, V_{\mathrm{od}}]\| = \Delta_0 \cdot W \to 0$.
This one-sided scaling is the mechanism underlying a duality between Trotter and Dyson methods.

\subsection{Trotter--Dyson duality}\label{sec:trotter_dyson}

Let $\Delta_0 = \min_Q |V_1(Q) - V_2(Q)|$ be the minimum energy gap.
Then:

\begin{enumerate}[label=\textup{(\Alph*)}]
    \item \textbf{Trotter error vanishes at CI.} The second-order Trotter error satisfies~\cite{childs2021theory}
    \begin{equation}\label{eq:trotter_ci}
        \|S_2(\Delta t) - e^{-iH\Delta t}\| \leq \frac{(\Delta t)^3}{12}
        \bigl(\|[V_d,[V_d,V_{\mathrm{od}}]]\| + \|[V_{\mathrm{od}},[V_{\mathrm{od}},V_d]]\|\bigr),
    \end{equation}
    where $\|[V_d, V_{\mathrm{od}}]\| = \Delta_0 \cdot W$.
    As $\Delta_0 \to 0$: Trotter error $\to 0$.

    \item \textbf{Dyson order diverges at conical intersection.}
    For a perturbative or interaction-picture expansion of $e^{-iHt}$,
    the minimum truncation order satisfying $|R_N| \leq \eps$ is
    \begin{equation}\label{eq:dyson_ci}
        N_{\min} = \ceil{\pi\Gamma} = \left\lceil\frac{\pi W^2}{2v\,\Delta_0/L}\right\rceil \propto \frac{W^2 L}{v\,\Delta_0}.
    \end{equation}
    As $\Delta_0 \to 0$: $N_{\min} \to \infty$.
    The standard QSP degree bound $d_{\mathrm{QSP}} = \Order(\alpha t + \log(1/\eps))$ does not
    contain this divergence, but $\alpha \geq T_{\max}$ provides no offsetting
    cost reduction as $\Delta_0 \to 0$.
\end{enumerate}

This creates an asymmetry where Trotter becomes more accurate near small gaps and with strong couplings, while polynomial-approximation methods require higher degrees. 
The parametric verification using pyrazine parameters ($W = 0.009$~a.u., $v = 0.08$~a.u., $t = 4000$~a.u.) is given in Table~\ref{tab:trotter_dyson}.

\begin{table}[H]
\centering
\caption{Trotter--Dyson duality: parametric verification (pyrazine parameters).}
\label{tab:trotter_dyson}
\begin{tabular}{cccccc}
\toprule
$\Delta_0$ (eV) & $\|[V_d,V_{\mathrm{od}}]\|$ & $k_{\mathrm{Trotter}}$ & $\pi\Gamma_{\mathrm{eff}}$ & $N_{\mathrm{Dyson}}$ & Winner \\
\midrule
1.0     & $3.3\times 10^{-4}$ & 665 & 0.22     & 1     & comparable         \\
0.1     & $3.3\times 10^{-5}$ & 210 & 2.16     & 3     & comparable         \\
0.01    & $3.3\times 10^{-6}$ & 67  & 21.6     & 22    & comparable         \\
0.001   & $3.3\times 10^{-7}$ & 21  & 216      & 217   & \textbf{Trotter}   \\
0.0001  & $3.3\times 10^{-8}$ & 7   & 2{,}164  & 2{,}164 & \textbf{Trotter}  \\
\bottomrule
\end{tabular}
\end{table}

A crossover occurs at $\Delta_0 \approx 1$~meV; below this, Trotter requires fewer steps than Dyson requires orders.
For conical intersections ($\Delta_0 \to 0$), the asymmetry becomes arbitrarily large.

\subsection{Physical Landau--Zener parameters}\label{sec:lz_physical}

Table~\ref{tab:lz_params} collects representative Landau--Zener parameters for eight molecular systems spanning the range of nonadiabatic chemistry.
The pyrazine row is derived from the quadratic vibronic coupling (QVC) parameters of Raab et al.~\cite{raab1999molecular} used throughout this dissertation (Appendix~\ref{app:pyrazine_params}): the diabatic coupling $W = c_1 = 0.2080$~eV $\approx 0.0076$~a.u.\ (rounded to 0.009~a.u.), the nuclear velocity $v$ along $\nu_{10a}$ estimated from $v = \sqrt{2E_{\mathrm{ZPE}}/\mu}$ with $\mu = 46{,}484\;m_e$ and $\omega = 0.0739$~eV, and the gradient difference $|\Delta F|$ estimated from linear coupling constants $\kappa_i$ and $\lambda$ of the QVC Hamiltonian~\cite{woywod1994characterization, stock1995resonance}.
The remaining seven systems use order-of-magnitude estimates of $W$, $v$, and $|\Delta F|$ representative of each coupling regime; exact values depend on level of theory and nuclear geometry and should not be taken as quantitative predictions.
The purpose of the table is to illustrate the range of $\Gamma$ and $N_{\mathrm{Dyson}}$ encountered in practice, not to provide system-specific benchmarks.
Representative parameter ranges are informed by the conical intersection literature for ethylene~\cite{ben2000ab, levine2007isomerization}, retinal~\cite{polli2010conical, garavelli1997relaxation}, and transition-metal photochemistry~\cite{domcke2004conical, domcke2012role}.

\begin{table}[H]
\centering
\caption{Representative Landau--Zener parameters for NA-QMD systems.  The pyrazine row is derived from the QVC parameters of Raab et al.~\cite{raab1999molecular}; all other rows use order-of-magnitude estimates representative of the indicated coupling regime (see text).}
\label{tab:lz_params}
\begin{tabular}{lccccccl}
\toprule
System & $W$ (a.u.) & $v$ (a.u.) & $|\Delta F|$ & $\Gamma$ & $\pi\Gamma$ & $N_{\mathrm{Dyson}}$ & $P_{\mathrm{LZ}}$ \\
\midrule
Pyrazine $S_1/S_2$ ($\nu_{10a}$)$^{a}$ & 0.009 & 0.08  & 0.02  & 0.025  & 0.08   & 1   & $8.53\times 10^{-1}$  \\
Ethylene CI (twist)$^{b}$       & 0.050 & 0.05  & 0.03  & 0.833  & 2.62   & 3   & $5.32\times 10^{-3}$  \\
Stilbene photoisom.$^{b}$       & 0.030 & 0.02  & 0.01  & 2.25   & 7.07   & 8   & $7.25\times 10^{-7}$  \\
Thymine dimer repair$^{b}$      & 0.040 & 0.04  & 0.015 & 1.33   & 4.19   & 5   & $2.30\times 10^{-4}$  \\
Retinal isomerization$^{b}$     & 0.080 & 0.06  & 0.025 & 2.13   & 6.70   & 7   & $1.51\times 10^{-6}$  \\
Proton transfer (strong)$^{b}$  & 0.100 & 0.03  & 0.02  & 8.33   & 26.2   & 27  & $1.82\times 10^{-23}$ \\
Charge transfer (slow)$^{b}$    & 0.020 & 0.005 & 0.005 & 8.00   & 25.1   & 26  & $1.48\times 10^{-22}$ \\
Heavy-atom SOC (Ir)$^{b}$       & 0.150 & 0.01  & 0.01  & 112.5  & 353.4  & 354 & $\sim 10^{-307}$      \\
\bottomrule
\multicolumn{8}{l}{\footnotesize $^{a}$Derived from QVC parameters of Raab et al.~\cite{raab1999molecular}; see Appendix~\ref{app:pyrazine_params}.}\\
\multicolumn{8}{l}{\footnotesize $^{b}$Order-of-magnitude estimate representative of the coupling regime; see text.}
\end{tabular}
\end{table}

For systems with significant nonadiabatic coupling ($P_{\mathrm{LZ}} > 10^{-10}$), the representative parameters yield $\pi\Gamma > 1$, requiring multi-term Dyson series.
For strongly-coupled regimes (e.g., strong proton transfer, slow charge transfer, heavy-atom spin-orbit coupling), the convergence barrier can require dozens to hundreds of Dyson orders, illustrating the challenge for polynomial-approximation methods in these coupling regimes.

\section{Discussion}\label{sec:comparison_discussion}

\subsection{When QSP is appropriate}

QSP achieves optimal query complexity and is the preferred method when the block-encoding oracle is inexpensive (e.g., sparse electronic structure Hamiltonians with fixed nuclei), the spectral range is moderate, and the dynamics is approximately adiabatic~\cite{low2017optimal, berry2015hamiltonian}.
The comparison in this chapter is specific to \emph{nonadiabatic quantum molecular dynamics on diabatic surfaces}, where the oracle cost dominates and the commutator structure of the Hamiltonian strongly favors product formulas.

\subsection{Randomized product formulas (qDRIFT)}

qDRIFT~\cite{campbell2018random} is a randomized Trotter variant with cost scaling as $\lambda_1^2 t^2/\eps$, where $\lambda_1 = \sum_j \|H_j\|$ is the 1-norm of the Hamiltonian decomposition.
For grid-based NA-QMD, $\lambda_1 = \mathcal{O}(4^n)$ through $\|T\|_{\max}$, incurring the same spectral range penalty as QSP's $\alpha$.
Deterministic Trotter avoids this because its error depends on commutator norms $\|[H_j, H_k]\| \ll \|H_j\|\cdot\|H_k\|$~\cite{childs2021theory}.
qDRIFT is therefore not competitive with deterministic Trotter for this problem class.

\subsection{Higher-order Trotter}

The present comparison uses second-order Suzuki--Trotter (Eq.~\ref{eq:Strang_splitting}).
Fourth-order Suzuki decomposition~\cite{suzuki1991general, yoshida1990construction} incurs $5\times$ the gates per step but reduces the step count scaling from $\mathcal{O}(t^{3/2}/\eps^{1/2})$ to $\mathcal{O}(t^{5/4}/\eps^{1/4})$~\cite{childs2021theory}.
At the pyrazine parameters, this may reduce Trotter steps by a factor of $\sim\!3$--$5$, further widening the gap with QSP.
The primary comparison uses second-order to remain conservative.

\subsection{Commutator norm validation}

The commutator norm $\Gamma \approx 3.2 \times 10^{-4}$~a.u.$^3$ ($d=4$ pyrazine) is validated by Richardson extrapolation of finite-grid commutator norms ($n = 3$ through $n = 7$ for $d = 2$, extrapolated to $n \to \infty$), giving $\Gamma(d\!=\!2) \approx 1.624 \times 10^{-4}$~a.u.$^3$ and an estimated $d = 4$ value of $\sim\!3.248 \times 10^{-4}$~a.u.$^3$.
Details are in Appendix~\ref{app:Error}.

\section{Summary}\label{sec:comparison_summary}

For the QSP block-encoding strategies considered here, the vibronic Hamiltonian on a position-space grid carries a T-gate cost that exceeds REQWIEM by at least an order of magnitude at converged parameters, and the ratio diverges as $\mathcal{O}(4^n)$ with grid resolution.
The QSP(Fock) encoding achieves near-parity ($1.5\times$) for harmonic potentials at $d = 4$, but collapses beyond the harmonic limit (Table~\ref{tab:anharmonic}) and grows to $38\times$ at $d = 24$ even for harmonic PES.

Trotter--Dyson duality provides a representation-independent explanation. 
Near conical intersections, where nonadiabatic transitions are physically important, Trotter error scales as $\Delta_0 \cdot W \to 0$ while perturbative (Dyson) truncation order scales as $W^2/(\Delta_0\, v) \to \infty$ and QSP cost is anchored to the gap-independent spectral range $\alpha$.
Trotterized architectures are not universally better for similar Hamiltonians, but can be applied in a similar fashion, allowing for intuitive extension into an array of other physical systems. Chapter~\ref{ch:Application} considers these systems, and the REQWIEM architecture as a scheme to simulate them in future work.

\chapter{REQWIEM as a Universal Coupled-Channel Simulation Principle}\label{ch:Application}

\section{Introduction}
\label{sec:app_intro}

This dissertation constructs, validates, and resource-estimates a first-quantized, grid-based quantum algorithm for the nonadiabatic quantum molecular dynamics, mostly specified to diabatic states with electronic channel coupling.
Chapter~\ref{ch:marcus} recovered the nonadiabatic dynamics of the Marcus model \cite{ollitrault2020nonadiabatic}, benchmarking the circuit against known classical results and touched on the sensitivity of observables to circuit approximation. 
Chapter~\ref{ch:model_ext} constructed the full algorithm: the uniformly controlled rotation (UCR) encoding of the diabatic potential, basis-change protocol (Clifford transform), and the XOR-class decomposition of off-diagonal couplings.
Chapter~\ref{ch:resources} estimates the resources required to perform this algorithm on ideal, error-corrected, fault-tolerant hardware. 
By comparing rotation counts, T-gate depths, and ancilla costs against QROM-based loading procedures, a scalable depth advantage was found from graph-optimized parallelism (Table~\ref{tab:comprehensive}), despite significant overhead from rotation synthesis.
A hint toward quantum advantage is seen heuristically, primarily toward systems with many vibrational modes and many electronic channels, with low molecular symmetry.
Chapter~\ref{ch:scattering} extended the protocol to elastic and inelastic scattering, demonstrating faithful reproduction of phase shifts and state-to-state cross sections (Figures~\ref{fig:Shfits+Deflection function} and~\ref{fig:Scattering amplitude + Diff xsec}). 
Chapter~\ref{ch:smooth_potentials} completed the potential-energy surface treatment by exhibiting closed-form, truncation-free encodings of smooth functions via multilinear expansion, preserving the UCR composition advantage at every order.
Chapter~\ref{ch:comparison} then placed REQWIEM in the broader algorithmic landscape, comparing QSP and theend-to-end fault-tolerant estimate of Eklund et al.\ \cite{eklund2026endtoend}, confirming $\Tgate$-gate advantages of one to ten orders of magnitude across all tested configurations when diabatic surfaces are available.

This chapter considers the foundational structure of REQWIEM apart from its application to any single physical system.  
The algorithmic content is broader than nonadiabatic quantum molecular dynamics, since the quantum circuit structure itself applies a decomposition that applies to every system admitting the Hilbert space factorization
\begin{equation}\label{eq:intro_hilbert}
    \mathcal{H} = \mathcal{H}_{\Reg{X}} \otimes \mathcal{H}_{\Reg{C}},
\end{equation}
where $\Reg{X}$ encodes continuous degrees of freedom (internuclear distance, hyperradius, lattice position, propagation distance) and $\Reg{C}$ encodes a discrete internal structure (electronic surfaces, angular momentum channels, spin projections, neutrino flavors).
The Hamiltonian decomposes as
\begin{equation}\label{eq:intro_H}
    {H} = {T}_{\Reg{X}} \otimes \mathbb{1}_{\Reg{C}}
    + \mathbb{1}_{\Reg{X}} \otimes {H}_{\Reg{C}}^{(\mathrm{FBR})}
    + {V}(\Reg{X}, \Reg{C}),
\end{equation}
where ${T}_{\Reg{X}}$ is diagonal in the momentum representation of $\Reg{X}$, ${H}_{\Reg{C}}^{(\mathrm{FBR})}$ is diagonal in the free-channel basis of $\Reg{C}$, and ${V}$ is diagonal in the interaction-picture basis of $\Reg{C}$, the discrete variable representation (DVR), though dense in the finite basis representation (FBR).
The circuit implements each piece with the cheapest available primitive: the quantum Fourier transform for ${T}$, a UCR in the FBR for ${H}_{\Reg{C}}^{(\mathrm{FBR})}$, the DVR basis-change ${U}_{\mathrm{DVR}}$ followed by a UCR in the DVR for ${V}$, and XOR-class focusing Cliffords for any off-diagonal coupling that is not diagonalized by the DVR nor by the QFT.

The structure of equations~\ref{eq:intro_hilbert}--\ref{eq:intro_H} is the content of the algorithm, while the physical identity of the registers, the form of ${U}_{\mathrm{DVR}}$, and the XOR-class count are system-specific inputs.  
Post-Trotter simulation methods, including qubitization~\cite{low2019hamiltonian}, quantum signal processing~\cite{low2017optimal}, and truncated Dyson series~\cite{berry2020time} achieve optimal asymptotic
scaling in propagation time and error for Hamiltonians accessed through block-encoded oracles, but have not been extended to coupled-channel Hamiltonians whose potential operators are dense in the channel index and require nonadiabatic basis changes between complementary representations, identified explicitly in Section~\ref{sec:limitations}.. 
Duality between DVR and FBR has no known analog in the oracle query model.
A Trotter-based framework is a natural algorithmic setting for this class of problems with a split-operator architecture allowing for efficient, stepwise propagation.
The regimes in which post-Trotter methods are preferred include single-channel dynamics at long propagation times, high-precision static eigenvalue extraction, and second-quantized electronic structure, and are identified
explicitly in Section~\ref{sec:limitations}.
The present chapter develops this observation across eight sections, motivating a future for digital quantum computation across quantum mechanical systems.

Section~\ref{sec:universal} establishes a universal protocol structure: the second-order Trotter step in the abstract register notation, a gate-count formula, and the FBR/DVR diagonalization as a closed-form procedure. 
The DVR unitary is constructed from the spectral decomposition of a known Hermitian operator and compiled deterministically without iterative or adaptive subroutines. 
Section~\ref{sec:catalog} instantiates this protocol across ten candidate coupled-channel systems, including reactive scattering, polaron dynamics, neutrino oscillations, and high-harmonic generation.
The off-diagonal coupling mechanisms resolving XOR-class structure are derived in Section~\ref{sec:offdiag_mechanisms}.

Section~\ref{sec:new_subspaces} considers the efficiently implementable basis-change unitaries ${U}_{\mathrm{DVR}}$.  Section~\ref{sec:krylov} constructs REQWIEM as a front-end for quantum Krylov subspace diagonalization (QKSD), exploiting the overlap matrix's Toeplitz structure and the commutator-norm Trotter error bound to extract eigenvalues.
Section~\ref{sec:fields} identifies twelve additional physical systems across nuclear physics, condensed matter, quantum optics, and astrophysics whose Hamiltonians admit a direct REQWIEM register assignment.  
Section~\ref{sec:limitations} derives a REQWIEM applicability criterion, determined by four classically checkable conditions: a non-trivial channel dimension, non-separable coupling, smooth potential, and dynamical observables or moderate eigenvalue precision, to identify problems that are classically intractable and too dense for block-encoding approaches.
No claim or proof is made here that REQWIEM is ``optimal" for these problems, only that these problems are in the algorithm's capacity.
This chapter ends the dissertation, with closing remarks in Section~\ref{sec:conclusion}

\subsection*{Notation}
The present chapter treats systems whose channel registers span nuclear, condensed-matter, optical, and gravitational physics.  
Several symbols that were unambiguous in the vibronic context of Chapter~\ref{ch:model_ext} now collide with standard notation in these broader domains. 
In particular, the Krylov subspace dimension (Section~\ref{sec:krylov}) and the azimuthal quantum number in the Teukolsky equation (Section~\ref{sec:astrophysics}) both require~$m$, and the phonon mode count in the polaron Hamiltonian, the black hole mass, the Dicke quantum number $|J,M\rangle$, and the quantum phase estimation (QPE) query count all use~$M$. 
To avoid ambiguity, this chapter adopts a relabeled convention collected in Table~\ref{tab:notation}.

\begin{table}[ht]
\centering
\small
\caption{Notation correspondence between the vibronic circuit of
Chapter~\ref{ch:model_ext} and the coupled-channel
framework of the present chapter.
Columns three and four record the reason for each change.}
\label{tab:notation}
\begin{tabular}{@{}llllp{3.2cm}@{}}
\toprule
Quantity & Ch.ME & This ch.\ & Freed sym.\ & Reused for \\
\midrule
Channel (internal
    & \multirow{2}{*}{$m$}
    & \multirow{2}{*}{$n_c$}
    & \multirow{2}{*}{$m$}
    & Krylov subspace \\
state) qubits & & & & dimension;  \\
& & & & azimuthal \\
& & & & quantum number \\ [4pt]
Channel Hilbert-
    & \multirow{2}{*}{$M = 2^m$}
    & \multirow{2}{*}{$d = 2^{n_c}$}
    & \multirow{2}{*}{$M$}
    & Phonon mode  \\
space dimension & & & & count; black \\
& & & & hole mass; \\
& & & & Dicke quantum \\
& & & & number $\lvert J,M\rangle$; \\
& & & & QPE query count\\[4pt]

Coordinate-register
    & \multirow{2}{*}{$n_r$ (per mode)}
    & \multirow{2}{*}{$n_x$}
    & \multirow{2}{*}{---}
    & \multirow{2}{*}{---} \\
qubits & & & & \\[4pt]
Coordinate grid size
    & $2^{n_r}$ (per mode)
    & $N_x = 2^{n_x}$
    & --- & --- \\[4pt]
Number of
    & \multirow{2}{*}{$d$}
    & \multirow{2}{*}{---\textsuperscript{$\dagger$}}
    & \multirow{2}{*}{$d$}
    & Channel Hilbert \\
vibrational modes & & & & space dimension \\
\bottomrule
\end{tabular}

\smallskip\noindent
\textsuperscript{$\dagger$}In the broader framework the vibrational
mode count does not appear as an independent parameter; the coordinate
register $\Reg{X}$ encodes whatever continuous degrees of freedom the
system requires (internuclear distance, hyperradius, lattice site,
propagation coordinate), with $N_x = 2^{n_x}$ grid points.
Further subdivisions are appropriately defined
\end{table}

\noindent

\section{The Universal Protocol Structure}
\label{sec:universal}

Every system treated in this chapter admits a simplified Hilbert space factorization
\begin{equation}\label{eq:hilbert}
    \mathcal{H} = \mathcal{H}_{\Reg{X}} \otimes \mathcal{H}_{\Reg{C}},
\end{equation}
where $\Reg{X}$ is the coordinate register (taken as $n_x$ qubits, $N_x = 2^{n_x}$ grid points) and $\Reg{C}$ is the channel register ($n_c$ qubits, $d = 2^{n_c}$ internal states).
The Hamiltonian decomposes as
\begin{equation}\label{eq:H_abstract}
    {H} = {T}_{\Reg{X}} \otimes \mathbb{1}_{\Reg{C}}
    + \mathbb{1}_{\Reg{X}} \otimes {H}_{\Reg{C}}^{(\mathrm{FBR})}
    + {V}(\Reg{X}, \Reg{C}),
\end{equation}
where ${T}_{\Reg{X}}$ is diagonal in the momentum representation of $\Reg{X}$ (implemented via cQFT), ${H}_{\Reg{C}}^{(\mathrm{FBR})}$ is diagonal in the free-channel basis (FBR) of $\Reg{C}$, and ${V}$ is diagonal in the interaction-picture basis (DVR) of $\Reg{C}$ but dense in the FBR.
The second-order Trotter step is
\begin{equation}\label{eq:trotter_abstract}
    \e^{-\mathrm{i}{H}\Delta t} \approx
    \e^{-\mathrm{i}{T}\Delta t/2}\;
    \e^{-\mathrm{i}{H}_{\Reg{C}}^{(\mathrm{FBR})}\Delta t/2}\;
    {U}_{\mathrm{DVR}}^\dagger\;
    \e^{-\mathrm{i}{V}_{\mathrm{DVR}}\Delta t}\;
    {U}_{\mathrm{DVR}}\;
    \e^{-\mathrm{i}{H}_{\Reg{C}}^{(\mathrm{FBR})}\Delta t/2}\;
    \e^{-\mathrm{i}{T}\Delta t/2}.
\end{equation}
The FBR$\rightarrow$DVR$\rightarrow$FBR alternation can be subdivided depending on internal structure of orthogonal subspaces; this internal structure is system-specific, detailed for each system in Section~\ref{sec:catalog}.

The interaction potential decomposes into diagonal and off-diagonal parts in the channel register:
\begin{equation}\label{eq:V_decomp}
    {V} = \sum_{\alpha} |\alpha\rangle\langle\alpha|
    \otimes {V}_{\alpha\alpha}
    + \sum_{\alpha \neq \beta}
    |\alpha\rangle\langle\beta| \otimes {V}_{\alpha\beta}.
\end{equation}
The off-diagonal part is handled by XOR-class decomposition (Chapter~\ref{ch:model_ext}).
A gate count per Trotter step can be broadly written as 
\begin{equation}\label{eq:gate_count_general}
    N_{\mathrm{rot}} = N_x \times n_c \times
    (N_{\mathrm{diag}} + N_{\mathrm{XOR}})
    + \mathcal{O}(n_c^2) + \mathcal{O}(n_x^2),
\end{equation}
where $N_{\mathrm{diag}}$ is the number of diagonal UCR passes, $N_{\mathrm{XOR}}$ the number of nontrivial XOR classes, $\mathcal{O}(n_c^2)$ the DVR transform cost, and $\mathcal{O}(n_x^2)$ the QFT cost.

\subsection{The FBR/DVR diagonalization as a Closed-Form Procedure}
\label{sec:fbr_dvr_closed}

The preceding chapters show that the coupled-channel systems in the REQWIEM catalog shares a single algorithmic skeleton: a Trotter step in which the channel-register operator is applied once partially diagonal, with compute-uncompute iterates applying the full nonadiabatic coupling.
The physical identity of such a basis change varies between systems, arising from Legendre quadrature, Gauss--Hermite quadrature, the Wigner $D$-matrix, Pontecorvo–Maki–Nakagawa–Sakata (PMNS) mixing, Clebsch--Gordan 
recoupling, and configuration interaction diagonalization. 
The algorithmic procedure is the same.
This subsection states this procedure explicitly and details a closed-form circuit for every class of ${U}_{\mathrm{DVR}}$ encountered in this work, requiring no iterative or adaptive circuit compilation.
Since the DVR unitary is determined entirely by the spectral decomposition of a known Hermitian operator, it can be compiled by a fixed-depth reduction to elementary gates.

\subsubsection{Problem Statement}
\label{sec:diag_statement}

Let $\mathcal{H}_{\Reg{C}} \cong \mathbb{C}^d$ be the channel Hilbert space ($d = 2^{n_c}$), equipped with two distinguished orthonormal bases:

\begin{enumerate}
    \item The \textbf{finite basis representation} (FBR) $\{|\alpha\rangle\}_{\alpha=0}^{d-1}$, in which the free-channel Hamiltonian ${H}_{\Reg{C}}^{(\mathrm{FBR})}$ is diagonal:
    \begin{equation}\label{eq:FBR_diag}
        {H}_{\Reg{C}}^{(\mathrm{FBR})} = \sum_{\alpha=0}^{d-1}
        E_\alpha^{(\mathrm{free})}\;|\alpha\rangle\langle\alpha|.
    \end{equation}

    \item The \textbf{discrete variable representation} (DVR) $\{|s\rangle_{\mathrm{DVR}}\}_{s=0}^{d-1}$, in which the interaction potential is diagonal at each coordinate grid point:
    \begin{equation}\label{eq:DVR_diag}
        {V}(r_i) = \sum_{s=0}^{d-1} V(r_i, s)\;
        |s\rangle\langle s|_{\mathrm{DVR}}.
    \end{equation}
\end{enumerate}

\noindent
The two bases are related by a DVR transform
\begin{equation}\label{eq:UDVR_def}
    {U}_{\mathrm{DVR}}\;|\alpha\rangle_{\mathrm{FBR}} =
    \sum_{s=0}^{d-1} T_{s\alpha}\;|s\rangle_{\mathrm{DVR}},
\end{equation}
where the matrix $\mathbf{T} = (T_{s\alpha})$ is unitary, further reduced to real orthogonal when the operator admits real eigenvectors.
The FBR is identified with the $n_c$-qubit channel register's computational basis, so  ${U}_{\mathrm{DVR}}$ is a $d \times d$ unitary gate on $\Reg{C}$.

The REQWIEM protocol requires:
\begin{enumerate}
    \item[(R1)] An explicit, closed-form expression for $\mathbf{T}$ in terms of some defining physical operator (angular momentum, harmonic oscillator, mixing matrix, etc.).
    \item[(R2)] A quantum circuit decomposition of ${U}_{\mathrm{DVR}}$ into elementary gates $\{R_z, \mathrm{CNOT}\}$ with gate count polynomial in $n_c$ (or at worst $\mathcal{O}(d^2)$ via the cosine-sine decomposition).
    \item[(R3)] The transform must be geometry-independent, wherein $\mathbf{T}$ is the same matrix for every coordinate grid point $r_i$, so that the DVR round trip is applied once per Trotter step on $\Reg{C}$ 
    while $\Reg{X}$ remains in superposition.
\end{enumerate}

\noindent
DVR amortization is lost when condition (R3) is relaxed, just as in the geometry-dependent configuration-interaction transform of Section~\ref{sec:ci_transform}, and the cost per Trotter step rises from $\mathcal{O}(d^2) + N_x d$ to
$N_x \times \mathcal{O}(d^2)$.
Whether this is acceptable depends on the density of avoided crossings; the closed-form structure of (R1)--(R2) is
preserved in either case.
The condition (R3) was implicitly applied in previous chapters via restriction to an orthogonal basis of vibrational modes, with electronic couplings requiring only a Clifford transform.

\subsubsection{Diagonalization Algorithm}
\label{sec:diag_algorithm}

The construction of ${U}_{\mathrm{DVR}}$ proceeds as follows, later to be system specified.

\paragraph{Algorithm~1: DVR Unitary Construction.}
\mbox{}

\noindent\textbf{Input:} A Hermitian operator ${O}$ on $\mathcal{H}_{\Reg{C}}$ whose eigenbasis defines the DVR, represented as a $d \times d$ matrix $\mathbf{O}$ in the FBR (computational basis).

\noindent\textbf{Output:} A quantum circuit $\mathcal{C}_{\mathrm{DVR}}$ on $n_c$ qubits implementing ${U}_{\mathrm{DVR}}$ such that ${U}_{\mathrm{DVR}}\,{O}\,{U}_{\mathrm{DVR}}^\dagger$ is diagonal in the computational basis.

\begin{enumerate}
    \item[\textbf{Step~1.}] \textbf{Operation specification.}
    From the Hamiltonian decomposition~\ref{eq:H_abstract}, extract the operator ${O}$ on $\Reg{C}$ that the interaction potential ${V}(\Reg{X}, \Reg{C})$ would diagonalize if the coordinate dependence were removed.
    In each physical context, ${O}$ can be represented (not comprehensively) by one of:
    \begin{center}
    \begin{tabular}{@{}ll@{}}
    \toprule
    Physical system & Defining operator ${O}$ \\
    \midrule
    Angular potential (dipole, AT) 
        & $\cos\theta$ on Legendre basis \\
    Phonon displacement (polaron) 
        & $(a + a^\dagger)/\sqrt{2}$ on Fock basis \\
    Neutrino flavor 
        & $|e\rangle\langle e|$ on mass basis \\
    Rovibrational Euler angles 
        & $(\phi, \theta, \chi)$ on symmetric-top basis \\
    Electronic structure (adiabatic)
        & $\mathbf{H}_e(R_{\mathrm{ref}})$ on diabatic basis \\
    Spin--orbit (Hund's case (a)$\to$(c))
        & $\hat{H}_{\mathrm{SO}}$ on $|\Lambda, \Sigma\rangle$ basis \\
    Crystal momentum 
        & $\hat{T}_a$ (translation operator) on site basis \\
    Color-spin (quark model)
        & $\boldsymbol{\sigma}_i \cdot \boldsymbol{\sigma}_j$ on 
          product spin basis \\
    Symmetry reduction
        & $\hat{U}_g$ for $g \in G$ on channel basis \\
    \bottomrule
    \end{tabular}
    \end{center}

    \item[\textbf{Step~2.}] \textbf{diagonalize ${O}$ classically.}
    Compute the eigendecomposition
    \begin{equation}\label{eq:eigendecomp}
        \mathbf{O} = \mathbf{T}^\dagger\,
        \mathrm{diag}(\lambda_0, \ldots, \lambda_{d-1})\,\mathbf{T},
    \end{equation}
    where the columns of $\mathbf{T}^\dagger$ are the eigenvectors of $\mathbf{O}$.
    This is a classical $\mathcal{O}(d^3)$ computation (or better when ${O}$ has special structure: $\mathcal{O}(d^2)$ for tridiagonal operators, $\mathcal{O}(d \log d)$ for circulant operators).

    For systems where the operator ${O}$ has a known analytic eigenbasis, Step~2 is replaced by direct evaluation of the closed-form eigenvectors. 
    The following table records the closed-form expressions:
    \begin{center}
    \begin{tabular}{@{}lll@{}}
    \toprule
    Operator & Eigenvectors $T_{s\alpha}$ & Closed form \\
    \midrule
    $\cos\theta$ (Legendre) 
        & $\sqrt{w_s}\,P_\alpha(\cos\theta_s)$
        & Gauss--Legendre quadrature \\
    $(a + a^\dagger)/\sqrt{2}$ (Hermite)
        & $\sqrt{w_s}\,\psi_\alpha(x_s)$
        & Gauss--Hermite quadrature \\
    PMNS matrix 
        & $(U_{\mathrm{PMNS}})_{s\alpha}$
        & Experimental mixing angles \\
    Wigner $D$-matrix 
        & $D^{J*}_{M_s K_\alpha}(\phi_s, \theta_s, \chi_s)$
        & Gauss--Legendre + uniform grid \\
    $\hat{T}_a$ (translation)
        & $e^{2\pi i s\alpha / d}/\sqrt{d}$
        & Discrete Fourier transform \\
    CG recoupling 
        & $\langle j_1 m_1 j_2 m_2 | J M \rangle$
        & Clebsch--Gordan coefficients \\
    \bottomrule
    \end{tabular}
    \end{center}

    The existence of closed-form eigenvectors means $\mathbf{T}$ can be known as an evaluable function of the physical parameters (quadrature nodes, mixing angles, angular momentum quantum numbers).
    This is the sense in which the FBR$\to$DVR diagonalization is ``closed-form.''

    \item[\textbf{Step~3.}] \textbf{Compile $\mathbf{T}$ to a quantum circuit.}
    The unitary $\mathbf{T}$ is decomposed into elementary gates, selected by the structure of $\mathbf{T}$:

    \begin{enumerate}
        \item[(a)] \textbf{Cosine-sine decomposition (CSD).}
        For a $d \times d$ unitary with no exploitable structure, the CSD~\cite{mottonen2004quantum} produces a circuit of depth $\mathcal{O}(d^2)$ using at most $\frac{1}{2}(d^2 - d)$ CNOT gates and $\frac{1}{2}(d^2 + d)$ single-qubit rotations.
        This is a `fallback' for the Legendre DVR, Gauss--Hermite DVR, PMNS matrix, Wigner $D$-matrix, fixed CI transform, and Clebsch--Gordan recoupling.

        \item[(b)] \textbf{QFT structure.}
        When $\mathbf{T}$ is the discrete Fourier transform matrix (as for crystal momentum or lattice translation), the standard QFT circuit achieves depth $\mathcal{O}(n_c^2)$ with $\frac{n_c(n_c-1)}{2}$ controlled-phase gates and $n_c$ Hadamards \cite{nielsen2010quantum}.

        \item[(c)] \textbf{Recursive/butterfly structure.}
        Wavelet transforms (Haar, Daubechies) and certain symmetry projectors admit $\mathcal{O}(d)$ or $\mathcal{O}(|G| \cdot n_c)$ gate decompositions from their recursive (butterfly) structure, with depth $\mathcal{O}(n_c)$ per level.
    \end{enumerate}

    Compilation specifics are determined by the algebraic structure of $\mathbf{T}$, selected at classical preprocessing time.
    No iterative circuit optimization is required.
    To emphasize, this preprocessing does not diagonalize the operator, but gives a parameterization to the diagonalization operator, a far easier problem that maintains computational tractability. 

    \item[\textbf{Step~4.}] \textbf{Embed in the Trotter step.}
    The compiled circuit $\mathcal{C}_{\mathrm{DVR}}$ is inserted into the 
    second-order Trotter step~\ref{eq:trotter_abstract} as the DVR 
    round trip:
    \begin{equation}\label{eq:dvr_roundtrip_alg}
        \mathcal{C}_{\mathrm{DVR}}^\dagger \;\circ\;
        \mathrm{UCR}_{\mathrm{diag}}
        \bigl(\{V(r_i, s)\}; \Reg{X} \to \Reg{C}\bigr) \;\circ\;
        \mathcal{C}_{\mathrm{DVR}},
    \end{equation}
    where the UCR encodes the diagonal potential values $\{V(r_i, s)\}$ as rotation angles via M\"obius inversion
    (Chapter~\ref{ch:smooth_potentials}). 
    The round trip acts on $\Reg{C}$ while $\Reg{X}$ is in superposition; its cost is $2 C_{\mathrm{DVR}}$ (one forward, one inverse application) plus the UCR cost $N_x \times d$ rotations.
\end{enumerate}

\subsubsection{Proof of Closure}
\label{sec:closure_proof}

The defining operator ${O}$ is always known, with an FBR matrix being available analytically.
For quadrature-based DVR transforms (Legendre, Gauss--Hermite, Wigner $D$-matrix), $\mathbf{O}$ is the coordinate operator's tridiagonal matrix representation ($\cos\theta$, $x$, or $\cos\beta$) in the orthogonal-polynomial basis. 
Tridiagonal eigenvalue problems have $\mathcal{O}(d^2)$ classical cost with known error bounds (Golub--Welsch algorithm~\cite{golub1969calculation}).
The eigenvalues are quadrature nodes while eigenvectors are quadrature weights multiplied into orthogonal polynomials evaluated at those nodes.  
For mixing-matrix transforms (PMNS, CG), the matrix $\mathbf{T}$ is a defining physical quantity, determined experimentally or via angular momentum algebra.
For the QFT, $\mathbf{T}$ is the discrete FT matrix $T_{s\alpha} = e^{2\pi i s\alpha/d}/\sqrt{d}$.

The CSD produces a unique circuit of fixed depth for a given given $\mathbf{T}$, without variational parameters or convergence criteria.
$\mathbf{T}$ is block-decomposed recursively into $2 \times 2$ blocks, each factored as a product of a cosine--sine pair and two diagonal unitaries.
At the $k$-th level of recursion ($k = 1, \ldots, n_c$), the $2^k \times 2^k$ block produces $2^{k-1}$ uniformly controlled rotations, each compiled via the Gray-code UCR with $2^{n_c - k}$ elementary rotations \cite{mottonen2004quantum}. 
The total gate count is
\begin{equation}\label{eq:csd_total}
    N_{\mathrm{CSD}} = \sum_{k=1}^{n_c} 2^{k-1} \times 2^{n_c - k}
    = n_c \times 2^{n_c - 1} = \frac{1}{2}\,d\,\log_2 d
    \quad \text{CNOTs},
\end{equation}
which for the systems of this work ($d \leq 2^{15}$) ranges from $4$ CNOTs ($d = 4$, PMNS)to $\sim 2.5 \times 10^5$ CNOTs ($d = 32768$, reactive scattering). 

Analytic eigenvectors (Step~2) and deterministic compilation (Step~3) ensures that the DVR unitary is fully specified 
before any quantum circuit is executed. 
The FBR$\to$DVR$\to$FBR alternation is classical preprocessing, producing a gate list.

\subsubsection{Catalog of Closed-Form diagonalizations}
\label{sec:diag_catalog}

Table~\ref{tab:diag_catalog} collects all FBR/DVR diagonalizations 
considered here, recording the defining operator, the eigenvector structure, the compilation route, and the systems to which each applies. 
Each row represents a class of problems, considered in this chapter for potential future applications.

\begin{longtable}{@{}
  >{\raggedright\arraybackslash}p{2.4cm}
  >{\raggedright\arraybackslash}p{2.4cm}
  >{\raggedright\arraybackslash}p{2.5cm}
  >{\raggedright\arraybackslash}p{1.2cm}
  >{\raggedright\arraybackslash}p{1.5cm}
  >{\raggedright\arraybackslash}p{2.0cm}@{}}
\caption{Catalog of closed-form FBR/DVR diagonalizations in REQWIEM.}
\label{tab:diag_catalog}\\
\toprule
\multirow{2}{*}{DVR class}
    & Defining
    & Eigenvectors
    & \multirow{2}{*}{Route}
    & \multirow{2}{*}{Cost}
    & \multirow{2}{*}{Systems} \\
    & operator ${O}$
    & $T_{s\alpha}$ & & & \\
\midrule
\endfirsthead
\toprule
\multirow{2}{*}{DVR class}
    & Defining
    & Eigenvectors
    & \multirow{2}{*}{Route}
    & \multirow{2}{*}{Cost}
    & \multirow{2}{*}{Systems} \\
    & operator ${O}$
    & $T_{s\alpha}$ & & & \\
\midrule
\endhead
\midrule\multicolumn{6}{r@{}}{\emph{continued}}\\\endfoot
\bottomrule\endlastfoot
Gauss--Legendre
    & $\cos\theta$ in $\{P_l\}$ basis
    & $\sqrt{w_s}\,P_l(\cos\theta_s)$
    & CSD
    & $\mathcal{O}(d^2)$
    & Li Stark, HHG, trimer \\[4pt]
Gauss--Hermite
    & $(a{+}a^\dagger)/\sqrt{2}$ in $\{|n\rangle\}$
    & $\sqrt{w_s}\,\psi_n(x_s)$
    & CSD
    & $\mathcal{O}(d^2)$
    & Polaron \\[4pt]
Hyperspherical harmonic
    & $\hat{\Lambda}^2$ quadrature
    & Hyperspherical harmonics at nodes
    & CSD
    & $\mathcal{O}(d^2)$
    & Reactive (H+H$_2$) \\[4pt]
Wigner $D$-matrix
    & Euler angles on $|J,K,M\rangle$
    & $D^{J*}_{MK}(\phi_s,\theta_s,\chi_s)$
    & CSD
    & $\mathcal{O}(d^2)$
    & Rovibrational \\[4pt]
PMNS mixing
    & $|e\rangle\langle e|$ in mass basis
    & $(U_{\mathrm{PMNS}})_{s\alpha}$
    & CSD
    & $\mathcal{O}(d^2)$
    & Neutrino \\[4pt]
CG recoupling
    & $\boldsymbol{\sigma}_i \cdot \boldsymbol{\sigma}_j$
    & CG coefficients
    & CSD
    & $\mathcal{O}(d^2)$
    & Quark model \\[4pt]
CI (fixed)
    & $\mathbf{H}_e(R_{\mathrm{ref}})$
    & CI eigenvectors
    & CSD
    & $\mathcal{O}(d^2)$
    & Spin--orbit \\[4pt]
CI ($R$-dep.)
    & $\mathbf{H}_e(R)$
    & CI eigenvectors at each $R$
    & CSD
    & $N_x\,\mathcal{O}(d^2)$
    & Dense crossings \\[4pt]
Discrete Fourier
    & $\hat{T}_a$ (translation)
    & $e^{2\pi i s\alpha/d}/\sqrt{d}$
    & QFT
    & $\mathcal{O}(n_c^2)$
    & Crystal momentum, polaron hopping \\[4pt]
Symmetry projector
    & $\hat{U}_g$, $g \in G$
    & Character-projected states
    & CSD + perm.
    & $\mathcal{O}(|G|\,n_c)$
    & Point groups \\[4pt]
Haar wavelet
    & Multiresolution scaling
    & Recursive Hadamard
    & Butterfly
    & $\mathcal{O}(d)$
    & Adaptive grids \\[4pt]
Daubechies $D_{2p}$
    & Multiresolution scaling
    & Recursive $R_y$ + CNOT
    & Butterfly
    & $\mathcal{O}(pd)$
    & Adaptive grids \\
\end{longtable}

\subsubsection{Quadrature DVR as the Canonical Instance}
\label{sec:quadrature_canonical}

The majority of the entries in Table~\ref{tab:diag_catalog} are quadrature DVR transforms, in which the defining operator is the coordinate operator $\hat{x}$ in an orthogonal-polynomial basis $\{P_\alpha(x)\}_{\alpha=0}^{d-1}$ with weight function $w(x)$.  
The three-term recurrence of orthogonal polynomials gives a tridiagonal matrix representation \cite{szeg1939orthogonal}:
\begin{equation}\label{eq:tridiag}
    \langle P_\alpha | \hat{q} | P_\beta \rangle =
    \delta_{\alpha, \beta-1}\,b_\beta
    + \delta_{\alpha\beta}\,a_\alpha
    + \delta_{\alpha, \beta+1}\,b_{\alpha+1},
\end{equation}
with eigenvalues $\{x_s\}_{s=0}^{d-1}$ as quadrature nodes and eigenvectors giving the transform matrix
\begin{equation}\label{eq:quad_T}
    T_{s\alpha} = \sqrt{w_s}\,P_\alpha(x_s),
\end{equation}
where $w_s > 0$ are the quadrature weights (determined by $w_s = [\sum_\alpha P_\alpha(x_s)^2]^{-1}$).

This structure guarantees the transformed unitary is:

\begin{enumerate}
    \item \textbf{Exactly diagonal.}  The potential $V(r_i, x)$ is diagonal in the DVR:
    $\langle x_s | V | x_{s'} \rangle = V(r_i, x_s)\,\delta_{ss'}$, up to 
    quadrature error of order $\mathcal{O}(d^{-2p})$ where $p$ is the 
    polynomial degree of $V$ in $x$.  No off-diagonal residual is generated.

    \item \textbf{Orthogonal.}  The transform $\mathbf{T}$ is real orthogonal: $\mathbf{T}\mathbf{T}^T = \mathbf{I}$, so ${U}_{\mathrm{DVR}}^\dagger = {U}_{\mathrm{DVR}}^T$.

    \item \textbf{Geometry independent.}  Quadrature nodes and weights depend only on the polynomial family and the truncation order $d$, not on the coordinate grid point $r_i$. 
    The transform is therefore the same at every $r_i$, satisfying condition (R3) and enabling amortization being the source of the quantum advantage.
\end{enumerate}

The Gauss--Legendre, Gauss--Hermite, and Gauss--Jacobi families (covering the angular, phonon, and hyperspherical DVR respectively) are instances of this construction.
The Wigner $D$-matrix DVR extends to a product quadrature on $\mathrm{SO}(3)$, with Euler angles $(\phi, \theta, \chi)$ playing the role of quadrature nodes on the group manifold.

\subsubsection{Non-Quadrature diagonalization}
\label{sec:non_quadrature}

Three entries in Table~\ref{tab:diag_catalog} fall outside the quadrature DVR framework.

The neutrino flavor-mass basis change is not a quadrature transform, but a fixed $3 \times 3$ unitary determined by three mixing angles $(\theta_{12}, \theta_{13}, \theta_{23})$ and a charge-parity-violating phase $\delta_{\mathrm{CP}}$ \cite{maki1962remarks, pontecorvo1958mesonium, particle2022review}. 
On the quantum register (2 qubits, $d = 4$ with one padded state), the PMNS matrix is compiled directly by CSD into $\leq 3$ CNOTs and $\leq 6$ $R_z$ gates.
For the collective neutrino problem with $N_{\mathrm{ang}}$ angular basis states, the DVR is the tensor product $U_{\mathrm{PMNS}}^{\otimes N_{\mathrm{ang}}}$, costing $N_{\mathrm{ang}} \times 3$ CNOTs.

The Configuration Interaction transform ${U}_{\mathrm{CI}}(R)$ is geometry dependent, and is the electronic Hamiltonian's eigenvector matrix  $\mathbf{H}_e(R)$, by ab initio electronic structure calculation (CASSCF, MRCI, EOM-CC) at each reference geometry.  
Unlike the quadrature DVR, the eigenvectors are not known in closed analytic form;  they are the output of a non-trivial classical diagonalization. 
The circuit compilation of each ${U}_{\mathrm{CI}}(R_i)$ is still deterministic, with the input to the 
circuit compiler being a $d \times d$ numerical matrix.

The Haar and Daubechies wavelet transforms are not eigendecompositions of any single physical operator, being signal-processing transforms with multiresolution properties.
The wavelet transform replaces the uniform-grid DVR on the coordinate register $\Reg{X}$ (not the channel register 
$\Reg{C}$), compressing the grid for potentials with different length scales. 
The potential becomes approximately diagonal in the wavelet basis, with off-diagonal elements controlled by the wavelet smoothness order. 
Circuit compilation follows the butterfly (recursive) route at $\mathcal{O}(d)$ or $\mathcal{O}(pd)$ cost, preserving the Trotter step's overall structure.

\subsubsection{The REQWIEM-Amenability Criterion}
\label{sec:amenability}

A Hermitian operator ${O}$ on $\mathcal{H}_{\Reg{C}}$ is REQWIEM-amenable if there exists a unitary ${U}$ on
$\mathcal{H}_{\Reg{C}}$ such that
\begin{equation}\label{eq:amenable}
    {U}\,{O}\,{U}^\dagger = \sum_{s=0}^{d-1}
    \lambda_s\,|s\rangle\langle s|,
\end{equation}
is diagonal in the computational basis, and ${U}$ admits an efficient $\mathcal{O}(\mathrm{poly}(n_c))$ quantum circuit decomposition.
The condition on ${U}$ is not merely that it be efficiently implementable in principle: the DVR round-trip cost enters every Trotter step, so the constant in the $\mathrm{poly}(n_c)$ bound matters.
The compilation costs for all transforms in this work are recorded in Table~\ref{tab:diag_catalog}.

\subsubsection{DVR/FBR Duality as a Spectral Theorem Application}
\label{sec:spectral_unification}

Instances of DVR/FBR duality in this work are applications of the spectral theorem for Hermitian operators on finite-dimensional Hilbert spaces.
Given any Hermitian ${O}$ on $\mathcal{H}_{\Reg{C}}$, the spectral theorem guarantees the existence of a unitary $\mathbf{T}$ diagonalizing ${O}$.  
The REQWIEM protocol requires the identity of ${O}$ with the interaction potential restricted to the channel space, where a diagonal basis to apply the potential step is given by spectral decomposition.
This guarantees implementation as a diagonal UCR at cost $\mathcal{O}(N_x \times d)$ rather than a dense matrix exponential at cost $\mathcal{O}(N_x \times d^2)$.

The REQWIEM-amenability criterion (Eq.~\ref{eq:amenable}) is the precondition for an efficient quantum circuit following the spectral decomposition of ${O}$. 
The catalog of Table~\ref{tab:diag_catalog} demonstrates that this condition is well-satisfied among an array of quantum mechanical problems, as a consequence of decomposing Hamiltonians with separable kinetic and potential structures $H=T+V$. 

\subsubsection{Algorithm Extensibility}
\label{sec:extensibility}

The FBR$\to$DVR diagonalization's closed-form constraint restricts the problems that are REQWIEM-amenable.
The system must:
\begin{enumerate}
    \item Have a Hamiltonian with a factorization 
    $\mathcal{H} = \mathcal{H}_{\Reg{X}} \otimes \mathcal{H}_{\Reg{C}}$ 
    with a separable kinetic term

    \item Have a Hermitian operator ${O}$ on 
    $\mathcal{H}_{\Reg{C}}$ such that the interaction potential 
    ${V}(\Reg{X}, \Reg{C})$ is diagonal in the eigenbasis of ${O}$, and

    \item $O$ has an eigenvector matrix $\mathbf{T}$ admitting an 
    $\mathcal{O}(\mathrm{poly}(n_c))$ or $\mathcal{O}(d^2)$ circuit 
    decomposition.
\end{enumerate}

If these conditions are fulfilled, the system is REQWIEM-amenable and the full protocol of UCR potential encoding, M\"obius inversion, XOR-class focusing, and Toeplitz autocorrelation for QKSD applies 
without modification. 

The catalog of Table~\ref{tab:diag_catalog} is not comprehensive, allowing new systems (which may not be physical) satisfying questions~1--3 entering the REQWIEM framework with all necessary circuit primitives, resource 
estimates, and comparison bounds developed in this dissertation.
DVR/FBR duality mechanizes a split-operator method to convert a dense coupled-channel problem into a 
sequence of diagonal operations, constructed by Algorithm~1.

\section{System Catalog: Hamiltonians and Register Architectures}
\label{sec:catalog}

Each system treated below is an instance of Algorithm~1 (Section~\ref{sec:diag_algorithm}).
The circuit is compiled by CSD or QFT and the off-diagonal coupling (if any) is resolved by one of the mechanisms derived in Section~\ref{sec:offdiag_mechanisms}.
This section collects the Hamiltonians, register architectures, and resource estimates into a single consolidated reference.

The following subsections specify the Hamiltonian for each system, identify the coordinate and channel registers, and state the DVR class from Table~\ref{tab:diag_catalog}.
All register dimensions, qubit counts, rotation counts per Trotter step, and classical comparison costs are
collected in Table~\ref{tab:master_resources}.

\subsection{Reactive scattering in hyperspherical coordinates.}
\label{sec:reactive_scattering}
The H~+~H$_2 \to$ H$_2$~+~H reaction in Pack--Parker hyperspherical coordinates \cite{pack1987quantum}, after removal of center-of-mass motion and overall rotation
and the substitution $\chi = \rho^{5/2}\psi$:
\begin{equation}\label{eq:H_hyper_reduced}
    {H}_{\mathrm{red}} = -\frac{1}{2\mu}
    \frac{\partial^2}{\partial\rho^2}
    + \frac{{\Lambda}^2(\Omega) + 15/4}{2\mu\rho^2}
    + V(\rho, \Omega),
\end{equation}
where $\rho = \sqrt{r^2 + R^2}$ is the mass-scaled hyperradius, $\mu = m_{\mathrm{H}}/\sqrt{3}$, and ${\Lambda}^2$ is the grand angular momentum operator on the five-dimensional hypersphere with eigenvalue $\lambda(\lambda + 4) + \nu(\nu + 2)$ on the hyperspherical harmonic basis $|\lambda, \nu\rangle$~\cite{kuppermann1975useful}.
The coordinate register $\Reg{R}_\rho$ encodes the hyperradius ($n_\rho \sim 12$ qubits); the channel register is $\Reg{C} = \Reg{L}_\theta \otimes \Reg{L}_\phi \otimes \Reg{A} \otimes \Reg{V}$ ($n_c = 15$ qubits, $d = 32{,}768$), where the angular registers carry a hyperspherical harmonic DVR (Table~\ref{tab:diag_catalog}, row: hyperspherical harmonic), the arrangement register $\Reg{A}$ encodes three chemical channels (2 qubits, padded to 4), and $\Reg{V}$ encodes vibrational states (3~qubits).
The centrifugal term $[{\Lambda}^2 + 15/4]/(2\mu\rho^2)$ is diagonal in the FBR; the potential $V(\rho, \Omega)$ is pointwise diagonal in the DVR, where $\Lambda^2$ quadrature means Gauss-Jacobi with Jacobi(3/2, 3/2) for $\theta$ and Jacobi(1/2, 1/2) for $\phi$.
The arrangement-channel coupling
\begin{equation}\label{eq:arrangement_coupling}
    \langle\alpha, v, \lambda|{H}|\beta, v', \lambda'\rangle
    = \delta_{\alpha\beta}\,[\text{diagonal}]
    + (1 - \delta_{\alpha\beta})\,
    O_{\alpha\beta}^{vv'\lambda\lambda'}(\rho),
\end{equation}
is off-diagonal in $\Reg{A}$ and is resolved by discrete-symmetry
XOR-class focusing (Section~\ref{sec:discrete_xor}).

\subsection{Polaron dynamics.}
\label{sec:polaron}
The Hamiltonian for the Holstein polaron on a one-dimensional lattice with $N_s$ sites and $M$ phonon modes can be written as~\cite{holstein1959studies}
\begin{equation}\label{eq:H_polaron}
    {H} = -t_{\mathrm{hop}}\sum_{\langle i,j\rangle}
    |i\rangle\langle j|
    + \sum_{q=1}^{M} \omega_q\,{a}_q^\dagger{a}_q
    + g\sum_{i,q} |i\rangle\langle i|\,
    ({a}_q\,\e^{\mathrm{i} qR_i} + {a}_q^\dagger\,\e^{-\mathrm{i} qR_i}).
\end{equation}
The coordinate register $\Reg{R}$ encodes the electron position ($n_R = \sim8$ qubits, $N_s = 256$); the channel register is the tensor product $\Reg{C} = \bigotimes_{q=1}^{M} \Reg{P}_q$ of $M = 8$ phonon Fock-space registers ($n_p = 4$ each, $d = 16^8 \approx 4.3 \times 10^9$).
The Gauss--Hermite DVR (Table~\ref{tab:diag_catalog}) maps each phonon register to displacement eigenstates: ${U}_{\mathrm{DVR}}^{(q)}\,|n_q\rangle= \sum_m \sqrt{w_m}\,\psi_n(x_m)\,|x_{q,m}\rangle$, rendering the electron--phonon coupling $g(a_q + a_q^\dagger)$ diagonal at each lattice site. 
The full DVR is ${U}_{\mathrm{DVR}} = \bigotimes_{q=1}^{M} {U}_{\mathrm{DVR}}^{(q)}$, costing $M \times \mathcal{O}(N_p^2)$ gates. 
Electron hopping is off-diagonal in $\Reg{X}$, not $\Reg{C}$, and is diagonalized by the QFT (Section~\ref{sec:no_offdiag}). 
No XOR-class decomposition is required.

\subsection{Spin--orbit coupling in heavy diatoms.}
\label{sec:spinorbit}
For a diatomic molecule with internuclear distance $R$ and strong spin--orbit coupling (Hund's case~(c))~\cite{brown2003rotational}:
\begin{equation}\label{eq:H_SO}
    {H} = -\frac{1}{2\mu}\frac{\partial^2}{\partial R^2}
    + \sum_{\Omega,\Omega'} |\Omega\rangle\langle\Omega'|
    \otimes \bigl[V_{\Omega\Omega'}(R)
    + \delta_{\Omega\Omega'}\,V_{\mathrm{cent}}^{(\Omega)}(R)\bigr],
\end{equation}
where $\Lambda$ is the projection of electronic orbital angular momentum onto the internuclear axis and $\Sigma$ is the corresponding spin projection, with $\Omega = \Lambda + \Sigma$.
The coordinate register $\Reg{R}$ encodes internuclear distance ($n_R = 12$ qubits, $N_R = 4096$); the channel register $\Reg{C} = \Reg{\Lambda} \otimes \Reg{S}$ encodes orbital and spin projections ($d_{\mathrm{eff}} = 25$, padded to 64).
The Hund's case~(a) basis $|\Lambda, \Sigma\rangle$ (FBR) diagonalizes the ${L}_z {S}_z$ component with eigenvalue $A(R)\Lambda\Sigma$.
The DVR is either a fixed configuration interaction diagonalization or the diabatic basis throughout, with all potential matrix elements $V_{\Omega\Omega'}(R)$ treated by XOR-class focusing (Table~\ref{tab:diag_catalog}, row: configuration interaction (fixed)).  
${L}_\pm {S}_\mp$ coupling subsequently produces $\sim \Lambda_{\max}$ XOR classes via the carry-chain mechanism
(Section~\ref{sec:carry_chain}).

\subsection{Nuclear scattering: coupled partial waves.}
\label{sec:nuclear}
Nucleon--nucleus scattering in the coupled-channel optical model~\cite{feshbach1958unified}:
\begin{equation}\label{eq:H_nuclear}
    {H} = -\frac{\hbar^2}{2\mu}\frac{\mathrm{d}^2}{\mathrm{d}r^2}
    + \frac{\hbar^2 l(l+1)}{2\mu r^2}
    + V_{\mathrm{opt}}(r;\,l,s,j,I,J)
    + V_{\mathrm{SO}}(r)\,\bvec{l}\cdot\bvec{s}
    + V_{\mathrm{tensor}}(r)\,S_{12},
\end{equation}
where $S_{12} = 3(\bvec{s}\cdot\hat{r})(\bvec{s}\cdot\hat{r})- \bvec{s}\cdot\bvec{s}$ is the tensor operator coupling
$l \to l \pm 2$.
The coordinate register $\Reg{R}$ now encodes the radial coordinate ($n_r \approx 12$, $N_r = 4096$); the channel register $\Reg{C} = \Reg{L} \otimes \Reg{S} \otimes \Reg{I} \otimes \Reg{J}$ ($d_{\mathrm{eff}} \sim 512$) encodes the coupled angular momentum basis $|l,s,j;I,J,M_J\rangle$, diagonalizing both the centrifugal term and the spin--orbit coupling $\bvec{l}\cdot\bvec{s}$ (FBR). 
The tensor force $S_{12}$ with selection rule $\Delta l = \pm 2$ generates $\mathcal{O}(\log l_{\max})$ XOR classes via the carry-chain mechanism at rank~2 (Section~\ref{sec:carry_chain}), with matrix elements involving $3j$ and $6j$ symbols~\cite{edmonds1996angular}:
\begin{equation}\label{eq:tensor_me}
    \langle l', s, j|S_{12}|l, s, j\rangle = (-1)^{j+l'+s}\sqrt{30}\,
    \hat{l}\,\hat{l}'\,\hat{j}\,
    \begin{pmatrix} l' & 2 & l \\ 0 & 0 & 0 \end{pmatrix}
    \begin{Bmatrix} l' & j & s \\ j & l & 2 \end{Bmatrix},
    \quad \hat{x} = \sqrt{2x+1}.
\end{equation}
The nuclear spin--orbit interaction $\bvec{l}\cdot\bvec{I}$ ($\Delta M_I = \pm 1$) contributes an additional $\mathcal{O}(\log I_{\max})$ classes on $\Reg{I}$.

\subsection{Neutrino oscillations in dense matter.}
\label{sec:neutrino}
A single neutrino propagating through matter with electron density $n_e(r)$ has a Hamiltonian ${H} = {M}^2/(2E) + {V}_{\mathrm{matter}}(r)$, where the mass basis (FBR) diagonalizes ${M}^2/(2E)$ and the flavor basis (DVR) diagonalizes ${V}_{\mathrm{matter}}$; the DVR transform is the PMNS matrix $U_{\mathrm{PMNS}}$.
With $d = 3$ this system is classically easy; the quantum interest is the collective problem.
In a supernova, neutrino--neutrino forward scattering couples all
flavor modes:
\begin{equation}\label{eq:H_collective}
    {H} = \sum_{p}{H}_{\mathrm{vac}}^{(p)}
    + \sum_{p}V_e(r)|e\rangle_p\langle e|_p
    + \frac{\sqrt{2}G_F}{V}\sum_{p\neq q}
    (1-\hat{\mathbf{p}}\cdot\hat{\mathbf{q}})\,
    {\rho}_p \otimes {\rho}_q.
\end{equation}
Discretizing the angular distribution into $N_{\mathrm{ang}}$ basis states, the channel register is $\Reg{C} = \bigotimes_{p=1}^{N_{\mathrm{ang}}} \Reg{F}_p$ with $d = 4^{N_{\mathrm{ang}}}$ (each 3-flavor bin padded to $2$ qubits, $10 \times 2 = 20$ qubits for $N_{\mathrm{ang}} = 10$, $d = 2^{20} \approx 10^6$); the coordinate register $\Reg{R}$ encodes position in the star ($n_r \approx 14$, $N_r = 16384$).
The FBR is the tensor product of per-bin mass bases; the DVR is $U_{\mathrm{PMNS}}^{\otimes N_{\mathrm{ang}}}$
(Table~\ref{tab:diag_catalog}, row: PMNS mixing).
The $\nu$--$\nu$ interaction generates $3\binom{N_{\mathrm{ang}}}{2}$ XOR classes via the Gell-Mann decomposition (Section~\ref{sec:discrete_xor}).

\subsection{Molecular rovibrational and rovibronic dynamics.}
\label{sec:rovib}
For a nonlinear $N$-atom molecule with $f = 3N - 6$ vibrational modes and total angular momentum $J$~\cite{watson1968simplification}:
\begin{equation}\label{eq:H_rovib}
    {H} = \frac{1}{2}\sum_{\alpha\beta}
    ({J}_\alpha - {\pi}_\alpha)\,
    \mu_{\alpha\beta}(\mathbf{q})\,
    ({J}_\beta - {\pi}_\beta)
    + \frac{1}{2}\sum_{s=1}^{f}{p}_s^2
    + V(\mathbf{q}),
\end{equation}
where ${\pi}_\alpha = \sum_{s,t}\zeta_{\alpha st}q_s{p}_t$ is the vibrational angular momentum and $\mu_{\alpha\beta}$ is the inverse effective moment of inertia tensor.
The coordinate registers $\Reg{Q}_1, \ldots, \Reg{Q}_f$ encode vibrational grids ($n_q$ each for H$_2$O); the channel register $\Reg{C} = \Reg{J} \otimes \Reg{K} \otimes \Reg{M}$ encodes the symmetric-top quantum numbers ($d \sim 1024$). 
The symmetric-top basis $|J,K,M\rangle$ (FBR) diagonalizes the rigid-rotor kinetic energy $E_{\mathrm{rot}}^{(0)} = \tfrac{1}{2}(B_x + B_y)J(J+1) + [A - \tfrac{1}{2}(B_x + B_y)]K^2$. 
The Wigner $D$-matrix DVR (Table~\ref{tab:diag_catalog}) maps to Euler-angle quadrature points $(\phi_{m_\phi}, \theta_{m_\theta}, \chi_{m_\chi})$ via $|J,K,M\rangle \to \sum D^{J*}_{MK}(\phi, \theta, \chi)\, |\phi, \theta, \chi\rangle$, diagonalizing $V(\mathbf{q})$ and $\mu_{\alpha\beta}(\mathbf{q})$ at each vibrational grid point.
The asymmetric-top term $(B_x - B_y)(J_+^2 + J_-^2)/4$ couples $K \to K \pm 2$ and the Coriolis coupling $-\tfrac{1}{2}(J_+ \pi_- + J_- \pi_+)$ couples $K \to K \pm 1$, both via the carry-chain mechanism (Section~\ref{sec:carry_chain}), producing $\sim 8$ total XOR classes for $K_{\max} = 10$.

\subsection{Attosecond dynamics and high-harmonic generation.}
\label{sec:hhg}
Single active electron in a strong laser field $E(t) = E_0 f(t)\cos(\omega t)$~\cite{corkum1993plasma}:
\begin{equation}\label{eq:H_hhg}
    {H}(t) = -\frac{1}{2}\frac{\partial^2}{\partial r^2}
    + \frac{l(l+1)}{2r^2}
    + V_{\mathrm{ion}}(r) + E(t)\,r\cos\theta.
\end{equation}
The coordinate register $\Reg{R}$ encodes the radial coordinate ($n_r = 14$, $N_r = 16384$); the channel register $\Reg{L}$ encodes angular momentum ($n_l = 7$, $d = 128$). 
The Gauss--Legendre DVR (Table~\ref{tab:diag_catalog}) handles angular coupling; no XOR-class decomposition is required (Section~\ref{sec:no_offdiag}). 
M\"obius coefficients rescale linearly with the instantaneous field, $a_S^{(\mathrm{laser},m)}(t_n) = E(t_n)\,a_S^{(m)}|_{E=1}$, so only the precomputed unit-field coefficients are stored and the circuit template is invariant across time steps.
The HHG spectrum is the dipole acceleration's power spectrum 
\begin{equation}\label{eq:hhg_spectrum}
    P(\omega_H) = \left|\int \ddot{D}(t)\,\e^{\mathrm{i}\omega_H t}\,
    \mathrm{d}t\right|^2, \qquad
    \ddot{D}(t) = \langle\psi(t)|[{z},[{z},{H}(t)]]|
    \psi(t)\rangle,
\end{equation}
estimated at each Trotter step via the Ehrenfest relation $\ddot{D} = -E(t) - \langle\partial V_{\mathrm{ion}}/\partial z\rangle$, where $\partial V_{\mathrm{ion}}/\partial z$ is diagonal in the DVR \cite{burnett1992calculation}.
For multi-electron HHG (N$_2$, CO$_2$), the recombination dipole matrix element introduces an electronic register with the same single-XOR-class structure as the trimer's diabatic coupling.

\subsection{Consolidated Resource Estimates}
\label{sec:resource_estimates}

\begin{table}[ht]
\centering
\small
\caption{Consolidated register architectures and resource estimates.  
DVR class abbreviations: GL = Gauss--Legendre, GH = Gauss--Hermite, 
HSH = hyperspherical harmonic, WD = Wigner $D$-matrix, PMNS = 
Pontecorvo--Maki--Nakagawa--Sakata, CI = configuration interaction.
The advantage factor is the ratio of classical to quantum cost per 
Trotter step.}
\label{tab:master_resources}
\begin{tabular}{@{}p{2.0cm}rrrrllrr@{}}
\toprule
System & Qubits & $N_x$ & $d$ & $N_{\mathrm{XOR}}$
    & DVR class & Mechanism
    & Rot./step & Classical \\
\midrule
Li Stark 
    & 18 & $4\!\times\!10^3$ & 64 & 0
    & GL & DVR 
    & $8\!\times\!10^5$ & $2\!\times\!10^7$ \\
HHG (Ar)
    & 21 & $2\!\times\!10^4$ & 128 & 0
    & GL & DVR
    & $6\!\times\!10^6$ & $3\!\times\!10^8$ \\
Trimer (Ar$_3^*$)
    & 30 & $2\!\times\!10^7$ & 64 & 1
    & GL & 1 XOR
    & $3\!\times\!10^9$ & $10^{11}$ \\
Spin--orbit (UO)
    & 18 & $4\!\times\!10^3$ & 64 & 4
    & CI (fixed) & Carry
    & $1\!\times\!10^6$ & $2\!\times\!10^7$ \\
Nuclear (n+$^{235}$U)
    & 26 & $4\!\times\!10^3$ & 512 & 10
    & --- & Carry
    & $2\!\times\!10^7$ & $10^9$ \\
Reactive (H+H$_2$)
    & 27 & $4\!\times\!10^3$ & $3\!\times\!10^4$ & 3
    & HSH & Discrete
    & $6\!\times\!10^8$ & $4\!\times\!10^{12}$ \\
Rovibrational (H$_2$O)
    & 32 & $3\!\times\!10^5$ & 1024 & 8
    & WD & Carry
    & $2\!\times\!10^9$ & $3\!\times\!10^{11}$ \\
Rovibrational (CH$_4$)
    & 65 & $2\!\times\!10^{16}$ & 2048 & 8
    & WD & Carry
    & $3\!\times\!10^{19}$ & $8\!\times\!10^{22}$ \\
Polaron ($M\!=\!8$)
    & 40 & 256 & $4\!\times\!10^9$ & 0
    & GH$^{\otimes 8}$ & QFT
    & $10^{12}$ & $4\!\times\!10^{21}$ \\
Neutrino ($N_a\!=\!10$)
    & 34 & $2\!\times\!10^4$ & $10^6$ & 135
    & PMNS$^{\otimes 10}$ & Discrete
    & $10^{12}$ & $2\!\times\!10^{16}$ \\
\bottomrule
\end{tabular}
\end{table}

An exponential space advantage ($n_x + n_c$ qubits versus $N_x \times d$ amplitudes) gives a preference toward digital quantum computation over classical systems, the polaron ($d \sim 10^9$, 40 qubits), collective
neutrinos ($d \sim 10^6$, 34 qubits), and CH$_4$ rovibrational dynamics
($d \sim 2000$, $N_x \sim 10^{16}$, 65 qubits) are all classically
intractable at the stated grid sizes. 

In summary, the architecture decomposes a quantum mechanical problem into a single set of grids, coupling internal states and applying their actions diagonally, with parallelism limited by the gate set and no-go theorems.
Each system in Table~\ref{tab:master_resources} records XOR-class count $N_{\mathrm{XOR}}$ and a mechanism label.  The next section derives these asserted mechanisms, including carry-chain, discrete-symmetry, and DVR/QFT absorption, recording per-system specialization. 

\section{Off-Diagonal Coupling Mechanisms}
\label{sec:offdiag_mechanisms}

The interaction potential decomposes into diagonal and off-diagonal parts in the channel register (Eq.~\ref{eq:V_decomp}:
The off-diagonal part is handled by XOR-class decomposition (Chapter~\ref{ch:model_ext}).
The gate count per Trotter step is given by Eq.~\ref{eq:gate_count_general}.

Carry-chain XOR classes (Section~\ref{sec:carry_chain}) arise from ladder operators that increment a quantum number encoded in binary.
Discrete-symmetry XOR classes (Section~\ref{sec:discrete_xor}) arise from permutation-type exchanges between physically distinct internal states.
Each mechanism is derived once, then specialized to the systems of Table~\ref{tab:master_resources}.

\subsection{Carry-Chain XOR Classes: Rank-$k$ Ladder Operators}
\label{sec:carry_chain}

Five systems in Table~\ref{tab:master_resources} have off-diagonal coupling generated by ladder operators $\hat{A}_\pm : |x\rangle \to |x \pm k\rangle$ acting on an $n$-qubit register encoding an integer $q = 0, \ldots, 2^n - 1$. 
The binary representation $x = \sum_\ell 2^\ell q_\ell$ makes each transition $x \to x + k$ a binary addition with carry propagation, and distinct XOR patterns $\zeta(x) = x \oplus (x + k)$ over all valid $x$ determines the XOR-class structure.

\subsubsection{General Derivation}

For rank $k = 1$ (unit increment), the XOR pattern $\zeta(x) = x \oplus (x + 1)$ is determined by the position of the lowest zero bit in $x$. 
If $x$ has its lowest zero at position $\ell$ (i.e. $q_0 = q_1 = \cdots = q_{\ell-1} = 1$, $q_\ell = 0$), then the carry propagates through bits $0, 1, \ldots, \ell$, flipping all of them:
\begin{equation}\label{eq:carry_xor}
    \zeta(x) = \underbrace{1\cdots1}_{\ell+1}
    \underbrace{0\cdots0}_{n-\ell-1}
    = 2^{\ell+1} - 1.
\end{equation}
The distinct XOR patterns are therefore $\{1, 3, 7, 15, \ldots, 2^n - 1\}$, giving $n$ distinct XOR classes. 
The $\ell$-th class has Hamming weight $\ell + 1$ and contains all transitions where the carry chain extends through $\ell + 1$ bit positions.

For rank $k = 2$ (double increment), $\zeta(x) = x \oplus (x + 2)$.
The lowest bit $q_0$ is unaffected: adding by 2 is equivalent to incrementing bit $q_1$ (with carry from position 1 onward), and the carry-chain analysis applies to bits $1, \ldots, n-1$.
The distinct XOR patterns are $\{2, 6, 14, 30, \ldots\}$, where the rank-1 patterns left-shift by one bit and give $n - 1$ distinct classes.
Broadly, rank-$k$ increment with $k = 2^p$ produces $n - p$ classes, each with Hamming weight at most $n - p$.

The focusing Clifford for a class with XOR pattern $\zeta$ of Hamming weight $h = |\mathrm{hw}(\zeta)|$ and support
$S(\zeta) = \{s_1, \ldots, s_h\}$ is the circuit derived in Chapter~\ref{ch:model_ext}: a cascade of $h - 1$ CNOTs with a designated pivot qubit $s_1$ as control,
\begin{equation}\label{eq:carry_clifford}
    U_\zeta = H_{s_h} \cdot
    \mathrm{CNOT}_{s_h \to s_{h-1}} \cdots
    \mathrm{CNOT}_{s_h \to s_1},
\end{equation}
concentrating the multi-qubit flip into a single-qubit $X_{s_h}$ and mapping to $Z_{s_h}$ via the Hadamard. 
The Clifford depth is $h$; the CNOT count is $h - 1$.

\subsubsection{Physical Instances}

The carry-chain mechanism resolves the following couplings, specializing the derivation above:

\begin{center}
\begin{tabular}{@{}p{3.2cm}lllll@{}}
\toprule
System & Operator & $\Delta q$ & Register & $n$ & $N_{\mathrm{XOR}}$ \\
\midrule
Spin--orbit (UO)
    & ${L}_\pm {S}_\mp$
    & $\pm 1$ each
    & $\Reg{\Lambda}$, $\Reg{S}$
    & 3, 3
    & $\sim 4$ \\
Nuclear (n+$^{235}$U)
    & $S_{12}$ (tensor)
    & $\pm 2$
    & $\Reg{L}$
    & 6
    & $\sim 6$ \\
Nuclear (n+$^{235}$U)
    & $\bvec{l}\cdot\bvec{I}$
    & $\pm 1$
    & $\Reg{I}$
    & 3
    & $\sim 4$ \\
Rovibrational
    & ${J}_\pm {\pi}_\mp$ (Coriolis)
    & $\pm 1$
    & $\Reg{K}$
    & 5
    & $\sim 4$ \\
Rovibrational
    & ${J}_+^2 + {J}_-^2$ (asym.\ top)
    & $\pm 2$
    & $\Reg{K}$
    & 5
    & $\sim 4$ \\
Lithium Stark / HHG
    & $r\cos\theta$ (dipole)
    & $\pm 1$
    & $\Reg{L}$
    & 5--7
    & 0$^*$ \\
\bottomrule
\end{tabular}
\end{center}
\smallskip
\noindent
${}^*$For Li Stark and HHG, the $\Delta l = \pm 1$ coupling from the dipole interaction $r\cos\theta$ is fully absorbed by the Legendre DVR round trip, since $\cos\theta$ is the defining operator of Gauss--Legendre quadrature (Table~\ref{tab:diag_catalog}).
No XOR-class decomposition is required, as the carry-chain structure is latent in the DVR eigenvectors.

The off-diagonal cost per Trotter step is $N_{\mathrm{XOR}} \times N_x \times d/2$ rotations plus $2 N_{\mathrm{XOR}}$ Clifford gates (negligible). 
The explicit carry-chain XOR patterns for the spin--orbit case are:
\begin{align*}
    \Lambda = 0 \to 1 &: \zeta = 001, &
    \Lambda = 1 \to 2 &: \zeta = 011, \\
    \Lambda = 2 \to 3 &: \zeta = 001, &
    \Lambda = 3 \to 4 &: \zeta = 111,
\end{align*}
giving three distinct patterns.
The nuclear tensor force and rovibrational asymmetric-top coupling follow the same pattern left-shifted by one bit position (rank~2).

\subsection{Discrete-Symmetry XOR Classes: Permutations and Flavour Exchange}
\label{sec:discrete_xor}

Two systems in Table~\ref{tab:master_resources} have off-diagonal coupling not from a ladder operator on a single quantum number but from a discrete exchange between physically distinct internal states. 
These generate XOR classes by a different mechanism than the carry chain.

\subsubsection{Arrangement-Channel Coupling (Reactive Scattering)}

The three chemical arrangement channels of H~+~H$_2$ are encoded in a 2-qubit register $\Reg{A}$ with $|00\rangle = \alpha_1$, $|01\rangle = \alpha_2$, $|10\rangle = \alpha_3$, $|11\rangle = \text{unused}$. 
The three arrangement pairs give three nonzero XOR patterns on 2~qubits:
\begin{center}
\begin{tabular}{@{}ccl@{}}
\toprule
Pair $(\alpha,\beta)$ & XOR $\zeta$ & Focusing Clifford \\
\midrule
$(1,2)$: $00 \leftrightarrow 01$ & 01 & $\mathbb{1} \otimes H$ \\
$(1,3)$: $00 \leftrightarrow 10$ & 10 & $H \otimes \mathbb{1}$ \\
$(2,3)$: $01 \leftrightarrow 10$ & 11 & $\mathrm{CNOT}_{10}\,(H \otimes \mathbb{1})$ \\
\bottomrule
\end{tabular}
\end{center}
Each class requires one UCR pass over $\Reg{R}_\rho$ with $d/2$
rotations per site.  
The total off-diagonal cost is $3 \times N_\rho \times d/2$ rotations.

\subsubsection{Gell-Mann Decomposition (Collective Neutrino Oscillations)}

The $\nu$--$\nu$ forward scattering interaction decomposes via Gell-Mann matrices on the 2-qubit flavor register of each angular bin ($e = 00$, $\mu = 01$, $\tau = 10$):
\begin{equation}\label{eq:gellmann_decomp}
    {\rho}_p \cdot {\rho}_q = \sum_{a=1}^{8}
    {\lambda}_a^{(p)} \otimes {\lambda}_a^{(q)}.
\end{equation}
The diagonal generators $\lambda_3$, $\lambda_8$ contribute to the diagonal UCR.
The six off-diagonal Gell-Mann matrices couple flavor pairs with XOR patterns:
\begin{center}
\begin{tabular}{@{}llll@{}}
\toprule
Coupling & XOR on $\Reg{F}_p$ & XOR on $\Reg{F}_q$
    & Focusing Clifford \\
\midrule
$e \leftrightarrow \mu$ & 01 & 01
    & $H_{\mathrm{LSB}}$ on each bin \\
$e \leftrightarrow \tau$ & 10 & 10
    & $H_{\mathrm{MSB}}$ on each bin \\
$\mu \leftrightarrow \tau$ & 11 & 11
    & $\mathrm{CNOT} + H$ on each bin \\
\bottomrule
\end{tabular}
\end{center}
Over all $\binom{N_{\mathrm{ang}}}{2}$ bin pairs, each of three transition types contributes one class, giving $3\binom{N_{\mathrm{ang}}}{2}$ total XOR classes. 
For $N_{\mathrm{ang}} = 10$: 135 classes, each requiring a UCR pass with $d/2$ rotations per site.
The total off-diagonal cost is $135 \times N_r \times d/2 \sim 10^{12}$ rotations, versus a classical cost $N_r \times d^2 \sim 2 \times 10^{16}$.

The XOR-class scaling $\mathcal{O}(N_{\mathrm{ang}}^2)$ reflects the all-to-all nature of the $\nu$--$\nu$ interaction and is the only example given here where $N_{\mathrm{XOR}}$ grows faster than logarithmically.
Mean-field or spectral truncation of the angular distribution can reduce this at the cost of physical fidelity.

\subsection{Systems Without Off-Diagonal Channel Coupling}
\label{sec:no_offdiag}

Two systems in Table~\ref{tab:master_resources} require no XOR-class decomposition.

\subsubsection{Polaron: QFT-diagonalized Hopping}

The electron hopping term $-t_{\mathrm{hop}}\sum_{\langle i,j\rangle} |i\rangle\langle j|$ is off-diagonal in the coordinate register $\Reg{R}$, not the channel register $\Reg{C}$.
The QFT on $\Reg{R}$ diagonalizes it:
\begin{equation}\label{eq:hopping_qft}
    -t_{\mathrm{hop}}\sum_{\langle i,j\rangle}|i\rangle\langle j|
    \;\xrightarrow{\;\mathrm{QFT}\;}\;
    \sum_k \varepsilon_k\,|k\rangle\langle k|, \qquad
    \varepsilon_k = -2t_{\mathrm{hop}}\cos(2\pi k/N_s).
\end{equation}
The electron--phonon coupling is fully diagonal in the Gauss--Hermite DVR of the phonon registers. 
The DVR's tensor-product structure ${U}_{\mathrm{DVR}} = \bigotimes_{q=1}^{M} {U}_{\mathrm{DVR}}^{(q)}$ arises from mutually commuting phonon modes $[a_q, a_{q'}^\dagger] = \delta_{qq'}$. 
Each per-mode DVR costs $\mathcal{O}(N_p^2)$ gates; the full channel-register transform costs $M \times \mathcal{O}(N_p^2)$ rather than $\mathcal{O}(d^2) = \mathcal{O}(N_p^{2M})$, exploiting the product structure.

\subsubsection{HHG: DVR-Absorbed Dipole Coupling and Time-Dependent Protocol}

The single-active-electron Hamiltonian~\ref{eq:H_hhg} contains only one coupling between angular momentum channels: the dipole interaction $E(t)\,r\cos\theta$.  
This coupling is diagonal in the DVR, defining the Gauss--Legendre DVR on $\Reg{L}$: the potential $V_{\mathrm{ion}}(r) + E(t)r\cos\theta$ is pointwise diagonal at each angular quadrature point $\theta_m$. 
No XOR-class decomposition is needed; the entire angular coupling is handled by the DVR round trip.

The time-dependent field is accommodated by the split-operator framework: the M\"obius coefficients rescale    classically at each step while the circuit template is invariant.
The M\"obius coefficients rescale with $E(t_n)$ at each Trotter step, so the circuit template is invariant and only classical angle rescaling is required.
In qubitization, each distinct field value requires a new PREPARE oracle, multiplying the oracle count by the number of distinct time steps ($\sim 23{,}000$ for a 10-cycle pulse at 800~nm)\cite{krausz2009attosecond}.

\subsection{Summary of Off-Diagonal Coupling}
\label{sec:offdiag_summary}

The off-diagonal cost is bounded by $N_{\mathrm{XOR}} \times N_x \times d/2$ rotations.
For most physical systems $N_{\mathrm{XOR}} \leq 10$. 
Tables~\ref{tab:taxonomy_condensed} and~\ref{tab:xor_cost} collect off-diagonal coupling taxonomy and the per-system
cost.

\begin{longtable}{@{}p{2.6cm}p{2.8cm}p{2.0cm}p{1.8cm}p{3.2cm}@{}}
\caption{Off-diagonal coupling taxonomy.}
\label{tab:taxonomy_condensed}\\
\toprule
System & Coupling operator & Selection rule & XOR classes
    & Mechanism \\
\midrule
\endfirsthead
\toprule
System & Coupling operator & Selection rule & XOR classes
    & Mechanism \\
\midrule
\endhead
\midrule\multicolumn{5}{r@{}}{\emph{continued}}\\\endfoot
\bottomrule\endlastfoot
Li Stark / HHG & $r\cos\theta$
    & $\Delta l = \pm 1$ & 0
    & DVR absorbs coupling \\
Trimer: electronic & $V_{01}\sigma_x$
    & $\Delta n = 1$ & 1 & Single Hadamard \\
Trimer: angular & dense via DVR
    & all $\Delta l$ & 0 & DVR absorbs coupling \\
Polaron: hopping & $\sum|i\rangle\langle j|$
    & $\Delta R = \pm 1$ & 0 & QFT absorbs coupling \\
Reactive: arrangement & $O_{\alpha\beta}$
    & $\alpha\!\leftrightarrow\!\beta$ & 3
    & Discrete symmetry \\
Spin--orbit & ${L}_+{S}_-$
    & $\Delta\Lambda\!=\!\pm 1$ & $\sim\!\Lambda_{\max}$
    & Carry chain (rank 1) \\
Nuclear: tensor & $S_{12}$
    & $\Delta l = \pm 2$ & $\mathcal{O}(\log l_{\max})$
    & Carry chain (rank 2) \\
Rovibrational: Coriolis & ${J}_\pm{\pi}_\mp$
    & $\Delta K = \pm 1$ & $\mathcal{O}(\log K_{\max})$
    & Carry chain (rank 1) \\
Rovibrational: asym.\ top & ${J}_+^2\!+\!{J}_-^2$
    & $\Delta K = \pm 2$ & $\mathcal{O}(\log K_{\max})$
    & Carry chain (rank 2) \\
Neutrino: $\nu$-$\nu$ & $\lambda_a^{(p)}\lambda_a^{(q)}$
    & 3 per bin pair
    & $3\binom{N_a}{2}$
    & Discrete (Gell-Mann) \\
\end{longtable}

\begin{table}[ht]
\centering
\caption{XOR-class cost per Trotter step across systems.}
\label{tab:xor_cost}
\begin{tabular}{@{}lrrr@{}}
\toprule
System & $N_{\mathrm{XOR}}$ & Max Clifford depth
    & Off-diag.\ rotations \\
\midrule
Li Stark / HHG & 0 & 0 & 0 \\
Trimer (2 states) & 1 & 1 & $N_x \times d/2$ \\
Reactive scattering & 3 & 3 & $3 N_x \times d/2$ \\
Spin--orbit (UO) & 4 & 3 & $4 N_x \times d/2$ \\
Nuclear (n+$^{235}$U) & 10 & 6 & $10 N_x \times d/2$ \\
Neutrino ($N_a = 10$) & 135 & 4
    & $135 N_x \times d/2$ \\
Rovibrational (H$_2$O) & 8 & 4 & $8 N_x \times d/2$ \\
\bottomrule
\end{tabular}
\end{table}

\section{Extension to Other Orthogonal Subspaces}
\label{sec:new_subspaces}

Section~\ref{sec:fbr_dvr_closed} translates DVR transforms to quantum circuits to satisfy the amenability criterion (Section~\ref{sec:amenability}), and Table~\ref{tab:diag_catalog} catalogs the closed-form diagonalizations for systems treated in Section~\ref{sec:catalog}. 
This section extends the catalog with four additional families not represented in those systems: symmetry-adapted bases derived from molecular point groups, Bloch crystal-momentum bases for periodic systems, wavelet multiresolution bases for potentials with disparate length scales, and geometry-dependent configuration-interaction transforms for strongly correlated electronic states.

\subsection{Symmetry-Adapted Bases and Point-Group Reduction}
\label{sec:symmetry_adapted}

\subsubsection{Block-diagonalization by Finite Symmetry Groups}

Let $G$ be a finite group of order $|G|$ acting on the channel register by a unitary representation $g \mapsto {U}_g$, with $[{H}, {U}_g] = 0$ for all $g \in G$. 
By Maschke's theorem, $\mathcal{H}_{\Reg{C}}$ decomposes into irreducible representations (irreps)~\cite{tinkham2003group}:
\begin{equation}\label{eq:irrep_decomp}
    \mathcal{H}_{\Reg{C}} = \bigoplus_{\Gamma} \mathcal{H}_\Gamma^{\oplus m_\Gamma},
\end{equation}
where the direct sum runs over all irreps $\Gamma$ of $G$, $m_\Gamma$ is the multiplicity, and $\dim\mathcal{H}_\Gamma = d_\Gamma$. 
The projector onto the $\Gamma$-sector is
\begin{equation}\label{eq:projector}
    {P}_\Gamma = \frac{d_\Gamma}{|G|}
    \sum_{g \in G} \chi_\Gamma(g)^*\,{U}_g,
\end{equation}
where $\chi_\Gamma(g) = \mathrm{Tr}[{U}_g|_{\mathcal{H}_\Gamma}]$ is the character of $g$ in irrep $\Gamma$~\cite{tinkham2003group}. 
Since $[{H},{U}_g] = 0$, the Hamiltonian is block-diagonal in the symmetry-adapted basis: ${H} = \bigoplus_\Gamma {H}|_{\mathcal{H}_\Gamma}$.
Dynamics initiated in a single symmetry sector remains in that sector, with an effective channel dimension for each sector $d_\Gamma \ll d$.

\subsubsection{Symmetry Projector Circuit Implementation}

Representation matrices ${U}_g$ for molecular point groups are permutations on the channel basis. 
A permutation matrix on $n_c$ qubits is a product of SWAP and controlled-NOT gates: if $g$ maps basis state $|s\rangle \mapsto |g(s)\rangle$, then ${U}_g$ is the circuit that, for each bit position $k$ where $g(s)_k \neq s_k$, applies a CNOT or SWAP conditioned on the appropriate control pattern. 
The circuit depth is $\mathcal{O}(n_c)$ per group element.

The symmetry-adapted basis change ${U}_{\mathrm{sym}}$ that simultaneously block-diagonalizes all ${U}_g$ is constructed from the character table of $G$ and the Gram--Schmidt orthogonalization of projected states $\{{P}_\Gamma|s\rangle\}$. 
This produces a real orthogonal matrix (for groups with real characters, e.g.\ all molecular
point groups) of dimension $d \times d$, compiles by CSD at cost $\mathcal{O}(d^2)$ (same as Legendre DVR).
Effective DVR within each irrep is $d_\Gamma \times d_\Gamma$, costing $\mathcal{O}(d_\Gamma^2)$.

\subsubsection{XOR-Class Structure Under Symmetry}

The permutation $g$ maps $|s\rangle \to |g(s)\rangle$, so its XOR pattern on $\Reg{C}$ is $\zeta_g(s) = s \oplus g(s)$. 
For a group element $g$ of cycle type $(c_1, c_2, \ldots)$ acting on the binary-encoded channel index, the set $\{\zeta_g(s) : s = 0,\ldots,d-1\}$ is fixed by the group multiplication table and can be classically precomputed. The number of distinct XOR patterns over all $g \in G$ is bounded by $|G| - 1$.
For molecular point groups relevant to polyatomic dynamics (e.g.\ $C_{2v}$ with $|G| = 4$, $T_d$ with $|G| = 24$, $O_h$ with $|G| = 48$), symmetry compresses the XOR catalog as well as the channel dimension.

\subsubsection{Example: $C_{2v}$ Symmetry in H$_2$O}

The water molecule belongs to $C_{2v} = \{E, C_2, \sigma_v, \sigma_v'\}$ with four one-dimensional irreps $A_1$, $A_2$, $B_1$, $B_2$. 
For the rovibrational register of Section~\ref{sec:rovib} ($n_c = 14$ qubits, $d_{\mathrm{eff}} \approx 231$, padded to 1024), the symmetry-adapted block structure reduces the effective channel dimension per sector to approximately 256. 
The per-Trotter-step UCR cost falls by a factor of four within each symmetry block, and the off-diagonal Coriolis
coupling (Section~\ref{sec:rovib}) connects only states within the same symmetry block: $C_{2v}$ selection rules forbid $B_1 \leftrightarrow B_2$ coupling, reducing the number of active XOR classes by half.

\subsection{Bloch Basis for Periodic and Lattice Systems}
\label{sec:bloch}

\subsubsection{Translational Symmetry and Crystal Momentum}

For a Hamiltonian with discrete translational symmetry under lattice translation by $a$,
\begin{equation}\label{eq:translation}
    {T}_a\,{H}\,{T}_a^\dagger = {H}, \qquad
    {T}_a|R_j\rangle = |R_{j+1}\rangle,
\end{equation}
the Bloch theorem states that eigenstates factor as $|\psi_{n,k}\rangle = e^{ikR}|u_{n,k}\rangle$, where $k \in
[-\pi/a, \pi/a)$ is the crystal momentum and $n$ is the band index \cite{ehrenreich1977solid}.
In the REQWIEM register architecture, the coordinate register $\Reg{X}$ encodes the $N_s = 2^{n_x}$ lattice sites $R_j = ja$. 
The QFT on $\Reg{X}$ implements the discrete Fourier transform
\begin{equation}\label{eq:lattice_qft}
    {F}_{N_s}|R_j\rangle = \frac{1}{\sqrt{N_s}}
    \sum_{k=0}^{N_s-1} e^{2\pi ijk/N_s}|k\rangle,
\end{equation}
mapping site basis to crystal momentum basis. 
This transform is already present as the kinetic-energy step of the REQWIEM Trotter sequence.

\subsubsection{Multi-Band Channel Register}

For a system with $N_b$ bands, the channel register $\Reg{C}$ encodes the band index $n = 0, \ldots, N_b - 1$, requiring $n_c = \lceil\log_2 N_b \rceil$ qubits.
The single-band tight-binding Hamiltonian of Section~\ref{sec:polaron} generalizes to
\begin{equation}\label{eq:multiband_H}
    {H} = \sum_{n,k} \varepsilon_n(k)\,|n,k\rangle\langle n,k|
    + \sum_{n \neq n'} \sum_k V_{nn'}(k)\,
    |n,k\rangle\langle n',k|
    + {H}_{\mathrm{phonon}},
\end{equation}
where $\varepsilon_n(k)$ is the band dispersion (diagonal in the crystal-momentum basis) and $V_{nn'}(k)$, the interband coupling from an external field or phonon-mediated hybridization.
The diagonal term $\sum_{n,k} \varepsilon_n(k)|n,k\rangle\langle n,k|$ is encoded as a UCR on $\Reg{C}$ with the band dispersions as angle vectors, evaluated after the QFT on $\Reg{X}$. 
Interband coupling $V_{nn'}(k)$ is off-diagonal in the band index and handled by the XOR-class protocol: the XOR pattern $\zeta = n \oplus n'$ on the $n_c$-qubit band register, focused by the same Clifford construction derived in Chapter~\ref{ch:model_ext}.

\subsubsection{Dispersion Encoding and Momentum-Dependent UCR}

Band dispersion $\varepsilon_n(k)$ is nonlinear in crystal momentum $k$.
The kinetic phase $e^{-i\varepsilon_n(k)\Delta t}$ in the Trotter step must be applied conditioned on both the band index $n \in \Reg{C}$ and the crystal momentum $k \in \Reg{X}$ (after QFT). 
This is a UCR on $\Reg{X}$ multiplexed by $\Reg{C}$:
\begin{equation}\label{eq:band_ucr}
    {U}_{\mathrm{kin}} = \sum_{n=0}^{N_b-1} |n\rangle\langle n|
    \otimes \exp\!\left(-i\Delta t\sum_{k=0}^{N_s-1}
    \varepsilon_n(k)|k\rangle\langle k|\right).
\end{equation}
This is the DVR/UCR step structure, with a band dispersion table $\{\varepsilon_n(k)\}$ playing the role of a potential energy surface. 
The M\"obius inversion of Chapter~\ref{ch:smooth_potentials} converts the tabulated dispersion to Gray-code UCR angles at cost $\mathcal{O}(N_b \times N_s)$ classically with zero additional gate overhead.

\subsubsection{Scope}

This construction encompasses multi-band Hubbard models (orbital and spin channels in $\Reg{C}$, site index in $\Reg{X}$), the Kondo lattice (localized spin register $\otimes$ conduction electron register), and topological insulators (Chern bands with Berry curvature encoded in the off-diagonal coupling $V_{nn'}(k)$), leaving the circuit template invariant with specific register assignment and M\"obius coefficient tables. 

\subsection{Wavelet Multiresolution Basis}
\label{sec:wavelet}

\subsubsection{Motivation}

A uniformly-spaced DVR allocates equal resolution across all of $\Reg{X}$.
For potentials with disparate length scales, such as deep core repulsion at $r \lesssim 1\,a_0$ and diffuse van der Waals attraction at $r \sim 10$--$100\,a_0$, most grid points lie in the smooth asymptotic region and
contribute negligible potential variation, inflating $N_x$ without improving accuracy. 
This is seen clearly in Chapter~\ref{ch:scattering}, where $n=13,\,N=8192$ was required for a single-channel He-He scattering event.
A multiresolution basis concentrates grid points where the potential varies rapidly, reducing effective $N_x$ for a fixed error target.

\subsubsection{Quantum Wavelet Transform}

The Haar wavelet transform on $n_x$ qubits is the recursion
\begin{align}\label{eq:haar_recursion}
    {W}^{(1)}&: \begin{pmatrix}|2j\rangle \\ |2j+1\rangle\end{pmatrix}
    \mapsto \begin{pmatrix}(|2j\rangle + |2j+1\rangle)/\sqrt{2} \\
    (|2j\rangle - |2j+1\rangle)/\sqrt{2}\end{pmatrix}, \quad
    j = 0, \ldots, 2^{n_x-1} - 1,
\end{align}
applied iteratively from the coarsest scale down.
On a quantum register, each level of recursion is a layer of Hadamard gates on alternating qubits, interleaved with CNOT gates that propagate the separation. 
The full $n_x$-qubit Haar transform requires
\begin{equation}\label{eq:haar_cost}
    N_{\mathrm{CNOT}}^{(\mathrm{Haar})} = 2^{n_x} - 1, \qquad
    N_H^{(\mathrm{Haar})} = 2^{n_x},
\end{equation}
gates, with depth $\mathcal{O}(n_x)$. 
The butterfly structure parallelizes across $2^{n_x}$ sites.
The Daubechies $D_4$ wavelet replaces each Hadamard with a $2\times 2$ orthogonal matrix $\begin{pmatrix}h_0 & h_1 \\ h_2 & h_3\end{pmatrix}$, implementable as a single $R_y$ rotation sandwiched between two CNOTs, adding a factor of
3 to the CNOT count while improving the approximation order from 1 (Haar) to 2 (D$_4$).

\subsubsection{Adaptive Potential Encoding}

Smooth potential regions in the wavelet basis are represented by low-frequency scaling coefficients while sharp features are captured by high-frequency detail coefficients at fine scales. 
The multilinear coefficient $a_S$ for a high-order subset $S$ (large $|S|$) is negligible when the potential is smooth at the corresponding scale, so most rotations are near zero and can be truncated below a threshold $\theta_{\mathrm{cut}}$ without accumulating Trotter error. 
The effective number of nontrivial rotations per Trotter step falls from $N_x \times d$ to $N_x^{(\mathrm{eff})} \times d$, where
\begin{equation}\label{eq:neff}
    N_x^{(\mathrm{eff})} = \#\left\{k : |\phi_k| \geq
    \theta_{\mathrm{cut}}\right\} \ll N_x,
\end{equation}
for potentials that are smooth on the coarse wavelet scale. 
Potential structure and truncation threshold dictate the sparsity ratio $N_x^{(\mathrm{eff})} / N_x$, estimated estimated system by system; the reduction can be substantial for long-range potentials with Coulomb or dispersion
tails.

\subsubsection{Limitation: Approximate DVR Diagonality}

A uniform-grid DVR has the property that the potential is diagonal at the quadrature points: $\langle\theta_m|V|\theta_{m'}\rangle = V(\theta_m)\delta_{mm'}$. 
In the wavelet basis, exactness is lost, where the potential operator ${V}$ is not diagonal in the wavelet basis, being approximately sparse with off-diagonal elements decaying as wavelet smoothness order increases.
For the Daubechies D$_{2p}$ family, off-diagonal elements in the wavelet basis decay as $\mathcal{O}(\Delta x^{2p})$ where $\Delta x$ is the fine-scale grid spacing~\cite{beylkin1991fast}.
Using the wavelet transform as ${U}_{\mathrm{DVR}}$ therefore introduces a truncation error proportional to the off-diagonal residual, balanced against the grid compression gained. 
For potentials satisfying the Nyquist condition of Chapter~\ref{ch:wavepacket_prop}, the $D_4$ wavelet at $p = 2$ achieves fourth-order accuracy, and the off-diagonal residual is smaller than the Trotter error for grid spacings satisfying the coherent propagation criteria.

\subsection{Geometry-Dependent Configuration-Interaction Transforms}
\label{sec:ci_transform}

\subsubsection{$R$-Dependent Electronic Basis Problem}

For a diatomic molecule with strong avoided crossings or dense conical intersection seams, neither a fixed adiabatic basis (singular at the crossing) nor a fixed diabatic basis (approximate everywhere) is optimal.
A geometry-dependent unitary ${U}_{\mathrm{CI}}(R)$ diagonalizing the $d \times d$ electronic Hamiltonian matrix $\mathbf{H}_e(R)$ at each internuclear distance can be written as
\begin{equation}\label{eq:ci_diag}
    {U}_{\mathrm{CI}}(R)\,\mathbf{H}_e(R)\,
    {U}_{\mathrm{CI}}^\dagger(R) = \mathrm{diag}
    \bigl(E_1(R), \ldots, E_d(R)\bigr),
\end{equation}
provides the adiabatic potential energy surfaces $E_k(R)$ as the diagonal entries.
The columns of ${U}_{\mathrm{CI}}(R)$ are the adiabatic electronic eigenstates at geometry $R$, obtained from ab initio electronic structure calculation (CASSCF, MRCI, EOM-CC) at each reference geometry.

\subsubsection{Circuit Implementation as a Coordinate-Multiplexed UCR}

The geometry-dependent transform on a quantum cirucit is a unitary acting on the channel register, controlled by the coordinat register:
\begin{equation}\label{eq:ci_circuit}
    {U}_{\mathrm{CI}} = \sum_{i=0}^{N_x-1}
    |r_i\rangle\langle r_i|_{\Reg{X}}
    \otimes {U}_{\mathrm{CI}}(r_i)_{\Reg{C}}.
\end{equation}
This is a uniformly controlled unitary (UCU) on $\Reg{C}$, parameterized by $\Reg{X}$.
Each ${U}_{\mathrm{CI}}(r_i)$ is a $d \times d$ unitary, compiled by CSD~\cite{mottonen2004quantum} into $\mathcal{O}(d^2)$ CNOT and $R_z$ gates.
The full UCU therefore costs
\begin{equation}\label{eq:ci_cost}
    C_{\mathrm{CI}} = N_x \times \mathcal{O}(d^2),
\end{equation}
gates, compared to the geometry-independent DVR which costs $\mathcal{O}(d^2)$ regardless of $N_x$. 
This is the loss of amortization identified in Table~\ref{tab:diag_catalog}: the geometry-independent DVR acts on $\Reg{C}$ while $\Reg{X}$ is in superposition, requiring the geometry-dependent UCU to branch over all $N_x$ coordinate values.

\subsubsection{Diabatization and Residual Off-Diagonal Structure}

After application of ${U}_{\mathrm{CI}}$, the potential is diagonal (adiabatic surfaces $E_k(R)$), but the kinetic energy acquires off-diagonal derivative coupling matrix elements (NACMEs):
\begin{equation}\label{eq:nacme_circuit}
    \mathbf{F}_{kl}(R) = \langle\phi_k(R)|\nabla_R|\phi_l(R)\rangle
    = \langle k|{U}_{\mathrm{CI}}(R)\,\nabla_R\,
    {U}_{\mathrm{CI}}^\dagger(R)|l\rangle,
\end{equation}
arising from $R$-dependence in the adiabatic basis. 
These are the nonadiabatic coupling matrix elements seen throughout this dissertation.
Derivative coupling in the split-operator method contributes an off-diagonal term to the potential step, entering through the commutator $[T, {U}_{\mathrm{CI}}]$.
This is handled by the XOR-class protocol with coupling matrix elements
\begin{equation}\label{eq:nacme_xor}
    V_{kl}^{(\mathrm{NAC})}(r_i) \approx
    -\frac{1}{2\mu}\frac{F_{kl}(r_i) + F_{kl}(r_{i+1})}
    {2\Delta r},
\end{equation}
estimated by finite difference on the coordinate grid. 
Each off-diagonal pair $(k, l)$ with $\mathbf{F}_{kl} \neq 0$ contributes one XOR class (with pattern $k \oplus l$) and one UCR pass.

A fully diabatic treatment avoids the NACMEs by choosing a fixed ${U}_{\mathrm{CI}}$ that approximately diagonalizes $\mathbf{H}_e$ over nuclear configuration space.
Residual off-diagonal elements $V_{kl}(R)$ are treated by the XOR-class protocol, as in Sections~\ref{sec:spinorbit}--\ref{sec:nuclear}. 
Geometry-dependent UCU becomes unnecessary, the DVR amortization is recovered, and the total cost returns to $\mathcal{O}(d^2) + N_x \times d \times N_{\mathrm{XOR}}$ per Trotter step. 
Whether the adiabatic UCU or the diabatic XOR approach is preferred depends on the density of avoided crossings: a dense seam (many $R$-values with strong coupling) favors the adiabatic UCU; an isolated crossing or a slowly varying coupling favors the diabatic approach.

\section{REQWIEM as a Front-End for Quantum Krylov Subspace diagonalization}
\label{sec:krylov}

Real-time propagation and eigenvalue extraction are complementary tasks.
REQWIEM is designed to evolves an initial state $|\psi_0\rangle$ under the coupled-channel Hamiltonian ${H}$ and produces the time-dependent state $|\psi(t)\rangle$ along with the expectation values of physical observables, channel populations, coordinate-space densities, and transition dipoles at each Trotter
step.
Eigenvalues are not a direct output of the propagation; extracting them requires additional structure.

The real-time variant of Quantum Krylov subspace diagonalization (QKSD)~\cite{parrish2019hybrid, stair2020multireference, cortes2022quantum} builds a projected representation of ${H}$ from a set of time-displaced states $\{|\psi(t_k)\rangle\}_{k=0}^{m-1}$, solving a classical generalized eigenvalue problem in the projected subspace. 
Extraction to low-lying energy eigenvalues never requires a full $d \times N_x$-dimensional Hamiltonian diagonalization. 
This section derives the overlap and Hamiltonian matrix elements needed for QKSD from the REQWIEM propagator, constructs the controlled-REQWIEM circuit required for their estimation on a fault-tolerant device, and analyzes conditions under which the combination is advantageous.

\subsection{The Real-Time Krylov Subspace}
\label{sec:krylov_subspace}

Let $|\psi_0\rangle \in \mathcal{H}_{\Reg{X}} \otimes \mathcal{H}_{\Reg{C}}$ be a normalized initial state with a Krylov time step $\Delta\tau > 0$.
The real-time Krylov basis of dimension $m$ is
\begin{equation}\label{eq:krylov_basis}
    \mathcal{K}_m = \mathrm{span}\bigl\{
    |\psi(t_k)\rangle\bigr\}_{k=0}^{m-1}, \qquad
    |\psi(t_k)\rangle = e^{-i{H}k\Delta\tau}|\psi_0\rangle.
\end{equation}
The overlap matrix $\mathbf{S} \in \mathbb{C}^{m \times m}$ and the Krylov--Hamiltonian matrix $\mathbf{H}_K \in \mathbb{C}^{m \times m}$ in this basis are
\begin{align}
    S_{jk} &= \langle\psi(t_j)|\psi(t_k)\rangle
    = \langle\psi_0|e^{i{H}j\Delta\tau}
    e^{-i{H}k\Delta\tau}|\psi_0\rangle
    = \langle\psi_0|e^{-i{H}(k-j)\Delta\tau}|\psi_0\rangle,
    \label{eq:overlap_matrix} \\
    (H_K)_{jk} &= \langle\psi(t_j)|{H}|\psi(t_k)\rangle
    = i\,\partial_{t_k} S_{jk}.
    \label{eq:krylov_H}
\end{align}
The projected eigenvalue problem
\begin{equation}\label{eq:gen_eig}
    \mathbf{H}_K\,\mathbf{c} = E\,\mathbf{S}\,\mathbf{c},
\end{equation}
is solved classically. 
Its eigenvalues $\{E_n^{(m)}\}$ converge to the true eigenvalues $\{E_n\}$ of ${H}$ as $m$ increases and as the
initial state $|\psi_0\rangle$ acquires overlap with more eigenstates~\cite{cortes2022quantum}.

\subsection{Overlap Matrix Toeplitz Structure}
\label{sec:toeplitz}

$S_{jk}$ depends only on
the time difference $\ell = k - j$, not $j$ and $k$ independently (Eq.~\ref{eq:overlap_matrix})
\begin{equation}\label{eq:toeplitz}
    S_{jk} = C(\ell\,\Delta\tau), \qquad \ell = k - j,
\end{equation}
where
\begin{equation}\label{eq:autocorrelation}
    C(t) = \langle\psi_0|e^{-i{H}t}|\psi_0\rangle,
\end{equation}
is the initial state's autocorrelation function.
The overlap matrix is Hermitian Toeplitz: $S_{jk} = C((k-j)\Delta\tau)$ and $S_{jk} = S_{kj}^*$. 

Being Hermitian Toeplitz, the $m \times m$ matrix $\mathbf{S}$ is determined by $m$ complex numbers $\{C(0), C(\Delta\tau), \ldots, C((m-1)\Delta\tau)\}$, the autocorrelation function sampled at integer multiples of $\Delta\tau$.
Rather than estimating $m(m-1)/2$ independent overlaps, only $m-1$ independent quantum circuits are needed (plus a known $C(0) = 1$).

Hamiltonian matrix elements follow from Eq.~\ref{eq:krylov_H} by finite difference:
\begin{equation}\label{eq:H_finite_diff}
    (H_K)_{jk} \approx i\,\frac{C((k-j+1)\Delta\tau)
    - C((k-j-1)\Delta\tau)}{2\Delta\tau},
\end{equation}
determining $\mathbf{H}_K$ by the same autocorrelation samples, at the cost of one additional point $C(m\Delta\tau)$. 

The autocorrelation function $C(\ell\Delta\tau)$ is a complex-valued
transition amplitude. 
Extraction on a fault-tolerant device requires the controlled-$U_\ell$ construction developed below. 
This overhead is specific to eigenvalue extraction via QKSD. 
Dynamical observables central to the preceding chapters (channel populations $P_\alpha(t) = \langle\psi(t)|{\Pi}_\alpha|\psi(t)\rangle$, coordinate-space densities) are real-valued expectation values obtainable by measuring the channel register $\Reg{C}$ after an uncontrolled REQWIEM propagation, with no ancilla qubit, no controlled gates, and no Hadamard test.  
The controlled-REQWIEM circuit of Section~\ref{sec:controlled_reqwiem} is therefore optional, invoked only when spectroscopic eigenvalue resolution is the target rather than time-dependent dynamics.

\subsection{Overlap Estimation via the Hadamard Test}
\label{sec:hadamard_test}

On a fault-tolerant device, $|\psi(t)\rangle$ is not directly readable. 
Each complex autocorrelation value $C(\ell\Delta\tau)$ is estimated by Hadamard test:
\begin{equation}\label{eq:hadamard_circuit}
\Qcircuit @C=1em @R=0.8em {
    \lstick{|0\rangle_{\Reg{A}}} & \gate{H} & \ctrl{1} & \gate{H} &
        \meter & \rstick{\mathrm{Re}[C]} \\
    \lstick{|\psi_0\rangle} & \qw & \gate{U_\ell} & \qw & \qw &
}
\end{equation}
where $U_\ell = e^{-i{H}\ell\Delta\tau}$ is the REQWIEM propagator for $\ell$ Krylov steps and $H$ denotes the Hadamard gate on the ancilla.
Measuring the ancilla in the $X$ basis (equation~\ref{eq:hadamard_circuit}) gives $\mathrm{Re}[C(\ell\Delta\tau)]$; replacing the final Hadamard with $S^\dagger H$ (i.e.\ measuring in the $Y$ basis) gives
$\mathrm{Im}[C(\ell\Delta\tau)]$. 
The probability of outcome $+1$ in each case is
\begin{align}
    \Pr(+1)_X &= \tfrac{1}{2}\bigl(1 +
        \mathrm{Re}[C(\ell\Delta\tau)]\bigr),
    \label{eq:prob_x} \\
    \Pr(+1)_Y &= \tfrac{1}{2}\bigl(1 +
        \mathrm{Im}[C(\ell\Delta\tau)]\bigr).
    \label{eq:prob_y}
\end{align}
Each estimate requires $\mathcal{O}(\eps^{-2})$ shots to achieve additive
precision $\eps$ by standard concentration bounds.
For an $m$-point Krylov expansion, the total shot count is $2(m-1) \times \mathcal{O}(\eps^{-2})$,
linear in $m$.

\subsection{Controlled-REQWIEM: Circuit Construction and Overhead}
\label{sec:controlled_reqwiem}

The Hadamard test requires the controlled-$U_\ell$ operation, where $U_\ell$ is applied if and only if the ancilla is in state $|1\rangle$.
Since $U_\ell = (U_{\Delta\tau})^\ell$ is a product of Trotter steps, it suffices to construct the controlled single-step operator $CU_{\Delta\tau}$ and repeat it $\ell$ times.

Each Trotter step consists of three types of operations on $\Reg{X} \otimes \Reg{C}$: QFT layers on $\Reg{X}$, UCR diagonal phases on $\Reg{X}$ or $\Reg{C}$, and Clifford XOR-class focusers on $\Reg{C}$.
The controlled versions are constructed as follows.

The QFT on $n_x$ qubits is a sequence of Hadamard and controlled-phase gates \cite{nielsen2010quantum}.
For the Hadamard gate, $CH$ costs 3 CNOTs and 4 $R_z$ gates~\cite{selinger2012efficient}. 
For a controlled phase $P(\theta)$, the controlled version $CP(\theta)$ is the Toffoli-phase gate costing 2 CNOTs and 3 $R_z$ gates. 
The total overhead factor for the controlled-QFT relative to the uncontrolled QFT is $\mathcal{O}(1)$ in gate count.

The UCR on register $\Reg{R}$ ($n$ qubits) is the Gray-code diagonal unitary $\mathrm{diag}(e^{i\theta_0}, \ldots, e^{i\theta_{2^n-1}})$.
Under ancilla control, this becomes the block-diagonal operator
\begin{equation}\label{eq:controlled_ucr}
    c\text{-}\mathrm{UCR} = |0\rangle\langle 0|_{\Reg{A}} \otimes
    \mathbb{1}_{\Reg{R}} + |1\rangle\langle 1|_{\Reg{A}} \otimes
    \mathrm{diag}(e^{i\theta_k}),
\end{equation}
diagonal on the $(n+1)$-qubit register $\Reg{A} \otimes \Reg{R}$:
\begin{equation}\label{eq:controlled_ucr_diagonal}
    c\text{-}\mathrm{UCR} = \mathrm{diag}\bigl(
    \underbrace{1, \ldots, 1}_{2^n},\;
    e^{i\theta_0}, \ldots, e^{i\theta_{2^n-1}}\bigr).
\end{equation}
Rather than converting each gate of the $n$-qubit UCR to its
controlled version, the controlled-UCR simply acts as itself on $n + 1$ qubits, with the ancilla $\Reg{A}$ entering the parity network as an ordinary qubit.

The angle vector for the $(n+1)$-qubit compilation is the
zero-padded extension
\begin{equation}\label{eq:padded_angles}
    \tilde{\bvec{\theta}} = (\underbrace{0, \ldots, 0}_{2^n},\;
    \theta_0, \ldots, \theta_{2^n - 1}) \in \mathbb{R}^{2^{n+1}},
\end{equation}
whose M\"obius-inverted coefficients are computed by the same classical inversion used for the uncontrolled UCR, applied to a vector twice as long with the first half zeroed. 
No change to the compilation pipeline is required. The $(n+1)$-qubit Gray-code
circuit costs
\begin{equation}\label{eq:controlled_ucr_cost}
    C_{c\text{-}\mathrm{UCR}} = 2^{n+1}\text{ CNOTs}
    + 2^{n+1}\,R_z,
\end{equation}
a factor of $2$ overhead in both CNOTs and $R_z$ gates relative to the uncontrolled $n$-qubit UCR ($2^n$ CNOTs, $2^n\,R_z$). 

The focusing Cliffords $U_\zeta$ are products of Hadamard and CNOT gates
on $\Reg{C}$. Their controlled versions are standard controlled-Clifford
circuits: each CNOT becomes a Toffoli ($6$ CNOTs) and each Hadamard becomes
a controlled-Hadamard ($3$ CNOTs)~\cite{barenco1995elementary}. 
This overhead is negligible relative to the UCR cost, with total Clifford depth scales as $\mathcal{O}(n_c)$ and $n_c \ll 2^{n_c}$ per XOR class.

The Hadamard test for a single autocorrelation value at lag $\ell$ requires $\ell$ applications of the controlled single-step circuit.
The total gate count for all $m - 1$ lags is
\begin{equation}\label{eq:krylov_total_cost}
    C_{\mathrm{QKSD}} = 2 \times C_{\mathrm{REQWIEM}} \times
    \sum_{\ell=1}^{m-1} \ell \times N_{\mathrm{shots}}
    = 2 \times C_{\mathrm{REQWIEM}} \times
    \frac{m(m-1)}{2} \times \mathcal{O}(\delta^{-2}),
\end{equation}
where $C_{\mathrm{REQWIEM}}$ is the per-step gate count from Table~\ref{tab:master_resources}. 
The dominant cost is the longest circuit at lag $\ell = m - 1$, which runs $(m-1)$ controlled Trotter steps.

\subsection{Error Budget for the Krylov Overlap and Optimal Step Size}
\label{sec:krylov_advantage}

QKSD eigenvalue approximation resolution improves as the Krylov window $t_{\max} = (m-1)\Delta\tau$ increases: eigenvalues separated by $\Delta E$ require $t_{\max} \gtrsim \pi / \Delta E$ to be resolved in the projected spectrum. 
Autocorrelation fidelity degrades by errors accumulated over the propagation. 
On an ideal device, three independent sources contribute to the overlap error at lag $\ell\Delta\tau$, following the error budget of Chapter~\ref{ch:resources}:

\begin{equation}\label{eq:overlap_error_budget}
    |C_{\mathrm{circuit}}(\ell\Delta\tau)
      - C_{\mathrm{exact}}(\ell\Delta\tau)|
    \;\leq\;
    \eps_{\mathrm{T}}(\ell\Delta\tau,\Delta t)
    + \eps_{\mathrm{disc}}(\ell\Delta\tau)
    + \eps_{\mathrm{synth}}(\ell\Delta\tau,\Delta t),
\end{equation}

For the second-order Strang splitting with per-step error
$\eps_{\mathrm{step}} \leq C_2\,\|[T,V]\|\,\Delta t^3$,
accumulated Trotter contribution at propagation time
$t = \ell\Delta\tau$ is
\begin{equation}\label{eq:trotter_overlap}
    \eps_{\mathrm{T}}(\ell\Delta\tau, \Delta t)
    \leq C_2\,\|[T,V]\|\,\ell\Delta\tau\,\Delta t^2,
\end{equation}
decreasing quadratically with $\Delta t$~\cite{suzuki1991general}.

The finite grid Hamiltonian ${H}_{\mathrm{grid}}$ differs from the continuum Hamiltonian by an operator-norm distance $\delta_H = \|{H}_{\mathrm{grid}} - {H}_{\mathrm{cont}}\|$, determined by the grid spacing $\Delta r = L / N_x$ and potential surface smoothness. 
Standard Hamiltonian perturbation theory gives
\begin{equation}\label{eq:disc_overlap}
    \eps_{\mathrm{disc}}(\ell\Delta\tau)
    \leq \delta_H\,\ell\Delta\tau,
\end{equation}
For potentials satisfying the aliasing bound of Chapter~\ref{ch:wavepacket_prop}, ($\|V\|_\infty / T_{\max} \leq 1$), $\delta_H$ decreases exponentially with $n_x$ on a well-resolved grid, and this term is subdominant.
It is nevertheless a hard floor on the overlap fidelity: no reduction of $\Delta t$ can improve it, and it imposes a ceiling on the achievable Krylov window,
\begin{equation}\label{eq:tmax_ceiling}
    t_{\max} < \eta / \delta_H,
\end{equation}
beyond which the discretization error alone exhausts the fidelity budget~$\eta$.

Each of the $N_{\mathrm{rot}}$ parameterized rotations per Trotter step (Eq.~\ref{eq:gate_count_general}) is synthesized from the Clifford$+T$ gate set to per-gate precision $\varepsilon_\theta$~\cite{ross2016optimal}.
Over $N_{\mathrm{steps}} = \ell\Delta\tau / \Delta t$ Trotter steps, coherent accumulation of synthesis errors gives
\begin{equation}\label{eq:synth_overlap}
    \eps_{\mathrm{synth}}(\ell\Delta\tau, \Delta t)
    \leq \frac{\ell\Delta\tau}{\Delta t}\,
    N_{\mathrm{rot}}\,\varepsilon_\theta,
\end{equation}
increasing as $\Delta t$ decreases, since more Trotter steps are required to cover the same propagation time.

After subtracting the discretization floor, the effective fidelity budget is
$\eta_{\mathrm{eff}} = \eta - \delta_H\,t_{\max}$, and the joint
constraint at the maximum lag $t_{\max} = (m-1)\Delta\tau$ is
\begin{equation}\label{eq:f_combined}
    f(\Delta t)
    \;\equiv\;
    \underbrace{C_2\,\|[T,V]\|\,t_{\max}}_{A}\,\Delta t^2
    \;+\;
    \underbrace{t_{\max}\,N_{\mathrm{rot}}\,
    \varepsilon_\theta}_{B}\;\frac{1}{\Delta t}
    \;\leq\; \eta_{\mathrm{eff}}.
\end{equation}
This function $f(\Delta t)$ is convex with a minimum at
$\Delta t_{\mathrm{opt}}
= \bigl(N_{\mathrm{rot}}\,\varepsilon_\theta
/ (2\,C_2\,\|[T,V]\|)\bigr)^{1/3}$,
with minimum value
\begin{equation}\label{eq:f_min}
    f_{\min}
    = \frac{3}{2^{2/3}}\,t_{\max}\,
    \bigl(C_2\,\|[T,V]\|\bigr)^{1/3}
    \bigl(N_{\mathrm{rot}}\,\varepsilon_\theta\bigr)^{2/3}.
\end{equation}
A solution to~\ref{eq:f_combined} exists if and only if $f_{\min} \leq \eta_{\mathrm{eff}}$, imposing a required
per-gate synthesis precision:
\begin{equation}\label{eq:eps_theta_crit}
    \varepsilon_\theta
    \;\leq\;
    \varepsilon_\theta^{(\mathrm{crit})}
    \;\equiv\;
    \frac{1}{N_{\mathrm{rot}}}
    \left(\frac{2^{2/3}\,\eta_{\mathrm{eff}}}
    {3\,t_{\max}}\right)^{\!3/2}
    \!\!\frac{1}{\bigl(C_2\,\|[T,V]\|\bigr)^{1/2}}.
\end{equation}
When this condition is satisfied, the maximum allowed Trotter step size is the larger positive root of the cubic
$A\,\Delta t^3 - \eta_{\mathrm{eff}}\,\Delta t + B = 0$.  
When synthesis precision comfortably satisfies equation~\ref{eq:eps_theta_crit} (the typical operating point), this root approaches the pure-Trotter bound
\begin{equation}\label{eq:dt_requirement}
    \Delta t \leq \left(\frac{\eta_{\mathrm{eff}}}
    {C_2\,\|[T,V]\|\,t_{\max}}\right)^{\!1/2},
\end{equation}
and the synthesis correction to the step count is modest.

The critical per-gate precision~\ref{eq:eps_theta_crit} scales as
$\|[T,V]\|^{-1/2}$: a smaller commutator norm relaxes synthesis
requirements because larger Trotter steps are possible, fewer steps are
taken, and fewer gates accumulate imprecisions.
This is a second, indirect benefit of the smooth-potential regime, where smooth potentials reduce both the Trotter error directly and the synthesis precision required to support the resulting step size.

For the Ar$_3^*$ trimer ($d = 64$, $N_x = 2^{24}$, $n_c = 6$, $N_{\mathrm{XOR}} = 1$, $N_{\mathrm{diag}} = 1$), the per-step rotation count from Eq.~\ref{eq:gate_count_general} is $N_{\mathrm{rot}} \approx N_x \times n_c \times 2 \approx 2 \times 10^8$.
With $\|[T,V]\| \sim 10^{-5}$~a.u., $m = 20$, $\Delta\tau = 10^3$~a.u.\ ($t_{\max} = 1.9 \times 10^4$~a.u.), and target overlap fidelity $\eta = 10^{-3}$:

The critical per-gate precision from equation~\ref{eq:eps_theta_crit} is $\varepsilon_\theta^{(\mathrm{crit})} \approx 7 \times 10^{-15}$, achievable with $C_T \approx 141$ $T$-gates per rotation via bare
Ross--Selinger synthesis ($\approx 28$ after optimization~\cite{ross2016optimal}).
At $\varepsilon_\theta = 10^{-14}$, which lies above this threshold, the minimum combined Trotter$+$synthesis error is
$f_{\min} \approx 1.2 \times 10^{-3} > \eta$: no step size can achieve the target fidelity.

The small value of $\|[T,V]\|$ relative to the LCU 1-norm $\alpha = d\,N_x\,\langle|V|\rangle$ permits Trotter step sizes of $\mathcal{O}(10^1)$~a.u.\ rather than $\mathcal{O}(10^{-3})$~a.u., keeping the number of steps moderate. 
A QKSD front-end based on qubitization, whose query count scales as $\mathcal{O}(\alpha t)$ with $\alpha \sim 10^6$~a.u., would require $\sim 10^{10}$ queries for the same Krylov window, a factor of $\sim 10^7$ more circuit evaluations than the REQWIEM-based approach, even after the
synthesis-corrected step count, assuming the nonadiabatic system can be implemented.

\subsection{Physical Observables from REQWIEM Within QKSD}
\label{sec:krylov_observables}

REQWIEM does generates the full state
$|\psi(t)\rangle \in \mathcal{H}_{\Reg{X}} \otimes \mathcal{H}_{\Reg{C}}$, from which
physical observables are extracted by partial measurement or by
operator expectation values estimated via additional Hadamard test circuits.
Channel-register populations are given as
\begin{equation}\label{eq:channel_populations}
    P_\alpha(t) = \|\langle\alpha|\psi(t)\rangle\|^2
    = \langle\psi(t)|{\Pi}_\alpha|\psi(t)\rangle, \qquad
    {\Pi}_\alpha = |\alpha\rangle\langle\alpha|_{\Reg{C}}
    \otimes \mathbb{1}_{\Reg{X}},
\end{equation}
where $\alpha$ labels an electronic surface, angular momentum channel, spin
projection, or neutrino flavor.
These populations are estimable by measuring the channel register $\Reg{C}$ directly after the propagation circuit. These are the physical observables most relevant to nonadiabatic dynamics, tracking population transfer between coupled surfaces, recurrence of wavepacket amplitude, and the time-dependent branching ratio between product channels.

Within QKSD, channel populations at the Krylov times provide observable-resolved spectral information.
The population autocorrelation in channel $\alpha$,
\begin{equation}\label{eq:channel_autocorrelation}
    C_\alpha(t) = \langle\psi_0|{\Pi}_\alpha\,
    e^{-i{H}t}|\psi_0\rangle,
\end{equation}
has a spectral decomposition $C_\alpha(t) = \sum_n \langle\psi_0|{\Pi}_\alpha|n\rangle\langle n|\psi_0\rangle e^{-iE_n t}$, where $|n\rangle$ are energy eigenstates.
Fourier transforming $C_\alpha(t)$ resolves the energy spectrum weighted by an overlap of each eigenstate with the $\alpha$-th channel, providing state-resolved partial densities of states without additional circuit modifications. The same Hadamard test circuit estimating $C(t)$ for QKSD is modified by inserting ${\Pi}_\alpha$ between the state preparation and the propagator:
\begin{equation}\label{eq:channel_hadamard}
    C_\alpha(t) = \langle\psi_0|{\Pi}_\alpha e^{-i{H}t}|\psi_0\rangle
    \approx \langle\psi_0|e^{i{H}t}{\Pi}_\alpha e^{-i{H}t}
    |\psi_0\rangle\big|_{t \to 0},
\end{equation}
estimated by a Hadamard test with the observable ${\Pi}_\alpha$ inserted into the circuit.

\subsection{Krylov Collapse and Optimal Subspace Dimension}
\label{sec:krylov_collapse}

The overlap matrix $\mathbf{S}$ approaches singularity with increasing Krylov dimension $m$, with time-displaced states becoming approximately linearly dependent as $t_{\max}$ exceeds the system's recurrence time.
This is the Krylov collapse problem~\cite{cortes2022quantum}. 
In the presence of shot noise in autocorrelation estimates, the overlap matrix's condition number $\mathbf{S}$ grows as $\mathcal{O}(\eps^{-1} e^{\kappa m})$ for some system-dependent rate $\kappa$, making the generalized eigenvalue problem~\ref{eq:gen_eig} numerically ill-conditioned.

A standard mitigation technique is to threshold the eigenvalues of $\mathbf{S}$: eigenvalues below a cutoff $\eps_S$ are discarded and the projected problem is solved in the reduced subspace. 
The optimal $m$ balances spectral resolution against conditioning, dependent on target precision and shot budget.
For coupled-channel systems, the symmetry-block structure of Section~\ref{sec:symmetry_adapted} reduces the effective Krylov dimension needed.
If the initial state $|\psi_0\rangle$ belongs to a single symmetry sector of dimension $d_\Gamma$, then the Krylov subspace is confined to that sector, and $m$ need only be large enough to resolve the spectrum within $\mathcal{H}_\Gamma$ rather than all of $\mathcal{H}_{\Reg{C}}$. 
Combined with a wide Krylov window accessible at small $\|[T,V]\|$, this makes the REQWIEM-QKSD combination well-suited to mid-size coupled-channel systems where classical Lanczos encounters memory limits but full-time quantum propagation to convergence is unnecessary.

With the protocol, its extensions, and its eigenvalue extraction capability established, the next section identifies physical domains to which the framework applies.

\section{Physics Subfields Amenable to REQWIEM}
\label{sec:fields}

The coupled-channel structure of equation~\ref{eq:intro_H} emerges wherever a quantum system decomposes into an (effectively) continuous degree of freedom (a coordinate, a propagation distance, a lattice site index) coupled to a discrete internal structure whose free evolution is diagonal in a basis different from the basis in which the coupling is diagonal.
This section identifies some of these classes of problems that admit a REQWIEM-amenable register assignment, derives the Hamiltonian decomposition and XOR-class structure for each, and records the register architecture and gate count in the unified format of Table~\ref{tab:taxonomy_condensed}.

The criterion for inclusion is the REQWIEM-amenability of Eq.~\ref{eq:amenable}: the interaction potential must be diagonalizable on the channel register by an efficiently implementable unitary.
Satisfactory systems inherit the full protocol: UCR potential encoding, M\"obius inversion, XOR-class focusing, and potential Toeplitz autocorrelation for QKSD.

\subsection{Nuclear and Hadronic Physics}
\label{sec:nuclear_fields}

Two extensions substantially broaden the scope of applicability within nuclear physics, being multi-nucleon transfer reactions and the no-core shell model for light nuclei.

\subsubsection{Multi-Nucleon Transfer Reactions}

Consider the deuteron-stripping reaction $d + {}^{A}X \to p + {}^{A+1}X^*$ in the distorted-wave Born approximation (DWBA) \cite{satchler1983direct}.
The post-form transition amplitude requires the simultaneous propagation of the proton--nucleus and neutron--nucleus relative motions, coupled by the residual $pn$ interaction $V_{pn}(r_{pn})$.
The Hamiltonian in the Jacobi coordinate system $(\mathbf{r}_p, \mathbf{r}_n)$ is
\begin{equation}\label{eq:deuteron_H}
    {H} = -\frac{1}{2\mu_p}\nabla_{r_p}^2
    - \frac{1}{2\mu_n}\nabla_{r_n}^2
    + U_p(r_p) + U_n(r_n)
    + V_{pn}(|\mathbf{r}_p - \mathbf{r}_n|),
\end{equation}
where $U_p$ and $U_n$ are the proton and neutron optical potentials.
Expanding in the coupled basis $|l_p, l_n, L, S, J, M_J\rangle$ of angular momenta for each Jacobi vector and their couplings, the Hamiltonian takes the form~\ref{eq:intro_H} with two coordinate registers:
\begin{equation}\label{eq:deuteron_regs}
    \Reg{X} = \Reg{R}_p \otimes \Reg{R}_n, \qquad
    \Reg{C} = \Reg{L}_p \otimes \Reg{L}_n \otimes
    \Reg{S} \otimes \Reg{J},
\end{equation}
where $\Reg{R}_p$ and $\Reg{R}_n$ encode the proton and neutron radial coordinates on $N_r$-point grids, and $\Reg{L}_p$, $\Reg{L}_n$, $\Reg{S}$, $\Reg{J}$ encode the angular momentum quantum numbers.
The kinetic energy is diagonal after QFT on each of $\Reg{R}_p$ and $\Reg{R}_n$ independently.
The optical potentials $U_p(r_p)$ and $U_n(r_n)$ are diagonal in the DVR of each radial register and are channel register independent, contributing a UCR on $\Reg{R}_p$ (or $\Reg{R}_n$) multiplexed by the identity on $\Reg{C}$.
Residual interactions $V_{pn}$ couple all partial waves of both particles, giving off-diagonal channel coupling.

In the partial-wave basis, matrix elements of $V_{pn}$ are
\begin{equation}\label{eq:vpn_matrix}
    \langle l_p', l_n', L'| V_{pn} |l_p, l_n, L\rangle
    = \sum_\lambda C_\lambda(r_p, r_n)\,
    \langle l_p', l_n', L'|{T}_\lambda|l_p, l_n, L\rangle,
\end{equation}
where ${T}_\lambda$ is a rank-$\lambda$ spherical tensor and the radial integral $C_\lambda(r_p, r_n)$ is separable in the DVR.
The selection rules $\Delta l_p = 0, \pm\lambda$ and $\Delta l_n = 0, \pm\lambda$ determine XOR-class structure on $\Reg{L}_p \otimes \Reg{L}_n$: each multipole $\lambda$ of $V_{pn}$ contributes $\mathcal{O}(\log l_{\max})$ XOR classes via carry-chain analysis (Section~\ref{sec:nuclear}), applied independently to $\Reg{L}_p$ and $\Reg{L}_n$. For $\lambda_{\max} = 2$ (quadrupole residual interaction) and $l_{\max} = 10$: approximately $2 \times 4 = 8$ XOR classes.

\subsubsection{No-Core Shell Model Dynamics}

The no-core shell model (NCSM) represents the nuclear many-body state in a harmonic oscillator (HO) basis truncated at $N_{\max}$ oscillator quanta~\cite{barrett2013ab}. 
For a nucleus with mass number $A$, the intrinsic Hamiltonian in Jacobi coordinates is
\begin{equation}\label{eq:ncsm_H}
    {H}_{\mathrm{int}} = \sum_{i=1}^{A-1}
    \frac{{p}_i^2}{2m} + \sum_{i<j} V_{ij}(r_{ij})
    + \sum_{i<j<k} V_{ijk}(r_{ij}, r_{ik}, r_{jk}),
\end{equation}
where $V_{ij}$ are nucleon--nucleon (NN) potentials and $V_{ijk}$ are three-nucleon forces (3NF).
In present register assignment, the center-of-mass Jacobi coordinate $\boldsymbol{\rho}_1$ provides the
coordinate register $\Reg{X}$ and all remaining Jacobi coordinates $(\boldsymbol{\rho}_2, \ldots, \boldsymbol{\rho}_{A-1})$ together with the spin-isospin quantum numbers $(m_{s_i}, m_{t_i})$ form the channel
register $\Reg{C}$. 
The HO basis for $\Reg{C}$ diagonalizes the center-of-mass kinetic term and the harmonic confining potential.
Residual NN and 3NF interactions are off-diagonal and handled by XOR-class decomposition on the spin-isospin register.

For ${}^4$He with $N_{\max} = 8$, the channel dimension is $d \sim \binom{N_{\max} + 3}{3} \times (2s+1)^A \times (2t+1)^A \sim 10^3$. 
The reduction factor $\sim d = 10^3$ over classical NCSM diagonalization is modest for light nuclei but grows rapidly: for ${}^{12}$C at $N_{\max} = 6$, $d \sim 10^6$ and the classical NCSM matrix dimension $\sim 10^9$ is already at the limit of present supercomputer memory.

\subsubsection{Color-Spin Interactions in Quark Models}

The non-relativistic quark model for baryon spectroscopy uses a Hamiltonian of the form
\begin{equation}\label{eq:quark_H}
    {H} = \sum_i \frac{{p}_i^2}{2m_q}
    + \sum_{i<j}\left[V_{\mathrm{conf}}(r_{ij})
    + V_{\mathrm{OGE}}(r_{ij})\,
    \boldsymbol{\lambda}_i^c \cdot \boldsymbol{\lambda}_j^c\,
    \boldsymbol{\sigma}_i \cdot \boldsymbol{\sigma}_j\right],
\end{equation}
where $\boldsymbol{\lambda}^c$ are the color Gell-Mann matrices, $\boldsymbol{\sigma}$ are spin Pauli matrices, and $V_{\mathrm{OGE}}$ is the one-gluon-exchange potential~\cite{de1975hadron}. 
The Jacobi coordinate $\boldsymbol{\rho}_1$ serves as $\Reg{X}$; the color-spin register $|c_1 c_2 c_3\rangle \otimes |s_1 s_2 s_3\rangle$ (3 color qubits, 3 spin qubits) serves as $\Reg{C}$ with $d = 8 \times 8 = 64$ (padded to 64). The color Casimir $\boldsymbol{\lambda}_i^c \cdot \boldsymbol{\lambda}_j^c$ is diagonal in the SU(3) color basis (singlet vs.\ octet), providing the FBR.
The spin operator $\boldsymbol{\sigma}_i \cdot \boldsymbol{\sigma}_j = 2{P}_{ij}^{\mathrm{spin}} - 1$ (where ${P}_{ij}^{\mathrm{spin}}$ is the spin-exchange operator~\cite{edmonds1996angular}) is diagonal in the coupled spin basis, providing the DVR. The DVR transform on $\Reg{C}$ is the Clebsch--Gordan recoupling unitary for three spin-$\frac{1}{2}$ particles, which decomposes into $\mathcal{O}(d^2)$ gates via CSD.

\subsection{Condensed Matter and Quantum Materials}
\label{sec:condensed_matter}

\subsubsection{Multiband Superconductors and Nambu Space}

In a two-band superconductor (e.g.\ MgB$_2$, Fe-based pnictides), the BCS mean-field Hamiltonian in Nambu notation is~\cite{suhl1959bardeen}
\begin{equation}\label{eq:bcs_H}
    {H}_{\mathrm{BCS}} = \sum_{k,n,\sigma}
    \xi_{nk}\,c_{nk\sigma}^\dagger c_{nk\sigma}
    + \sum_{k,n,n'}\Delta_{nn'}(k)\,
    c_{nk\uparrow}^\dagger c_{n'\bar{k}\downarrow}^\dagger
    + \mathrm{h.c.},
\end{equation}
where $n, n' \in \{1, 2\}$ are band indices, $\xi_{nk} = \varepsilon_{nk} - \mu$ is the quasiparticle dispersion, and $\Delta_{nn'}(k)$ is the $k$-dependent interband gap matrix.
In Nambu space, the four-component spinor is defined as
$\psi_{k} = (c_{1k\uparrow}, c_{2k\uparrow},c_{1\bar{k}\downarrow}^\dagger, c_{2\bar{k}\downarrow}^\dagger)^T$, so that
\begin{equation}\label{eq:nambu_BdG}
    {H}_{\mathrm{BdG}} = \frac{1}{2}\sum_k
    \psi_k^\dagger\,\mathcal{H}_{\mathrm{BdG}}(k)\,\psi_k, \quad
    \mathcal{H}_{\mathrm{BdG}}(k) =
    \begin{pmatrix} \xi_{nk}\delta_{nn'} & \Delta_{nn'}(k) \\
    \Delta_{nn'}^\dagger(k) & -\xi_{nk}\delta_{nn'} \end{pmatrix}.
\end{equation}
In the REQWIEM register assignment, the crystal momentum $k$ is encoded in $\Reg{X}$ (QFT on the real-space site register), and the Nambu--band channel $|n, \tau\rangle$ (band index $n \in \{1,2\}$, particle--hole index $\tau \in \{+,-\}$) is encoded in $\Reg{C}$ with $d = 4$ ($n_c = 2$ qubits). 
The quasiparticle dispersion $\xi_{nk}\tau_z$ is diagonal in the Nambu basis (FBR); the gap $\Delta_{nn'}(k)\tau_+$ is off-diagonal in both band and particle--hole indices.
The gap matrix contributes one XOR class per off-diagonal band pair: for a two-band system, $|\tau = +, n = 1\rangle \leftrightarrow |\tau = -, n = 2\rangle$ gives XOR pattern $\zeta = 11$ on the 2-qubit channel register, focused by a CNOT followed by a Hadamard. 
Interband scattering $V_{01}(k)$ (pair breaking between bands) gives a second XOR class $\zeta = 01$, focused by a Hadamard on the band qubit. 
The total XOR count is 2, and the Clifford depth is $\leq 2$ gates. 
The $k$-dependent gap $\Delta_{nn'}(k)$ is encoded by the M\"obius inversion of Chapter~\ref{ch:smooth_potentials}, treating the gap as an arbitrary smooth function on the $N_k$-point momentum grid.

\subsubsection{Topological Insulators and Chern Bands}

The Qi--Wu--Zhang model for a two-dimensional Chern insulator on a square lattice is~\cite{qi2006topological}
\begin{equation}\label{eq:qwz_H}
    {H}(\mathbf{k}) = \sin k_x\,\sigma_x
    + \sin k_y\,\sigma_y
    + (m_0 + \cos k_x + \cos k_y)\,\sigma_z,
\end{equation}
where $\sigma_{x,y,z}$ act on the two-band channel register ($n_c = 1$ qubit, $d = 2$) and $m_0$ is the Dirac mass parameter.
The coordinate register encodes the two-dimensional crystal momentum $\mathbf{k} = (k_x, k_y)$ using two $n_x$-qubit registers $\Reg{K}_x \otimes \Reg{K}_y$, reached by QFT on the real-space position registers. 
The $\sigma_z$ term is diagonal in the channel (FBR) basis, contributing a UCR on $\Reg{K}_x \otimes \Reg{K}_y$ multiplexed by $\sigma_z$. 
The $\sigma_x$ and $\sigma_y$ terms are off-diagonal with XOR pattern $\zeta = 1$ on the single channel qubit; both
are focused by a Hadamard ($\sigma_x$ term) or a phase-then-Hadamard ($\sigma_y$ term).
The Chern number
\begin{equation}\label{eq:chern}
    \mathcal{C} = \frac{1}{2\pi}\int_{\mathrm{BZ}}
    \Omega(\mathbf{k})\,d^2k, \qquad
    \Omega(\mathbf{k}) = -2\,\mathrm{Im}
    \langle\partial_{k_x}u_-|\partial_{k_y}u_-\rangle,
\end{equation}
is accessible from the REQWIEM simulation as the time-averaged Hall conductance $\sigma_{xy} = \mathcal{C} e^2/h$ via the Kubo formula~\cite{kubo1957statistical} applied to the time-dependent current--current correlator. 
The current-current correlator by itself is a channel-register observable of the type described in Section~\ref{sec:krylov_observables}.

\subsubsection{Quantum Spin Liquids and Emergent Gauge Fields}

The Kitaev honeycomb model~\cite{kitaev2006anyons} is
\begin{equation}\label{eq:kitaev_H}
    {H}_{\mathrm{Kit}} = -J_x\sum_{\langle ij\rangle_x}
    \sigma_i^x\sigma_j^x
    - J_y\sum_{\langle ij\rangle_y}\sigma_i^y\sigma_j^y
    - J_z\sum_{\langle ij\rangle_z}\sigma_i^z\sigma_j^z.
\end{equation}
By the Jordan--Wigner transformation on the honeycomb lattice, this maps to a free Majorana fermion Hamiltonian in a static $\mathbb{Z}_2$ gauge field $u_{ij} = \pm 1$ on each bond~\cite{kitaev2006anyons}. 
For each gauge sector (fixed configuration of $\{u_{ij}\}$), the fermionic Hamiltonian is quadratic and can be solved classically.
The quantum problem simultes the full gauge-unfixed dynamics including the flux excitations (visons).  
In the REQWIEM assignment, sublattice site indices are encoded in $\Reg{X}$ and the gauge-sector label $|\{u_{ij}\}\rangle$ is encoded in $\Reg{C}$, with $d = 2^{N_{\mathrm{bonds}}}$.
The free-fermion Hamiltonian for each gauge sector is diagonal in the Bogoliubov quasiparticle basis, providing the FBR; the vison hopping (flux pair creation and annihilation) is off-diagonal in the gauge register and contributes
XOR classes whose patterns are determined by the honeycomb lattice's bond connectivity. 
Total XOR count is bounded by $3N_{\mathrm{unit cells}}$ (one per bond type $x$, $y$, $z$ per unit cell), and each class is focused by a Hadamard on the relevant bond qubit.

\subsection{Quantum Optics and Cavity Quantum Electrodynamics}
\label{sec:cavity_qed}

\subsubsection{Jaynes--Cummings and Tavis--Cummings Models}

The Jaynes--Cummings Hamiltonian for a single two-level atom (transition
frequency $\omega_0$) coupled to a single cavity mode (frequency
$\omega_c$, annihilation operator ${a}$) is~\cite{jaynes2005comparison}
\begin{equation}\label{eq:jc_H}
    {H}_{\mathrm{JC}} = \omega_c\,{a}^\dagger{a}
    + \frac{\omega_0}{2}\,\sigma_z
    + g\,({a}^\dagger\sigma_- + {a}\,\sigma_+),
\end{equation}
where $g$ is the vacuum Rabi coupling.
In the REQWIEM assignment, the cavity Fock states $|n\rangle$ ($n = 0, \ldots, N_{\mathrm{Fock}} - 1$) are encoded in $\Reg{X}$ using $n_x = \lceil\log_2 N_{\mathrm{Fock}} \rceil$ qubits, and the atomic state $|\uparrow\rangle$, $|\downarrow\rangle$ is encoded in $\Reg{C}$ with $n_c = 1$ qubit.
The free-field term $\omega_c\,{a}^\dagger{a}$ is diagonal in the Fock basis (FBR on $\Reg{X}$) and the atomic term $\omega_0\sigma_z/2$ is diagonal in the spin-$z$ basis (FBR on $\Reg{C}$).
The Jaynes--Cummings interaction $g\,({a}^\dagger\sigma_- + {a}\,\sigma_+)$ is simultaneously off-diagonal in $\Reg{X}$ (by $\pm 1$ Fock excitation) and off-diagonal in $\Reg{C}$ (by $\pm 1$ spin flip). 
The Gauss--Hermite DVR maps the Fock register to the quadrature displacement basis (as in Section~\ref{sec:polaron}), in which ${a} + {a}^\dagger = \sqrt{2}\,{x}$ is diagonal. 
The interaction then becomes
\begin{equation}\label{eq:jc_dvr}
    g\,({a}^\dagger\sigma_- + {a}\,\sigma_+)
    \;\xrightarrow{\;\mathrm{DVR}\;}\;
    g\sqrt{2}\,x_m\,\sigma_x + \mathcal{O}({n}^{-1/2}),
\end{equation}
where $x_m$ is the DVR quadrature point and the correction $\mathcal{O}({n}^{-1/2})$ vanishes in the large-$N_{\mathrm{Fock}}$ limit. 
The $\sigma_x$ coupling is off-diagonal with XOR pattern $\zeta = 1$ on the spin qubit, focused by a Hadamard. 
The full Trotter step is:
\begin{equation}\label{eq:jc_trotter}
    e^{-i{H}_{\mathrm{JC}}\Delta t} \approx
    e^{-i\omega_c{n}\Delta t/2}\,
    e^{-i\omega_0\sigma_z\Delta t/4}\,
    {U}_{\mathrm{HO}}^\dagger\,
    e^{-ig\sqrt{2}x_m\sigma_x\Delta t}\,
    {U}_{\mathrm{HO}}\,
    e^{-i\omega_0\sigma_z\Delta t/4}\,
    e^{-i\omega_c{n}\Delta t/2},
\end{equation}
where ${U}_{\mathrm{HO}}$ is the Gauss--Hermite DVR transform on $\Reg{X}$.
The Tavis--Cummings generalization to $N_a$ atoms replaces the single-spin channel register with the collective Dicke register $|J, M\rangle$ ($J = 0, \ldots, N_a/2$, $M = -J, \ldots, J$), encoded on $n_c = \lceil\log_2(N_a + 1)\rceil$ qubits with $d = N_a + 1$.
The coupling $g\sqrt{N_a}\,({a}^\dagger J_- + {a} J_+)/\sqrt{J}$ produces $\mathcal{O}(\log N_a)$ XOR classes from the $\Delta M = \pm 1$ transitions in the Dicke ladder, following the same carry-chain analysis as the molecular spin--orbit case of Section~\ref{sec:spinorbit}.

\subsubsection{Ultrastrong Coupling Regime}

In the ultrastrong coupling (USC) regime, $g/\omega_c \geq 0.1$, the rotating-wave approximation that justifies equation~\ref{eq:jc_H} breaks down and the full quantum Rabi Hamiltonian must be used~\cite{forn2019ultrastrong}:
\begin{equation}\label{eq:rabi_H}
    {H}_{\mathrm{Rabi}} = \omega_c\,{a}^\dagger{a}
    + \frac{\omega_0}{2}\,\sigma_z
    + g\,({a}^\dagger + {a})\,\sigma_x.
\end{equation}
The counter-rotating terms ${a}^\dagger\sigma_+$ and ${a}\,\sigma_-$ (absent in the Jaynes--Cummings model) are now
present.
In the DVR, $({a}^\dagger + {a}) \to \sqrt{2}\,x_m$, so the full coupling becomes $g\sqrt{2}\,x_m\,\sigma_x$,
the same diagonal-times-Pauli structure as the rotating-wave case.
Counter-rotating terms generate no additional XOR classes and add no circuit complexity beyond the Jaynes--Cummings
case, giving a structural advantage of the DVR approach where the approximation useful in classical numerics has no benefit.

\subsubsection{Multi-Mode Cavity Arrays and Photon Hopping}

For a one-dimensional array of $N_{\mathrm{cav}}$ cavities with nearest-neighbor photon hopping amplitude $J_c$:
\begin{equation}\label{eq:cavity_array_H}
    {H}_{\mathrm{array}} = \sum_{i=1}^{N_{\mathrm{cav}}}
    {H}_{\mathrm{JC}}^{(i)}
    - J_c\sum_{\langle ij\rangle}
    ({a}_i^\dagger{a}_j + \mathrm{h.c.}).
\end{equation}
The cavity site index $i$ is encoded in $\Reg{X}$ and the Fock state of each cavity is distributed across $N_{\mathrm{cav}}$ phonon-mode registers in $\Reg{C}$, much like the Holstein polaron of Section~\ref{sec:polaron} with the identification photon $\leftrightarrow$ phonon, cavity site $\leftrightarrow$ lattice site. 
The hopping term ${a}_i^\dagger{a}_j$ is off-diagonal in the cavity site index; after QFT on $\Reg{X}$, it becomes diagonal in the crystal photon-momentum basis with dispersion $-2J_c\cos(2\pi k / N_{\mathrm{cav}})$. 
The per-cavity atom--photon coupling maps to one UCR per cavity mode.
Between-cavity hopping requires no XOR classes. 
For $N_{\mathrm{cav}} = 64$ cavities each with $N_{\mathrm{Fock}} = 16$ levels and $N_a = 2$ atoms per
cavity, the channel dimension is $d = 16^{64} \times 4^{64} \approx 10^{154}$, being classically intractable and in the regime where the $n_x + n_c$ qubit encoding provides an exponential space advantage.

\subsection{Astrophysics and Relativistic Wave Equations}
\label{sec:astrophysics}

\subsubsection{Collective Neutrino Problem Extensions}

Section~\ref{sec:neutrino} treated the Mikheyev-Smirnov-Wolfenstein (MSW) system and collective neutrino oscillation problems with three active flavors \cite{mikheev1985resonance, wolfenstein2018neutrino}. 
Two extensions are immediate within the existing circuit template.

Sterile neutrinos add a fourth flavor state $|\nu_s\rangle$ that does not interact via the weak force.
The flavor register $\Reg{F}$ extends to $n_f = 2$ qubits ($d = 4$, all four states occupied), and the PMNS matrix is replaced by the $4 \times 4$ unitary $U_{\mathrm{PMNS}}^{(4)}$ parameterized by three additional mixing
angles and two additional charge-parity (CP)-violating phases. 
The DVR is $U_{\mathrm{PMNS}}^{(4)}$, again compiled by CSD at cost $\mathcal{O}(d^2) = \mathcal{O}(16)$ gates. 
Additional XOR classes from active--sterile transitions are $\nu_e \leftrightarrow \nu_s$, $\nu_\mu \leftrightarrow \nu_s$, $\nu_\tau \leftrightarrow \nu_s$, each with XOR pattern $\zeta = 11$ on the 2-qubit flavor register, this is three additional XOR classes, each focused by CNOT + Hadamard.

CP-violating phases in the PMNS matrix enter as complex entries in the DVR unitary $U_{\mathrm{PMNS}}$, requiring no modification to the XOR-class decomposition, being absorbed into UCR complex rotation angles via the M\"obius inversion applied to the real and imaginary parts of the potential matrix elements separately.
The Jarlskog invariant $J_{\mathrm{CP}} = \mathrm{Im}[U_{e1}U_{\mu 2} U_{e2}^* U_{\mu 1}^*]$~\cite{jarlskog1985commutator}, controlling the magnitude of CP violation in neutrino oscillations, is accessible as a post-processed function of the channel-register density matrix elements $\rho_{\alpha\beta}(t) = \langle\psi(t)|{\Pi}_\alpha
{\Pi}_\beta|\psi(t)\rangle$.

\subsubsection{Neutron Star Crust and Superfluid Vortex Dynamics}

The inner crust of a neutron star contains a neutron superfluid coexisting with a crystalline nuclear lattice. 
The coupled superfluid--lattice dynamics is governed by scattering equations for neutron quasiparticles scattering off lattice nuclei, taking the same form as the nucleon--nucleus optical model of Section~\ref{sec:nuclear}~\cite{nicolas2008physics}.
The coordinate register $\Reg{R}$ encodes the neutron radial position within a Wigner--Seitz cell of radius $R_{\mathrm{WS}} \sim 50$~fm, and the channel register $\Reg{C}$ encodes the partial-wave and band index $(l, s, j, n_{\mathrm{band}})$.
The Hamiltonian is
\begin{equation}\label{eq:crust_H}
    {H}_{\mathrm{crust}} =
    -\frac{\hbar^2}{2m_n^*(\mathbf{r})}\nabla^2
    + V_{\mathrm{nuc}}(r) + V_{\mathrm{pair}}(r)
    + \frac{\hbar^2 l(l+1)}{2m_n^* r^2},
\end{equation}
where $m_n^*(\mathbf{r})$ is the position-dependent effective mass from the superfluid-lattice entrainment coupling, and $V_{\mathrm{pair}}(r)$ is the pairing gap potential.
The position-dependent effective mass contributes an additional coupling term to the kinetic energy:
\begin{equation}\label{eq:effmass_coupling}
    -\nabla\cdot\frac{\hbar^2}{2m_n^*(\mathbf{r})}\nabla
    = -\frac{\hbar^2}{2m_n^*}\nabla^2
    + \frac{\hbar^2}{2}\left(\nabla\frac{1}{m_n^*}\right)\cdot\nabla,
\end{equation}
where the second term is a momentum-dependent potential diagonal in the DVR.
The M\"obius expansion of Chapter~\ref{ch:smooth_potentials} encodes $1/m_n^*(r_i)$ as an arbitrary smooth function on the radial grid, with no additional gate primitives required. 
The XOR-class structure is identical to Section~\ref{sec:nuclear}: tensor force ($\Delta l = \pm 2$,
$\mathcal{O}(\log l_{\max})$ classes) and spin--orbit ($\Delta l = \pm 1$, $\mathcal{O}(\log l_{\max})$ classes).

\subsubsection{Black Hole Quasi-Normal Modes: Teukolsky Equation}

Gravitational perturbations of a Kerr black hole (mass $M$, spin $a$) are governed by the Teukolsky equation~\cite{teukolsky1973perturbations}.
In Boyer--Lindquist coordinates, separating $\psi_s(t, r, \theta,\phi) = e^{-i\omega t + im\phi}{}_{s}S_{lm}(\theta)\,{}_{s}R_{lm}(r)$, the radial and angular equations are
\begin{align}
    \Delta^{-s}\frac{d}{dr}\left[\Delta^{s+1}
    \frac{d\,{}_sR_{lm}}{dr}\right]
    &+ \left[\frac{K^2 - 2is(r-M)K}{\Delta}
    + 4is\omega r - {}_s\lambda_{lm}\right]
    {}_sR_{lm} = 0,
    \label{eq:teukolsky_radial} \\
    \frac{1}{\sin\theta}\frac{d}{d\theta}
    \left[\sin\theta\frac{d\,{}_sS_{lm}}{d\theta}\right]
    &+ \left[a^2\omega^2\cos^2\theta
    - \frac{m^2 + s^2 + 2ms\cos\theta}{\sin^2\theta}
    + {}_s A_{lm}\right]{}_sS_{lm} = 0,
    \label{eq:teukolsky_angular}
\end{align}
where $K = (r^2 + a^2)\omega - am$, $\Delta = r^2 - 2Mr + a^2$, $s$ is the spin weight ($s = -2$ for gravitational waves), and ${}_s\lambda_{lm}$ is the angular separation constant. 
The angular equation~\ref{eq:teukolsky_angular} is a spin-weighted spheroidal harmonic equation, and its eigenvalues ${}_s A_{lm}$ couple different values of $l$ for $a\omega \neq 0$. 
This is precisely the structure of the coupled-channel centrifugal problem: the angular register $\Reg{L}$ encodes $l = |s|, |s|+1, \ldots, l_{\max}$, the angular eigenvalue matrix ${}_s A_{lm}(a\omega)$ is dense in the $l$-basis (FBR) and diagonal in the spheroidal harmonic basis (DVR), and the DVR transform is the matrix that diagonalizes equation~\ref{eq:teukolsky_angular} at the given frequency $\omega$.
The radial coordinate $r$ encoded in $\Reg{R}$ on a tortoise coordinate grid $r^* = r + 2M\ln|r/(2M) - 1|$ that compresses the near-horizon region.
The spin weight $s$ and azimuthal index $m$ label distinct, non-coupled problems (no XOR class between different $s$ or $m$); within a fixed $(s, m)$ sector, the $l$-channel coupling has selection rule $\Delta l = \pm 2$ from the $a^2\omega^2\cos^2\theta$ term (rank-2 in $l$, identical to the nuclear tensor force of Section~\ref{sec:nuclear}).

Quasi-normal mode frequencies $\omega_{\mathrm{QNM}}$ are the Teukolsky operator's complex eigenvalues. 
They are directly accessible via the REQWIEM-QKSD combination of Section~\ref{sec:krylov}: propagate an initial wavepacket in the tortoise coordinate, form the autocorrelation function $C(t) = \langle\psi_0|e^{-i{H}t}|\psi_0\rangle$, and extract the complex poles of its Laplace transform by QKSD.
The complex part of the QNM frequency (the decay rate $1/\tau_{\mathrm{QNM}}$) is accessible because the outgoing-wave boundary conditions on the tortoise grid impose a non-Hermitian effective Hamiltonian (via a complex absorbing potential at the grid boundary), shifting eigenvalues of $\mathbf{H}_K$ off the real axis in direct proportion
to the ringdown rate.

\subsection{Strong-Field Atomic Physics and Quantum Chaos}
\label{sec:strong_field}

An atom in an external static or oscillatory electric field, at energies near the ionization threshold, focuses on chaotic quantum dynamics. 
The hydrogen atom in a uniform field $F\hat{z}$ separates in parabolic coordinates $(\xi,\eta)$; the phase space is
foliated by invariant tori labeled by a parabolic separation constant $\beta$. 
For any non-hydrogenic atom, the ionic core breaks the dynamical $\mathrm{SO}(4)$ symmetry of the Coulomb problem, producing a mixed regular--chaotic phase space with a quantum defect ($\delta_l$)-dependent structure and energy scaling $\varepsilon = E F^{-1/2}$~\cite{courtney1994long, courtney2021comprehensive}.

Experimentally, the chaotic dynamics manifests in recurrence spectroscopy: the Fourier transform of the photoabsorption cross section at fixed $\varepsilon$ is plotted versus the scaling variable $w = F^{-1/4}$ and identifies peaks at scaled actions $\tilde{S}_k$ of the classical closed orbits, starting and ending at the nucleus~\cite{du1988effect}.
Core scattering produces combination recurrences at $\tilde{S} = \tilde{S}_{k_1} + \tilde{S}_{k_2}$ absent in hydrogen, providing experimental signatures of non-integrability. 
Reproducing these spectra requires propagating a Rydberg wavepacket through $\sim 10^4$--$10^5$ atomic time units while resolving $d \sim 64$--$128$ angular momentum channels coupled by the Stark interaction and differentiated by the $l$-dependent core potential.

\subsubsection{Register Assignment and Hamiltonian Decomposition}

The single-active-electron Hamiltonian in the reduced radial representation ($\chi = r\psi$) is
\begin{equation}\label{eq:H_stark}
    {H} = -\frac{1}{2}\frac{\partial^2}{\partial r^2}
    + \frac{{l}^2}{2r^2}
    - \frac{1}{r} + V_c(r) + Fr\cos\theta,
\end{equation}
where $V_c(r)$ encodes the non-hydrogenic core.
This maps directly onto the REQWIEM register decomposition:
\begin{equation}\label{eq:stark_registers}
    \Reg{X} = \Reg{R}\;(n_r\text{ qubits, radial grid}),
    \qquad
    \Reg{C} = \Reg{L}\;(n_l\text{ qubits, angular momentum}).
\end{equation}
The free-channel Hamiltonian in the FBR is ${H}_{\Reg{C}}^{(\mathrm{FBR})} = l(l+1)/(2r^2)$, diagonal in $|l\rangle$. 
The Stark interaction $Fr\cos\theta$ is tridiagonal in $l$ via dipole selection rule $\langle l|\cos\theta|l'\rangle \propto \delta_{l',l\pm 1}$.  
Since $\cos\theta$ is defines the Gauss--Legendre quadrature (Table~\ref{tab:diag_catalog}), the DVR round trip absorbs the full $l$--$l'$ coupling without an XOR-class decomposition: $N_{\mathrm{XOR}} = 0$.

The Trotter step is
\begin{equation}\label{eq:stark_trotter}
    \e^{-\mathrm{i}{H}\Delta t} \approx
    \e^{-\mathrm{i}{T}_r\Delta t/2}\;
    \e^{-\mathrm{i}{V}_{\mathrm{cent}}\Delta t/2}\;
    {U}_{\mathrm{DVR}}^\dagger\;
    \e^{-\mathrm{i}{V}_{\mathrm{DVR}}\Delta t}\;
    {U}_{\mathrm{DVR}}\;
    \e^{-\mathrm{i}{V}_{\mathrm{cent}}\Delta t/2}\;
    \e^{-\mathrm{i}{T}_r\Delta t/2},
\end{equation}
where $V_{\mathrm{DVR}}(r_i, \theta_m) = -1/r_i + V_c(r_i) +
Fr_i\cos\theta_m$ is diagonal at each DVR grid point.
The single radial coordinate and absence of an electronic register or off-diagonal XOR step make this the simplest non-trivial instance of the coupled-channel framework.

\subsubsection{Core Scattering as an Interferometric Mechanism}

The DVR round trip acts as a multipath interferometer on the angular register.
At each Trotter step, ${U}_{\mathrm{DVR}}$ splits each $|l\rangle$ into $d$ angular paths $|\theta_m\rangle$, each path accumulates a phase from the diagonal potential $V(r_i, \theta_m)$, and ${U}_{\mathrm{DVR}}^\dagger$ recombines them.
In hydrogen ($V_c = 0$), a hidden parabolic symmetry causes the multi-step interferometer to preserve the parabolic quantum number $n_1$: the DVR phases eventually returnn the $l$-distribution to its initial pattern.
The core potential $V_c(r)$ perturbs this, modifying the radial wavepacket shape between Trotter steps so $r$-values operated on by the angular interferometer no longer align with parabolic eigenstates.
The $l$-distribution drifts stroboscopically (with synchronized oscillations), with drift rate controlled by quantum defects $\delta_l$~\cite{courtney1995classical}.

The quantum circuit encodes this mechanism without an explicit core-scattering operator is required.  
The $l$-dependent sensitivity arises from DVR transform weights $T_{ml} = \sqrt{w_m}\,\tilde{P}_l(\cos\theta_m)$ interacting with the radial wavepacket shape at the grid points where $V_c(r)$ is appreciable ($r \lesssim 5\,a_0$).

\subsubsection{Gate Count and Computational Advantage}

For the reference parameters of the lithium system ($n_r = 12$, $N_x = 4096$; $n_l = 6$, $d = 64$), the per-step rotation count is
\begin{equation}\label{eq:stark_rotations}
    N_{\mathrm{rot}} = 3\,N_x \times d + 2\,N_x
    = 3 \times 4096 \times 64 + 8192
    \approx 8 \times 10^5,
\end{equation}
comprising two centrifugal UCR half-steps ($2 \times N_x \times d$), one DVR potential step ($N_x \times d$), and two kinetic half-steps ($2 \times N_x$).  
The DVR transform itself costs $\mathcal{O}(d^2)$ elementary gates (concretely, $990$ CNOTs for the parity-optimized Legendre CSD at $d = 64$), applied twice per step, and is negligible relative to the UCR cost ($\sim 1\%$).

The classical cost of the same Trotter step is the cost of $d \times d$ matrix--vector multiplies at each of $N_x$ grid points: $C_{\mathrm{classical}} = N_x \times d^2 \approx 1.7 \times 10^7$
operations per step on a matrix of approximate dimension $N_x\times d = 2^{12}\times2^6 \approx 2.62\times10^5$. 
The per-step reduction factor is
\begin{equation}\label{eq:stark_advantage}
    \frac{C_{\mathrm{classical}}}{N_{\mathrm{rot}}}
    = \frac{N_x \times d^2}{3\,N_x \times d + 2\,N_x}
    \approx \frac{d}{3} \approx 21.
\end{equation}
Complexity reduction is linear in $d$, deriving from DVR amortization: the $\mathcal{O}(d^2)$ basis-change cost is applied once on $\Reg{L}$ while $\Reg{R}$ remains in superposition, rather than at each of the $N_x$ grid points.

\subsubsection{Scaled-Energy Scans and Time-Dependent Fields}

Recurrence spectroscopy requires scanning the field strength $F$ across $N_w \sim 100$--$300$ values at fixed scaled energy $\eps$, with each scan point requiring a full time propagation of $N_{\mathrm{steps}} \sim 6 \times 10^4$ Trotter steps (to resolve Rydberg states near $n \sim 40$). 
The total circuit cost is
\begin{equation}\label{eq:stark_total}
    C_{\mathrm{total}} \sim N_w \times N_{\mathrm{steps}} \times
    N_{\mathrm{rot}} \sim 300 \times 6 \times 10^4 \times
    8 \times 10^5 \sim 10^{13}\text{ rotations}.
\end{equation}
Each $w$-point is an independent propagation.
The circuit template is invariant across the scan, only requiring a rescaling of rotation coefficients of the Stark contribution. 
This scaling by $F(w) = w^{-4}$ at each scan point, while the Coulomb, core, and centrifugal coefficients are precomputed once and reused.
For time-dependent fields (ramp functions, half-cycle terahertz pulses, or Floquet driving $F(t) = F_0 + F_1\cos(\omega t)$, the same rescaling applies at each Trotter step, with no change to the gate
sequence.

\subsubsection{Extensions Within the Class}

The register architecture generalizes without modification to several related strong-field problems:
\begin{enumerate}
    \item \textbf{Diamagnetic Stark problem.~\cite{courtney1994long}}  Adding a magnetic field $B\hat{z}$ introduces a term $(B^2/8)\,r^2\sin^2\theta$ with Legendre components $\ell = 0$ and $\ell = 2$. 
    This adds two nontrivial Walsh sectors to the DVR potential UCR but requires no XOR classes and no change to the circuit template.

    \item \textbf{Multi-channel quantum defect theory.~\cite{dubau1984quantum}}  For atoms with multiple ionization thresholds (e.g.\ alkaline earths), the ionic core states are encoded in an electronic register $\Reg{E}$ with $n_e$ qubits.  
    The core S-matrix enters as off-diagonal coupling, handled by the XOR-class protocol. The register cost grows to $n_r + n_l + n_e$.

    \item \textbf{Two-electron atoms.}  For helium in a strong field, the inter-electron repulsion $1/|\mathbf{r}_1 -
          \mathbf{r}_2|$ couples two radial and two angular registers.
          The channel dimension grows to $d_1 \times d_2$ and the quantum advantage factor scales as $d_1 d_2 / 3$, placing this firmly in the classically intractable regime. 
\end{enumerate}

A broader catalog of amenable scientific domains, with register encodings and XOR-class counts, is given in Table~\ref{tab:new_fields}.

\begin{table}[ht]
\centering
\small
\caption{Register architectures and XOR-class counts for new
scientific fields amenable to REQWIEM.}
\label{tab:new_fields}
\begin{tabular}{@{}lllrrl@{}}
\toprule
System & $\Reg{X}$ encodes & $\Reg{C}$ encodes
    & $d$ & $N_{\mathrm{XOR}}$ & DVR \\
\midrule
Deuteron transfer & $r_p, r_n$ (2D) & $l_p, l_n, S, J$
    & $\sim 10^3$ & $\sim 8$ & CG recoupling \\
NCSM (${}^4$He) & $\rho_1$ & $\rho_2, \rho_3$, spin-isospin
    & $\sim 10^3$ & $\sim N_a^2$ & HO basis \\
Quark model & Jacobi $r_{12}$ & color-spin
    & 64 & 3 & CG (SU(3)) \\
BCS two-band & $k$ & Nambu-band
    & 4 & 2 & identity \\
Chern insulator & $\mathbf{k}$ (2D) & band
    & 2 & 2 & identity \\
Kitaev honeycomb & sublattice site & gauge sector
    & $2^{N_b}$ & $3N_{\mathrm{uc}}$ & Bogoliubov \\
Jaynes--Cummings & Fock $n$ & atom spin
    & 2 & 1 & Gauss--Hermite \\
Tavis--Cummings & Fock $n$ & Dicke $|J,M\rangle$
    & $N_a+1$ & $\mathcal{O}(\log N_a)$ & Gauss--Hermite \\
Cavity array & site $i$ & Fock $\otimes$ spin
    & $16^{N_{\mathrm{cav}}}$ & 0 & Gauss--Hermite \\
Sterile neutrino & $r$ (solar) & 4 flavors
    & 4 & 6 & PMNS$^{(4)}$ \\
Neutron star crust & $r$ (WS cell) & $l, s, j$, band
    & $\sim 512$ & $\sim 10$ & Legendre \\
Teukolsky (Kerr) & $r^*$ (tortoise) & $l$-channel
    & $\sim 64$ & $\mathcal{O}(\log l_{\max})$ & spheroidal \\
\bottomrule
\end{tabular}
\end{table}

\section{The Applicability Criterion: Identifying the Natural Domain}
\label{sec:limitations}

This section gives a classical preprocessing test to help answer the question if REQWIEM is the appropriate tool for a given problem, a resource cost heuristic, and what method is preferred. 
Inputs to this test are the potential's multilinear coefficient matrix, the potential smoothness on the proposed grid, and the target precision criterion partitions the space of coupled-channel problems into classes, organized below by the partition's specific properties.

\subsection{The Regime of Post-Trotter Methods}
\label{sec:qubitize_wins}

\subsubsection{Single-Channel Dynamics and the Finite Crossover Time}

For $d = 1$, assuming grid-based dynamics, the channel register $\Reg{C}$ is absent and the Hamiltonian reduces to ${H} = {T} + {V}(r)$. 
A Trotter step is a QFT, a diagonal phase, an inverse QFT, and the diagonal potential term, costing from $\mathcal{O}(n_x^2)$ gates per step (harmonic) to $\mathcal{O}(2^{n_x} + n_x^2)$ (smooth potential) with no $d$-dependent amortization. 
The qubitization block encoding of ${H}$ has 1-norm
\begin{equation}\label{eq:alpha_d1}
    \alpha = \|{T}\| + \sum_{i=0}^{N_x-1}|V(r_i)|
    \leq \frac{\pi^2}{2\Delta r^2} + N_x\langle|V|\rangle,
\end{equation}
with QROAM cost $\mathcal{O}(\sqrt{N_x})$ T-gates per query~\cite{berry2019qubitization}.
The total propagation costs for time $t$ at precision $\eps$ are:
\begin{align}
    C_{\mathrm{Trotter}}^{(d=1)} &= \mathcal{O}\!\left(n_x^2\right)
    \times \left(\frac{\|[T,V]\|\,t^3}{\eps}\right)^{1/2},
    \label{eq:trotter_d1} \\
    C_{\mathrm{qubitize}}^{(d=1)} &= \mathcal{O}\!\left(\alpha\,t\right)
    \times \mathcal{O}\!\left(\sqrt{N_x}\right).
    \label{eq:qubitize_d1}
\end{align}
Setting these equal and solving for the crossover time $t_\times^{(d=1)}$:
\begin{equation}\label{eq:crossover_d1}
    t_\times^{(d=1)} \sim \left(\frac{n_x^4\,\|[T,V]\|}
    {\alpha^2\,N_x\,\eps}\right)^{1/3}.
\end{equation}
Substituting $\alpha \sim N_x\langle|V|\rangle$ and
$\|[T,V]\| \sim \|V\|_{\max}/\ell_V^2$:
\begin{equation}\label{eq:crossover_d1_explicit}
    t_\times^{(d=1)} \sim \frac{n_x^{4/3}}{N_x^{1/3}\,\ell_V^{2/3}}
    \times \left(\frac{\|V\|_{\max}}
    {\langle|V|\rangle^2\,\eps}\right)^{1/3},
\end{equation}
For a single-channel scattering problem with $n_x = 12$, $N_x = 4096$,
$\ell_V = 5\,a_0$, $\|V\|_{\max} = 1$~a.u.,
$\langle|V|\rangle = 10^{-2}$~a.u., and $\eps = 10^{-3}$:
$t_\times^{(d=1)} \approx 3\times10^3$~a.u.,
which is physically accessible and well within the regime of post-trotter methods.

At propagation times $t \gg t_\times^{(d=1)}$, qubitization's
$\mathcal{O}(t)$ scaling overtakes Trotter's $\mathcal{O}(t^{3/2})$.
High-order Trotter ($p = 2, 3, \ldots$) extends the crossover time
by a factor $\mathcal{O}(\eps^{-1/(2p+1)})$ but does not eliminate it.

\subsubsection{High-Precision Static Eigenvalue Extraction}

Quantum phase estimation (QPE) combined with qubitization achieves Heisenberg-limited precision: $M = \mathcal{O}(\alpha/\eps)$ queries to the walk operator $W$ resolve an eigenvalue to precision $\eps$~\cite{low2019hamiltonian}.
The total gate cost is
\begin{equation}\label{eq:qpe_cost}
    C_{\mathrm{QPE}} = \mathcal{O}\!\left(\frac{\alpha}{\eps}\right)
    \times C_{\mathrm{query}}.
\end{equation}
REQWIEM extracts eigenvalues via QKSD (Section~\ref{sec:krylov}), with precision bounded by
\begin{equation}\label{eq:krylov_precision}
    \eps_{\mathrm{QKSD}} \gtrsim
    C_2\,\|[T,V]\|\,t_{\max}\,\Delta t^2
    + \frac{\pi}{t_{\max}}.
\end{equation}
Minimizing over $\Delta t$:
\begin{equation}\label{eq:krylov_optimum}
    \eps_{\mathrm{QKSD}}^{(\min)} \sim
    \left(\frac{C_2\,\|[T,V]\|}{t_{\max}}\right)^{1/3}.
\end{equation}
This precision floor decays as $t_{\max}^{-1/3}$, compared to QPE's $1/t_{\max}$ Heisenberg scaling. 
REQWIEM-QKSD achieves chemical accuracy ($\sim 10^{-3}$~a.u.) efficiently and provides time-resolved channel populations simultaneously.
For sub-wavenumber spectroscopic precision ($\eps < 10^{-6}$~a.u.) with a static, time-independent Hamiltonian and no dynamical observables, QPE combined with qubitization may be preferred.

\subsubsection{Second-Quantized Electronic Structure}

For a molecular electronic structure problem with $N$ spin-orbitals,
the second-quantized Hamiltonian is
\begin{equation}\label{eq:sq_H}
    {H} = \sum_{pq} h_{pq}\,a_p^\dagger a_q
    + \frac{1}{2}\sum_{pqrs} g_{pqrs}\,
    a_p^\dagger a_q^\dagger a_r a_s,
\end{equation}
with $\mathcal{O}(N^4)$ two-electron integrals $g_{pqrs}$~\cite{szabo2012modern}.
With tensor hypercontraction~\cite{lee2021even}, the QPE query complexity is $\mathcal{O}(N^2\lambda_{\max}/\eps)$ T-gates.
REQWIEM on a coordinate-space grid for $N_{\mathrm{nuc}} = 10$ atoms would require $n_x = 192$ qubits
encoding $N_x = 2^{192}$ grid points. 
The first-quantized grid does not exploit fermionic antisymmetry or two-electron integral
tensor sparsity. 
Second quantization in the molecular orbital basis is preferred for electronic structure, while REQWIEM's encoding advantage relies on precomputed potential energy surfaces.

\subsection{Classically Efficient Regimes}
\label{sec:classical_easy}

\subsubsection{Separable Hamiltonians}

If the Hamiltonian factorizes as
\begin{equation}\label{eq:separable_H}
    {H} = \sum_{r=1}^{f} {H}_r(x_r, p_r),
\end{equation}
the wavefunction remains a product state at all times and the classical
cost of independent propagation is
\begin{equation}\label{eq:separable_classical}
    C_{\mathrm{classical}}^{(\mathrm{sep})} = f\times\mathcal{O}(2^n n),
\end{equation}
while the REQWIEM cost on the full $f$-mode grid is
\begin{equation}\label{eq:separable_quantum}
    C_{\mathrm{REQWIEM}}^{(\mathrm{sep})} = \mathcal{O}\!\left(2^{fn}
    \times d \times N_{\mathrm{steps}}\right).
\end{equation}
The ratio of classical to quantum cost is $\mathcal{O}(f/2^{(f-1)n})$:
for $f = 9$ modes at $n = 6$ qubits each, $2^{(f-1)n} \approx
10^{14}$. 
Separability is classically detected in $\mathcal{O}(d^2)$ operations by inspecting the M\"obius coefficient matrix: if $a_S$ vanishes for all cross-mode index subsets $S$, the dynamics is separable and modes should be propagated independently. 
When cross-mode coefficients are non-negligible, the dynamics require a full coupled-grid treatment, and REQWIEM becomes preferred. 

\subsubsection{Gaussian States and Quadratic Hamiltonians}

A Hamiltonian at most quadratic in the mode operators
\begin{equation}\label{eq:quadratic_H}
    {H}_{\mathrm{quad}} = \frac{1}{2}\sum_{pq}
    A_{pq}\,{x}_p{x}_q
    + B_{pq}\,{p}_p{p}_q
    + C_{pq}\,({x}_p{p}_q + {p}_q{x}_p),
\end{equation}
generates dynamics that preserves Gaussian states via the symplectic map
\begin{equation}\label{eq:symplectic}
    \sigma(t) = S(t)\,\sigma(0)\,S(t)^T, \qquad
    \langle{\xi}(t)\rangle = S(t)\langle{\xi}(0)\rangle,
\end{equation}
at classical cost $\mathcal{O}(f^3)$, independent of grid size~\cite{weedbrook2012gaussian}.
Systems whose dynamics reduces to a Bogoliubov transformation (harmonic oscillator chains, normal-mode decoupled vibrations, BCS mean-field quasiparticle evolution) are mostly efficient in classical computation.
REQWIEM's advantage becomes apparent with anharmonic corrections generating inter-mode coefficients $a_S$ with $|S| \geq 2$, making the state non-Gaussian. 
A negative-valued Wigner function becomes a necessary condition for a quantum algorithm-preferred problem in continuous-variable systems, marking the boundary where REQWIEM can scalably outperform classical propagation \cite{mari2012positive}.

\subsubsection{Dynamics Tractable by Tensor Network Methods}

When the bipartite entanglement entropy $\mathcal{S}(\Reg{X}:\Reg{C})$ grows slowly, the wavefunction admits an efficient matrix product state (MPS) representation with bond dimension $\chi$, and TEBD costs~\cite{vidal2004efficient}
\begin{equation}\label{eq:mps_cost}
    C_{\mathrm{MPS}} = \mathcal{O}\!\left(N_x\,\chi^3\right)
    \text{ per Trotter step.}
\end{equation}
This identifies the ``slow scramblers."
When $\chi \ll \sqrt{N_x \times d}$, MPS is cheaper than REQWIEM. 
For nonadiabatic dynamics, the Schmidt rank grows to $\sim d$ within a few recurrence times, placing these problems in the volume-law entanglement regime where REQWIEM is $\mathcal{O}(d^2)$ cheaper than MPS per step.
The diagnostic can be the bipartite entanglement entropy $\mathcal{S} = -\mathrm{Tr}[\rho_{\Reg{X}}
\log\rho_{\Reg{X}}]$: if $\mathcal{S} \ll n_x + n_c$ throughout the propagation, MPS remains tractable. 
This condition can be checked on a classical simulation of the early-time dynamics before committing to a full quantum circuit.

\subsection{Structural Encoding Mismatch}\label{sec:wrong_encoding}

\subsubsection{Categorically Different Problems}
The REQWIEM template requires a continuous coordinate $\Reg{X}$ that admits a kinetic-energy operator diagonal under the QFT, and a discrete channel register $\Reg{C}$ that admits a DVR basis change.
Structurally, its similar to a Level 2 Basic Linear Algebra Subroutine (BLAS) for GPU computing, applied to a narrow but physically relevant class of Hamiltonians.
Combinatorial problems such as MaxCut, portfolio optimization, constraint satisfaction, and graph isomorphism do not satisfy these structures. 
As categorically different problems, any mapping to this subroutine produces large $\|[T,V]\|$ and eliminates the amortization advantage.
The current approaches to solve these problems include quantum approximate optimization algorithm QAOA~\cite{farhi2014quantum}, quantum annealing, or variational algorithms built for combinatorial structure.

\subsubsection{State-Dependent Potentials and Nonlinear Dynamics}

The REQWIEM protocol encodes the potential energy surface as a set of rotation angles precomputed classically by M\"obius inversion and loaded into the circuit before execution.
This assumes the potential is a known function of the coordinates $V(r_i, \theta_m, t)$ is fixed at each grid point independently of the quantum state $|\psi(t)\rangle$.
Smoothe particle hydrodynamics (SPH), and more broadly any system governed by nonlinear equations of motion, violates this assumption.
In SPH the kernel-smoothed density $\rho(\mathbf{r}) = \sum_j m_j\,W(|\mathbf{r} - \mathbf{r}_j|, h)$
determines the pressure, viscosity, and self-gravitational forces at each grid point, so the effective potential depends on the propagated state's local amplitude: $V = V[\rho] = V[|\psi|^2]$.
The rotation angles encoding this potential cannot be precomputed because they are functions of state amplitudes, estimated by measurement statistics.

In principle, quantum arithmetic circuits could evaluate the density functional $V[\rho]$ coherently: amplitude estimation extracts $|\psi(r_i)|^2$ into an ancilla register, a reversible arithmetic
circuit computes the kernel convolution and the resulting forces, and the output is used to set rotation angles via a phase-gradient or QROM lookup.
Each self-consistent evaluation would require $\mathcal{O}(N_x / \eps_{\mathrm{obs}})$ queries to the state (for amplitude estimation at $N_x$ grid points to precision $\eps_{\mathrm{obs}}$), followed by $\mathcal{O}(N_x^2)$ arithmetic operations for the pairwise kernel sum, all executed
coherently and uncomputed before the next Trotter step. 
The resulting circuit depth per step would scale as $\mathcal{O}(N_x^2 / \eps_{\mathrm{obs}})$, being quickly intractable with typical SPH requiring $N_x \sim 10^6$ grid points.

\subsection{Resource-Regime Considerations}
\label{sec:resource_limits}

\subsubsection{Fault-Tolerant Hardware Requirements}

REQWIEM is designed for fault-tolerant quantum hardware. 
All gate counts are logical operation counts on ideal, error-corrected qubits; on a surface-code architecture with physical error rate $p$ and code distance $d_{\mathrm{code}} \sim \log(C_{\mathrm{total}}/\eps_{\mathrm{logical}})$, each logical qubit requires $\mathcal{O}(d_{\mathrm{code}}^2)$ physical qubits~\cite{fowler2012surface}. 
For the Ar$_3^*$ trimer at $p = 10^{-3}$: $d_{\mathrm{code}} \sim 15$, giving $\sim 450$ physical qubits per logical qubit and a total of $\sim 13{,}500$ physical qubits for the 30-logical-qubit data register.
Systems at larger logical qubit counts (CH$_4$ at 65 qubits, collective neutrinos at 34 qubits) require $\sim 10^4$--$10^5$ physical qubits at the same code distance. Current platforms with $\sim 10^3$ physical
qubits at $p \sim 10^{-3}$ are approaching but have not yet reached this threshold; the trajectory of purported fault-tolerant quantum computing roadmaps places these device scales within the $10$--$15$ year horizon.
Further requirements include rotation precision of $\eps_\theta \sim10^{-14}$, rapid magic state factories and magic state injection, and coherence times that are hours/days. 
Although there are rapid advancements in the field, this extends the timeline prior to applicability, leaving REQWIEM as a specified subroutine for a fault-tolerant future.

\subsection{The REQWIEM Applicability Criterion}

\begin{enumerate}
\item \textbf{Non-trivial channel dimension}: $d \gtrsim 10$,
    so that the UCR's $\mathcal{O}(N_x \times d)$ scaling provides a decisive
    factor over classical $\mathcal{O}(N_x \times d^2)$. \emph{Check}: count
    the number of coupled internal states.

    \item \textbf{Non-separable coupling}: cross-mode rotation
    coefficients $a_S$ with $|S| \geq 2$ across distinct register
    pairs are non-negligible, so dynamics generate $\Reg{X}$--$\Reg{C}$ entanglement. \emph{Check}: compute the coefficient matrix and inspect the off-diagonal blocks.

    \item \textbf{Smooth potential}: $\|[T,V]\| \ll \|V\|_{\max}$,
    so the commutator-norm Trotter step advantage holds. \emph{Check}:
    apply the aliasing diagnostic $\|V\|_\infty/T_{\max} \leq 1$ on
    the proposed grid.

    \item \textbf{Dynamical observables or moderate eigenvalue
    precision}: the goal is wavepacket dynamics with channel populations
    as outputs, or eigenvalue precision $\eps \geq 10^{-6}$~a.u.\
    \emph{Check}: identify whether sub-wavenumber static eigenvalues
    are the sole output and whether no time-resolved observable is
    desired.
\end{enumerate}

\subsection{Open Problems and Opportunities}

The work presented in this dissertation opens several concrete directions, each of which addresses a technical gap and enables a field-level scientific program that is currently inaccessible.

\subsubsection{Geometry-dependent conical intersection \\
transforms and transition-metal photochemistry} 
The $R$-dependent unitary ${U}_{\mathrm{CI}}(R)$ of
Section~\ref{sec:ci_transform} costs $N_x \times \mathcal{O}(d^2)$ per
Trotter step, breaking DVR amortization.
Whether a smooth interpolation of ${U}_{\mathrm{CI}}(R)$ can recover partial amortization is open.
This is a primary barrier to first-principles non-Born-Oppenheimer dynamics in transition-metal photochemistry, where conical intersections among spin-orbit-coupled surfaces drive photocatalysis, light harvesting, and intersystem crossing in Fe, Ru, and Ir complexes. 
Classical MCTDH scales exponentially with the number of coupled states~\cite{meyer1990multi}; a partial-amortization CI transform would make 10--20 coupled surface dynamics tractable on approximately 50 logical qubits, bringing quantum simulation to the class of problems where classical wavepacket propagation is already at its practical limit.

\subsubsection{Wavelet basis error analysis \\ and multi-scale molecular dynamics}
The Daubechies $D_{2p}$ wavelet transform of Section~\ref{sec:wavelet} introduces off-diagonal potential elements decaying as $\mathcal{O}(\Delta x^{2p})$, and truncation at threshold
$\theta_{\mathrm{cut}}$ introduces systematic bias in the Trotter error bound. 
Analyzing how wavelet truncation error and Trotter error combine, and how $\theta_{\mathrm{cut}}$ should be chosen as a function of $\Delta x$, $\Delta t$, and $\eps$, is needed before the wavelet basis can be deployed in a controlled simulation. 
The scientific opportunity is multi-scale molecular dynamics: potentials with simultaneous structure at covalent-bond scales ($\sim 1 \,\text{a}_0$) and van der Waals scales
($\sim 20\,\text{a}_0$) currently require grid oversampling. 
This makes classical split-operator propagation prohibitive for systems with more than five or six modes. 
A wavelet-basis REQWIEM protocol with a verified error bound would make quantum simulation of protein-ligand binding and reactive dynamics in solvent tractable without grid oversampling.

\subsubsection{REQWIEM-QKSD demonstration on the Ar$_3^*$ trimer}
The Ar$_3^*$ trimer has a vibronic spectrum has been computed classically~\cite{horn1991vibrational, beck1988melting, bohmer1989stability} and provides a benchmark for QKSD demonstrability. 
Its channel dimension $d = 64$ is large enough for the QKSD advantage to manifest, and its Trotter step parameters can be characterized.
Demonstrating sub-chemical-accuracy eigenvalue extraction from a REQWIEM autocorrelation function, and comparing the Krylov window width achievable at fixed circuit depth against a qubitization-based QKSD baseline, would provide the first end-to-end resource comparison for coupled-channel eigenvalue problems on fault-tolerant hardware.
Successful demonstration here would validate the QKSD machinery for the broader catalog of systems in
Section~\ref{sec:fields}.

\subsubsection{Kerr black hole quasi-normal modes and gravitational wave science}
Section~\ref{sec:astrophysics} established that the Teukolsky equation for gravitational perturbations of a Kerr black hole maps directly onto the REQWIEM register architecture.
Tortoise coordinates $r^*$ in $\Reg{R}$, angular $l$-channel in $\Reg{C}$ ($d \sim 64$,
$\mathcal{O}(\log l_{\max})$ XOR classes from the $\Delta l = \pm 2$ spheroidal coupling), require approximately the same logical qubit count as the Ar$_3^*$ trimer.
Quasi-normal mode (QNM) spectra for near-extremal Kerr black holes ($a/M \to 1$) are directly extractable by applying REQWIEM-QKSD to the tortoise-coordinate propagator . 
This is where the spheroidal-spherical coupling matrix becomes ill-conditioned and the near-degenerate mode spectrum is difficult to resolve classically. 

The second-order Teukolsky equation governing nonlinear perturbations of Kerr spacetime has quadratic source terms that couple the $l$-channels with a density that scales as $\mathcal{O}(l_{\max}^4)$ classically; REQWIEM's linear-in-$d$ XOR structure handles this at $\mathcal{O}(N_{\mathrm{XOR}} \times N_{r^*} \times d/2)$ cost, providing access to second-order ringdown waveforms that the Laser Interferometer Space Antenna (LISA) will require to match observed signals at signal-to-noise ratios exceeding $10^3$~\cite{amaro2017laser}. 

Superradiant instability timescales $\tau_{\mathrm{inst}}$ for ultralight bosons (axions, dark photons) around Kerr black holes, as a gravitational-wave dark matter detection channel, requires
long-time evolution in the Kerr background at high precision, distinguishing signal from noise in pulsar timing arrays.
The REQWIEM-QKSD combination provides this capability. 
Together, these three problems constitute a meaningful path in quantum-assisted gravitational wave astronomy that the present framework makes quantitatively tractable for the first time.

\subsubsection{Ab initio nuclear reactions for astrophysical nucleosynthesis}
Section~\ref{sec:nuclear_fields} established the register architecture for multi-nucleon transfer and no-core shell model dynamics. 
The channel dimension is $d \sim 10^3$--$10^4$ for p-shell nuclei (Li, Be, B, C) at $N_{\max} = 8$, and the logical qubit count is $\sim 40$--$50$.
Big Bang nucleosynthesis and the p-process fall into this regime of dynamics, with cross sections at stellar energies currently determined by extrapolation from laboratory measurements or R-matrix fits. Ab initio NCSM reaction calculations at $N_{\max} \geq 8$ exceed classical computational capacity for nuclei heavier than ${}^4$He, opening the problem for quantum-assisted computational work. 
REQWIEM dynamics on the NCSM basis would provide reaction cross sections directly from first-principles nuclear structure, without the extrapolation uncertainties that limit current nucleosynthesis network calculations. 
The nucleosynthetic origin of elements up to carbon is encoded in these cross sections, and is a question that fault-tolerant quantum simulation, built on the architecture developed in this dissertation, is positioned to address.

\section{Conclusion}
\label{sec:conclusion}

The Real-Time Evolution of Quantum Wavepackets In Explicit Modularity architecture is not complicated on its face, reducing the problem to a first-quantized grid, applying operations controlled on internal state populations. 
The central claim made in this dissertation is that REQWIEM encodes the physical principle that the dynamics of a coupled-channel system decomposes, at every instant of time, into kinetic propagation, free-channel phase accumulation, and interaction-picture potential application, and that each of these three pieces is diagonal in a different, efficiently accessible basis. 
The QFT diagonalizes the kinetic energy, the FBR diagonalizes the free-channel Hamiltonian, the DVR diagonalizes the interaction potential, and the XOR-class focusing Clifford diagonalizes every off-diagonal coupling that neither the DVR nor the QFT can reach.
A M\"obius inversion converts arbitrary smooth functions on the grid into Gray-code rotation angles at zero gate overhead. 
The grid can represent any dimensionality, and the multilinear coefficients will expose an efficient decomposition in classical preprocessing if one exists.

\subsection{What Has Been Established}

Chapter~\ref{ch:model_ext} constructed the algorithm for the vibronic Hamiltonians: encoding the diabatic potential via UCR, XOR-class decomposition of off-diagonal couplings, and the DVR/FBR basis-change protocol using simple Hadamard transforms. 
At this stage, REQWIEM was a circuit for a specific model with polynomial potential energy surfaces, coupled electronic states, and bound-state dynamics.
Chapter~\ref{ch:resources} placed this construction on a quantitative fault-tolerant footing, demonstrating that the rotation-count scaling is competitive with QROM-based alternatives and that graph-optimized parallelism yields a depth advantage despite rotation-synthesis overhead.
Chapter~\ref{ch:comparison} extended this comparison beyond QROM methods to the broader class of oracle-based algorithms.
Trotter--Dyson duality shows that Trotter error vanishes near conical intersections while the perturbative (Dyson) truncation order diverges and QSP cost remains anchored to the gap-independent spectral range, providing a one-sided structural advantage for the split-operator approach in the nonadiabatic regime.

Chapter~\ref{ch:scattering} removed the first restriction: the confinement to bound-state dynamics. 
Extending the protocol to elastic scattering the extraction of phase shifts and
state-to-state cross sections exhibits that REQWIEM is a propagation method for continuum states, not merely a vibronic eigenvalue tool.
Without this extension, the reactive scattering, nuclear scattering, and neutrino oscillation examples
of Section~\ref{sec:catalog} would have no methodological foundation, each requiring propagation of wavepackets into asymptotic channels where the wavefunction is not square-integrable.
The scattering chapter finds that the split-operator circuit is capable of handling this regime faithfully, with the same Trotter error bounds and the same UCR cost formula derived for bound states.

Chapter~\ref{ch:smooth_potentials} removed the second restriction: the confinement to polynomial potentials.
Multilinear M\"obius expansion provides a closed-form encoding of arbitrary smooth functions on the binary grid, the smooth-potentials chapter completed the potential energy surface treatment.
Without this result, the algorithm could encode only quadratic and low-order polynomial surfaces, excluding ab initio and highly structured potentials that appear in Section~\ref{sec:catalog}. 
The M\"obius inversion converts the potential values at the $N_x \times d$ grid--channel points into
Gray-code UCR rotation angles at zero gate overhead, preserving the $\mathcal{O}(N_x \times d)$ scaling regardless of functional form.

With both restrictions removed, the present chapter considered the consistence of the completed framework, independent of the physical system that motivated its construction.

The DVR/FBR diagonalization is a closed-form procedure: Algorithm~1 constructs the DVR unitary from the spectral decomposition of a known Hermitian operator, compiles it deterministically via CSD, QFT, or butterfly decomposition, and embeds it in the Trotter step without iterative or adaptive subroutines (Section~\ref{sec:fbr_dvr_closed}).
Section~\ref{sec:new_subspaces} then extended the class of amenable transforms (Eq.~\ref{eq:amenable}) beyond those used in the scattering and smooth-potential chapters to include symmetry-adapted bases, Bloch crystal-momentum representations, wavelet multiresolution
bases, and geometry-dependent configuration interaction transforms. 
The circuit template (QFT, FBR UCR, DVR round trip, XOR-class passes) is invariant. 
Only the coefficient table, the explicit form of ${U}_{\mathrm{DVR}}$, and the XOR-class catalog change.

Section~\ref{sec:krylov} indicates that REQWIEM is a natural front-end for quantum Krylov subspace diagonalization.
The overlap matrix Toeplitz structure  $S_{jk} = C((k-j)\Delta\tau)$ reduces the eigenvalue extraction problem to estimation of a single autocorrelation function sampled at $m - 1$ time lags.  The controlled-REQWIEM circuit adds a single ancilla qubit and a factor of $\sim 2$ in CNOT count.
The commutator-norm Trotter error bound (Appendix~\ref{app:Error}) guarantees that the overlap fidelity
constraint
\begin{equation}\label{eq:conclusion_dt}
    \Delta t \leq \left(\frac{\eta}{C_2\,\|[T,V]\|\,
    (m-1)\Delta\tau}\right)^{1/2},
\end{equation}
is satisfied with step sizes orders of magnitude larger than the LCU 1-norm alone would require.
For the Ar$_3^*$ trimer, this translates to a factor of $\sim 10^7$ fewer circuit evaluations than a qubitization-based QKSD front-end. 
The channel-register populations generated at each Trotter step provide observable-resolved spectral information via their Fourier transforms, connecting the eigenvalue extraction to the scattering cross sections and transition rates computed in Chapter~\ref{ch:scattering}.

Sections~\ref{sec:fields} and~\ref{sec:limitations} delineate the framework boundary.
Section~\ref{sec:fields} identified additional physical systems across nuclear physics, condensed matter, quantum optics, and astrophysics whose Hamiltonians admit a direct REQWIEM register
assignment, each requiring the continuum-state propagation of Chapter~\ref{ch:scattering} and the non-polynomial potential encoding
of Chapter~\ref{ch:smooth_potentials}. 
The scope becomes broader than any single application domain, and that breadth is a direct consequence of the completeness results of Chapters~\ref{ch:scattering} -- ~\ref{ch:smooth_potentials}.

\subsection{Conditions for Structural Advantage}

Analysis of Section~\ref{sec:limitations} yields a classically checkable applicability criterion.
For REQWIEM to provide a structural advantage, four conditions must be simultaneously
satisfied: (1)~non-trivial channel dimension ($d \geq \mathcal{O}(10)$),
so that the UCR's linear scaling in $d$ dominates the classical quadratic
scaling; (2)~non-separable coupling, so that dynamics generate genuine
$\Reg{X}$--$\Reg{C}$ entanglement that classical mode-by-mode propagation
cannot exploit; (3)~smooth potential on the grid scale, so that the
commutator-norm Trotter step-size advantage holds; and (4)~dynamical
observables or moderate eigenvalue precision, so that the $t^{-1/3}$
resolution scaling of QKSD suffices.
The formal statement, with the classical check for each condition, is given in
Section~\ref{sec:limitations} (the REQWIEM Applicability Criterion).

These conditions are not satisfied for single-channel potential scattering at long times, purely harmonic dynamics, separable molecular Hamiltonians, or for electronic structure in the
second-quantized molecular orbital basis.
Identifying which regime a given problem occupies is a classical preprocessing task that requires only the rotation coefficient matrix, the potential smoothness, and the target precision. 

\subsection{DVR/FBR Duality}

A quantum algorithm simulating $H=T+V$ achieves efficiency when its circuit decomposition respects Hamiltonian structure. 
Each piece of ${H}$ is assigned to the primitive for which it is cheapest to implement.
For coupled-channel Hamiltonians, the structure is DVR/FBR duality, being two complementary orthonormal bases.
One diagonalizes the free-channel evolution and one diagonalizes the interaction potential, related by
an efficiently implementable unitary ${U}_{\mathrm{DVR}}$. 
When this duality holds, the split-operator Trotter sequence amortizes the $\mathcal{O}(d^2)$ basis-change cost across the entire coordinate grid at no additional gate expense, reducing the per-site potential step from $\mathcal{O}(d^2)$ (classical matrix-vector multiply) to $\mathcal{O}(d)$
(UCR diagonal). 
This amortization gives scalable advantage.

The DVR/FBR duality was introduced into coupled-channel quantum chemistry by Light, Hamilton, and Lill in 1985~\cite{light1985generalized} as a classical numerical technique.
The kinetic and potential energy operators of a molecular Hamiltonian are simultaneously sparse in complementary representations, and that alternating between those representations makes the classical split-operator method tractable. 
This alternation on a quantum circuit becomes a unitary acting in a superposition over the $2^{n_x}$ coordinate grid points, converting a dense $d \times d$ matrix-vector multiplyinto a diagonal phase accumulation.
The forty-year-old insight of Light, Hamilton, and Lill becomes, the mechanism that reduces this subclass of \#P problems to BQP-complete on fault-tolerant quantum hardware.

\subsection{Closing Remarks}

Construction of quantum algorithms can mimic the Cartesian ``mind-body" problem where the Res Cogitans (the thinking substance) becomes increasingly detached from the Res Extensa (physical extension)~\cite{descartes1996meditations}.
The LCU protocol guarantees asymptotic accuracy in a query-optimal idealism. 
This presents an idea of error bounds vanishing with logarithmic resources. 
In doing so, the volume is structured and a sparsity is imposed on the interaction picture, using PREPARE and SELECT subroutines to bridge a ``black-box" interface with a register of qubits that represent the system itself. 
In this way, qubitization may be epistemically similar to the Cartesian mind `thinking' the simulation into existence.
The consequent is the overhead of oracle loading, leading to deep, serialized circuits that result in a geometric, physical impracticability. 
It treats the Hamiltonian not as a physical machine to be run, but as an abstract monolith to be queried. 
Embedding $H$ into a quantum walk operator $W = e^{i \arccos(H/\lambda)}$, it converts energy (a physical quantity) into phase (information). 
In this way, it is a method for extracting knowledge (eigenvalues via Phase Estimation) about the system without necessarily replicating the physical flow of the system.
The mind perceives a concept like ``Simultaneity" instantly. LCU mimics this by representing the Hamiltonian as a simultaneous sum $\sum c_l U_l$.
The algorithm assumes access to an oracle (PREPARE/SELECT) that can instantaneously ``recall" any part of this sum, hiding the physical reality of implementation behind the black box of the oracle, much like the mind ignores the firing of individual neurons to produce a coherent thought.

A strong motivator for quantum computing is mapping a quantum system onto another quantum system. 
Take by contrast an algorithm that is designed for this mapping. 
Interactions are translated to connectivity constraints and a finite gate set.
Data is not encoded as a concept, but is distributed across resources as a physical ``load," potentially allowing easier handling of chaotic, entropically saturated systems. 
This is the Res Extensa, where Cartesian mechanics are applied by nature of the constructed Hamiltonian rather than the Hamiltonian's spectrum. 

The ``mind" (LCU) proposes algorithms that cannot inhabit the ``body" (Hardware/Hamiltonian), limited by Hamiltonian density brought by nonadiabatic coupling. 
Functionality of qubitization becomes limited to sparse or truncated basis sets. 

Descartes ultimately failed to join the abilities of mind and the causative action on the body. 
Quantum chemistry wonders similarly if LCU can drive dense chemical simulation without a prohibitively expensive translation layer.
If nonadiabatic dynamics does not sacrifice its full basis and does not assume sparsity, the Hamiltonian admits itself as a machine.
These systems are not unstructured data to be loaded and can be encoded via interactions on the quantum register without will or thought. 
The particle is dragged (propagated) through a potential, instead of an operator being processed. 
A molecule does not check its kinetic energy, then its potential energy, and then its coupling in a tragedy of discretization into basis states and a universal gate set.
It must take the holistic reality of nature and ``disperse" it.
Just as the human body is constrained by the leverage of joints and the speed of nerve impulses, REQWIEM is constrained by the native gate set and the connectivity graph between the qubits. 
The algorithm is a struggle to map the infinite, continuous degrees of freedom onto the finite, discrete chip geometry.

Break the ``whole" into parts via Trotterization:$$e^{-i(T+V)t} \approx \left( e^{-iT \frac{t}{n}} e^{-iV \frac{t}{n}} \right)^n$$ introduces a machine ``stutter." 
The REQWIEM algorithm admits that the ``Body" can only approximate the fluid motion of the ``Soul" (the wavefunction). 
Trotter error is the physical manifestation of the gap between the holistic reality and the simulation.

LCU (The Mind) tries to solve the problem by staying in the realm of pure information. 
It aims for the ``Continuum Limit" and ``Arbitrary Basis Sets," striving for a perfect mathematical representation of the wavefunction. 
But without a physical body (an efficient oracle implementation), it remains a ghost, as a query-optimal theory viable for Hamiltonians that admit some sparse representation.
REQWIEM (The Body) accepts the ``fallen" state of the hardware, building the Hamiltonian out of ``dust" (CNOTs and rotations). 
It does not possess the Mind's asymptotic purity (it has Trotter error), but it possesses existence by creating a physical analog of the chemical system within chip connectivity.

Process philosophy and instrumentalism consider the becoming and constructing alongside the description and the performance, taking for example the Aristotelian Dynamis (Potentiality) versus Energeia (Actuality)~\cite{aristotle2025complete}.
In standard quantum algorithms (like LCU), the Hamiltonian $H$ is treated as a static list of coefficients, being a mathematical object existing in potentiality.
It is data to be interpreted, with LCU observing the Hamiltonian's norm to predict resource costs.
The performance transmutes dynamis (possibility) into energeia (actuality). 
Such an algorithm would therefore act as a propagator, quantum mechanically dragging an initialized state through all paths.

Heidegger distinguished between objects that are merely present to be observed (Vorhanden) and objects that are engaged with as equipment (Zuhanden)~\cite{heidegger1962being}.
A study of the Hamiltonian contemplates the Heideggerian hammer, and enacts its approximate effect based on sparsity, norm, and spectrum, becoming exponentially more difficult with entropy saturation. 
Such a study is LCU, as Vorhandenheit.
An alterative is the engine's disappearance into its function. 
The compression and expansion of gasses to drive the pistons is forgotten, and you simply drive to your destination, as Zuhandenheit. 
The Cartesian body lets the user focus on the chemistry, embodying Technea. 

LCU will typically take a sparse representation with hierarchical interaction topology.
This breaks under strong correlation, introducing a philosophical Rhizome to be traversed, as sorting and arranging becomes intractable~\cite{gilles1976rhizome}. 
Such is the problem found in ML-MCTDH \cite{wang2015multilayer}.

Giambattista Vico argued that we can only truly know that which we have made ourselves: \emph{Verum ipsum factum}~\cite{vico1988most}.
Typically considered in epistemic constructivism in the truth manufactured from history, it can be related to science in the replication and accurate prediction of systems. 
Mathematics models represent physical reality, and ``truth" comes with aligning theory and observation. 
Physics does not determine what a system is, but focuses on what it does. 
The dynamics of a molecules cannot be ``known" by solving a differential equation as a system's description, rather, it can only be known when a machine replicating the system is built. 
An algorithm acting as a Heideggerian hammer is thereby argued to be a path to ontological ``truth," with the dual Vorhandenheit an approximation of that which is unknown. 

If such an algorithm can be meaningfully constructed, it would act as a ``ground truth," from which further optimization can grow. 
This dissertation argues for REQWIEM as a candidate for such an algorithm.
If \emph{Verum ipsum factum} is followed, true understanding of dense, chaotic dynamics at conical intersections follows after a mechanical assemblage of its Hamiltonian out of a quantum processor's gate set. 
Such is the target of this thesis, being the construction of a quantum circuit that develops a baseline. 
Future algorithms may trade precision for speed with T-gate budgeting or interaction picture approximation, to be compared against a Rhizome~\cite{gilles1976rhizome}.

An algorithm acting as propagator may be thought like a dedicated GPU Kernel or Shader.
In classical computing, a CPU is designed for complex logic and branching (if/then/else), much like the LCU/oracle approach which relies on conditional logic ("if index is $l$, apply $U_l$"). 
This is slow for dense data because of the overhead of checking conditions.
A GPU uses Single Instruction, Multiple Data (SIMD) to achieve parallelism.
Likewise, a layer of rotations may apply a potential energy phase factor ($e^{-iV(x)\Delta t}$) to a particular Hilbert space hyperplane, minimizing the ``instruction fetch" cost.

As a linear algebra tool, an algorithm like this may be classified as a BLAS Level 2 (Matrix-Vector) routine.
It performs the operation $|\psi_{t+1}\rangle = e^{-iH\Delta t} |\psi_t\rangle$ on the position register itself.
A standard solver (like in MATLAB or NumPy) is a software library sitting on top of generic hardware, whereas this would be closer to an Application-Specific Integrated Circuit (ASIC).

The potential for quantum advantage with REQWIEM is structural and bounded.
The compression of $N_x \times d$ classical amplitudes into $n_x + n_c$ qubits, combined with the UCR's linear scaling in $d$ produces an advantage factor of $\sim d$ per site over classical DVR methods \cite{light1989theoretical} that no classical algorithm can replicate
without exploiting the same quantum superposition. 
It derives from the Hamiltonian decomposition, allowing extensibility beyond electronic nonadiabaticity.
REQWIEM requires non-trivial channel coupling, smooth potentials \cite{childs2021theory},
non-separable dynamics, and fault-tolerant hardware that does not yet
exist at the required scale \cite{fowler2012surface}.

The algorithm described in this dissertation is not a near-term device proposal.
It is a specification for a physically motivated quantum simulation protocol for the class of
coupled-channel Hamiltonians, valid for any system that satisfies the aforementioned preconditions, executable on fault-tolerant hardware in the $10^3$--$10^6$ physical qubit regime, and extendable without modification to nuclear reactions, polaron transport, topological phases, cavity QED, and gravitational perturbation theory.
Future theoretical contributions aim to further advantage these Hamiltonian properties, go beyond Hermitian systems, and extend digital quantum simulation to simulate many more quantum systems, just as Feynman intended \cite{feynman1982simulating}.

\bibliography{refs} 

\appendices

\chapter{REQWIEM Algorithm}\label{app:reqwiem-algorithm}

Algorithm~\ref{alg:xor_step} assembles the complete Trotter step, operating with Algorithm~\ref{alg:single_frag} for off-diagonal terms.
On-diagonal and off-diagonal blocks use different circuit primitives optimized to their respective structures: quantum multiplexing with a three-tier hybrid decomposition for the on-diagonal potential (Section~\ref{sec:mux_alg}), and per-fragment Clifford diagonalization with UCR$_Z^{(m-1)}$ for the off-diagonal coupling (Section~\ref{sec:offdiag_alg}).

\begin{breakablealgorithm}
\caption{One Trotter step: REQWIEM with multiplexed on-diagonal potential and XOR-class-fragmented off-diagonal coupling}
\label{alg:xor_step}
\begin{algorithmic}[1]
\Require Channel register $\ket{s_\text{channel}}$ ($m$ qubits), vibrational registers $\{\ket{x_r}\}$ ($n_r$ qubits each), timestep $\tau$
\Require Classically precomputed: on-diagonal surface coefficients $\{V_{ss}\}$ and correction angles; per-fragment UCR angles $\{\phi_{\tilde\sigma}^{(\zeta,r,\ell)}\}$ for all XOR classes $\zeta$
\Ensure $|\psi\rangle \mapsto e^{-iH_{\text{vib}}\tau}|\psi\rangle + \mathcal{O}(\tau^3)$ (second-order Trotter)
\Statex
\Statex \textbf{ Kinetic half-step }
\For{each mode $r = 1,\ldots,d$ \textbf{in parallel}}
    \State $\text{cQFT}(\ket{x_r})$ \Comment{Position $\to$ momentum basis}
    \State Apply $R_z$ phase kicks for $p_r^2/2\mu_r$ \Comment{$n_r + \binom{n_r}{2}$ rotations, no channel involvement}
    \State $\text{cIQFT}(\ket{x_r})$ \Comment{Momentum $\to$ position basis}
\EndFor
\Statex
\Statex \textbf{ On-diagonal potential half-step (quantum multiplexing, Sec.~\ref{sec:mux_protocol}) }
\Statex
\State \textit{Phase~1: Reference surface $V_{00}$ (unconditional single-qubit terms):}
\For{each mode $r = 1,\ldots,d$ \textbf{in parallel}}
    \State Apply $R_z$ gates for constant, linear, and diagonal-quadratic
           terms of $V_{00}^{(r)}$ \Comment{$n + 1$ rotations per mode}
\EndFor
\Statex
\State \textit{Phase~2: Truth-value computation:}
\State Compute truth-value ancillas $\{|w_s\rangle\}_{s=1}^{M-1}$
        via Toffoli cascades on $\ket{s_\text{channel}}$
        \Comment{$T_{\text{truth}}$ Toffoli; Eq.~\ref{eq:mux_toffoli_overhead}}
\Statex
\State \textit{Phase~3: Fan-out:}
\State Fan-out $m$ channel qubits and $(M{-}1{-}m)$ truth-value ancillas
        to $d$ copies via binary-tree CNOTs
        \Comment{$(M{-}1)(d{-}1)$ CNOTs, depth $\lceil\log_2 d\rceil$}
\Statex
\State \textit{Tier~1: Constant and linear corrections (parallel across modes, serial across surfaces):}
\For{each correction surface $s = 1,\ldots,M{-}1$ \textbf{(serial within each mode)}}
    \For{each mode $r = 1,\ldots,d$ \textbf{in parallel via fan-out copies}}
        \State Apply $\mathrm{CR}_z$ gates for $\Delta V_s^{(r)}$,
               controlled on local copy of $|w_s\rangle$
               \Comment{$n + 1$ rotations per mode per surface}
    \EndFor
\EndFor
\Statex
\State \textit{Tier~2: Intra-mode quadratic cross-terms ($\mathrm{UCR}_Z^{(m)}$ on original channel register):}
\For{each mode $r = 1,\ldots,d$}
    \For{$(\ell,\ell') \in \binom{n_r}{2}$ intra-mode cross sites}
        \State $\text{CNOT}_{\ell'\to\ell}$;\;
               $\text{UCR}_Z^{(m)}(q_\ell^{(r)},\,\mathcal{C}^{(r)});\;$ \Comment{mode-parallel via fan-out copies}
               $\text{CNOT}_{\ell'\to\ell}$
               \Comment{$M$ rotations, $M{+}1$ CNOTs per site}
    \EndFor
\EndFor
\Statex
\State \textit{Tier~3: Inter-mode bilinear terms ($\mathrm{UCR}_Z^{(m)}$, chromatic scheduling):}
\For{each mode pair $(r,s)\in\mathcal{G}_2^{<}$ \textbf{serialized over $\chi(\mathcal{G}_2^{<})$ colors}}
    \Comment{On-diagonal bilinear}
    \For{$\ell = 0,\ldots,n_r-1$;\; $p = 0,\ldots,n_s-1$}
        \State $\text{CNOT}_{q_p^{(s)}\to q_\ell^{(r)}}$;\;
               $\text{UCR}_Z^{(m)}(q_\ell^{(r)},\,\mathcal{C}^{(r)});\;$ \Comment{mode-parallel via fan-out copies}
               $\text{CNOT}_{q_p^{(s)}\to q_\ell^{(r)}}$
               \Comment{$M$ rotations, $M{+}1$ CNOTs per site}
    \EndFor
\EndFor
\Statex
\State \textit{Phase~4: Fan-in and uncomputation:}
\State Fan-in all $(M{-}1)(d{-}1)$ ancilla copies
        (reverse CNOT trees)
        \Comment{$(M{-}1)(d{-}1)$ CNOTs, depth $\lceil\log_2 d\rceil$}
\State Uncompute truth-value ancillas
        (reverse Toffoli cascades)
        \Comment{$T_{\text{truth}}$ Toffoli}
\Statex
\Statex \textbf{ Off-diagonal potential full step (XOR-class symmetric product) }
\Statex \textit{Forward sweep} ($\zeta = 1, 2, \ldots, M-1$):
\For{$\zeta = 1$ \textbf{to} $M-1$}
    \State \Call{ApplyFragment}{$\zeta,\;\tau/2$} \Comment{Algorithm~\ref{alg:single_frag}}
\EndFor
\Statex \textit{Backward sweep} ($\zeta = M-1, M-2, \ldots, 1$):
\For{$\zeta = M-1$ \textbf{downto} $1$}
    \State \Call{ApplyFragment}{$\zeta,\;\tau/2$} \Comment{Algorithm~\ref{alg:single_frag}}
\EndFor
\Statex
\Statex \textbf{ On-diagonal potential half-step (quantum multiplexing) }
\State (Repeat on-diagonal half-step above with angle $\tau/2$)
\Statex
\Statex \textbf{ Kinetic half-step }
\State (Repeat kinetic half-step above)
\end{algorithmic}
\end{breakablealgorithm}

\begin{breakablealgorithm}
\caption{\textsc{ApplyFragment}: single off-diagonal fragment at half-angle}\label{alg:single_frag}
\begin{algorithmic}[1]
\Require Fragment index $\zeta$ with support $S(\zeta) = \{s_1,\ldots,s_h\}$, half-timestep $\tau/2$, channel register $\mathcal{C}$ ($m$ qubits), vibrational registers $\mathcal{V}_1,\ldots,\mathcal{V}_d$ ($n$ qubits each, total of $m+dn$)
\Require Classically precomputed UCR angles $\{\phi_{\tilde\sigma}^{(\zeta,r,\ell)}\}$ encoding $M/2$ coupling values per site
\Statex
\Statex \textbf{ Clifford focusing ($U_\zeta$) }
\For{$k = h$ \textbf{downto} $2$} \Comment{CNOT cascade: $s_1$ controls, $s_k$ targets}
    \State $\text{CNOT}_{s_1 \to s_k}$
\EndFor
\State $H_{s_1}$ \Comment{$X_{s_1} \to Z_{s_1}$; depth $h(\zeta)$, $h(\zeta){-}1$ CNOTs}
\Statex
\Statex \textbf{ Fan-out channel register }
\State Fan-out $\mathcal{C}$ to $d$ copies $\mathcal{C}^{(1)},\ldots,\mathcal{C}^{(d)}$
       via binary-tree CNOTs
       \Comment{$m(d{-}1)$ CNOTs, depth $\lceil\log_2 d\rceil$}
\Statex
\Statex \textbf{ Mode-parallel UCR block }
\For{each mode $r = 1,\ldots,d$ \textbf{in parallel}} \Comment{Using local copy $\mathcal{C}^{(r)}$}
    \State \textit{Constant-phase site:}
           $\text{CNOT}_{s_1^{(r)}\to\ell_0^{(r)}}$;\;
           $\text{UCR}_Z^{(m-1)}(\ell_0^{(r)},\;\tilde{S^c}^{(r)})$;\;
           $\text{CNOT}_{s_1^{(r)}\to\ell_0^{(r)}}$
           \Comment{$M/2$ rot., $M/2{+}1$ CNOTs}
    \For{$\ell = 0,\ldots,n_r - 1$} \Comment{Single-qubit (linear) sites}
        \State $\text{CNOT}_{s_1^{(r)}\to q_\ell^{(r)}}$;\;
               $\text{UCR}_Z^{(m-1)}(q_\ell^{(r)},\;\tilde{S^c}^{(r)})$;\;
               $\text{CNOT}_{s_1^{(r)}\to q_\ell^{(r)}}$
               \Comment{$M/2$ rot., $M/2{+}1$ CNOTs per site}
    \EndFor
    \For{$(\ell,\ell') \in \binom{n_r}{2}$ intra-mode cross sites}
        \State $\text{CNOT}_{\ell'\to\ell}$;\;
               $\text{CNOT}_{s_1^{(r)}\to q_\ell^{(r)}}$;\;
               $\text{UCR}_Z^{(m-1)}(q_\ell^{(r)},\;\tilde{S^c}^{(r)})$;\;
               $\text{CNOT}_{s_1^{(r)}\to q_\ell^{(r)}}$;\;
               $\text{CNOT}_{\ell'\to\ell}$
               \Comment{$M/2$ rot., $M/2{+}3$ CNOTs per site}
    \EndFor
\EndFor
\For{$(r,s) \in \mathcal{G}_4$ \textbf{serialized over $\chi(\mathcal{G}_4)$ colors}} \Comment{Off-diagonal bilinear cross sites}
    \For{$\ell = 0,\ldots,n_r-1$;\; $p = 0,\ldots,n_s-1$}
        \State $\text{CNOT}_{q_p^{(s)}\to q_\ell^{(r)}}$;\;
               $\text{CNOT}_{s_1^{(r)}\to q_\ell^{(r)}}$;\;
               $\text{UCR}_Z^{(m-1)}(q_\ell^{(r)},\;\tilde{S^c}^{(r)})$;\; \newline
               $\text{CNOT}_{s_1^{(r)}\to q_\ell^{(r)}}$;\;
               $\text{CNOT}_{q_p^{(s)}\to q_\ell^{(r)}}$
               \Comment{$M/2$ rot., $M/2{+}3$ CNOTs per site}
    \EndFor
\EndFor
\Statex
\Statex \textbf{ Fan-in channel register }
\State Fan-in $\mathcal{C}^{(1)},\ldots,\mathcal{C}^{(d)}$ back to $\mathcal{C}$
       (reverse CNOT tree)
       \Comment{$m(d{-}1)$ CNOTs, depth $\lceil\log_2 d\rceil$}
\Statex
\Statex \textbf{ Unfocusing ($U_\zeta^\dagger$) }
\State $H_{s_1}$
\For{$k = 2$ \textbf{to} $h$}
    \State $\text{CNOT}_{s_1 \to s_k}$ \Comment{Reverse cascade; depth $h(\zeta)$, $h(\zeta){-}1$ CNOTs}
\EndFor
\end{algorithmic}
\end{breakablealgorithm}

\chapter{Binary Arithmetic Encoding of Position-Register Gates}\label{app:binary_decomp}

This appendix derives the binary-expansion structure of the rotation angles used
throughout the REQWIEM circuit, establishes how these angles interact with the
channel-register UCR formalism, and demonstrates the absence of position-register
Walsh truncation.

Throughout, $x_r$ denotes the vibrational coordinate of mode~$r$, encoded in an
$n_r$-qubit register ($n$ when uniform across modes); $M = 2^m$ is the number of
padded electronic states; $d$ is the number of vibrational modes; $k$ is the
number of Trotter steps; and $\tau$ is the Trotter timestep.

Coupling-constant notation follows Eq.~\ref{eq:vibronic_ham_full}:
$a_r^{(s)}$, $a_{rs}^{(s)}$ for intrastate (on-diagonal) linear and bilinear
coefficients on surface~$s$, and $c_r^{(ss')}$, $c_{rs}^{(ss')}$ for interstate
(off-diagonal) coupling coefficients between surfaces~$s$ and~$s'$.
The total curvature on surface~$j$ for mode~$r$ is defined as
\begin{equation}\label{eq:app_total_curvature}
    \kappa_j^{(r)} \;=\; \tfrac{1}{2}\mu_r\omega_r^2 + a_{rr}^{(j)},
\end{equation}
combining the shared harmonic contribution $\tfrac{1}{2}\mu_r\omega_r^2$ with the
surface-dependent correction $a_{rr}^{(j)}$, following
Section~\ref{sec:mux_tier2}.

\section{Binary Encoding of the Position Register}
\label{app:binary_encoding}

Each vibrational coordinate $x_r$ is discretized on a uniform grid of
$N_r = 2^{n_r}$ points spanning the simulation domain
$[-L_r/2,\, L_r/2]$:
\begin{equation}\label{eq:app_binary_encoding}
    x_r = \Delta_r\!\left(\sum_{\ell=0}^{n_r-1} 2^\ell\, q_\ell^{(r)}
          - \frac{N_r}{2}\right),
    \qquad
    \Delta_r = \frac{L_r}{N_r},
\end{equation}
where $q_\ell^{(r)} \in \{0,1\}$ is the $\ell$-th qubit of mode~$r$.
The operator $x_r$ is diagonal in the computational basis, with
eigenvalues determined by the binary string $\{q_\ell^{(r)}\}$.
Substituting Eq.~\ref{eq:app_binary_encoding} into any polynomial
function of $x_r$ produces a multilinear polynomial in the individual
qubits, since $q_\ell^2 = q_\ell$ for binary variables.
Each resulting monomial corresponds to a diagonal operator in the
computational basis, implementable directly as a phase gate.

\section{Monomial Decompositions}\label{app:monomial_decomp}

The three classes of position-dependent terms appearing in the vibronic
Hamiltonian (Eq.~\ref{eq:vibronic_ham_full}) each yield a characteristic
set of rotation sites.
All rotation angles include the Trotter timestep~$\tau$, so that the
angles given below are those appearing directly in the compiled circuit.

\subsection{Linear Terms}\label{app:linear_decomp}

A linear coupling $a_r^{(j)}\, x_r$ (intrastate, $r \in \mathcal{G}_1$)
or $c_r^{(ss')}\, x_r$ (interstate, $r \in \mathcal{G}_3$) expands as
\begin{equation}\label{eq:app_linear_expand}
    a_r^{(j)}\, x_r = a_r^{(j)} \Delta_r
    \left(\sum_{\ell=0}^{n_r-1} 2^\ell\, q_\ell^{(r)}
          - \frac{N_r}{2}\right).
\end{equation}
The constant offset $-a_r^{(j)} \Delta_r N_r/2$ contributes a
state-dependent global phase, handled by quantum multiplexing
(on-diagonal, Section~\ref{sec:mux_alg}) or by the UCR constant-phase
site (off-diagonal, Section~\ref{sec:ucr_offdiag}).
The variable part yields $n_r$ single-qubit $R_z$ rotations with
circuit angles
\begin{equation}\label{eq:app_angle_linear}
    \theta_\ell^{(j)} = \tau\, a_r^{(j)}\, \Delta_r\, 2^\ell,
    \qquad \ell = 0, \ldots, n_r - 1,
\end{equation}
and analogously $\theta_\ell^{(ss')} = \tau\, c_r^{(ss')}\, \Delta_r\,
2^\ell$ for the off-diagonal coupling constant.
Each rotation is applied to qubit $q_\ell^{(r)}$ on the vibrational
register.

\subsection{Intra-Mode Quadratic Terms}
\label{app:quadratic_decomp}

An intra-mode quadratic term on surface~$j$ combines the shared harmonic
contribution and the surface-dependent correction into the total
curvature $\kappa_j^{(r)}$ (Eq.~\ref{eq:app_total_curvature}).
Expanding $\kappa_j^{(r)}\, x_r^2$ gives
\begin{equation}\label{eq:app_Q2_decomp}
    \kappa_j^{(r)}\, x_r^2 = \kappa_j^{(r)}\, \Delta_r^2
    \left[\sum_{\ell=0}^{n_r-1} 2^{2\ell}\, q_\ell^{(r)}
    + 2\!\sum_{\ell < p} 2^{\ell+p}\, q_\ell^{(r)} q_p^{(r)}
    - N_r \sum_\ell 2^\ell\, q_\ell^{(r)}
    + \frac{N_r^2}{4}\right].
\end{equation}
The terms organize into three classes:
\begin{enumerate}
    \item \textbf{Single-qubit diagonal} ($q_\ell^2 = q_\ell$):
    The contributions $2^{2\ell} q_\ell$ and the linear correction
    $-N_r\, 2^\ell q_\ell$ are absorbed into the rotation angles of the
    single-qubit gates already present from the linear term, modifying
    Eq.~\ref{eq:app_angle_linear} by additive shifts.
    No additional gates are required.

    \item \textbf{Intra-mode cross pairs}
    ($q_\ell^{(r)} q_p^{(r)}$, $\ell < p$):
    Each of the $\binom{n_r}{2}$ pairs produces a two-qubit $ZZ$
    interaction with circuit angle
    \begin{equation}\label{eq:app_angle_quadratic}
        \theta_{\ell p}^{(j)} = \tau\,\kappa_j^{(r)}\,
        \Delta_r^2\, 2^{\ell+p+1},
        \qquad 0 \leq \ell < p \leq n_r - 1,
    \end{equation}
    where $\kappa_j^{(r)} = \tfrac{1}{2}\mu_r\omega_r^2 + a_{rr}^{(j)}$
    is the full curvature on surface~$j$
    (Eq.~\ref{eq:app_total_curvature}), consistent with the folded
    formulation of Tier~2 (Section~\ref{sec:mux_tier2}).
    Since both qubits reside on the same mode register, the $ZZ$
    interaction is implemented by the standard CNOT sandwich:
    \begin{equation}\label{eq:app_ZZ_intramode}
        \mathrm{CNOT}_{p \to \ell} \;\longrightarrow\;
        R_z(q_\ell;\, \theta_{\ell p}^{(j)}) \;\longrightarrow\;
        \mathrm{CNOT}_{p \to \ell},
    \end{equation}
    requiring 1 rotation and 2 CNOTs per pair, with zero Toffoli gates.

    \item \textbf{Constant} ($N_r^2/4$):
    A state-dependent global phase, absorbed into the constant-phase
    rotation site (Section~\ref{app:constant_phase}).
\end{enumerate}

\subsection{Bilinear Cross-Mode Terms}\label{app:bilinear_decomp}

A bilinear coupling $a_{rt}^{(j)}\, x_r x_t$ (intrastate,
$(r,t) \in \mathcal{G}_2^{<}$) or $c_{rt}^{(ss')}\, x_r x_t$
(interstate, $(r,t) \in \mathcal{G}_4$), with $r \neq t$, expands as
\begin{equation}\label{eq:app_bilinear_expand}
    a_{rt}^{(j)}\, x_r x_t = a_{rt}^{(j)}\, \Delta_r \Delta_t
    \left[\sum_{\ell=0}^{n_r-1}\sum_{p=0}^{n_t-1}
          2^{\ell+p}\, q_\ell^{(r)} q_p^{(t)}
    - \frac{N_t}{2}\sum_\ell 2^\ell\, q_\ell^{(r)}
    - \frac{N_r}{2}\sum_p 2^p\, q_p^{(t)}
    + \frac{N_r N_t}{4}\right].
\end{equation}
Note that $r$ and $t$ are both mode indices, while $j$ (on-diagonal) or
$(s,s')$ (off-diagonal) labels the electronic state(s).
The terms organize as follows:
\begin{enumerate}
    \item \textbf{Single-qubit corrections}:
    The terms $-(N_t/2)\, 2^\ell\, q_\ell^{(r)}$ and
    $-(N_r/2)\, 2^p\, q_p^{(t)}$ shift the rotation angles of
    existing single-qubit gates on modes~$r$ and~$t$ respectively.
    No additional gates are required.

    \item \textbf{Inter-mode cross pairs} ($q_\ell^{(r)} q_p^{(t)}$):
    Each of the $n_r \times n_t$ pairs produces a two-qubit $ZZ$
    interaction across different mode registers, with circuit angle
    \begin{equation}\label{eq:app_angle_bilinear}
        \theta_{\ell p}^{(j)} = \tau\, a_{rt}^{(j)}\,
        \Delta_r \Delta_t\, 2^{\ell+p},
        \qquad
        \ell \in \{0, \ldots, n_r{-}1\},\;
        p \in \{0, \ldots, n_t{-}1\},
    \end{equation}
    and analogously $\theta_{\ell p}^{(ss')} = \tau\, c_{rt}^{(ss')}\,
    \Delta_r \Delta_t\, 2^{\ell+p}$ for off-diagonal couplings.
    Despite bridging separate registers, each inter-mode cross pair is
    implemented by the same CNOT sandwich used for intra-mode pairs:
    \begin{equation}\label{eq:app_ZZ_intermode}
        \mathrm{CNOT}_{q_p^{(t)} \to q_\ell^{(r)}} \;\longrightarrow\;
        R_z(q_\ell^{(r)};\, \theta_{\ell p}^{(j)}) \;\longrightarrow\;
        \mathrm{CNOT}_{q_p^{(t)} \to q_\ell^{(r)}},
    \end{equation}
    requiring 1 rotation and 2 CNOTs per pair, with
    {zero Toffoli gates}.

    \item \textbf{Constant} ($N_r N_t/4$):
    A state-dependent global phase, absorbed into the constant-phase
    rotation site (Section~\ref{app:constant_phase}).
\end{enumerate}

The Toffoli-free implementation of the bilinear $ZZ$ interaction
deserves emphasis.
In a na\"ive approach, the product $q_\ell^{(r)} q_p^{(t)}$ of qubits
on different registers would be computed into a work ancilla via a
Toffoli gate, rotated, and uncomputed.
The CNOT sandwich avoids this entirely: the CNOT
$q_p^{(t)} \to q_\ell^{(r)}$ maps the target qubit to
$q_\ell^{(r)} \oplus q_p^{(t)}$, and a subsequent $R_z$ on this
target followed by the reverse CNOT produces the net unitary
\begin{equation}\label{eq:app_ZZ_derivation}
    e^{-i\theta\, Z_\ell^{(r)} Z_p^{(t)} / 2}
    = \mathrm{CNOT}_{p \to \ell}\;
      R_z(\theta)\;
      \mathrm{CNOT}_{p \to \ell},
\end{equation}
which is the required conditioned phase $e^{\pm i\theta/2}$ depending
on the parity $q_\ell \oplus q_p$.
This identity holds regardless of whether the two qubits reside on
the same register (intra-mode) or different registers (inter-mode),
and is the reason the off-diagonal Toffoli count
$T_{\mathrm{od}} = 0$ despite the presence of
bilinear coupling terms.

\section{Constant-Phase Rotation Site}\label{app:constant_phase}

Each monomial decomposition produces constant terms that depend on the
electronic state index but not on the qubit variables.
For a given surface~$j$ and mode~$r$, these constants arise from:
\begin{enumerate}
    \item the vertical state energy $E_j^{(0)}$,
    \item the linear offset $-a_r^{(j)}\,\Delta_r\,N_r/2$,
    \item the diagonal-quadratic constant
          $\kappa_j^{(r)}\,\Delta_r^2\,N_r^2/4$
          (including the shared harmonic contribution
          $\tfrac{1}{2}\mu_r\omega_r^2$),
    \item the bilinear constant
          $a_{rt}^{(j)}\,\Delta_r\Delta_t\,N_r N_t/4$ for each
          bilinear mode pair $(r,t) \in \mathcal{G}_2^{<}$.
\end{enumerate}
These are collected into a single constant-phase $R_z$ gate per mode,
with angle
\begin{equation}\label{eq:app_constant_phase}
    \theta_{\mathrm{const}}^{(j,r)} = \tau\!\left[
    \frac{E_j^{(0)}}{d}
    - a_r^{(j)}\,\Delta_r\,\frac{N_r}{2}
    + \kappa_j^{(r)}\,\Delta_r^2\,\frac{N_r^2}{4}
    + \sum_{\substack{t:\,(r,t)\in\mathcal{G}_2^{<}}}
      a_{rt}^{(j)}\,\Delta_r\Delta_t\,\frac{N_r N_t}{4}
    \right],
\end{equation}
where the vertical energy is distributed equally across $d$ modes.
In the on-diagonal block, this constant-phase site is handled by the
Phase~1 and Tier~1 structure of Section~\ref{sec:mux_alg}; in the
off-diagonal block, it appears as the first UCR site within each
fragment (Algorithm~\ref{alg:single_frag}).

\section{Summary of Rotation Sites per Coupling Term}
\label{app:site_summary}

Collecting the results of
Sections~\ref{app:linear_decomp}--\ref{app:bilinear_decomp}, the
number of rotation sites generated by each class of potential term is:

\begin{center}
\renewcommand{\arraystretch}{1.25}
\begin{tabular}{l c c c}
\hline\hline
Term type & Sites per term & CNOT per site & Toffoli per site \\
\hline
Constant phase & $1$ & $0$ & $0$ \\
Linear ($x_r$) & $n_r$ & $0$ & $0$ \\
Intra-mode quadratic ($x_r^2$) & $\binom{n_r}{2}$ & $2$ & $0$ \\
Bilinear ($x_r x_t$, $r \neq t$) & $n_r n_t$ & $2$ & $0$ \\
\hline\hline
\end{tabular}
\end{center}

For uniform register size $n_r = n$, the {total} number of
rotation sites across all $d$ modes and all bilinear pairs in the
$C_1$ (no symmetry) limit is
\begin{equation}\label{eq:app_Ssites}
    S_{\mathrm{sites}}^{(\mathrm{total})}
    = d\!\left(1 + n + \binom{n}{2}\right)
      + \binom{d}{2}\, n^2,
\end{equation}
where the first term sums the constant, linear, and intra-mode
quadratic sites over all $d$ modes, and the second term counts the
$n^2$ inter-mode cross-pair sites for each of the $\binom{d}{2}$
bilinear mode pairs.
This matches the expression
$d(1{+}n) + d\binom{n}{2} + \binom{d}{2}n^2$ used in the main text.
For molecules with non-trivial symmetry, $\binom{d}{2}$ is replaced
by the number of symmetry-allowed bilinear pairs
$|\mathcal{G}_2^{<}|$ (on-diagonal) or $|\mathcal{G}_4|$
(off-diagonal).

Every site produces exactly one $R_z$ rotation per electronic-state
value addressed by the UCR\@: $M$ rotations per site for on-diagonal
UCR$_Z^{(m)}$ and $M/2$ rotations per site for off-diagonal
UCR$_Z^{(m-1)}$.
The connection between the position-register binary weights and the
channel-register UCR angles is derived in the following section.

\section{Interaction with the Channel-Register UCR}
\label{app:ucr_factorisation}

The multichannel extension of the REQWIEM circuit requires that each
rotation site carry a {state-dependent} angle: on the on-diagonal
block, the curvature $\kappa_j^{(r)}$ varies across the $M$ diabatic
surfaces; on the off-diagonal block, the coupling strength
$c_{rt}^{(ss')}$ varies across the $M/2$ pairs addressed within each
XOR-class fragment.
This state dependence is encoded by a uniformly controlled rotation
UCR$_Z^{(p)}$ acting at each site, with $p = m$ (on-diagonal) or
$p = m - 1$ (off-diagonal).
The binary weight factor from the position-register
decomposition passes through the UCR compilation as a site-independent
multiplicative constant, so that the UCR acts exclusively on the
coupling-constant vector.

\subsection{UCR Compilation via the M\"ott\"onen Decomposition}
\label{app:ucr_compilation}

A uniformly controlled rotation UCR$_Z^{(p)}$ with $P = 2^p$ input
angles $\{\theta_0, \ldots, \theta_{P-1}\}$ is compiled via the
M\"ott\"onen decomposition~\cite{mottonen12006decompositions} into $P$
rotations $R_z(\phi_k)$ interleaved with $P - 1$ CNOT gates on the
control qubits, following a Gray-code ordering.
The circuit angles $\{\phi_k\}$ are related to the input angles by the
Walsh--Hadamard transform:
\begin{equation}\label{eq:app_walsh_transform}
    \phi_k = \frac{1}{P}\sum_{j=0}^{P-1}
    (-1)^{\langle k,j\rangle}\,\theta_j,
\end{equation}
where $\langle k,j\rangle = \sum_\alpha k_\alpha j_\alpha$ is the
bitwise inner product and the $1/P$ prefactor reflects the non-unitary
convention standard in the M\"ott\"onen
decomposition~\cite{mottonen12006decompositions}.
The computational cost of the UCR ($P$ rotations and $P - 1$
CNOTs) is determined by $P$ alone, independent of the magnitudes of
the~$\phi_k$.

\subsection{Factorization: On-Diagonal Block}
\label{app:ucr_ondiag}

Consider an intra-mode cross-pair site $(\ell, p)$ within the
on-diagonal quadratic term of mode~$r$.
The $M$ state-dependent angles at this site are, from
Eq.~\ref{eq:app_angle_quadratic},
\begin{equation}\label{eq:app_ucr_input_ondiag}
    \theta_{\ell p}^{(j)} = \tau\,\kappa_j^{(r)}\,
    \Delta_r^2\, 2^{\ell+p+1},
    \qquad j = 0, \ldots, M - 1,
\end{equation}
where $\kappa_j^{(r)} = \tfrac{1}{2}\mu_r\omega_r^2 + a_{rr}^{(j)}$
is the full curvature on surface~$j$.
Note that the $j = 0$ entry carries the reference surface's curvature
$\kappa_0^{(r)}$ (generally nonzero), consistent with the folded
formulation of Section~\ref{sec:mux_tier2}.
Substituting into the Walsh
transform~\ref{eq:app_walsh_transform} with $P = M$:
\begin{equation}\label{eq:app_walsh_ondiag}
    \phi_{k,\ell p} = \frac{1}{M}\sum_{j=0}^{M-1}
    (-1)^{\langle k,j\rangle}\,\theta_{\ell p}^{(j)}
    = \underbrace{\frac{1}{M}\left(\sum_{j=0}^{M-1}
      (-1)^{\langle k,j\rangle}\,
      \kappa_j^{(r)}\right)}_{\displaystyle \hat{\kappa}_{r,k}}
      \;\cdot\;
      \underbrace{\tau\,\Delta_r^2\, 2^{\ell+p+1}}_{\displaystyle
        \text{binary weight}},
\end{equation}
where $\hat{\kappa}_{r,k}$ is the $k$-th Walsh coefficient of the
curvature vector
$\boldsymbol{\kappa}_r = (\kappa_0^{(r)}, \ldots,
\kappa_{M-1}^{(r)})$.
The binary weight factor $\tau\,\Delta_r^2\, 2^{\ell+p+1}$ passes
through the Walsh transform as a site-independent multiplicative
constant because it does not depend on the electronic state index~$j$.

The identical factorization holds for every site type.
The following table lists the input angle $\theta^{(j)}$, the
corresponding coupling-constant vector entering the Walsh transform,
and the binary weight for each site:

\begin{center}
\renewcommand{\arraystretch}{1.35}
\begin{tabular}{l l l}
\hline\hline
Site type & Input angle $\theta^{(j)}$ & Binary weight \\
\hline
Constant phase
  & $\theta_{\mathrm{const}}^{(j,r)}$
    (Eq.~\ref{eq:app_constant_phase})
  & (state-dependent; no binary factor) \\[4pt]
Linear ($q_\ell^{(r)}$)
  & $\tau\bigl[a_r^{(j)}\,\Delta_r\,2^\ell
    + \text{diag.\ quad.\ shift}\bigr]$
  & $\tau\,\Delta_r\, 2^\ell$ (leading) \\[4pt]
Intra-mode quad.\
  ($q_\ell^{(r)} q_p^{(r)}$)
  & $\tau\,\kappa_j^{(r)}\,\Delta_r^2\, 2^{\ell+p+1}$
  & $\tau\,\Delta_r^2\, 2^{\ell+p+1}$ \\[4pt]
Bilinear
  ($q_\ell^{(r)} q_p^{(t)}$)
  & $\tau\, a_{rt}^{(j)}\,\Delta_r \Delta_t\, 2^{\ell+p}$
  & $\tau\,\Delta_r \Delta_t\, 2^{\ell+p}$ \\
\hline\hline
\end{tabular}
\end{center}

In every case, the UCR circuit angle factorizes as
\begin{equation}\label{eq:app_angle_factorisation}
    \phi_{k,\boldsymbol\ell} =
    \hat{c}_k \;\times\;
    \tau\, 2^{f(\boldsymbol\ell,\, \mathbf{n})},
\end{equation}
where $\hat{c}_k$ is the $k$-th Walsh coefficient of the
coupling-constant vector (computed with the $1/P$ normalization of
Eq.~\ref{eq:app_walsh_transform}) and
$f(\boldsymbol\ell, \mathbf{n})$ is a linear function of the qubit
indices and register widths.
The Walsh transform acts entirely within the $M$-dimensional space of
electronic states; the $2^{n_r}$-dimensional position register is not
spectral-transformed.

\subsection{Factorization: Off-Diagonal Block (Per Fragment)}
\label{app:ucr_offdiag}

Within XOR-class fragment~$\zeta$, the focusing Clifford $U_\zeta$
reduces the off-diagonal operator to diagonal form with $M/2$ distinct
coupling values
$\{c_{\tilde\sigma}^{(\zeta)}\}_{\tilde\sigma \in \{0,1\}^{m-1}}$,
addressed by the $(m{-}1)$-qubit extended spectator configuration
(Section~\ref{sec:clifford_diag}, Appendix~\ref{app:offdiag}).
The UCR$_Z^{(m-1)}$ has $P = M/2$ input angles per site.

For a bilinear cross-pair site $(\ell, p)$ coupling modes~$r$ and~$t$
within the off-diagonal block, the input angles are
\begin{equation}\label{eq:app_ucr_input_offdiag}
    \theta_{\ell p}^{(\tilde\sigma)}
    = \tau\, c_{rt}^{(\tilde\sigma,\zeta)}\,
    \Delta_r \Delta_t\, 2^{\ell+p},
    \qquad \tilde\sigma = 0, \ldots, M/2 - 1,
\end{equation}
where $c_{rt}^{(\tilde\sigma,\zeta)}$ is the bilinear coupling
constant for the electronic pair addressed by spectator configuration
$\tilde\sigma$ in fragment~$\zeta$.
The Walsh transform with $P = M/2$ yields
\begin{equation}\label{eq:app_walsh_offdiag}
    \phi_{k,\ell p}^{(\zeta)}
    = \underbrace{\frac{1}{M/2}\left(\sum_{\tilde\sigma=0}^{M/2-1}
      (-1)^{\langle k,\tilde\sigma\rangle}\,
      c_{rt}^{(\tilde\sigma,\zeta)}\right)
      }_{\displaystyle \hat{c}_{rt,k}^{(\zeta)}}
      \;\cdot\;
      \underbrace{\tau\,\Delta_r \Delta_t\, 2^{\ell+p}}_{\displaystyle
        \text{binary weight}}.
\end{equation}
The factorization structure is identical to the on-diagonal case: the
binary weight depends only on the position-register site, while the
Walsh coefficient depends only on the coupling constants and the
XOR class.
The Walsh coefficients $\hat{c}_{rt,k}^{(\zeta)}$ correspond to the
classically precomputed UCR angles
$\phi_{\tilde\sigma}^{(\zeta,r,\ell)}$ of
Algorithm~\ref{alg:single_frag} after the binary weight has been
applied.

The target-qubit sign from the CNOT sandwich
(Section~\ref{sec:ucr_offdiag}) is incorporated as follows.
The block-diagonal structure established in
Eq.~\ref{eq:block_diag_main} assigns eigenvalue
$(-1)^{k_{s_1}} c_{\tilde\sigma}^{(\zeta)}$ to each computational
basis state of the channel register, where $s_1$ is the focusing
target qubit.
The outer CNOT pair in the per-site circuit
(Eq.~\ref{eq:ucr_offdiag_site})
\begin{equation}\label{eq:app_cnot_sign}
    \mathrm{CNOT}_{s_1 \to \ell} \;\longrightarrow\;
    \mathrm{UCR}_Z^{(m-1)}(\ell;\, \widetilde{S^c}) \;\longrightarrow\;
    \mathrm{CNOT}_{s_1 \to \ell},
\end{equation}
encodes the sign $(-1)^{k_{s_1}}$ by XORing the target qubit $s_1$
into the position-register qubit $\ell$ before and after the UCR\@.
This effectively replaces the coupling-constant vector entering the
Walsh transform by
$\{(-1)^{k_{s_1}} c_{\tilde\sigma}^{(\zeta)}\}$, absorbing the sign
into the input angles before the binary weight is applied.
The factorization~\ref{eq:app_walsh_offdiag} therefore holds as
written, with the sign already incorporated into
$c_{rt}^{(\tilde\sigma,\zeta)}$.

For bilinear sites in the off-diagonal block, the complete per-site
circuit nests the CNOT sandwich for the $ZZ$ interaction
(Eq.~\ref{eq:app_ZZ_intermode}) inside the CNOT sandwich for the
$Z_{s_1}$ sign (Eq.~\ref{eq:app_cnot_sign}):
\begin{equation}\label{eq:app_bilinear_offdiag_circuit}
    \mathrm{CNOT}_{q_p^{(t)} \to q_\ell^{(r)}} \;\longrightarrow\;
    \mathrm{CNOT}_{s_1 \to q_\ell^{(r)}} \;\longrightarrow\;
    \mathrm{UCR}_Z^{(m-1)}(q_\ell^{(r)};\, \widetilde{S^c})
    \;\longrightarrow\;
    \mathrm{CNOT}_{s_1 \to q_\ell^{(r)}} \;\longrightarrow\;
    \mathrm{CNOT}_{q_p^{(t)} \to q_\ell^{(r)}}.
\end{equation}
The outermost CNOT pair computes
$q_\ell^{(r)} \oplus q_p^{(t)}$ into $q_\ell^{(r)}$, encoding the
bilinear $ZZ$ interaction; the inner CNOT pair computes
$q_\ell^{(r)} \oplus q_p^{(t)} \oplus k_{s_1}$, encoding the
target-qubit sign.
The net effect is a conditioned phase
$e^{-i\theta\, Z_\ell^{(r)} Z_p^{(t)} Z_{s_1} / 2}$, with the angle
$\theta$ selected by the $(m{-}1)$-qubit UCR according to the extended
spectator configuration.
The five-gate sequence of two outer CNOTs, two inner CNOTs, and one
UCR$_Z^{(m-1)}$ is the largest per-site circuit for the KDC Hamiltonian, contributing $M/2$ rotations and $M/2 + 3$ CNOTs per bilinear site per fragment, with zero Toffoli gates.

\section{Comparison with Per-Term CNOT Cascade}
\label{sec:per_term_comparison}

If each multilinear term is implemented independently via a CNOT cascade that reduces $\prod_{j \in S} Z_j$ to a single $Z_j$, followed by an $R_z$ rotation and the reverse cascade, the cost per term is $2(|S| - 1)$ CNOT gates and one $R_z$.
The total CNOT count across all $2^n$ terms is
\begin{equation}\label{eq:total_cnot_naive}
    N_{\mathrm{CNOT}}^{(\mathrm{cascade})} = 2\sum_{k=2}^{n} (k-1)\binom{n}{k} = n\cdot 2^{n} - 2^{n+1} + 2\,,
\end{equation}
scaling as $\mathcal{O}(n \cdot 2^n)$.

\begin{center}
\renewcommand{\arraystretch}{1.3}
\begin{tabular}{l c c c}
\hline\hline
Implementation & CNOT count & $R_z$ count & Scaling \\
\hline
Per-term cascade & $n\cdot 2^n - 2^{n+1} + 2$ & $2^n$ & $\mathcal{O}(n \cdot 2^n)$ \\
Gray-code UCR$_Z^{(n)}$ & $2^n - 1$ & $2^n$ & $\mathcal{O}(2^n)$ \\
\hline
Improvement factor & $\approx n$ & $1$ & $n\times$ \\
\hline\hline
\end{tabular}
\end{center}

The $R_z$ count is identical in both approaches, confirming that the improvement is purely in the CNOT overhead.
The Gray-code UCR achieves the information-theoretic lower bound: $2^n - 1$ CNOTs are the minimum needed to traverse all $2^n$ computational basis states via single-bit transitions.

\chapter{Off-Diagonal Potential: XOR-Class Fragmentation}\label{app:offdiag}

This appendix derives the circuit implementation of the off-diagonal vibronic coupling operator
\begin{equation}\label{eq:Vod_def_app}
    V_{\text{od}} = \sum_{\substack{i,j=0 \\ i<j}}^{M-1} c_{ij}(\mathbf{x})\,G_{ij},
    \qquad
    G_{ij} \;=\; |i\rangle\langle j| + |j\rangle\langle i|,
\end{equation}
where $M = 2^m$ is the number of diabatic electronic states encoded in $m = \lceil\log_2 M\rceil$ channel qubits, $d$ vibrational modes are each discretized on $n$-qubit grids, and $c_{ij}(\mathbf{x})$ are position-dependent coupling functions.
The treatment proceeds through six stages: XOR-class partitioning of the coupling pairs (Section~\ref{app:xor_partition}), spectator-qubit structure of individual generators (Section~\ref{app:spectator}), per-fragment Clifford diagonalization (Section~\ref{app:clifford_diag}), the fan-out protocol within each fragment (Section~\ref{app:fanout_frag}), the second-order symmetric Trotter integration that preserves interferometric population transfer (Section~\ref{app:symmetric_trotter}), and the treatment of non-power-of-$2$ electronic states (Section~\ref{app:padding}).
Exact gate counts and circuit depths are collected in Section~\ref{app:offdiag_resources}.

\subsection{XOR-Class Partition}
\label{app:xor_partition}

For each nonzero $m$-bit string $\zeta \in \{1,\ldots,M-1\}$, the {XOR class} $C_\zeta$ is the set of ordered pairs
\begin{equation}\label{eq:Czeta_def}
    C_\zeta \;=\; \bigl\{(i,j) : 0 \le i < j \le M-1,\; i \oplus j = \zeta\bigr\},
\end{equation}
where $\oplus$ denotes bitwise XOR.
The {support} of $\zeta$ is $S(\zeta) = \{q \in \{0,\ldots,m-1\} : \zeta_q = 1\}$, with Hamming weight $h(\zeta) = |S(\zeta)| = \operatorname{popcount}(\zeta)$.
The complement $S^c(\zeta) = \{0,\ldots,m-1\}\setminus S(\zeta)$ has cardinality $\kappa(\zeta) = m - h(\zeta)$ and is called the {spectator set}.

The $\binom{M}{2}$ coupling pairs partition into $M-1$ disjoint XOR classes:
\begin{equation}\label{eq:pair_partition}
    \bigl\{(i,j) : 0 \le i < j \le M-1\bigr\} \;=\; \bigsqcup_{\zeta=1}^{M-1} C_\zeta.
\end{equation}
Each class has exactly $|C_\zeta| = M/2 = 2^{m-1}$ pairs.

This is shown via the following argument. 
For any $\zeta \neq 0$, the map $i \mapsto i \oplus \zeta$ is a fixed-point-free involution on $\{0,\ldots,M-1\}$, partitioning the $M$ states into $M/2$ transpositions $\{i, i\oplus\zeta\}$.
Each transposition yields exactly one ordered pair $(i,j)$ with $i < j$, so $|C_\zeta| = M/2$.
Summing: $(M-1)\cdot M/2 = \binom{M}{2}$.

The {off-diagonal fragment Hamiltonian} for class $\zeta$ is
\begin{equation}\label{eq:Hzeta_def}
    H_\zeta \;=\; \sum_{(i,j)\in C_\zeta} c_{ij}(\mathbf{x})\,G_{ij}.
\end{equation}
By the partition property, $V_{\text{od}} = \sum_{\zeta=1}^{M-1} H_\zeta$.

\section{Spectator-Qubit Structure of Individual Generators}
\label{app:spectator}

Let $(i,j) \in C_\zeta$ with $\zeta = i\oplus j$, support $S = S(\zeta)$, and spectator set $S^c$.
Write $\sigma = i|_{S^c} = j|_{S^c}$ for the common spectator configuration.
Then
\begin{equation}\label{eq:Gij_decomposition}
    G_{ij} \;=\; |\sigma\rangle\langle\sigma|_{S^c} \;\otimes\; \bigotimes_{q \in S} X_q,
\end{equation}
where the projector acts on the spectator qubits and the $X$ product acts on the support qubits.

Since $i_q = j_q$ for all $q \in S^c$ and $i_q \neq j_q$ for $q \in S$, the transition $|i\rangle\langle j|$ factors as $|\sigma\rangle\langle\sigma|_{S^c} \otimes |i_S\rangle\langle j_S|$ where $j_S = i_S \oplus \mathbf{1}_S$ (all support bits flipped).
Adding the Hermitian conjugate, $|i_S\rangle\langle j_S| + |j_S\rangle\langle i_S|$ is the operator that swaps $|i_S\rangle \leftrightarrow |j_S\rangle$.
Since $|S|$-qubit states $|i_S\rangle$ and $|j_S\rangle = |i_S \oplus \mathbf{1}_S\rangle$ differ on every bit, the swap is $\bigotimes_{q\in S} X_q$ restricted to the two-dimensional subspace $\{|i_S\rangle, |j_S\rangle\}$.
But $\bigotimes_{q\in S} X_q$ acts identically on {all} computational basis states of the support qubits (it flips every bit in $S$), and $|\sigma\rangle\langle\sigma|_{S^c}$ projects onto the correct spectator subspace.

The fragment Hamiltonian decomposes as
\begin{equation}\label{eq:Hzeta_structure}
    H_\zeta \;=\; \underbrace{\left[\sum_{\sigma \in \{0,1\}^{\kappa(\zeta)}} c_\sigma(\mathbf{x})\,|\sigma\rangle\langle\sigma|_{S^c}\right]}_{\text{spectator-addressed coupling}} \;\otimes\; \underbrace{\bigotimes_{q \in S(\zeta)} X_q}_{\equiv\; X_S},
\end{equation}
where $c_\sigma(\mathbf{x}) = c_{i(\sigma),\,j(\sigma)}(\mathbf{x})$ is the coupling function for the unique pair $(i,j) \in C_\zeta$ whose spectator bits are $\sigma$.

If all $c_\sigma$ were equal, the spectator sum would reduce to $I_{S^c}$ and $H_\zeta = c(\mathbf{x})\,X_S$, which conjugates under $H^{\otimes m}$ to $c(\mathbf{x})\,Z_S$ (diagonal).
For non-uniform $c_\sigma$, the spectator-addressed operator is
\begin{equation}\label{eq:spectator_hadamard}
    H_{S^c}\!\left[\sum_\sigma c_\sigma |\sigma\rangle\langle\sigma|\right]\!H_{S^c}
    \;=\;
    \frac{1}{2^{\kappa}}\sum_{\mu \in \{0,1\}^{\kappa}} \left(\sum_\sigma (-1)^{\sigma\cdot\mu}\,c_\sigma\right) \bigotimes_{q \in S^c}\!\sigma_q^{(\mu_q)},
\end{equation}
where $\sigma_q^{(0)} = I_q$ and $\sigma_q^{(1)} = X_q$.
The terms with $\mu \neq \mathbf{0}$ contain Pauli $X$ operators on spectator qubits and are off-diagonal in the computational basis.
Hence $H^{\otimes m} V_{\text{od}}\,H^{\otimes m}$ is {not} diagonal for generic coupling functions when $M > 2$.

The support is $S = \{0\}$ (qubit 0), the spectator is $S^c = \{1\}$ (qubit 1), and $\kappa = 1$.
The two pairs are $(0,1)$ with $\sigma = 0$ and $(2,3)$ with $\sigma = 1$.
The fragment Hamiltonian is
\begin{equation}
    H_{01} = c_{01}\,|0\rangle\langle 0|_1 \otimes X_0 \;+\; c_{23}\,|1\rangle\langle 1|_1 \otimes X_0.
\end{equation}
Conjugating by $H^{\otimes 2}$:
\begin{align}
    H^{\otimes 2}\,H_{01}\,H^{\otimes 2}
    &= \frac{c_{01}+c_{23}}{2}\,I_1 \otimes Z_0 \;+\; \frac{c_{01}-c_{23}}{2}\,X_1\otimes Z_0. \label{eq:M4_zeta01_conjugated}
\end{align}
The second term is off-diagonal whenever $c_{01} \neq c_{23}$.

\subsection{Per-Fragment Clifford Diagonalization}\label{app:clifford_diag}

Since the global Hadamard does not diagonalize $V_{\text{od}}$, each fragment $H_\zeta$ is diagonalized individually by a fragment-specific Clifford circuit $U_\zeta$.

\subsubsection{Focusing Operation}
\label{app:focusing}

For a class $\zeta$ with support $S = \{s_1, s_2, \ldots, s_{h}\}$ ordered so that $s_1$ is the {target} qubit, the focusing Clifford is
\begin{equation}\label{eq:Uzeta_def}
    U_\zeta \;=\; H_{s_1}\;\prod_{k=2}^{h(\zeta)} \mathrm{CNOT}_{s_1 \to s_k},
\end{equation}
where the CNOT cascade has qubit $s_1$ as {control} and each $s_k$ ($k \geq 2$) as target, and $H_{s_1}$ is the Hadamard on the target qubit.
The order of the CNOT product is right-to-left.

Under conjugation by $\mathrm{CNOT}_{c \to t}$, the Pauli operators transform as $X_c \mapsto X_c X_t$ and $X_t \mapsto X_t$.
Consequently, the product $X_S = \prod_{q \in S}X_q$ is mapped as follows: each $\mathrm{CNOT}_{s_1 \to s_k}$ sends $X_{s_1} \to X_{s_1}X_{s_k}$, so the product $X_{s_1}\prod_{k\ge2}X_{s_k}$ is replaced by $X_{s_1}X_{s_k}\cdot X_{s_k}\cdot\prod_{j\neq 1,k}X_{s_j} = X_{s_1}\prod_{j\neq k}X_{s_j}$.
After all $(h-1)$ CNOTs, the multi-qubit flip $X_S$ is reduced to $X_{s_1}$.
The reversed direction ($s_k \to s_1$) would spread $X_{s_1}$ to all support qubits rather than concentrating it, and would fail to diagonalize the fragment.

Under the focusing Clifford, the non-target support qubits $s_2, \ldots, s_h$ become additional spectators.
Defining the {extended spectator configuration} $\tilde\sigma(i) = \bigl(\sigma(i),\;\alpha(i)\bigr)$ where $\sigma(i) = i|_{S^c}$ are the original spectator bits and $\alpha(i)_k = i_{s_k}\oplus i_{s_1}$ for $k = 2,\ldots,h$ are the new spectator bits, the transformed fragment is:
\begin{equation}\label{eq:block_diag}
    U_\zeta\,H_\zeta\,U_\zeta^\dagger
    \;=\;
    \left[\sum_{(i,j)\in C_\zeta} c_{ij}(\mathbf{x})\;|\tilde\sigma(i)\rangle\langle\tilde\sigma(i)|_{\tilde{S^c}}\right]
    \;\otimes\;
    Z_{s_1},
\end{equation}
where $\tilde{S^c} = \{0,\ldots,m-1\}\setminus\{s_1\}$ comprises all channel qubits except the target.
This operator is diagonal in the computational basis of all $m$ channel qubits.

Consider a single generator $G_{ij} = |\sigma\rangle\langle\sigma|_{S^c}\otimes(|i_S\rangle\langle j_S| + |j_S\rangle\langle i_S|)$ where $j_S = i_S \oplus \mathbf{1}_S$.

\emph{Step 1 (CNOT focusing).}
Each $\mathrm{CNOT}_{s_1 \to s_k}$ XORs the value of bit $s_1$ into bit $s_k$.
After the full cascade, support bits of $|i_S\rangle$ become: target bit $s_1$ retains $i_{s_1}$; each non-target bit $s_k$ becomes $i_{s_k}\oplus i_{s_1}$.
Similarly for $|j_S\rangle$: target bit becomes $j_{s_1} = i_{s_1}\oplus 1$; each non-target bit becomes $j_{s_k}\oplus j_{s_1} = (i_{s_k}\oplus 1)\oplus(i_{s_1}\oplus 1) = i_{s_k}\oplus i_{s_1}$.

Non-target support bits of $|i_S\rangle$ and $|j_S\rangle$ are now {identical}: both equal $\alpha(i)_k = i_{s_k}\oplus i_{s_1}$.
Only the target bit differs.
Hence after the CNOT cascade:
\begin{equation}
    G_{ij}' = |\sigma\rangle\langle\sigma|_{S^c}\;\otimes\;|\alpha(i)\rangle\langle\alpha(i)|_{S\setminus\{s_1\}}\;\otimes\; X_{s_1}.
\end{equation}

\emph{Step 2 (Hadamard on target).}
$H_{s_1}\,X_{s_1}\,H_{s_1} = Z_{s_1}$.

\emph{Step 3 (Combining).}
Different pairs $(i,j) \in C_\zeta$ yield distinct extended spectator configurations $\tilde\sigma(i)$, so their projectors are mutually orthogonal.
Summing over all pairs gives Eq.~\ref{eq:block_diag}.

In the computational basis $|k\rangle = |k_{m-1}\cdots k_0\rangle$ of the channel register,
\begin{equation}\label{eq:diagonal_eigenvalue}
    \langle k|\;U_\zeta\,H_\zeta\,U_\zeta^\dagger\;|k\rangle
    \;=\;
    (-1)^{k_{s_1}}\;c_{\tilde\sigma(k)}(\mathbf{x}),
\end{equation}
where $\tilde\sigma(k)$ determines the unique pair $(i,j) \in C_\zeta$ addressed by the extended spectator configuration, and $c_{\tilde\sigma(k)}$ is its coupling function.
The $m-1$ non-target bits of $k$ select which pair, and the target bit $k_{s_1}$ provides the sign $\pm 1$.

Regardless of the Hamming weight $h(\zeta)$, the diagonalized fragment $U_\zeta H_\zeta U_\zeta^\dagger$ is addressed by $m - 1$ spectator qubits.
The $\mathrm{UCR}_Z^{(m-1)}$ that implements the diagonal evolution therefore has a uniform cost of $M/2$ rotations per site across all fragments.

\subsubsection{Circuit Structure of $U_\zeta$}
\label{app:Uzeta_circuit}

The focusing Clifford $U_\zeta$ consists of:
\begin{enumerate}
    \item $h(\zeta) - 1$ CNOT gates (the focusing cascade), each targeting qubit $s_1$,
    \item $1$ Hadamard gate on qubit $s_1$.
\end{enumerate}
The CNOT depth is $h(\zeta)-1$ (sequential, since all share the same target).
The total depth of $U_\zeta$ is $h(\zeta)$.
The adjoint $U_\zeta^\dagger$ has the same cost (reverse the gate order, noting $H^\dagger = H$ and $\mathrm{CNOT}^\dagger = \mathrm{CNOT}$).

\subsection{Per-Fragment Exponentiation with Fan-Out}
\label{app:fanout_frag}

After the Clifford $U_\zeta$, the fragment Hamiltonian is diagonal in the channel register's computational basis.
The channel register serves as a {read-only control} for position-dependent phase rotations on the vibrational register, enabling fan-out parallelization across the $d$ vibrational modes.

\subsubsection{Algorithm for a Single Fragment}
\label{app:single_frag_alg}

\begin{algorithm}
\textbf{Input:} Fragment index $\zeta$, timestep $\tau$, channel register $\mathcal{C}$ ($m$ qubits), vibrational registers $\mathcal{V}_1, \ldots, \mathcal{V}_d$ ($n$ qubits each).

\begin{enumerate}
    \item \textbf{Clifford focusing} ($U_\zeta$):
    \begin{enumerate}
        \item Apply $\mathrm{CNOT}_{s_1 \to s_k}$ for $k = h(\zeta), h(\zeta)-1, \ldots, 2$.
        \item Apply $H_{s_1}$.
    \end{enumerate}
    Depth: $h(\zeta)$. CNOTs: $h(\zeta)-1$.

    \item \textbf{Fan-out} the channel register $\mathcal{C}$ to $d$ copies $\mathcal{C}^{(1)}, \ldots, \mathcal{C}^{(d)}$ via a binary CNOT tree.

    Depth: $\lceil\log_2 d\rceil$. CNOTs: $m(d-1)$.

    \item \textbf{Mode-parallel UCR circuits:} For each mode $r = 1,\ldots,d$ in parallel, implement the controlled phase
    \begin{equation}\label{eq:per_mode_phase}
        \exp\!\Big(-i\tau\;\langle k^{(r)}|\,U_\zeta H_\zeta U_\zeta^\dagger\,|k^{(r)}\rangle\Big)
        = \exp\!\Big(-i\tau\,(-1)^{k_{s_1}^{(r)}}\,c_{\sigma(k^{(r)})}(x_r)\Big),
    \end{equation}
    using copy $\mathcal{C}^{(r)}$ as read-only control.

    The implementation uses $\mathrm{UCR}_Z$ gates:
    \begin{enumerate}
        \item For each rotation site, apply factored circuit: the outer CNOT pair (if cross-term) creates the $R_{ZZ}$ interaction, the inner CNOT pair encodes the sign $(-1)^{k_{s_1}}$, and the $\mathrm{UCR}_Z^{(m-1)}$ is controlled by $m-1$ extended spectator qubits from $\mathcal{C}^{(r)}$.
        \item Each $\mathrm{UCR}_Z^{(m-1)}$ decomposes into $2^{m-1}$ rotations, yielding $M/2$ rotation angles per site.
    \end{enumerate}

    \item \textbf{Fan-in} the $d$ copies back to $\mathcal{C}$ (reverse CNOT tree).

    Depth: $\lceil\log_2 d\rceil$. CNOTs: $m(d-1)$.

    \item \textbf{Unfocusing} ($U_\zeta^\dagger$):
    \begin{enumerate}
        \item Apply $H_{s_1}$.
        \item Apply $\mathrm{CNOT}_{s_1 \to s_k}$ for $k = 2, 3, \ldots, h(\zeta)$.
    \end{enumerate}
    Depth: $h(\zeta)$. CNOTs: $h(\zeta)-1$.
\end{enumerate}
\end{algorithm}

\subsubsection{UCR Structure Within a Fragment}
\label{app:ucr_frag}

After the Clifford $U_\zeta$, the diagonal operator on the channel register for fragment $\zeta$ at position $\mathbf{x}$ takes the form:
\begin{equation}\label{eq:diag_operator_frag}
    D_\zeta(\mathbf{x}) = \sum_{k=0}^{M-1} |k\rangle\langle k|\;\cdot\;(-1)^{k_{s_1}}\;c_{\tilde\sigma(k)}(\mathbf{x}).
\end{equation}
The eigenvalue has a factored structure: the target bit $k_{s_1}$ contributes a sign $\pm 1$, and the $m-1$ non-target bits select one of $M/2 = 2^{m-1}$ coupling functions $c_{\tilde\sigma}$.
This factored structure enables an efficient implementation.

The phase $\exp(-i\tau\,(-1)^{k_{s_1}}\,c_{\tilde\sigma(k)}\,q_\ell^{(r)}\,q_{\ell'}^{(r)})$ for a two-qubit (cross-term) site $(\ell,\ell')$ is implemented as:
\begin{equation}\label{eq:factored_ucr_circuit}
    \mathrm{CNOT}_{\ell'\to\ell}\;\;\mathrm{CNOT}_{s_1\to\ell}\;\;\mathrm{UCR}_Z^{(m-1)}(\ell;\;\tilde{S^c})\;\;\mathrm{CNOT}_{s_1\to\ell}\;\;\mathrm{CNOT}_{\ell'\to\ell},
\end{equation}
where $\mathrm{UCR}_Z^{(m-1)}$ is a uniformly controlled rotation on qubit $\ell$ with $m-1$ control qubits from the extended spectator set $\tilde{S^c} = \{0,\ldots,m-1\}\setminus\{s_1\}$.
The outer CNOT pair creates $q_\ell q_{\ell'}$ interaction (standard $R_{ZZ}$ sandwich), and the inner CNOT pair creates $Z_{s_1}\,Z_\ell$ interaction that encodes the sign $(-1)^{k_{s_1}}$.

For single-qubit (linear) sites, only the inner CNOT pair and UCR are needed:
\begin{equation}\label{eq:factored_ucr_single}
    \mathrm{CNOT}_{s_1\to\ell}\;\;\mathrm{UCR}_Z^{(m-1)}(\ell;\;\tilde{S^c})\;\;\mathrm{CNOT}_{s_1\to\ell},
\end{equation}

Since the extended spectator set always has $m-1$ qubits regardless of $\zeta$, the cost per site is {independent of $h(\zeta)$}:
\begin{enumerate}
    \item Cross-term (two-qubit) site: $M/2 = 2^{m-1}$ rotations, $2^{m-1}+3$ CNOTs, depth $2^m + 3$.
    \item Single-qubit site: $2^{m-1}$ rotations, $2^{m-1}+1$ CNOTs, depth $2^m + 1$.
\end{enumerate}
The depth per site follows from the Gray-code $\mathrm{UCR}_Z^{(m-1)}$ decomposition: $2^{m-1}$ rotations interleaved with $2^{m-1}-1$ CNOTs contribute depth $2(2^{m-1}) - 1 = 2^m - 1$; the outer CNOT pair (cross-term) or inner CNOT pair (single-qubit) adds $+4$ or $+2$ respectively.

\subsubsection{Total UCR Rotation Count Per Fragment}
\label{app:rotation_count_frag}

For the general polynomial vibronic coupling model with $d$ modes on $n$-qubit grids, each pair generates $d(1+n) + d\binom{n}{2} + \binom{d}{2}n^2$ rotation sites in total.
For the linear vibronic model ($c_{ij}$ linear in $\mathbf{x}$), only $|\mathcal{G}_3| + |\mathcal{G}_4|$ modes carry nonzero coupling gradients, generating $|\mathcal{G}_{\text{od}}|\cdot n$ single-qubit sites per pair.

The total rotation count for fragment $\zeta$ across all $d$ modes is independent of $\zeta$:
\begin{equation}\label{eq:Rzeta}
    R_\zeta = 2^{m-1}\left[d\,(1+n) + d\binom{n}{2} + \binom{d}{2}n^2\right] = \frac{M}{2}\,S_{\text{sites}},
\end{equation}
where $S_{\text{sites}} = d(1+n) + d\binom{n}{2} + \binom{d}{2}n^2$ is the total number of rotation sites per pair.

For the linear vibronic model:
\begin{equation}\label{eq:Rzeta_linear}
    R_\zeta^{(\text{lin})} = 2^{m-1}\cdot |\mathcal{G}_{\text{od}}|\cdot n = \frac{M}{2}\,|\mathcal{G}_{\text{od}}|\,n.
\end{equation}

\subsection{Second-Order Symmetric Trotter Decomposition}
\label{app:symmetric_trotter}

\subsubsection{Motivation: Trotter Order Preservation}
\label{app:trotter_order}

The off-diagonal potential is the sum of $M-1$ non-commuting fragments:
\begin{equation}
    V_{\text{od}} = \sum_{\zeta=1}^{M-1} H_\zeta.
\end{equation}
A first-order product formula $\prod_\zeta e^{-iH_\zeta\tau}$ incurs error $\mathcal{O}(\tau^2)$, which would dominate the $\mathcal{O}(\tau^3)$ error of the global second-order Trotter step.
To maintain overall second-order accuracy, the fragments must be symmetrized.

The second-order symmetric decomposition of $e^{-iV_{\text{od}}\tau}$ is
\begin{equation}\label{eq:symmetric_product}
    \mathcal{V}_2(\tau) \;=\; \prod_{\zeta=1}^{M-1} e^{-iH_\zeta\tau/2}\;\cdot\;\prod_{\zeta=M-1}^{1} e^{-iH_\zeta\tau/2},
\end{equation}
satisfying $\|\mathcal{V}_2(\tau) - e^{-iV_{\text{od}}\tau}\| = \mathcal{O}(\tau^3)$.

The full second-order Trotter step for the complete Hamiltonian $H = T + V_{\text{d}} + V_{\text{od}}$ is:
\begin{equation}\label{eq:full_trotter_step}
    \mathcal{U}(\tau) = e^{-iT\tau/2}\,e^{-iV_{\text{d}}\tau/2}\,\mathcal{V}_2(\tau)\,e^{-iV_{\text{d}}\tau/2}\,e^{-iT\tau/2} + \mathcal{O}(\tau^3).
\end{equation}
Alternatively, each $H_\zeta$ may be promoted to a separate term in the global symmetric product over all $(M+1)$ terms $\{T, V_{\text{d}}, H_1, \ldots, H_{M-1}\}$.

\subsubsection{Circuit for One Trotter Step}
\label{app:trotter_circuit}

The off-diagonal contribution to one Trotter step consists of $2(M-1)$ fragment applications (each at half-angle):

\begin{algorithm}
\textbf{Off-diagonal block of one Trotter step:}
\begin{enumerate}
    \item \textbf{Forward sweep} ($\zeta = 1, 2, \ldots, M-1$): For each $\zeta$ in order, execute the single-fragment algorithm (Section~\ref{app:single_frag_alg}) with timestep $\tau/2$.
    \item \textbf{Backward sweep} ($\zeta = M-1, M-2, \ldots, 1$): For each $\zeta$ in reverse order, execute the single-fragment algorithm with timestep $\tau/2$.
\end{enumerate}
\end{algorithm}

\subsection{Verification of Interferometric Population Transfer}
\label{app:mz_verification}

This subsection verifies explicitly that the per-fragment protocol produces the correct channel-population dynamics.

\subsubsection{Population Transfer Kernel}
\label{app:pop_transfer_general}

Given the full quantum state $|\Psi(t)\rangle$ on channel $\otimes$ vibrational registers, the population of diabatic state $a$ is
\begin{equation}
    P_a(t) = \operatorname{Tr}_{\text{vib}}\!\left[\langle a|\,\rho(t)\,|a\rangle\right], \qquad \rho(t) = |\Psi(t)\rangle\langle\Psi(t)|,
\end{equation}

For a single fragment $\zeta$ applied for time $\tau$, the unitary $e^{-iH_\zeta\tau}$ transfers population between states within each pair $(i,j) \in C_\zeta$, leaving all other states unchanged.
Specifically, if $\zeta = i \oplus j$, the transition amplitude between diabatic states $a$ and $b$ due to fragment $\zeta$ alone is:
\begin{equation}\label{eq:transition_amplitude}
    \langle b|\,e^{-iH_\zeta\tau}\,|a\rangle_{\text{chan}} =
    \begin{cases}
    \cos\bigl(c_{a,a\oplus\zeta}(\mathbf{x})\,\tau\bigr)\,\delta_{ab} - i\sin\bigl(c_{a,a\oplus\zeta}(\mathbf{x})\,\tau\bigr)\,\delta_{b,a\oplus\zeta}, & \text{if } a\oplus b \in \{0, \zeta\}, \\[4pt]
    \delta_{ab}, & \text{otherwise.}
    \end{cases}
\end{equation}

The fragment Hamiltonian $H_\zeta = \sum_\sigma c_\sigma |\sigma\rangle\langle\sigma|_{S^c} \otimes X_S$ is block-diagonal in the spectator basis.
For each spectator configuration $\sigma$, the restriction to the $2$-dimensional subspace $\{|i(\sigma)\rangle, |j(\sigma)\rangle\}$ is $c_\sigma X_S$, which generates the rotation $\exp(-ic_\sigma X_S \tau) = \cos(c_\sigma\tau)\,I - i\sin(c_\sigma\tau)\,X_S$.
Within this subspace, $X_S$ swaps $|i(\sigma)\rangle \leftrightarrow |j(\sigma)\rangle$, producing the claimed Rabi-like transfer.
States outside all such $2$-dimensional subspaces (i.e., those with $a\oplus b \notin \{0,\zeta\}$) are unaffected.

The population transfer from state $a$ to state $b = a\oplus\zeta$ due to fragment $\zeta$ at time $\tau$ is:
\begin{equation}\label{eq:pop_transfer}
    \Delta P_{a \to b}^{(\zeta)} = \sin^2\!\bigl(c_{ab}(\mathbf{x})\,\tau\bigr),
\end{equation}
integrated over the vibrational coordinate with appropriate wavepacket weighting.
This is the standard Mach-Zehnder interference pattern, with the coupling function $c_{ab}$ controlling the beamsplitter reflectivity.

\subsubsection{Equivalence of Clifford-Sandwiched Circuit}
\label{app:clifford_equivalence}

The circuit implementation
\begin{equation}\label{eq:circuit_implementation}
    U_\zeta^\dagger \;\exp\!\bigl(-i\,U_\zeta H_\zeta U_\zeta^\dagger\,\tau\bigr)\;U_\zeta
    \;=\;
    \exp(-iH_\zeta\tau),
\end{equation}
produces {exactly} the same population transfer as the direct exponentiation of $H_\zeta$.
Moreover, the operator $U_\zeta H_\zeta U_\zeta^\dagger$ is diagonal in the computational basis, so $\exp(-iU_\zeta H_\zeta U_\zeta^\dagger\,\tau)$ can be implemented exactly using $\mathrm{UCR}_Z$ gates with no approximation.

The first equality is the spectral mapping theorem: for any unitary $U$ and Hermitian $A$, $U^\dagger e^{-iUAU^\dagger \tau} U = e^{-iA\tau}$.
Since the operator is diagonal, its exponential is the entrywise exponential of the diagonal, which is realized exactly by $Z$-rotations controlled on the channel qubits.
The Clifford conjugation acts only on the channel register and commutes with the partial trace over the vibrational register, so any observable computed from the vibrational-register reduced density matrix is identical whether computed from $e^{-iH_\zeta\tau}|\Psi\rangle$ or from the circuit output.
After the unfocusing Clifford $U_\zeta^\dagger$, the channel register returns to the diabatic computational basis; each fragment's Clifford sandwich is self-contained and no Clifford frame propagation is required between fragments.

\subsubsection{Composite Population Transfer After All Fragments}
\label{app:composite_transfer}

After the full symmetric product $\mathcal{V}_2(\tau)$, the channel register returns to the diabatic basis (every $U_\zeta^\dagger$ undoes its $U_\zeta$).
The composite propagator is:
\begin{equation}\label{eq:composite_propagator}
    \mathcal{V}_2(\tau) = \prod_{\zeta=1}^{M-1} e^{-iH_\zeta\tau/2} \cdot \prod_{\zeta=M-1}^{1} e^{-iH_\zeta\tau/2} = e^{-iV_{\text{od}}\tau} + \mathcal{O}(\tau^3).
\end{equation}

The diabatic populations after one Trotter step $\mathcal{U}(\tau)$ satisfy:
\begin{equation}\label{eq:pop_update}
    P_a(\tau) = P_a(0) - \tau^2\sum_{b\neq a}\int|c_{ab}(\mathbf{x})|^2\bigl(|\psi_a(\mathbf{x})|^2-|\psi_b(\mathbf{x})|^2\bigr)\,d\mathbf{x} + \mathcal{O}(\tau^3),
\end{equation}
where the sum runs over all states $b$ coupled to $a$.
This agrees with exact short-time expansion of $e^{-iV_{\text{od}}\tau}$ to second order.

The symmetric product reproduces the ideal evolution to second order, then extracts the population update.

\emph{Step 1 (Symmetric product accuracy).}
Expand each fragment exponential to second order:
\begin{equation}
    e^{-iH_\zeta\tau/2} = I - i\frac{\tau}{2}H_\zeta - \frac{\tau^2}{8}H_\zeta^2 + \mathcal{O}(\tau^3).
\end{equation}
The forward product gives:
\begin{equation}\label{eq:forward_product}
    \prod_{\zeta=1}^{M-1}e^{-iH_\zeta\tau/2} = I - i\frac{\tau}{2}V_{\text{od}} - \frac{\tau^2}{8}\sum_\zeta H_\zeta^2 - \frac{\tau^2}{4}\sum_{\zeta<\zeta'}H_\zeta H_{\zeta'} + \mathcal{O}(\tau^3),
\end{equation}
where the coefficient $\tau^2/8$ on $\sum_\zeta H_\zeta^2$ arises from the second-order term of each individual exponential, while the cross-term coefficient $\tau^2/4$ arises from the product of first-order contributions of distinct factors: $(-i\tau/2)H_\zeta \cdot (-i\tau/2)H_{\zeta'} = -\tau^2 H_\zeta H_{\zeta'}/4$ for $\zeta < \zeta'$.
The backward product similarly gives:
\begin{equation}\label{eq:backward_product}
    \prod_{\zeta=M-1}^{1}e^{-iH_\zeta\tau/2} = I - i\frac{\tau}{2}V_{\text{od}} - \frac{\tau^2}{8}\sum_\zeta H_\zeta^2 - \frac{\tau^2}{4}\sum_{\zeta<\zeta'}H_{\zeta'} H_\zeta + \mathcal{O}(\tau^3),
\end{equation}
where the reversed ordering produces $H_{\zeta'}H_\zeta$ rather than $H_\zeta H_{\zeta'}$.
Multiplying forward $\times$ backward and collecting terms through $\mathcal{O}(\tau^2)$, three contributions arise at second order: the second-order terms $A_2$ and $B_2$ from each factor individually, plus the product $A_1 B_1$ of their first-order terms:
\begin{align}
    \mathcal{V}_2(\tau)
    &= I - i\tau\,V_{\text{od}}
    \;\underbrace{- \frac{\tau^2}{4}\sum_\zeta H_\zeta^2
    - \frac{\tau^2}{4}\sum_{\zeta<\zeta'}\bigl(H_\zeta H_{\zeta'} + H_{\zeta'}H_\zeta\bigr)}_{A_2 + B_2}
    \;\underbrace{- \frac{\tau^2}{4}\,V_{\text{od}}^2}_{A_1 B_1}
    + \mathcal{O}(\tau^3) \notag\\
    &= I - i\tau\,V_{\text{od}}
    - \frac{\tau^2}{2}\,V_{\text{od}}^2
    + \mathcal{O}(\tau^3) \label{eq:S2_expansion}\\
    &= e^{-iV_{\text{od}}\tau} + \mathcal{O}(\tau^3). \notag
\end{align}
where the second equality uses $V_{\text{od}}^2 = \sum_\zeta H_\zeta^2 + \sum_{\zeta \neq \zeta'} H_\zeta H_{\zeta'}$ and $\sum_{\zeta<\zeta'}(H_\zeta H_{\zeta'} + H_{\zeta'}H_\zeta) = \sum_{\zeta \neq \zeta'} H_\zeta H_{\zeta'}$: the $A_2 + B_2$ contribution equals $-(\tau^2/4)\,V_{\text{od}}^2$, combining with the $A_1 B_1 = (-i\tau/2)^2 V_{\text{od}}^2 = -(\tau^2/4)\,V_{\text{od}}^2$ term to give $-(\tau^2/2)\,V_{\text{od}}^2$.
The commutator part $\sum_{\zeta<\zeta'}[H_\zeta, H_{\zeta'}]$, which distinguishes the forward and backward products, drops out of the symmetric combination: $H_\zeta H_{\zeta'} + H_{\zeta'} H_\zeta = \{H_\zeta, H_{\zeta'}\}$ is symmetric under exchange.

\emph{Step 2 (Population update).}
The ideal evolution of the state $|\Psi\rangle = \sum_s |s\rangle|\psi_s\rangle$ gives:
\begin{equation}
    |\Psi(\tau)\rangle = \left(I - i\tau V_{\text{od}} - \frac{\tau^2}{2}V_{\text{od}}^2 + \mathcal{O}(\tau^3)\right)|\Psi\rangle.
\end{equation}
The first-order contribution to $P_a$ vanishes because $\langle a|V_{\text{od}}|a\rangle = 0$ (the off-diagonal potential has no diagonal matrix elements).
The second-order contribution is:
\begin{align}
    P_a(\tau) - P_a(0)
    &= -\tau^2\,\mathrm{Re}\!\left[\langle\Psi|\, V_{\text{od}}\,|a\rangle\langle a|\, V_{\text{od}}\,|\Psi\rangle\right] + \mathcal{O}(\tau^3) \notag\\
    &= -\tau^2\sum_{b \neq a}\int |c_{ab}(\mathbf{x})|^2\bigl(|\psi_a(\mathbf{x})|^2 - |\psi_b(\mathbf{x})|^2\bigr)\,d\mathbf{x} + \mathcal{O}(\tau^3), \label{eq:pop_second_order}
\end{align}
where $\langle b|V_{\text{od}}|a\rangle = c_{ab}(\mathbf{x})$ for $a \neq b$ (real coupling).
Since $\mathcal{V}_2(\tau) = e^{-iV_{\text{od}}\tau} + \mathcal{O}(\tau^3)$ by Step 1, and the per-fragment Clifford conjugations are each exact, the circuit reproduces the same population update to second order.
The only error source is the inter-fragment Trotter splitting, which enters at $\mathcal{O}(\tau^3)$ by the palindromic cancellation established in Eq.~\ref{eq:S2_expansion}.

\paragraph{Consistency check: $M = 2$.}
For $M = 2$ ($m = 1$), there is a single XOR class $\zeta = 1$ with $h = 1$.
The focusing Clifford reduces to a single Hadamard gate (no CNOT cascade, since $|S| = 1$).
The symmetric product has $2(M-1) = 2$ fragment applications, but since there is only one fragment, the forward and backward passes compose trivially: $e^{-iH_1\tau/2} \cdot e^{-iH_1\tau/2} = e^{-iH_1\tau} = e^{-iV_{\text{od}}\tau}$ (exact, with no inter-fragment Trotter error).
The UCR$_Z^{(m-1)} = \mathrm{UCR}_Z^{(0)}$ reduces to a bare $R_z$ gate (one rotation per site, no control qubits), recovering the original two-state circuit.

\subsection{Non-Power-of-$2$ Electronic States}
\label{app:padding}

When the physical number of electronic states $N < M = 2^m$, the channel register is padded to $M$ states by setting all coupling coefficients involving unphysical states to zero:
\begin{equation}\label{eq:padding_condition}
    c_{ij}(\mathbf{x}) = 0 \qquad \text{whenever } i \geq N \text{ or } j \geq N.
\end{equation}

The zero-padding condition~\ref{eq:padding_condition} ensures the following:
\begin{enumerate}
    \item \textbf{XOR-class partition.}
    The partition proceeds identically over all $M-1$ classes.
    Fragments $C_\zeta$ in which every pair $(i,j)$ involves at least one unphysical state ($i \geq N$ or $j \geq N$) have $c_{ij} = 0$ for all their pairs, so their fragment Hamiltonian $H_\zeta = 0$.
    These empty fragments contribute only the identity to the evolution and can be detected at compile time and omitted from the circuit, saving $2$ fragment applications per empty class.

    \item \textbf{UCR angles.}
    Within each non-empty fragment, the UCR angles for spectator configurations $\tilde\sigma$ addressing unphysical pairs are zero.
        The Walsh coefficients $\hat{c}_k^{(\zeta)}$ (Appendix~\ref{app:binary_decomp}) are computed from fewer nonzero inputs, but the UCR$_Z^{(m-1)}$ circuit structure is unchanged ($M/2$ rotations per site), with many rotation angles equal to zero or simplified.

    \item \textbf{Amplitude confinement.}
    The wavefunction $|\Psi\rangle$ has zero amplitude on unphysical states $|s\rangle$ for $s \geq N$ by construction: they are not populated by the initial state, and the padded Hamiltonian does not couple physical states to them (since $c_{ij} = 0$ whenever $i \geq N$ or $j \geq N$).
    This property is preserved exactly by each stage of the circuit.
    The Clifford focusing $U_\zeta$ only relabels computational basis states via CNOT and Hadamard gates; since the wavefunction has zero amplitude on every basis state with $s \geq N$, the transformed state in the focused basis inherits this zero-amplitude structure (mapped to the corresponding focused-basis labels).
    The diagonal evolution applies zero-valued phases to spectator configurations addressing unphysical pairs.
    The unfocusing Clifford $U_\zeta^\dagger$ restores the diabatic labelling.
    At no stage does amplitude leak into the unphysical subspace.

    \item \textbf{Population update.}
    The population update restricts to the physical subspace: the sum $\sum_{b \neq a}$ in Eq.~\ref{eq:pop_update} runs only over $b < N$, since $c_{ab} = 0$ for $b \geq N$.
    No probability leaks into the padded states, and the second-order accuracy of the symmetric product is unaffected.
\end{enumerate}

Claims (i) and (ii) follow directly from the zero-padding condition and the definitions.
Claim (iii) follows by induction over fragment applications: if $\langle s|\Psi\rangle = 0$ for all $s \geq N$ before a fragment application, then $\langle s|e^{-iH_\zeta\tau}|\Psi\rangle = 0$ for $s \geq N$, since $\langle s|H_\zeta|s'\rangle = 0$ whenever $s \geq N$ or $s' \geq N$ (all coupling coefficients involving unphysical states vanish).
The Clifford sandwich therefore reduces as $U_\zeta^\dagger e^{-iU_\zeta H_\zeta U_\zeta^\dagger \tau} U_\zeta = e^{-iH_\zeta\tau}$.
Claim (iv) then follows from (iii) and the second-order expansion.

For the benchmark systems, the worst case is $N = 5$ padded to $M = 8$ ($m = 3$), introducing $3$ unphysical states and $\binom{8}{2} - \binom{5}{2} = 18$ zero-coupling pairs distributed across the $M - 1 = 7$ XOR classes.
The number of empty fragments depends on the binary labels assigned to the physical states; an optimal assignment minimizes the number of non-empty fragments, reducing circuit depth by eliminating fragment applications.
For $N = 5$, $M = 8$, the minimum number of non-empty fragments is $4$, compared to $7$ for the worst-case assignment.

\subsection{Resource Counts}
\label{app:offdiag_resources}

Exact gate counts and depths for the off-diagonal block are given below.
Throughout, $M = 2^m$, $d$ is the number of vibrational modes, $n$ is the number of grid qubits per mode, $h(\zeta) = \operatorname{popcount}(\zeta)$, and $\kappa(\zeta) = m - h(\zeta)$.

\subsubsection{Per-Fragment Costs}
\label{app:per_frag_costs}

\begin{center}
\renewcommand{\arraystretch}{1.4}
\begin{tabular}{l c c}
\hline\hline
Operation & CNOT count & Depth \\
\hline
$U_\zeta$ (focusing + $H$) & $h(\zeta)-1$ & $h(\zeta)$ \\
Fan-out ($m$ qubits $\to$ $d$ copies) & $m(d-1)$ & $\lceil\log_2 d\rceil$ \\
UCR per cross-term site & $2^{m-1}+3$ & $2^m+3$ \\
UCR per single-qubit site & $2^{m-1}+1$ & $2^m+1$ \\
UCR per constant-phase site & $2^{m-1}-1$ & $2^m-1$ \\
Fan-in & $m(d-1)$ & $\lceil\log_2 d\rceil$ \\
$U_\zeta^\dagger$ (unfocusing) & $h(\zeta)-1$ & $h(\zeta)$ \\
\hline\hline
\end{tabular}
\end{center}

\noindent

The mode-parallel UCR depth (after fan-out, all $d$ modes execute simultaneously) is uniform across fragments:
\begin{equation}\label{eq:ucr_depth_frag}
    D^{(\mathrm{UCR})} = \bigl(2^{m}+3\bigr)\cdot\binom{n}{2} + \bigl(2^{m}+1\bigr)\cdot n + \bigl(2^{m}-1\bigr),
\end{equation}
where the three terms correspond to intra-mode cross sites, single-qubit sites, and constant-phase sites (one per mode), all executed sequentially within a single mode but in parallel across modes.
The independence from $\zeta$ follows from the uniform size $m-1$ of the extended spectator set.
For the linear vibronic model (coupling linear in $\mathbf{x}$), only the single-qubit sites contribute from the coupling itself, and the formula simplifies to $D^{(\mathrm{UCR,lin})} = (2^{m}+1)\cdot |\mathcal{G}_{\text{od}}|\cdot n$.

\chapter{Circuit Derivations}
\label{app:circuits}

Circuits are given in the binary-encoded basis as multi-controlled operations.
A series of operations $\alpha \cdot q_{n-1}q_{n-2}\cdots q_1q_0$ imparts a rotation $\alpha$ skewing the measurement axis $R_z$ on the target qubit $q_0$, controlled by all other qubits $q_{n-1}\cdots q_1$.

\section{Polynomial Functions}
\label{sec:poly-functions}

\subsection{Harmonic Oscillator}

The explicit Trotterization follows the expansion of an exponentiated quadratic.
Using the harmonic potential operator $V$ as an example,
\begin{equation}\label{eq:BenentiDerivation}
    \begin{split}
        e^{-iV\tau}\ket{\psi} & = e^{-i\tau\frac{m\omega^2}{2}(x-x_0)^2 -i\tau\delta_0}\ket{\psi} \\
        & = e^{-i\tau\frac{m\omega^2}{2}(x^2 - 2xx_0 + x_0^2) -i\tau\delta_0}\ket{\psi} \\
        & = e^{-i\tau\frac{m\omega^2}{2}x^2}e^{-i\tau\frac{m\omega^2}{2}(-2xx_0)}e^{-i\tau\frac{m\omega^2}{2}x_0^2 -i\tau\delta_0}\ket{\psi} \\
        & = e^{-i\tau\frac{m\omega^2\Delta^2}{2}\sum_{\ell=0}^{n-1}\sum_{j=0}^{n-1}2^jq_j2^\ell q_\ell}\\
        & \times e^{-i\tau\frac{m\omega^2}{2}(-2 x_0\Delta\sum_{j=0}^{n-1}2^jq_j)}\times e^{-i\tau(\frac{m\omega^2}{2}x_0^2 + \delta_0)}\ket{\psi} \\
        & = e^{2i\tau\eta \Delta^2 \sum_{\ell=0}^{n-1}\sum_{j=0}^{n-1}2^jq_j2^\ell q_\ell} \\
        & \times e^{-i\tau\eta x_0 \Delta\sum_{j=0}^{n-1}2^jq_j}\times e^{-i\tau(\eta x_0^2 + \delta_0)}\ket{\psi},
    \end{split}
\end{equation}
for $x$ defined by a binary basis $\ket{q_{n-1}\ldots q_1 q_0}$, with $2^n$ discrete positions spanning $x=0$ to $x=L$.

The $R_z$ gates act as
\begin{equation}
    R_z(\alpha) \otimes \mathbb{1} = \ket{0}\bra{0}\otimes \mathbb{1} + e^{i\alpha}\ket{1}\bra{1}\otimes \mathbb{1}\,,
\end{equation}
and therefore commute with other $R_z$ gates:
\begin{equation}
    [R_z(\alpha)\otimes\mathbb{1},\; R_z(\beta)\otimes\mathbb{1}] = 0\,.
\end{equation}
Likewise, controlled-$R_z$ gates ($\mathrm{CR}_z$) act as
\begin{equation}
    \mathrm{CR}_z(\alpha) \otimes\mathbb{1} = \ket{0}\bra{0}\otimes\mathbb{1}\otimes\mathbb{1} + \ket{1}\bra{1}\otimes R_z(\alpha) \otimes \mathbb{1}\,.
\end{equation}
Two $\mathrm{CR}_z$ gates that act on different tensor spaces therefore commute:
\begin{equation}
    [\mathrm{CR}_z(\alpha) \otimes\mathbb{1},\; \mathrm{CR}_z(\beta) \otimes\mathbb{1}] = 0\,.
\end{equation}
The quadratic term from Eq.~\ref{eq:BenentiDerivation} can then be reformed as
\begin{equation}
    \begin{split}
        e^{-i\tau\frac{m\omega^2}{2}x^2} & = e^{-i\tau\phi_2 \left(\sum_j^{n-1}2^jq_j\right)^2} \\
        & = e^{-i\tau\phi_2 \left(\sum_j^{n-1}2^{2j}q_j^2 + 2\sum_{j<\ell}2^{j+\ell}q_jq_\ell\right)} \\
        & = e^{-i\tau\phi_2 \left(\sum_j^{n-1}2^{2j}q_j + 2\sum_{j<\ell}2^{j+\ell}q_jq_\ell\right)} \\
        & = e^{-i\tau\phi_2 \left(\sum_j^{n-1}2^{2j}q_j\right)}\times e^{-i\tau\phi_2\left(2\sum_{j<\ell}2^{j+\ell}q_jq_\ell\right)}\,.
    \end{split}
\end{equation}
Combining with the first-order term,
\begin{equation}
\begin{split}
    & e^{-i\tau\phi_2 \sum_j^{n-1}2^{2j}q_j}\times e^{-i\tau\phi_2\left(2\sum_{j<\ell}2^{j+\ell}q_jq_\ell\right)}\times e^{-i\tau\phi_1\sum_j^{n-1}2^jq_j} \\
    & = e^{-i\tau\phi_2\left(2\sum_{j<\ell}2^{j+\ell}q_jq_\ell\right)}\times e^{-i\tau\phi_2 \sum_j^{n-1}2^{2j}q_j}\times e^{-i\tau\phi_1\sum_j^{n-1}2^jq_j} \\
    & = e^{-i\tau\phi_2\left(2\sum_{j<\ell}2^{j+\ell}q_jq_\ell\right)}\times e^{-i\tau(\phi_2 \sum_j^{n-1}2^{2j}q_j + \phi_1\sum_j^{n-1}2^jq_j)}\,.
\end{split}
\end{equation}

\subsection{Cubic}

For a cubic term $\phi_3 x^3$:
\begin{equation}
    \begin{split}
        e^{-i\tau\phi_3x^3} & = e^{-i\tau\phi_3 \left(\sum_j^{n-1}2^jq_j\right)^3} \\
        & = e^{-i\tau\phi_3 \left(\sum_j^{n-1}2^{3j}q_j^3 + 3\sum_{j=\ell\neq m}2^{2j+m}q_jq_m + 6\sum_{j<\ell<m}2^{j+\ell+m}q_jq_\ell q_m\right)}\,.
    \end{split}
\end{equation}
The first-order terms $\sum_j^{n-1}2^{3j}q_j^3 = \sum_j^{n-1}2^{3j}q_j$ commute with operators from lower degrees ($x^2$, $x$) and can be absorbed into the single-qubit sublayer of $R_z$ gates.
The second-order terms $3\sum_{j=\ell\neq m}2^{2j+m}q_jq_m$ commute with operators from $x^2$ and can be absorbed into the two-qubit $\mathrm{CR}_z$ gates.
The third-order terms $6\sum_{j<\ell<m}2^{j+\ell+m}q_jq_\ell q_m$ act as $ZZZ$ gates, with two control qubits and one target qubit.

\subsection{Quartic}

A quartic term $\phi_4 x^4$ decomposes as
\begin{equation}
    \begin{split}
        e^{-i\tau\phi_4x^4} & = e^{-i\tau\phi_4 \left(\sum_j^{n-1}2^jq_j\right)^4} \\
        & = e^{-i\tau\phi_4 \left(\sum_j^{n-1}2^{4j}q_j\right)}\\
        & \times e^{-i\tau\phi_4 \left(4\sum_{j=\ell=m\neq p}2^{3j+p}q_jq_p\right)} \\
        & \times e^{-i\tau\phi_4\left(3\sum_{j=\ell\neq m = p}2^{2j+2m}q_jq_m\right)} \\
        & \times e^{-i\tau\phi_4\left(6\sum_{j= \ell\neq m\neq p}2^{2j+m+p}q_jq_mq_p\right)} \\
        & \times e^{-i\tau\phi_4\left(24\sum_{j<\ell<m<p}2^{j + \ell + m + p}q_jq_\ell q_mq_p\right)}\,.
    \end{split}
\end{equation}

To compress this circuit, each term is ordered to be commutable via tensor products, so that second-order terms have the condition $j<\ell$, third-order terms have $j<\ell<m$, and so on.

For the first second-order term $\phi_4 \left(4\sum_{j=\ell=m\neq p}2^{3j+p}q_jq_p\right)$, two orientations for the indices arise: $j=\ell=m <p$ and $p<j=\ell=m$. Reindexing $p \rightarrow \ell$ yields
\begin{equation}
\begin{split}
     e^{-i\tau\phi_4 \left(4\sum_{j=\ell=m\neq p}2^{3j+p}q_jq_p\right)} &= e^{-i\tau\phi_4 \left(4\sum_{j<\ell}2^{3j+\ell}q_jq_\ell + 4\sum_{\ell<j}2^{3j+\ell}q_jq_\ell\right)} \\
     & = e^{-i\tau\phi_4 \left(4\sum_{j<\ell}(2^{3j+\ell}+2^{3\ell+j})q_jq_\ell\right)}\,.
     \end{split}
\end{equation}

The second second-order term $e^{-i\tau\phi_4\left(3\sum_{j=\ell\neq m = p}2^{2j+2m}q_jq_m\right)}$ has two equivalent combinations, $j = \ell < m=p$ and $m = p < j=\ell$, gaining a factor of two upon reindexing: $e^{-i\tau\phi_4\left(6\sum_{j<\ell}2^{2j+2\ell}q_jq_\ell\right)}$.

The third-order term has six orientations. Since $m$ and $p$ are interchangeable indices, these reduce to three orientations with an additional factor of 2 in the exponential. After reindexing,
\begin{equation}
    \begin{split}
        e^{-i\tau\phi_4\left(6\sum_{j= \ell\neq m\neq p}2^{2j+m+p}q_jq_mq_p\right)} & = e^{-i\tau\phi_4\left(12\sum_{j<\ell<m}\left(2^{2j+\ell+m} + 2^{j + 2\ell + m} + 2^{j + \ell + 2m}\right)q_jq_\ell q_m\right)}\,.
    \end{split}
\end{equation}

Collecting all terms, the quartic operator becomes
\begin{equation}
    \begin{split}
        e^{-i\tau\phi_4x^4} & = e^{-i\tau\phi_4 \left(\sum_j^{n-1}2^{4j}q_j\right)}\\
        & \times e^{-i\tau\phi_4 \left(4\sum_{j<\ell}(2^{3j+\ell}+2^{3\ell+j})q_jq_\ell\right)} \\
        & \times e^{-i\tau\phi_4\left(6\sum_{j<\ell}2^{2j+2\ell}q_jq_\ell\right)} \\
        & \times e^{-i\tau\phi_4\left(12\sum_{j<\ell<m}\left(2^{2j+\ell+m} + 2^{j + 2\ell + m} + 2^{j + \ell + 2m}\right)q_jq_\ell q_m\right)} \\
        & \times e^{-i\tau\phi_4\left(24\sum_{j<\ell<m<p}2^{j + \ell + m + p}q_jq_\ell q_mq_p\right)}\,.
    \end{split}
\end{equation}

Combining all terms in $f(x) = \phi_4x^4 + \phi_3x^3 + \phi_2x^2+ \phi_1x + \phi_0$ yields a circuit with depth $\sum_{i=1}^k {\binom{n}{i}} + 4$, ordered starting with the global phase and increasing with degree. Each additional degree requires an additional control qubit, producing $\mathrm{CR}_z$, $C\mathrm{CR}_z$, $C^3R_z$, etc.\ gates. Parameters for the full quartic circuit are given in Table~\ref{tab:quartic-params}.

\begin{table}[ht]
\centering
\caption{Gate parameters for the quartic polynomial circuit.}
\label{tab:quartic-params}
\begin{tabular}{clccc}
 \toprule
 Degree & Parameter & Control(s) & Target & Gate Type\\
 \midrule
 0 & $\phi_0$ & N/A & $q_0$ & $R_z$, X \\
 1 & $2^j(\phi_1 + \phi_22^{j} +\phi_32^{2j} +\phi_42^{3j})$ & N/A & $q_j$ & $R_z$ \\
 2 & $2^{j+\ell} (2\phi_2 + 3\phi_3(2^{j} + 2^{\ell})+4\phi_4(2^{2j}+2^{2\ell}) + 6\phi_4 2^{j + \ell})$ & $q_j$ & $q_\ell$ & $\mathrm{CR}_z$\\
 3 & $2^{j+\ell+m}(6\phi_3 + 12\phi_4(2^j + 2^\ell +2^m))$ & $q_j$, $q_\ell$ & $q_m$ & $C\mathrm{CR}_z$ \\
 4 & $2^{j+\ell+m+p}(24\phi_4)$ & $q_j$, $q_\ell$, $q_m$ & $q_p$ & $C^3R_z$ \\
 \bottomrule
\end{tabular}
\end{table}

\begin{figure}[ht]
    \centering
    \includegraphics[width=\textwidth]{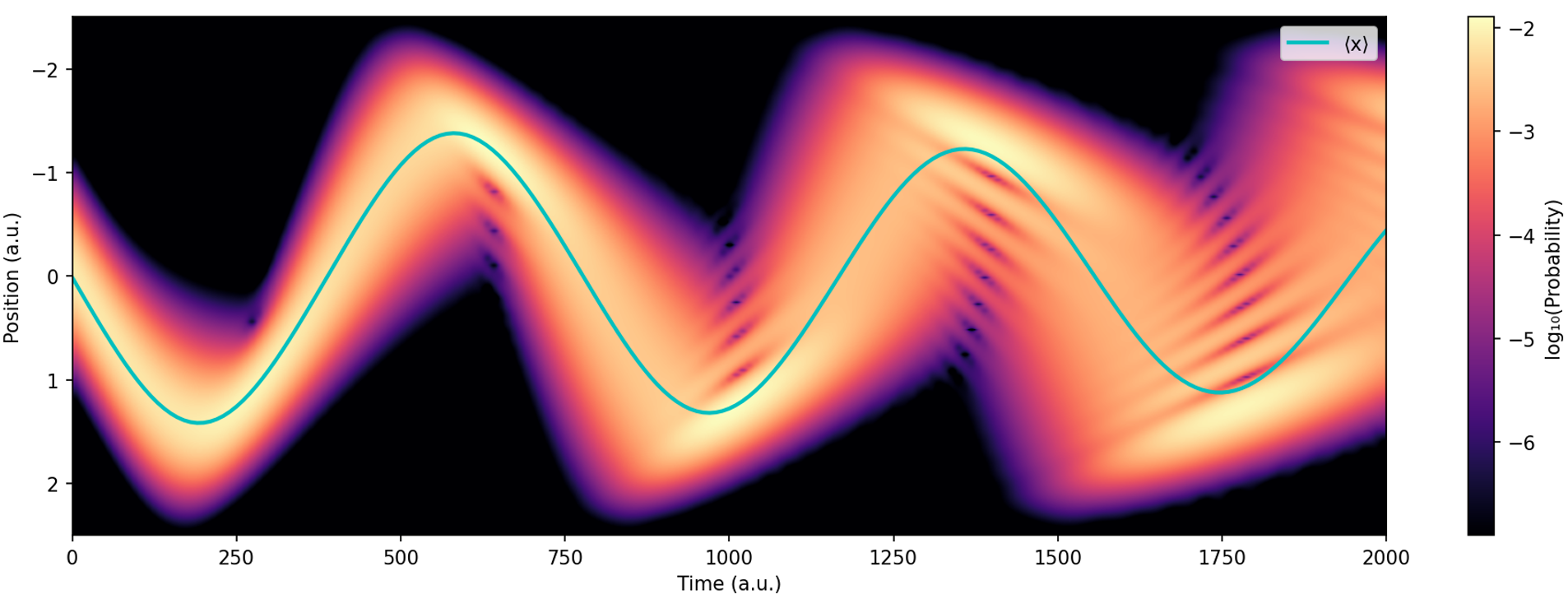}
    \caption{Anharmonic oscillator dynamics. Simulation parameters: 10 qubits, $m=1836$~a.u., $x_0=0.0\,\text{a.u.}$, $p_0 = 5.0\,\text{a.u.}$,  $\sigma = 0.25\,\text{a.u.}$, $\tau=1.0\,\text{a.u.}$, $t=2000\,\text{a.u.}$, $V(x) = 0.003x^4 + 0.02x^2$, $L=5.0\,\text{a.u.}$}
    \label{fig:AnharmonicOscillator}
\end{figure}

\begin{figure}[ht]
    \centering
    \includegraphics[width=\textwidth]{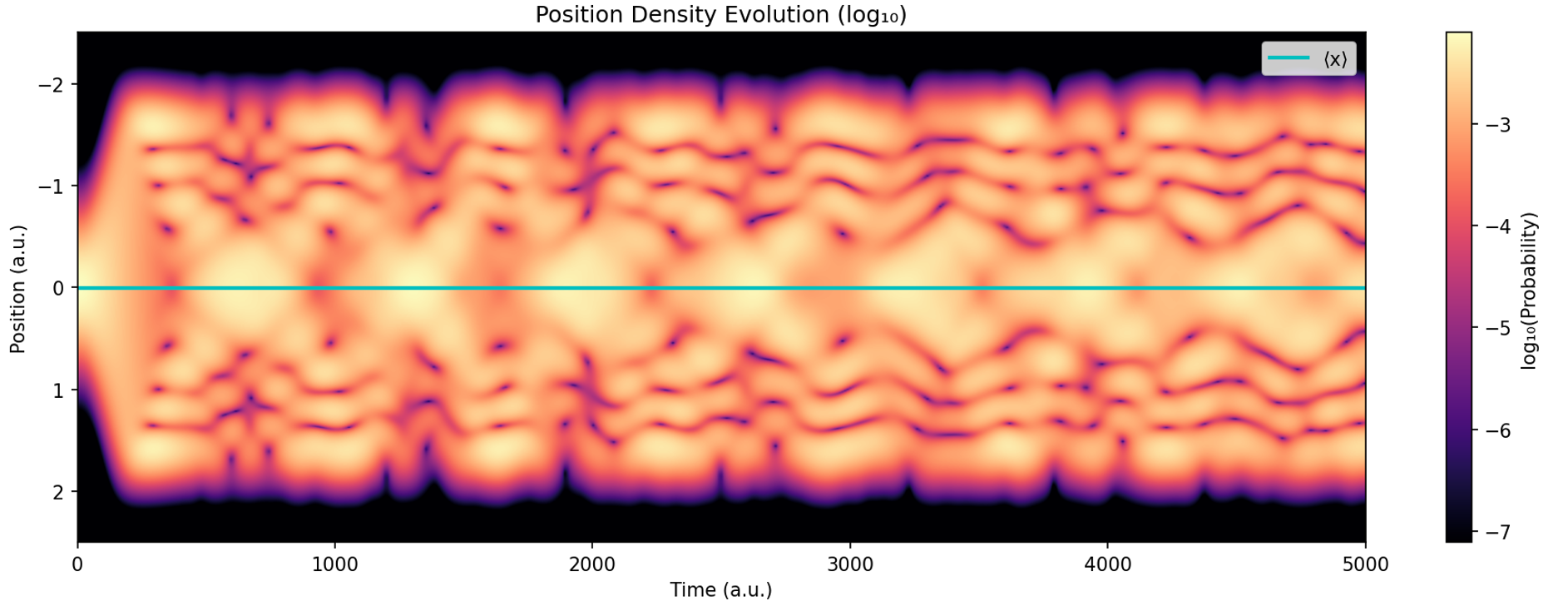}
    \caption{Symmetric double-well dynamics. Simulation parameters: 10 qubits, $m=1836$~a.u., $x_0=0.0\,\text{a.u.}$, $p_0 = 0.0\,\text{a.u.}$,  $\sigma = 0.25\,\text{a.u.}$, $\tau=1.0\,\text{a.u.}$, $t=5000\,\text{a.u.}$, $V(x) = 0.02x^4 - 0.06x^2$, $L=5.0\,\text{a.u.}$}
    \label{fig:SymmDouble}
\end{figure}

\begin{figure}[ht]
    \centering
    \includegraphics[width=\textwidth]{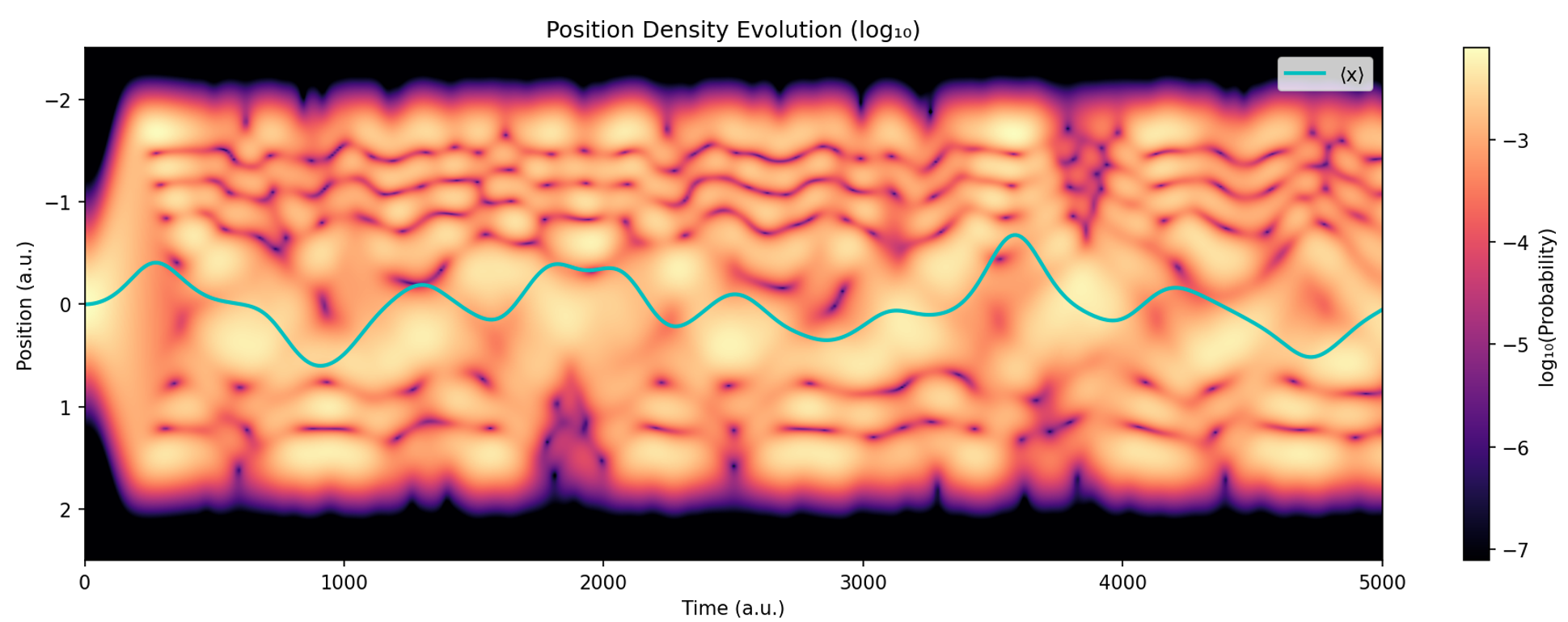}
    \caption{Asymmetric double-well dynamics. Simulation parameters: 10 qubits, $m=1836$~a.u., $x_0=0.0\,\text{a.u.}$, $p_0 = 0.0\,\text{a.u.}$,  $\sigma = 0.25\,\text{a.u.}$, $\tau=1.0\,\text{a.u.}$, $t=5000\,\text{a.u.}$, $V(x) = 0.02x^4 - 0.06x^2+0.012x$, $L=5.0\,\text{a.u.}$}
    \label{fig:AsymmDouble}
\end{figure}

Three common potential shapes for the quartic potential are the anharmonic oscillator, the symmetric double-well, and the asymmetric double-well. Simulations of Trotterized Hamiltonians with a wavepacket propagating in each of these potentials are shown in Figures~\ref{fig:AnharmonicOscillator}, \ref{fig:SymmDouble}, and~\ref{fig:AsymmDouble}.

For an 8-qubit simulation, this method requires $162 + 4 = 166$ rotations prior to two-qubit decomposition of multi-qubit gates. This stands in contrast to the previous $\mathcal{O}(n^k) \approx 4000$ rotations accepted as ``efficient.''
Likewise for a 10-qubit simulation, gate count reduces to $1023+4=1027$ rotations, compared to $\mathcal{O}(n^k)\approx 10^4$ rotations.

\begin{figure}[ht]
    \centering
    \includegraphics[width=\textwidth]{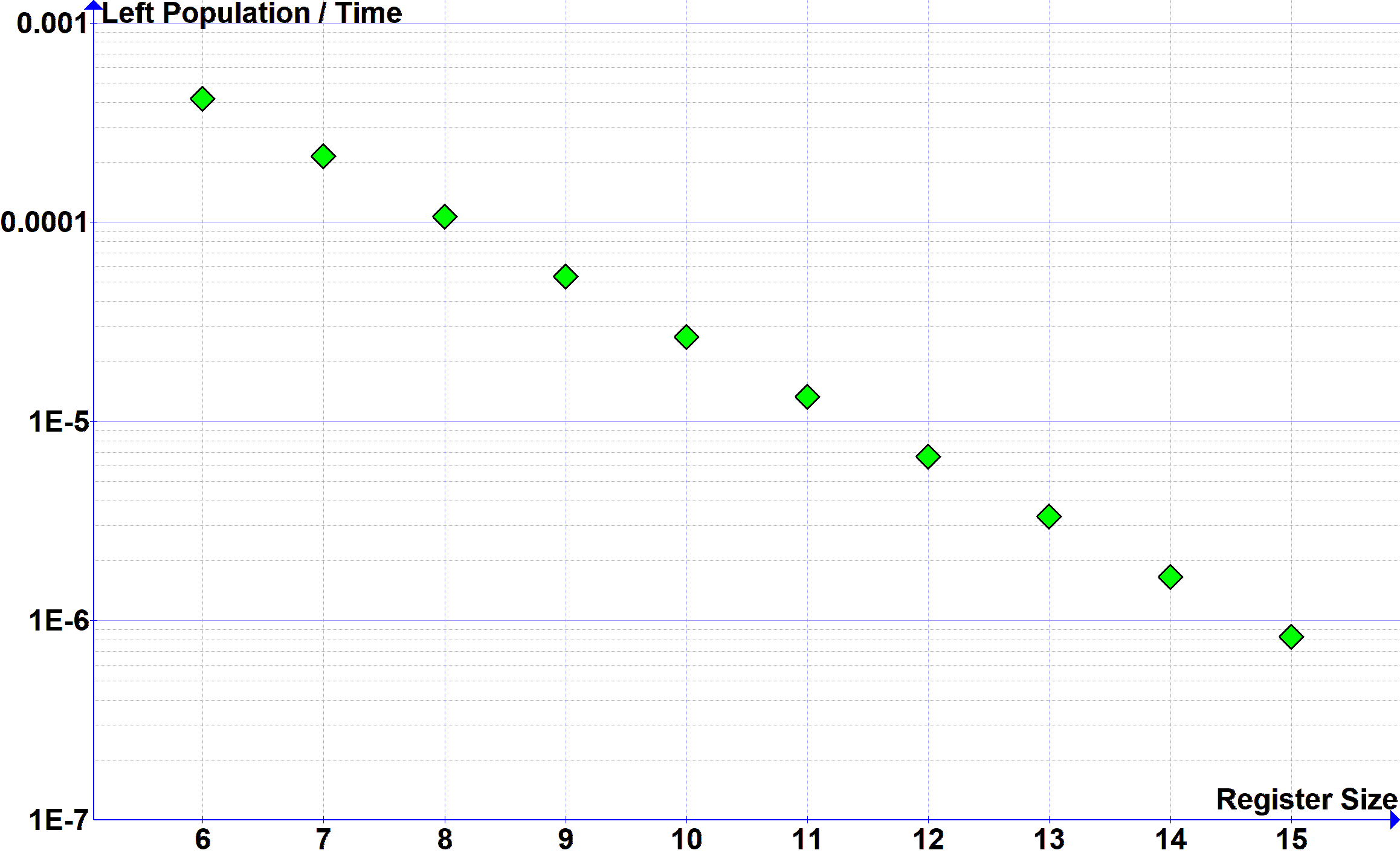}
    \caption{For a total propagation time of $100$~a.u.\ (divided into Trotter steps of $10$~a.u.), the slope of the left well's population was calculated. This is plotted against the size of the position register, showing exponential decrease as $(1/2^n)$ for all values of $n$ where the momentum distribution did not exceed the Nyquist frequency.}
    \label{fig:Instability}
\end{figure}

To assess finite-grid effects, a wavepacket was initialized at the center of a symmetric double-well with $p_0 = 0.0$, where any symmetry breaking would arise solely from numerical instability and the finite grid. A quartic potential $V(x) = 0.001x^4 - 0.05x^2 + 0.625$ was constructed at the center of a box of size $L=20\,\text{a.u.}$. A Gaussian wavepacket $\psi(x) \propto \exp{[-(x - x_0)^2 / (4\delta^2)]} \times \exp{[ip_0(x-x_0)]}$ with $\delta = 1/3$, $x_0 = L/2$, $p_0 = 0.0$ was propagated over ten Trotter steps of size $10$~a.u.\ with mass $m = 1818.18$~a.u. The deviation of the left-well population from $0.5$ was recorded for qubit register sizes $n=6$ through $n=15$. This deviation was found to scale as $\propto 0.500452^n$ ($R^2 = 0.9997$). When register sizes too small to capture all momentum eigenvalues were excluded, the scaling improved to $\propto 0.500000^n$ ($R^2 > 0.9999$). All rotation angles and numerical operations were performed with floating-point precision. These results are presented in Figure~\ref{fig:Instability}.

\subsection{Quintic and Sextic}

With higher-degree polynomials, a general algorithm for the compression emerges.
To build a circuit for a sextic polynomial, the quintic decomposition $\phi_5x^5$ is needed first. The bitwise decomposition is
\begin{equation}
    \begin{split}
        \phi_5 x^5 & = \phi_5 \sum_j2^{5j}q_j +5\phi_5\sum_{j=\ell=m=p\neq q}2^{4j +q} + 10\phi_5\sum_{j=\ell=m \neq p =q}2^{3j + 2p} \\
        & + 20\phi_5\sum_{j=\ell=m \neq p \neq q}2^{3j +p + q} + 30\phi_5\sum_{j=\ell\neq m =p \neq q}2^{2j + 2m + q} \\
        & + 60\phi_5\sum_{j=\ell \neq m \neq p \neq q}2^{2j + m + p+q} + 120\phi_5\sum_{j \neq \ell\neq m\neq p \neq q}2^{j+\ell+m+p+q}\,.
    \end{split}
\end{equation}
Index ordering and renaming gives
\begin{equation}
    \begin{split}
        \phi_5 x^5 & = \phi_5 \sum_j2^{5j}q_j +5\phi_5\sum_{j<\ell}(2^{4j +\ell}+2^{j+4\ell}) + 10\phi_5\sum_{j<\ell}(2^{3j + 2\ell}+2^{2j + 3\ell}) \\
        & + 20\phi_5\sum_{j<\ell<m}(2^{3j +\ell + m} + 2^{j +3\ell + m} + 2^{j +\ell + 3m})\\
        & + 30\phi_5\sum_{j<\ell<m}(2\cdot 2^{2j + 2\ell + m} + 2^{2j + \ell + 2m} + 2^{j + 2\ell + 2m}) \\
        & + 60\phi_5\sum_{j<\ell<m<p}(2^{2j + \ell+m+p} +2^{j + 2\ell+m+p} + 2^{j + \ell+2m+p} + 2^{j + \ell+m+2p})\\
        & + 120\phi_5\sum_{j<\ell<m<p<q}2^{j+\ell+m+p+q}\,,
    \end{split}
\end{equation}

For a sextic term $\phi_6x^6$, the bitwise decomposition follows similarly.
The leading coefficients are determined by equivalence classes characterizing how many indices are equal to one another $[n_a, \ldots, n_i]$. The leading coefficient is given by $k!/(n_a!\times n_b! \times \ldots \times n_i!)$. For example, with $j=\ell=m \neq p = q \neq r$, $k=6$ and the equivalence class $[3,2,1]$, yielding $6!/(3!\times2!\times 1!) = 60$.

After imposing ordered indices and collecting all contributions, the combined parameters for a sextic polynomial are given in Table~\ref{tab:sextic-params}.

\begin{table}[ht]
\centering
\caption{Gate parameters for each Walsh order of a sextic polynomial, including contributions from all lower-degree terms.}
\label{tab:sextic-params}
\renewcommand{\arraystretch}{1.4}
\begin{tabular}{cp{0.85\textwidth}}
\toprule
Degree & Parameter \\
\midrule
0 & $\phi_0$ \\
1 & $2^j(\phi_1 + \phi_2 2^j + \phi_3 2^{2j} + \phi_4 2^{3j} + \phi_5 2^{4j} + \phi_6 2^{5j})$ \\
2 & $2\phi_2 \cdot 2^{j+\ell} + 3\phi_3(2^{2j+\ell} + 2^{j+2\ell}) + 4\phi_4(2^{3j+\ell} + 2^{j+3\ell}) + 6\phi_4 \cdot 2^{2j+2\ell}
+ 5\phi_5(2^{4j+\ell} + 2^{j+4\ell}) + 10\phi_5(2^{3j+2\ell} + 2^{2j+3\ell})
+ 6\phi_6(2^{5j+\ell} + 2^{j+5\ell}) + 15\phi_6(2^{4j+2\ell} + 2^{2j+4\ell}) + 20\phi_6 \cdot 2^{3j+3\ell}$ \\
3 & $6\phi_3 \cdot 2^{j+\ell+m} + 12\phi_4(2^{2j+\ell+m} + 2^{j+2\ell+m} + 2^{j+\ell+2m})
+ 20\phi_5(2^{3j+\ell+m} + 2^{j+3\ell+m} + 2^{j+\ell+3m})
+ 30\phi_5(2 \cdot 2^{2j+2\ell+m} + 2^{2j+\ell+2m} + 2^{j+2\ell+2m})
+ 30\phi_6(2^{4j+\ell+m} + 2^{j+4\ell+m} + 2^{j+\ell+4m})
+ 60\phi_6(2^{3j+2\ell+m} + 2^{3j+\ell+2m} + 2^{2j+3\ell+m} + 2^{2j+\ell+3m} + 2^{j+2\ell+3m} + 2^{j+3\ell+2m})
+ 90\phi_6 \cdot 2^{2j+2\ell+2m}$ \\
4 & $24\phi_4 \cdot 2^{j+\ell+m+p} + 60\phi_5(2^{2j+\ell+m+p} + 2^{j+2\ell+m+p} + 2^{j+\ell+2m+p} + 2^{j+\ell+m+2p})
+ 120\phi_6(2^{3j+\ell+m+p} + 2^{j+3\ell+m+p} + 2^{j+\ell+3m+p} + 2^{j+\ell+m+3p})
+ 180\phi_6(2^{2j+2\ell+m+p} + 2^{j+\ell+2m+2p} + 2^{j+2\ell+2m+p} + 2^{j+2\ell+m+2p} + 2^{2j+\ell+2m+p} + 2^{2j+\ell+m+2p})$ \\
5 & $120\phi_5 \cdot 2^{j+\ell+m+p+q} + 360\phi_6(2^{2j+\ell+m+p+q} + 2^{j+2\ell+m+p+q} + 2^{j+\ell+2m+p+q} + 2^{j+\ell+m+2p+q} + 2^{j+\ell+m+p+2q})$ \\
6 & $720\phi_6 \cdot 2^{j+\ell+m+p+q+r}$ \\
\bottomrule
\end{tabular}
\end{table}

\begin{figure}[ht]
    \centering
    \includegraphics[width=\textwidth]{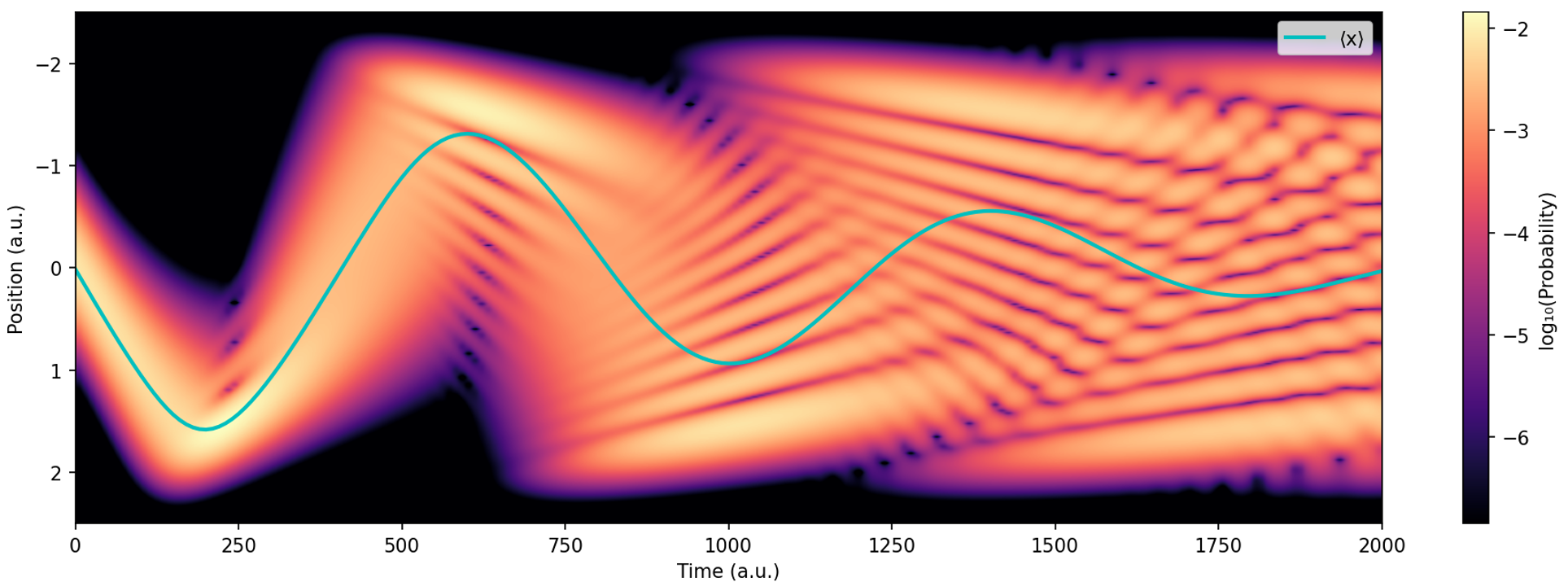}
    \caption{Anharmonic motion for a wavepacket in a sextic potential. Simulation parameters: 10 qubits, $m=1836$~a.u., $x_0=0.0\,\text{a.u.}$, $p_0 = 5.0\,\text{a.u.}$,  $\sigma = 0.25\,\text{a.u.}$, $\tau=1.0\,\text{a.u.}$, $t=2000\,\text{a.u.}$, $V(x) = 0.002x^6-0.004x^4 + 0.015x^2$, $L=5.0\,\text{a.u.}$}
    \label{fig:SexticAnharmonic}
\end{figure}

Simulation of a sextic anharmonic oscillator with potential $V(x) = 0.0001x^6 - 0.0005x^4 + 0.001x^2$ is shown in Figure~\ref{fig:SexticAnharmonic}. The wavepacket was initialized with $n=10$ qubits, $m=1836$~a.u., $x_0=0.0\,\text{a.u.}$, $p_0 = 5.0\,\text{a.u.}$,  $\sigma = 0.25\,\text{a.u.}$, $\tau=1.0\,\text{a.u.}$, $t=2000\,\text{a.u.}$, $V(x) = 0.002x^6-0.004x^4 + 0.015x^2$, $L=5.0\,\text{a.u.}$

The circuit depth is reduced by tensor product space commutativity, with complexity in the calculation limited to the computation of coefficients. The gate count is $4+\sum_{i=1}^6 {\binom{n}{i}}$: $250$ gates for $n=8$ qubits and $469$ gates for $n=9$ qubits, prior to two-qubit decomposition of multi-controlled gates. This is in contrast to the $\mathcal{O}(8^6) \approx 262{,}000$ and $\mathcal{O}(9^6) \approx 531{,}000$ gates when extending older polynomial-complexity methods~\cite{benenti2008quantum}.

\subsection{Septic, Octic, and the General Algorithm}
\label{subsec:septic-octic}

The general algorithm for compressing higher-degree polynomial circuits may be summarized as follows:
\begin{enumerate}
    \item Coefficients are defined by equivalence classes characterizing how many indices are equal to one another (e.g., $j = \ell \neq m \rightarrow [2,1] \rightarrow 3!/(2! \cdot 1!) = 3$).
    \item The permutations of equivalence classes between indices as powers of 2 give the necessary prefactor for an ordered operator (e.g., $j=\ell\neq m \rightarrow j<m$, $\sum_{j<m}(2^{2j+m}+2^{j+2m})$).
    \item Calculated coefficients and prefactors are multiplied with the initial coefficient of the polynomial degree $\phi_k$.
    \item Each resulting term is included in the corresponding Walsh order (e.g., 2 controls and 1 target corresponds to a 3-qubit operation in the third-order Walsh term) together with all preceding factors from lower degrees.
\end{enumerate}
The resulting operator is diagonal in the qubit basis and follows a fast-forwardable structure. The total number of operations is $4+\sum_{i=1}^k{\binom{n}{i}}$ for a register of $n$ qubits, with $\binom{n}{i}$ $i$-qubit operations per Walsh order.

The septic and octic parameters, which follow the same algorithm, are listed in Tables~\ref{tab:septic-params} and~\ref{tab:octic-params}.

\begin{table}   
\centering
\caption{Contributions to each Walsh order from the septic term $\phi_7 x^7$.}
\label{tab:septic-params}
\renewcommand{\arraystretch}{1.4}
\begin{tabular}{cp{0.85\textwidth}}
\toprule
Degree & Parameter \\
\midrule
0 & N/A \\
\hline
1 & $\phi_7 \cdot2^{7j}$ \\
\hline
2 & $\phi_7 \cdot (7\cdot(2^{6j+l} +2^{j+6l})+21\cdot(2^{5j+2l}+2^{2j+5l}) +35\cdot(2^{4j+3l}+2^{3j+4l}))$ \\
\hline
3 & $\phi_7 \cdot (42\cdot(2^{5j+l+m} +2^{j+5l+m} +2^{j+l+5m})$ \\
& $+105\cdot(2^{4j+2l+m}+2^{4j+l+2m}+2^{2j+4l+m}+2^{2j+l+4m}+2^{j+2l+4m}+2^{j+4l+2m})$ \\
&$+140\cdot(2^{3j+3l+m}+2^{3j+l+3m}+2^{j+3l+3m})+210\cdot(2^{3j+2l+2m}+2^{2j+3l+2m}+2^{2j+2l+3m}))$\\
\hline
4 & $\phi_7 \cdot (210\cdot(2^{4j+l+m+p} +2^{j+4l+m+p} +2^{j+l+4m+p} +2^{j+l+m+4p})$ \\
& $+420\cdot(2^{3j + 2l +m+p}+2^{3j + l +2m+p}+2^{3j + l +m+2p}+2^{2j + 3l +m+p}+2^{j + 3l +2m+p}$\\
&$+2^{j + 3l +m+2p}+2^{2j + l +3m+p}+2^{j + 2l +3m+p}+2^{j + l +3m+2p}+2^{2j + l +m+3p}$\\
&$+2^{j + 2l +m+3p}+2^{j + l +2m+3p})$\\
&$+630\cdot(2^{2j + 2l + 2m + p}+2^{j + 2l + 2m + 2p}+2^{2j + l + 2m + 2p}+2^{2j + 2l + m + 2p}))$\\
\hline
5 & $\phi_7 \cdot (840\cdot(2^{3j+l+m+p+q} +2^{j+3l+m+p+q} +2^{j+l+3m+p+q} +2^{j+l+m+3p+q} +2^{j+l+m+p+3q})$ \\
& $1260\cdot(2^{2j + 2l+ m +p+q}+2^{j + 2l+ 2m +p+q}+2^{j + 2l+ m +2p+q}+2^{j + 2l+ m +p+2q}+2^{2j + l+ 2m +p+q}$ \\
& $+2^{2j + l+ m +2p+q}+2^{2j + l+ m +p+2q}+2^{j + l+ 2m +2p+q}+2^{j + l+ 2m +p+2q}+2^{j + l+ m +2p+2q}))$\\
\hline
6 & $\phi_7 \cdot (2520\cdot(2^{2j + l + m + p+q+r} + 2^{j + 2l + m + p+q+r} +2^{j + l + 2m + p+q+r} +2^{j + l + m + 2p+q+r} $ \\
& $+2^{j + l + m + p+2q+r}+2^{j + l + m + p+q+2r}))$ \\
\hline
7 & $\phi_7 \cdot5040\cdot 2^{j+l+m+p+q+r+s}$\\
\bottomrule
\end{tabular}
\end{table}

The octic term parameters is given in Table~\ref{tab:octic-params}.
\FloatBarrier

\begin{footnotesize}
\begin{longtable}{cp{0.82\textwidth}}
\caption{Contributions to each Walsh order from the octic term $\phi_8 x^8$.}
\label{tab:octic-params} \\
\toprule
Degree & Parameter \\
\midrule
\endfirsthead

\multicolumn{2}{c}{\tablename\ \thetable{} -- \textit{continued from previous page}} \\
\toprule
Degree & Parameter \\
\midrule
\endhead

\midrule
\multicolumn{2}{r}{\textit{Continued on next page}} \\
\endfoot

\bottomrule
\endlastfoot

0 & N/A \\
\hline
1 & $\phi_8 \cdot2^{8j}$ \\
\hline
2 & $\phi_8 \cdot (8\cdot(2^{7j+l}+2^{j+7l}) + 28 \cdot (2^{6j+2l}+2^{2j+6l}) + 56 \cdot (2^{5j+3l}+2^{3j+5l}) +70\cdot (2^{4j+4l}))$ \\
\hline
3 & $\phi_8 \cdot (56\cdot(2^{6j+l+m}+2^{j+6l+m}+2^{j+l+6m})$
$+ 168 \cdot (2^{5j+2l+m}+2^{5j+l+2m}+2^{2j+5l+m}+2^{2j+l+5m}+2^{j+2l+5m}+2^{j+5l+2m})$
$+ 280 \cdot (2^{4j+3l+m}+2^{4j+l+3m}+2^{3j+4l+m}+2^{3j+l+4m}+2^{j+3l+4m}+2^{j+4l+3m})$
$+ 420 \cdot (2^{4j+2l+2m}+2^{2j+4l+2m}+2^{2j+2l+4m})$
$+560\cdot (2^{3j+3l+2m}+2^{3j+2l+3m}+2^{2j+3l+3m}))$ \\
\hline
4 & $\phi_8 \cdot (336\cdot(2^{5j+l+m+p}+2^{j+5l+m+p}+2^{j+l+5m+p}+2^{j+l+m+5p})$
$+ 840 \cdot (2^{4j+2l+m+p}+2^{4j+l+2m+p}+2^{4j+l+m+2p}+2^{2j+4l+m+p}$
$+2^{j+4l+2m+p}+2^{j+4l+m+2p}+2^{2j+l+4m+p}+2^{j+2l+4m+p}$
$+2^{j+l+4m+2p}+2^{2j+l+m+4p}+2^{j+2l+m+4p}+2^{j+l+2m+4p})$
$+ 1120 \cdot (2^{3j+3l+m+p}+2^{3j+l+3m+p}+2^{3j+l+m+3p}+2^{j+3l+3m+p}+2^{j+3l+m+3p}+2^{j+l+3m+3p})$
$+1680\cdot (2^{3j+2l+2m+p}+2^{3j+2l+m+2p}+2^{3j+l+2m+2p}+2^{2j+3l+2m+p}$
$+2^{2j+3l+m+2p}+2^{j+3l+2m+2p}+2^{2j+2l+3m+p}+2^{2j+l+3m+2p}$
$+2^{j+2l+3m+2p}+2^{2j+2l+m+3p}+2^{2j+l+2m+3p}+2^{j+2l+2m+3p})
+ 2520\cdot(2^{2j + 2l + 2m + 2p}))$ \\
\hline
5 & $\phi_8 \cdot (1680\cdot(2^{4j+l+m+p+q}+2^{j+4l+m+p+q}+2^{j+l+4m+p+q}+2^{j+l+m+4p+q}+2^{j+l+m+p+4q})$
$+ 3360 \cdot (2^{3j + 2l + m + p + q}+2^{3j + l + 2m + p + q}+2^{3j + l + m + 2p + q}+2^{3j + l + m + p + 2q}$
$+2^{2j + 3l + m + p + q}+2^{j + 3l + 2m + p + q}+2^{j + 3l + m + 2p + q}+2^{j + 3l + m + p + 2q}$
$+2^{2j + l + 3m + p + q}+2^{j + 2l + 3m + p + q}+2^{j + l + 3m + 2p + q}+2^{j + l + 3m + p + 2q}$
$+2^{2j + l + m + 3p + q}+2^{j + 2l + m + 3p + q}+2^{j + l + 2m + 3p + q}+2^{j + l + m + 3p + 2q}$
$+2^{2j + l + m + p + 3q}+2^{j + 2l + m + p + 3q}+2^{j + l + 2m + p + 3q}+2^{j + l + m + 2p + 3q})$
$+ 5040 \cdot (2^{2j + 2l + 2m + p+q}+2^{2j + 2l + m + 2p+q}+2^{2j + 2l + m + p+2q}$
$+2^{2j + l + 2m + 2p+q}+2^{2j + l + 2m + p+2q}+2^{2j + l + m + 2p+2q}$
$+2^{j + 2l + 2m + 2p+q}+2^{j + 2l + 2m + p+2q}+2^{j + 2l + m + 2p+2q}+2^{j + l + 2m + 2p+2q}))$ \\
\hline
6 & $\phi_8 \cdot (6720\cdot(2^{3j + l + m + p + q +r}+2^{j + 3l + m + p + q +r}+2^{j + l + 3m + p + q +r}+2^{j + l + m + 3p + q +r}$
$+2^{j + l + m + p + 3q +r}+2^{j + l + m + p + q +3r})$
$+ 10080 \cdot (2^{2j + 2l + m + p + q +r}+2^{2j + l + 2m + p + q +r}+2^{2j + l + m + 2p + q +r}+2^{2j + l + m + p + 2q +r}$
$+2^{2j + l + m + p + q +2r}+2^{j + 2l + 2m + p + q +r}+2^{j + 2l + m + 2p + q +r}+2^{j + 2l + m + p + 2q +r}$
$+2^{j + 2l + m + p + q +2r}+2^{j + l + 2m + 2p + q +r}+2^{j +l +2 m + p + 2q +r}+2^{j + l + 2m + p + q +2r}$
$+2^{j + l + m + 2p + 2q +r}+2^{j + l + m + 2p + q +2r} + 2^{j + l + m + p + 2q +2r}))$ \\
\hline
7 & $\phi_8 \cdot 20160\cdot(2^{2j+l+m+p+q+r+s}+2^{j+2l+m+p+q+r+s}+2^{j+l+2m+p+q+r+s}+2^{j+l+m+2p+q+r+s}$
$+2^{j+l+m+p+2q+r+s}+2^{j+l+m+p+q+2r+s}+2^{j+l+m+p+q+r+2s})$ \\
\hline
8 & $\phi_8 \cdot 40320 \cdot 2^{j+l+m+p+q+r+s+t}$ \\

\end{longtable}
\end{footnotesize}

For a $k$-degree polynomial on an $n$-qubit register (including the global phase), the circuit requires $2$ Pauli $X$ gates, $2+\binom{n}{1}$ $R_z$ gates, $\binom{n}{2}$ $\mathrm{CR}_z$ gates, and so on up to $\binom{n}{k}$ $C^{k-1}R_z$ gates with $k-1$ control qubits.

\begin{figure}[ht]
    \centering
    \includegraphics[width=\textwidth]{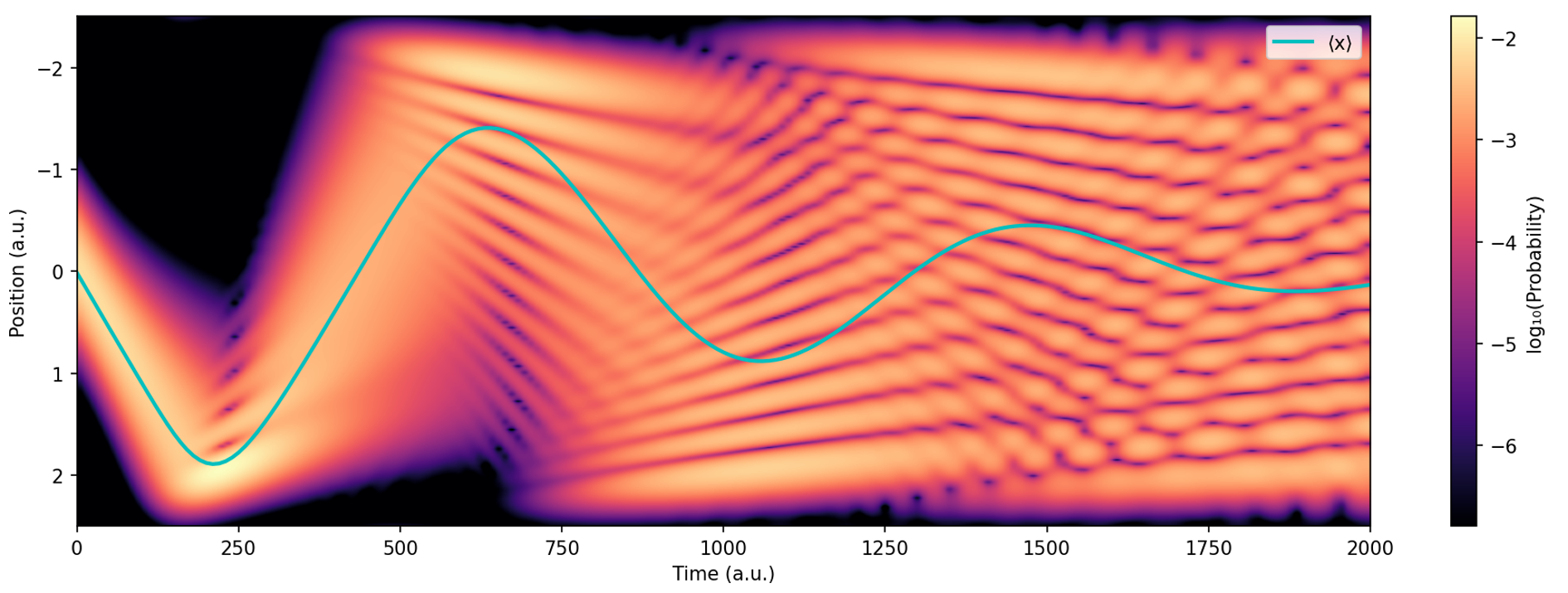}
    \caption{Density plot for position and momentum evolution of a wavepacket in an octic anharmonic oscillator. Signatures of breathing-mode oscillations are visible, as found in anharmonic crystals and in the formation of polarons.}
    \label{fig:OcticAnharmonic}
\end{figure}

\begin{figure}[ht]
    \centering
    \includegraphics[width=\textwidth]{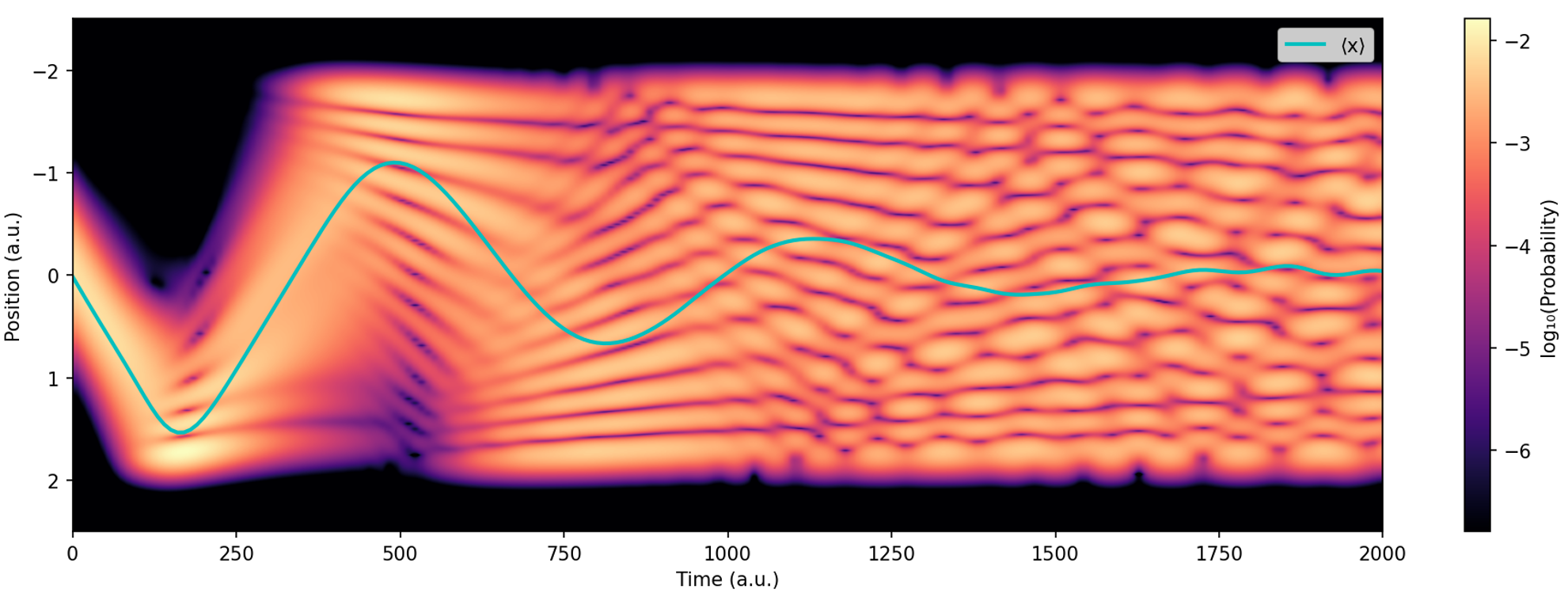}
    \caption{Triple-well potential dynamics. In addition to complex vibrational and excitational landscapes, the triple-well potential supports state confinement relevant to flux qubits and multi-level quantum dots. Features of real-time decoherence from the central well via coupling to the symmetric side wells are visible.}
    \label{fig:TripleWell}
\end{figure}

Octic polynomials are of particular importance in quantum chemistry, where they model additional anharmonic corrections ($\phi_8$) and can represent triple-well potentials. Dynamics for a symmetric octic anharmonic oscillator with $V(x) = 0.0002x^8 + 0.0003x^6 -0.003x^4 + 0.005x^2$ are shown in Figure~\ref{fig:OcticAnharmonic}. A triple-well potential with $V(x) = 0.001x^8 + 0.012x^6 -0.06x^4 +0.05x^2$ is shown in Figure~\ref{fig:TripleWell}.

\begin{figure}[ht]
    \centering
    \includegraphics[width=\textwidth]{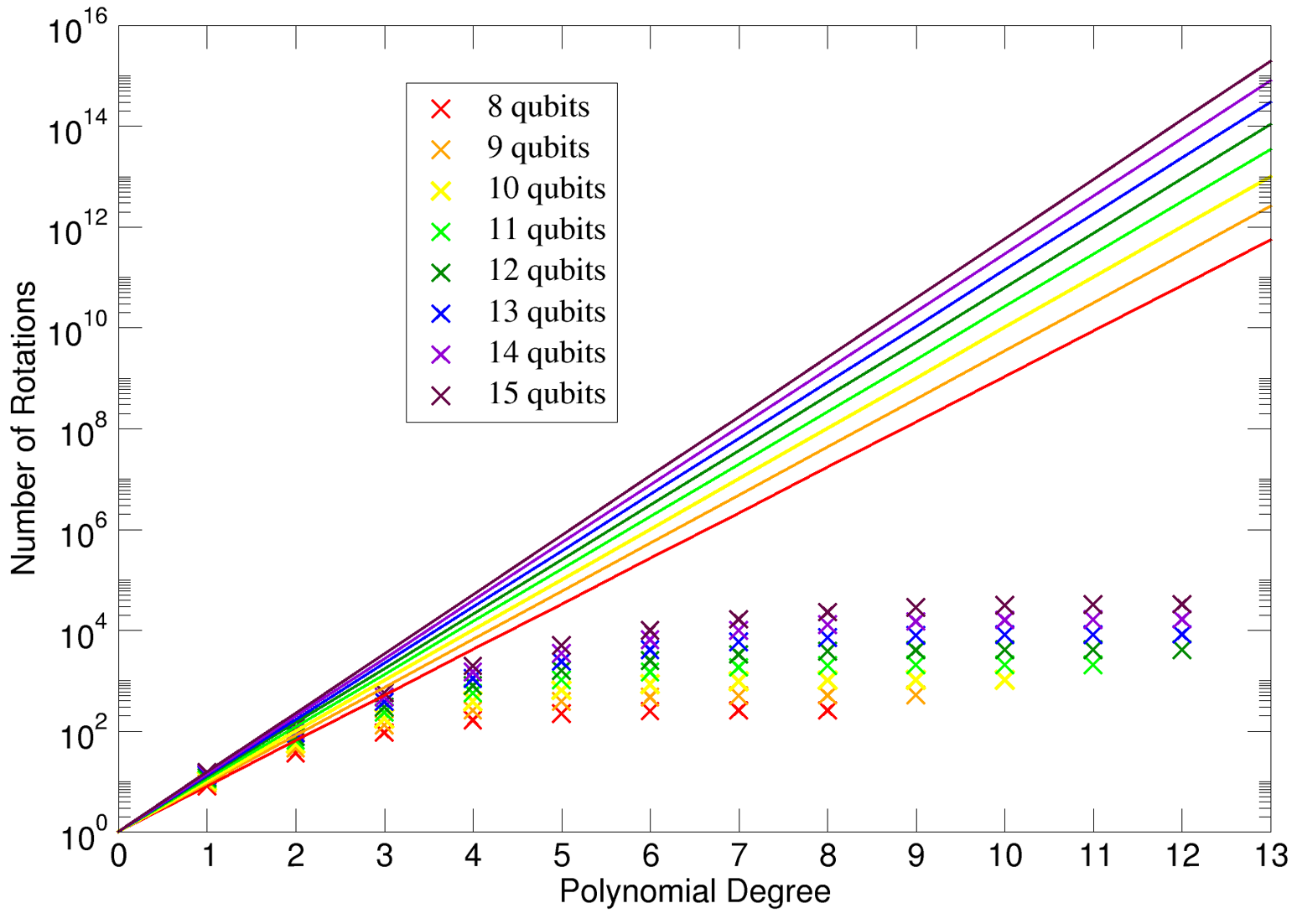}
    \caption{Gate scaling comparison between previous polynomial expansion (Extending from Benenti and Strini~\cite{benenti2008quantum} (lines) and the present compression (Xs). Left panel: gate count vs.\ polynomial degree. Right panel: gate count vs.\ register size.}
    \label{fig:CompressionScaling}
\end{figure}

Extending to octic (degree-8) polynomials, the compressed circuit requires 259, 514, and 1016 gates for $n=8,9,10$ qubits, respectively, compared to $1.68\times10^7$, $4.30\times10^7$, and $10^8$ gates with the older polynomial-complexity method. The scaling of this compression is highlighted in Figure~\ref{fig:CompressionScaling}.

\section{Trigonometric Functions}
\label{sec:trig-functions}

\subsection{Cosine Function}

\subsubsection{Complex Exponential Method}
Using Euler's formula,
\begin{equation}
\cos(x) = \mathrm{Re}(e^{ix}) = \mathrm{Re}\left(\exp\left(i\sum_{j=0}^{n-1} 2^j q_j\right)\right)\,.
\end{equation}
This can be expressed as
\begin{equation}
\cos(x) = \mathrm{Re}\left(\prod_{j=0}^{n-1} \exp(i \cdot 2^j \cdot q_j)\right)\,.
\end{equation}
When $q_j = 0$, the factor is $\exp(i \cdot 2^j \cdot q_j) = 1$; when $q_j = 1$, it becomes $\cos(2^j) + i\sin(2^j)$. Therefore,
\begin{equation}
\cos(x) = \mathrm{Re}\left(\prod_{j=0}^{n-1} [1 + q_j(\cos(2^j) + i\sin(2^j) - 1)]\right)\,.
\end{equation}

\subsubsection{Explicit Form}
\begin{equation}
\begin{split}
    \cos(x) &= \sum_{S \subseteq \{0,\ldots,n-1\}} (-1)^{\lfloor|S|/2\rfloor} \left[\prod_{j \in S_{\mathrm{even}}} \sin(2^j)\right] \\
    &\quad\times\left[\prod_{j \in S_{\mathrm{odd}}} \cos(2^j)\right] \left[\prod_{j \in S} q_j\right] \left[\prod_{j \notin S} (1-q_j)\right]\,.
\end{split}
\end{equation}

\subsection{Sine Function}

Using the imaginary part of Euler's formula,
\begin{equation}
\sin(x) = \mathrm{Im}(e^{ix}) = \mathrm{Im}\left(\prod_{j=0}^{n-1} \exp(i \cdot 2^j \cdot q_j)\right)\,.
\end{equation}
Expanding,
\begin{equation}
\sin(x) = \mathrm{Im}\left(\prod_{j=0}^{n-1} [1 + q_j(\cos(2^j) + i\sin(2^j) - 1)]\right)\,.
\end{equation}

\subsubsection{Explicit Form}
\begin{equation}
\begin{split}
 \sin(x) &= \sum_{S \subseteq \{0,\ldots,n-1\}} (-1)^{\lfloor(|S|-1)/2\rfloor} \left[\prod_{j \in S_{\mathrm{odd}}} \sin(2^j)\right] \\
 &\quad\times\left[\prod_{j \in S_{\mathrm{even}}} \cos(2^j)\right] \left[\prod_{j \in S} q_j\right] \left[\prod_{j \notin S} (1-q_j)\right]\,.
\end{split}
\end{equation}

\section{Finite Difference Operators}
\label{sec:finite-diff}

For more general functions, the finite difference operator provides a systematic framework. For a single bit $j$,
\begin{equation}
\Delta_j f(x) = f(x + 2^j) - f(x)\,,
\end{equation}
For a subset $S \subseteq \{0, 1, \ldots, n-1\}$, the multi-index finite difference operator is
\begin{equation}
\Delta_S f(x) = \prod_{j \in S} \Delta_j f(x)\,.
\end{equation}
Any function $f$ on the finite grid can then be written as
\begin{equation}\label{eq:finite-diff-expansion}
f(x) = \sum_{S \subseteq \{0,1,\ldots,n-1\}} (\Delta_S f)(0) \prod_{j \in S} q_j\,.
\end{equation}

\subsection{Tangent Function}

Expanding the product of difference operators for the tangent,
\begin{equation}
(\Delta_S \tan)(0) = \sum_{T \subseteq S} (-1)^{|S|-|T|} \tan\left(\sum_{j \in T} 2^j\right)\,,
\end{equation}
where $|S|$ and $|T|$ are the respective cardinalities of subsets $S$ and $T$. Therefore,
\begin{equation}
\tan(x) = \sum_{S \subseteq \{0,1,\ldots,n-1\}} \left[\sum_{T \subseteq S} (-1)^{|S|-|T|} \tan\left(\sum_{j \in T} 2^j\right)\right] \prod_{j \in S} q_j\,,
\end{equation}
For any subset $S = \{j_1, j_2, \ldots, j_m\}$, the coefficient is
\begin{equation}
c_S = \sum_{k=0}^{m} (-1)^{m-k} \sum_{\substack{T \subseteq S \\ |T| = k}} \tan\left(\sum_{j \in T} 2^j\right)\,.
\end{equation}

\subsection{Cosine and Sine Squared Functions}

Using the double-angle identity $\cos^2(x) = \frac{1+\cos(2x)}{2}$,
\begin{equation}
\cos^2(x) = \frac{1}{2}\left(1 + \mathrm{Re}\left(\prod_{j=0}^{n-1} [1 + q_j(\cos(2^{j+1}) + i\sin(2^{j+1}) - 1)]\right)\right)\,.
\end{equation}
Similarly, $\sin^2(x) = \frac{1-\cos(2x)}{2}$ gives
\begin{equation}
\sin^2(x) = \frac{1}{2}\left(1 - \mathrm{Re}\left(\prod_{j=0}^{n-1} [1 + q_j(\cos(2^{j+1}) + i\sin(2^{j+1}) - 1)]\right)\right)\,.
\end{equation}

\subsection{Tangent Squared Function}

\begin{equation}
\tan^2(x) = \sum_{S \subseteq \{0,1,\ldots,n-1\}} \left[\sum_{T \subseteq S} (-1)^{|S|-|T|} \tan^2\left(\sum_{j \in T} 2^j\right)\right] \prod_{j \in S} q_j\,.
\end{equation}

\subsection{Pythagorean Identity}

The identity $\cos^2(x) + \sin^2(x) = 1$ implies that the finite difference coefficients satisfy
\begin{equation}
(\Delta_S \cos^2)(0) + (\Delta_S \sin^2)(0) = \begin{cases}
1 & \text{if } S = \emptyset\,, \\
0 & \text{if } S \neq \emptyset\,.
\end{cases}
\end{equation}
Through the double-angle formulas, these relate to the cosine finite differences as
\begin{align}
(\Delta_S \cos^2)(0) &= \tfrac{1}{2}[(\Delta_S 1)(0) + (\Delta_S \cos(2\,\cdot\,))(0)]\,, \\
(\Delta_S \sin^2)(0) &= \tfrac{1}{2}[(\Delta_S 1)(0) - (\Delta_S \cos(2\,\cdot\,))(0)]\,.
\end{align}

\section{Hyperbolic Functions}
\label{sec:hyperbolic}

\subsection{Hyperbolic Cosine}

Using $\cosh(x) = \frac{e^x + e^{-x}}{2}$ with the exponential decomposition
\begin{equation}
e^{\pm x} = \prod_{j=0}^{n-1} [1 + q_j(e^{\pm 2^j} - 1)]\,,
\end{equation}
the hyperbolic cosine becomes
\begin{equation}
\cosh(x) = \frac{1}{2}\left(\prod_{j=0}^{n-1} [1 + q_j(e^{2^j} - 1)] + \prod_{j=0}^{n-1} [1 + q_j(e^{-2^j} - 1)]\right)\,.
\end{equation}
Using $\cosh^2(x) = \frac{1+\cosh(2x)}{2}$ and $e^{\pm 2x} = \prod_{j=0}^{n-1} [1 + q_j(e^{\pm 2^{j+1}} - 1)]$,
\begin{equation}
\cosh^2(x) = \frac{1}{2} + \frac{1}{4}\left(\prod_{j=0}^{n-1} [1 + q_j(e^{2^{j+1}} - 1)] + \prod_{j=0}^{n-1} [1 + q_j(e^{-2^{j+1}} - 1)]\right)\,.
\end{equation}

\subsection{Hyperbolic Sine}

Using $\sinh(x) = \frac{e^x - e^{-x}}{2}$,
\begin{equation}
\sinh(x) = \frac{1}{2}\left(\prod_{j=0}^{n-1} [1 + q_j(e^{2^j} - 1)] - \prod_{j=0}^{n-1} [1 + q_j(e^{-2^j} - 1)]\right)\,.
\end{equation}
Using $\sinh^2(x) = \frac{\cosh(2x) -1}{2}$,
\begin{equation}
\sinh^2(x) = \frac{1}{4}\left(\prod_{j=0}^{n-1} [1 + q_j(e^{2^{j+1}} - 1)] + \prod_{j=0}^{n-1} [1 + q_j(e^{-2^{j+1}} - 1)]\right) - \frac{1}{2}\,.
\end{equation}

\section{Exponential and Gaussian Functions}
\label{sec:exponential}

\subsection{Basic Exponential}

\begin{equation}
e^x = \prod_{j=0}^{n-1} e^{2^j q_j} = \prod_{j=0}^{n-1} [1 + q_j(e^{2^j} - 1)]\,.
\end{equation}

\subsection{Gaussian Functions}

For the function $f(x) = e^{x^2}$, the finite difference expansion gives
\begin{equation}
e^{x^2} = \sum_{S \subseteq \{0,1,\ldots,n-1\}} \left[\sum_{T \subseteq S} (-1)^{|S|-|T|} \exp\left(\left(\sum_{j \in T} 2^j\right)^{\!2}\right)\right] \prod_{j \in S} q_j\,,
\end{equation}
For the negative Gaussian $e^{-x^2}$,
\begin{equation}
e^{-x^2} = \sum_{S \subseteq \{0,1,\ldots,n-1\}} \left[\sum_{T \subseteq S} (-1)^{|S|-|T|} \exp\left(-\left(\sum_{j \in T} 2^j\right)^{\!2}\right)\right] \prod_{j \in S} q_j\,.
\end{equation}

Alternatively, using the quadratic decomposition $x^2 = \sum_{j=0}^{n-1}2^{2j}q_j + 2\sum_{j<\ell}2^{j+\ell}q_jq_\ell$, the Gaussian can be factored as
\begin{equation}
    e^{-x^2} = \prod_{j=0}^{n-1}\left[1+q_j(e^{-2^{2j}}-1)\right]\prod_{j<\ell}\left[1+q_jq_\ell(e^{-2^{j+\ell+1}}-1)\right]\,.
\end{equation}

\section{Algebraic Functions}
\label{sec:algebraic}

\subsection{Square Root}
\begin{equation}
\sqrt{x} = \sum_{S \subseteq \{0,1,\ldots,n-1\}} \left[\sum_{T \subseteq S} (-1)^{|S|-|T|} \sqrt{\sum_{j \in T} 2^j}\right] \prod_{j \in S} q_j\,,
\end{equation}
For example, with two qubits: $\sqrt{x} = q_0 + \sqrt{2}\, q_1 + (\sqrt{3} - \sqrt{2} - 1)\, q_0 q_1$.

\subsection{Square Root of $(1+x)$ and $(1+x^2)$}
\begin{align}
\sqrt{1+x} &= \sum_{S \subseteq \{0,1,\ldots,n-1\}} \left[\sum_{T \subseteq S} (-1)^{|S|-|T|} \sqrt{1 + \sum_{j \in T} 2^j}\right] \prod_{j \in S} q_j\,, \\
\sqrt{1+x^2} &= \sum_{S \subseteq \{0,1,\ldots,n-1\}} \left[\sum_{T \subseteq S} (-1)^{|S|-|T|} \sqrt{1 + \left(\sum_{j \in T} 2^j\right)^{\!2}}\right] \prod_{j \in S} q_j\,.
\end{align}

\subsection{Inverse Square Root}
For $x \neq 0$,
\begin{equation}
\frac{1}{\sqrt{x}} = \sum_{S \subseteq \{0,1,\ldots,n-1\}} \left[\sum_{T \subseteq S,\, T \neq \emptyset} (-1)^{|S|-|T|} \frac{1}{\sqrt{\sum_{j \in T} 2^j}}\right] \prod_{j \in S} q_j\,.
\end{equation}
The variants $1/\sqrt{1+x}$ and $1/\sqrt{1+x^2}$ follow analogously with the appropriate argument substitutions.

\subsection{Two-Variable Euclidean Norm}

For $f(x,y) = \sqrt{x^2 + y^2}$ with $x = \sum_{j=0}^{n-1} 2^j q_j^{(x)}$ and $y = \sum_{i=0}^{m-1} 2^i q_i^{(y)}$,
\begin{equation}
f(x,y) = \sum_{S_x \subseteq \{0,\ldots,n-1\}} \sum_{S_y \subseteq \{0,\ldots,m-1\}} c_{S_x,S_y} \prod_{j \in S_x} q_j^{(x)} \prod_{i \in S_y} q_i^{(y)}\,,
\end{equation}
where
\begin{equation}
c_{S_x,S_y} = \sum_{T_x \subseteq S_x} \sum_{T_y \subseteq S_y} (-1)^{|S_x|-|T_x|+|S_y|-|T_y|} \sqrt{\left(\sum_{j \in T_x} 2^j\right)^{\!2} + \left(\sum_{i \in T_y} 2^i\right)^{\!2}}\,.
\end{equation}

\subsection{Quantum Circuit Implementation}

All functions presented above share the common form
\begin{equation}
f(x) = \sum_{S} c_S \prod_{j \in S} q_j\,,
\end{equation}
suitable for multi-controlled rotations in which each term maps to a single gate, linear combination of unitaries (LCU) for the sum, and amplitude or phase encoding for the output.

\section{Non-Integer Discretization}
\label{sec:non-integer}

\subsection{General Discretization with $\Delta x = L/2^n$}

When discretizing a box of size $L$ with $2^n$ points, the grid positions become
\begin{equation}
x = \Delta x \cdot \sum_{j=0}^{n-1} 2^j q_j = \frac{L}{2^n} \sum_{j=0}^{n-1} 2^j q_j\,,
\end{equation}
giving $x \in \{0, \Delta x, 2\Delta x, \ldots, (2^n-1)\Delta x\}$ with $\Delta x = L/2^n$.

The finite difference operator with respect to bit $j$ becomes
\begin{equation}
\Delta_j f(x) = f(x + 2^j \Delta x) - f(x)\,,
\end{equation}
and the general formula for any function $f(x)$ on the discretized grid is
\begin{equation}\label{eq:general-disc}
f(x) = \sum_{S \subseteq \{0,1,\ldots,n-1\}} \left[\sum_{T \subseteq S} (-1)^{|S|-|T|} f\left(\frac{L}{2^n} \sum_{j \in T} 2^j\right)\right] \prod_{j \in S} q_j\,.
\end{equation}

\subsection{Example: $\cos(x)$ with $L = 2\pi$, $n = 3$}

Here $\Delta x = \pi/4$, so $x \in \{0, \pi/4, \pi/2, 3\pi/4, \pi, 5\pi/4, 3\pi/2, 7\pi/4\}$. The coefficients are
\begin{align}
(\Delta_\emptyset \cos)(0) &= \cos(0) = 1\,, \\
(\Delta_{\{0\}} \cos)(0) &= \cos(\pi/4) - \cos(0) = \frac{\sqrt{2}}{2} - 1\,, \\
(\Delta_{\{1\}} \cos)(0) &= \cos(\pi/2) - \cos(0) = -1\,, \\
(\Delta_{\{2\}} \cos)(0) &= \cos(\pi) - \cos(0) = -2\,, \\
(\Delta_{\{0,1\}} \cos)(0) &= \cos(3\pi/4) - \cos(\pi/2) - \cos(\pi/4) + \cos(0) = 1 - \sqrt{2}\,.
\end{align}

\subsection{Scaling and Transformations}

For a combined transformation $f(\alpha x + \beta)$ on a grid with spacing $\Delta x = L/2^n$, the coefficients become
\begin{equation}
c_S = \sum_{T \subseteq S} (-1)^{|S|-|T|} f\left(\beta + \alpha \frac{L}{2^n} \sum_{j \in T} 2^j\right)\,.
\end{equation}
Only the classically precomputed coefficients $c_S$ change with discretization and scaling; the quantum circuit structure remains unchanged.

\subsection{Sum of Functions}

For a sum of functions $f(x) = \sum_if_i(x)$, the circuit depth does not increase. By the commutativity of tensor product spaces,
\begin{equation}
    \begin{split}
        f(x) &= \sum_i\left(\sum_{S}{c_S}_i\prod_{j\in S}q_j\right) = \sum_S\left(\sum_{i}{c_S}_i\right)\prod_{j\in S}q_j\,.
    \end{split}
\end{equation}
It suffices to sum the coefficients for each gate. The number of gates remains $\sum_{i=0}^n \binom{n}{i} = 2^n$, with the number of layers scaling polynomially.

\section{Inverse Polynomial Potentials}
\label{sec:inverse-poly}

\subsection{Construction of Scattering Surfaces in a Binary Basis}

The primary challenge for a function such as $f(x) = 1/x$ is that direct substitution of the binary expansion into the denominator,
\begin{equation}
    f(x) = \frac{1}{x} = \frac{1}{\sum_j^{n-1}q_j2^j}\,,
\end{equation}
does not lend itself to a product-of-qubit-variable form. Instead, the multilinear expansion is asserted:
\begin{equation}
    \frac{1}{x} = \sum_{S\subseteq \{0,1,\ldots,n-1\}}a_S\prod_{j \in S}q_j\,,
\end{equation}
where the coefficients are given by M\"obius inversion,
\begin{equation}
    a_S = \sum_{T \subseteq S}(-1)^{|S| - |T|}\cdot\frac{1}{\sum_{j\in T}2^j}\,,
\end{equation}
with $\sum_{j \in \varnothing}2^j = 0$ by convention.

These products are exponentiated as phase rotations of a unitary $U = e^{2\pi i\alpha/x}$, equivalently represented as controlled $R_z$ rotations:
\begin{equation}
    U = \prod_{S\subseteq \{0, 1, \ldots, n-1\}}C^{|S|}R_z(2\pi \alpha\, a_S)\,,
\end{equation}

For $n=2$ qubits, the coefficients are
\begin{equation}
    \begin{split}
        a_\varnothing & = 0\,, \qquad
        a_{\{0\}} = 1\,, \qquad
        a_{\{1\}} = \frac{1}{2}\,, \\
        a_{\{0,1\}} & = 0 - 1 - \frac{1}{2} + \frac{1}{3} = -\frac{7}{6}\,.
    \end{split}
\end{equation}

The unitary for the Rutherford potential $V(x) = 1/x$ is therefore
\begin{equation}
    U=e^{-\frac{i}{\hbar} V(x)} = e^{-\frac{i}{\hbar}\sum_{S\subseteq \{0, 1, \ldots, n-1\}}\left(\sum_{T \subseteq S}(-1)^{|S|-|T|}\cdot \frac{1}{\sum_{j \in T}2^j}\right)\prod_{j\in S}q_j}\,.
\end{equation}
The coefficients $a_S$ are computed combinatorially, with the number of gates scaling exponentially with register size. Depth compression may allow $\lfloor n/k\rfloor$ $k$-qubit operations to be performed in parallel.

\subsection{Centrifugal Potentials}

The centrifugal potential $V_{\mathrm{cent}} = \frac{\ell(\ell+1)}{2\mu r^2}$ requires a circuit for $1/x^2$. Squaring the multilinear expansion for $1/x$ and using $q_j^2 = q_j$ gives
\begin{equation}
\left(\frac{1}{x}\right)^{\!2} = \sum_{T \subseteq \{0,1,\ldots,n-1\}} b_T^{(2)} \prod_{j \in T} q_j\,,
\end{equation}
where
\begin{equation}
b_T^{(2)} = \sum_{\substack{S_1, S_2 \subseteq \{0,1,\ldots,n-1\} \\ S_1 \cup S_2 = T}} a_{S_1} \cdot a_{S_2}\,,
\end{equation}
Since all coefficients $a_S$ and $a_{S'}$ are absorbed into the same $|S|$-fold controlled $R_z$ gate when $S = S'$, an $x^{-2}$ circuit requires no more gates than $x^{-1}$.

The unitary for the centrifugal potential is therefore
\begin{equation}
    U = e^{-i\tau \ell(\ell+1)/(2\mu r^2)} = e^{-i\tau (\ell(\ell+1)/(2\mu))  \sum_{T} b_T^{(2)} \prod_{j \in T} q_j}\,.
\end{equation}

\section{Lennard-Jones and Higher Inverse Powers}
\label{sec:lennard-jones}

By induction, for any positive integer $k$,
\begin{equation}
\left(\frac{1}{x}\right)^{\!k} = \sum_{T \subseteq \{0,1,\ldots,n-1\}} b_T^{(k)} \prod_{j \in T} q_j\,,
\end{equation}
where
\begin{equation}
b_T^{(k)} = \sum_{\substack{S_1, \ldots, S_k \subseteq \{0,1,\ldots,n-1\} \\ S_1 \cup \cdots \cup S_k = T}} a_{S_1} \cdots a_{S_k}\,.
\end{equation}

\subsection{Repeated Squaring Algorithm}

For efficiency, the power $k$ is decomposed into powers of 2: $k = \sum_{i=0}^{\lfloor \log_2 k \rfloor} b_i 2^i$, $b_i \in \{0,1\}$. The coefficients for each power of two are computed iteratively:
\begin{align}
(1/x)^1 &= \sum_{S} a_S \prod_{j \in S} q_j\,, \\
(1/x)^{2^{i+1}} &= \left[(1/x)^{2^{i}}\right]^2\,,
\end{align}
where squaring produces new coefficients via the convolution
\begin{equation}
d_T = \sum_{\substack{S_1, S_2 \subseteq \{0,1,\ldots,n-1\} \\ S_1 \cup S_2 = T}} c_{S_1} \cdot c_{S_2}\,,
\end{equation}
For example, $(1/x)^{12}$ with decomposition $12 = 2^3 + 2^2$ requires computing $(1/x)^4$ and $(1/x)^8$ via repeated squaring and then multiplying these two expansions together. A Lennard-Jones potential $V(r) = 4\epsilon [(\sigma/r)^{12}-(\sigma/r)^{6}]$ can therefore be constructed by computing $b_T^{(6)}$ and $b_T^{(12)}$ and combining them with appropriate scaling.

\section{One-Dimensional Finite Difference Expansion}
\label{sec:1d-expansion}

Any function $f$ on the $n$-bit grid admits the exact expansion
\begin{equation}\label{eq:1d-expansion}
    f(x) = \sum_{S\subseteq\{0,\dots,n-1\}} (\Delta_S f)(0) \prod_{j\in S} q_j\,,
\end{equation}
where the coefficients are given by inclusion--exclusion:
\begin{equation}\label{eq:1d-coeff}
    (\Delta_S f)(0) = \sum_{A\subseteq S} (-1)^{|S\setminus A|}\, f\!\left(\sum_{j\in A} 2^j\right)\,.
\end{equation}
Each term in Eq.~\ref{eq:1d-expansion} corresponds to a multi-controlled $R_z$ rotation in the quantum circuit: the coefficient $(\Delta_S f)(0)$ sets the rotation angle, the qubits $\{q_j : j\in S\setminus\{\min S\}\}$ serve as controls, and $q_{\min S}$ is the target. There are at most $2^n$ such gates.

\section{Two-Variable Expansion: The Inseparable Case}
\label{sec:two-var-inseparable}

\subsection{Register Setup}

Let $f\colon \{0,\dots,2^n-1\}\times\{0,\dots,2^m-1\}\to\mathbb{R}$ be an arbitrary function of two discrete variables, encoded as:
\begin{itemize}
    \item $x$ in an $n$-qubit register $\{q_0,\dots,q_{n-1}\}$, so $x = \sum_{j=0}^{n-1} q_j\, 2^j$,
    \item $y$ in an $m$-qubit register $\{r_0,\dots,r_{m-1}\}$, so $y = \sum_{k=0}^{m-1} r_k\, 2^k$.
\end{itemize}

\subsection{Two-Variable Finite Difference Operators}

    The shifts in each variable are defined independently:
    \begin{align}
        \Delta_j^{(x)} f(x,y) &= f(x+2^j,\, y) - f(x,y)\,, \qquad j\in\{0,\dots,n{-}1\}\,,\label{eq:delta-x}\\[4pt]
        \Delta_k^{(y)} f(x,y) &= f(x,\, y+2^k) - f(x,y)\,, \qquad k\in\{0,\dots,m{-}1\}\,.\label{eq:delta-y}
    \end{align}

Since these operators act on different arguments, they commute: $\Delta_j^{(x)}\Delta_k^{(y)} = \Delta_k^{(y)}\Delta_j^{(x)}$ for all $j,k$. This allows the mixed multi-index difference to be defined unambiguously.

    For subsets $S\subseteq\{0,\dots,n{-}1\}$ and $T\subseteq\{0,\dots,m{-}1\}$, a mixed multi-index difference is given by:
    \begin{equation}\label{eq:mixed-diff-2d}
        \Delta_S^{(x)}\Delta_T^{(y)} f(x,y) = \prod_{j\in S}\Delta_j^{(x)}\;\prod_{k\in T}\Delta_k^{(y)}\; f(x,y)\,.
    \end{equation}

\subsection{Explicit Coefficient Formula}

    Expanding each $\Delta_j = (\mathrm{shift}_j - \mathrm{id})$ as a signed sum over subsets and evaluating at the origin, the coefficients are given as:
    \begin{equation}\label{eq:app_2d-coeff}
        c_{S,T} \;\equiv\; \bigl(\Delta_S^{(x)}\Delta_T^{(y)} f\bigr)(0,0) \;=\; \sum_{A\subseteq S}\sum_{B\subseteq T}(-1)^{|S\setminus A|+|T\setminus B|}\; f\!\left(\sum_{j\in A}2^j,\;\sum_{k\in B}2^k\right).
    \end{equation}

    Expanding $\Delta_S^{(x)}$ first,
    \begin{equation}
        \Delta_S^{(x)} = \prod_{j\in S}(\mathrm{shift}_j^{(x)} - \mathrm{id}) = \sum_{A\subseteq S}(-1)^{|S\setminus A|}\,\mathrm{shift}_A^{(x)}\,,
    \end{equation}
    where $\mathrm{shift}_A^{(x)}$ translates $x\mapsto x+\sum_{j\in A}2^j$. Similarly for $\Delta_T^{(y)}$. Since the shifts in different variables commute, the composition gives:
    \begin{equation}
        \Delta_S^{(x)}\Delta_T^{(y)}f(x,y) = \sum_{A\subseteq S}\sum_{B\subseteq T}(-1)^{|S\setminus A|+|T\setminus B|}\;f\!\left(x+\sum_{j\in A}2^j,\; y+\sum_{k\in B}2^k\right)\,.
    \end{equation}
    Setting $x=y=0$ yields Eq.~\ref{eq:app_2d-coeff}.

\subsection{Full Two-Variable Expansion}

    Any function $f$ on the grid $\{0,\dots,2^n{-}1\}\times\{0,\dots,2^m{-}1\}$ admits the exact expansion:
    \begin{equation}\label{eq:app_2d-expansion}
        f(x,y) = \sum_{\substack{S\subseteq\{0,\dots,n-1\}\\[2pt]T\subseteq\{0,\dots,m-1\}}} c_{S,T}\;\prod_{j\in S}q_j\;\prod_{k\in T}r_k,
    \end{equation}
    with $c_{S,T}$ given by Eq.~\ref{eq:app_2d-coeff}. There are $2^n\cdot 2^m = 2^{n+m}$ terms.

The set of monomials $\{\prod_{j\in S}q_j\prod_{k\in T}r_k\}_{S,T}$ forms a basis for all functions on $\{0,1\}^{n+m}$ (the multilinear polynomials over GF(2) extended to $\mathbb{R}$). Since the binary encoding maps the grid bijectively to $\{0,1\}^{n+m}$, any function on the grid has a unique expansion in this basis. The coefficients are determined by the requirement that the expansion reproduce $f$ at all grid points, equivalent to Eq.~\ref{eq:app_2d-coeff} by the M\"obius inversion formula on the Boolean lattice.

\subsection{Circuit Interpretation}

Each monomial $\prod_{j\in S}q_j\prod_{k\in T}r_k$ maps to a single multi-controlled $R_z(c_{S,T})$ gate with controls drawn from both registers. The total gate count is at most $2^{n+m}$.

\subsection{Verification: $n=m=1$}

\begin{example}[$n=m=1$]\label{ex:2d-1bit}
    With single-bit registers, there are four function values and four coefficients:

    \begin{center}
    \begin{tabular}{cccc}
        \toprule
        $S$ & $T$ & $c_{S,T}$ & Monomial \\
        \midrule
        $\emptyset$ & $\emptyset$ & $f(0,0)$ & $1$ \\[3pt]
        $\{0\}$ & $\emptyset$ & $f(1,0)-f(0,0)$ & $q_0$ \\[3pt]
        $\emptyset$ & $\{0\}$ & $f(0,1)-f(0,0)$ & $r_0$ \\[3pt]
        $\{0\}$ & $\{0\}$ & $f(1,1)-f(0,1)-f(1,0)+f(0,0)$ & $q_0\, r_0$ \\
        \bottomrule
    \end{tabular}
    \end{center}

    Substituting $(q_0,r_0)=(1,1)$:
    \begin{align}
        &f(0,0) + [f(1,0)-f(0,0)]\cdot 1 + [f(0,1)-f(0,0)]\cdot 1 \notag\\
        &\quad+ [f(1,1)-f(0,1)-f(1,0)+f(0,0)]\cdot 1\cdot 1 = f(1,1)\,. \quad\checkmark
    \end{align}
\end{example}

\subsection{Inseparable Example}

\begin{example}[Inseparable function, $n=m=1$]\label{ex:2d-inseparable}
    Let $f(x,y) = xy + x + 2y + 3$, evaluated on the grid:
    \begin{equation}
        f(0,0)=3,\quad f(1,0)=4,\quad f(0,1)=5,\quad f(1,1)=7\,.
    \end{equation}
    Computing the coefficients:
    \begin{align}
        c_{\emptyset,\emptyset} &= f(0,0) = 3\,,&
        c_{\{0\},\emptyset} &= f(1,0)-f(0,0) = 1\,,\\
        c_{\emptyset,\{0\}} &= f(0,1)-f(0,0) = 2\,,&
        c_{\{0\},\{0\}} &= f(1,1)-f(0,1)-f(1,0)+f(0,0) = 1\,.
    \end{align}
    The expansion is $f(x,y) = 3 + q_0 + 2r_0 + q_0 r_0$.
    The nonzero cross-term $c_{\{0\},\{0\}}=1$ reflects the inseparable $xy$ interaction. The circuit requires four gates: one unconditional $R_z(3)$, two singly controlled gates $R_z(1)$ and $R_z(2)$, and one doubly controlled gate $R_z(1)$ with controls on both $q_0$ and $r_0$.
\end{example}

\section{Two-Variable Separable Case and Recovery}
\label{sec:two-var-separable}

For a separable function $f(x,y)=f(x)g(y)$, each factor has its own single-variable expansion:
\begin{align}
    f(x) &= \sum_{S\subseteq\{0,\dots,n-1\}} (\Delta_S f)(0)\prod_{j\in S}q_j\,,\\
    g(y) &= \sum_{T\subseteq\{0,\dots,m-1\}} (\Delta_T g)(0)\prod_{k\in T}r_k\,.
\end{align}
Taking the product directly:
\begin{equation}\label{eq:2d-separable}
    f(x)\,g(y) = \sum_{S,T}\bigl[(\Delta_S f)(0)\cdot(\Delta_T g)(0)\bigr]\;\prod_{j\in S}q_j\;\prod_{k\in T}r_k\,.
\end{equation}
    If $f(x,y) = f(x)\,g(y)$, then the general coefficients Eq.~\ref{eq:app_2d-coeff} factor:
    \begin{equation}\label{eq:2d-factor}
        c_{S,T} = (\Delta_S f)(0)\cdot(\Delta_T g)(0)\,.
    \end{equation}

    Substituting $f(x,y)\to f(x)\,g(y)$ in Eq.~\ref{eq:app_2d-coeff}, the sign factors as $(-1)^{|S\setminus A|+|T\setminus B|} = (-1)^{|S\setminus A|}\cdot(-1)^{|T\setminus B|}$. Since each factor in the summand depends on only one summation variable, the double sum separates:
    \begin{equation}
        c_{S,T} = \underbrace{\left[\sum_{A\subseteq S}(-1)^{|S\setminus A|}\,f\!\left(\sum_{j\in A}2^j\right)\right]}_{=\,(\Delta_S f)(0)}\;\underbrace{\left[\sum_{B\subseteq T}(-1)^{|T\setminus B|}\,g\!\left(\sum_{k\in B}2^k\right)\right]}_{=\,(\Delta_T g)(0)}\,.
    \end{equation}

\begin{example}[Separable function, $n=m=1$]\label{ex:2d-separable}
    Let $f(x,y)=(x+1)(y+2)$ with $f(x)=x+1$, $g(y)=y+2$. Function values:
    \begin{equation}
        f(0,0)=2,\quad f(1,0)=4,\quad f(0,1)=3,\quad f(1,1)=6\,.
    \end{equation}
    Single-variable coefficients: $(\Delta_\emptyset f)(0) = 1$, $(\Delta_{\{0\}} f)(0) = 1$, $(\Delta_\emptyset g)(0) = 2$, $(\Delta_{\{0\}} g)(0) = 1$.
    The expansion $f(x,y)= 2+2q_0+r_0+q_0r_0$ factors as $(1+q_0)(2+r_0)$.\quad$\checkmark$
\end{example}

\section{Three-Variable Expansion: The Inseparable Case}
\label{sec:three-var-inseparable}

\subsection{Register Setup}

Let $f\colon \{0,\dots,2^n{-}1\}\times\{0,\dots,2^m{-}1\}\times\{0,\dots,2^\ell{-}1\}\to\mathbb{R}$. The variables are encoded as:
\begin{itemize}
    \item $x$ in $n$ qubits $\{q_0,\dots,q_{n-1}\}$,
    \item $y$ in $m$ qubits $\{r_0,\dots,r_{m-1}\}$,
    \item $z$ in $\ell$ qubits $\{s_0,\dots,s_{\ell-1}\}$.
\end{itemize}

\subsection{Three-Variable Finite Difference Operators}

A partial finite difference in three variables can be written as:
    \begin{align}
        \Delta_j^{(x)}f(x,y,z) &= f(x+2^j,y,z)-f(x,y,z)\,, \quad j\in\{0,\dots,n{-}1\}\,,\\
        \Delta_k^{(y)}f(x,y,z) &= f(x,y+2^k,z)-f(x,y,z)\,, \quad k\in\{0,\dots,m{-}1\}\,,\\
        \Delta_p^{(z)}f(x,y,z) &= f(x,y,z+2^p)-f(x,y,z)\,, \quad p\in\{0,\dots,\ell{-}1\}\,.
    \end{align}

All operators commute pairwise since they act on different arguments.

\subsection{Explicit Coefficient Formula}

    Evaluating the mixed difference at the origin:
    \begin{equation}\label{eq:3d-coeff}
        c_{S,T,U} = \sum_{A\subseteq S}\sum_{B\subseteq T}\sum_{C\subseteq U}(-1)^{|S\setminus A|+|T\setminus B|+|U\setminus C|}\;f\!\left(\sum_{j\in A}2^j,\;\sum_{k\in B}2^k,\;\sum_{p\in C}2^p\right).
    \end{equation}

The construction is similar to the two-variable expansion, expanding each product of shift-minus-identity operators and using their commutativity.

\subsection{Full Three-Variable Expansion}

    Any function $f$ on the grid $\{0,\dots,2^n{-}1\}\times\{0,\dots,2^m{-}1\}\times\{0,\dots,2^\ell{-}1\}$ admits:
    \begin{equation}\label{eq:3d-expansion}
        {f(x,y,z) = \sum_{\substack{S\subseteq\{0,\dots,n-1\}\\[2pt]T\subseteq\{0,\dots,m-1\}\\[2pt]U\subseteq\{0,\dots,\ell-1\}}} c_{S,T,U}\;\prod_{j\in S}q_j\;\prod_{k\in T}r_k\;\prod_{p\in U}s_p},
    \end{equation}
    with $c_{S,T,U}$ given by Eq.~\ref{eq:3d-coeff}. There are $2^{n+m+\ell}$ terms.

\subsection{Classification of Interactions}

The coefficients decompose naturally by their {interaction order}:

\begin{center}
\begin{tabular}{lll}
    \toprule
    Non-empty subsets & Type & Continuous analog \\
    \midrule
    Only one of $S,T,U$ & Single-variable & $\partial^{|S|}f/\partial x_S$ etc.\ \\[3pt]
    Two of $S,T,U$ & Pairwise interaction & $\partial^2 f/\partial x_S\,\partial y_T$ etc.\ \\[3pt]
    All three $S,T,U\neq\emptyset$ & Genuine three-body & $\partial^3 f/\partial x_S\,\partial y_T\,\partial z_U$ \\
    \bottomrule
\end{tabular}
\end{center}

The three-body terms measure genuine trilinear correlation that cannot be decomposed into any pairwise product structure.

\subsection{Verification: $n=m=\ell=1$}

\begin{example}[$n=m=\ell=1$]\label{ex:3d-1bit}
    Writing $f_{abc}\equiv f(a,b,c)$, the eight coefficients are:

    \begin{center}
    \begin{tabular}{cccll}
        \toprule
        $S$ & $T$ & $U$ & $c_{S,T,U}$ & Monomial \\
        \midrule
        $\emptyset$ & $\emptyset$ & $\emptyset$ & $f_{000}$ & $1$ \\[3pt]
        $\{0\}$ & $\emptyset$ & $\emptyset$ & $f_{100}-f_{000}$ & $q_0$ \\[3pt]
        $\emptyset$ & $\{0\}$ & $\emptyset$ & $f_{010}-f_{000}$ & $r_0$ \\[3pt]
        $\emptyset$ & $\emptyset$ & $\{0\}$ & $f_{001}-f_{000}$ & $s_0$ \\[3pt]
        $\{0\}$ & $\{0\}$ & $\emptyset$ & $f_{110}-f_{100}-f_{010}+f_{000}$ & $q_0 r_0$ \\[3pt]
        $\{0\}$ & $\emptyset$ & $\{0\}$ & $f_{101}-f_{100}-f_{001}+f_{000}$ & $q_0 s_0$ \\[3pt]
        $\emptyset$ & $\{0\}$ & $\{0\}$ & $f_{011}-f_{010}-f_{001}+f_{000}$ & $r_0 s_0$ \\[3pt]
        $\{0\}$ & $\{0\}$ & $\{0\}$ & $f_{111}-f_{110}-f_{101}-f_{011}$ & $q_0 r_0 s_0$ \\
             &       &       & $\quad+f_{100}+f_{010}+f_{001}-f_{000}$ & \\
        \bottomrule
    \end{tabular}
    \end{center}

    Substituting any $(q_0,r_0,s_0)\in\{0,1\}^3$ and summing the active terms recovers the corresponding function value.
\end{example}

\subsection{Inseparable Example}

\begin{example}[Inseparable function, $n=m=\ell=1$]\label{ex:3d-inseparable}
    Let $f(x,y,z) = xyz + xy + z + 1$, evaluated on the vertices of the unit cube:
    \begin{equation}
        \begin{aligned}
            f_{000}&=1\,,\quad f_{100}=1\,,\quad f_{010}=1\,,\quad f_{001}=2\,,\\
            f_{110}&=2\,,\quad f_{101}=2\,,\quad f_{011}=2\,,\quad f_{111}=4\,.
        \end{aligned}
    \end{equation}
    Computing all coefficients yields the expansion
    \begin{equation}
        f(x,y,z) = 1 + s_0 + q_0 r_0 + q_0 r_0 s_0\,,
    \end{equation}
    matching the algebraic form $1 + z + xy + xyz$ upon identifying $q_0=x$, $r_0=y$, $s_0=z$. The nonzero three-body term $c_{\{0\},\{0\},\{0\}}=1$ confirms that the function is inseparable.
\end{example}

\section{Three-Variable Separable Case and Recovery}
\label{sec:three-var-separable}

    If $f(x,y,z)=f(x)\,g(y)\,h(z)$, then:
    \begin{equation}\label{eq:3d-factor}
        {c_{S,T,U} = (\Delta_S f)(0)\;\cdot\;(\Delta_T g)(0)\;\cdot\;(\Delta_U h)(0)}.
    \end{equation}

    Substituting the product form $f(x,y,z)\to f(x)\,g(y)\,h(z)$ into Eq.~\ref{eq:3d-coeff} and using the additive decomposition $(-1)^{|S\setminus A|+|T\setminus B|+|U\setminus C|} = (-1)^{|S\setminus A|}\cdot(-1)^{|T\setminus B|}\cdot(-1)^{|U\setminus C|}$, the triple sum separates since each factor depends on only one summation variable:
    \begin{equation}
        c_{S,T,U} = \underbrace{\left[\sum_{A\subseteq S}(-1)^{|S\setminus A|}\,f\!\left(\sum_{j\in A}2^j\right)\right]}_{(\Delta_S f)(0)}\;\underbrace{\left[\sum_{B\subseteq T}(-1)^{|T\setminus B|}\,g\!\left(\sum_{k\in B}2^k\right)\right]}_{(\Delta_T g)(0)}\;\underbrace{\left[\sum_{C\subseteq U}(-1)^{|U\setminus C|}\,h\!\left(\sum_{p\in C}2^p\right)\right]}_{(\Delta_U h)(0)}\,.
    \end{equation}

\subsection{Recovery of the Factored Expansion}

Substituting Eq.~\ref{eq:3d-factor} back into Eq.~\ref{eq:3d-expansion}:
\begin{equation}
    f(x)g(y)h(z) = \sum_{S,T,U}(\Delta_S f)(0)\,(\Delta_T g)(0)\,(\Delta_U h)(0)\;\prod_{j\in S}q_j\prod_{k\in T}r_k\prod_{p\in U}s_p\,,
\end{equation}
Since the sums over $S$, $T$, $U$ are independent, this rearranges as
\begin{equation}
    = \left[\sum_S (\Delta_S f)(0)\prod_{j\in S}q_j\right]\left[\sum_T (\Delta_T g)(0)\prod_{k\in T}r_k\right]\left[\sum_U (\Delta_U h)(0)\prod_{p\in U}s_p\right] = f(x)\cdot g(y)\cdot h(z)\,,
\end{equation}
where each factor is the single-variable expansion Eq.~\ref{eq:1d-expansion}.

\subsection{Explicit Verification: $n=m=\ell=1$}

\begin{example}[Full separable example, $n=m=\ell=1$]\label{ex:3d-separable-full}
    Let $f(x,y,z) = (x+1)(2y+1)(z+3)$ with $f(a)=a+1$, $g(b)=2b+1$, $h(c)=c+3$. The factored coefficients are:

    \begin{center}
    \begin{tabular}{ccccccc}
        \toprule
        $S$ & $T$ & $U$ & $\Delta_S f$ & $\Delta_T g$ & $\Delta_U h$ & $c_{S,T,U}$ \\
        \midrule
        $\emptyset$ & $\emptyset$ & $\emptyset$ & 1 & 1 & 3 & 3 \\
        $\{0\}$ & $\emptyset$ & $\emptyset$ & 1 & 1 & 3 & 3 \\
        $\emptyset$ & $\{0\}$ & $\emptyset$ & 1 & 2 & 3 & 6 \\
        $\emptyset$ & $\emptyset$ & $\{0\}$ & 1 & 1 & 1 & 1 \\
        $\{0\}$ & $\{0\}$ & $\emptyset$ & 1 & 2 & 3 & 6 \\
        $\{0\}$ & $\emptyset$ & $\{0\}$ & 1 & 1 & 1 & 1 \\
        $\emptyset$ & $\{0\}$ & $\{0\}$ & 1 & 2 & 1 & 2 \\
        $\{0\}$ & $\{0\}$ & $\{0\}$ & 1 & 2 & 1 & 2 \\
        \bottomrule
    \end{tabular}
    \end{center}

    Cross-checking the three-body term, it is seen that $c_{\{0\},\{0\},\{0\}} = 24-18-8-12+6+9+4-3 = 2$, agreeing with $1\cdot 2\cdot 1 = 2$.

    The full expansion $f(x,y,z) = 3 + 3q_0 + 6r_0 + s_0 + 6q_0r_0 + q_0s_0 + 2r_0s_0 + 2q_0r_0s_0$ factors as $(1+q_0)(1+2r_0)(3+s_0)$.
\end{example}

\section{Generalization to d-Dimensions}
\label{sec:d-dimensions}

The pattern established in two and three dimensions extends naturally. For $d$ variables $x_1,\dots,x_d$ encoded in registers of sizes $n_1,\dots,n_d$ with qubit sets $\{q_j^{(i)}\}_{j=0}^{n_i-1}$:

 For subsets $S_i\subseteq\{0,\dots,n_i{-}1\}$, $i=1,\dots,d$, define:
    \begin{equation}
        c_{S_1,\dots,S_d} = \sum_{A_1\subseteq S_1}\cdots\sum_{A_d\subseteq S_d}(-1)^{\sum_{i=1}^d |S_i\setminus A_i|}\;f\!\left(\sum_{j\in A_1}2^j,\;\dots,\;\sum_{j\in A_d}2^j\right)\,.
    \end{equation}
    Then:
    \begin{equation}\label{eq:app_d-expansion}
        f(x_1,\dots,x_d) = \sum_{S_1,\dots,S_d} c_{S_1,\dots,S_d}\;\prod_{i=1}^d\prod_{j\in S_i}q_j^{(i)}\,.
    \end{equation}

The circuit consists of at most $2^{n_1+\cdots+n_d}$ multi-controlled $R_z$ gates, each with controls drawn from up to $d$ registers simultaneously.

If $f(x_1,\dots,x_d) = \prod_{i=1}^d f_i(x_i)$, then:
    \begin{equation}
        c_{S_1,\dots,S_d} = \prod_{i=1}^d (\Delta_{S_i}f_i)(0)\,.
    \end{equation}

    For a general inseparable function, each of the $2^{n_1+\cdots+n_d}$ coefficients requires evaluation of $f$ at up to $2^{|S_1|+\cdots+|S_d|}$ grid points. For a separable function, one computes $\sum_{i=1}^d 2^{n_i}$ single-variable coefficients and multiplies, reducing the cost from exponential in the total qubit count to linear in the number of variables.

\section{Summary}
\label{sec:summary}

\begin{center}
\renewcommand{\arraystretch}{1.5}
\begin{tabular}{lccc}
    \toprule
    & \textbf{1D} & \textbf{2D} & \textbf{3D} \\
    \midrule
    Registers & $\{q_j\}$ & $\{q_j\},\{r_k\}$ & $\{q_j\},\{r_k\},\{s_p\}$ \\
    Subset indices & $S$ & $S,T$ & $S,T,U$ \\
    Gate count & $2^n$ & $2^{n+m}$ & $2^{n+m+\ell}$ \\
    Separable cost & --- & $2^n + 2^m$ coeff. & $2^n+2^m+2^\ell$ coeff. \\
    Interaction types & single & single, pairwise & single, pairwise, three-body \\
    \bottomrule
\end{tabular}
\end{center}

\noindent In all cases, the expansion Eq.~\ref{eq:d-expansion} provides:
\begin{enumerate}
    \item A {closed-form} representation of any function on the discrete grid as a multilinear polynomial in qubit variables.
    \item A {direct mapping} from each monomial term to a multi-controlled $R_z$ rotation gate.
    \item A {natural decomposition} of the coefficients into interaction orders, generalizing mixed partial derivatives to the discrete setting.
    \item {Automatic recovery} of the factored form whenever the function is separable, reducing the computational cost from exponential in the total qubit count to linear in the number of variables.
\end{enumerate}

\chapter{Greedy Search}\label{app:greedy}
\begin{figure*}
    \centerline{
    \includegraphics[width=140mm,scale=1.0]{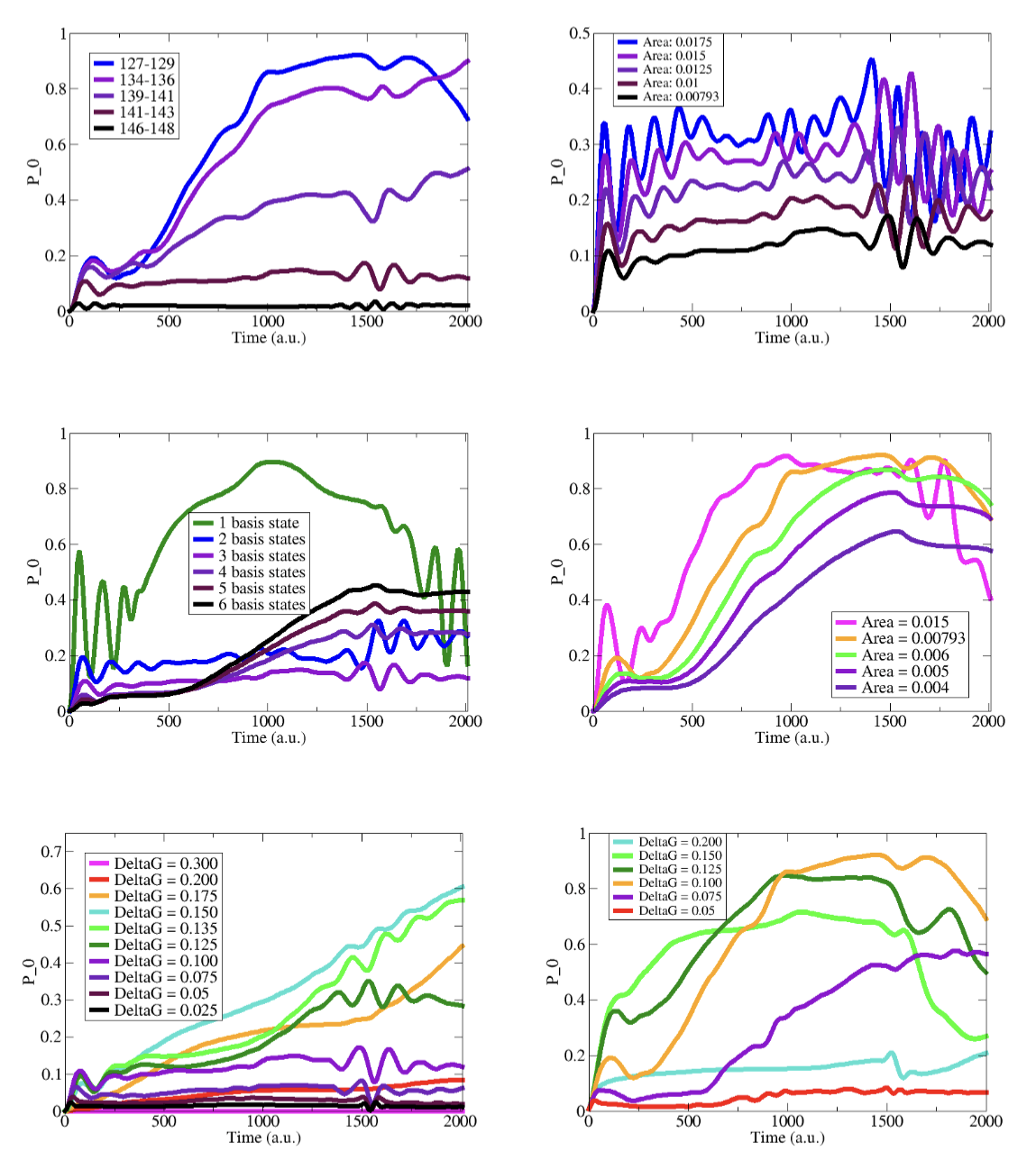}}
    \caption{\label{fig:greedy-search} Greedy search of Marcus Model parameters.
    }
\end{figure*}

A greedy search of Marcus Model parameters are given in Figure~\ref{fig:greedy-search}.
    (Top left) Coupling is area preserving with an offset $\Delta G = 0.1$, spanning three basis states, including the midpoint (127-129) and the intersection (134-136).
    (Top right) Coupling centered at the intersection between potential surfaces with varying area with constant offset $\Delta G = 0.1$, spanning three states.
    Area of 0.00793 is the area-preserving result to an approximate Gaussian coupling.
    (Middle left) Area preserving coupling with varying width with constant offset $\Delta G = 0.1$.
    (Middle Right) Coupling centered at the midpoint (basis states 127-129) between potential surfaces with varying area with constant offset $\Delta G = 0.1$, spanning three states. 
    (Bottom Left) Coupling centered at the intersection between potential surfaces for varying energy offsets, in contrast to varying energy offset for coupling between surfaces occurring at the midpoint (bottom right).
    The closest resemblance to \cite{ollitrault2020nonadiabatic} occurs with a step coupling spanning three basis states, being area-preserving to the exact Gaussian form, and always centered at the midpoint of the qubit basis.

\chapter{Symmetry Constraints on Vibronic Coupling Terms}\label{app:symmetry_appendix}

This appendix derives symmetry-allowed vibronic coupling terms for each benchmark system
and quantifies the reduction in independent coupling parameters relative to the generic
($C_1$, no symmetry) case.

Vibrational mode coordinates are denoted $x_i$ and the symmetry index sets
$\mathcal{G}_1$--$\mathcal{G}_4$ follow the definitions of the main text.
The bilinear-only subset of the on-diagonal index set is denoted
$\mathcal{G}_2^{<} = \{(\alpha,\beta) \in \mathcal{G}_2 : \alpha < \beta\}$, distinguishing
it from the full set $\mathcal{G}_2$ which also includes the diagonal
$(\alpha,\alpha)$ pairs.

For non-degenerate point groups the diagonal
quadratic operator $x_\alpha^2$ transforms as $\Gamma(x_\alpha^2) = \Gamma_\alpha \otimes
\Gamma_\alpha = \Gamma_{\mathrm{sym}}$ unconditionally, so intrastate quadratic terms
$a_{\alpha\alpha}^{(s)} x_\alpha^2$ are always allowed on every diabatic surface.
However, interstate quadratic terms $c_{\alpha\alpha}^{(ss')} x_\alpha^2$ require
$\Gamma_s \otimes \Gamma_{\mathrm{sym}} \otimes \Gamma_{s'} = \Gamma_s \otimes \Gamma_{s'}
\supset \Gamma_{\mathrm{sym}}$, which is satisfied only when $\Gamma_s = \Gamma_{s'}$.
Since all benchmark systems considered here have $\Gamma_s \neq \Gamma_{s'}$ for the
dynamically dominant electronic state pairs, interstate quadratic terms vanish for these
pairs and are omitted from the off-diagonal circuit.

\section{NO4-Anthracene}\label{sec:app_NO4_anth}

$(NO)_4$-anthracene retains $D_{2h}$ symmetry, which has eight one-dimensional
irreps: 

$A_g, B_{1g}, B_{2g}, B_{3g}, A_u, B_{1u}, B_{2u}, B_{3u}$.
The vibronic model of Pradhan and Zeng~\cite{pradhan2022design} describes 
intramolecular singlet fission (iSF) process $S_2 \to S_1$, proceeding through a conical
intersection driven by pseudo-Jahn--Teller coupling between the near-degenerate $S_2$
($B_{2u}$) and $S_3$ ($B_{1g}$) states via $B_{3u}$ modes, on a timescale of approximately
16~fs.
The five lowest electronic states ($N = 5$, padded to $M = 8$ with $m = 3$
channel qubits) transform as~\cite{pradhan2022design}:
\begin{equation}
    D_1\!(S_0): A_g, \quad D_2\!(S_1): A_g, \quad D_3\!(S_2): B_{2u}, \quad D_4\!(S_3): B_{1g}, \quad D_5\!(S_4): B_{3u}.
\end{equation}
Here $D_1$ ($S_0$) is the closed-shell ground state with diradical/tetraradical admixture,
$D_2$ ($S_1$) is the singlet-coupled triplet-pair state,
$D_3$ ($S_2$) is the bright state with charge-transfer character ($f = 0.049$),
$D_4$ ($S_3$) is an antisymmetric charge-transfer/tetraradical configuration, and
$D_5$ ($S_4$) is a higher charge-transfer state.
The vibronic Hamiltonian uses linear plus quadratic diagonal couplings and linear
off-diagonal (interstate) couplings, with modes selected based on large linear vibronic
constants from the reference $S_0$-optimised $D_{2h}$ geometry (B3LYP/cc-pVDZ).
Four of the five states carry distinct representations, but
$\Gamma_{D_1} = \Gamma_{D_2} = A_g$,
so the $D_1$--$D_2$ pair satisfies the interstate quadratic selection rule
$\Gamma_s = \Gamma_{s'}$, permitting nonzero $c_{\alpha\alpha}^{(D_1 D_2)}$ terms.
All remaining $\binom{5}{2} - 1 = 9$ pairs have $\Gamma_s \neq \Gamma_{s'}$, and their interstate
quadratic coupling terms vanish identically, eliminating $9 \times 19 = 171$ potential terms
from the off-diagonal block.

The $d = 19$ active vibrational modes are distributed across multiple irreducible
representations of $D_{2h}$, not restricted to two symmetry classes.
The distribution of modes across $D_{2h}$ irreps is given in
Table~\ref{tab:NO4_anth_modes}; the precise assignment follows from the
normal-mode analysis of Pradhan and Zeng~\cite{pradhan2023triplet}.

\begin{table}[htbp]
\centering
\caption{Distribution of the 19 active vibrational modes of $(NO)_4$-anthracene
across the irreps of $D_{2h}$.
The on-diagonal bilinear count per irrep is $\binom{d_\Gamma}{2}$; their sum gives
$|\mathcal{G}_2^{<}| = 44$.
Mode assignments follow from the vibronic model of
Pradhan and Zeng~\cite{pradhan2022design}:
only four of the eight $D_{2h}$ irreps are populated;
all modes are in-plane below 3000~cm$^{-1}$ (no C--H stretching).
The $B_{3u}$ modes drive the critical pseudo-Jahn--Teller coupling
between $S_2$ ($B_{2u}$) and $S_3$ ($B_{1g}$) responsible for the 16~fs iSF.}
\label{tab:NO4_anth_modes}
\begin{tabular}{lccl}
\hline
Irrep ($\Gamma$) & $d_\Gamma$ & $\binom{d_\Gamma}{2}$ & Coupling role \\
\hline
$A_g$    & 8 & 28 & Diagonal tuning $+$ $H_{12}$ off-diagonal \\
$B_{1g}$ & 3 &  3 & Off-diagonal: $H_{14}$, $H_{24}$, $H_{35}$ \\
$B_{2u}$ & 3 &  3 & Off-diagonal: $H_{13}$, $H_{23}$, $H_{45}$ \\
$B_{3u}$ & 5 & 10 & Off-diagonal: $H_{15}$, $H_{25}$, $H_{34}$ (pJT) \\
\hline
Total    & 19 & 44 & \\
\hline
\end{tabular}
\end{table}

Because modes span more than two irreps, the on-diagonal bilinear
conflict graph $G_{\mathrm{d}} = \bigsqcup_\Gamma K_{d_\Gamma}$ is a disjoint union of
complete graphs whose largest component has chromatic index
$\chi'(\mathcal{G}_2^{<}) = 7$,
indicating that the largest same-symmetry block contains at most $8$ modes.

For the intrastate blocks, the term counts per surface reflect the multi-irrep distribution:
\begin{center}
\renewcommand{\arraystretch}{1.3}
\begin{tabular}{lcccl}
\hline
Term & $C_1$ count & $D_{2h}$ count & Reduction & Detail \\
\hline
Linear ($a_\alpha^{(s)}\, x_\alpha$) & 19 & 8 & 58\% & $A_g$ modes only ($\mathcal{G}_1$) \\
Quadratic ($a_{\alpha\alpha}^{(s)}\, x_\alpha^2$) & 19 & 19 & 0\% & Always allowed ($\mathcal{G}_2^{=}$) \\
Bilinear ($a_{\alpha\beta}^{(s)}\, x_\alpha x_\beta$) & 171 & 44 & 74\% & Same-irrep pairs ($\mathcal{G}_2^{<}$) \\
\hline
\end{tabular}
\end{center}

For the interstate blocks, the $\binom{5}{2} = 10$ pairwise electronic couplings generate
product symmetries from the set
$\{A_g \otimes A_g,\; A_g \otimes B_{2u},\; A_g \otimes B_{1g},\; A_g \otimes B_{3u},\;
B_{2u} \otimes B_{1g},\; B_{2u} \otimes B_{3u},\; B_{1g} \otimes B_{3u}\}
= \{A_g,\; B_{2u},\; B_{1g},\; B_{3u}\}$.
These four product symmetries coincide exactly with the four irreps populated by vibrational
modes.
The four irreps $\{A_g, B_{1g}, B_{2u}, B_{3u}\}$ form a closed Klein-four subgroup under the $D_{2h}$ multiplication table, so every product of any two populated irreps is itself a member of the set.
Consequently, every bilinear mode pair $(\alpha,\beta)$ with $\alpha < \beta$ satisfies the
off-diagonal selection rule for at least one electronic coupling, and the union of all off-diagonal conflict graphs across the ten pairs covers the full $\binom{19}{2} = 171$ mode pairs.

The worst-case off-diagonal chromatic index across all electronic pairs is
\begin{equation}
    \chi'(\mathcal{G}_4) = 19.
\end{equation}
This value equals $d$ (the complete graph $K_{19}$ chromatic index for odd~$d$),
indicating that the union of bilinear mode-pair selections across all ten electronic
couplings effectively covers all mode pairs. 
Systems with fewer electronic states or fewer populated
irreps exhibit smaller $\chi'(\mathcal{G}_4)$ values (cf.\ pyrazine,
Section~\ref{sec:app_pyrazine}).

\section{NO4-Anthracene Dimer}\label{sec:app_NO4_dimer}

$(NO)_4$-anthracene dimer models the triplet-pair separation process
in cofacial $(NO)_4$-anthracene aggregates, as studied by
Pradhan and Zeng~\cite{pradhan2023triplet}.
In the reference vibronic model, the dimer geometry at the
$S_3^d$-optimised cofacial arrangement (intermonomer separation $R = 3.2$~\AA)
possesses $D_{2h}$ symmetry.
The six diabatic states ($N = 6$, padded to $M = 8$ with $m = 3$) transform under
$D_{2h}$ as
\begin{equation}
    D_1^d: A_g, \quad D_2^d: B_{1u}, \quad D_3^d: A_g, \quad
    D_4^d: B_{1u}, \quad D_5^d: B_{1g}, \quad D_6^d: A_g,
\end{equation}
representing the intermolecular triplet-pair state $T_1 T_1$ ($D_1^d$), the
intramolecular triplet-pair states $S_1 S_0 \mp S_0 S_1$ ($D_2^d, D_3^d$),
the intermolecular triplet-pair states $T_1 T_2 \mp T_2 T_1$ ($D_4^d, D_5^d$),
and the higher intermolecular state $T_2 T_2$ ($D_6^d$).
Note that $A_g$ appears three times ($D_1^d, D_3^d, D_6^d$) and $B_{1u}$ twice
($D_2^d, D_4^d$), introducing pairs with $\Gamma_s = \Gamma_{s'}$ for which
interstate quadratic terms are symmetry-allowed.

For the resource analysis in Section~\ref{sec:direct_comparison}, the effective symmetry
is taken as $C_{2v}$ rather than the full $D_{2h}$ of the reference model.
This symmetry lowering is physically motivated by the inclusion of two special
coordinates in the 21-mode model that break $D_{2h}$: the $B_{2g}$ conrotation
mode ($44ib_{2m}$), reducing symmetry from $D_{2h}$ to at most $C_{2h}$
along this large-amplitude coordinate,
and the intermonomer distance $R$, which is modelled as a Morse potential with
exponential off-diagonal scaling $f(R) = 6.05 \exp(-3.33\;\text{\AA}^{-1}(R - 2.66\;\text{\AA}))$.
The $d = 21$ modes in the effective $C_{2v}$ treatment span all four irreducible
representations: $A_1, A_2, B_1, B_2$.
Under $C_{2v}$, the six electronic states transform as
\begin{equation}
    \tilde{X}: B_1, \quad \tilde{A}: A_2, \quad \tilde{B}: B_2, \quad \tilde{C}: A_1, \quad \tilde{D}: A_2, \quad \tilde{E}: B_1.
\end{equation}
Unlike $(NO)_4$-anthracene, the representation assignments are not all unique:
$\Gamma_{\tilde{A}} = \Gamma_{\tilde{D}} = A_2$ and
$\Gamma_{\tilde{X}} = \Gamma_{\tilde{E}} = B_1$.
For these two pairs, the interstate quadratic selection rule $\Gamma_s = \Gamma_{s'}$ is
satisfied, permitting nonzero diagonal quadratic interstate coupling:
\begin{equation}
    c_{\alpha\alpha}^{(\tilde{A}\tilde{D})} \neq 0, \quad c_{\alpha\alpha}^{(\tilde{X}\tilde{E})} \neq 0 \quad \text{(symmetry-allowed for all } \alpha\text{)},
\end{equation}
introducing $2 \times 21 = 42$ additional interstate quadratic terms absent in the
$(NO)_4$-anthracene model.
The corresponding circuit cost is 42 additional UCR sequences in the off-diagonal block,
each contributing $M/2$ rotations per site within the relevant XOR-class fragment
(Section~\ref{sec:ucr_offdiag}).
All remaining 13 pairs satisfy $\Gamma_s \neq \Gamma_{s'}$, and their interstate quadratic
terms vanish.

The distribution of the $d = 21$ modes across $C_{2v}$ irreps is given in
Table~\ref{tab:NO4_dimer_modes}.
Under the parent $D_{2h}$ symmetry of the reference model, the approximate distribution
is $A_g \approx 6$--8, $B_{1g} \approx 2$--3, $B_{2u} \approx 2$--3,
$B_{3u} \approx 3$--5, $B_{2g} = 1$ (conrotation), and $R = 1$
(Pradhan and Zeng~\cite{pradhan2023triplet}; exact counts require the
Supporting Information of that reference).
These $D_{2h}$ irreps map to $C_{2v}$ under the $D_{2h} \to C_{2v}$ subduction
appropriate to the dimer geometry.

\begin{table}[htbp]
\centering
\caption{Approximate distribution of the 21 active modes of the $(NO)_4$-anthracene
dimer.
\textbf{Left:} the parent $D_{2h}$ irreps from the vibronic model of
Pradhan and Zeng~\cite{pradhan2023triplet}, with approximate mode counts (exact values
require the Supporting Information of that reference).
\textbf{Right:} the effective $C_{2v}$ irreps used in the chapter's resource analysis
after $D_{2h} \to C_{2v}$ subduction.
The on-diagonal bilinear count per $C_{2v}$ irrep is $\binom{d_\Gamma}{2}$; their sum
gives $|\mathcal{G}_2^{<}|$.
}
\label{tab:NO4_dimer_modes}
\begin{tabular}{lcc|lcc}
\hline
\multicolumn{3}{c|}{$D_{2h}$ (reference)} & \multicolumn{3}{c}{$C_{2v}$ (effective)} \\
Irrep & $d_\Gamma$ & Coupling role & Irrep & $d_\Gamma$ & $\binom{d_\Gamma}{2}$ \\
\hline
$A_g$    & $\sim\!6$--8 & Diag.\ tuning $+$ inter/intra TP
    & $A_1$  & $\sim\!7$--9 & -- \\
$B_{1u}$ & 0  & (off-diag.\ only)
    & \multicolumn{3}{l}{\footnotesize$(A_g$ and $B_{1u}$ merge into $A_1)$} \\
$B_{1g}$ & $\sim\!2$--3 & $H_{15}^d$, $H_{35}^d$, $H_{56}^d$
    & $A_2$  & $\sim\!2$--3 & -- \\
$B_{2g}$ & 1 & Conrotation ($44ib_{2m}$)
    & $B_1$  & $\sim\!4$--6 & -- \\
$B_{3u}$ & $\sim\!3$--5 & $H_{12}^d$, $H_{23}^d$, $H_{26}^d$, $H_{34}^d$, $H_{46}^d$
    & \multicolumn{3}{l}{\footnotesize$(B_{2g}$ and $B_{3u}$ merge into $B_1)$} \\
$B_{2u}$ & $\sim\!2$--3 & $H_{25}^d$, $H_{34}^d$ (mostly weak)
    & $B_2$  & $\sim\!2$--3 & -- \\
$R$ (special) & 1 & Morse diag.; exp.\ off-diag.
    & \multicolumn{3}{l}{\footnotesize(maps to $A_1$; tot.\ sym.\ stretch)} \\
\hline
Total    & 21 & & Total & 21 & $|\mathcal{G}_2^{<}|$ \\
\hline
\end{tabular}
\end{table}

Because modes span all four $C_{2v}$ irreps, the on-diagonal bilinear conflict graph
$G_{\mathrm{d}} = \bigsqcup_\Gamma K_{d_\Gamma}$ has chromatic index
\begin{equation}
    \chi'(\mathcal{G}_2^{<}) = 9,
\end{equation}
indicating that the largest same-symmetry block
contains at most $10$ modes (even) or $9$ modes (odd).
In the approximate $D_{2h} \to C_{2v}$ mapping of Table~\ref{tab:NO4_dimer_modes},
the $A_1$ block (merging $A_g$ and $B_{1u}$ modes plus $R$) is the largest,
consistent with $d_{A_1} \approx 7$--9 and $\chi'(K_9) = 9$ or $\chi'(K_{10}) = 9$.

For the intrastate blocks, the term counts per surface are:
\begin{center}
\renewcommand{\arraystretch}{1.3}
\begin{tabular}{lcccl}
\hline
Term & $C_1$ count & $C_{2v}$ count & Reduction & Detail \\
\hline
Linear ($a_\alpha^{(s)}\, x_\alpha$) & 21 & $d_{A_1}$ & -- & $A_1$ modes only ($\mathcal{G}_1$) \\
Quadratic ($a_{\alpha\alpha}^{(s)}\, x_\alpha^2$) & 21 & 21 & 0\% & Always allowed ($\mathcal{G}_2^{=}$) \\
Bilinear ($a_{\alpha\beta}^{(s)}\, x_\alpha x_\beta$) & 210 & $|\mathcal{G}_2^{<}|$ & -- & Same-irrep pairs ($\mathcal{G}_2^{<}$) \\
\hline
\end{tabular}
\end{center}

For interstate blocks, the $\binom{6}{2} = 15$ pairwise electronic couplings each have a product symmetry $\Gamma_s \otimes \Gamma_{s'}$.
Since the six electronic states span all four $C_{2v}$ representations (with $A_2$ and $B_1$ each appearing twice), the fifteen product representations sample all four irreps of $C_{2v}$,
each selecting a different subset of bilinear mode pairs.
The overlap of representation assignments
($\Gamma_{\tilde{A}} = \Gamma_{\tilde{D}}$ and
$\Gamma_{\tilde{X}} = \Gamma_{\tilde{E}}$) enriches the set of distinct product symmetries
beyond what five unique representations would provide, contributing to a large worst-case
off-diagonal chromatic index.

Under the parent $D_{2h}$ classification of Pradhan and Zeng~\cite{pradhan2023triplet},
symmetry-allowed vibronic couplings show that eight of the fifteen
electronic pairs have non-negligible coupling and span mode symmetries
$A_g$, $B_{1g}$, $B_{1u}$, and $B_{2u}$.
Triplet-pair separation proceeds primarily through avoided crossings rather than conical
intersections, with the very small $|g|$ at $\mathrm{MECI}_{12}^d$ creating a large
vibrational space of small $S_1^d$--$S_2^d$ gaps.
The separation efficiency is approximately 80\% per monomer collision on a 1.1~ps timescale.

The worst-case off-diagonal chromatic index across all electronic pairs is
\begin{equation}
    \chi'(\mathcal{G}_4) = 15.
\end{equation}
This value exceeds the on-diagonal chromatic index ($\chi'(\mathcal{G}_2^{<}) = 9$)
and reflects the fact that the union of bilinear mode-pair selections across all fifteen
electronic couplings covers a substantial fraction of all $\binom{21}{2} = 210$ possible mode pairs.
The four-irrep partition of $C_{2v}$ provides less fine-grained mode separation than the
eight-irrep partition of $D_{2h}$ available to the isolated monomer, so the off-diagonal
conflict graph retains more connectivity than in the $(NO)_4$-anthracene system.

\section{Anthracene-C60}\label{sec:app_dyad}

The anthracene-fullerene dyad nominally derives from components with $C_s$ symmetry
(anthracene) and $I_h$ symmetry (C$_{60}$), but the covalent linkage between the two
moieties breaks all nontrivial symmetry elements.
The combined dyad therefore belongs to the trivial point group $C_1$, with no
symmetry-imposed selection rules on the vibronic coupling terms.
All $\binom{d}{2}$ bilinear mode pairs are symmetry-allowed for both intrastate and
interstate couplings, and the conflict graph is the complete graph $K_d$ for both the
on-diagonal and off-diagonal blocks.

The four relevant electronic states ($M = 4$, $m = 2$) are the locally excited singlet states
on each moiety and the charge-transfer states:
\begin{equation}
    \mathrm{LE}_{\mathrm{Anth}}, \quad \mathrm{LE}_{C_{60}}, \quad \mathrm{CT}_1, \quad \mathrm{CT}_2,
\end{equation}
Under $C_1$ symmetry, all four states transform as the trivial representation, and all
interstate coupling terms (linear, quadratic, and bilinear) are symmetry-allowed for every
electronic pair.

For the full $d = 246$ model, the term counts are:
\begin{center}
\renewcommand{\arraystretch}{1.3}
\begin{tabular}{lccl}
\hline
Term & $C_1$ count & $C_1$ count & Detail \\
\hline
Linear ($a_\alpha^{(s)}\, x_\alpha$) & 246 & 246 & All modes ($\mathcal{G}_1$) \\
Quadratic ($a_{\alpha\alpha}^{(s)}\, x_\alpha^2$) & 246 & 246 & Always allowed ($\mathcal{G}_2^{=}$) \\
Bilinear ($a_{\alpha\beta}^{(s)}\, x_\alpha x_\beta$) & 30{,}135 & 30{,}135 & $\binom{246}{2}$ ($\mathcal{G}_2^{<}$) \\
\hline
\end{tabular}
\end{center}
The on-diagonal conflict graph is $K_{246}$ with chromatic index
$\chi'(\mathcal{G}_2^{<}) = 245$ (even $d$, so $\chi'(K_d) = d - 1$).

For the interstate blocks (all six pairwise couplings):
\begin{center}
\renewcommand{\arraystretch}{1.3}
\begin{tabular}{lccc}
\hline
Term & $C_1$ count & Reduction & Detail \\
\hline
Linear ($c_\alpha^{(ss')}\, x_\alpha$) & 246 & 0\% & All modes ($\mathcal{G}_3$) \\
Bilinear ($c_{\alpha\beta}^{(ss')}\, x_\alpha x_\beta$) & 30{,}135 & 0\% & $\binom{246}{2}$ ($\mathcal{G}_4$) \\
\hline
\end{tabular}
\end{center}
The off-diagonal conflict graph is likewise $K_{246}$, with chromatic index
$\chi'(\mathcal{G}_4) = 245$,
confirming that no symmetry reduction is available for this system.

The complete-graph chromatic indices for both blocks are:
\begin{equation}\label{eq:chi_C60}
    \chi'(\mathcal{G}_2^{<}) = \chi'(\mathcal{G}_4) = \chi'(K_{246}) = 245,
\end{equation}

For the reduced $d = 11$ model, the mode selection is driven by maximal coupling strength
from a subset of the 246 modes.
Under $C_1$ symmetry, all $\binom{11}{2} = 55$ mode pairs are allowed for both intrastate
and interstate bilinear coupling, and the conflict graph is $K_{11}$ with chromatic index
\begin{equation}
    \chi'(\mathcal{G}_2^{<}) = \chi'(\mathcal{G}_4) = \chi'(K_{11}) = 11.
\end{equation}

\section{Pyrazine}\label{sec:app_pyrazine}

Pyrazine belongs to $D_{2h}$ with the two relevant electronic states ($M = 2$, $m = 1$):
\begin{equation}
    S_1 \sim B_{3u}, \quad S_2 \sim B_{2u},
\end{equation}
giving the product symmetry $\Gamma_{S_1} \otimes \Gamma_{S_2} = B_{3u} \otimes B_{2u} = B_{1g}$.
Since $\Gamma_{S_1} \neq \Gamma_{S_2}$, all interstate quadratic terms
$c_{\alpha\alpha}^{(S_1 S_2)} x_\alpha^2$ vanish, consistent with the Hamiltonian in
Eq.~\ref{eq:PyrazineHamiltonian}.

For the two-mode model of Section~\ref{sec:Pyrazine}, the tuning mode $\nu_{6a}$ transforms
as $A_g$ and the coupling mode $\nu_{10a}$ transforms as $B_{1g}$.
The interstate linear coupling $c_1\, x_{10a}$ is allowed since
$B_{1g} = \Gamma_{S_1} \otimes \Gamma_{S_2}$ ($\nu_{10a} \in \mathcal{G}_3$);
the intrastate linear term $a_1\, x_{6a}$ is allowed since
$A_g = \Gamma_{\mathrm{sym}}$ ($\nu_{6a} \in \mathcal{G}_1$).
The single bilinear term $c_2\, x_{6a} x_{10a}$ transforms as
$A_g \otimes B_{1g} = B_{1g}$, satisfying the interstate selection rule
($(\nu_{6a}, \nu_{10a}) \in \mathcal{G}_4$), which is consistent with the Hamiltonian in
Eq.~\ref{eq:PyrazineHamiltonian}.
The intrastate bilinear coupling between these two modes is forbidden since
$A_g \otimes B_{1g} = B_{1g} \neq A_g$, reflecting the fact that the tuning and coupling
modes belong to orthogonal symmetry classes and do not undergo Duschinsky mixing on a single
surface.

In the full 24-mode model of Raab et al.~\cite{raab1999molecular}, the modes distribute
across the eight representations of $D_{2h}$ as shown in
Table~\ref{tab:mode_distribution}.
This model refines the earlier Hamiltonian of Stock et al.~\cite{stock1995resonance}
by including bilinear off-diagonal coupling terms
$c_{ij}\, Q_i Q_j$ for mode pairs whose product symmetry equals $B_{1g}$.
Raab et al.\ determined the 102 coupling constants via CIS calculations at the
MP2-optimised ground-state geometry, with selected coefficients adjusted to reproduce the
experimental $S_2$ absorption spectrum.
The MCTDH treatment of all 24 modes achieves good agreement with experiment using
only a realistic phenomenological broadening ($\tau = 150$~fs), in contrast to the
4-mode model which requires a strong artificial broadening ($\tau = 30$~fs),
demonstrating that the 20 additional modes act as a symmetry-respecting internal heat bath.

\begin{table}[htbp]
\centering
\caption{Distribution of the 24 active vibrational modes of pyrazine across the eight
irreps of $D_{2h}$, following Raab et al.~\cite{raab1999molecular},
Table~I.
The on-diagonal bilinear count per irrep is $\binom{d_\Gamma}{2}$; their sum gives
$|\mathcal{G}_2^{<}| = 31$.
All 24 modes enter both the diagonal and the off-diagonal parts of the Hamiltonian.
Mode nomenclature follows Innes et al.~\cite{innes1988electronic}.}
\label{tab:mode_distribution}
\begin{tabular}{lccl}
\hline
Irrep ($\Gamma$) & $d_\Gamma$ & $\binom{d_\Gamma}{2}$ & Modes \\
\hline
$A_g$    & 5 & 10 & $\nu_{6a}$, $\nu_{1}$, $\nu_{9a}$, $\nu_{8a}$, $\nu_{2}$ \\
$B_{1g}$ & 1 &  0 & $\nu_{10a}$ \\
$B_{2g}$ & 2 &  1 & $\nu_{4}$, $\nu_{5}$ \\
$B_{3g}$ & 4 &  6 & $\nu_{6b}$, $\nu_{3}$, $\nu_{8b}$, $\nu_{7b}$ \\
$A_u$    & 2 &  1 & $\nu_{16a}$, $\nu_{17a}$ \\
$B_{1u}$ & 4 &  6 & $\nu_{12}$, $\nu_{18a}$, $\nu_{19a}$, $\nu_{13}$ \\
$B_{2u}$ & 4 &  6 & $\nu_{18b}$, $\nu_{14}$, $\nu_{19b}$, $\nu_{20b}$ \\
$B_{3u}$ & 2 &  1 & $\nu_{16b}$, $\nu_{11}$ \\
\hline
Total    & 24 & 31 & \\
\hline
\end{tabular}
\end{table}

\begin{figure}
    \centerline{
    \includegraphics[width=120mm,scale=0.8]{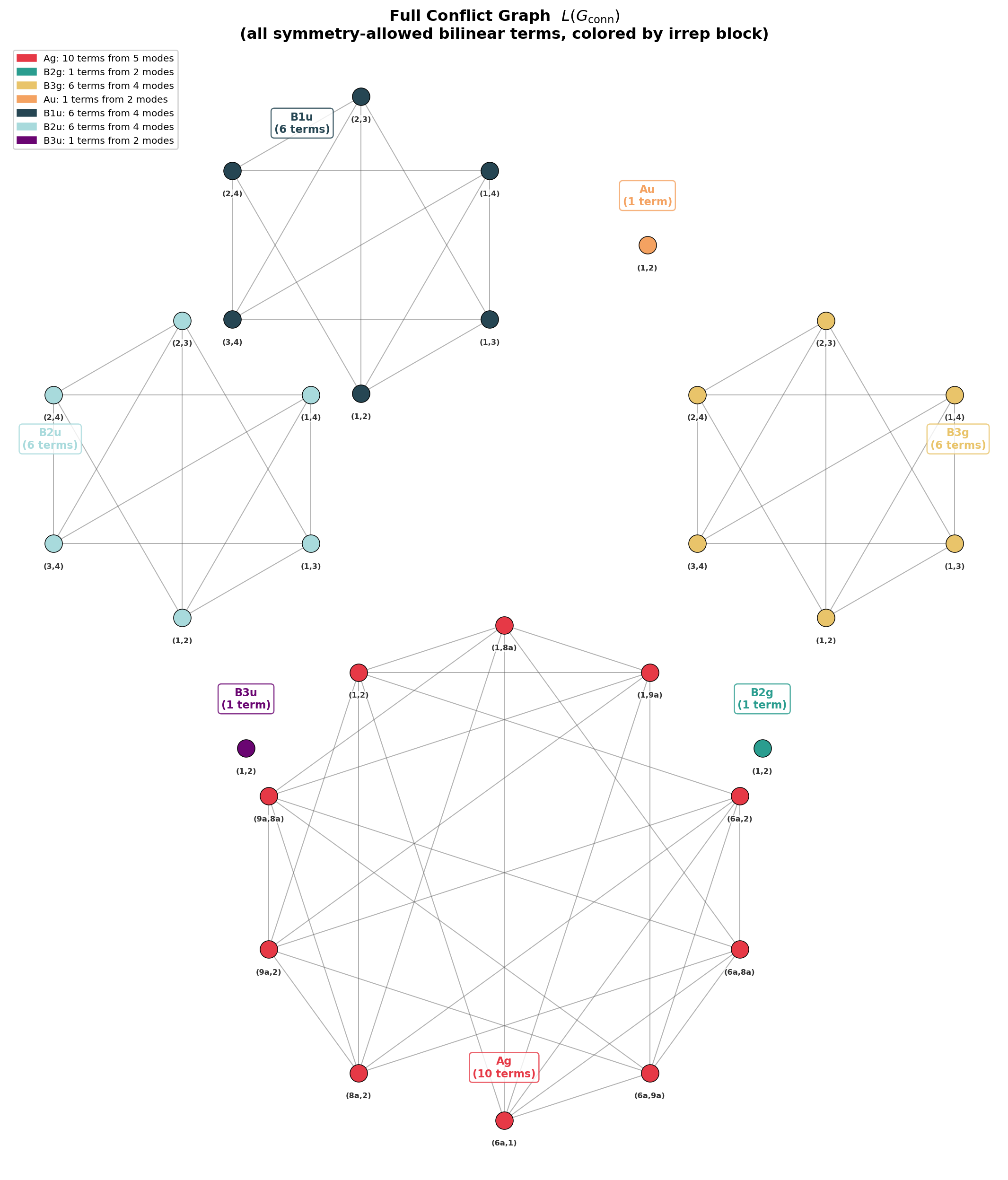}
    }
    \caption{\label{fig:ag_conflict_full} Full conflict graph for on-diagonal bilinear terms in 24-mode pyrazine.
    The $B_{1g}$ symmetry group is not present, as it contains only the single $\nu_{10a}$ mode and contributes no bilinear pairs.}
\end{figure}

The five $A_g$ modes form the largest same-symmetry clique, $K_5$, whose chromatic index
$\chi'(K_5) = 5$ dominates the on-diagonal conflict graph
$G_{\mathrm{d}} = \bigsqcup_\Gamma K_{d_\Gamma}$ and sets
$\chi'(\mathcal{G}_2^{<}) = 5$.
Only the subset with $B_{1g}$ symmetry
participates in the interstate linear coupling ($|\mathcal{G}_3| = 1$).
The interstate bilinear terms are restricted to mode pairs satisfying
$\Gamma_\alpha \otimes \Gamma_\beta = B_{1g}$, selecting complementary pairs:
$A_g \otimes B_{1g}$, $B_{2g} \otimes B_{3g}$, $A_u \otimes B_{1u}$, and
$B_{2u} \otimes B_{3u}$, giving $|\mathcal{G}_4| = 29$.\footnote{The mode assignment from
Raab et al.\ yields $5\!\times\!1 + 2\!\times\!4 + 2\!\times\!4 + 4\!\times\!2 = 29$.
This one-pair difference does not affect $\chi'(\mathcal{G}_4) = 5$ or any
resource estimate.}
The off-diagonal conflict graph is correspondingly the disjoint union of four complete
bipartite graphs,
$G_{\mathrm{od}} = K_{5,1} \sqcup K_{2,4}
 \sqcup K_{2,4} \sqcup K_{4,2}$,
with chromatic index $\chi'(\mathcal{G}_4) = 5$, reflecting the fine-grained mode partitioning of $D_{2h}$.
The full conflict graph is shown in Figure~\ref{fig:ag_conflict_full}.

\chapter{Phase Gradient Resource Recalculation}\label{app:Motlagh_Resource}
The resource estimates reported by Motlagh et al.~\cite{motlagh2025quantum} for their benchmark systems are computed under simplifications that significantly understate the true cost of simulating vibronic dynamics with their phase gradient algorithm.
Each vibrational mode is discretizeed onto $N = 2^n = 16$ grid points ($n = 4$ qubits per mode) without justification from a convergence or Nyquist analysis; the concrete gate counts include only single-mode terms ($x_r$ and $x_r^2$), omitting the bilinear cross-mode terms $x_r x_s$ that their own formalism introduces; and the Trotter step cost conflates the on-diagonal and off-diagonal potential fragments, obscuring the non-commutativity of the three-way splitting.
In this section, corrected resource estimates are derived by adopting a physically justified grid resolution of $n = 10$ ($N = 1024$), increasing the phase gradient register to $b = 45$ bits, including all bilinear vibronic coupling terms, and properly accounting for the second-order Trotter structure with kinetic merging across timesteps.

\subsubsection{Grid resolution}\label{sec:app_grid_resolution}

The Motlagh et al.\ proposal discretizes each vibrational mode onto a real-space grid of $N = 2^n$ points with spacing $\Delta = \sqrt{2\pi/K}$, representing position eigenvalues $Q|x\rangle = \Delta(x - K/2)|x\rangle$ for $x \in \{0,\ldots,N-1\}$.
The choice $n = 4$ ($N = 16$, $\Delta \approx 0.63$) is taken without analysis.

The harmonic ground state (Motlagh et al., Eq.~14) has momentum-space width $\sigma_P = \sqrt{2\pi/K}$ in the grid representation.
For the wavefunction to be adequately sampled, the grid spacing must resolve the highest significant momentum component: $\Delta \leq \pi/p_{\max}$.
Nonadiabatic dynamics populate vibrational levels up to $n_{\max} \sim 10$--$15$ through linear displacement and vibrational redistribution~\cite{raab1999molecular,worth2004beyond}.
With $N = 16$ and $n_{\max} = 10$, the number of grid points per oscillator period is $N/(2n_{\max}+1) \approx 0.76$.
This is well below Nyquist sampling, and the discretization error is $\mathcal{O}(1)$.

Converged MCTDH calculations require 32--64 primitive basis functions per mode for 2-state models~\cite{raab1999molecular} and 8--32 per mode for 24-mode models~\cite{worth2004beyond}, depending on displacement and coupling strength.
For the singlet fission systems of Pradhan and Zeng~\cite{pradhan2022design,pradhan2023triplet}, which involve larger displacements, $n = 10$ ($N = 1024$) is adopted as a conservative choice, consistent with the first-quantization grid resolutions of Su et al.~\cite{su2021fault}.

\subsubsection{Phase gradient precision}\label{sec:phase_precision}

The phase gradient resource state accumulates phases with precision $2\pi/2^b$.
For a bilinear term $\gamma_{rs} x_r x_s$ at $n = 10$, the maximum phase magnitude is $\theta_{\max}^{(\textrm{BVC})} = \pi\tau K|\gamma_{rs}|/2$.
With $N = 1024$, $\tau \sim 1$~a.u., and $|\gamma| \sim 0.01$--$0.05$~a.u., $\theta_{\max} \sim 16$--$80$~rad.
Requiring per-step phase error $\varepsilon_{\textrm{phase}} \lesssim 10^{-10}$ (negligible over $\sim\!10^4$ Trotter steps at $\varepsilon = 0.01$ target) gives
\begin{equation}\label{eq:b_requirement}
    b \geq \log_2\!\left(\frac{2\pi \theta_{\max}}{\varepsilon_{\textrm{phase}}}\right) \approx 43.
\end{equation}
A higher-precision accumulator is taken with $b = 45$, providing precision $2\pi/2^{45} \approx 1.8 \times 10^{-13}$, giving a comfortable margin for accumulation across $\mathcal{O}(M^2)$ terms per step.
Motlagh et al.\ use $b = 20$ ($\textrm{precision} \approx 6 \times 10^{-6}$), which is marginally adequate at $n = 4$ but becomes insufficient at $n = 10$ where phase magnitudes grow by $N_{\textrm{new}}/K_{\textrm{orig}} = 64$ for quadratic terms.

\subsubsection{Per-term Toffoli costs}\label{sec:per_term}

Following the circuit structure of Motlagh et al.\ (their Section~3 and companion notebook~\cite{motlagh2025quantum}), each potential energy term proceeds as QROM $\to$ arithmetic $\to$ phase accumulation $\to$ uncompute.
The costs of the individual primitives at $n = 10$, $b = 45$ are:
\begin{alignat}{2}
    C_{\textrm{sq}} &= \tfrac{k(k-1)}{2} &&= 45 \;\text{Toffoli} \quad\text{(out-of-place square~\cite{su2021fault})}, \label{eq:Csq}\\
    C_{\textrm{mult}} &= n^2 &&= 100 \;\text{Toffoli} \quad\text{(out-of-place multiply)}, \label{eq:Cmult}\\
    C_{\textrm{semi}}(n_b) &= n_b b - \tfrac{n_b(n_b\!-\!1)}{2} &&\quad\text{(controlled semi-adders into $b$-bit register~\cite{gidney2018halving})}, \label{eq:Csemi}
\end{alignat}
giving $C_{\textrm{semi}}(10) = 405$ and $C_{\textrm{semi}}(20) = 710$.

A dirty-ancilla QROM selecting among $2^n$ bitstrings costs $2^n$ Toffoli~\cite{babbush2018encoding}.
For the diagonal fragment ($\zeta = 0$), the QROM addresses all $N$ electronic states with cost $C_{\textrm{QROM}}^{(0)} = 2^{\lceil\log_2 N\rceil}$.
For off-diagonal XOR class $\zeta$ with Hamming weight $h(\zeta)$, the focusing CNOTs (Motlagh et al., Section~3.2) reduce the effective address space, giving
\begin{equation}\label{eq:qrom_focused}
    C_{\textrm{QROM}}^{(\zeta)} = 2^{m - h(\zeta)}, \qquad m = \lceil\log_2 N\rceil.
\end{equation}
This distinction provides a modest per-term reduction for off-diagonal terms.

Each term type requires one forward and one inverse QROM ($2C_{\textrm{QROM}}$), the appropriate arithmetic, and phase accumulation:
\begin{align}
    C_{Q}^{(\zeta)} &= 2C_{\textrm{QROM}}^{(\zeta)} + C_{\textrm{semi}}(k), \label{eq:CQ}\\
    C_{Q^2}^{(\zeta)} &= 2C_{\textrm{QROM}}^{(\zeta)} + 2C_{\textrm{sq}} + C_{\textrm{semi}}(2k), \label{eq:CQQ}\\
    C_{x_r x_s}^{(\zeta)} &= 2C_{\textrm{QROM}}^{(\zeta)} + 2C_{\textrm{mult}} + C_{\textrm{semi}}(2k). \label{eq:CBVC}
\end{align}
The bilinear term~\ref{eq:CBVC} has the same circuit topology as the diagonal quadratic~\ref{eq:CQQ}, with \texttt{OutOfPlaceSquare} replaced by \texttt{OutOfPlaceMultiply}.
The cost difference per term is $C_{x_r x_s} - C_{Q^2} = 2(n^2 - k(k-1)/2) = k(k+1) = 110$ Toffoli.
Representative per-term costs are shown in Table~\ref{tab:per_term}.

\begin{table}[ht]
\centering
\caption{Per-term Toffoli costs at original ($n\!=\!4$, $b\!=\!20$) and corrected ($n\!=\!10$, $b\!=\!45$) parameters.  QROM values shown for the diagonal fragment ($\zeta\!=\!0$) at $N\!=\!6$ ($C_{\textrm{QROM}}=8$).}\label{tab:per_term}
\renewcommand{\arraystretch}{1.15}
\begin{tabular}{l c c c | c c c}
\hline\hline
& \multicolumn{3}{c|}{$n = 4$, $b = 20$} & \multicolumn{3}{c}{$n = 10$, $b = 45$} \\
Term type & Arith. & Semi-add. & Total & Arith. & Semi-add. & Total \\
\hline
$x_r$ (linear)       & $0$  & $74$  & $90$  & $0$   & $405$ & $421$  \\
$x_r^2$ (diag.\ quad.) & $12$ & $132$ & $160$ & $90$  & $710$ & $816$  \\
$x_r x_s$ (bilinear) & ---  & ---   & ---   & $200$ & $710$ & $926$  \\
\hline\hline
\end{tabular}
\end{table}

\subsubsection{Trotter structure and kinetic merging}\label{sec:trotter_struct}

The vibronic Hamiltonian decomposes into three non-commuting groups: the kinetic energy $H_T = \sum_r(\omega_r/2)P_r^2$, the on-diagonal potential $V_{\textrm{d}}$ (XOR class $\zeta=0$, containing the tuning terms $\kappa_r^{(s)}x_r$, diagonal quadratic $\gamma_{rr}^{(s)}x_r^2$, and on-diagonal bilinear $\gamma_{rs}^{(s)}x_rx_s$), and the off-diagonal coupling $V_{\textrm{od}}$ (XOR classes $\zeta \neq 0$, containing coupling terms $\lambda_r^{(ss')}x_r$ and bilinear couplings $\mu_{rs}^{(ss')}x_rx_s$).
Motlagh et al.\ apply a Clifford transform to diagonalize each off-diagonal fragment before applying the phase gradient protocol; this does not change the Trotterization structure, which remains a three-way second-order Strang splitting:
\begin{equation}\label{eq:trotter_step}
    U(\tau) = e^{iH_T\tau/2}\, e^{iV_{\textrm{d}}\tau/2}\, e^{iV_{\textrm{od}}\tau}\, e^{iV_{\textrm{d}}\tau/2}\, e^{iH_T\tau/2} + \mathcal{O}(\tau^3).
\end{equation}

\paragraph{Kinetic merging.}
When chaining $n$ Trotter steps, the adjacent half-kinetic operators $e^{iH_T\tau/2}$ from the end of step $\ell$ and the beginning of step $\ell+1$ commute (both are functions of $P_r$ only) and merge into a single full kinetic operator $e^{iH_T\tau}$:
\begin{equation}\label{eq:merged_chain}
\begin{split}
    U(t) &\approx \;e^{iH_T\tau/2}\!\left[\prod_{\ell=1}^{n}\left(e^{iV_{\textrm{d}}\tau/2}\,e^{iV_{\textrm{od}}\tau}\,e^{iV_{\textrm{d}}\tau/2}\right)\! \cdot\! e^{iH_T\tau}\right]^{n-1}\!\\
    &\qquad\times\; e^{iV_{\textrm{d}}\tau/2}\,e^{iV_{\textrm{od}}\tau}\,e^{iV_{\textrm{d}}\tau/2}\;e^{iH_T\tau/2}.
\end{split}
\end{equation}
The circuit cost of each kinetic, diagonal, or off-diagonal evaluation is independent of the timestep fraction (the Toffoli count is determined by register widths and QROM sizes, not by the magnitude of rotation angles).
The operation counts for $n$ steps are therefore:
\begin{equation}\label{eq:op_counts}
    \underbrace{(n+1)}_{\textrm{kinetic}} \;\text{evaluations}, \quad
    \underbrace{2n}_{\textrm{diag.\ potential}} \;\text{evaluations}, \quad
    \underbrace{n\vphantom{2}}_{\textrm{off-diag.\ potential}} \;\text{evaluations}.
\end{equation}
Note that the $V_{\textrm{d}}(\tau/2)$ terms cannot merge across the kinetic or off-diagonal barriers, since $[V_{\textrm{d}}, H_T] \neq 0$ and $[V_{\textrm{d}}, V_{\textrm{od}}] \neq 0$.
For large $n$, the per-step cost is
\begin{equation}\label{eq:step_cost}
    \Omega_{\textrm{step}} = C_{\textrm{kin}} + 2\,C_{V_{\textrm{d}}} + C_{V_{\textrm{od}}},
\end{equation}
where the three cost components are defined below.

\subsubsection{Fragment costs}\label{sec:frag_costs}

\paragraph{Kinetic fragment.}
Each of $M$ modes requires a forward QFT, a squaring with phase accumulation (using the mode-specific frequency $\omega_r/2$ as a classical coefficient, so no QROM is needed), and an inverse QFT.
Using the approximate QFT of Nam and Maslov~\cite{nam2020approximate}:
\begin{equation}\label{eq:kin_cost}
    C_{\textrm{kin}} = M\!\left[
    2\!\cdot\!2k(\lceil\log_2 k\rceil\!-\!1) + 2C_{\textrm{sq}} + C_{\textrm{semi}}(2k)
    \right].
\end{equation}
At $n = 10$: AQFT$(10) = 60$ Toffoli, giving $C_{\textrm{kin}} = 920M$.

\paragraph{Diagonal potential ($\zeta = 0$).}
The on-diagonal fragment contains all terms $V_{ss}$ for each electronic state $s$.
In the worst case ($C_1$ symmetry, all coupling constants nonzero), the term counts are $M$ linear, $M$ diagonal quadratic, and $\binom{M}{2}$ on-diagonal bilinear:
\begin{equation}\label{eq:Vd_cost}
    C_{V_{\textrm{d}}} = M\!\left(C_Q^{(0)} + C_{Q^2}^{(0)}\right) + \binom{M}{2}C_{x_rx_s}^{(0)},
\end{equation}
where the superscript $(0)$ indicates the diagonal QROM size $C_{\textrm{QROM}}^{(0)} = 2^m$.

\paragraph{Off-diagonal potential ($\zeta \neq 0$).}
The $2^m - 1$ off-diagonal XOR classes each contain coupling terms for state pairs $(s,s')$ with $s \oplus s' = \zeta$.
After the focusing CNOTs reduce the electronic register, the QROM cost for class $\zeta$ is $C_{\textrm{QROM}}^{(\zeta)} = 2^{m-h(\zeta)}$.
In the worst case:
\begin{equation}\label{eq:Vod_cost}
    C_{V_{\textrm{od}}} = \sum_{\zeta=1}^{2^m-1}\left[
    M\, C_Q^{(\zeta)} + \binom{M}{2}C_{x_rx_s}^{(\zeta)}
    \right],
\end{equation}
For molecules with nontrivial point group symmetry, the vibronic selection rules (Section~\ref{sec:symmetry}) reduce the number of nonzero terms.
The diagonal coupling modes ($\mathcal{G}_1$) must have $\Gamma(x_r) = \Gamma_{\textrm{sym}}$; the off-diagonal coupling modes ($\mathcal{G}_3$) must satisfy $\Gamma(x_r) = \Gamma_s \otimes \Gamma_{s'}$; and the bilinear pairs are restricted to $\mathcal{G}_2$ (on-diagonal) and $\mathcal{G}_4$ (off-diagonal).
However, the Motlagh et al.\ benchmark systems have low molecular symmetry ($C_s$ or $C_1$), providing minimal reduction.
The term counts for each benchmark are listed in Table~\ref{tab:term_counts}; estimates presented here use the worst-case $C_1$ counts throughout; for higher-symmetry molecules, the bilinear term counts can be reduced by the factors tabulated in Section~\ref{sec:symmetry}.

\begin{table}[ht]
\centering
\caption{Term counts and XOR class structure for the Motlagh et al.\ benchmark systems.
``Existing'' counts the terms actually costed in their Table~1 (no bilinear terms).
Off-diagonal bilinear counts are summed over all $2^m\!-\!1$ XOR classes.}\label{tab:term_counts}
\renewcommand{\arraystretch}{1.15}
\begin{tabular}{lccc rr rr c}
\hline\hline
& & & & \multicolumn{2}{c}{On-diagonal} & \multicolumn{2}{c}{Off-diagonal} & \\
System & $N$ & $M$ & $m$ & $N_Q + N_{Q^2}$ & $N_{\textrm{BVC}}^{(\textrm{d})}$ & $N_Q^{(\textrm{od})}$ & $N_{\textrm{BVC}}^{(\textrm{od})}$ & Existing \\
\hline
(NO)$_4$-Anth           & 5  & 19  & 3 & 38    & 171       & 133   & 1\,197    & $\sim\!190$ \\
(NO)$_4$-Anth Dimer     & 6  & 21  & 3 & 42    & 210       & 147   & 1\,470    & $\sim\!252$ \\
Anth/C$_{60}$ ($M\!=\!11$)  & 4  & 11  & 2 & 22    & 55        & 33    & 165       & $\sim\!88$  \\
Anth/C$_{60}$ ($M\!=\!246$) & 4  & 246 & 2 & 492   & 30\,135   & 738   & 90\,405   & $\sim\!1968$ \\
\hline\hline
\end{tabular}
\end{table}

\subsubsection{Trotter error recalibration}\label{sec:trotter_recal}

The empirical Trotter errors reported by Motlagh et al.\ were calibrated at $N = 16$ without bilinear terms.
Two corrections are required.

\paragraph{Grid-induced norm increase.}
The leading Trotter error for the second-order formula~\cite{childs2021theory} is governed by nested commutators $\|[H_a,[H_a,H_b]]\|$.
The dominant contributions are the kinetic--potential commutators.
In the grid representation, $\|x_r\| = \|P_r\| = \sqrt{\pi K/2}$, and the nested commutator norms for each term type are:
\begin{align}
    \|[T,[T,\kappa x_r]]\| &\sim |\kappa|\omega_r^2 \sqrt{\tfrac{\pi K}{2}}, \label{eq:nested_lin}\\
    \|[T,[T,\gamma x_r^2]]\| &\sim |\gamma|\omega_r^2 \cdot \tfrac{\pi K}{2}, \label{eq:nested_diag}\\
    \|[T,[T,\gamma x_r x_s]]\| &\sim |\gamma|(\omega_r^2\!+\!\omega_s^2)\cdot\tfrac{\pi K}{2}. \label{eq:nested_bilin}
\end{align}
The dominant quadratic and bilinear terms~\ref{eq:nested_diag}--\ref{eq:nested_bilin} scale linearly in $N$.
Increasing $N$ from 16 to 1024 therefore amplifies the commutator sum by a factor of
\begin{equation}\label{eq:beta_grid}
    \beta_{\textrm{grid}} = \frac{K_{\textrm{new}}}{K_{\textrm{orig}}} = \frac{1024}{16} = 64.
\end{equation}
The nested commutator $[T,[T,\gamma Q^2]]$ produces terms proportional to $\|P_r\|\!\cdot\!\|x_r\| = \pi K/2$, not $(\pi K/2)^2$.
The error scales as $N^1$, not $N^2$, because each commutation with $T = (\omega/2)P^2$ lowers the polynomial degree by one via $[P^2, Q^n] \sim nQ^{n-1}P$.

\paragraph{Bilinear term contributions.}
\subsubsection{On-Diagonal Bilinear terms}
\begin{figure}
\centerline{
\Qcircuit @C=0.1em @R=1.0em {
\lstick{\ket{q_{0_A}}}& \ctrl{3} & \ctrl{4} & \ctrl{5} & \qw& \qw & \qw & \qw& \qw & \qw & \qw \\
\lstick{\ket{q_{1_A}}}& \qw & \qw & \qw & \ctrl{2} & \ctrl{3} & \ctrl{4}& \qw & \qw & \qw & \qw\\
\lstick{\ket{q_{2_A}}}& \qw & \qw & \qw & \qw & \qw & \qw & \ctrl{1} & \ctrl{2} & \ctrl{3} & \qw\\
\lstick{\ket{q_{0_B}}}& \gate{\phi 2^{0+0}} & \qw & \qw & \gate{\phi 2^{1+0}}& \qw & \qw& \gate{\phi 2^{2+0}} & \qw& \qw & \qw & \qw\\
\lstick{\ket{q_{1_B}}}& \qw & \gate{\phi 2^{0+1}} & \qw & \qw & \gate{\phi 2^{1+1}}& \qw & \qw & \gate{\phi 2^{2+1}}& \qw & \qw\\
\lstick{\ket{q_{2_B}}}& \qw & \qw & \gate{\phi 2^{0+2}} & \qw & \qw & \gate{\phi 2^{1+2}}& \qw & \qw & \gate{\phi 2^{2+2}} & \qw
}
}
\caption{\label{fig:Bilinear_sample} Quantum circuit for bilinear term (on-diagonal).}
\end{figure}
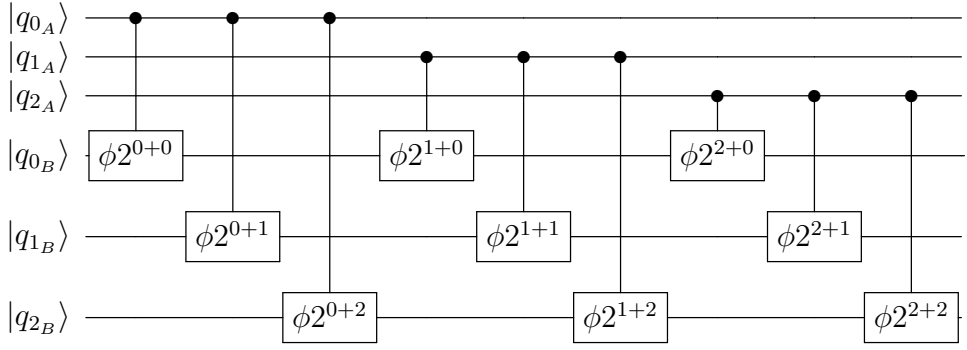

Consider an operator $U$ 
\begin{equation}
    U = e^{-i\phi x_0x_1}.
\end{equation}
With a bitwise decomposition, it is seen that
\begin{equation}
        x_0 = \sum_j^{n-1}2^jq_{j_A}, \quad x_1 = \sum_\ell^{m-1}2^\ell q_{\ell _B},
\end{equation}
where qubits $n_A$ define a position register $A$ and qubits $n_B$ define a position register $B$, as independent coordinate axes, and $\phi$ is some phase rotation. The product $x_0x_1$ can expand as 

\begin{equation}
\begin{split}
    x_0x_1 &= \sum_j^{n-1}2^jq_{j_A}\sum_\ell^{m-1}2^\ell q_{\ell _B}  \\
    &= \sum_j^{n-1}\sum_\ell^{m-1}2^{j+\ell}q_{j_A}q_{\ell _B}.
\end{split}
\end{equation}
This results in phase rotations with weight $2^{j+\ell}\phi$ on qubit $n_{\ell _B}$ controlled by qubit $n_{j_A}$, as illustrated in Figure~\ref{fig:Bilinear_sample}.. 

The bilinear terms add $N_{\textrm{BVC}}^{(\textrm{d})} + N_{\textrm{BVC}}^{(\textrm{od})}$ additional commutator contributions, each of comparable magnitude to the diagonal quadratic terms.
Parameterizfing the relative coupling strength as $\eta = |\bar\gamma_{\textrm{BVC}}|/|\bar\gamma_{\textrm{QVC}}|$, the bilinear amplification factor is
\begin{equation}\label{eq:beta_bvc}
    \beta_{\textrm{BVC}} = 1 + \eta^2 \frac{N_{\textrm{BVC}}^{(\textrm{d})} + N_{\textrm{BVC}}^{(\textrm{od})}}{N_{Q^2}},
\end{equation}
where $N_{Q^2} = M$ is the diagonal quadratic count.
The combined error amplification is
\begin{equation}\label{eq:beta_total}
    \beta = \beta_{\textrm{grid}} \times \beta_{\textrm{BVC}} = 64\!\left(1 + \eta^2\frac{N_{\textrm{BVC}}^{(\textrm{tot})}}{M}\right).
\end{equation}

\paragraph{Adjusted step count.}
The empirical calibration gives Trotter error $\varepsilon_1$ at timestep $\Delta t_1 = 0.1$~fs.
The total error for $n$ steps at timestep $\tau = t/n$ scales as $\varepsilon \sim n\tau^3 W \sim t\tau^2 W$, where $W$ is the commutator sum.
To maintain target error $\varepsilon_{\textrm{req}}$:
\begin{equation}\label{eq:n_adjusted}
    n_{\textrm{steps}} = \left\lceil \frac{t}{\Delta t_1}\sqrt{\frac{\varepsilon_1\,\beta}{\varepsilon_{\textrm{req}}}}\;\right\rceil.
\end{equation}
The grid correction alone yields $\sqrt{\beta_{\textrm{grid}}} = \sqrt{64} = 8$, so even without bilinear terms the step count increases by $8\times$.

\subsubsection{Assembled resource estimates}\label{sec:assembled}

The total Toffoli count is
\begin{equation}\label{eq:total_cost}
    C_{\textrm{total}} = n_{\textrm{steps}}\,\Omega_{\textrm{step}} + C_{\textrm{kin}},
\end{equation}
where the additional $C_{\textrm{kin}}$ accounts for the $(n+1)$th kinetic evaluation from the boundary half-steps.
Estimates are reported at two bilinear coupling strengths: moderate ($\eta = 1/3$, bilinear couplings one-third the diagonal quadratic strength) and worst case ($\eta = 1$, equal strengths).
The corrected resource estimates are presented in Table~\ref{tab:extended_resources}.

\begin{table}[ht]
\centering
\small
\caption{Corrected resource estimates at $n = 10$, $b = 45$, with second-order Trotter structure $T/2 \to V_{\textrm{d}}/2 \to V_{\textrm{od}} \to V_{\textrm{d}}/2 \to T/2$ and kinetic merging.
All entries: $\varepsilon = 1\%$, $t = 100$~fs.
``Paper'' reproduces Motlagh et al., Table~1 ($n\!=\!4$, $b\!=\!20$, no bilinear terms).
}\label{tab:extended_resources}
\renewcommand{\arraystretch}{1.2}
\begin{tabular}{l c c c c c}
\hline\hline
System & Qubits & $\Omega_{\textrm{step}}$ & $n_{\textrm{steps}}$ & Total Toffoli & vs.\ Paper \\
\hline
\multicolumn{6}{l}{\textit{Paper (Motlagh et al., $n=4$, $b=20$, no BVC):}} \\
(NO)$_4$-Anth           & 146  & $3.37 \times 10^4$ & 513    & $1.73 \times 10^7$ & --- \\
(NO)$_4$-Anth Dimer     & 154  & $2.48 \times 10^4$ & 111    & $2.76 \times 10^6$ & --- \\
Anth/C$_{60}$ ($M\!=\!11$)  & 113  & $1.20 \times 10^4$ & 55     & $6.62 \times 10^5$ & --- \\
Anth/C$_{60}$ ($M\!=\!246$) & 1053 & $1.16 \times 10^6$ & 55     & $2.66 \times 10^7$ & --- \\[6pt]
\hline
\multicolumn{6}{l}{\textit{Corrected, grid only ($n=10$, $b=45$, no BVC):}} \\
(NO)$_4$-Anth           & 308  & $1.19 \times 10^5$ & 4\,097     & $4.88 \times 10^{8}$  & $28\times$ \\
(NO)$_4$-Anth Dimer     & 328  & $1.32 \times 10^5$ & 888        & $1.17 \times 10^{8}$  & $42\times$ \\
Anth/C$_{60}$ ($M\!=\!11$)  & 227  & $5.05 \times 10^4$ & 440        & $2.22 \times 10^{7}$  & $34\times$ \\
Anth/C$_{60}$ ($M\!=\!246$) & 2577 & $1.13 \times 10^6$ & 440        & $4.97 \times 10^{8}$  & $19\times$ \\[6pt]
\hline
\multicolumn{6}{l}{\textit{Corrected, moderate ($n=10$, $b=45$, $\eta = 1/3$):}} \\
(NO)$_4$-Anth           & 308  & $1.53 \times 10^6$ & 12\,290    & $1.88 \times 10^{10}$ & $1\,090\times$ \\
(NO)$_4$-Anth Dimer     & 328  & $1.87 \times 10^6$ & 2\,792     & $5.21 \times 10^{9}$  & $1\,890\times$ \\
Anth/C$_{60}$ ($M\!=\!11$)  & 227  & $3.02 \times 10^5$ & 790        & $2.39 \times 10^{8}$  & $361\times$ \\
Anth/C$_{60}$ ($M\!=\!246$) & 2577 & $1.39 \times 10^8$ & 3\,274     & $4.55 \times 10^{11}$ & $17\,100\times$ \\[6pt]
\hline
\multicolumn{6}{l}{\textit{Corrected, worst case ($n=10$, $b=45$, $\eta = 1$):}} \\
(NO)$_4$-Anth           & 308  & $1.53 \times 10^6$ & 35\,000    & $5.36 \times 10^{10}$ & $3\,100\times$ \\
(NO)$_4$-Anth Dimer     & 328  & $1.87 \times 10^6$ & 7\,989     & $1.49 \times 10^{10}$ & $5\,400\times$ \\
Anth/C$_{60}$ ($M\!=\!11$)  & 227  & $3.02 \times 10^5$ & 2\,015     & $6.09 \times 10^{8}$  & $920\times$ \\
Anth/C$_{60}$ ($M\!=\!246$) & 2577 & $1.39 \times 10^8$ & 9\,742     & $1.35 \times 10^{12}$ & $50\,900\times$ \\
\hline\hline
\end{tabular}
\end{table}

\subsubsection{Circuit depth and serialization}\label{sec:depth}

The Motlagh et al.\ paper reports only Toffoli count.
For this algorithm, circuit depth is essentially equal to Toffoli count because there is almost no opportunity for parallelism.
Within each term, the operations QROM $\to$ arithmetic $\to$ phase accumulation $\to$ uncompute form a single data-dependency chain.
Across terms within a fragment, all terms share the phase gradient register $|R\rangle$ and must be applied in sequence; the caching protocol (Motlagh et al., Section~3.1) makes this explicitly serial.
Across fragments within a Trotter step, the three-way splitting~\ref{eq:trotter_step} is inherently sequential.
Across Trotter steps, each step depends on the output of the previous one.

For fault-tolerant execution at $t_{\textrm{cycle}} \sim 1$~$\mu$s per logical Toffoli~\cite{litinski2019game}, the wall-clock time is $t_{\textrm{wall}} = C_{\textrm{total}} \times t_{\textrm{cycle}}$.
To maintain overall logical error $\delta = 0.01$ across depth $D = C_{\textrm{total}}$, the per-gate logical error must satisfy $p_L < \delta/D$, requiring surface code distances $d \geq 2\log_{10}(D/\delta) - 1$ and $\sim\!2d^2$ physical qubits per logical qubit.
The resulting fault-tolerant overhead is summarized in Table~\ref{tab:ft_overhead}.

\begin{table}[ht]
\centering
\small
\caption{Fault-tolerant overhead for corrected estimates (moderate scenario, $\delta = 0.01$, $t_{\textrm{cycle}} = 1$~$\mu$s, $10^4$ shots for $1\%$ population accuracy).}\label{tab:ft_overhead}
\renewcommand{\arraystretch}{1.15}
\begin{tabular}{l c c c c c}
\hline\hline
System & Total Toffoli & $t_{\textrm{wall}}$ (1 shot) & $t_{\textrm{wall}}$ ($10^4$ shots) & $d$ & Phys.\ qubits \\
\hline
(NO)$_4$-Anth         & $1.88 \times 10^{10}$ & $5.2$~h    & $6.0$~yr   & 25 & $\sim\!385\,000$ \\
Dimer                  & $5.21 \times 10^{9}$  & $1.4$~h    & $1.7$~yr   & 23 & $\sim\!347\,000$ \\
Anth/C$_{60}$ ($M\!=\!11$)    & $2.39 \times 10^{8}$  & $4.0$~min  & $27.6$~d   & 21 & $\sim\!200\,000$ \\
Anth/C$_{60}$ ($M\!=\!246$)   & $4.55 \times 10^{11}$ & $5.3$~d    & $144$~yr   & 27 & $\sim\!3\,760\,000$ \\
\hline\hline
\end{tabular}
\end{table}

\noindent
Even the smallest system (Anth/C$_{60}$, $M = 11$) requires $\sim\!28$~days at the moderate coupling scenario with $10^4$ shots, or $\sim\!3$~days with amplitude estimation ($\sim\!100$ rounds at doubled depth).
The largest system ($M = 246$) is categorically infeasible.
These wall-time estimates exclude the use of the Motlagh et al.\ algorithm in the iterative materials discovery pipeline depicted in their Figures~1 and~4.

\subsubsection{Decomposition of the cost increase}\label{sec:decomposition}

Table~\ref{tab:decomposition} separates the correction into three multiplicative factors.

\begin{table}[ht]
\centering
\caption{Decomposition of the correction factor into independent multiplicative contributions (moderate, $\eta = 1/3$).}\label{tab:decomposition}
\renewcommand{\arraystretch}{1.15}
\begin{tabular}{l c c c c}
\hline\hline
System & \shortstack{Per-term cost \\ ($n,b$ increase)} & \shortstack{BVC terms \\ ($\mathcal{O}(M^2)$)} & \shortstack{Trotter steps \\ ($\sqrt\beta$)} & Combined \\
\hline
(NO)$_4$-Anth        & $5.1\times$  & $\sim\!13\times$  & $24\times$ & $1\,090\times$ \\
Dimer                 & $5.1\times$  & $\sim\!14\times$  & $25\times$ & $1\,890\times$ \\
Anth/C$_{60}$ ($M\!=\!11$)  & $5.3\times$  & $\sim\!6\times$   & $14\times$ & $361\times$ \\
Anth/C$_{60}$ ($M\!=\!246$) & $5.3\times$  & $\sim\!123\times$ & $60\times$ & $17\,100\times$ \\
\hline\hline
\end{tabular}
\end{table}

\noindent
The three factors compound multiplicatively.
The Trotter step correction (dominated by the $8\times$ factor from $\sqrt{\beta_{\textrm{grid}}} = \sqrt{64}$) is the largest single contributor for small systems.
For large $M$, the $\mathcal{O}(M^2)$ bilinear term count dominates, driving the $M = 246$ correction to $17\,100\times$ even at moderate coupling.

\chapter{Pyrazine Parameters}\label{app:pyrazine_params}
\section{2-mode Pyrazine}

The pyrazine molecule has the point group $D_{2h}$, with the $S_1$ state having $B_{3u}$ symmetry and $S_2$ having $B_{2u}$ symmetry. 
An expansion of the vibrational Hamiltonian has nonzero terms if the off-diagonal couplings have $B_{1g}$ symmetry and on-diagonal couplings have $A_g$ symmetry. 
Knowing this, the non-zero terms in the Hamiltonian are
\begin{equation}
\begin{split}
    \hat{H} & = \frac{p_x^2}{2\mu_x}+\frac{p_y^2}{2\mu_y}+\begin{pmatrix}
        \frac{1}{2}\mu_x\omega_x^2 & 0 \\ 0 & \frac{1}{2}\mu_x\omega_x^2
    \end{pmatrix}x^2 +\begin{pmatrix}
        \frac{1}{2}\mu_y\omega_y^2 & 0 \\ 0 & \frac{1}{2}\mu_y\omega_y^2
    \end{pmatrix}y^2  \\
    &+\begin{pmatrix}
        -\Delta E & 0 \\ 0 & \Delta E
    \end{pmatrix}
    + \begin{pmatrix}
        a_1 & 0 \\ 0 & b_1
    \end{pmatrix}x + \begin{pmatrix}
        a_2 & 0 \\ 0 & b_2
    \end{pmatrix}x^2\\&+\begin{pmatrix}
        a_3 & 0 \\ 0 & b_3
    \end{pmatrix}y^2+\begin{pmatrix}
        0 & c_1 \\ c_1 & 0
    \end{pmatrix}y + \begin{pmatrix}
        0 & c_2 \\ c_2 & 0
    \end{pmatrix}xy.
    \end{split}
\end{equation}

Parameters for the simulation are given in the following table, with vibrational frequencies taken from the experimentally-derived figures from Raab et al. \cite{raab1999molecular, yamazaki1983intramolecular}, and couplings taken from the single-excitation configuration interaction method \cite{raab1999molecular, foresman1992toward}.

\begin{center}
\begin{tabular}{lcc}
\toprule
\textbf{Description} & \textbf{Symbol} & \textbf{Value} \\
\midrule
Trotter steps & $k$ & 2500 \\
Trotter step size & $\tau$ & $1.0\,\text{a.u.}$ \\
Evolution time & $t$ & $2500\,\text{a.u.}$\\
Qubits on $x$-axis ($v6a$ mode) & $n_x$ & 9 \\
Qubits on $y$-axis ($v10a$ mode) & $n_y$ & 9 \\
Number of vibrational modes & $d$ & 2 \\
Number of electronic channels & $M$ & 2 \\
Channel register size & $m$ & 1\\
Box size & $L_x$, $L_y$ & 8.0 a.u., 8.0 a.u. \\
$S_1$/$S_2$ initial population & $P_0$/$P_1$ & 0.75/0.25 \\
Reduced mass & $\mu_x$, $\mu_y$ & 46484.0 $m_e$, 25520.0 $m_e$\\
Harmonic frequency & $\omega_x$, $\omega_y$ & 0.0739 eV, 0.1139 eV\\
Initial position & $x_0$, $y_0$ & 0.25 a.u., 0.0 a.u.\\
Initial momentum & $p_{x,0}$, $p_{y,0}$ & 0.0 a.u., 0.0 a.u.\\
Gaussian width & $\sigma_x,\,\sigma_y$ & 0.2109 a.u.,0.2310 a.u.\\
Energy shift & $\Delta E$ & 0.4617 eV\\
&$a_1$, $b_1$ & -0.0981 eV, 0.1355 eV\\
&$a_2$, $b_2$ & 0.00002 eV, -0.00917 eV\\
&$a_3$, $b_3$ & -0.01159 eV, -0.01159 eV\\
&$c_1$, $c_2$ & 0.2080 eV, 0.00628 eV\\
\bottomrule
\end{tabular}
\end{center}

\section{4-mode Pyrazine}
The inclusion of two more modes is easily handled once the parameters are given.
The 4-mode pyrazine model parameters are given using the same resources as the two-mode model.

\begin{longtable}{lcc}
\caption{Simulation parameters for the 4-mode, 2-channel vibronic coupling model.}
\label{tab:sim-params} \\
\toprule
\textbf{Description} & \textbf{Symbol} & \textbf{Value} \\
\midrule
\endfirsthead

\multicolumn{3}{c}{\tablename\ \thetable{} -- \textit{continued from previous page}} \\
\toprule
\textbf{Description} & \textbf{Symbol} & \textbf{Value} \\
\midrule
\endhead

\midrule
\multicolumn{3}{r}{\textit{Continued on next page}} \\
\endfoot

\bottomrule
\endlastfoot

Trotter steps & $k$ & 2500 \\
Trotter step size & $\tau$ & $1.0\,\text{a.u.}$ \\
Evolution time & $t$ & $2500\,\text{a.u.}$ \\
Qubits on grid & $n_{x_0}, n_{x_1}, n_{x_2}, n_{x_3}$ & (8, 8, 8, 8) \\
$\#$ vibrational modes & $d$ & 4 \\
$\#$ electronic channels & $M$ & 2 \\
Box size on all axes & $L_{x_{\{0,1,2,3\}}}$ & 4.0 a.u. \\
$S_1$/$S_2$ initial population & $P_0$/$P_1$ & 0.0/1.0 \\
Reduced masses ($m_e$) & $(\mu_{x_0}, \mu_{x_1}, \mu_{x_2}, \mu_{x_3})$ & (46484.0, $\mu_{x_0}$, $\mu_{x_0}$, 25520.0) \\
Harmonic frequencies (eV) & ($\omega_{x_0}, \omega_{x_1}, \omega_{x_2}, \omega_{x_3}$) & (0.0739, 0.1258, 0.1525, 0.1139) \\
Initial position & ($x_{0_0}, x_{1_0}, x_{2_0}, x_{3_0}$) & (0.2, 0, 0, 0) \\
Initial momentum & ($p_{0_0}, p_{1_0}, p_{2_0}, p_{3_0}$) & (0, 0, 0, 0) \\
Gaussian widths ($a_0$) & ($\sigma_{x_0}, \sigma_{x_1}, \sigma_{x_2}, \sigma_{x_3}$) & (0.2109, 0.1847, 0.1760, 0.2310) \\
Energy shift & $\Delta E$ & 0.4617 eV \\
\midrule
\multicolumn{3}{c}{\textbf{On-Diagonal Parameters}} \\
\midrule
Linear parameters ($v6a$) & ($a_0$, $b_0$) & (-0.0981 eV, 0.1355 eV) \\
Linear parameters ($v1$) & ($a_1$, $b_1$) & (-0.0503 eV, -0.1629 eV) \\
Linear parameters ($v9a$) & ($a_2$, $b_2$) & (0.1452 eV, 0.0375 eV) \\
Linear parameters ($v10a$) & ($a_3$, $b_3$) & (-0.0116 eV, -0.0116 eV) \\
Quadratic parameters ($v6a$) & ($a_{00}, b_{00}$) & (0.00002 eV, -0.00917 eV) \\
Quadratic parameters ($v1$) & ($a_{11}, b_{11}$) & (-0.00810 eV, -0.00488 eV) \\
Quadratic parameters ($v9a$) & ($a_{22}, b_{22}$) & (-0.00116 eV, 0.00022 eV) \\
Quadratic parameters ($v10a$) & ($a_{33}, b_{33}$) & (-0.01159 eV, -0.01159 eV) \\
Bilinear parameters ($v6a, v1$) & ($a_{01}, b_{01}$) & (0.00108 eV, -0.00298 eV) \\
Bilinear parameters ($v6a, v9a$) & ($a_{02}, b_{02}$) & (-0.00204 eV, -0.00189 eV) \\
Bilinear parameters ($v1, v9a$) & ($a_{12}, b_{12}$) & (0.00474 eV, 0.00155 eV) \\
\midrule
\multicolumn{3}{c}{\textbf{Off-Diagonal Parameters}} \\
\midrule
Linear coupling ($v10a$) & ($c_{3}, c_3$) & 0.2080 eV \\
Bilinear coupling ($v6a, v10a$) & ($c_{03}, c_{30}$) & 0.00628 eV \\
Bilinear coupling ($v1, v10a$) & ($c_{13}, c_{31}$) & -0.00551 eV \\
Bilinear coupling ($v9a, v10a$) & ($c_{23}, c_{32}$) & 0.00127 eV \\

\end{longtable}

\section{Landau-Zener Bowtie Convergence}

\begin{figure}[t]
\centering
\includegraphics[width=\textwidth]{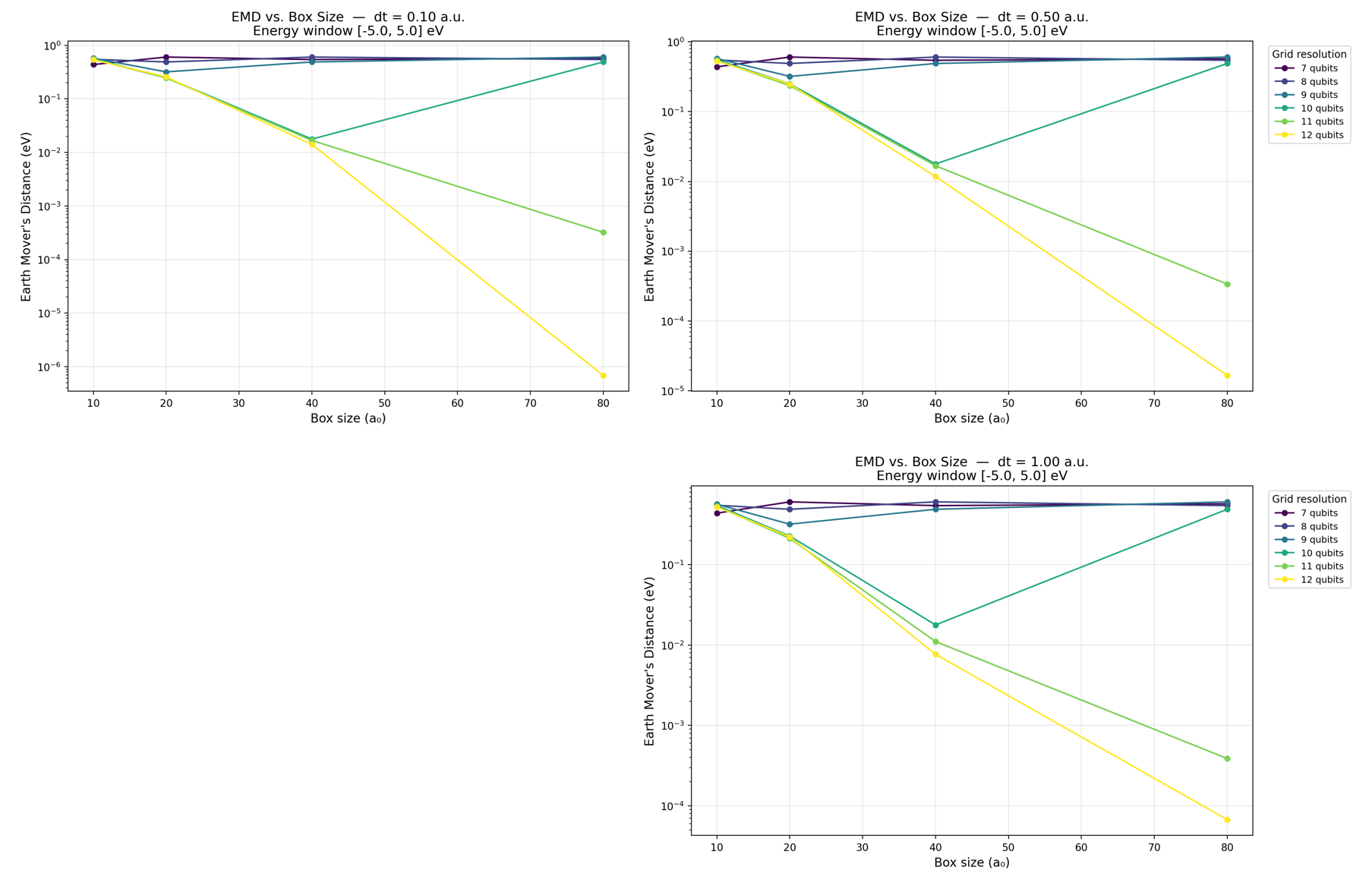}
\caption{\label{fig:BoxSizeLZConv}
Convergence for Landau-Zener bowtie model for changing box size. Earth mover's distance (Wasserstein-1 distance) acts as a measure for convergence, with coherent wavefunction propagation and spectral peak instenties requiring a sufficient timestep, register size, and box size.}
\end{figure}

\begin{figure}[t]
\centering
\includegraphics[width=\textwidth]{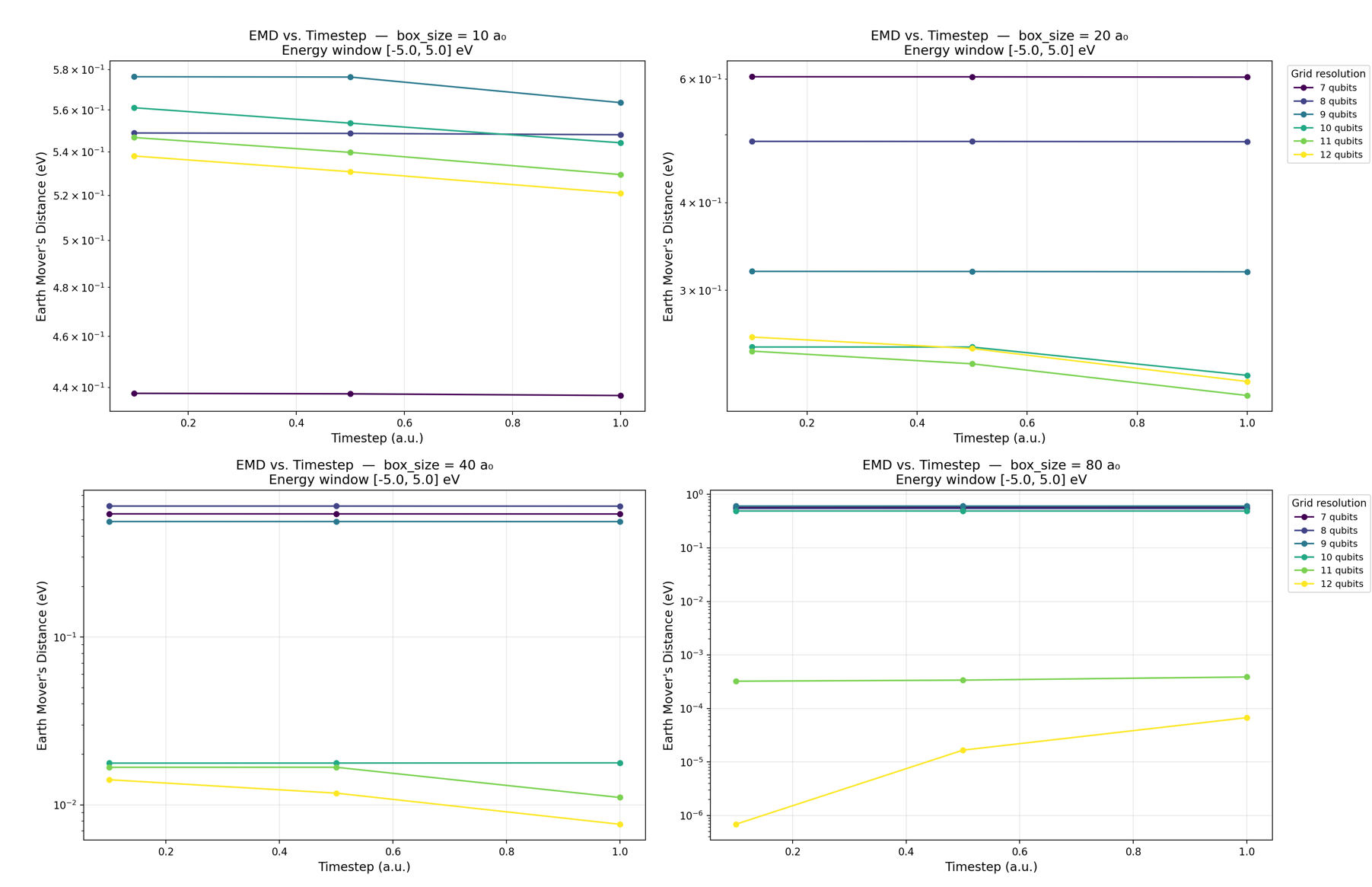}
\caption{\label{fig:TimestepLZConv}
Convergence for Landau-Zener bowtie model for changing time step.}
\end{figure}

Figures~\ref{fig:BoxSizeLZConv} and~\ref{fig:TimestepLZConv} demonstrate convergence of the Landau–Zener bowtie model with respect to box size and timestep, respectively, quantified by the Wasserstein-1 distance.


\chapter{Suboptimal Scaling for Pairwise Coupling}\label{app:basic_coupling}
\section{Basic Nonadiabatic Coupling}

This appendix reviews the nonadiabatic coupling scheme of
Ollitrault et al.~\cite{ollitrault2020nonadiabatic}, showing that it
achieves suboptimal scaling of $\mathcal{O}(M^2\log M)$ Toffoli gates
for $M$ electronic channels.

\subsection{Coupling Primitives for $M=2$}\label{sec:coupling_M2}

The two-channel case ($m=1$) has the channel register as a single qubit $\ket{s}$.  These primitives form the building blocks
for the general construction and are originally due to Ollitrault
et al.~\cite{ollitrault2020nonadiabatic}.

\subsubsection{Linear Vibronic Coupling (LVC)}

\begin{figure}[htbp]
\centerline{
\Qcircuit @C=0.1em @R=1.0em {
\lstick{\ket{q_{0}}}& \qw & \ctrl{3} & \qw & \qw & \qw \\
\lstick{\ket{q_{1}}}& \qw & \qw & \ctrl{2} & \qw & \qw \\
\lstick{\ket{q_{2}}}& \qw & \qw & \qw & \ctrl{1} & \qw \\
\lstick{\ket{s}} & \gate{R_x(b\tau)} & \gate{R_x(2^0a\tau)}
& \gate{R_x(2^1a\tau)}& \gate{R_x(2^2a\tau)} & \qw
}
}
\caption{\label{fig:LVC}Quantum circuit for a linear vibronic coupling
between one mode register and one channel qubit, with timestep~$\tau$.
The constant offset $b$ is applied unconditionally; the linear
coefficient $a$ is weighted by binary factors $2^i$.}
\end{figure}
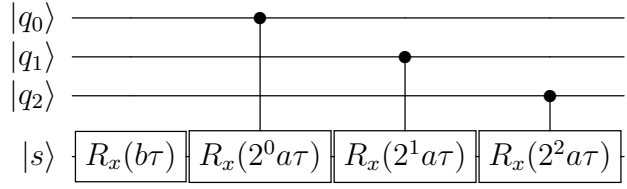

A linear coupling applies
\begin{equation}
    R_x\!\bigl(f(x)\bigr)\ket{s}
    = R_x(a\,x + b)\ket{s},
\end{equation}
for a given channel state $\ket{s}$, where the position variable $x$ is
encoded in the binary-weighted values of the $n$ grid qubits.  The
result is $n$ singly controlled $R_x$ gates (with binary weights
$2^0 a,\,2^1 a,\ldots,2^{n-1}a$) plus one unconditional rotation by
$b$, as shown in Figure~\ref{fig:LVC}.

For $d$ vibrational modes, LVC requires $dn$ singly controlled $R_x$
gates and $2dn$ CNOTs (from their standard decomposition), with no
multi-qubit gates beyond two-body.

\subsubsection{Quadratic Vibronic Coupling (QVC)}

\begin{figure}[htbp]
\centerline{
\Qcircuit @C=0.1em @R=1.0em {
\lstick{\ket{q_{0}}}& \qw & \ctrl{3} & \qw & \qw
  & \ctrl{3} & \qw & \ctrl{3}& \qw\\
\lstick{\ket{q_{1}}}& \qw & \qw & \ctrl{2} & \qw
  & \ctrl{2} & \ctrl{2} & \qw& \qw\\
\lstick{\ket{q_{2}}}& \qw & \qw & \qw & \ctrl{1}
  & \qw & \ctrl{1} & \ctrl{1}& \qw\\
\lstick{\ket{s}}& \gate{R_x(c\tau)} & \gate{R_x(2^0(a+b)\tau)}
  & \gate{R_x((2^{2}a+2^1b)\tau)}
  & \gate{R_x((2^{4}a+2^2b)\tau)}
  & \gate{R_x(2^{0+1}a\tau)}
  & \gate{R_x(2^{1+2}a\tau)}
  & \gate{R_x(2^{0+2}a\tau)} & \qw\\
}
}
\caption{\label{fig:QVC_sample}Quantum circuit for a quadratic vibronic
coupling (off-diagonal) with timestep~$\tau$.}
\end{figure}
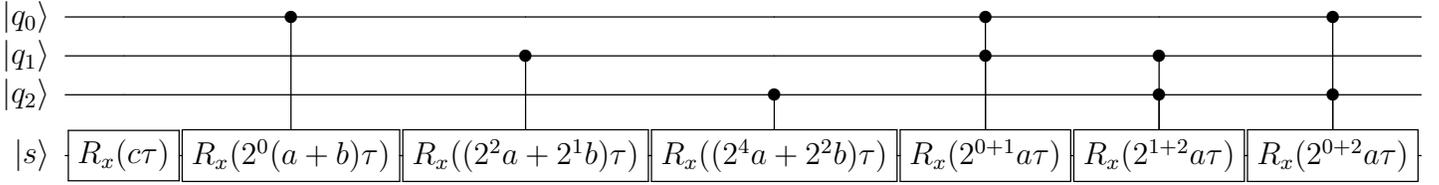

A quadratic coupling applies
\begin{equation}
    R_x\!\bigl(f(x)\bigr)\ket{s}
    = R_x(a\,x^2 + b\,x + c)\ket{s}.
\end{equation}
The binary expansion of $x^2$ generates
$\sum_{k=1}^{2}\binom{n}{k}=n+\binom{n}{2}$ terms: $n$ singly
controlled gates (from the linear-plus-diagonal piece) and
$\binom{n}{2}$ doubly controlled gates (from the cross terms
$2^{i+j}$), plus one constant rotation.  An example for $n=3$ is
shown in Figure~\ref{fig:QVC_sample}.

Doubly controlled gates are implemented as $CCX$--$R_x$--$CCX$
sequences, contributing $dn(n-1)$ Toffoli gates for $d$ modes.

\subsubsection{Bilinear Vibronic Coupling (BVC)}

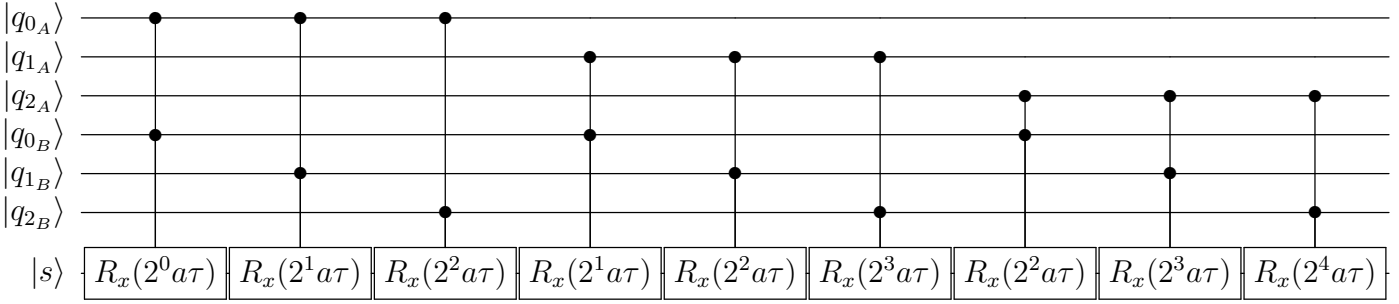
\begin{figure}[htbp]
\centerline{
\Qcircuit @C=0.1em @R=1.0em {
\lstick{\ket{q_{0_A}}}& \ctrl{6} & \ctrl{6} & \ctrl{6}
  & \qw & \qw & \qw & \qw & \qw & \qw & \qw \\
\lstick{\ket{q_{1_A}}}& \qw & \qw & \qw
  & \ctrl{5} & \ctrl{5} & \ctrl{5} & \qw & \qw & \qw & \qw  \\
\lstick{\ket{q_{2_A}}}&\qw & \qw & \qw
  & \qw & \qw & \qw & \ctrl{4} & \ctrl{4} & \ctrl{4} & \qw\\
\lstick{\ket{q_{0_B}}}& \ctrl{3} & \qw & \qw
  & \ctrl{3} & \qw & \qw & \ctrl{3} & \qw & \qw & \qw\\
\lstick{\ket{q_{1_B}}}& \qw & \ctrl{2} & \qw
  & \qw & \ctrl{2} & \qw & \qw & \ctrl{2} & \qw & \qw\\
\lstick{\ket{q_{2_B}}}& \qw & \qw & \ctrl{1}
  & \qw & \qw & \ctrl{1} & \qw & \qw & \ctrl{1} & \qw\\
\lstick{\ket{s}}& \gate{R_x(2^0a\tau)} & \gate{R_x(2^1a\tau)}
  & \gate{R_x(2^2a\tau)}& \gate{R_x(2^1a\tau)}
  & \gate{R_x(2^2a\tau)} & \gate{R_x(2^3a\tau)}
  & \gate{R_x(2^2a\tau)} & \gate{R_x(2^3a\tau)}
  & \gate{R_x(2^4a\tau)} & \qw
}
}
\caption{\label{fig:BVC_sample}Quantum circuit for a bilinear vibronic
coupling (off-diagonal) with timestep~$\tau$.  Prefactors $2^{i+j}$ are
weighted by qubit indices $i$ and $j$ from registers $A$ and $B$.}
\end{figure}

An off-diagonal bilinear coupling between mode registers $A$ and $B$
applies
\begin{equation}
    R_x\!\bigl(f(x_A,x_B)\bigr)
    = R_x(a\,x_A\,x_B),
\end{equation}
where the binary expansion produces $n^2$ terms $2^{i+j}k_{i_A}k_{j_B}$
with one qubit from each register as control.  An example for $n=3$
per mode is shown in Figure~\ref{fig:BVC_sample}.

Each of the $n^2$ terms requires a Toffoli pair ($CCX$--$R_x$--$CCX$),
giving $2n^2$ Toffoli gates per mode pair.  For $\binom{d}{2}$ mode
pairs, the bilinear terms alone contribute $d(d-1)n^2$ Toffoli gates.

\subsubsection{Summary of Two-Channel Gate Costs}

For $M=2$ ($m=1$), with $d$ vibrational modes and $n$ grid qubits per
mode, the gate counts for each coupling type are:

\begin{center}
\begin{tabular}{lccc}
\toprule
\textbf{Term} & \textbf{Count} & \textbf{Rotations} & \textbf{Toffoli gates} \\
\midrule
LVC (linear)   & $d$                   & $dn$            & 0 \\
QVC (quadratic) & $d$                   & $d\bigl(n+\binom{n}{2}\bigr)$ & $dn(n-1)$ \\
BVC (bilinear)  & $\binom{d}{2}$        & $\binom{d}{2}n^2$ & $d(d-1)n^2$ \\
\midrule
\textbf{Total}  & ---                   & ---             & $d(d-1)n^2 + dn(n-1)$ \\
\bottomrule
\end{tabular}
\end{center}

\subsection{Information-Theoretic Lower Bound}\label{sec:lower_bound}

Off-diagonal coupling is constrained to be real and symmetric:
\begin{equation}
    V^{(ij)} = c^{(ij)}\bigl(\ket{i}\!\bra{j}+\ket{j}\!\bra{i}\bigr),
    \qquad c^{(ij)}\in\mathbb{R},
\end{equation}
where $c^{(ij)}$ depends on the nuclear coordinates (via LVC, QVC, or
BVC terms).  For each nuclear-coordinate site (i.e., each term in the
binary expansion of the coupling function), the coupling matrix is an
$M\times M$ real symmetric matrix with vanishing diagonal, having
$\binom{M}{2}=M(M-1)/2$ independent parameters.

Each generator $G_{ij}=\ket{i}\!\bra{j}+\ket{j}\!\bra{i}$ is a real
symmetric matrix generating the Lie algebra $\mathfrak{so}(M)$, with
\begin{equation}\label{eq:dim_soM}
    \dim\bigl(\mathfrak{so}(M)\bigr)=\frac{M(M-1)}{2}.
\end{equation}
Any scheme that independently addresses all $M(M-1)/2$ couplings must
therefore use at least $M(M-1)/2=\mathcal{O}(M^2)$ independent
rotation parameters. $\mathcal{O}(M^2)$ is taken as the target
complexity for an optimal scheme.

\subsection{Pair-Isolation Approach}\label{sec:pair_isolation}

The natural extension of the Ollitrault et
al.~\cite{ollitrault2020nonadiabatic} scheme to $M>2$ channels couples
each pair $(i,j)$ independently via multi-controlled rotations (Equation S7 of Ollitrault et al. \cite{ollitrault2020nonadiabatic}).  

\subsubsection{Construction}

For a given pair $(i,j)$ with $i\neq j$, the unitary that couples only
channels $\ket{i}$ and $\ket{j}$ with angle $\theta$ is
\begin{equation}\label{eq:Uij}
    U_{ij}(\theta) = \mathbb{1}
    + (\cos\tfrac{\theta}{2}-1)
      \bigl(\ket{i}\bra{i}+\ket{j}\bra{j}\bigr)
    - i\sin\tfrac{\theta}{2}
      \bigl(\ket{i}\!\bra{j}+\ket{j}\!\bra{i}\bigr).
\end{equation}
On the $m$-qubit channel register, this can be implemented by the
following subroutine:
\begin{enumerate}
    \item Compute $\zeta = i\oplus j$ (classically).
    \item Select a target qubit $t$ such that $\zeta_t=1$.
    \item \textbf{For} each qubit $k\neq t$: \textbf{if} bit $k$ of $i$
          is $0$, apply $X_k$ (to condition on $\ket{0}$).
    \item Apply $C^{m-1}\!R_x(\theta_{ij})$ with all $m-1$ non-target
          qubits as controls and qubit $t$ as target.
    \item Reverse the conditioning $X$ gates from step~3.
\end{enumerate}
The $C^{m-1}\!R_x$ gate is the bottleneck: it requires computing a
truth value (AND of $m-1$ control qubits), implemented as a linear
chain of Toffoli gates.

\subsubsection{Toffoli Cost}\label{sec:pair_toffoli}

Using one ancilla qubit, a $C^{m-1}\!R_x$ gate decomposes as:
\begin{itemize}
    \item $m-2$ Toffoli gates to compute the $(m-1)$-bit AND onto an
          ancilla,
    \item one singly controlled $CR_x$ gate (Toffoli-free),
    \item $m-2$ Toffoli gates to uncompute the ancilla,
\end{itemize}
for a total of $2(m-2)$ Toffoli gates per pair.  Summing over all
$M(M-1)/2$ pairs:
\begin{equation}\label{eq:toffoli_pair}
    T_{\text{pair}} = \frac{M(M-1)}{2}\cdot 2(m-2)
    = (m-2)\,M(M-1)
    = \mathcal{O}(M^2\log M).
\end{equation}
This exceeds the optimal $\mathcal{O}(M^2)$ target by a factor of
$\Theta(\log M)$.

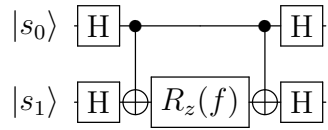
\begin{figure}[htbp]
\centerline{
\Qcircuit @C=0.1em @R=1.0em {
\lstick{\ket{s_0}}& \gate{\text{H}} & \ctrl{1} & \qw
  & \ctrl{1} & \gate{\text{H}} & \qw \\
\lstick{\ket{s_1}}& \gate{\text{H}} & \targ & \gate{R_z(f)}
  & \targ & \gate{\text{H}} & \qw \\
}
}
\caption{\label{fig:Ollitrault_S7}Quantum circuit to couple the
$\ket{00}$ and $\ket{11}$ electronic channels via
$\exp(i\sigma_x^{(0)}\sigma_x^{(1)}f)$, from the supplemental material
of Ollitrault et al.~\cite{ollitrault2020nonadiabatic}.  For $m=2$ this
requires no Toffoli gates, but the extension to $m\geq 3$ does.}
\end{figure}

An example of one such pairwise coupling circuit for $M=4$ is shown in
Figure~\ref{fig:Ollitrault_S7}.

\subsection{XOR Class Decomposition}\label{sec:xor_classes}

A natural optimization groups pairs that share the same XOR pattern,
allowing simultaneous coupling via a single generator.
Independently, Motlagh et al.~\cite{motlagh2025quantum} employ an
equivalent XOR fragmentation of the potential Hamiltonian, loading
state-dependent coefficients via quantum read-only memory (QROM).

\subsubsection{Definitions}

For two channel-register basis states $\ket{i}$ and $\ket{j}$, define
the {XOR pattern} $\zeta=i\oplus j\in\{0,1\}^m\setminus\{0^m\}$
and the {Hamming weight} $h=\mathrm{wt}(\zeta)$.  The $2^m-1$
nonzero patterns partition all $M(M-1)/2$ pairs into disjoint
{XOR classes}~$\mathcal{C}_\zeta$.

\paragraph{Class size.}
For any nonzero $\zeta$, the class $\mathcal{C}_\zeta$ contains exactly
\begin{equation}\label{eq:class_size}
    |\mathcal{C}_\zeta| = 2^{m-1},
\end{equation}
pairs, independent of $h$.  To see this, note that for each
$i\in\{0,\ldots,2^m-1\}$ there is a unique partner $j=i\oplus\zeta$,
and the ordering constraint $i<j$ selects exactly half of the $2^m$
values of~$i$.

Summing over all nonzero $\zeta$:
\begin{equation}
    \sum_{\zeta\neq 0}|\mathcal{C}_\zeta|
    = (2^m-1)\cdot 2^{m-1}
    = \frac{M(M-1)}{2},
\end{equation}
confirming the partition is complete.

\paragraph{Active and spectator qubits.}
Within class $\mathcal{C}_\zeta$ (weight $h$), the $h$ qubit positions
where $\zeta$ has a~$1$ are {active}: these qubits flip between
$i$ and $j$.  The remaining $m-h$ positions are {spectator}: they
have the same value in $i$ and $j$.  Two pairs in the same class are
distinguished by:
\begin{itemize}
    \item the $m-h$ spectator-qubit values ($2^{m-h}$ configurations),
    \item the $h-1$ active-qubit values of the lower-indexed state on
          the non-target active qubits ($2^{h-1}$ configurations, after
          fixing the ordering $i<j$ on the most significant active bit).
\end{itemize}
The product is $2^{m-h}\cdot 2^{h-1}=2^{m-1}$, consistent with
Eq.~\ref{eq:class_size}.

\subsubsection{Generator Structure}

The generator $X_{S(\zeta)}=\prod_{k\in S(\zeta)}X_k$, where
$S(\zeta)$ is the set of active qubit positions, simultaneously couples
all $2^{m-1}$ pairs in $\mathcal{C}_\zeta$ with the {same}
rotation angle.  To assign {independent} coupling strengths to
each pair, one must further condition on the distinguishing bits.

As an illustrative example, applying $R_x(\theta)$ to the lowest
channel qubit $\ket{s_0}$ in a 3-qubit register ($m=3$) simultaneously
couples four pairs:
\begin{equation}
    \ket{000}\leftrightarrow\ket{001},\quad
    \ket{010}\leftrightarrow\ket{011},\quad
    \ket{100}\leftrightarrow\ket{101},\quad
    \ket{110}\leftrightarrow\ket{111}.
\end{equation}
This class has $h=1$, so all distinguishing bits are spectator bits
($m-h=2$ spectators), and all four pairs can be independently
addressed by conditioning on $\ket{s_1 s_2}$.  However, for the class
$\zeta=011$ ($h=2$), the four pairs are:
\begin{equation}
    \ket{000}\leftrightarrow\ket{011},\quad
    \ket{001}\leftrightarrow\ket{010},\quad
    \ket{100}\leftrightarrow\ket{111},\quad
    \ket{101}\leftrightarrow\ket{110}.
\end{equation}
Here there is only $m-h=1$ spectator qubit, giving $2^1=2$ spectator
configurations, while the class contains $2^{m-1}=4$ pairs.  The
remaining factor of $2^{h-1}=2$ comes from the non-target active qubit,
which the spectator-only scheme cannot distinguish.

\subsection{Spectator-Qubit Multiplexing}\label{sec:multiplexed}

The multiplexed approach attempts to achieve
independent control within each XOR class by conditioning on spectator
qubits only.

\subsubsection{Construction}

For a class $\mathcal{C}_\zeta$ with weight $h$, select a target active
qubit $t$ and apply:
\begin{enumerate}
    \item \textbf{Clifford diagonalization:} CNOT from
          $h-1$ non-target active qubits to $t$, followed by a Hadamard
          on $t$.  This conjugates $X_{S(\zeta)}$ into $Z_t$, reducing
          the multi-qubit coupling to a $Z$-rotation on qubit $t$.
    \item \textbf{Spectator-controlled rotations:} Apply $2^{m-h}$
          rotations $R_{z_t}(\tilde{c}_a)$, indexed by
          $a\in\{0,1\}^{m-h}$, where rotation $a$ is controlled on 
          spectator qubits at positions where $a$ has a~$1$.
    \item \textbf{Uncompute:} Reverse the Clifford diagonalization.
\end{enumerate}

The $2^{m-h}$ transformed coefficients $\{\tilde{c}_a\}$ are obtained
from physical couplings $\{c_b\}_{b\in\{0,1\}^{m-h}}$ by M\"obius
inversion on the Boolean lattice:
\begin{equation}\label{eq:mobius}
    \tilde{c}_a = \sum_{b:\,a\leq b}(-1)^{|b|-|a|}\,c_b,
\end{equation}
where $a\leq b$ denotes componentwise bit comparison and $|\cdot|$ is
the Hamming weight.  This ensures that the total rotation angle
accumulated by spectator configuration $b$ is exactly $c_b$, by the
inclusion-exclusion principle:
\begin{equation}
    \theta_{\mathrm{tot}}(b)
    = \sum_{a:\,a\leq b}\tilde{c}_a
    = \sum_{a\leq b}\sum_{k\leq a}(-1)^{|a|-|k|}\,c_k
    = c_b,
\end{equation}
where the last equality follows because the inner sums cancel for all
$k\neq b$.  The inversion~\ref{eq:mobius} is computed classically in
$\mathcal{O}(2^{m-h})$ operations.

\subsubsection{Degrees of Freedom}\label{sec:dof_deficit}

The multiplexed scheme assigns $2^{m-h}$ independent rotation angles
per class (one per spectator configuration), but each class contains
$2^{m-1}$ pairs (Eq.~\ref{eq:class_size}).  Since the Clifford
diagonalization step treats all $2^{h-1}$ active-qubit configurations
identically, pairs that share the same spectator values but differ in
their non-target active-qubit values receive the same coupling strength.

\begin{center}
\begin{tabular}{cccc}
\toprule
\textbf{Hamming weight $h$}
  & \textbf{Degrees of freedom}
  & \textbf{Pairs per class}
  & \textbf{Fraction addressed} \\
\midrule
$1$ & $2^{m-1}$ & $2^{m-1}$ & $1$ \\
$2$ & $2^{m-2}$ & $2^{m-1}$ & $1/2$ \\
$3$ & $2^{m-3}$ & $2^{m-1}$ & $1/4$ \\
$h$ & $2^{m-h}$ & $2^{m-1}$ & $2^{1-h}$ \\
\bottomrule
\end{tabular}
\end{center}

For classes with $h=1$ (all $m$ such classes), the scheme is fully
expressive: every pair is independently addressable.  For $h\geq 2$,
each spectator-controlled rotation conflates $2^{h-1}$ pairs, so only a
$2^{1-h}$ fraction of the couplings can be set independently.

The total number of independent rotation parameters across all classes
is
\begin{equation}\label{eq:total_dof}
    \sum_{h=1}^{m}\binom{m}{h}\cdot 2^{m-h}
    = \sum_{h=0}^{m}\binom{m}{h}\cdot 2^{m-h}\cdot 1^h - 2^m
    = 3^m - 2^m,
\end{equation}
using the binomial theorem.

Comparing to the required $M(M-1)/2 = 2^{m-1}(2^m-1)$:

\begin{center}
\begin{tabular}{cccc}
\toprule
$m$ & $3^m-2^m$ (scheme d.o.f.)
    & $2^{m-1}(2^m-1)$ (required)
    & Deficit factor \\
\midrule
1 & 1    & 1     & $1.0\times$ \\
2 & 5    & 6     & $1.2\times$ \\
3 & 19   & 28    & $1.5\times$ \\
4 & 65   & 120   & $1.8\times$ \\
5 & 211  & 496   & $2.4\times$ \\
6 & 665  & 2{,}016  & $3.0\times$ \\
\bottomrule
\end{tabular}
\end{center}

For $m\geq 2$, the scheme has strictly fewer degrees of freedom than
independent couplings.  Asymptotically,
\begin{equation}
    \frac{3^m-2^m}{2^{m-1}(2^m-1)}
    \;\xrightarrow{m\to\infty}\;
    2\!\left(\frac{3}{4}\right)^{\!m}\;\to\;0,
\end{equation}
so the deficit grows exponentially.  The spectator-only multiplexing
scheme is fundamentally underparameterized and cannot represent an
arbitrary real symmetric coupling matrix for $M>2$.

\subsection{Restoring Full Expressivity via Truth-Value
Computation}\label{sec:full_expressive}

To independently address all $2^{m-1}$ pairs within each XOR class, the
rotation within each class must be conditioned not only on the $m-h$
spectator qubits but also on the $h-1$ non-target active qubits.  This
requires $m-1$ total control qubits per rotation, restoring the
truth-value computation overhead.

\subsubsection{Construction}

Within class $\mathcal{C}_\zeta$ (weight $h$), after Clifford
diagonalization, each of the $2^{m-1}$ pairs is uniquely identified by
an $(m-1)$-bit string: the $m-h$ spectator values concatenated with the
$h-1$ non-target active-qubit values.  To apply an independent rotation
to pair $b\in\{0,1\}^{m-1}$, one must compute the truth value ``all
$m-1$ control qubits match pattern~$b$,'' which requires:
\begin{itemize}
    \item conditioning $X$ gates on the control qubits where $b$ has
          a~$0$ (to convert the pattern to all-ones),
    \item computing the $(m-1)$-way AND via a Toffoli chain: $m-2$
          Toffoli gates to compute, $m-2$ to uncompute, totaling
          $2(m-2)$ Toffoli gates per pair,
    \item one $\mathrm{CR}_z$ gate (Toffoli-free),
    \item reversing the conditioning $X$ gates.
\end{itemize}

\subsubsection{Toffoli Count}

The total Toffoli cost over all classes is:
\begin{equation}\label{eq:toffoli_full}
    T_{\text{full}}
    = \underbrace{(2^m-1)}_{\text{classes}}
      \times\underbrace{2^{m-1}}_{\text{pairs/class}}
      \times\underbrace{2(m-2)}_{\text{Toffoli/pair}}
    = (m-2)\,M(M-1)
    = \mathcal{O}(M^2\log M),
\end{equation}
identical to the pair-isolation cost~\ref{eq:toffoli_pair}.  The XOR
class grouping and Clifford diagonalization reduce the Clifford overhead
(fewer single-qubit gates and CNOTs) but do not reduce the Toffoli
count, because the number of control qubits per rotation remains $m-1$.

\subsubsection{Why XOR Grouping Alone Cannot Eliminate the Overhead}

The $\Theta(\log M)$ Toffoli overhead is intrinsic to truth-value-based
pair isolation: identifying a specific $(m-1)$-bit pattern among $m-1$
control qubits requires computing an AND, which costs $\Theta(m)$
Toffoli gates regardless of how the pairs are organized into classes.

The spectator-only scheme (Section~\ref{sec:multiplexed}) avoids
Toffoli gates for high-weight classes precisely by surrendering degrees
of freedom: it conditions on only $m-h < m-1$ qubits, so each rotation
addresses $2^{h-1}$ pairs simultaneously rather than one.  The savings
in Toffoli gates come at the cost of an exponentially growing
expressivity deficit.

The decision to partially diagonalize with XOR fragmentation follows with a potentially optimal gate scaling, given in Motlagh et al. \cite{motlagh2025quantum}, used in the main text.

\chapter{T-Gate Reduction Pipeline for REQWIEM}
\label{app:tgate_pipeline}

This appendix provides a self-contained treatment of the $T$-gate optimization
pipeline applicable to the REQWIEM circuit architecture. 
Each synthesis and optimization stage is defined, with a composite reduction factor calculated, including estimates for the fraction of rotations eligible for Hamming weight phasing in representative molecular systems.

Throughout, the {empirical} linear fits to the numerical benchmark data reported in \cite{kliuchnikov2023shorter} are used rather than the asymptotic theoretical bounds, more accurately capturing the per-rotation $T$-count at the finite precisions relevant to molecular simulation.

\subsection{Stage 1: Single-Gate Synthesis}
\label{app:synthesis}

\subsubsection{Ross--Selinger Synthesis (Baseline)}
\label{app:ross_selinger}

The Ross--Selinger algorithm \cite{ross2016optimal} provides an asymptotically
optimal decomposition of an arbitrary rotation $\Rz{\theta}$ into the
Clifford+$T$ gate set.

For any $\theta \in \mathbb{R}$ and synthesis precision $\eps > 0$, there
exists a Clifford+$T$ circuit $U$ with
\begin{equation}
    \|U - \Rz{\theta}\| \leq \eps \qquad \text{and} \qquad
    \Tcount(U) = 3\log_2\!\left(\frac{1}{\eps}\right)
    + \BigO\!\left(\log\log\frac{1}{\eps}\right).
    \label{eq:rs_tcount}
\end{equation}

\noindent
This result is {gate-optimal} in the sense that no single-gate synthesis protocol over Clifford+$T$ can achieve a smaller leading coefficient than~$3$.
The numerical benchmarks of \cite{kliuchnikov2023shorter} confirm the theoretical scaling and yield the empirical linear fit
\begin{equation}
    T_{\mathrm{RS}}(\eps) \;=\; 3.02\,\log_2(1/\eps) + 1.77,
    \label{eq:trs_def}
\end{equation}
for random rotation angles drawn uniformly from $[0, 2\pi]$, where the additive constant absorbs the lower-order correction from
\ref{eq:rs_tcount}.

\subsubsection{Mixed-Fallback Synthesis over Clifford+\texorpdfstring{$\sqrt{T}$}{sqrt(T)}}
\label{app:mixed_fallback}

Kliuchnikov et al. \cite{kliuchnikov2023shorter} introduced a refined protocol that operates over the extended gate set
Clifford+$\sqrt{T}$, where
\begin{equation}
    \sqrt{T} = \begin{pmatrix} 1 & 0 \\ 0 & e^{i\pi/8} \end{pmatrix}.
\end{equation}
A {mixed-fallback} strategy probabilistically selects among multiple candidate unitaries using ancilla measurements,
achieving shorter expected gate sequences.

For random rotation angles $\Rz{\theta}$ and precision $\eps > 0$,
mixed-fallback synthesis over Clifford+$\sqrt{T}$ produces a circuit with
expected $T$-count fitting the empirical linear relation
\begin{equation}
    T_{\mathrm{MF}}(\eps) = 0.53\,\log_2\!\left(\frac{1}{\eps}\right) + 4.86.
    \label{eq:tmf_def}
\end{equation}
\noindent
Every rotation in the REQWIEM circuit is eligible for mixed-fallback synthesis: the UCR decomposition produces $\Rz{\theta}$ gates on diagonal blocks and $\Rx{\theta} = H\,\Rz{\theta}\,H$ gates on off-diagonal blocks, both of which are Clifford conjugates of $R_z$ rotations.

The {mixed-fallback reduction factor} at precision $\eps$ is
\begin{equation}
    R_{\mathrm{MF}}(\eps) \;=\; \frac{T_{\mathrm{RS}}(\eps)}{T_{\mathrm{MF}}(\eps)}
    \;=\; \frac{3.02\,\log_2(1/\eps) + 1.77}{0.53\,\log_2(1/\eps) + 4.86}.
    \label{eq:rmf}
\end{equation}

\noindent
The asymptotic ratio of leading coefficients is $3.02/0.53 \approx 5.70$, which is approached in the limit $\eps \to 0$.  
At finite precisions relevant to molecular simulation, the additive constants reduce the realized improvement.  
Evaluating Eq.~\ref{eq:rmf} at representative precisions (\ref{tab:rmf_values}) yields $R_{\mathrm{MF}} \approx 4.5\times$ at
$\eps = 10^{-10}$, increasing slowly toward the asymptotic limit at higher precisions.

\begin{table}[h]
\centering
\caption{Per-rotation $T$-counts (empirical fits) and mixed-fallback reduction
factor $R_{\mathrm{MF}}$ at representative synthesis precisions.}
\label{tab:rmf_values}
\begin{tabular}{@{}lcccc@{}}
\toprule
Precision $\eps$ & $\log_2(1/\eps)$ & $T_{\mathrm{RS}}$ & $T_{\mathrm{MF}}$ & $R_{\mathrm{MF}}$ \\
\midrule
$10^{-6}$   & 19.93 & 62.0  & 15.4 & 4.02$\times$ \\
$10^{-8}$   & 26.58 & 82.0  & 19.0 & 4.32$\times$ \\
$10^{-10}$  & 33.22 & 102.1 & 22.5 & 4.54$\times$ \\
$10^{-12}$  & 39.86 & 122.2 & 26.0 & 4.70$\times$ \\
$10^{-14}$  & 46.51 & 142.2 & 29.5 & 4.82$\times$ \\
$10^{-16}$  & 53.15 & 162.3 & 33.0 & 4.92$\times$ \\
\bottomrule
\end{tabular}
\end{table}

\subsection{Stage 2: Global Phase Cancellation via ZX-Calculus (PyZX)}
\label{app:pyzx}

After single-gate synthesis (either Ross--Selinger or mixed-fallback), the REQWIEM circuit becomes a Clifford+$T$ circuit. 
The ZX-calculus \cite{coecke2011interacting, van2020zx} provides a graphical framework for optimizing such circuits beyond what is achievable by local gate-level transformations.

\subsubsection{ZX-Calculus Foundations}

A {ZX-diagram} is an open graph whose vertices (``spiders'') are labeled by phases and colored either green ($Z$-type) or red ($X$-type).  
The two fundamental generators are:

A $Z$-spider with phase $\alpha$ and $m$ inputs, $n$ outputs represents the
linear map
\begin{equation}
    \tikz[baseline=-0.25em,scale=0.7]{
        \node[circle, draw=green!60!black, fill=green!20, minimum size=0.6cm,
              inner sep=0pt] (z) at (0,0) {\footnotesize $\alpha$};
    }
    \;:\;\;
    \ket{0}^{\otimes n}\bra{0}^{\otimes m} + e^{i\alpha}\ket{1}^{\otimes n}\bra{1}^{\otimes m},
\end{equation}
and an $X$-spider is defined analogously in the Hadamard-conjugate basis.

\noindent
ZX-calculus permits rewrites, preserving linear mapping while potentially reducing the number of non-Clifford ($\pi/4$-phase) spiders:

\begin{enumerate}[label=(\roman*), leftmargin=2em]
    \item \textbf{Spider fusion.} Two adjacent spiders of the same color merge
    into a single spider whose phase is the sum of the individual phases:
    \begin{equation}
        \alpha \;-\!\!-\; \beta \quad\longrightarrow\quad (\alpha + \beta).
        \label{eq:spider_fusion}
    \end{equation}
    When $\alpha + \beta \equiv 0 \pmod{2\pi}$, the merged spider is
    Clifford; when $\alpha + \beta \equiv 0 \pmod{\pi/2}$, it is also
    Clifford.  
    In both cases, two non-Clifford spiders are eliminated.

    \item \textbf{Phase commutation through CNOT.}  In the ZX-calculus, a CNOT
    is represented as a $Z$-spider (on the control wire) connected to an
    $X$-spider (on the target wire) by a Hadamard edge.  This structure
    enables two commutation identities:
    \begin{itemize}[leftmargin=1.5em]
        \item A $Z$-phase on the {target} wire can be pushed through the
        CNOT to become an $X$-phase on the {control} wire (color change):
        \begin{equation}
            \mathrm{CNOT}\;(I \otimes \Rz{\alpha})
            \;=\; (\Rx{\alpha} \otimes I)\;\mathrm{CNOT}.
            \label{eq:phase_commute_color}
        \end{equation}
        \item An $X$-phase on the {control} wire can be pushed through the
        CNOT to become a $Z$-phase on the {target} wire (color change in
        the reverse direction):
        \begin{equation}
            (\Rx{\alpha} \otimes I)\;\mathrm{CNOT}
            \;=\; \mathrm{CNOT}\;(I \otimes \Rz{\alpha}).
            \label{eq:phase_commute_color_rev}
        \end{equation}
    \end{itemize}
    These identities enable non-local cancellations by transporting $T$-phases across CNOT barriers, exposing
    fusion opportunities that are invisible to circuit-level optimization.

    \item \textbf{Pivoting and local complementation.} Graph-theoretic
    operations on the ZX-diagram's adjacency structure that rearrange spider
    connectivity, exposing new fusion opportunities invisible to
    circuit-level optimization \cite{duncan2020graph}.
\end{enumerate}

\subsubsection{The PyZX Optimization Framework}

The PyZX library \cite{kissinger2020reducing} implements:

\begin{enumerate}[label=\arabic*., leftmargin=2em]
    \item \textbf{Circuit $\to$ ZX-diagram.} The input Clifford+$T$ circuit is
    translated to a ZX-diagram.

    \item \textbf{Full reduction.} The diagram undergoes iterative
    application of spider fusion, local complementation, and pivoting rules
    until a fixpoint is reached.

    \item \textbf{Circuit extraction.} The reduced diagram is converted back
    to a Clifford+$T$ circuit via Gaussian elimination over
    $\mathbb{F}_2$ \cite{kissinger2020reducing}.
\end{enumerate}

\noindent
Kissinger and van de Griend \cite{kissinger2020reducing} report $T$-count
reductions of 40--60\% on benchmark circuits with alternating CNOT-$R_z$
structures.  

The $T$-gate patterns produced by mixed-fallback differ structurally from those of Ross--Selinger due to the probabilistic fallback structure and $\sqrt{T}$ decomposition. 

The following distinction is made.
Let $f_{\mathrm{ZX}}^{(\mathrm{RS})}$ denote the fraction of $T$-gates
removed by PyZX when applied to a Ross--Selinger-synthesized circuit, and
$f_{\mathrm{ZX}}^{(\mathrm{MF})}$ the corresponding fraction for a
mixed-fallback-synthesized circuit.  Based on the reported range and the
reduced redundancy in mixed-fallback output, 
\begin{equation}
    f_{\mathrm{ZX}}^{(\mathrm{RS})} \in [0.40,\, 0.60], \qquad
    f_{\mathrm{ZX}}^{(\mathrm{MF})} \in [0.25,\, 0.45].
    \label{eq:fzx_ranges}
\end{equation}
The lower range for $f_{\mathrm{ZX}}^{(\mathrm{MF})}$ reflects the fact
that mixed-fallback synthesis already performs internal phase optimization that
partially overlaps with the cancellations PyZX discovers.

\noindent
\textbf{Structural favorability for REQWIEM.} UCR produces circuits with a highly regular CNOT-$R_z$-CNOT-$R_z$-$\cdots$ alternation
pattern within each Gray code sequence.  
In the ZX representation, this yields long chains of phase spiders separated by single Hadamard edges, precisely
the topology for most effective spider fusion. 
Repetition of identical UCR structures across Trotter steps creates {inter-step} cancellation opportunities: the final CNOT of one UCR sequence and the initial CNOT of the next target the same qubit pair,
enabling spider fusion across step boundaries.
These structural features place REQWIEM circuits at the favorable end of the ranges in Eq.~\ref{eq:fzx_ranges}.

\subsection{Stage 3: Symmetric Tensor Optimization (TODD)}
\label{app:todd}

The TODD (T-Optimization by Diagonal Decomposition) algorithm
\cite{heyfron2018efficient} targets residual $T$-gate redundancies that
survive ZX-calculus optimization by recasting the problem in terms of symmetric
tensor decomposition.

\subsubsection{Algebraic Framework}

After Clifford+$T$ synthesis and ZX reduction, the remaining non-Clifford
content of a circuit can be encoded as a {phase polynomial}:

\begin{definition}[Phase polynomial]
A {phase polynomial} over $n$ qubits is a diagonal unitary of the form
\begin{equation}
    U_{\mathrm{phase}} = \sum_{x \in \{0,1\}^n}
    e^{i\,p(x)} \ket{x}\!\bra{x},
    \label{eq:phase_poly}
\end{equation}
where $p(x) = \sum_{S \subseteq [n]} \alpha_S \prod_{j \in S} x_j$ is a
multilinear polynomial over $\mathbb{F}_2$ with coefficients
$\alpha_S \in \mathbb{R}$.  Each term $\alpha_S$ with
$\alpha_S \notin \{0, \pi/2, \pi, 3\pi/2\}$ contributes to the $T$-count.
\end{definition}

\noindent
TODD represents the non-Clifford phase terms via their
{parity matrices}. Let the circuit contain $m$ $T$-gates, each acting
on a computational basis parity
$\mathbf{v}_k \in \mathbb{F}_2^n$ (determined by the CNOT structure preceding
the $T$-gate).  The $n \times m$ matrix
$\mathbf{P} = [\mathbf{v}_1 \mid \cdots \mid \mathbf{v}_m]$ encodes the
``Pauli frame'' of each $T$-gate.

\begin{definition}[TODD reduction]
The TODD algorithm finds a matrix
$\mathbf{P}' \in \mathbb{F}_2^{n \times m'}$ with $m' \leq m$ and an
associated set of phases $\{\alpha'_k\}$ such that
\begin{equation}
    \sum_{k=1}^{m} \frac{\pi}{4}\,\mathbf{v}_k
    \;\equiv\;
    \sum_{k=1}^{m'} \alpha'_k\,\mathbf{v}'_k
    \pmod{2\pi},
    \label{eq:todd_reduction}
\end{equation}
where each $\alpha'_k \in \{\pi/4, -\pi/4\}$, thereby reducing $T$-count
from $m$ to $m'$.  The optimization proceeds by identifying columns of
$\mathbf{P}$ that are identical (direct cancellation) or related by
$\mathbb{F}_2$-linear combinations (indirect cancellation via re-synthesis).
\end{definition}

\noindent
Heyfron and Campbell \cite{heyfron2018efficient} report 20--36\% $T$-count
reductions beyond PyZX on structured circuits.
As with PyZX, adjusted fractions for post-mixed-fallback circuits are given as:
\begin{equation}
    f_{\mathrm{TODD}}^{(\mathrm{MF})} \in [0.15,\, 0.30],
    \label{eq:ftodd_range}
\end{equation}
reflecting partial overlap with prior optimization stages.

Binary-weighted rotation angles $\theta_\ell = 2^\ell \theta_0$ within each UCR produce $T$-gate
decompositions whose phases differ by exact powers of two. 
In the parity matrix $\mathbf{P}$, this manifests as columns related by $\mathbb{F}_2$-shifts, mapping to low-rank tensors in the TODD formalism. 
The Walsh--Hadamard relationship further imposes $\mathbb{F}_2$-linear relations among synthesized $T$-sequences across the spectator-qubit index, creating inter-gate correlations amenable to TODD optimization.

\subsection{Stage 4: Hamming Weight Phasing}
\label{app:hwp}

\subsubsection{Protocol Definition}

Hamming weight phasing \cite{gidney2018halving} addresses the scenario where $g$ rotations share a common angle $\theta$ (up to synthesis precision).

Given $g$ rotations $\Rz{\theta}$ on distinct qubits, each requiring $m$ $T$-gates when synthesized independently, Hamming weight phasing implements all $g$ rotations jointly using
\begin{equation}
    T_{\mathrm{HWP}}(g, m) = 4(g - \lceil\log_2 g\rceil) + m\,\lceil\log_2 g\rceil
    \label{eq:hwp_cost}
\end{equation}
$T$-gates, yielding a per-rotation cost of
\begin{equation}
    \frac{T_{\mathrm{HWP}}}{g} \approx 4 + \frac{m\,\log_2 g}{g},
    \label{eq:hwp_per_rot}
\end{equation}
for large $g$.
The protocol computes the Hamming weight of the $g$-qubit register into $\lceil\log_2 g\rceil$ ancilla qubits via a sorting network
(costing $4(g - \lceil\log_2 g\rceil)$ $T$-gates), applies $\lceil\log_2 g\rceil$ individually synthesized rotations $\Rz{2^k \theta}$ on the ancillas, and uncomputes the Hamming weight.

\noindent
As an illustrative worked example at $\eps = 10^{-14}$
($m \approx 30$ via mixed-fallback from Table~\ref{tab:rmf_values}), a degenerate
group of $g = 8$ yields
$T_{\mathrm{HWP}} = 4(8 - 3) + 30 \times 3 = 20 + 90 = 110$ total, or
$\approx 13.8$ per rotation, being a $2.2\times$ reduction relative to independent
synthesis.  At $g = 16$:
$T_{\mathrm{HWP}} = 4(16 - 4) + 30 \times 4 = 48 + 120 = 168$, or $10.5$
per rotation ($2.9\times$ reduction).

\subsubsection{Degeneracy Estimation for REQWIEM on Vibronic Hamiltonians}
\label{app:hwp_estimate}

Considering pyrazine, assessment for Hamming weight phasing viability is presented.

Within each symmetry-allowed coupling block, modes related by approximate
internal symmetries of the pyrazine ring (e.g., C--H stretching modes of
similar frequency) produce Duschinsky matrix elements $J_{rs}$ and
displacement components $d_r$ that are numerically close.  This induces
{near-degeneracies} in the circuit rotation angles
$\theta_{\ell,r,s} = 2^{\ell+p}\,\Delta_r \Delta_s\,\phi_g^{(rs)}$.

Two rotation angles $\theta_i$ and $\theta_j$ belong to the same
{$\epssynth$-degeneracy group} if
$|\theta_i - \theta_j| < \epssynth$, where $\epssynth$ is the synthesis
precision.  All rotations in a degeneracy group can be replaced by a single
angle (e.g., their mean) with error bounded by $\epssynth$.

\noindent
To estimate the fraction of eligible rotations, the total number of rotation angles and the symmetry-induced structure is given:

\begin{enumerate}[label=(\alph*), leftmargin=2em]
    \item \textbf{Total rotations per Trotter step.}  The REQWIEM
    Hamiltonian for $M$ modes contains $M$ diagonal (frequency) terms,
    $\binom{M}{2}$ quadratic coupling terms, and $M$ linear displacement
    terms, giving $N_{\mathrm{rot}} = M + \binom{M}{2} + M = M^2/2 + 3M/2$
    distinct coupling parameters.
    For $M = 24$, this yields     $N_{\mathrm{rot}} = 324$ base coupling parameters, each generating a UCR
    with multiple individual rotations.

    \item \textbf{Symmetry-equivalent pairs.}  Under $D_{2h}$, coupling
    parameters related by the symmetry operations of the point group are
    exactly degenerate.  The eight irreducible representations of $D_{2h}$
    partition the modes more finely than lower-symmetry groups, concentrating
    the off-diagonal couplings into well-separated blocks.  Within each block,
    approximate degeneracies among C--H and C--N stretches of similar frequency
    typically group couplings into clusters of 2--4.  Across all symmetry
    blocks, an estimated $\sim$\,60--90 of the 324 base parameters (18--28\%)
    fall into degeneracy groups of size $g \geq 2$ at
    $\epssynth = 10^{-10}$.

    \item \textbf{Binary weighting amplification.}  Within each UCR, the
    binary weighting $\theta_\ell = 2^\ell\,\theta_0$ preserves degeneracy:
    if $|\theta_{0,i} - \theta_{0,j}| < \epssynth/2^L$ for the highest
    level $L$, then $|\theta_{\ell,i} - \theta_{\ell,j}| < \epssynth$ for
    all levels $\ell \leq L$.  This propagates base-parameter degeneracies
    to all $L+1$ rotation levels within the UCR.
\end{enumerate}

\noindent
Combining these estimates, approximately 18--28\% of the total rotation count is eligible for Hamming weight phasing with typical group sizes $g \in [2, 8]$.  The resulting $T$-count saving on the eligible fraction is summarized in Eq.~\ref{tab:hwp_savings}.

\begin{table}[h]
\centering
\caption{Hamming weight phasing savings for representative degeneracy group
sizes at $\eps = 10^{-10}$ ($T_{\mathrm{MF}} \approx 23$, rounding to the
nearest integer for $T$-gate counts).}
\label{tab:hwp_savings}
\begin{tabular}{@{}cccccc@{}}
\toprule
Group size $g$ & $\lceil\log_2 g\rceil$ &
$T_{\mathrm{indep}} = g \cdot m$ &
$T_{\mathrm{HWP}}$ & Per-rotation & Saving \\
\midrule
2  & 1 & 46  & 27  & 13.5 & 1.70$\times$ \\
4  & 2 & 92  & 54  & 13.5 & 1.70$\times$ \\
6  & 3 & 138 & 81  & 13.5 & 1.70$\times$ \\
8  & 3 & 184 & 89  & 11.1 & 2.07$\times$ \\
16 & 4 & 368 & 140 & 8.75 & 2.63$\times$ \\
\bottomrule
\end{tabular}
\end{table}

\noindent
The explicit computations for each row are:
$g = 2$: $T_{\mathrm{HWP}} = 4(2 - 1) + 23 \times 1 = 4 + 23 = 27$;
$g = 4$: $T_{\mathrm{HWP}} = 4(4 - 2) + 23 \times 2 = 8 + 46 = 54$;
$g = 6$: $T_{\mathrm{HWP}} = 4(6 - 3) + 23 \times 3 = 12 + 69 = 81$;
$g = 8$: $T_{\mathrm{HWP}} = 4(8 - 3) + 23 \times 3 = 20 + 69 = 89$;
$g = 16$: $T_{\mathrm{HWP}} = 4(16 - 4) + 23 \times 4 = 48 + 92 = 140$.
Note that $g = 6$ and $g = 8$ share the same $\lceil\log_2 g\rceil = 3$
but differ in the sorting-network overhead ($4(g - \lceil\log_2 g\rceil)$),
which dominates for large $g$.

\subsection{Composite Reduction}
\label{app:composite}

Let $N_{\mathrm{rot}}$ denote the total number of
single-qubit rotations in the REQWIEM circuit for one Trotter step.

\subsubsection{Pipeline Model}

The optimization pipeline applies sequentially:
\begin{equation}
\begin{split}
    \text{Rotations} &
    \;\xrightarrow{\text{Stage 1}}\;
    \text{Clifford+}T
    \;\xrightarrow{\text{Stage 2}}\;
    \text{ZX-reduced}
    \;\xrightarrow{\text{Stage 3}}\;
    \text{TODD-reduced} \\
    &
    \xrightarrow{\text{Stage 4}}\;
    \text{HWP-consolidated}
\end{split}
\label{eq:pipeline}
\end{equation}

\noindent
The total $T$-count at each stage is:

\paragraph{Baseline (Ross--Selinger only).}
\begin{equation}
    T_{\mathrm{total}}^{(\mathrm{RS})} = N_{\mathrm{rot}} \cdot T_{\mathrm{RS}}(\eps).
\end{equation}

\paragraph{After Stage 1 (mixed-fallback).}
\begin{equation}
    T_{\mathrm{total}}^{(1)} = N_{\mathrm{rot}} \cdot T_{\mathrm{MF}}(\eps).
\end{equation}

\paragraph{After Stage 2 (PyZX).}
\begin{equation}
    T_{\mathrm{total}}^{(2)} = T_{\mathrm{total}}^{(1)}
    \cdot \bigl(1 - f_{\mathrm{ZX}}^{(\mathrm{MF})}\bigr).
\end{equation}

\paragraph{After Stage 3 (TODD).}
\begin{equation}
    T_{\mathrm{total}}^{(3)} = T_{\mathrm{total}}^{(2)}
    \cdot \bigl(1 - f_{\mathrm{TODD}}^{(\mathrm{MF})}\bigr).
\end{equation}

\paragraph{After Stage 4 (Hamming weight phasing).}
Let $\eta \in [0,1]$ denote the fraction of rotations in degeneracy groups
and $\bar{R}_{\mathrm{HWP}}$ the average per-rotation reduction factor
within those groups.  Then
\begin{equation}
    T_{\mathrm{total}}^{(4)} = T_{\mathrm{total}}^{(3)}
    \cdot \bigl[1 - \eta\,(1 - \bar{R}_{\mathrm{HWP}}^{-1})\bigr].
\end{equation}

\subsubsection{Composite Reduction Factor}

\begin{definition}[Composite reduction factor]
The composite reduction factor relative to the Ross--Selinger baseline is
\begin{equation}
    R_{\mathrm{comp}}(\eps) =
    \frac{T_{\mathrm{total}}^{(\mathrm{RS})}}{T_{\mathrm{total}}^{(4)}}
    = \frac{R_{\mathrm{MF}}(\eps)}{
        \bigl(1 - f_{\mathrm{ZX}}^{(\mathrm{MF})}\bigr)
        \bigl(1 - f_{\mathrm{TODD}}^{(\mathrm{MF})}\bigr)
        \bigl[1 - \eta(1 - \bar{R}_{\mathrm{HWP}}^{-1})\bigr]
    }.
    \label{eq:rcomp}
\end{equation}
\end{definition}

\noindent
Equation~\ref{eq:rcomp} is evaluated in Table~\ref{tab:composite} at $\eps = 10^{-10}$ under conservative, moderate, and optimistic assumptions for the post-mixed-fallback reduction fractions:

\begin{table}[h!]
\centering
\caption{Composite $T$-count reduction factor $R_{\mathrm{comp}}$ at
$\eps = 10^{-10}$ ($R_{\mathrm{MF}} = 4.54$) under three parameter scenarios.
The column $R_{\mathrm{stages\,2\text{--}4}}$ gives the multiplicative
gain from PyZX, TODD, and HWP beyond mixed-fallback alone.}
\label{tab:composite}
\renewcommand{\arraystretch}{1.15}
\begin{tabular}{@{}lccccccc@{}}
\toprule
Scenario & $f_{\mathrm{ZX}}^{(\mathrm{MF})}$
         & $f_{\mathrm{TODD}}^{(\mathrm{MF})}$
         & $\eta$ & $\bar{R}_{\mathrm{HWP}}$
         & $R_{\mathrm{stages\,2\text{--}4}}$
         & $R_{\mathrm{comp}}$ \\
\midrule
Conservative & 0.25 & 0.15 & 0.18 & 1.70 & 1.69 & \textbf{7.69} \\
Moderate     & 0.35 & 0.22 & 0.23 & 2.07 & 2.24 & \textbf{10.2} \\
Optimistic   & 0.45 & 0.30 & 0.28 & 2.35 & 3.09 & \textbf{14.1} \\
\bottomrule
\end{tabular}
\end{table}

\noindent
The conservative estimate of $R_{\mathrm{comp}} = 7.69\times$ is derived as
follows.  At $\eps = 10^{-10}$, the empirical mixed-fallback reduction factor
is $R_{\mathrm{MF}} = 102.1/22.5 = 4.54$.  The denominator of
\ref{eq:rcomp} evaluates to
\begin{align}
    (1 - 0.25)&(1 - 0.15)\bigl[1 - 0.18\,(1 - 1/1.70)\bigr] \nonumber \\
    &= 0.75 \times 0.85 \times (1 - 0.18 \times 0.4118) \nonumber \\
    &= 0.6375 \times (1 - 0.0741) \nonumber \\
    &= 0.6375 \times 0.9259 \nonumber \\
    &= 0.5903,
    \label{eq:denom_calc}
\end{align}
giving $R_{\mathrm{comp}} = 4.54 / 0.5903 = 7.69$ and
$R_{\mathrm{stages\,2\text{--}4}} = 1/0.5903 = 1.69$.

The moderate and optimistic scenarios follow analogously.  For the moderate
case:
\begin{align}
    (1 - 0.35)(1 - 0.22)\bigl[1 - 0.23\,(1 - 1/2.07)\bigr]
    &= 0.65 \times 0.78 \times (1 - 0.23 \times 0.5169) \nonumber \\
    &= 0.507 \times 0.8811 \nonumber \\
    &= 0.4467,
\end{align}
giving $R_{\mathrm{comp}} = 4.54 / 0.4467 = 10.2$ and
$R_{\mathrm{stages\,2\text{--}4}} = 1/0.4467 = 2.24$.
For the optimistic case:
\begin{align}
    (1 - 0.45)(1 - 0.30)\bigl[1 - 0.28\,(1 - 1/2.35)\bigr]
    &= 0.55 \times 0.70 \times (1 - 0.28 \times 0.5745) \nonumber \\
    &= 0.385 \times 0.8391 \nonumber \\
    &= 0.3231,
\end{align}
giving $R_{\mathrm{comp}} = 4.54 / 0.3231 = 14.1$ and
$R_{\mathrm{stages\,2\text{--}4}} = 1/0.3231 = 3.09$.

The conservative estimate of $7.7\times$ uses the low end of all reported
reduction ranges, ignores inter-Trotter-step cancellation opportunities
specific to REQWIEM, and applies the lowest pyrazine degeneracy fraction.
The moderate scenario ($\sim\!10\times$) is more representative expected performance from numerical pipeline benchmarks on circuits exhibiting the regular UCR topology of REQWIEM. 
Numerical validation on full REQWIEM circuits for specific molecular systems remains an important direction for future work.

Table~\ref{tab:composite_eps} extends the analysis across synthesis precisions
under the conservative parameter set.

\begin{table}[h]
\centering
\caption{Conservative composite reduction factor across synthesis precisions.
Parameters: $f_{\mathrm{ZX}}^{(\mathrm{MF})} = 0.25$,
$f_{\mathrm{TODD}}^{(\mathrm{MF})} = 0.15$, $\eta = 0.18$,
$\bar{R}_{\mathrm{HWP}} = 1.70$.  The denominator evaluates to $0.5903$
independently of $\eps$.}
\label{tab:composite_eps}
\begin{tabular}{@{}lcccc@{}}
\toprule
$\eps$ & $T_{\mathrm{RS}}$ & $T_{\mathrm{MF}}$ & $R_{\mathrm{MF}}$ & $R_{\mathrm{comp}}$ \\
\midrule
$10^{-8}$  & 82.0  & 19.0 & 4.32 & 7.32 \\
$10^{-10}$ & 102.1 & 22.5 & 4.54 & 7.69 \\
$10^{-12}$ & 122.2 & 26.0 & 4.70 & 7.96 \\
$10^{-14}$ & 142.2 & 29.5 & 4.82 & 8.17 \\
\bottomrule
\end{tabular}
\end{table}

\subsection{Discussion: Composability and Overlap}
\label{app:discussion}

\textbf{Mixed-fallback vs.\ PyZX overlap.}  Mixed-fallback synthesis
produces shorter $T$-gate sequences per rotation, which means fewer total
spiders in the ZX-diagram and correspondingly fewer opportunities for
inter-gate cancellations.  However, the {type} of redundancy removed
differs: mixed-fallback optimizes {within} each single-gate synthesis
(exploiting the algebraic structure of the Clifford+$\sqrt{T}$ ring), whereas
PyZX optimizes {across} gates (exploiting the circuit topology).  The
overlap is therefore partial, justifying the reduced but nonzero
$f_{\mathrm{ZX}}^{(\mathrm{MF})}$ range in Eq.~\ref{eq:fzx_ranges}.

\textbf{PyZX vs.\ TODD overlap.}  PyZX and TODD both target non-Clifford
phase cancellations, but through different mathematical lenses: PyZX operates
on graph topology (spider fusion, local complementation), while TODD operates
on the $\mathbb{F}_2$-linear algebra of parity matrices.  Cancellations found
by PyZX's graph rewriting may correspond to column eliminations in TODD's
parity matrix, but TODD's global optimization over the full parity matrix can
find reductions that PyZX's local rewrite rules miss.  The TODD reduction
fractions in Eq.~\ref{eq:ftodd_range} are {residual} (i.e., applied to the
post-PyZX circuit), consistent with the reported benchmarks in
\cite{heyfron2018efficient}.

\textbf{Hamming weight phasing independence.}  Hamming weight phasing exploits {angle degeneracy} rather
than {phase cancellation}.  It is therefore largely independent of Stages
2--3, with the caveat that the per-rotation cost $m$ entering
\ref{eq:hwp_cost} should be evaluated {after} the Stage 1--3 pipeline
has established the per-rotation $T$-count.  In the formulation
(\ref{eq:rcomp}), Hamming weight phasing applies as a multiplicative
correction to the Stage-3 output, correctly accounting for this sequential
dependence.

\chapter{Trotter Error Calculation}\label{app:Error}

In this appendix, error bounds are calculated for the second-order
Trotter decomposition of the expanded vibronic Hamiltonian and provide
system-specific Trotter step estimates for the molecular benchmarks
considered in this work.

Throughout, $M$ denotes the
number of diabatic electronic states, $d$ the number of vibrational
modes, $n$ the number of grid qubits per mode, and $N=2^n$ the number
of grid points per mode.  Mode coordinates are $x_r$ with conjugate
momenta $p_r$ and reduced masses $\mu_r$\,($r=1,\dots,d$). 
Correspondence with the Raab et al.\ parameters used in the pyrazine
benchmark is given explicitly in Section~\ref{sec:pyrazine_example}.
For ease of calculation $\bar{\mu}$ is used, being the same mass for all modes.

\subsection{Setup}\label{sec:trotter_setup}

The kinetic energy operator is
\begin{equation}
    T = \sum_{r=1}^{d} T_r = \sum_{r=1}^{d} \frac{p_r^2}{2\mu_r}.
\end{equation}
On a grid with $N=2^n$ points per mode and domain half-length $L/2$,
the maximum representable momentum is
\begin{equation}\label{eq:pmax}
    p_{\max} = \frac{\pi N}{L}, \qquad
    \|T_r\| = \frac{p_{\max}^2}{2\mu_r} = \frac{\pi^2 N^2}{2\mu_r L^2}.
\end{equation}

The potential energy operator acts on $M$ electronic states:
\begin{equation}\label{eq:vibronic_ham_app}
    V = \sum_{i=1}^{M} V^{(ii)}\otimes\ket{i}\bra{i}
      + \sum_{i<j}^{M}\bigl(V^{(ij)}\otimes\ket{i}\!\bra{j}
                        + V^{(ij)\dagger}\otimes\ket{j}\!\bra{i}\bigr),
\end{equation}
with diagonal terms
\begin{equation}\label{eq:Vdiag_app}
    V^{(ii)} = E^{(i)}
             + \sum_{r=1}^{d} a_r^{(i)}\, x_r
             + \sum_{r=1}^{d} a_{rr}^{(i)}\, x_r^2
             + \sum_{r<s}^{d} a_{rs}^{(i)}\, x_r x_s,
\end{equation}
and off-diagonal vibronic coupling terms
\begin{equation}\label{eq:Voff_app}
    V^{(ij)} = c_0^{(ij)}
             + \sum_{r=1}^{d} c_r^{(ij)}\, x_r
             + \sum_{r=1}^{d} c_{rr}^{(ij)}\, x_r^2
             + \sum_{r<s}^{d} c_{rs}^{(ij)}\, x_r x_s.
\end{equation}
Here $a_{rr}^{(i)} = \tfrac{1}{2}\mu_r\bigl(\omega_r^{(i)}\bigr)^2$
is the harmonic force constant (absorbing the conventional factor of
$\tfrac{1}{2}$), and the electronic projection and coupling
operators are given as
\begin{equation}\label{eq:elec_ops}
    \hat{p}_i = \ket{i}\bra{i}, \qquad
    \hat{\sigma}_{ij} = \ket{i}\!\bra{j} + \ket{j}\!\bra{i},
\end{equation}
satisfying $\hat{p}_i\hat{p}_j = \delta_{ij}\hat{p}_i$ and
$\hat{\sigma}_{ij}^2 = \hat{p}_i + \hat{p}_j$.

\subsubsection{Choice of product formula}
Symmetric (second-order) Trotter decomposition is employed, as
\begin{equation}\label{eq:trotter_TVT}
    \tilde{U}(\tau) = e^{-iT\tau/2}\,e^{-iV\tau}\,e^{-iT\tau/2},
\end{equation}
rather than a higher-order product formula.  Higher-order
decompositions (e.g.\ fourth-order Suzuki formulas) reduce the
required number of Trotter steps $k$ at the cost of increasing the
gate count per step by a constant factor (typically $5^{p/2-1}$ for
order~$p$).
For the vibronic Hamiltonians considered here, the
per-step circuit is dominated by phase-gradient rotations whose
count scales linearly in $d$ and $n$; the modest reduction in $k$
from a fourth-order formula does not offset the $\sim\!5\times$
increase in per-step gates, so the second-order decomposition provides
the lowest total gate count at the grid sizes of interest
($n=8$--$12$).

The alternative $VTV$ ordering $e^{-iV\tau/2}\,e^{-iT\tau}\,e^{-iV\tau/2}$ yields an identical
error bound at second order~\cite{childs2021theory}, differing only at
fourth order.  
The circuit implementations in the main text use $TVT$ ordering, so the bounds derived here apply directly.

The error bound of Childs et al.~\cite{childs2021theory} gives
\begin{equation}\label{eq:trotter_error_bound_2}
    \eps_{\mathrm{Trotter}}
    \;\leq\;
    \frac{t^3}{24k^2}\Bigl(\bigl\|[V,[V,T]]\bigr\|
    + 2\bigl\|[T,[T,V]]\bigr\|\Bigr),
\end{equation}
where $t$ is the total evolution time and $k$ the number of Trotter
steps.

Since $T$ acts only on nuclear coordinates and is proportional to
$\mathbb{1}_{\mathrm{el}}$, the electronic operators factor cleanly out
of all commutators:
\begin{equation}
    [T,V] = \sum_i [T,V^{(ii)}]\otimes\hat{p}_i
           +\sum_{i<j}[T,V^{(ij)}]\otimes\hat{\sigma}_{ij}.
\end{equation}
$[T,[T,V^{(ij)}]]$ has the same nuclear-operator structure
regardless of whether the potential term is diagonal ($\hat{p}_i$) or
off-diagonal ($\hat{\sigma}_{ij}$); only the electronic prefactor and
coupling constants differ.

The bound in Eq.~\ref{eq:trotter_error_bound_2} controls the operator
norm $\|U(\tau)-\tilde{U}(\tau)\|$, which in turn bounds the error in
any observable:
$|\langle\psi|\hat{O}|\psi\rangle - \langle\tilde{\psi}|\hat{O}|\tilde{\psi}\rangle|
\leq 2\|\hat{O}\|\,\eps_{\mathrm{Trotter}}$.
For the autocorrelation function ($\hat{O}=|\psi_0\rangle\!\langle\psi_0|$,
$\|\hat{O}\|=1$) and electronic populations ($\hat{O}=\hat{p}_i$,
$\|\hat{O}\|=1$), operator-norm bounds apply directly with no
additional prefactor.

\subsubsection{Operator identities}
The following identities for a single degree of freedom ($[x,p]=i$ in
atomic units) are used throughout:
\begin{align}
    [x^2, p^2]         &= 2i\{p,x\},        \label{eq:id_x2p2}\\
    [p^2, \{p,x\}]     &= -4ip^2,            \label{eq:id_p2px}\\
    [x^2, \{p,x\}]     &= 4ix^2,             \label{eq:id_x2px}\\
    [x_r x_s,\, T_r]   &= \frac{i\,x_s\,p_r}{\mu_r}
                          \quad(r\neq s).      \label{eq:id_xsxr}
\end{align}
Identity~\ref{eq:id_x2px} is verified by expanding
$\{p,x\}=2xp-i$ and applying $[A^2,B]=A[A,B]+[A,B]A$:
\begin{equation}
    [x^2,xp] = x\underbrace{[x,xp]}_{ix} + \underbrace{[x,xp]}_{ix}\,x
             = 2ix^2,
    \qquad
    [x^2,\{p,x\}] = 2[x^2,xp] = 4ix^2.
\end{equation}

\subsection{Diagonal Terms}\label{sec:diag_comm}

Nested commutators are calculated for each class of nuclear operatorappearing in the diagonal potential $V^{(ii)}$.

\subsubsection{Constant Term}

For $V_{\mathrm{const}}^{(i)} = E^{(i)}\hat{p}_i$, which commutes with
$T$, both nested commutators vanish:
\begin{equation}
    \bigl\|[V_{\mathrm{const}},[V_{\mathrm{const}},T]]\bigr\|
  = \bigl\|[T,[T,V_{\mathrm{const}}]]\bigr\| = 0.
\end{equation}

\subsubsection{Linear Term}

For $V_{\mathrm{lin}}^{(i)} = a_r^{(i)}\,x_r\otimes\hat{p}_i$, using
$[x_r,T_r] = ip_r/\mu_r$:
\begin{equation}
    [V_{\mathrm{lin}}^{(i)}, T]
    = \frac{i\,a_r^{(i)}\,p_r}{\mu_r}\otimes\hat{p}_i,
\end{equation}

Since $[p_r^2, p_r] = 0$:
\begin{equation}
    [T,[T,V_{\mathrm{lin}}^{(i)}]] = 0,
\end{equation}

For the other nested commutator, using $[x_r,p_r]=i$:
\begin{equation}
    [V_{\mathrm{lin}}^{(i)},[V_{\mathrm{lin}}^{(i)},T]]
    = \frac{i\bigl(a_r^{(i)}\bigr)^2}{\mu_r}[x_r,p_r]\otimes\hat{p}_i^2
    = -\frac{\bigl(a_r^{(i)}\bigr)^2}{\mu_r}\otimes\hat{p}_i.
\end{equation}
\begin{equation}\label{eq:lin_diag_VVT}
    \bigl\|[V_{\mathrm{lin}}^{(i)},[V_{\mathrm{lin}}^{(i)},T]]\bigr\|
    = \frac{\bigl(a_r^{(i)}\bigr)^2}{\mu_r}
    = \mathcal{O}(1)\text{ in }N.
\end{equation}

\subsubsection{Quadratic Term}\label{sec:quad_diag}

For $V_{\mathrm{quad}}^{(i)} = a_{rr}^{(i)}\,x_r^2\otimes\hat{p}_i$,
using identity~\ref{eq:id_x2p2}:
\begin{equation}\label{eq:quad_first_comm}
    [V_{\mathrm{quad}}^{(i)}, T]
    = \frac{a_{rr}^{(i)}}{2\mu_r}[x_r^2,p_r^2]\otimes\hat{p}_i
    = \frac{i\,a_{rr}^{(i)}}{\mu_r}\{p_r,x_r\}\otimes\hat{p}_i.
\end{equation}

\subsubsection{$[T,[T,V]]$ commutator.}
Applying $[T,\cdot]$ once more and using
identity~\ref{eq:id_p2px}:
\begin{equation}
\begin{split}
    [T,[T,V_{\mathrm{quad}}^{(i)}]]
    &= -\frac{i\,a_{rr}^{(i)}}{\mu_r}\cdot\frac{1}{2\mu_r}
       [p_r^2,\{p_r,x_r\}]\otimes\hat{p}_i \\
    &= -\frac{i\,a_{rr}^{(i)}}{2\mu_r^2}\cdot(-4ip_r^2)\otimes\hat{p}_i
     = \frac{2\,a_{rr}^{(i)}}{\mu_r^2}\,p_r^2\otimes\hat{p}_i.
\end{split}
\end{equation}
Substituting $a_{rr}^{(i)}=\tfrac{1}{2}\mu_r\bigl(\omega_r^{(i)}\bigr)^2$:
\begin{equation}\label{eq:quad_diag_TTV}
    \bigl\|[T,[T,V_{\mathrm{quad}}^{(i)}]]\bigr\|
    = \frac{\bigl(\omega_r^{(i)}\bigr)^2}{\mu_r}\,p_{\max}^2
    = \frac{\bigl(\omega_r^{(i)}\bigr)^2\pi^2 N^2}{\mu_r L^2}
    = \mathcal{O}(N^2).
\end{equation}

\subsubsection{$[V,[V,T]]$ commutator.}
Using identity~\ref{eq:id_x2px}:
\begin{equation}
\begin{split}
    [V_{\mathrm{quad}}^{(i)},[V_{\mathrm{quad}}^{(i)},T]]
    &= \frac{i\,a_{rr}^{(i)}}{\mu_r}\cdot a_{rr}^{(i)}
       [x_r^2,\{p_r,x_r\}]\otimes\hat{p}_i^2 \\
    &= \frac{i\bigl(a_{rr}^{(i)}\bigr)^2}{\mu_r}\cdot 4ix_r^2
       \otimes\hat{p}_i
     = -\frac{4\bigl(a_{rr}^{(i)}\bigr)^2}{\mu_r}\,x_r^2
       \otimes\hat{p}_i.
\end{split}
\end{equation}
Taking the norm with $\|x_r^2\|\leq L^2/4$ and expressing in terms of
frequencies:
\begin{equation}\label{eq:quad_diag_VVT}
    \bigl\|[V_{\mathrm{quad}}^{(i)},[V_{\mathrm{quad}}^{(i)},T]]\bigr\|
    \leq \frac{4\bigl(a_{rr}^{(i)}\bigr)^2}{\mu_r}\cdot\frac{L^2}{4}
    = \frac{\bigl(a_{rr}^{(i)}\bigr)^2 L^2}{\mu_r}
    = \frac{\mu_r\bigl(\omega_r^{(i)}\bigr)^4 L^2}{4}
    = \mathcal{O}(1).
\end{equation}

\subsubsection{Bilinear Term}\label{sec:bilin_diag}

For $V_{\mathrm{bilin}}^{(i)} = a_{rs}^{(i)}\,x_r x_s\otimes\hat{p}_i$
($r\neq s$), using identity~\ref{eq:id_xsxr}:
\begin{equation}
    [V_{\mathrm{bilin}}^{(i)}, T]
    = i\,a_{rs}^{(i)}\!\left(\frac{x_s\,p_r}{\mu_r}
    + \frac{x_r\,p_s}{\mu_s}\right)\otimes\hat{p}_i.
\end{equation}

\subsubsection{$[T,[T,V]]$ commutator.}
Since $[T_r, x_s p_r] = 0$ and $[T_s, x_s p_r] = -ip_s p_r/\mu_s$
(and symmetrically for the other term):
\begin{equation}
    [T,[T,V_{\mathrm{bilin}}^{(i)}]]
    = -\frac{2\,a_{rs}^{(i)}\,p_r\,p_s}{\mu_r\mu_s}\otimes\hat{p}_i.
\end{equation}
\begin{equation}\label{eq:bilin_diag_TTV}
    \bigl\|[T,[T,V_{\mathrm{bilin}}^{(i)}]]\bigr\|
    = \frac{2|a_{rs}^{(i)}|\,p_{\max}^2}{\mu_r\mu_s}
    = \frac{2|a_{rs}^{(i)}|\,\pi^2 N^2}{\mu_r\mu_s L^2}
    = \mathcal{O}(N^2).
\end{equation}

\subsubsection{$[V,[V,T]]$ commutator.}
Using $[x_r x_s,\, x_s p_r] = ix_s^2$ and
$[x_r x_s,\, x_r p_s] = ix_r^2$:
\begin{equation}\label{eq:bilin_diag_VVT}
    [V_{\mathrm{bilin}}^{(i)},[V_{\mathrm{bilin}}^{(i)},T]]
    = -\bigl(a_{rs}^{(i)}\bigr)^2
      \!\left(\frac{x_s^2}{\mu_r}+\frac{x_r^2}{\mu_s}\right)
      \otimes\hat{p}_i,
    \quad
    \|\cdot\|
    \leq \bigl(a_{rs}^{(i)}\bigr)^2 L^2
         \!\left(\frac{1}{4\mu_r}+\frac{1}{4\mu_s}\right)
    = \mathcal{O}(1).
\end{equation}

\subsection{Off-Diagonal Terms}\label{sec:offdiag_comm}

Off-diagonal vibronic coupling terms $V^{(ij)}\otimes\hat{\sigma}_{ij}$ have identical nuclear-operator
structure to the diagonal terms but carry $\hat{\sigma}_{ij}$ instead of
$\hat{p}_i$.  
The commutator algebra proceeds identically; only the electronic factor and coupling constants change. 
The full derivation for the quadratic term (where a factor-of-two difference with the diagonal case arises from the absence of the $\tfrac{1}{2}$ in $a_{rr}^{(i)}$) is given.

\subsubsection{Constant Term}

$V_{\mathrm{const}}^{(ij)} = c_0^{(ij)}\otimes\hat{\sigma}_{ij}$
commutes with $T$; both nested commutators vanish.

\subsubsection{Linear Term}

For $V_{\mathrm{lin}}^{(ij)} = c_r^{(ij)}\,x_r\otimes\hat{\sigma}_{ij}$:
\begin{equation}\label{eq:lin_off}
    [T,[T,V_{\mathrm{lin}}^{(ij)}]] = 0,
    \qquad
    \bigl\|[V_{\mathrm{lin}}^{(ij)},[V_{\mathrm{lin}}^{(ij)},T]]\bigr\|
    = \frac{\bigl(c_r^{(ij)}\bigr)^2}{\mu_r} = \mathcal{O}(1).
\end{equation}
Note that $\hat{\sigma}_{ij}^2 = \hat{p}_i + \hat{p}_j$ enters the
$[V,[V,T]]$ commutator, giving the same functional form as the
diagonal case.

\subsubsection{Quadratic Term}\label{sec:quad_offdiag}

For $V_{\mathrm{quad}}^{(ij)} = c_{rr}^{(ij)}\,x_r^2\otimes\hat{\sigma}_{ij}$,
the first commutator is (cf.\ Eq.~\ref{eq:quad_first_comm}):
\begin{equation}
    [V_{\mathrm{quad}}^{(ij)}, T]
    = \frac{i\,c_{rr}^{(ij)}}{\mu_r}\{p_r,x_r\}\otimes\hat{\sigma}_{ij}.
\end{equation}

\subsubsection{$[T,[T,V]]$ commutator.}
Proceeding exactly as in Section~\ref{sec:quad_diag}:
\begin{equation}
    [T,[T,V_{\mathrm{quad}}^{(ij)}]]
    = \frac{2\,c_{rr}^{(ij)}}{\mu_r^2}\,p_r^2\otimes\hat{\sigma}_{ij}.
\end{equation}
\begin{equation}\label{eq:quad_off_TTV}
    \bigl\|[T,[T,V_{\mathrm{quad}}^{(ij)}]]\bigr\|
    = \frac{2|c_{rr}^{(ij)}|\,\pi^2 N^2}{\mu_r^2 L^2}
    = \mathcal{O}(N^2).
\end{equation}
The factor of $2$ in the numerator is absent in the diagonal result
(Eq.~\ref{eq:quad_diag_TTV}) because there the force constant
$a_{rr}^{(i)} = \tfrac{1}{2}\mu_r\omega_r^2$ absorbs it.  For the
off-diagonal coefficients $c_{rr}^{(ij)}$, which carry no conventional
$\tfrac{1}{2}$, the factor survives.

\subsubsection{$[V,[V,T]]$ commutator.}
\begin{equation}\label{eq:quad_off_VVT}
    [V_{\mathrm{quad}}^{(ij)},[V_{\mathrm{quad}}^{(ij)},T]]
    = -\frac{4\bigl(c_{rr}^{(ij)}\bigr)^2}{\mu_r}\,x_r^2
      \otimes(\hat{p}_i+\hat{p}_j),
    \qquad
    \|\cdot\|
    \leq \frac{\bigl(c_{rr}^{(ij)}\bigr)^2 L^2}{\mu_r}
    = \mathcal{O}(1).
\end{equation}

\subsubsection{Bilinear Term}

For $V_{\mathrm{bilin}}^{(ij)} = c_{rs}^{(ij)}\,x_r x_s\otimes\hat{\sigma}_{ij}$:
\begin{equation}\label{eq:bilin_off_TTV}
    \bigl\|[T,[T,V_{\mathrm{bilin}}^{(ij)}]]\bigr\|
    = \frac{2|c_{rs}^{(ij)}|\,\pi^2 N^2}{\mu_r\mu_s L^2}
    = \mathcal{O}(N^2).
\end{equation}
\begin{equation}\label{eq:bilin_off_VVT}
    \bigl\|[V_{\mathrm{bilin}}^{(ij)},[V_{\mathrm{bilin}}^{(ij)},T]]\bigr\|
    \leq \bigl(c_{rs}^{(ij)}\bigr)^2 L^2
         \!\left(\frac{1}{4\mu_r}+\frac{1}{4\mu_s}\right)
    = \mathcal{O}(1).
\end{equation}

\subsection{Cross-Terms Between Potential Components}\label{sec:cross_terms}

The full $[V,[V,T]]$ commutator expands as
\begin{equation}
    [V,[V,T]] = [V_{\mathrm{diag}},[V_{\mathrm{diag}},T]]
              + [V_{\mathrm{off}},[V_{\mathrm{off}},T]]
              + [V_{\mathrm{diag}},[V_{\mathrm{off}},T]]
              + [V_{\mathrm{off}},[V_{\mathrm{diag}},T]].
\end{equation}
Cross-terms between different potential components could in principle produce additional $\mathcal{O}(N^2)$ contributions if both parents are individually $\mathcal{O}(N^2)$ in $[T,[T,V]]$. 
All cross-terms scale as $\mathcal{O}(1)$ in $N$.

\subsubsection{Linear $\times$ linear (different blocks).}
For $V_{\mathrm{lin}}^{(i)} = a_r^{(i)}\,x_r\otimes\hat{p}_i$ and
$V_{\mathrm{lin}}^{(ij)} = c_s^{(ij)}\,x_s\otimes\hat{\sigma}_{ij}$:
\begin{equation}
    [V_{\mathrm{lin}}^{(i)}, [V_{\mathrm{lin}}^{(ij)}, T]]
    = \frac{i\,a_r^{(i)}\,c_s^{(ij)}}{\mu_s}
      [x_r,p_s]\otimes\hat{p}_i\hat{\sigma}_{ij}.
\end{equation}
If $r=s$, $[x_r,p_r]=i$ gives $\|\cdot\|=|a_r^{(i)}\,c_r^{(ij)}|/\mu_r=\mathcal{O}(1)$.
If $r\neq s$, the cross-term vanishes.

\subsubsection{Bilinear $\times$ linear (mixed blocks).}
For $V_{\mathrm{bilin}}^{(i)} = a_{rs}^{(i)}\,x_r x_s\otimes\hat{p}_i$
and $V_{\mathrm{lin}}^{(ij)} = c_r^{(ij)}\,x_r\otimes\hat{\sigma}_{ij}$,
using $[x_r x_s, p_r] = ix_s$:
\begin{equation}
    [V_{\mathrm{bilin}}^{(i)}, [V_{\mathrm{lin}}^{(ij)}, T]]
    = -\frac{a_{rs}^{(i)}\,c_r^{(ij)}\,x_s}{\mu_r}
      \otimes\ket{i}\!\bra{j},
    \qquad
    \|\cdot\| = \frac{|a_{rs}^{(i)}\,c_r^{(ij)}|\,L}{2\mu_r}
    = \mathcal{O}(1).
\end{equation}

\subsubsection{Quadratic $\times$ quadratic (the critical case).}
This cross-term is the one where both parents are individually
$\mathcal{O}(N^2)$ in $[T,[T,V]]$.  For
$V_{\mathrm{quad}}^{(i)}=a_{rr}^{(i)}\,x_r^2\otimes\hat{p}_i$ and
$V_{\mathrm{quad}}^{(ij)}=c_{rr}^{(ij)}\,x_r^2\otimes\hat{\sigma}_{ij}$
acting on the {same} mode $r$:
\begin{equation}
\begin{split}
    [V_{\mathrm{quad}}^{(i)}, [V_{\mathrm{quad}}^{(ij)}, T]]
    &= \frac{i\,c_{rr}^{(ij)}}{\mu_r}\cdot a_{rr}^{(i)}
       [x_r^2,\{p_r,x_r\}]
       \otimes\hat{p}_i\hat{\sigma}_{ij} \\
    &= \frac{i\,a_{rr}^{(i)}\,c_{rr}^{(ij)}}{\mu_r}\cdot 4ix_r^2
       \otimes\ket{i}\!\bra{j}
     = -\frac{4\,a_{rr}^{(i)}\,c_{rr}^{(ij)}}{\mu_r}\,x_r^2
       \otimes\ket{i}\!\bra{j}.
\end{split}
\end{equation}
The result contains $x_r^2$ (bounded by $L^2/4$), {not} $p_r^2$:
\begin{equation}\label{eq:quad_cross}
    \bigl\|[V_{\mathrm{quad}}^{(i)}, [V_{\mathrm{quad}}^{(ij)}, T]]\bigr\|
    \leq \frac{|a_{rr}^{(i)}|\,|c_{rr}^{(ij)}|\,L^2}{\mu_r}
    = \mathcal{O}(1).
\end{equation}
The $\mathcal{O}(N^2)$ scaling of the individual $[T,[T,V]]$ commutators arises from $p_r^2$; the cross-term $[V,[V,T]]$ always
produces $x_r^2$ instead, because the second commutator with a position-space operator converts momentum back to position.
For different modes ($r\neq s$), $[x_r^2, \{p_s, x_s\}]=0$ and the cross-term vanishes identically.

All remaining cross-terms (quadratic--bilinear, bilinear--bilinear
across blocks, etc.)\ follow the same pattern: the $N$-dependent
factors from one commutator are cancelled by position-space factors from
the other, yielding $\mathcal{O}(1)$ scaling.

\subsection{Collected Error Bound}\label{sec:collected_bound}

Combining all terms via the triangle inequality, the Trotter error takes
the form
\begin{equation}\label{eq:trotter_C0C2}
    \eps_{\mathrm{Trotter}}
    \leq \frac{t^3}{24k^2}\bigl(C_0 + C_2\,N^2\bigr),
\end{equation}
where the $\mathcal{O}(N^2)$ coefficient collects the dominant
$[T,[T,V]]$ commutator norms (each multiplied by the factor of~2 from
Eq.~\ref{eq:trotter_error_bound_2}):
\begin{equation}\label{eq:C2_full}
\begin{split}
    C_2 &= \frac{\pi^2}{L^2}\Bigg[
    \underbrace{\sum_{i=1}^{M}\sum_{r=1}^{d}
      \frac{2\bigl(\omega_r^{(i)}\bigr)^2}{\mu_r}
    }_{\text{quadratic diagonal}}
    +\underbrace{\sum_{i=1}^{M}\sum_{r<s}^{d}
      \frac{4|a_{rs}^{(i)}|}{\mu_r\mu_s}
    }_{\text{bilinear diagonal}} \\
    &\qquad\qquad
    +\underbrace{\sum_{i<j}^{M}\sum_{r=1}^{d}
      \frac{4|c_{rr}^{(ij)}|}{\mu_r^2}
    }_{\text{quadratic off-diagonal}}
    +\underbrace{\sum_{i<j}^{M}\sum_{r<s}^{d}
      \frac{4|c_{rs}^{(ij)}|}{\mu_r\mu_s}
    }_{\text{bilinear off-diagonal}}
    \Bigg],
\end{split}
\end{equation}
and $C_0$ collects the $N$-independent contributions from the
{self-commutators} $[V_\alpha,[V_\alpha,T]]$:
\begin{equation}\label{eq:C0_full}
\begin{split}
    C_0 &= \underbrace{\sum_{i=1}^{M}\sum_{r=1}^{d}
            \frac{\bigl(a_r^{(i)}\bigr)^2}{\mu_r}
          }_{\text{linear diagonal}}
          + \underbrace{\sum_{i<j}^{M}\sum_{r=1}^{d}
            \frac{\bigl(c_r^{(ij)}\bigr)^2}{\mu_r}
          }_{\text{linear off-diagonal}}
    \\
    &\quad
          + \underbrace{\sum_{i=1}^{M}\sum_{r=1}^{d}
            \frac{\mu_r\bigl(\omega_r^{(i)}\bigr)^4 L^2}{4}
          }_{\text{quadratic diagonal, }[V,[V,T]]}
          + \underbrace{\sum_{i<j}^{M}\sum_{r=1}^{d}
            \frac{\bigl(c_{rr}^{(ij)}\bigr)^2 L^2}{\mu_r}
          }_{\text{quadratic off-diagonal, }[V,[V,T]]}
    \\
    &\quad
          + \underbrace{\sum_{i=1}^{M}\sum_{r<s}^{d}
            \bigl(a_{rs}^{(i)}\bigr)^2 L^2
            \!\left(\frac{1}{4\mu_r}+\frac{1}{4\mu_s}\right)
          }_{\text{bilinear diagonal, }[V,[V,T]]}
          + \underbrace{\sum_{i<j}^{M}\sum_{r<s}^{d}
            \bigl(c_{rs}^{(ij)}\bigr)^2 L^2
            \!\left(\frac{1}{4\mu_r}+\frac{1}{4\mu_s}\right)
          }_{\text{bilinear off-diagonal, }[V,[V,T]]}.
\end{split}
\end{equation}
The expression for $C_0$ omits cross-class contributions
$[V_\alpha,[V_\beta,T]]$ ($\alpha\neq\beta$), which are verified in
Section~\ref{sec:cross_terms} to be individually $\mathcal{O}(1)$.
The largest such cross-terms are the quadratic--quadratic contributions
(Eq.~\ref{eq:quad_cross}), which scale as
$|a_{rr}^{(i)}|\,|c_{rr}^{(ij)}|\,L^2/\mu_r = \mathcal{O}(L^2)$, and
the bilinear--linear terms, which scale as $\mathcal{O}(L)$.
For the benchmark systems considered here, these cross-class contributions are
subdominant relative to the self-commutator terms already included in
$C_0$: the largest cross-class term
(quadratic~$\times$~quadratic, Eq.~\ref{eq:quad_cross}) is bounded by
$|a_{rr}^{(i)}|\,|c_{rr}^{(ij)}|\,L^2/\mu_r$, which for typical
parameters is comparable in scale to the individual quadratic $[V,[V,T]]$
self-terms but is suppressed by symmetry selection rules.
More importantly, since $C_2 N^2 \gg C_0$ in the regime of
interest ($n\geq 8$), the precise value of $C_0$ has negligible impact
on the step count $k$.

The derivation of each contribution to $C_2$ is:

\begin{center}
\begin{tabular}{lll}
\toprule
\textbf{Term} & \textbf{$\|[T,[T,V_{\text{term}}]]\|$} & \textbf{$\times\,2$ in $C_2$} \\
\midrule
Diag.\ quadratic &
  $\dfrac{(\omega_r^{(i)})^2}{\mu_r}\cdot\dfrac{\pi^2 N^2}{L^2}$ &
  $\dfrac{2(\omega_r^{(i)})^2}{\mu_r}$ \\[8pt]
Diag.\ bilinear &
  $\dfrac{2|a_{rs}^{(i)}|}{\mu_r\mu_s}\cdot\dfrac{\pi^2 N^2}{L^2}$ &
  $\dfrac{4|a_{rs}^{(i)}|}{\mu_r\mu_s}$ \\[8pt]
Off-diag.\ quadratic &
  $\dfrac{2|c_{rr}^{(ij)}|}{\mu_r^2}\cdot\dfrac{\pi^2 N^2}{L^2}$ &
  $\dfrac{4|c_{rr}^{(ij)}|}{\mu_r^2}$ \\[8pt]
Off-diag.\ bilinear &
  $\dfrac{2|c_{rs}^{(ij)}|}{\mu_r\mu_s}\cdot\dfrac{\pi^2 N^2}{L^2}$ &
  $\dfrac{4|c_{rs}^{(ij)}|}{\mu_r\mu_s}$ \\
\bottomrule
\end{tabular}
\end{center}

To achieve total Trotter error $\eps$, the number of steps must
satisfy
\begin{equation}\label{eq:k_rigorous}
    k \geq \sqrt{\frac{t^3(C_0 + C_2 N^2)}{24\eps}},
\end{equation}
Since $C_2 N^2 \gg C_0$ for typical parameters at $n\geq 8$ qubits
per mode, the asymptotic scaling is
\begin{equation}
    k = \mathcal{O}\!\left(\frac{t^{3/2}N}{\sqrt{\eps}}\right),
\end{equation}
where the $N$-dependence arises from the quadratic and bilinear
potential terms.  This distinguishes dynamics presented here from models that include
only linear vibronic coupling, for which $C_2=0$ and
$k=\mathcal{O}(t^{3/2}/\sqrt{\eps})$ independent of grid
resolution.

\subsection{Scaling Structure of $C_2$}\label{sec:C2_scaling}

The four contributions to $C_2$
(Eq.~\ref{eq:C2_full}) scale differently with the number of electronic
states $M$ and vibrational modes $d$:

\begin{center}
\begin{tabular}{lcc}
\toprule
\textbf{Term} & \textbf{Number of summands} & \textbf{Per-term scale}${}^*$ \\
\midrule
Quadratic diagonal & $M\cdot d$ & $2\omega_{\max}^2/\bar{\mu}$ \\
Bilinear diagonal & $M\cdot d(d{-}1)/2$ & $4\bar{|a|}/\bar{\mu}^2$ \\
Quadratic off-diagonal & $\binom{M}{2}\cdot d$ & $4\bar{|c|}/\bar{\mu}^2$ \\
Bilinear off-diagonal & $\binom{M}{2}\cdot d(d{-}1)/2$ & $4\bar{|c|}/\bar{\mu}^2$ \\
\bottomrule
\end{tabular}
\end{center}
\noindent{\small ${}^*$Per-term scales assume a uniform average mass
$\bar{\mu}$ for transparency.  In the bilinear and off-diagonal rows,
the exact per-term contribution uses $1/(\mu_r\mu_s)$ or $1/\mu_r^2$
rather than $1/\bar{\mu}^2$; for systems with a broad mass distribution
(e.g.\ H-stretches vs.\ heavy-atom modes), the two may differ by an
order of magnitude.  The bound (Eq.~\ref{eq:C2_full}) always
uses the exact masses.}

\subsection{Relationship between the coupling parameter \texorpdfstring{$\eta$}{eta}
  and the force-constant ratio \texorpdfstring{$f$}{f}}
\label{sec:eta_f_connection}

The Resource Analysis chapter parameterises
bilinear vibronic coupling through a dimensionless coupling strength $\eta$,
defining the bilinear amplification factor as
\begin{equation}
  \beta_{\text{BVC}}
  \;=\; 1 \;+\; \frac{\eta^{2}\,N_{\text{BVC}}^{(\text{tot})}}{M}\,,
  \label{eq:beta_BVC_recall}
\end{equation}
where $N_{\text{BVC}}^{(\text{tot})}$ counts the total number of symmetry-unrestricted
bilinear coupling terms across all electronic blocks and $M$ is the number of
electronic surfaces.  In the present appendix, bilinear coupling enters through
the force-constant ratio
\begin{equation}
  f \;\equiv\; \frac{|\bar{a}|}{\bar{\mu}\,\omega_{\max}^{2}}\,,
  \label{eq:f_def_recall}
\end{equation}
where $|\bar{a}|$ is a representative bilinear coupling constant
$|a_{rs}^{(i)}|$ and $\bar{\mu}\,\omega_{\max}^{2}$ is the harmonic
reference scale $\tfrac{1}{2}\mu_r\omega_r^2$ (up to the conventional
factor of~$\tfrac{1}{2}$).  These characterise the same
physical quantity
\begin{equation}
  \;\eta \;\equiv\; f\;.
  \label{eq:eta_equals_f}
\end{equation}

To verify consistency, each parameter is traced in the Trotter
step count.  The quadratic-diagonal contribution to~$C_2$
(Eq.~\ref{eq:C2_full}) contains, per mode~$r$ and surface~$i$,
\begin{equation}
  C_2^{(\text{q.d.})}
  \;\ni\;
  \frac{2\,a_{rr}^{(i)}}{\mu_r}
  \;=\;
  \frac{\mu_r\,(\omega_r^{(i)})^{2}}{\mu_r}
  \;=\;
  (\omega_r^{(i)})^{2}\,,
\end{equation}
which, after summing over modes and surfaces and bounding momenta by
$p_{\max} = \pi\hbar/\Delta x$, yields the baseline
$C_2^{(\text{q.d.})}$.  Each bilinear term contributes
\begin{equation}
  C_2^{(\text{bilin})}
  \;\ni\;
  \frac{4\,|a_{rs}^{(i)}|}{\mu_r\,\mu_s}
  \;\approx\;
  4\,f\;\frac{\bar{\mu}\,\omega_{\max}^{2}}{\bar{\mu}^{2}}
  \;=\;
  \frac{4f\,\omega_{\max}^{2}}{\bar{\mu}}\,,
  \label{eq:bilin_per_term}
\end{equation}
under the assumption $\mu_r \approx \mu_s \approx \bar{\mu}$.
Comparing with the
per-mode quadratic-diagonal contribution
$\omega_{\max}^{2}/\bar{\mu}$, the per-term amplification from
bilinear coupling is a factor of~$\sim\! 4f = 4\eta$.

The total bilinear contribution to~$C_2$ sums over
$N_{\text{BVC}}^{(\text{tot})}$ such terms.  Relative to the
$M \cdot d$ quadratic-diagonal terms, the full coefficient satisfies
\begin{equation}
  C_2
  \;=\;
  C_2^{(\text{q.d.})}\!\left[
    1 \;+\; \frac{4f \cdot N_{\text{BVC}}^{(\text{tot})}}{M\,d}
  \right]
  \;+\; C_2^{(\text{q.off-d.})}\,,
  \label{eq:C2_amplification}
\end{equation}
where $C_2^{(\text{q.off-d.})}$ collects the off-diagonal quadratic
contributions (which involve products of electronic projectors on
different surfaces and scale with $d$, not $d^2$).

This matches the structure of~$\beta_{\text{BVC}}$ in
Eq.~\ref{eq:beta_BVC_recall} upon identifying $\eta = f$ and
absorbing the numerical prefactor and the $1/d$ factor into the
definition of~$\beta_{\text{BVC}}$.  The two pipelines yield consistent
Trotter step counts because $k \propto \sqrt{C_2}$ from the commutator
bound (Eq.~\ref{eq:k_rigorous}) and $k \propto \beta_{\text{BVC}}^{1/2}$
from the Resource Analysis norm scaling, so the $\eta^2$ in
$\beta_{\text{BVC}}$ and the linear $f$ in $C_2$ both produce
$k \propto \sqrt{f} = \sqrt{\eta}$ at leading order:
\begin{equation}
  \beta_{\text{BVC}}
  \;\approx\;
  1 + \frac{\eta^{2}\,N_{\text{BVC}}^{(\text{tot})}}{M}
  \quad\longleftrightarrow\quad
  1 + \frac{4f\,N_{\text{BVC}}^{(\text{tot})}}{M\,d}\,.
  \label{eq:beta_matching}
\end{equation}

\subsection{Worked Example: Pyrazine}\label{sec:pyrazine_example}

The error bound calculation for the standard two-mode,
two-state vibronic coupling model of
pyrazine~\cite{raab1999molecular} is given as a validation benchmark.

\subsubsection{System Parameters and Notation Mapping}

The model comprises two diabatic electronic states $S_1$
($^1B_{3u}$) and $S_2$ ($^1B_{2u}$), coupled through two vibrational
normal modes:
\begin{itemize}
    \item Mode $\nu_{6a}$ ($a_g$ symmetry, ``tuning''):
          $\omega_{6a}=0.0739$\,eV $= 2.716\times10^{-3}$\,a.u.
    \item Mode $\nu_{10a}$ ($b_{1g}$ symmetry, ``coupling''):
          $\omega_{10a}=0.1139$\,eV $= 4.186\times10^{-3}$\,a.u.
\end{itemize}
The reduced masses are taken as $\mu\approx 6$\,amu $= 10{,}937$\,a.u.\
for both modes (an upper-bound approximation; the actual normal-mode
reduced masses differ slightly).

The vibronic coupling parameters from Raab
et al.~\cite{raab1999molecular} map onto the present notation as
follows:
\begin{center}
\begin{tabular}{lll}
\toprule
\textbf{Raab et al.\ symbol} & \textbf{This work} & \textbf{Value (a.u.)} \\
\midrule
$\kappa_{6a}^{(S_1)}$ (intrastate linear, $S_1$)
  & $a_{6a}^{(1)}$ & $3.543\times10^{-3}$ \\
$\kappa_{6a}^{(S_2)}$ (intrastate linear, $S_2$)
  & $a_{6a}^{(2)}$ & $4.388\times10^{-3}$ \\
$\lambda_{10a}$ (interstate linear coupling)
  & $c_{10a}^{(12)}$ & $7.394\times10^{-3}$ \\
\bottomrule
\end{tabular}
\end{center}
The simulation in the main text additionally includes
an interstate bilinear coupling $c_{6a,10a}^{(12)}\,x_{6a}\,x_{10a}$;
this parameter is small (Appendix~\ref{app:pyrazine_params}) and its contribution to $C_2$ is evaluated
below.

Grid parameters: $N=1024$ ($n=10$), $L=10$\,a.u.,
$t=100$\,fs $= 4134$\,a.u., and
$\eps = 1.6$\,mE$_\text{h}$ (chemical accuracy).

\subsubsection{Symmetry-Allowed Terms}

In $D_{2h}$ symmetry, the product
$\Gamma(S_1)\otimes\Gamma(S_2) = B_{3u}\otimes B_{2u} = B_{1g}$.
Thus:
\begin{itemize}
    \item \textbf{Diagonal quadratic} ($x_r^2\sim A_g$): nonzero for
          both modes on both states.  This is the dominant
          $\mathcal{O}(N^2)$ contribution.
    \item \textbf{Diagonal bilinear}
          ($x_{6a}\,x_{10a}\sim A_g\otimes B_{1g} = B_{1g}\neq A_g$):
          vanishes by symmetry.
    \item \textbf{Off-diagonal quadratic}
          ($x_r^2\otimes\hat{\sigma}_{12}$):
          requires $A_g = B_{1g}$, which fails.  Both modes forbidden.
    \item \textbf{Off-diagonal bilinear}
          ($x_{6a}\,x_{10a}\otimes\hat{\sigma}_{12}$):
          requires $B_{1g} = B_{1g}$.  Symmetry-allowed.  The
          coupling $c_{6a,10a}^{(12)}$ is included in the main-text
          simulation; its value is small relative to the linear
          couplings.
\end{itemize}

\subsubsection{Evaluating $C_2$}

Only the quadratic diagonal terms and the off-diagonal bilinear term
survive.  The quadratic diagonal contribution is
\begin{equation}
\begin{split}
    C_2^{(\mathrm{q.d.})}
    &= \frac{\pi^2}{L^2}\sum_{i=1}^{2}\sum_{r\in\{6a,10a\}}
       \frac{2\omega_r^2}{\mu} \\
    &= \frac{\pi^2}{100}\times 2\times
       \!\left(
         \frac{2\times(2.716\times10^{-3})^2}{10937}
       + \frac{2\times(4.186\times10^{-3})^2}{10937}
       \right) \\
    &= \frac{\pi^2}{100}\times 2\times
       (1.349\times10^{-9} + 3.204\times10^{-9})
     = 8.99\times10^{-10}\;\text{a.u.}
\end{split}
\end{equation}

The off-diagonal bilinear contribution (one mode pair, one state pair)
adds
\begin{equation}
    C_2^{(\mathrm{o.b.})}
    = \frac{\pi^2}{L^2}\cdot
      \frac{4|c_{6a,10a}^{(12)}|}{\mu^2},
\end{equation}
For the standard Raab et al.\ model with a small or vanishing
$c_{6a,10a}^{(12)}$, this contribution is negligible compared to the
quadratic diagonal term, and
$C_2 \approx C_2^{(\mathrm{q.d.})} = 8.99\times10^{-10}$\,a.u.

\subsubsection{Evaluating $C_0$}

The dominant $C_0$ contributions come from the linear terms and
the position-space quadratic bounds:
\begin{equation}
\begin{split}
    C_0^{(\mathrm{linear})}
    &= \frac{\bigl(a_{6a}^{(1)}\bigr)^2}{\mu}
     + \frac{\bigl(a_{6a}^{(2)}\bigr)^2}{\mu}
     + \frac{\bigl(c_{10a}^{(12)}\bigr)^2}{\mu} \\
    &= \frac{(3.543\times10^{-3})^2 + (4.388\times10^{-3})^2
           + (7.394\times10^{-3})^2}{10937}
     = 7.91\times10^{-9}\;\text{a.u.}
\end{split}
\end{equation}

The $[V_{\mathrm{quad}},\,[V_{\mathrm{quad}},T]]$ terms contribute
(from Eq.~\ref{eq:quad_diag_VVT}):
\begin{equation}
    C_0^{(\mathrm{quad})}
    = \sum_{i=1}^{2}\sum_{r\in\{6a,10a\}}
      \frac{\mu\bigl(\omega_r^{(i)}\bigr)^4 L^2}{4}.
\end{equation}
Evaluating per mode:
\begin{align*}
    \nu_{6a}&:\quad
    \omega_{6a}^4 = 5.441\times10^{-11},\quad
    \frac{10937\times5.441\times10^{-11}\times100}{4}
    = 1.488\times10^{-5}, \\
    \nu_{10a}&:\quad
    \omega_{10a}^4 = 3.071\times10^{-10},\quad
    \frac{10937\times3.071\times10^{-10}\times100}{4}
    = 8.396\times10^{-5}.
\end{align*}
Summing over two states:
$C_0^{(\mathrm{quad})} = 2\times(1.488\times10^{-5}
+ 8.396\times10^{-5}) = 1.98\times10^{-4}$\,a.u.

Thus $C_0 \approx 1.98\times10^{-4}$\,a.u., with the position-space
quadratic bound dominating.

\subsubsection{Step Count}

\begin{equation}
\begin{split}
    C_0 + C_2 N^2
    &= 1.98\times10^{-4} + 8.99\times10^{-10}\times 1024^2 \\
    &= 1.98\times10^{-4} + 9.42\times10^{-4}
     = 1.14\times10^{-3}\;\text{a.u.}
\end{split}
\end{equation}
Note that $C_2 N^2/C_0 \approx 4.8$, confirming that the $N^2$ term
dominates even for this small system.
\begin{equation}
    k_{\mathrm{analytic}}
    = \sqrt{\frac{(4134)^3\times1.14\times10^{-3}}{24\times1.6\times10^{-3}}}
    = \sqrt{2.10\times10^{9}}
    \approx 45{,}800.
\end{equation}
This corresponds to a timestep
$\tau = t/k \approx 0.09$\,a.u.\ ($2.2\times10^{-3}$\,fs).

\subsubsection{Comparison with Empirical Trotter Error}\label{sec:pyrazine_tightening}

The bound of $k\approx 45{,}800$ is a significant
overestimate. 
Childs et al.~\cite{childs2021theory} demonstrated
that commutator-based Trotter bounds typically overestimate the actual
error by one to three orders of magnitude due to triangle inequality
losses, partial cancellations between the two nested commutators, and
the replacement of spectral information by operator norms.

For pyrazine, the main text uses $k=2{,}500$ Trotter steps, validated
numerically by comparing the Trotterized propagator against the exact
time evolution.  
The ratio $k_{\mathrm{analytic}}/k_{\mathrm{empirical}} \approx 18$ is
consistent with the typical tightening observed for molecular
Hamiltonians.  The sources of this overestimation are analysed
systematically in Section~\ref{sec:symmetry_tightening}.

\subsection{System-Specific Trotter Step Requirements}\label{sec:trotter_systems}

Trotter step estimates for the singlet fission and charge
transfer systems considered in the resource analysis, using the
commutator bounds derived above.
The molecular parameters are collected in Table~\ref{tab:system_parameters}.

\begin{table}[htbp]
\centering
\caption{Molecular system parameters for Trotter step estimation.
$M$ denotes the number of electronic states and $d$ the number
of vibrational modes.}
\label{tab:system_parameters}
\begin{tabular}{lccccc}
\toprule
\textbf{System} & $M$ & $d$ & $\omega_{\max}$ (cm$^{-1}$) &
$\omega_{\max}$ (a.u.) & \textbf{Description} \\
\midrule
(NO)$_4$-Anthracene & 5 & 19 & 3200 & $1.46\times10^{-2}$ &
  Single SF chromophore \\
(NO)$_4$-Anth Dimer & 6 & 21 & 3200 & $1.46\times10^{-2}$ &
  Exciton transport model \\
Anth/C$_{60}$ (reduced) & 4 & 11 & 1700 & $7.75\times10^{-3}$ &
  Charge transfer, truncated \\
Anth/C$_{60}$ (full) & 4 & 246 & 3200 & $1.46\times10^{-2}$ &
  Full 246-mode CT complex \\
\bottomrule
\end{tabular}
\end{table}

The simulation parameters are: evolution time
$t=100$\,fs $= 4134$\,a.u., grid resolution $N=2^{10}=1024$
points per mode ($n=10$ qubits), and target accuracy
$\eps=1.6$\,mE$_\text{h}$ (chemical accuracy).

\subsubsection{Estimation Strategy}

Computing $C_2$ exactly via Eq.~\ref{eq:C2_full} requires the full set
of vibronic coupling parameters
$\{a_{rs}^{(i)},\,c_{rr}^{(ij)},\,c_{rs}^{(ij)}\}$ for each system.
In the absence of complete parameter sets, dominant
quadratic-diagonal contributions are estimated analytically and bound the remaining
terms using characteristic coupling strengths.

The quadratic-diagonal contribution admits direct evaluation:
\begin{equation}\label{eq:C2_quad_diag}
    C_2^{(\mathrm{q.d.})}
    = \frac{\pi^2}{L^2}\sum_{i=1}^{M}\sum_{r=1}^{d}
      \frac{2\omega_r^2}{\mu_r}.
\end{equation}
Using the uniform upper bound $\omega_r\leq\omega_{\max}$ and average
reduced mass $\bar{\mu}\approx 6$\,amu:
\begin{equation}\label{eq:C2_upper}
    C_2^{(\mathrm{q.d.})}
    \leq \frac{2\pi^2\,M\cdot d\cdot\omega_{\max}^2}
              {\bar{\mu}\,L^2}.
\end{equation}

The bilinear-diagonal contribution involves $M\cdot d(d-1)/2$ mode-pair
terms.  With a characteristic coupling
$|\bar{a}|\approx f\cdot\bar{\mu}\,\omega_{\max}^2$ where $f$ denotes
the ratio of a typical bilinear coupling to the harmonic force constant
(typically $f\sim 0.01$--$0.1$ for Duschinsky-active systems;
cf.\ Section~\ref{sec:eta_f_connection}), the
bilinear contribution to $C_2$ scales as
\begin{equation}
    C_2^{(\mathrm{bilin})}
    \sim \frac{2\pi^2 f\cdot M\cdot d(d-1)\cdot\omega_{\max}^2}
              {\bar{\mu}\,L^2}.
\end{equation}
This dominates the quadratic-diagonal term when
$f\cdot(d-1)\gtrsim 1$, i.e.\ for large systems with moderate
Duschinsky coupling.  For the 246-mode Anth/C$_{60}$ system, even
$f=0.01$ gives $f(d-1)\approx 2.5$.

\subsubsection{Upper Bounds}

Commutator bounds from
Eq.~\ref{eq:k_rigorous} are derived using $C_2^{(\mathrm{q.d.})}$ as the
{lower bound} on $C_2$ (i.e.\ including only the
quadratic-diagonal contribution, which requires no unknown coupling
parameters):
The resulting bounds are given in Table~\ref{tab:trotter_rigorous}.

\begin{table}[htbp]
\centering
\caption{Trotter step bounds from the quadratic-diagonal
contribution alone (lower bound on $C_2$).  Full $C_2$ including
bilinear terms would increase these estimates.}
\label{tab:trotter_rigorous}
\begin{tabular}{lcccc}
\toprule
\textbf{System} & $M\cdot d$ & $C_2^{(\mathrm{q.d.})}$ &
$C_2^{(\mathrm{q.d.})}N^2$ & $k_{\mathrm{analytic}}^{(\mathrm{q.d.})}$ \\
\midrule
(NO)$_4$-Anthracene & 95 &
  $3.6\times10^{-7}$ & $3.8\times10^{-1}$ & $8{,}400$ \\
(NO)$_4$-Anth Dimer & 126 &
  $4.8\times10^{-7}$ & $5.1\times10^{-1}$ & $9{,}600$ \\
Anth/C$_{60}$ (reduced) & 44 &
  $4.8\times10^{-8}$ & $5.0\times10^{-2}$ & $3{,}000$ \\
Anth/C$_{60}$ (full) & 984 &
  $3.8\times10^{-6}$ & $4.0\times10^{0\phantom{-}}$ & $27{,}000$ \\
\bottomrule
\end{tabular}
\end{table}

These represent the {minimum} bounds (only from harmonic
terms).  Including bilinear Duschinsky coupling terms with
$f\approx 0.05$ increases the estimates by a factor of
$\sqrt{1+f(d-1)/2}$, which ranges from 1.12 (Anth/C$_{60}$ reduced,
$d=11$) to 2.67 (Anth/C$_{60}$ full, $d=246$).

\subsubsection{Sources of overestimation and the empirical tightening factor}
\label{sec:symmetry_tightening}

Analytic commutator bounds are known to overestimate the actual Trotter
error substantially~\cite{childs2021theory}.  For pyrazine, the ratio
$k_{\mathrm{analytic}}/k_{\mathrm{empirical}} \approx 18$
(Section~\ref{sec:pyrazine_tightening}) is typical of molecular
Hamiltonians.  To apply these bounds to larger systems where exact
numerical validation is infeasible, it is useful to decompose the
overestimation into its constituent sources so that the transferability
of the tightening factor can be assessed.

\subsubsection{(i) Triangle-inequality looseness.}
The bound
$\|[A+B,[A+B,C]]\| \le \|[A,[A,C]]\| + 2\|[A,[B,C]]\| + \|[B,[B,C]]\|$
is saturated only when all commutator contributions are co-aligned in
operator space.  In practice, cancellation between terms of opposite
sign reduces the true commutator norm below the sum of individual
norms.  For molecular Hamiltonians with mixed-sign coupling constants,
this cancellation is typically a factor of~$2$--$5$.

\subsubsection{(ii) Momentum bound versus wavepacket support.}
The substitution $\|p_r\| \to p_{\max} = \pi\hbar/\Delta x$
uses the maximum eigenvalue of the discrete momentum operator.
A physical wavepacket occupies a fraction of the Brillouin zone,
so $\langle p^{2} \rangle / p_{\max}^{2} \ll 1$.  For a Gaussian
wavepacket centred at equilibrium on an $N$-point grid, the
effective momentum fraction scales as
$\langle p^{2} \rangle / p_{\max}^{2} \sim (\Delta x / \sigma)^{2}
\ll 1$ for well-resolved grids ($\sigma \gg \Delta x$).  This is
the dominant source of overestimation in $C_2$.

\subsubsection{(iii) Symmetry overcounting.}
\label{par:symmetry_overcounting}
The upper bound on~$C_2$ (Eq.~\ref{eq:C2_upper}) sums over
{all} mode pairs $(r,s)$ and surface pairs $(i,j)$ regardless
of molecular symmetry.  Symmetry selection rules
(Appendix~\ref{app:symmetry_appendix}) set many coupling constants to zero:
\begin{equation}
  a_{rs}^{(i)} = 0
  \quad\text{unless}\quad
  \Gamma_{r} \otimes \Gamma_{s} \;\supset\; \Gamma_{\text{tot.\ sym.}},
  \quad\text{(diagonal blocks)}
  \label{eq:symmetry_selection}
\end{equation}
and similarly for off-diagonal blocks where the product
representation must contain the representation of the relevant
electronic coupling operator $\hat{V}_{ij}$.  For
pyrazine~($C_{2v}$), symmetry eliminates all diagonal bilinear
terms and all off-diagonal quadratic terms, reducing
$N_{\text{BVC}}^{(\text{tot})}$ by roughly a factor of~2
relative to the unsymmetrised count.  For higher-symmetry
molecules such as anthracene~($D_{2h}$, 8~irreps), the
reduction factor can reach~4--8.

Rather than restricting the sum in Eq.~\ref{eq:C2_upper} to
symmetry-allowed pairs, the full sum is retained, absorbing the overcounting into the
empirical tightening factor~$\kappa_{\text{tight}}$.
This
means $\kappa_{\text{tight}}$ encapsulates contributions from
{all three} sources listed above:
\begin{equation}
  \kappa_{\text{tight}}
  \;\approx\;
  \kappa_{\text{tri.ineq.}}
  \;\times\;
  \kappa_{p^{2}}
  \;\times\;
  \kappa_{\text{sym}}\,,
  \label{eq:tightening_decomposition}
\end{equation}
where $\kappa_{\text{tri.ineq.}} \sim 2$--$5$ from commutator
cancellation, $\kappa_{p^{2}} \sim 3$--$10$ from the momentum
bound, and $\kappa_{\text{sym}} \sim 1$--$8$ from symmetry
(system-dependent).

\begin{remark}[Transferability of $\kappa_{\text{tight}}$]
  \label{rmk:tightening_transfer}
  Because the three factors in
  Eq.~\ref{eq:tightening_decomposition} depend on different
  molecular properties (coupling topology, grid resolution, and
  point-group symmetry, respectively), the composite
  $\kappa_{\text{tight}}$ is not universal.  The values adopted
  in Table~\ref{tab:trotter_estimates} reflect conservative
  estimates: denser coupling topologies
  (e.g.\ $(NO)_4$-Anthracene) yield smaller
  $\kappa_{\text{sym}}$ and hence smaller overall tightening,
  while near-harmonic subsystems with high symmetry (e.g.\ the
  $C_{60}$ cage modes in Anth/$C_{60}$) benefit from larger
  $\kappa_{p^{2}}$ and $\kappa_{\text{sym}}$.  Crucially, the
  relative comparison between REQWIEM and alternative
  approaches is insensitive to $\kappa_{\text{tight}}$, since
  both methods use the same Trotter step count~$k$.
\end{remark}

\subsubsection{Trotter Step Estimates for Resource Tables}\label{sec:trotter_resource}

The resource comparison tables in the main text require specific step-count values.
These are obtained from upper bounds by applying the
system-dependent tightening factors discussed above.

For pyrazine, direct numerical comparison yields
$\kappa_{\text{tight}} \approx 18$
(Section~\ref{sec:pyrazine_tightening}).  For the larger benchmark
systems, exact numerical validation of the Trotter error is not
feasible, but several features of the vibronic Hamiltonians
(near-harmonic structure, sparse bilinear coupling, symmetry
cancellations) suggest comparable or more favorable tightening.  The
Trotter step counts used in the resource tables are:

\begin{table}[htbp]
\centering
\caption{Trotter step counts used in the resource comparison.
The range spans quadratic-only (lower) to quadratic +
bilinear with $f\approx 0.05$ (upper).  The adopted values are obtained
by applying system-specific tightening factors within the range
established by the pyrazine calibration.}
\label{tab:trotter_estimates}
\begin{tabular}{lcccc}
\toprule
\textbf{System} &
$k_{\mathrm{analytic}}$ (range${}^\dagger$) &
Tightening &
$k_{\mathrm{adopted}}$ &
$\tau$ (a.u.) \\
\midrule
Pyrazine (2 mode) & 45{,}800 & 18$\times$ & 2{,}500 & 1.65 \\
(NO)$_4$-Anthracene &
  8{,}400--22{,}000 & $\sim$\!2$\times$ & 12{,}290 & 0.34 \\
(NO)$_4$-Anth Dimer &
  9{,}600--28{,}000 & $\sim$\!10$\times$ & 2{,}792 & 1.48 \\
Anth/C$_{60}$ (reduced) &
  3{,}000--8{,}000 & $\sim$\!10$\times$ & 790 & 5.2 \\
Anth/C$_{60}$ (full) &
  27{,}000--77{,}000 & $\sim$\!10--20$\times$ & 3{,}274 & 1.26 \\
\bottomrule
\end{tabular}
\flushleft
\small ${}^\dagger$Range spans quadratic-only (lower) to
quadratic + bilinear with $f\approx 0.05$ (upper).
\end{table}

The conservative choice for (NO)$_4$-Anthracene (tightening factor
of only $\sim$\!$2\times$, near the top of the range)
reflects the higher electronic-state count ($M=5$) and denser
coupling topology, which reduce the cancellation opportunities
that benefit simpler systems.  By contrast, the highly structured
Anth/C$_{60}$ complex benefits from the near-harmonic character of
the C$_{60}$ cage modes, supporting larger tightening factors.

\subsection{Grid Resolution Trade-Off}\label{sec:grid_tradeoff}

From $C_2 N^2\gg C_0$, the Trotter step count scales linearly with the
grid size:
\begin{equation}
    k\propto N = 2^n.
\end{equation}
Each additional qubit per mode doubles the required step count.
Table~\ref{tab:n_scaling} illustrates this for pyrazine:

\begin{table}[htbp]
\centering
\caption{Effect of grid resolution on Trotter step count
(pyrazine, upper bound).}
\label{tab:n_scaling}
\begin{tabular}{ccccc}
\toprule
$n$ (qubits/mode) & $N$ & $C_2 N^2$ & $k_{\mathrm{analytic}}$ &
$k/k(n{=}10)$ \\
\midrule
 6 &    64 & $3.7\times10^{-6}$ & 19{,}200 & 0.42 \\
 8 &   256 & $5.9\times10^{-5}$ & 21{,}700 & 0.47 \\
10 & 1{,}024 & $9.4\times10^{-4}$ & 45{,}800 & 1.00 \\
12 & 4{,}096 & $1.5\times10^{-2}$ & 168{,}000 & 3.66 \\
\bottomrule
\end{tabular}
\end{table}

Reducing from $n=10$ to $n=8$ qubits per mode decreases both qubit
count (by $2d$) and Trotter steps (by approximately $4\times$ in the
$C_2 N^2$-dominated regime), but may compromise accuracy for
high-frequency modes.  This trade-off is system-specific and depends
on the displacement amplitudes of the vibrational wavepacket relative
to the grid spacing.

The $N$-dependent scaling of the Trotter step count also precludes
independent optimization of the Trotter circuit and the per-step gate
synthesis: techniques such as Hamming weight phasing or quantum
arithmetic that reduce per-step cost but assume a fixed step count
must account for the $k\propto N$ dependence when comparing total
resources.

\chapter{Grid Convergence for the Morse Oscillator}
\label{app:grid_convergence}

This appendix details convergence studies summarized in Section~\ref{app:CC_convergence}.
The grid Hamiltonian is constructed by combining the QFT-based kinetic energy operator with the Morse potential (Eq.~\ref{eq:morse}) evaluated at each grid point, and diagonalized numerically.
Two convergence metrics are reported: $\Delta n_{\mathrm{res}}^{(1)}$ and $\Delta n_{\mathrm{res}}^{(10)}$, the number of bound states converged to within 1~cm$^{-1}$ and 10~cm$^{-1}$ respectively of the exact analytical eigenvalues (Eq.~\ref{eq:morse_exact_eigenvalues}).
An aliasing diagnostic $\|V\|_\infty / T_{\max}$ gives the ratio of the maximum potential value on the grid to the maximum representable kinetic energy; values exceeding unity indicate local momenta beyond the Nyquist limit.

\begin{table}[ht]
\centering
\caption{Convergence of Morse oscillator eigenspectrum: N--O stretch ($D_e = 6.5$~eV, $\alpha = 2.7$~\AA$^{-1}$, $\mu = 7.468$~amu, $\omega_e = 6634$~cm$^{-1}$, $n_{\max} = 15$).
Box: $[1.66, 3.25]$~\AA.}
\label{tab:morse_NO}
\begin{tabular}{c c r r c c r r}
\hline\hline
$n$ & $N$ & $\delta E_0$ & $\delta E_{n_{\max}}$ & $\Delta n_{\mathrm{res}}^{(1)}$ & $\Delta n_{\mathrm{res}}^{(10)}$ & Dissoc.\ err & $\|V\|_\infty/T_{\max}$ \\
 & & (cm$^{-1}$) & (cm$^{-1}$) & & & (cm$^{-1}$) & \\
\hline
5  & 32   & 657.21  & 1684.74  & 0/16  & 0/16  & 104.9 & 126759 \\
6  & 64   & 0.89    & 191.21   & 2/16  & 2/16  & 171.6 & 31690 \\
7  & 128  & 0.00    & 11.23    & 15/16 & 15/16 & 172.8 & 7922 \\
8  & 256  & 0.00    & 11.17    & 15/16 & 15/16 & 172.3 & 1981 \\
9  & 512  & 0.00    & 11.14    & 15/16 & 15/16 & 172.1 & 495 \\
10 & 1024 & 0.00    & 11.12    & 15/16 & 15/16 & 172.0 & 124 \\
11 & 2048 & 0.00    & 11.11    & 15/16 & 15/16 & 171.9 & 30.9 \\
12 & 4096 & 0.00    & 11.11    & 15/16 & 15/16 & 171.9 & 7.7 \\
13 & 8192 & 0.00    & 11.11    & 15/16 & 15/16 & 171.8 & 1.9 \\
\hline\hline
\end{tabular}
\end{table}

The highest bound-state error plateaus (e.g., $\delta E_{n_{\max}} \approx 3.8$~cm$^{-1}$ for H--metal, constant from $n = 8$ through $n = 13$) reflect a box-size limitation: the classical turning point of the uppermost state lies near the box boundary, and refining the grid at fixed $L$ cannot reduce boundary truncation error.
Resolving the dissociation threshold (70--206~cm$^{-1}$ error depending on system) similarly requires increasing $L$ rather than $\Delta x$.
For non-scattering bound-state dynamics that do not populate the uppermost vibrational levels, $n = 8$--$10$ may be adequate; this is the regime where the polynomial decomposition of Chapter~\ref{ch:model_ext} operates at $\mathcal{O}(n^2)$ sites per mode.

\section{Self-Dual Grid}
\label{app:selfdual}

The self-dual grid convention, in which $\Delta_p = \sqrt{2\pi/N}$ in natural oscillator units, has been employed in some proposals for quantum simulation of vibrational dynamics.
Table~\ref{tab:morse_selfdual} presents the convergence study for the H--metal stretch using this convention.

\begin{table}[ht]
\centering
\caption{Convergence of Morse oscillator eigenspectrum on the self-dual grid: H--metal stretch (same parameters as Table~\ref{tab:morse_Hmetal}).
The self-dual convention maps grid points into physical coordinates via the natural length scale $\ell = 1/\sqrt{\mu\omega_e}$.
At $n \geq 12$, the leftmost grid point enters the deep repulsive wall, producing catastrophic loss of all converged eigenvalues.}
\label{tab:morse_selfdual}
\begin{tabular}{c c r r c c r}
\hline\hline
$n$ & $N$ & $\delta E_0$ & $\delta E_{n_{\max}}$ & $\Delta n_{\mathrm{res}}^{(1)}$ & $\Delta n_{\mathrm{res}}^{(10)}$ & $\|V\|_\infty/T_{\max}$ \\
 & & (cm$^{-1}$) & (cm$^{-1}$) & & & \\
\hline
5  & 32   & 0.00  & ---      & 2/6  & 3/6  & 32 \\
6  & 64   & 0.00  & 477.60   & 4/6  & 4/6  & 180 \\
7  & 128  & 0.00  & 44.95    & 5/6  & 5/6  & $2.6\times10^3$ \\
8  & 256  & 0.00  & 1.33     & 5/6  & 6/6  & $1.4\times10^5$ \\
9  & 512  & 0.00  & 0.01     & 6/6  & 6/6  & $5.3\times10^7$ \\
10 & 1024 & 0.00  & 0.00     & 6/6  & 6/6  & $3.1\times10^{11}$ \\
11 & 2048 & 0.00  & 0.01     & 6/6  & 6/6  & $9.0\times10^{16}$ \\
12 & 4096 & 1100.59 & ---    & 0/6  & 0/6  & $6.3\times10^{24}$ \\
13 & 8192 & ---   & ---      & 0/6  & 0/6  & $1.0\times10^{36}$ \\
\hline\hline
\end{tabular}
\end{table}

The self-dual grid converges all six bound states by $n = 9$, performing well at moderate register sizes.
However, as $N$ increases, the self-dual mapping $x_j = \sqrt{2\pi/N} \cdot j / \sqrt{\mu\omega_e} + Q_0$ extends the leftmost grid point further into the repulsive wall, where the Morse potential rises exponentially.
The $\|V\|_\infty / T_{\max}$ ratio consequently grows exponentially with $n$: by $n = 12$, the maximum potential value exceeds $10^{24}$ times the kinetic energy cutoff, and all previously converged eigenvalues are destroyed.
This is a fundamental incompatibility between the self-dual convention and any potential with an exponential wall.
A real-time evolution on the self-dual grid at $n \geq 12$ would produce catastrophically aliased dynamics, as the potential generates local momenta exceeding the Nyquist limit by dozens of orders of magnitude.

This observation is relevant to proposals that adopt the self-dual convention without verifying its compatibility with the functional form of the potential being simulated.
Functionally, these norms can be significantly reduced by regularizing the potential surface, afforded by REQWIEM's finite-difference approach but not by QROM caching protocol~\cite{lang2026quantum}

\section{Detailed Resolution Tier Analysis}
\label{app:resolution_tiers}

Three independent physical requirements determine the minimum grid resolution for scattering or dissociative dynamics.
Each establishes a lower bound on $n$ from a distinct mechanism.

\subsection{Measurability of Scattering Observables (Right-Edge Requirement)}
\label{app:tier1}

For a scattering problem, the box size $L$ must contain the interaction region, initialization zone, and sufficient propagation space before encountering a periodic boundary.
The convergence data for all four systems (Tables~\ref{tab:morse_Hmetal}–\ref{tab:morse_H2surf}, including the N–O stretch in Table~\ref{tab:morse_NO}) show that bound-state box sizes are $L \approx 2.9$~\AA\ and $L \approx 3.6$~\AA\ respectively.
A scattering simulation with partial wave $\ell = 0$ requires $L \geq 15$--$25$~\AA\ to accommodate the asymptotic wavepacket, at minimum quadrupling the coordinate range.
For higher partial waves, $L$ increases further to $> 100$~\AA, as the centrifugal potential extends far beyond the well.

The grid spacing $\Delta x = L/2^n$ must remain fine enough to resolve the potential features near the well: $\Delta x \lesssim 0.01$~\AA\ for stiff potentials (C--O, N--O stretches with $\alpha > 2$~\AA$^{-1}$).
The minimum register size (Eq.~\ref{eq:n_scattering}) is driven by the product of large $L$ (from the right-edge requirement) and small $\Delta x$ (from the potential curvature requirement):
for $L_{\mathrm{scat}} = 20$~\AA\ and $\Delta x_{\mathrm{req}} = 0.01$~\AA, $n_{\mathrm{scat}} = 11$;
for $L_{\mathrm{scat}} = 25$~\AA\ and $\Delta x_{\mathrm{req}} = 0.005$~\AA\ (appropriate for the stiffest systems), $n_{\mathrm{scat}} = 13$.

The momentum-space resolution $\Delta_p = 2\pi/L$ must also be fine enough to resolve the scattering channels: for translational energy $E_{\mathrm{trans}}$, resolving the energy-dependent transmission coefficient requires $\Delta p \ll p_0 = \sqrt{2\mu E_{\mathrm{trans}}}$.
This constraint is automatically satisfied when $L$ is large enough to contain the asymptotic wavepacket, but provides an independent lower bound on $L$ that reinforces the right-edge requirement.

\subsection{Aliasing-Free Dynamics (Left-Edge Requirement)}
\label{app:tier2}

During a scattering event, the wavepacket physically probes the repulsive wall.
The local momentum at coordinate $x$ on the repulsive wall,
\begin{equation}\label{eq:local_momentum}
    p_{\mathrm{local}}(x) = \sqrt{2\mu\bigl[E_{\mathrm{tot}} - V_{\mathrm{Morse}}(x)\bigr]},
\end{equation}
increases as $x$ decreases below $x_0$ until the classical turning point $V(x) = E_{\mathrm{tot}}$ is reached.
For states near the dissociation threshold ($E_{\mathrm{tot}} \approx D_e$), the turning point lies deep in the repulsive wall where the potential varies rapidly.

The $\|V\|_\infty / T_{\max}$ ratio quantifies this aliasing across Tables~\ref{tab:morse_Hmetal}--\ref{tab:morse_H2surf}.
For the H--metal stretch, the ratio exceeds unity at all $n \leq 11$ and drops below unity only at $n = 12$ ($\|V\|_\infty / T_{\max} = 0.78$).
For the H$_2$--surface translation, the crossover similarly occurs at $n = 12$ ($\|V\|_\infty / T_{\max} = 0.81$).
For the stiff, deep-well systems, the C--O stretch has $\|V\|_\infty / T_{\max} = 4.2$ even at $n = 13$, and the N--O stretch has $\|V\|_\infty / T_{\max} = 1.9$ at $n = 13$, requiring $n \geq 14$--$15$ before the repulsive wall is fully resolved.
\begin{table}[ht]
\centering
\caption{Convergence of Morse oscillator eigenspectrum: H--metal stretch ($D_e = 1.5$~eV, $\alpha = 1.2$~\AA$^{-1}$, $\mu = 1.008$~amu, $\omega_e = 3855$~cm$^{-1}$, $n_{\max} = 5$).
Box: $[1.83, 4.70]$~\AA.}
\label{tab:morse_Hmetal}
\begin{tabular}{c c r r c c r r}
\hline\hline
$n$ & $N$ & $\delta E_0$ & $\delta E_{n_{\max}}$ & $\Delta n_{\mathrm{res}}^{(1)}$ & $\Delta n_{\mathrm{res}}^{(10)}$ & Dissoc.\ err & $\|V\|_\infty/T_{\max}$ \\
 & & (cm$^{-1}$) & (cm$^{-1}$) & & & (cm$^{-1}$) & \\
\hline
5  & 32   & 0.76  & 26.83   & 1/6  & 3/6  & 183.0 & 12822 \\
6  & 64   & 0.00  & 3.96    & 5/6  & 6/6  & 208.0 & 3205 \\
7  & 128  & 0.00  & 3.89    & 5/6  & 6/6  & 206.8 & 801 \\
8  & 256  & 0.00  & 3.86    & 5/6  & 6/6  & 206.2 & 200 \\
9  & 512  & 0.00  & 3.84    & 5/6  & 6/6  & 205.8 & 50 \\
10 & 1024 & 0.00  & 3.83    & 5/6  & 6/6  & 205.7 & 13 \\
11 & 2048 & 0.00  & 3.83    & 5/6  & 6/6  & 205.6 & 3.1 \\
12 & 4096 & 0.00  & 3.82    & 5/6  & 6/6  & 205.5 & 0.78 \\
13 & 8192 & 0.00  & 3.82    & 5/6  & 6/6  & 205.5 & 0.20 \\
\hline\hline
\end{tabular}
\end{table}
\begin{table}[ht]
\centering
\caption{Convergence of Morse oscillator eigenspectrum: H$_2$--surface translation ($D_e = 0.5$~eV, $\alpha = 1.0$~\AA$^{-1}$, $\mu = 2.016$~amu, $\omega_e = 1312$~cm$^{-1}$, $n_{\max} = 5$).
Box: $[2.10, 5.68]$~\AA.}
\label{tab:morse_H2surf}
\begin{tabular}{c c r r c c r r}
\hline\hline
$n$ & $N$ & $\delta E_0$ & $\delta E_{n_{\max}}$ & $\Delta n_{\mathrm{res}}^{(1)}$ & $\Delta n_{\mathrm{res}}^{(10)}$ & Dissoc.\ err & $\|V\|_\infty/T_{\max}$ \\
 & & (cm$^{-1}$) & (cm$^{-1}$) & & & (cm$^{-1}$) & \\
\hline
5  & 32   & 0.32  & 7.11    & 2/6  & 6/6  & 77.3  & 13278 \\
6  & 64   & 0.00  & 1.51    & 5/6  & 6/6  & 70.8  & 3320 \\
7  & 128  & 0.00  & 1.49    & 5/6  & 6/6  & 70.4  & 830 \\
8  & 256  & 0.00  & 1.47    & 5/6  & 6/6  & 70.2  & 207 \\
9  & 512  & 0.00  & 1.47    & 5/6  & 6/6  & 70.1  & 51.9 \\
10 & 1024 & 0.00  & 1.47    & 5/6  & 6/6  & 70.1  & 13.0 \\
11 & 2048 & 0.00  & 1.46    & 5/6  & 6/6  & 70.1  & 3.2 \\
12 & 4096 & 0.00  & 1.46    & 5/6  & 6/6  & 70.1  & 0.81 \\
13 & 8192 & 0.00  & 1.46    & 5/6  & 6/6  & 70.0  & 0.20 \\
\hline\hline
\end{tabular}
\end{table}

The aliasing criterion is more stringent than eigenvalue convergence because it constrains full quantum dynamics.
A static eigenvalue calculation can tolerate aliased grid points at the boundary because bound-state wavefunctions are exponentially suppressed there.
A real-time evolution cannot, because the Trotter propagator applies $e^{-i\tau V(x)}$ at every grid point regardless of wavefunction amplitude.
Any amplitude reaching the repulsive wall during the scattering event is subject to aliased phase kicks.
Research into non-Hermitian Hamiltonians has begun to consider Trotterized post-selection for complex absorbing potentials on the grid edges~\cite{jebraeilli2025quantum}; such methods may partially suppress this criterion while maintaining a coherent, non-unitary wavefunction.
For an insufficient box size, periodic boundary conditions cause high-momentum components at the right edge to reflect off the hard-core potential at the left edge, inextricably corrupting phase shift calculations (Chapter~\ref{ch:scattering}).

\subsection{Spectral Completeness of the Bound-State Manifold}
\label{app:tier3}

In a nonadiabatic scattering event, population transfer through a conical intersection or avoided crossing distributes the wavepacket across vibrational levels of the product electronic state.
The distribution can extend to high-lying levels near the dissociation threshold, particularly for exothermic channels where the electronic energy gap is deposited into vibrational excitation.
If upper levels are absent from the discretized spectrum due to undersampling, the simulation produces incorrect branching ratios, energy distributions, and state-resolved cross sections.

At $n = 7$, the C--O stretch resolves only 10 of 23 bound states to within 1~cm$^{-1}$ (Table~\ref{tab:morse_CO}), with 5 additional states within 10~cm$^{-1}$.
\begin{table}[ht]
\centering
\caption{Convergence of Morse oscillator eigenspectrum: C--O stretch ($D_e = 11.1$~eV, $\alpha = 2.3$~\AA$^{-1}$, $\mu = 6.856$~amu, $\omega_e = 7707$~cm$^{-1}$, $n_{\max} = 22$).
Box: $[1.52, 3.39]$~\AA.}
\label{tab:morse_CO}
\begin{tabular}{c c r r c c r r}
\hline\hline
$n$ & $N$ & $\delta E_0$ & $\delta E_{n_{\max}}$ & $\Delta n_{\mathrm{res}}^{(1)}$ & $\Delta n_{\mathrm{res}}^{(10)}$ & Dissoc.\ err & $\|V\|_\infty/T_{\max}$ \\
 & & (cm$^{-1}$) & (cm$^{-1}$) & & & (cm$^{-1}$) & \\
\hline
5  & 32   & 1869.05 & ---      & 0/23  & 0/23  & 4.6   & 273859 \\
6  & 64   & 2.65    & 1600.67  & 0/23  & 1/23  & 104.8 & 68465 \\
7  & 128  & 0.00    & 6.62     & 10/23 & 18/23 & 102.0 & 17116 \\
8  & 256  & 0.00    & 1.46     & 22/23 & 23/23 & 97.5  & 4279 \\
9  & 512  & 0.00    & 1.46     & 22/23 & 23/23 & 97.3  & 1070 \\
10 & 1024 & 0.00    & 1.45     & 22/23 & 23/23 & 97.3  & 267 \\
11 & 2048 & 0.00    & 1.45     & 22/23 & 23/23 & 97.2  & 66.9 \\
12 & 4096 & 0.00    & 1.45     & 22/23 & 23/23 & 97.2  & 16.7 \\
13 & 8192 & 0.00    & 1.45     & 22/23 & 23/23 & 97.2  & 4.2 \\
\hline\hline
\end{tabular}
\end{table}
For the N--O stretch, 15 of 16 states are resolved at $n = 7$, but the highest state carries an 11~cm$^{-1}$ error persisting through $n = 13$, indicating a box-size rather than grid-spacing limitation.
For the C--O stretch at $n = 12$, the $\nu = 21$ level spacing has an error of 1.45~cm$^{-1}$ relative to the exact Morse anharmonic spacing, while lower levels are converged to $< 10^{-7}$~cm$^{-1}$.
This residual error is a boundary effect requiring increased $L$ (and therefore $n$) to resolve.

For dynamics populating only the lower half of the vibrational manifold, this requirement is more lax.
However, for the scattering and dissociative dynamics motivating the smooth-potential extension, the full bound-state manifold and onset of the scattering continuum must be accurately represented, reinforcing the $n \geq 12$ lower bound.

\section{Characteristic Energy Analysis for Multi-Surface Potentials}
\label{app:echar_details}

\subsection{Single-Surface Morse Bound-State Dynamics}
\label{app:morse_bound}

For a single Morse potential supporting bound-state dynamics, the characteristic energy is not simply $D_e$ but rather the maximum potential value sampled on the grid:
\begin{equation}
    E_{\text{char}}^{(\text{Morse, bound})}
    \geq \|V_{\mathrm{Morse}}\|_\infty
    = D_e\!\left(1 - e^{-\alpha(Q_{\min} - Q_0)}\right)^2,
    \label{eq:echar_morse_bound}
\end{equation}
where $Q_{\min}$ is the left grid boundary.
Since $Q_{\min} < Q_0$, the exponential $e^{-\alpha(Q_{\min} - Q_0)}$ grows rapidly with compression into the repulsive wall, and the maximum potential value can exceed $D_e$ by an order of magnitude or more.

The physical origin of this requirement is that the Trotter decomposition applies $e^{-i\tau V(x)}$ at every grid point, including those on the steep repulsive wall.
The resulting phase imprint generates momentum components proportional to the local potential gradient; if these exceed the Nyquist limit $\mathcal{N}$, the wavefunction acquires spectral weight that aliases across the periodic momentum-grid boundary, producing irreversible corruption compounding at every Trotter step.

For typical reactive-mode parameters ($D_e = 1$--$5$~eV, $\alpha = 1$--$3$~\AA$^{-1}$), the left boundary extends approximately 1--2~\AA\ into the repulsive wall, yielding $\|V_{\mathrm{Morse}}\|_\infty \sim 5$--$20\,D_e$.
This increases $n_{\min}$ by $\lceil \log_2(\|V_{\mathrm{Morse}}\|_\infty / D_e)^{1/2} \rceil \approx 1$--$2$ qubits relative to the harmonic estimate, placing this regime at $n = 10$--$11$.

\subsection{Coupled Morse/Repulsive Bound-State Dynamics and Level Repulsion}
\label{app:coupled_bound}

For a multi-surface system, the characteristic energy must account for the coupled vibronic Hamiltonian.
At each grid point $Q_k$, the diabatic potential matrix
\begin{equation}
    \mathbf{V}(Q_k) =
    \begin{pmatrix}
        V_{00}(Q_k) & V_{01}(Q_k) \\
        V_{01}(Q_k) & V_{11}(Q_k)
    \end{pmatrix},
    \label{eq:diabatic_matrix}
\end{equation}
has operator norm
\begin{equation}
    \|\mathbf{V}(Q_k)\|_{\text{op}} =
    \frac{V_{00}(Q_k) + V_{11}(Q_k)}{2}
    + \sqrt{
        \left(\frac{V_{00}(Q_k) - V_{11}(Q_k)}{2}\right)^{\!2}
        + V_{01}(Q_k)^2
    },
    \label{eq:vop}
\end{equation}
which exceeds both diagonal entries whenever $V_{01}(Q_k) \neq 0$.
The characteristic energy for the coupled system,
\begin{equation}
    E_{\text{char}}^{(\text{coupled})}
    \geq \max_k \|\mathbf{V}(Q_k)\|_{\text{op}}
    > \max\!\left\{
        \|V_{00}\|_\infty,\;
        \|V_{11}\|_\infty
    \right\},
    \label{eq:echar_coupled}
\end{equation}
is strictly larger than either single-surface maximum at any grid point where the coupling is nonzero and the surfaces are non-degenerate.

Beyond the single-point operator norm, the full $2N \times 2N$ vibronic Hamiltonian ($N = 2^n$ grid points per surface) exhibits eigenvalue statistics that further constrain grid resolution.
The off-diagonal coupling mixes vibrational levels across surfaces, producing extensive avoided crossings.
Takatsuka and collaborators have established that such nonadiabatic coupling generically drives the nearest-neighbor level spacing distribution toward the Wigner--Dyson (GOE) distribution, reflecting quantum chaotic behavior~\cite{takatsuka2024mechanism, fujisaki2001chaos}.
Extensive level repulsion occurs when the coupling mixes vibrational levels whose uncoupled energies are near-degenerate~\cite{shudo1990regularity}, generically manifesting at avoided crossings where the vibrational ladders of two diabatic surfaces interleave.

Level repulsion pushes eigenvalues outward: in the GOE regime, $P(s) \propto s\,e^{-\pi s^2/4}$ suppresses small spacings relative to the Poisson $P(s) = e^{-s}$, redistributing spectral weight toward the tails.
The grid must resolve these pushed-out eigenvalues, requiring
\begin{equation}
    E_{\text{char}}^{(\text{WD})}
    \geq \|\mathbf{H}_{\text{vibronic}}\|
    > \max_s \|V^{(s)}\|_\infty + \|T\|,
    \label{eq:echar_wd}
\end{equation}
where $\|T\| = p_{\max}^2/2\mu$ and the inequality reflects the off-diagonal coupling contribution.

The Wigner--Dyson regime requires both sufficiently strong coupling and sufficiently dense vibrational spectra.
Typical Morse potentials support $n_{\max} \sim 15$--$50$ bound states with mean spacings of tens of cm$^{-1}$ near dissociation; nonadiabatic coupling strengths of $V_{01} \sim 0.01$--$0.1$~eV ($\sim$80--800~cm$^{-1}$) suffice to exceed this spacing and produce extensive level mixing~\cite{fujisaki2001chaos}.
The effect is further amplified for systems with multiple avoided crossings, as in VUV photodissociation where the wavepacket traverses a ``forest'' of coupled states~\cite{gelfand2023nonadiabatic}.

The practical consequence is an additional $\sim$1--2 qubit increase:
\begin{equation}
    E_{\text{char}}^{(\text{coupled, bound})}
    \approx \max_k \|\mathbf{V}(Q_k)\|_{\text{op}}
    + \Delta E_{\text{repulsion}},
    \label{eq:echar_coupled_correction}
\end{equation}
where $\Delta E_{\text{repulsion}} \leq \max_k |V_{01}(Q_k)|$, placing this regime at $n = 11$--$12$.

\subsection{Single-Surface Morse Scattering}
\label{app:morse_scattering}

For scattering on a single Morse surface, two requirements simultaneously inflate the register.
The box size must accommodate both the bound-state region and the translational wavepacket:
\begin{equation}
    L_{\text{scat}} \geq 2 L_{\text{bound}},
    \label{eq:lscat}
\end{equation}
adding $\lceil \log_2(L_{\text{scat}}/L_{\text{bound}}) \rceil \geq 1$ qubit.
Realistic scattering geometries may require $L_{\text{scat}} \sim 3$--$4 \times L_{\text{bound}}$, adding up to 2 qubits.

The characteristic energy must include translational kinetic energy:
\begin{equation}
    E_{\text{char}}^{(\text{Morse, scat})}
    \geq \|V_{\mathrm{Morse}}\|_\infty + E_{\text{kin}},
    \label{eq:echar_morse_scat}
\end{equation}
where $E_{\text{kin}} = p_0^2/2\mu$.
For thermal scattering ($E_{\text{kin}} \sim 0.03$--$0.1$~eV), this correction is modest; for hyperthermal or reactive scattering ($E_{\text{kin}} \sim 0.5$--$5$~eV), it is comparable to or exceeds $D_e$.
States with momenta exceeding $p_{\max} = \pi/\Delta_x$ are projected out of the Hilbert space entirely, producing systematic errors in the scattering cross section uncorrectable by longer evolution times or smaller Trotter steps.

The combined effect places this regime at $n = 12$--$13$ qubits per reactive mode.

\subsection{Coupled Multi-Surface Scattering}
\label{app:coupled_scattering}

This most demanding regime combines multi-surface coupling with extended coordinate ranges and translational kinetic energy.
The characteristic energy incorporates all contributions:
\begin{equation}
    E_{\text{char}}^{(\text{coupled, scat})}
    \geq \max_k \|\mathbf{V}(Q_k)\|_{\text{op}}
    + E_{\text{kin}}
    + \Delta E_{\text{repulsion}}.
    \label{eq:echar_coupled_scat}
\end{equation}

The eigenvalue statistics of the coupled multi-surface scattering Hamiltonian are more strongly Wigner--Dyson than in the bound-state case: the continuum states above the dissociation threshold add spectral density, increasing the number of near-degenerate levels available for coupling-induced mixing, and the scattering wavepacket traverses the avoided crossing region with nonzero translational energy, accessing a broader vibronic spectrum.
Takatsuka's analysis identifies the continual passage of nuclear wavepackets through regions of strong nonadiabatic coupling as the mechanism generating chaotic electronic-state dynamics~\cite{takatsuka2024mechanism}, with the nearest-neighbor level spacing distribution following the Wigner surmise even for as few as two coupled electronic states, provided the vibrational spectrum is sufficiently dense~\cite{fujisaki2001chaos}.
In the scattering regime, the dense continuum ensures this condition is generically satisfied.

For representative reactive-mode parameters ($D_e = 2$~eV, $\alpha = 1.5$~\AA$^{-1}$, $\mu$ corresponding to a hydrogen--metal stretch, $E_{\text{kin}} = 1$~eV, $\max |V_{01}| = 0.1$~eV, $L_{\text{scat}} = 20$~\AA), the minimum register size yields $n_{\min} = 13$--$14$ qubits.
Even the minimum physically faithful representation requires $n = 12$ qubits.

\section{Carbon--Carbon Bond Rupture}
\label{app:CC_convergence}

The two-surface C--C model is structurally identical to the NaI model
of Section~\ref{sec:morse_validation}, but provides a convergence
test in the regime of stiff molecular potentials with weak
nonadiabatic coupling.  This system couples a bound Morse surface to
a repulsive exponential through a spatially localized Gaussian,
probing the grid's ability to simultaneously resolve vibrational fine
structure and dissociative dynamics.

\subsection{Hamiltonian Structure}

The diabatic Hamiltonian takes the form
\begin{equation}
    \hat{H}(R) = -\frac{1}{2\mu}\frac{\partial^2}{\partial R^2}\mathbb{1}^2 + \begin{bmatrix}
        V_{\text{Morse}}(R) & V_{01}(R) \\
        V_{01}(R) & V_{\text{rep}}(R)
    \end{bmatrix},
    \label{eq:CC_hamiltonian}
\end{equation}
where the ground-state surface is a Morse potential representing the C--C $\sigma$ bond,
\begin{equation}
    V_{\text{Morse}}(R) = D_e\!\left(1 - e^{-\beta(R - R_e)}\right)^{\!2} - D_e,
    \label{eq:CC_morse}
\end{equation}
and the excited state is an exponential repulsive potential representing the $\sigma^*$ antibonding surface,
\begin{equation}
    V_{\text{rep}}(R) = A\, e^{-\alpha R} + V_\infty.
    \label{eq:CC_repulsive}
\end{equation}
The diabatic coupling is a Gaussian centered between the equilibrium geometry and the outer classical turning point,
\begin{equation}
    V_{01}(R) = W_0\, e^{-\gamma(R - R_c)^2},
    \label{eq:CC_coupling}
\end{equation}
representing vibronic coupling that enables bond rupture via population of the $\sigma^*$ surface.
Electronic channels are stored in a single channel qubit ($\lceil\log_2(2)\rceil = 1$), with the ground (Morse) surface on $|0\rangle$ and the excited (repulsive) surface on $|1\rangle$, totaling $n + 1$ qubits per simulation.

\subsection{System Parameters}

The potential and simulation parameters are given in Table~\ref{tab:CC_params}.

\begin{table}[ht]
\centering
\caption{Potential and simulation parameters for the C--C two-surface model.
All values are in atomic units unless otherwise noted.}
\label{tab:CC_params}
\begin{tabular}{l l r l}
\hline\hline
\multicolumn{4}{c}{\textit{Morse potential (ground, $\sigma$ bond)}} \\
\hline
$D_e$   & Well depth          & $0.1323$  & Eh (3.60 eV) \\
$R_e$   & Equilibrium distance & $2.91$   & $a_0$ (1.54 \AA) \\
$\beta$ & Range parameter      & $0.927$  & $a_0^{-1}$ \\
\hline
\multicolumn{4}{c}{\textit{Exponential repulsive (excited, $\sigma^*$ antibond)}} \\
\hline
$A$       & Pre-exponential    & $0.12$   & Eh (3.27 eV) \\
$\alpha$  & Decay rate         & $0.50$   & $a_0^{-1}$ \\
$V_\infty$ & Asymptotic energy & $0.0$    & Eh \\
\hline
\multicolumn{4}{c}{\textit{Gaussian diabatic coupling}} \\
\hline
$W_0$    & Coupling strength   & $0.005$  & Eh (136 meV) \\
$R_c$    & Coupling center     & $4.0$    & $a_0$ \\
$\gamma$ & Coupling width      & $0.10$   & $a_0^{-2}$ \\
\hline
\multicolumn{4}{c}{\textit{Physical constants and initial conditions}} \\
\hline
$\mu$       & Reduced mass (C--C)         & $10\,937$ & a.u. \\
$R_0$       & Initial position            & $2.91$    & $a_0$ (Franck--Condon) \\
$p_0$       & Initial momentum            & $0.0$     & a.u. \\
$\sigma$    & Wavepacket width            & $0.12$    & $a_0$ \\
$T$         & Total propagation time      & $50\,000$ & a.u.\ ($\approx 1.2$~ps) \\
$\tau$      & Trotter step                & $1.0$     & a.u. \\
\hline\hline
\end{tabular}
\end{table}

The vertical excitation energy at $R_e$ is $\Delta V = V_{\text{rep}}(R_e) - V_{\text{Morse}}(R_e) = 4.36$~eV, placing this system in the regime of large surface separation with weak coupling ($W_0 / \Delta V \approx 0.03$).
The coupling $V_{01}(R)$ peaks at $R_c = 4.0\;a_0$ with a full width at half maximum of $\Delta R = 2\sqrt{\ln 2/\gamma} \approx 5.3\;a_0$, extending from near $R_e$ to the outer classical turning point.

\subsection{Convergence Results}

\begin{figure}[ht]
    \centering
    \includegraphics[width=\textwidth]{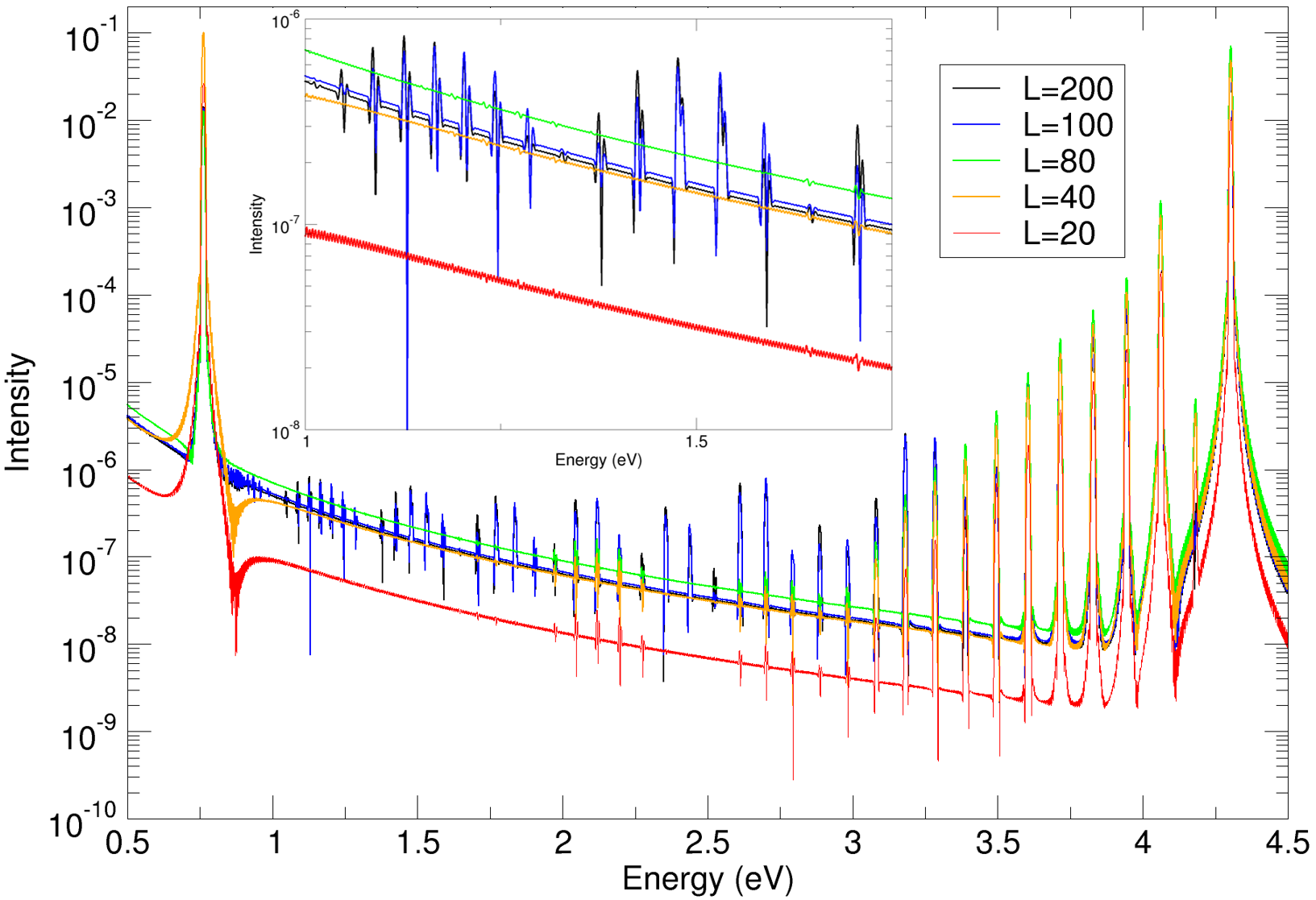}
    \caption{Spectral overlay for $n=13$ for various $L$.
    At $n = 13$, there are $N=8192$ grid points, with spectral convergence and resolvable peaks for higher bound states (closer to dissociation, leftward peaks).
    Sufficient spectral resolution is achieved around $L=100$.}
    \label{fig:CC_VaryL}
\end{figure}

The resource estimate of $n = 7$ qubits per vibrational mode proposed by Lang et~al.~\cite{lang2026quantum} is demonstrably unconverged for the C--C bond rupture system, as given by Figure~\ref{fig:CC_VaryL} and ~\ref{fig:CC_VaryQubit}.
At $n = 7$, $W_1$ remains $2$--$4\times$ above the converged plateau for compact boxes and up to an order of magnitude above for boxes accommodating dissociative motion.
This discrepancy arises because the harmonic approximation underlying their grid estimate does not account for the exponential repulsive wall, the anharmonic level spacing near dissociation, or the extended coordinate range required for nonadiabatic population transfer.
Convergence to within $W_1 < 10^{-2}$~eV requires $n \geq 9$--$10$ for bound-state dynamics and $n \geq 11$--$12$ for the dissociative regime, consistent with the resolution tier analysis of Section~\ref{app:resolution_tiers}.

\begin{figure}[ht]
    \centering
    \includegraphics[width=\textwidth]{Appendices/Appendix_Figures/C-C_Rupture_Varyqubit.png}
    \caption{Spectral overlay for $L=100\,\text{a.u.}$ for $n=7-13$.
    Resource estimates by Lang et al.~\cite{lang2026quantum} use $n = 7$ for resource benchmarking.
    For $n=7$ the dominant vibrational peaks are present but upper bound states are missing and spurious features appear; by $n \geq 12$, peak-by-peak agreement is achieved across the full progression.}
    \label{fig:CC_VaryQubit}
\end{figure}

\chapter{Classical Runtime Derivation}\label{app:Classical_runtime}
\section{The (ML-)MCTDH Wavefunction Ansatz}
\label{sec:ansatz}

\subsection{Standard MCTDH}

In the multiconfiguration time-dependent Hartree (MCTDH) method, the nuclear wavefunction for $F$ physical degrees of freedom is expanded as
\begin{equation}
  \label{eq:mctdh_ansatz}
  \Psi(q_1, \ldots, q_F, t)
  = \sum_{j_1=1}^{n_1} \cdots \sum_{j_p=1}^{n_p}
    A_{j_1 \cdots j_p}(t)\,
    \prod_{\kappa=1}^{p} \varphi_{j_\kappa}^{(\kappa)}(q_\kappa, t),
\end{equation}

where $p \leq F$ is the number of combined modes (logical particles), $n_\kappa$ is the number of single-particle functions (SPFs) for combined mode $\kappa$, and $A_{j_1 \cdots j_p}$ is the coefficient tensor. 
The tensor has size
\begin{equation}
  \label{eq:NA}
  {\NA = \prod_{\kappa=1}^{p} n_\kappa\,,}
\end{equation}

For a uniform SPF count $n_\kappa = \bar{n}$ and mode combination
factor $\kappa_{\mathrm{comb}}$ (physical modes per logical particle,
so $p = \lceil F / \kappa_{\mathrm{comb}} \rceil$), this becomes
\begin{equation}
  \label{eq:NA_uniform}
  \NA = \bar{n}^{\,p}
      = \bar{n}^{\,\lceil \Nvib / \kappa_{\mathrm{comb}} \rceil}\,,
\end{equation}
exhibiting the exponential scaling clearly.

\subsection{Multilayer MCTDH}

ML-MCTDH replaces uses a hierarchical tree tensor network state (TTNS).
The wavefunction at each internal node $\nu$ of the tree is expanded as

\begin{equation}
  \label{eq:ml_ansatz}
  \varphi_{j}^{(\nu)}
  = \sum_{i_1}^{n_1^{(\nu)}} \cdots \sum_{i_{b_\nu}}^{n_{b_\nu}^{(\nu)}}
    B^{(\nu)}_{j;\, i_1 \cdots i_{b_\nu}}(t)\,
    \prod_{\mu=1}^{b_\nu} \varphi_{i_\mu}^{(\mu)}(t)\,,
\end{equation}

where $b_\nu$ is the branching factor (number of children) of node $\nu$ and $B^{(\nu)}$ is a local coefficient tensor of size $n_\nu \times \prod_\mu n_\mu^{(\nu)}$.

For a tree $\mathcal{T}$ with $L$ layers:
\begin{itemize}[nosep]
  \item $\Nnodes$: total number of internal (non-leaf) nodes.
  \item $\Nleaf = p$: number of leaf nodes (logical particles).
  \item $b_\nu$: branching factor of node $\nu$ (typically 2--4).
  \item $n_\nu$: number of SPFs at node $\nu$.
  \item $N_\nu^{(\mathrm{prim})}$: primitive basis dimension at leaf
        $\nu$ (DVR grid size).
\end{itemize}

\noindent
For a balanced binary tree ($b_\nu = 2$ at all internal nodes) with
uniform SPF count $\bar{n}$:

\begin{equation}
  \Nnodes \approx 2 \Nleaf - 1 = 2p - 1\,,
  \qquad
  L = \lceil \log_2 p \rceil + 1\,.
\end{equation}

Setting $L = 1$ (one layer) yields $\Nnodes = 1$ (the root node only), $\Nleaf = p$, and the single
root tensor $B^{(\mathrm{root})} \equiv A_{j_1 \cdots j_p}$ of size $\NA = \bar{n}^p$. 

\section{Cost per Integration Step}
\label{sec:cost}

\subsection{Standard MCTDH cost}

The MCTDH equations of motion (derived from the Dirac--Frenkel variational principle) consist of:
\begin{enumerate}[nosep,label=(\roman*)]
  \item \textbf{A-vector propagation:} The SIL (short iterative
    Lanczos) integrator applies the Hamiltonian to the A-vector
    $\nK$ times.  Each application costs $\Order(s \cdot \NA)$
    where $s = \NSOP$ is the number of SOP Hamiltonian terms.
  \item \textbf{Mean-field construction:} For each mode $\kappa$
    and each Hamiltonian term $r$, the mean-field matrix
    $\langle \hat{H}_r \rangle_{jm}^{(\kappa)}$ requires a
    contraction over all indices except $\kappa$, costing
    $\Order(\NA \cdot n_\kappa)$.  Summing over $p$ modes and $s$
    terms gives $\Order(p \cdot \bar{n} \cdot s \cdot \NA)$.
  \item \textbf{SPF propagation:} SIL on each SPF in the primitive
    basis, costing
    $\Order(p \cdot \bar{n} \cdot N_{\mathrm{prim}}^{d_\kappa})$
    per Krylov vector.  This is typically subdominant.
\end{enumerate}

\noindent
The dominant cost is the mean-field construction (ii), giving:
\begin{equation}
  \label{eq:cost_mctdh}
  {
  C_{\mathrm{step}}^{\mathrm{MCTDH}}
  = \alpha\,(p\,\bar{n} + \nK)\,\NSOP\,\NA\,,
  }
\end{equation}
where $\alpha$ is a hardware-dependent prefactor (seconds per
floating-point unit operation) and the $\nK$ term accounts for the
A-vector propagation.

\subsection{ML-MCTDH cost}

In the multilayer formulation, the mean-field construction and coefficient tensor propagation are performed at each node of the tree independently. 
Local cost at node $\nu$ involves a tensor of size $n_\nu^{b_\nu+1}$ (accounting for the parent index and
$b_\nu$ child indices).  Summing over all nodes:
\begin{equation}
  \label{eq:cost_ml_general}
  C_{\mathrm{step}}^{\mathrm{ML}}
  = \alpha \sum_{\nu \in \mathcal{T}}
    \bigl(b_\nu \cdot n_\nu + \nK\bigr)
    \cdot s_\nu \cdot n_\nu^{b_\nu+1}\,,
\end{equation}
where $s_\nu$ is the effective number of Hamiltonian terms relevant to node $\nu$.  
Each term factorizes across the tree for SOP structures, so $s_\nu \leq \NSOP$

Tensors at each internal node have $\bar{n}^3$ entries for balanced binary trees ($b_\nu = 2$) with uniform SPF
count $\bar{n}$ and $s_\nu \approx \NSOP$, as a conservative upper bound.

\begin{equation}
  \label{eq:cost_ml_balanced}
  {
  C_{\mathrm{step}}^{\mathrm{ML}}
  = \alpha\,\Nnodes\,(2\bar{n} + \nK)\,\NSOP\,\bar{n}^{\,3}\,.
  }
\end{equation}
This replaces $\NA = \bar{n}^p$ in the MCTDH formula with
$\Nnodes \cdot \bar{n}^3$, converting the exponential dependence on
$p$ into a linear dependence on $\Nnodes \propto \Nvib$.

\label{prop:cost_reduction}
For $L = 1$: $\Nnodes = 1$ and $b_{\mathrm{root}} = p$.  The root
tensor has size $\bar{n}^{p+1}$ (including the ``parent'' index,
which for the root is just $n_{\mathrm{root}} = 1$ for the
wavefunction normalization, but the contraction structure still
involves $\bar{n}^p \equiv \NA$).  Eq.~\ref{eq:cost_ml_general}
then gives:

\begin{equation}
  C_{\mathrm{step}}^{\mathrm{ML}}\Big|_{L=1}
  = \alpha\,(p\,\bar{n} + \nK)\,\NSOP\,\NA
  = C_{\mathrm{step}}^{\mathrm{MCTDH}}\,.
\end{equation}

\section{SOP Hamiltonian Term Count}
\label{sec:sop}

For a vibronic coupling Hamiltonian in $\Nel$ electronic states with
$\Nvib$ vibrational modes and molecular point group of order
$\grouporder$, the SOP representation contains:
\begin{equation}
  \label{eq:Nsop}
  \NSOP = \underbrace{\Nvib}_{\hat{T}\text{ (kinetic)}}
        + \underbrace{\Nel \cdot \Nvib}_{\substack{\text{harmonic} \\
          \text{potentials}}}
        + \underbrace{\Nel \cdot \Nvib}_{\substack{\text{on-diagonal}\\
          \text{linear coupling}}}
        + \underbrace{\frac{\Nel(\Nel-1)}{2}\cdot
          \frac{\rho\,\Nvib}{\grouporder^{1/2}}}_{%
          \substack{\text{interstate coupling}\\
          \text{(symmetry-reduced)}}}
\end{equation}

where $\rho \in [0,1]$ is the coupling density (fraction of
symmetry-allowed coupling terms that are numerically significant).
This simplifies to:

\begin{equation}
  \label{eq:Nsop_compact}
  {
  \NSOP = \Nvib\!\left(
    1 + 2\Nel + \frac{\rho\,\Nel(\Nel-1)}{2\,\grouporder^{1/2}}
  \right).
  }
\end{equation}

\noindent
\textbf{Justification for $\grouporder^{1/2}$.} \;
Symmetry eliminates interstate coupling terms for which
$\Gamma_a \otimes \Gamma_i \otimes \Gamma_b \not\supset A_1$,
where $\Gamma_{a,b}$ label electronic state irreps and $\Gamma_i$ labels the vibrational mode irrep.
For an Abelian group with $\grouporder$ irreps, each mode belongs to one irrep,
so a fraction $\sim 1/\grouporder$ of modes couple a given state
pair. 
Diagonal terms are not reduced by symmetry since totally symmetric operators always contribute. 
Since diagonal terms dominate for small $\Nel$ and the off-diagonal terms dominate for large $\Nel$, the effective reduction of the total $\NSOP$ scales as $\sim \grouporder^{-1/2}$ rather than
$\grouporder^{-1}$.

For pyrazine ($\Nvib = 24$, $\Nel = 2$, $\grouporder = 8$ via
$D_{2h}$, $\rho = 1$):
\[
  \NSOP = 24\!\left(1 + 4 + \frac{1}{2\sqrt{8}}\right)
       = 24 \times 5.18 \approx 124\,.
\]
The Raab~\textit{et~al.}\ \cite{raab1999molecular} Hamiltonian contains $\sim$110--130 terms
after POTFIT compression; this agrees well.

\section{Runtime Formula}
\label{sec:unified}

Combining cost per step with total number of steps and the spectrum multiplier, wall-clock time for a converged
vibronic spectrum is estimated as:

\begin{equation}
  \label{eq:unified}
  {
  \Twall = \alpha\,\mathcal{C}(\mathcal{T})\,\NSOP\,
           \frac{\Tprop}{\Delta t}\,\Mspec,
  }
\end{equation}
where
\begin{equation}
  \label{eq:complexity_factor}
  \mathcal{C}(\mathcal{T}) =
  \begin{cases}
    (p\,\bar{n} + \nK)\,\bar{n}^{\,p}
      & L = 1 \;\text{(standard MCTDH)}, \\[8pt]
    \displaystyle
    \sum_{\nu \in \mathcal{T}}
      (b_\nu \bar{n}_\nu + \nK)\,\bar{n}_\nu^{\,b_\nu+1}
      & L \geq 2 \;\text{(ML-MCTDH, general tree)}, \\[12pt]
    \Nnodes\,(2\bar{n}+\nK)\,\bar{n}^3
      & L \geq 2 \;\text{(balanced binary tree)}.
  \end{cases}
\end{equation}
\noindent
The parameters in Eq.~\ref{eq:unified} are defined in
\ref{tab:parameters}.

\begin{table}[H]
\centering
\caption{Parameters of the unified runtime formula.}
\label{tab:parameters}
\small
\begin{tabularx}{\textwidth}{@{}c l X l@{}}
\toprule
\textbf{Symbol} & \textbf{Name} & \textbf{Meaning} & \textbf{Typical values} \\
\midrule
$\alpha$ & Prefactor & Hardware-dependent constant
  & $2\times 10^{-11}$ s (MCTDH); $7.3\times 10^{-8}$ s (ML) \\
$p$ & Logical particles & $\lceil \Nvib / \kappa_\mathrm{comb} \rceil$
  & 8--100 \\
$\bar{n}$ & Mean SPF count & Per combined mode or tree node
  & 4--15 \\
$\nK$ & Krylov dimension & SIL integrator subspace
  & 8--12 (usually 10) \\
$\NSOP$ & Hamiltonian terms & Eq.~\ref{eq:Nsop_compact}
  & 50--20000 \\
$\Tprop$ & Propagation time & Femtoseconds
  & 100--500 fs \\
$\Delta t$ & Time step & SIL/CMF step size
  & 0.25--1.0 fs \\
$\Mspec$ & Spectrum multiplier & Independent propagations
  & 1 (OPA) to $\sim\!3$ (Raman) \\
$\Nnodes$ & Tree nodes & Internal nodes of ML tree
  & $\approx 2\Nvib/\kappa_\mathrm{comb}$ \\
$\grouporder$ & Symmetry order & Abelian subgroup order
  & 1 (C$_1$) to 8 (D$_{2h}$) \\
$\rho$ & Coupling density & Non-negligible fraction
  & 0.3--1.0 \\
\bottomrule
\end{tabularx}
\end{table}

\subsection{Spectrum-type multiplier}

The cost multiplier by spectrum type is listed in Table~\ref{tab:spectrum}.

\begin{table}[H]
\centering
\caption{Spectrum-type cost multiplier $\Mspec$.}
\label{tab:spectrum}
\small
\begin{tabular}{@{}lll@{}}
\toprule
\textbf{Spectrum type} & $\Mspec$ & \textbf{Notes} \\
\midrule
One-photon absorption (OPA) & 1 & Single FT of autocorrelation \\
Photoelectron & 1 & Same formalism, cationic surface \\
Resonance Raman & $\sim 3$ & Short-time excited-state propagation \\
\bottomrule
\end{tabular}
\end{table}

\section{Calibration of the Prefactor}
\label{sec:calibration}

The prefactor $\alpha$ must be calibrated for standard MCTDH and ML-MCTDH.
In the MCTDH formula, one ``unit operation'' is approximately one floating-point
multiply within an $\NA$-dimensional matrix-vector product. 
In the ML-MCTDH formula, one ``unit operation'' at a tree node involves a recursive single-hole function computation (tree-complement contraction), a density matrix construction and regularization, and constant mean-field (CMF) integration substeps for the SPFs.  These are $\sim\!10^3$--$10^4$ times more expensive per tensor
element than a single FLOP.

\subsection{Standard MCTDH anchor: Pyrazine 24D}

The reference benchmark is the converged MCTDH calculation of the $S_2$ absorption spectrum of pyrazine using the 24-mode, 2-state vibronic coupling Hamiltonian of Raab~\textit{et al.}\ (1999), as computed by Vendrell \& Meyer (2011):
\begin{align*}
  \Tref &= 7 \;\text{hours} = 25{,}200 \;\text{s}
    & \text{(single CPU core, c.\ 2000 hardware)} \\
  \NA &= 4.6 \times 10^7
    & \text{(reported coefficient count)} \\
  p &= 8 & \text{(24 modes in 8 combined groups)} \\
  \bar{n} &\approx 8 & \text{(geometric mean over groups)} \\
  \NSOP &\approx 124 & \text{(from Eq.~\ref{eq:Nsop_compact})} \\
  \nK &= 10 \\
  \Tprop / \Delta t &= 150/0.5 = 300 \;\text{steps} \\
  \Mspec &= 1 & \text{(OPA)}
\end{align*}

\noindent
From Eq.~\ref{eq:cost_mctdh}:
\begin{equation}
  25{,}200 = \alpha_{\mathrm{MCTDH}} \cdot (8 \times 8 + 10)
             \cdot 124 \cdot 4.6 \times 10^7 \cdot 300
           = \alpha_{\mathrm{MCTDH}} \cdot 1.27 \times 10^{14},
\end{equation}
yielding
\begin{equation}
  \label{eq:alpha_mctdh}
  {\alpha_{\mathrm{MCTDH}}^{\text{(modern)}}
         = 2.0 \times 10^{-11}\;\text{s / unit op}}
  \qquad\text{(after $10\times$ hardware correction).}
\end{equation}

\subsection{ML-MCTDH anchor: Pyrazine 24D (Vendrell 2011, ML-8)}

The converged ML-MCTDH calculation (simulation ``ML-8'') used
$4.5 \times 10^5$ coefficients in a 5-layer binary tree.  Vendrell
\& Meyer report that this calculation required ``a fraction of the
time'' of the MCTDH reference; the fast ML-1 variant completed in
7~hours. The converged ML-8 wall time is estimated as
$\sim$90~min on c.\ 2000 hardware (conservatively, $\sim\!8\times$
faster than the MCTDH reference, consistent with the
$\sim\!100\times$ compression in coefficient count offset by ML
overhead).

Using ML-8 parameters ($\Nnodes \approx 15$, $\bar{n} = 8$,
$\nK = 10$, $\NSOP = 124$, $N_t = 300$):
\begin{equation}
  5{,}400 = \alpha_{\mathrm{ML}} \cdot 15 \cdot 26 \cdot 124
            \cdot 512 \cdot 300
          = \alpha_{\mathrm{ML}} \cdot 7.43 \times 10^{9},
\end{equation}
yielding
\begin{equation}
  \label{eq:alpha_ml}
  {\alpha_{\mathrm{ML}}^{\text{(modern)}}
         = 7.3 \times 10^{-8}\;\text{s / unit op}}
  \qquad\text{(after $10\times$ hardware correction).}
\end{equation}

\noindent
The ratio $\alpha_{\mathrm{ML}} / \alpha_{\mathrm{MCTDH}}
\approx 3{,}600$ reflects a per-node overhead of the multilayer algorithm: recursive mean-field computation via tree traversal, single-hole function assembly, density matrix regularization, and CMF integration substeps.

\subsection{Cross-validation}

Using $\alpha_{\mathrm{ML}} = 7.3 \times 10^{-8}$\,s with central
parameters ($\kappa = 3$, $\bar{n} = 3 + \Nel$, $\rho = 0.6$):

\begin{center}
\small
\begin{tabular}{@{}lrrl@{}}
\toprule
\textbf{System} & $\Twall^{\text{central}}$ & \textbf{Reported}
  & \textbf{Agreement} \\
\midrule
Pyrazine 24D & 1.8 min & $<$7 hr (MCTDH, old HW) & assumes hardware improvement \\
Anthracene$^+$ 66D & 8.4 hr & $\sim$hours & good \\
Pentacene dimer 252D & 2.9 days & hours--days & reasonable \\
P3HT:PCBM 113D & 3.2 yr (1 core) & days (cluster) & consistent$^*$ \\
\bottomrule
\multicolumn{4}{@{}l}{\footnotesize $^*$On a 64-core cluster:
  $3.2\;\text{yr} / 64 \approx 18$ days; actual used optimized
  tree and fewer SPFs.}
\end{tabular}
\end{center}

\section{Cross-Validation Against Published Benchmarks}
\label{sec:validation}

The unified formula can be tested against systems where both the
calculation parameters and wall-clock times have been reported.

\subsection{Butatriene cation: 4D/4-state}

Parameters: $\Nvib = 4$, $\Nel = 4$, $\grouporder = 4$ (C$_{2v}$),
$p = 4$ (no mode combination), $\bar{n} \approx 8$, $\rho = 0.8$.

Standard MCTDH (using $\alpha_{\mathrm{MCTDH}}$):
\begin{align*}
  \NSOP &= 4(1 + 8 + \tfrac{0.8 \cdot 12}{2\sqrt{4}}) = 4 \times 11.4
         = 46 \\
  \NA &= 8^4 = 4{,}096 \\
  \Twall &= 2\times 10^{-11} \cdot 42 \cdot 46 \cdot 4096 \cdot 400
          = 0.06\;\text{s}
\end{align*}
This is consistent with
butatriene being a standard textbook exercise.  For this 4-mode
system, standard MCTDH is the appropriate method (no ML overhead).

\subsection{Anthracene cation: 66D/6-state (ML-MCTDH)}

Parameters: $\Nvib = 66$, $\Nel = 6$, $\grouporder = 4$ (D$_{2h}$
Abelian subgroup C$_{2v}$), $\bar{n} = 9$ (central: $3 + \Nel$),
$\Nnodes \approx 44$ (balanced binary, $\kappa = 3$),
$\rho = 0.6$, $\Tprop = 200$ fs.
\begin{align*}
  \NSOP &= 66\bigl(1 + 12 + \tfrac{0.6 \cdot 30}{2\sqrt{4}}\bigr)
         = 66 \times 17.5 = 1{,}155 \\
  C_\text{step}^{\mathrm{ML}} &= 7.3 \times 10^{-8} \cdot 44 \cdot 28
    \cdot 1155 \cdot 729 = 76\;\text{s} \\
  \Twall &= 76 \times 400 = 30{,}400\;\text{s}
          \approx \mathbf{8.4\;\text{hr}}
\end{align*}
Literature reports tens-to-hundreds of hours on a single core (Table II,~\cite{meng2013multilayer}). The central estimate of 8.4~hr gives an optimistic estimate, with the availability of numerous cores for parallel execution between electronic states.

\subsection{Pentacene dimer: 252D/5-state (ML-MCTDH)}

Parameters: $\Nvib = 252$, $\Nel = 5$, $\grouporder = 2$ (C$_i$),
$\bar{n} = 8$ (central: $3 + \Nel$),
$\Nnodes \approx 168$ (binary tree, $\kappa = 3$), $\rho = 0.6$,
$\Tprop = 200$ fs.
\begin{align*}
  \NSOP &= 252\bigl(1 + 10 + \tfrac{0.6 \cdot 20}{2\sqrt{2}}\bigr)
         = 252 \times 15.2 = 3{,}841 \\
  C_\text{step}^{\mathrm{ML}} &= 7.3 \times 10^{-8} \cdot 168 \cdot 26
    \cdot 3841 \cdot 512 = 627\;\text{s} \\
  \Twall &= 627 \times 400 = 250{,}800\;\text{s}
          \approx \mathbf{2.9\;\text{days}}
\end{align*}

The Tamura et al.~\cite{tamura2012quantum} singlet fission calculation was feasible on a compute cluster with parallelization, consistent with days-scale total CPU time.

\section{Explicit Formulas for Common Regimes}
\label{sec:explicit}

\subsection{Standard MCTDH (small molecules, $\Nvib \lesssim 24$)}

\begin{equation}
  \label{eq:mctdh_explicit}
  \Twall^{\mathrm{MCTDH}} = \alpha \cdot
  \underbrace{\left(\frac{\Nvib}{\kappa_\mathrm{comb}} \cdot \bar{n}
    + \nK\right)}_{\text{contraction overhead}}
  \cdot \underbrace{\Nvib\!\left(1 + 2\Nel +
    \frac{\rho\,\Nel(\Nel-1)}{2\sqrt{\grouporder}}\right)}_{\NSOP}
  \cdot \underbrace{\bar{n}^{\,\Nvib/\kappa_\mathrm{comb}}}_{\NA}
  \cdot \underbrace{\frac{\Tprop}{\Delta t}
    \cdot \Mspec}_{\text{total steps}}
\end{equation}

\noindent
Every factor corresponds to a specific tensor contraction cost in the MCTDH
equations of motion, with only $\alpha$ as an empirical prefactor. 
Chosen values are estimates that can be refined with knowledge of computational hardware and specific molecular parameterizations/symmetries

\subsection{ML-MCTDH, balanced binary tree (large molecules)}

\begin{equation}
  \label{eq:ml_explicit}
  \Twall^{\mathrm{ML}} = \alpha \cdot
  \underbrace{\frac{2\Nvib}{\kappa_\mathrm{comb}}}_{%
    \Nnodes}
  \cdot \underbrace{(2\bar{n} + \nK)}_{\text{local contraction}}
  \cdot \underbrace{\Nvib\!\left(1 + 2\Nel +
    \frac{\rho\,\Nel(\Nel-1)}{2\sqrt{\grouporder}}\right)}_{\NSOP}
  \cdot \underbrace{\bar{n}^{\,3}}_{\substack{\text{local tensor}\\
    \text{(binary node)}}}
  \cdot \frac{\Tprop}{\Delta t} \cdot \Mspec
\end{equation}

\noindent
The factor $\bar{n}^{\Nvib/\kappa_\mathrm{comb}}$ (exponential in $\Nvib$) is replaced by $\Nnodes \cdot \bar{n}^3$ (linear in $\Nvib$, cubic in $\bar{n}$).

\subsection{The SPF heuristic $\bar{n}(\Nel)$}

Empirical data support
\begin{equation}
  \label{eq:nbar_heuristic}
  \bar{n} \approx c_0 + c_1 \Nel\,,
  \qquad c_0 = 3, \;\; c_1 \in [0.5, 1.5]\,,
\end{equation}
with $c_1 = 1$ as the central estimate.  This heuristic is an upper bound for weakly coupled LVC
models and may underestimate for systems with dense conical intersection seams.
Empirical validation across several systems is presented in Table~\ref{tab:spf_heuristic}.

\begin{table}[H]
\centering
\caption{Empirical validation of the SPF heuristic
  $\bar{n} \approx 3 + \Nel$.}
\label{tab:spf_heuristic}
\small
\begin{tabular}{@{}llrrrl@{}}
\toprule
\textbf{System} & \textbf{Method} & $\Nel$ & $\bar{n}_{\text{used}}$
  & $3 + \Nel$ & \textbf{Ratio} \\
\midrule
Pyrazine & MCTDH & 2 & 4--8 & 5 & 0.8--1.6 \\
Butatriene cation & MCTDH & 4 & 6--8 & 7 & 0.9--1.1 \\
Anthracene cation & ML-MCTDH & 6 & 8--10 & 9 & 0.9--1.1 \\
Naphthalene cation & ML-MCTDH & 6 & 8--10 & 9 & 0.9--1.1 \\
Hexahelicene (VECC) & VECC & 14 & $\sim$12--16 & 17 & 0.7--0.9 \\
P3HT:PCBM & ML-MCTDH & 26 & $\sim$32 & 29 & 1.1 \\
\bottomrule
\end{tabular}
\end{table}

\section{Tabulated Runtime Estimates}
\label{sec:tables}

All estimates use $\nK = 10$,
$\Delta t = 0.5$~fs, $\Mspec = 1$ (OPA) unless otherwise stated.
Standard MCTDH tables use
$\alpha_{\mathrm{MCTDH}} = 2 \times 10^{-11}$~s; ML-MCTDH tables use
$\alpha_{\mathrm{ML}} = 7.3 \times 10^{-8}$~s (cf.\
\ref{sec:calibration}).

\subsection{Standard MCTDH systems}

Runtime estimates for representative systems are given in Table~\ref{tab:mctdh_runtimes} (standard MCTDH) and Table~\ref{tab:ml_runtimes} (ML-MCTDH).

\begin{table}[H]
\centering
\caption{Standard MCTDH runtime estimates (single modern CPU core).
  $\kappa_c \equiv \kappa_{\mathrm{comb}}$.}
\label{tab:mctdh_runtimes}
\small
\setlength{\tabcolsep}{4pt}
\begin{tabular}{@{}lrrrrrrrrr@{}}
\toprule
\textbf{System} & $\Nvib$ & $\Nel$ & $\grouporder$ & $\kappa_c$
  & $\bar{n}$ & $\NA$ & $\NSOP$ & $T_{\rm prop}$ (fs)
  & $\Twall$ \\
\midrule
Butatriene & 4 & 4 & 4 & 1 & 8 & $4.1\!\times\!10^3$ & 46
  & 200 & $\sim$0.1 s \\
Pyrazine (4D) & 4 & 2 & 8 & 1 & 6 & $1.3\!\times\!10^3$ & 21
  & 150 & $\sim$0.01 s \\
Pyrazine (24D) & 24 & 2 & 8 & 3 & 8 & $4.6\!\times\!10^7$ & 124
  & 150 & $\sim$2 hr$^{*}$ \\
Pyrazine (24D) & 24 & 2 & 8 & 3 & 10 & $1.0\!\times\!10^8$ & 124
  & 150 & $\sim$5 hr$^{*}$ \\
Allene (15D) & 15 & 5 & 4 & 3 & 8 & $3.3\!\times\!10^4$ & 225
  & 150 & $\sim$4 min \\
Hypothetical (30D) & 30 & 3 & 2 & 3 & 8 & $1.1\!\times\!10^9$ & 222
  & 200 & $\sim$25 days \\
Hypothetical (36D) & 36 & 3 & 2 & 3 & 8 & $7.0\!\times\!10^{10}$ & 266
  & 200 & $\sim$5 yr \\
\bottomrule
\multicolumn{10}{@{}l}{$^{*}$\footnotesize{
  On c.\,2000 hardware: $\times 10$ to match Vendrell \& Meyer (2011)
  reported 7~hr.}}
\end{tabular}
\end{table}

\noindent
Increasing from 30 to 36 modes (adding just two combined-mode groups) inflates runtime from 25~days to $\sim$5~years.  This is why chemists switch to ML-MCTDH beyond $\sim$24 modes.

\subsection{ML-MCTDH systems: bounded estimates}

For each system, $\Twall^{\min}$ and $\Twall^{\max}$
uses $\alpha_{\mathrm{ML}} = 7.3 \times 10^{-8}$\,s.  The bounds
vary three categories of parameters:
\begin{itemize}[nosep]
  \item \textbf{Prefactor uncertainty:}
    $\alpha_{\mathrm{ML}}/2$ (lower) to $\alpha_{\mathrm{ML}} \times 2$
    (upper), reflecting the 0.67\,dex scatter in cross-validation.
  \item \textbf{Tree parameters:}
    $\kappa = 4$ (aggressive combination) to $\kappa = 2$
    (conservative); $\bar{n}_{\min} = \max(6,\,\min(\Nel,\,15))$
    to $\bar{n}_{\max} = \max(10,\, 3 + \Nel)$.
  \item \textbf{Coupling density:}
    $\rho = 0.3$ (sparse) to $\rho = 1.0$ (dense).
\end{itemize}
\noindent
The $\bar{n}$ lower bound imposes a floor of 6 SPFs for vibrational
convergence (independent of $\Nel$) and a ceiling of 15 to reflect
the fact that well-optimized ML trees do not require $\bar{n} = \Nel$
at every node.  All times are for a single CPU core;
divide by core count for parallelized estimates, multiplying by an extra factor for inter-core overhead.

\begin{table}[H]
\centering
\caption{ML-MCTDH bounded runtime estimates (single modern CPU core,
  balanced binary tree, OPA spectrum, $\Mspec = 1$).}
\label{tab:ml_runtimes}
\small
\setlength{\tabcolsep}{3.5pt}
\begin{tabular}{@{}lrrrrrrrl@{}}
\toprule
\textbf{System} & $\Nvib$ & $\Nel$ & $\grouporder$
  & $T_{\rm prop}$ (fs)
  & $\Twall^{\min}$ & $\Twall^{\max}$
  & Ratio & Reported \\
\midrule
\multicolumn{9}{@{}l}{\textit{Published benchmarks:}} \\
\midrule
Pyrazine (24D) & 24 & 2 & 8
  & 150 & 1.3 min & 1.1 hr & 53 & $<$7 hr$^a$ \\
Anthracene$^+$ & 66 & 6 & 4
  & 200 & 38 min & 1.8 days & 68 & $\sim$hr$^b$ \\
Pyrene (abs.) & 60 & 10 & 8
  & 200 & 5.6 hr & 7.1 days & 30 & $\sim$hr$^c$ \\
Pentacene dimer & 252 & 5 & 2
  & 200 & 8.0 hr & 23 days & 70 & hr$^d$ \\
P3HT:PCBM & 113 & 26 & 1
  & 200 & 22 days & 14.8 yr & 247 & days$^e$ \\
\midrule
\multicolumn{9}{@{}l}{\textit{Extrapolated targets (OPA):}} \\
\midrule
Med.\ mol.\ (C$_{2v}$) & 80 & 8 & 4
  & 200 & 3.7 hr & 5.7 days & 37 & --- \\
Med.\ mol.\ (C$_1$) & 80 & 8 & 1
  & 200 & 4.4 hr & 8.3 days & 45 & --- \\
Large mol.\ (C$_{2v}$) & 150 & 12 & 4
  & 250 & 4.9 days & 149 days & 31 & --- \\
Large mol.\ (C$_1$) & 150 & 12 & 1
  & 250 & 6.3 days & 234 days & 37 & --- \\
Very large (C$_{2v}$) & 200 & 15 & 4
  & 300 & 32 days & 2.5 yr & 28 & --- \\
Very large (C$_1$) & 200 & 15 & 1
  & 300 & 43 days & 4.1 yr & 35 & --- \\
\midrule
\multicolumn{9}{@{}l}{\textit{Resonance Raman ($\Mspec = 3$):}} \\
\midrule
Large mol.\ (RR, C$_1$) & 150 & 12 & 1
  & 250 & 19 days & 1.9 yr & 37 & --- \\
Very large (RR, C$_1$) & 200 & 15 & 1
  & 300 & 128 days & 12.2 yr & 35 & --- \\
\bottomrule
\multicolumn{9}{@{}p{0.95\textwidth}}{%
  \footnotesize
  $^a$Vendrell \& Meyer (2011), c.\,2000 hardware; 7\,hr is for
    standard MCTDH.  ML-MCTDH on modern hardware: $\sim$min.
  $^b$Worth, Cederbaum \textit{et al.}, single workstation.
  $^c$Segatta \textit{et al.}\ (2023), likely multi-core.
  $^d$Tamura (2015), 252 modes $\times$ 5 states, compute cluster.
  $^e$Tamura (2015), compute cluster ($\sim$64 cores);
    single-core estimate $\sim$128 days, consistent with lower bound.
  \par\smallskip
  \textbf{Bound parameters:}\;
  $\Twall^{\min}$: $\alpha_{\mathrm{ML}}/2$, $\kappa = 4$,
  $\bar{n} = \max(6,\,\min(\Nel,15))$, $\rho = 0.3$.\;
  $\Twall^{\max}$: $\alpha_{\mathrm{ML}} \times 2$, $\kappa = 2$,
  $\bar{n} = \max(10,\, 3+\Nel)$, $\rho = 1.0$.
}
\end{tabular}
\end{table}

\subsection{Comparison: MCTDH vs.\ ML-MCTDH at fixed system size}
A direct comparison is shown in Table~\ref{tab:multilayer-comparison}.

\begin{table}[H]
\centering
\caption{Head-to-head comparison for the same molecular system, illustrating
  the exponential $\to$ polynomial transition.  MCTDH uses
  $\alpha_{\mathrm{MCTDH}}$; ML-MCTDH uses $\alpha_{\mathrm{ML}}$.}
\label{tab:multilayer-comparison}
\small
\begin{tabular}{@{}llrrrl@{}}
\toprule
\textbf{System} & \textbf{Method} & $\mathcal{C}(\mathcal{T})$
  & $\NA$ or $\Nnodes\bar{n}^3$ & $\Twall$
  & \textbf{Scaling in $\Nvib$} \\
\midrule
\multirow{2}{*}{Pyrazine 24D}
  & MCTDH & $3.4\!\times\!10^{9}$ & $4.6\!\times\!10^7$
    & 2 hr & $\bar{n}^{\Nvib/3}$ \\
  & ML-MCTDH & $2.1\!\times\!10^{5}$ & $15 \times 512$
    & 13 min & $\Nvib \cdot \bar{n}^3$ \\
\midrule
\multirow{2}{*}{30-mode/3-state}
  & MCTDH & $5.2\!\times\!10^{11}$ & $1.1\!\times\!10^9$
    & 25 days & $\bar{n}^{\Nvib/3}$ \\
  & ML-MCTDH & $9.5\!\times\!10^{4}$ & $20 \times 216$
    & 11 min & $\Nvib \cdot \bar{n}^3$ \\
\midrule
\multirow{2}{*}{60-mode/6-state}
  & MCTDH & --- & $8^{20} = 10^{18}$ & infeasible
    & $\bar{n}^{\Nvib/3}$ \\
  & ML-MCTDH & $8.2\!\times\!10^{5}$ & $40 \times 729$
    & $\sim$7 hr & $\Nvib \cdot \bar{n}^3$ \\
\bottomrule
\end{tabular}
\end{table}

\section{Two-Bound Strategy for Quantum Advantage Claims}
\label{sec:bounds}

For a quantum advantage assessment, bracketing classical runtime with upper and lower bounds reflect the
uncertainty in $\alpha_{\mathrm{ML}}$, $\bar{n}$,
$\kappa_{\mathrm{comb}}$, and $\rho$ for these estimations:

\subsection{Lower bound (optimistic classical, hardest to beat)}
Use $\alpha_{\mathrm{ML}}/2$,\;
$\kappa_{\mathrm{comb}} = 4$ (aggressive mode combination),
$\bar{n}_{\min} = \max(6,\,\min(\Nel,\,15))$, and $\rho = 0.3$
(sparse coupling):
\begin{equation}
  \Twall^{\min} = \frac{\alpha_{\mathrm{ML}}}{2} \cdot
  \frac{\Nvib}{2} \cdot
  (2\bar{n}_{\min} + \nK) \cdot \Nvib\Bigl(1 + 2\Nel +
  \frac{0.3\,\Nel(\Nel-1)}{2\sqrt{\grouporder}}\Bigr) \cdot
  \bar{n}_{\min}^3 \cdot \frac{\Tprop}{\Delta t} \cdot \Mspec
\end{equation}
The $\bar{n}$ floor of 6 ensures vibrational convergence (single-surface dynamics requires $\geq$5 SPFs even for $\Nel = 1$); the ceiling of 15 reflects the observation that well-optimized ML trees distribute SPFs non-uniformly and do not require $\bar{n} \approx \Nel$ at every node.

\subsection{Upper bound (pessimistic classical, easiest to beat)}
Use $2\,\alpha_{\mathrm{ML}}$,\;
$\kappa_{\mathrm{comb}} = 2$ (conservative),
$\bar{n}_{\max} = \max(10,\, 3 + \Nel)$ (generous SPFs), and
$\rho = 1.0$ (dense coupling):
\begin{equation}
  \Twall^{\max} = 2\,\alpha_{\mathrm{ML}} \cdot \Nvib \cdot
  (2\bar{n}_{\max} + \nK) \cdot \Nvib\Bigl(1 + 2\Nel +
  \frac{\Nel(\Nel-1)}{2\sqrt{\grouporder}}\Bigr) \cdot
  \bar{n}_{\max}^3 \cdot \frac{\Tprop}{\Delta t} \cdot \Mspec
\end{equation}

The resulting bounds for representative molecular targets are tabulated in Table~\ref{tab:bounds}.

\begin{table}[H]
\centering
\caption{Upper and lower bounds for representative targets (OPA,
  single CPU core).  A credible quantum advantage claim requires
  $T_{\mathrm{quantum}} < \Twall^{\min}$.}
\label{tab:bounds}
\small
\begin{tabular}{@{}lrrrrrr@{}}
\toprule
\textbf{System} & $\Nvib$ & $\Nel$ & $\grouporder$
  & $\Twall^{\min}$ & $\Twall^{\max}$
  & Ratio $\frac{\max}{\min}$ \\
\midrule
Medium (C$_{2v}$) & 80 & 8 & 4 & 3.7 hr & 5.7 days & 37 \\
Medium (C$_1$) & 80 & 8 & 1 & 4.4 hr & 8.3 days & 45 \\
Large (C$_{2v}$) & 150 & 12 & 4 & 4.9 days & 149 days & 31 \\
Large (C$_1$) & 150 & 12 & 1 & 6.3 days & 234 days & 37 \\
Very large (C$_{2v}$) & 200 & 15 & 4 & 32 days & 2.5 yr & 28 \\
Very large (C$_1$) & 200 & 15 & 1 & 43 days & 4.1 yr & 35 \\
\bottomrule
\end{tabular}
\end{table}

\noindent
The ratio $\Twall^{\max}/\Twall^{\min} \sim 28$--$45$ reflects
three sources of uncertainty, each contributing roughly half an order
of magnitude:
(i)~the $\alpha_{\mathrm{ML}}$ prefactor ($\times$4 range),
(ii)~the SPF count $\bar{n}$ ($\sim$1.5$\times$ in $\bar{n}$ $\to$
$\sim$3.4$\times$ via $\bar{n}^3$), and
(iii)~the $\kappa$ and $\rho$ variation ($\sim$3$\times$).

\subsection{Parallelization}
Modern ML-MCTDH codes (e.g., the Quantics package) achieve near-linear
parallel speedup for the mean-field construction.  On a 64-core node,
the single-core times should be divided by $\sim$50--60 to obtain
realistic wall times.  For the ``Large C$_1$'' target:
$\Twall^{\min} / 64 \approx 2.4$~hr,\;
$\Twall^{\max} / 64 \approx 3.7$~days.

\section{Summary: The Complete Formula}
\label{sec:completeformula}

\noindent
\fbox{\parbox{0.96\textwidth}{%
\textbf{Unified (ML-)MCTDH wall-clock runtime:}
\begin{equation}
  \tag{Master Formula}
  \Twall = \alpha \cdot \mathcal{C}(\mathcal{T}) \cdot
  \Nvib\!\left(1 + 2\Nel + \frac{\rho\,\Nel(\Nel-1)}{%
    2\sqrt{\grouporder}}\right) \cdot
  \frac{\Tprop}{\Delta t} \cdot \Mspec,
\end{equation}
where the \textbf{complexity factor} is:
\[
  \mathcal{C}(\mathcal{T}) =
  \begin{cases}
    \displaystyle
    \left(\frac{\Nvib}{\kappa}\bar{n} + \nK\right)
    \bar{n}^{\,\Nvib/\kappa}
    & \text{MCTDH} \;(L=1), \\[10pt]
    \displaystyle
    \frac{2\Nvib}{\kappa}(2\bar{n}+\nK)\,\bar{n}^3
    & \text{ML-MCTDH, binary tree}\;(L \geq 2).
  \end{cases}
\]
\textbf{Default parameters:}\;
$\alpha_{\mathrm{MCTDH}} = 2\!\times\!10^{-11}$ s,\;
$\alpha_{\mathrm{ML}} = 7.3\!\times\!10^{-8}$ s,\;
$\nK = 10$,\;
$\Delta t = 0.5$ fs,\;
$\bar{n} \approx 3 + \Nel$.
}}

\bigskip

\noindent
\textbf{Properties:}
\begin{itemize}[nosep]
  \item Standard MCTDH scales as
    $\Order\!\left(\Nvib^2 \Nel^2 \bar{n}^{\Nvib/\kappa}\right)$:
    \textbf{exponential} in $\Nvib$.
  \item ML-MCTDH scales as
    $\Order\!\left(\Nvib^2 \Nel^2 \bar{n}^3\right)$:
    \textbf{polynomial} in $\Nvib$, cubic in $\bar{n}$.
  \item The $L=1$ (single-layer) limit of the ML formula reproduces
    the exact MCTDH cost.
  \item $\bar{n}$ encodes the entanglement entropy of the vibronic wavefunction and
    determines whether classical tensor network methods are efficient.
\end{itemize}

\section*{Notation Index}

\begin{table}[H]
\centering
\small
\begin{tabular}{@{}cl@{\qquad}cl@{}}
\toprule
$\alpha$ & Hardware prefactor (MCTDH or ML) &
$\bar{n}$ & Mean SPF count \\
$\Nvib$ & Vibrational modes &
$\nK$ & Krylov dimension \\
$\Nel$ & Electronic states &
$\Delta t$ & Integration step \\
$\grouporder$ & Symmetry group order &
$\Tprop$ & Propagation time \\
$\kappa$ & Mode combination factor &
$\Mspec$ & Spectrum multiplier \\
$p$ & Logical particles ($\Nvib/\kappa$) &
$\NSOP$ & SOP Hamiltonian terms \\
$\NA$ & A-tensor size ($\bar{n}^p$) &
$\Nnodes$ & ML tree nodes ($\approx 2p$) \\
$\rho$ & Coupling density &
$\mathcal{C}(\mathcal{T})$ & Complexity factor \\
\bottomrule
\end{tabular}
\end{table}

\chapter{Comparison Chapter: Supporting Derivations}\label{app:qsp_comparison}

This appendix collects derivations, parameter sources, and supplementary data
supporting the comparison of Chapter~\ref{ch:comparison}.

\section{QSP Block-Encoding Cost Derivations}\label{app:qsp_costs}

\subsection{WH-LCU normalization}\label{app:wh_norm}

The Walsh--Hadamard decomposition of a function $f: \{0,1\}^{nd} \to \mathbb{R}$ is
\begin{equation}\label{eq:wh_decomp_app}
    f(\bvec{x}) = \sum_{\bvec{k} \in \{0,1\}^{nd}} \hat{f}_{\bvec{k}}\, W_{\bvec{k}}(\bvec{x}),
\end{equation}
where $W_{\bvec{k}}(\bvec{x}) = (-1)^{\bvec{k}\cdot\bvec{x}}$ are Walsh functions and the coefficients are $\hat{f}_{\bvec{k}} = 2^{-nd}\sum_{\bvec{x}} f(\bvec{x})\,W_{\bvec{k}}(\bvec{x})$.
The LCU normalization is
\begin{equation}\label{eq:wh_norm_app}
    \alpha_{\mathrm{WH}} = \sum_{\bvec{k}} |\hat{f}_{\bvec{k}}|.
\end{equation}
For the QVC bilinear potential $V_{ss'}(\bvec{x}) = \sum_{r<r'} a_{rr'}^{(ss')} x_r x_{r'}$, the Walsh coefficients concentrate on multi-index pairs with $|\bvec{k}| = 2$, yielding
\begin{equation}\label{eq:alpha_wh_explicit}
    \alpha_{\mathrm{WH}} \approx \binom{d}{2} |a_{rr'}|_{\max}\, \Delta^2\, 2^{2(n-1)}
    = \mathcal{O}(d^2 \cdot 4^n).
\end{equation}
Numerical evaluation at the pyrazine parameters (Appendix~\ref{app:pyrazine_params}):
\begin{center}
\begin{tabular}{ccc}
\toprule
$n$ & $\alpha_{\mathrm{WH}}$ & $d_{\mathrm{QSP}}$ \\
\midrule
8  & $\sim 2.2 \times 10^3$ & $\sim 8.8 \times 10^6$  \\
10 & $3.45 \times 10^4$     & $1.38 \times 10^8$      \\
12 & $5.52 \times 10^5$     & $2.21 \times 10^9$      \\
\bottomrule
\end{tabular}
\end{center}

\subsection{$T{+}V$ split per-query cost}\label{app:tv_per_query}

Each QSP query to the $T{+}V$ block encoding involves:
\begin{enumerate}[label=(\alph*)]
    \item $2d$ QFT applications, each costing $n(n{-}1)/2$ controlled rotations and $n$ Hadamards~(Eq.~\ref{eq:QFT}); for $d=4$, $n=10$: $4 \times 90 = 360$ rotations.
    \item Diagonal phase kicks for $p_r^2$ using $\mathcal{O}(n^2)$ Toffoli gates per mode for arithmetic; for $d=4$, $n=10$: $\sim\!400$ Toffoli.
    \item SELECT/PREPARE networks for the 2-term LCU: $\sim\!1{,}000$ additional Toffoli from comparators and controlled swaps.
\end{enumerate}
Total per query: $\sim\!1{,}800$ Toffoli + $360$ rotations.
At $C_T^{(\mathrm{Toff})} = 4$ and $C_T^{(\mathrm{rot})} = 47$:
\begin{equation}\label{eq:tgate_per_query}
    \Tgate\text{-gates/query} = 1{,}800 \times 4 + 360 \times 47 = 7{,}200 + 16{,}920 = 24{,}120.
\end{equation}
Total: $d_{\mathrm{QSP}} \times 24{,}120 = 6.93 \times 10^4 \times 24{,}120 \approx 1.68 \times 10^9$ $\Tgate$-gates.

\subsection{Fock-basis sparsity and normalization}\label{app:fock_sparsity}

In the Fock basis, the QVC Hamiltonian has nonzero entries from:
\begin{itemize}
    \item Diagonal: $\omega_k(n_k + 1/2)$ --- 1 entry per row per mode.
    \item Off-diagonal (linear coupling): $\kappa_k \sqrt{n_k+1}$ connects $|n_k\rangle \leftrightarrow |n_k \pm 1\rangle$ --- 2 entries per row per mode.
    \item Bilinear: $\gamma_{kk'}(a_k + a_k^\dagger)(a_{k'} + a_{k'}^\dagger)$ --- 4 entries per mode pair.
\end{itemize}
The sparsity per row is bounded by
\begin{equation}\label{eq:fock_sparsity}
    s \leq 1 + 2d + 4\binom{d}{2} = 2d^2 + 1.
\end{equation}
For $d = 4$: $s = 33$.
This counts only on-diagonal electronic-block connections.
Off-diagonal vibronic coupling $V_{ij}$ adds $\mathcal{O}(d)$ entries per state pair, making reported sparsity a lower bound.

The block-encoding normalization for sparse qubitization~\cite{babbush2018encoding} is
\begin{equation}\label{eq:alpha_fock}
    \alpha_{\mathrm{2Q}} = s \cdot \max_{i,j} |H_{ij}| \leq (2d^2+1) \cdot d\, N_{\max}\, \max(|\kappa|, |\gamma|, \ldots).
\end{equation}
At $N_{\max} = 30$, $d = 4$, $\max|\kappa| = 0.00904$~a.u.: $\alpha_{\mathrm{2Q}} = 22.0$.

\section{Arbitrary Smooth PES: QROM Scaling}\label{app:qrom_scaling}

For an arbitrary smooth potential $V(Q_1,\ldots,Q_d)$ that does not factorize into a polynomial of low degree, the Fock-basis sparse oracle must store the full $d$-dimensional tensor of matrix elements.
The number of nonzero entries per row becomes
\begin{equation}\label{eq:seff_arbitrary}
    s_{\mathrm{eff}} = \mathcal{O}(N_{\max}^d),
\end{equation}
and the QROM oracle cost scales accordingly.

REQWIEM avoids this by the multilinear expansion (Eq.~\ref{eq:multilinear_expansion}), which decomposes $V$ into a sum over subsets $S \subseteq \{1,\ldots,d\}$ with M\"obius-inverted coefficients (Eq.~\ref{eq:mobius_coefficients}).
Each subset contributes a UCR circuit with $2^{|S|} - 1$ CNOTs and $2^{|S|}$ $R_z$ gates (Eq.~\ref{eq:graycode_gates}), keeping the total cost polynomial in $d$ for bounded interaction order.

\section{Dyson Series Convergence: Extended Derivation}\label{app:dyson_derivation}

\subsection{Connection to the Taylor series of $\exp(-x)$}

The interaction-picture Dyson series for the Landau--Zener Hamiltonian (Eq.~\ref{eq:H_LZ}) produces an asymptotic series whose $N$th term has magnitude
\begin{equation}\label{eq:cN_magnitude}
    |c_N| \sim \frac{(\pi\Gamma)^N}{N!}.
\end{equation}
This is precisely the $N$th term of the Taylor series of $\exp(-x)$ evaluated at $x = \pi\Gamma$.
By Stirling's approximation ($N! \approx \sqrt{2\pi N}(N/\e)^N$),
\begin{equation}\label{eq:stirling_bound}
    \frac{x^N}{N!} \approx \frac{1}{\sqrt{2\pi N}}\left(\frac{\e\, x}{N}\right)^{\!N}.
\end{equation}
This quantity is:
\begin{itemize}
    \item \emph{Increasing} for $N < x$ (since $\e x/N > 1$);
    \item \emph{Maximum} at $N = \floor{x}$, with value $\sim (2\pi x)^{-1/2}$;
    \item \emph{Decreasing} for $N > x$, with super-exponential decay for $N > \e x$.
\end{itemize}

The $N$th partial sum $S_N = \sum_{n=0}^{N} (-x)^n/n!$ alternates in sign.
During the growth phase ($N < \floor{x}$), successive partial sums overshoot and undershoot the true value $\exp(-x)$ by amounts far exceeding $\exp(-x)$ itself.
Only after $N \gtrsim \e x$ do the terms become small enough for the alternating series to converge, being the Stokes phenomenon, where the exponentially small quantity $\exp(-\pi\Gamma)$ is ``hidden'' behind the factorial growth of the asymptotic series and cannot be resolved by any finite truncation with $N < \floor{\pi\Gamma}$ terms~\cite{berry1989uniform, dingle1973asymptotic}.

For algorithms approximating $e^{-iHt}$ via a truncated Dyson series or interaction-picture perturbation expansion~\cite{berry2020time}, the truncation order must satisfy $N \geq \floor{\pi\Gamma}$ to resolve the non-perturbative transition amplitude.

QSP/QSVT approximates $e^{-iHt}$ by a polynomial $P_d(x)$ of degree $d$ in the eigenvalue $x = E/\alpha$, using Chebyshev-type polynomial approximation rather than a perturbative expansion.
The standard degree bound $d_{\mathrm{QSP}} = \Order(\alpha t + \log(1/\eps))$~\cite{low2017optimal, gilyen2019quantum} is a uniform bound in the eigenvalue argument and does not contain an additive gap-dependent term.
The Dyson convergence barrier therefore does not directly impose an additional polynomial degree requirement on QSP.
However, QSP receives no cost reduction from gap closure ($\Delta_0 \to 0$) because the block-encoding normalization $\alpha \geq T_{\max}$ is gap-independent, while Trotter's commutator-norm cost decreases as $\|[V_d, V_{\mathrm{od}}]\| = \Delta_0 \cdot W \to 0$.
This one-sided scaling causes the gap in implementation resource overhead.

Trotter error depends on commutator norms $\|[T,V]\|$, which decrease near conical intersections where the diagonal potential surfaces approach each other (Section~\ref{sec:trotter_dyson}).
The duality is exact: as $\Delta_0 \to 0$,
\begin{align}
    \text{Trotter error} &\propto \Delta_0 \cdot W \to 0, \label{eq:trotter_limit}\\
    \text{Dyson order}   &\propto W^2/(\Delta_0\, v) \to \infty. \label{eq:dyson_limit}
\end{align}

\chapter{List of Initialisms and Acronyms}
\label{app:acronyms}

This appendix provides a comprehensive list of all initialisms, acronyms, and abbreviations used throughout this dissertation, organized alphabetically.
Where an abbreviation carries multiple meanings (e.g., CI for both Conical Intersection and Configuration Interaction), both definitions are listed.
The rightmost column indicates the chapter or section of first appearance.

\vspace{1em}

\begin{longtable}{@{} >{\ttfamily}l p{0.52\textwidth} l @{}}
\toprule
\textnormal{\textbf{Abbreviation}} & \textbf{Definition} & \textbf{First Use} \\
\midrule
\endfirsthead
\toprule
\textnormal{\textbf{Abbreviation}} & \textbf{Definition} & \textbf{First Use} \\
\midrule
\endhead
\midrule
\multicolumn{3}{r}{\textit{Continued on next page\ldots}} \\
\endfoot
\bottomrule
\endlastfoot
%
%
AIMS & Ab Initio Multiple Spawning & Ch.\ \ref{ch:Introduction} \\
AIMSWISS & Ab Initio Multiple Spawning with Informed Stochastic Selections & Ch.\ \ref{ch:Introduction} \\
Adam & Adaptive Moment Estimation (optimizer) & Ch.\ \ref{ch:wavepacket_prop} \\
AMD & Advanced Micro Devices, Inc. & Ch.\ \ref{ch:smooth_potentials} \\
ASIC & Application-Specific Integrated Circuit & Ch.\ \ref{ch:Application} \\
AQS & Analog Quantum Simulation & Ch.\ \ref{ch:Introduction} \\
%
%
BCH & Baker--Campbell--Hausdorff & Ch.\ \ref{ch:wavepacket_prop} \\
BCS & Bardeen--Cooper--Schrieffer & Ch.\ \ref{ch:Application} \\
BdG & Bogoliubov--de Gennes & Ch.\ \ref{ch:Application} \\
BLAS & Basic Linear Algebra Subprograms & Ch.\ \ref{ch:Introduction} \\
BQP & Bounded-error Quantum Polynomial (complexity class) & Ch.\ \ref{ch:Introduction} \\
BRGC & Binary Reflected Gray Code & Ch.\ \ref{ch:wavepacket_prop} \\
BVC & Bilinear Vibronic Coupling & Ch.\ \ref{ch:model_ext} \\
%
%
CASSCF & Complete Active Space Self-Consistent Field & Ch.\ \ref{ch:Introduction} \\
CASPT2 & Complete Active Space Second-Order Perturbation Theory & Ch.\ \ref{ch:smooth_potentials} \\
CC & Coupled Cluster & Ch.\ \ref{ch:Introduction} \\
CCS & Coupled-Coherent States & Ch.\ \ref{ch:Introduction} \\
CCSD & Coupled Cluster Singles and Doubles & Ch.\ \ref{ch:Introduction} \\
CCSD(T) & Coupled Cluster Singles, Doubles, and Perturbative Triples & Ch.\ \ref{ch:Introduction} \\
CCSDT & Coupled Cluster Singles, Doubles, and Triples & Ch.\ \ref{ch:Introduction} \\
CG & Clebsch--Gordan & Ch.\ \ref{ch:Application} \\
CI & Conical Intersection / Configuration Interaction & Ch.\ \ref{ch:Introduction} \\
CIS & Configuration Interaction Singles & App.\ \ref{app:symmetry_appendix}\\
cIQFT & Centered Inverse Quantum Fourier Transform & Ch.\ \ref{ch:marcus} \\
CMF & Constant Mean-Field (integration scheme) & App.\ \ref{app:Classical_runtime} \\
CNOT & Controlled-NOT (gate) & Ch.\ \ref{ch:wavepacket_prop} \\
CODATA & Committee on Data for Science and Technology & Ch.\ \ref{ch:smooth_potentials} \\
CP & Charge-Parity (symmetry) & Ch.\ \ref{ch:Application} \\
CPU & Central Processing Unit & Ch.\ \ref{ch:wavepacket_prop} \\
cQFT & Centered Quantum Fourier Transform & Ch.\ \ref{ch:wavepacket_prop} \\
CSD & Cosine-Sine Decomposition & Ch.\ \ref{ch:Application} \\
CW & Continuous Wave & Ch.\ \ref{ch:Introduction} \\
%
%
DMRG & Density Matrix Renormalization Group & Ch.\ \ref{ch:Introduction} \\
DNA & Deoxyribonucleic Acid & Ch.\ \ref{ch:Introduction} \\
DOF & Degree(s) of Freedom & Ch.\ \ref{ch:Introduction} \\
DFT & Density Functional Theory & App.\ \ref{app:symmetry_appendix} \\
DPEM & Diabatic Potential Energy Matrix & Ch.\ \ref{ch:Introduction} \\
DPW & Dual Plane Waves & Ch.\ \ref{ch:Introduction} \\
DQS & Digital Quantum Simulation & Ch.\ \ref{ch:Introduction} \\
DVR & Discrete Variable Representation & Ch.\ \ref{ch:Introduction} \\
DWBA & Distorted-Wave Born Approximation & Ch.\ \ref{ch:Application} \\
%
%
EOM & Equation(s) of Motion & Ch.\ \ref{ch:Introduction} \\
EOM-CC & Equation-of-Motion Coupled Cluster & Ch.\ \ref{ch:Introduction} \\
ESIPT & Excited-State Intramolecular Proton Transfer & Ch.\ \ref{ch:Introduction} \\
%
%
FBR & Finite Basis Representation & Ch.\ \ref{ch:Introduction} \\
FCI & Full Configuration Interaction & Ch.\ \ref{ch:Introduction} \\
FD & Finite Difference & Ch.\ \ref{ch:smooth_potentials} \\
FFT & Fast Fourier Transform & Ch.\ \ref{ch:model_ext} \\
FLOP & Floating-Point Operation & App.\ \ref{app:Classical_runtime} \\
FSSH & Fewest Switches Surface Hopping & Ch.\ \ref{ch:Introduction} \\
FTQC & Fault-Tolerant Quantum Computing & Ch.\ \ref{ch:Introduction} \\
%
%
GF & Galois Field & Ch.\ \ref{ch:smooth_potentials} \\
GH & Gauss--Hermite (quadrature) & Ch.\ \ref{ch:Application} \\
GHZ & Greenberger--Horne--Zeilinger (state) & Ch.\ \ref{ch:model_ext} \\
GL & Gauss--Legendre (quadrature) & Ch.\ \ref{ch:Application} \\
GOE & Gaussian Orthogonal Ensemble & App.\ \ref{app:grid_convergence} \\
GPU & Graphics Processing Unit & Ch.\ \ref{ch:Introduction} \\
%
%
HBT & 2-(2$'$-hydroxyphenyl)benzothiazole & Ch.\ \ref{ch:Introduction} \\
HEA & Hardware-Efficient Ansatz & Ch.\ \ref{ch:marcus} \\
HHG & High-Harmonic Generation & Ch.\ \ref{ch:Application} \\
HHL & Harrow--Hassidim--Lloyd (algorithm) & Ch.\ \ref{ch:scattering} \\
HO & Harmonic Oscillator & Ch.\ \ref{ch:Application} \\
HSH & Hyperspherical Harmonic & Ch.\ \ref{ch:Application} \\
HWP & Hamming Weight Phasing & App.\ \ref{app:tgate_pipeline} \\
%
%
IBM & International Business Machines & Ch.\ \ref{ch:wavepacket_prop} \\
IQAE & Iterative Quantum Amplitude Estimation & Ch.\ \ref{ch:resources} \\
irrep & Irreducible Representation & Ch.\ \ref{ch:model_ext} \\
iSF & Intramolecular Singlet Fission & App.\ \ref{app:symmetry_appendix} \\
IVR & Intramolecular Vibrational (Energy) Redistribution & Ch.\ \ref{ch:model_ext} \\
%
%
JC & Jaynes-Cummings (model) & Ch.\ \ref{ch:Application} \\
JCP & Jarlskog Invariant (CP-violation measure) & Ch.\ \ref{ch:Application} \\
%
%
L-BFGS-B & Limited-Memory Broyden--Fletcher--Goldfarb--Shanno, Bound-Constrained & Ch.\ \ref{ch:wavepacket_prop} \\
LCU & Linear Combination of Unitaries & Ch.\ \ref{ch:Introduction} \\
LHST & Local Hilbert-Schmidt Test & Ch.\ \ref{ch:marcus} \\
LISA & Laser Interferometer Space Antenna & Ch.\ \ref{ch:Application} \\
LJ & Lennard-Jones (potential) & Ch.\ \ref{ch:scattering} \\
LVC & Linear Vibronic Coupling & Ch.\ \ref{ch:Introduction} \\
LZ & Landau--Zener (model) & Ch.\ \ref{ch:model_ext} \\
%
%
MATLAB & Matrix Laboratory (software) & Ch.\ \ref{ch:Application} \\
MCTDH & Multiconfigurational Time-Dependent Hartree & Ch.\ \ref{ch:Introduction} \\
MECI & Minimum Energy Conical Intersection & App.\ \ref{app:symmetry_appendix} \\
MLAE & Maximum Likelihood Amplitude Estimation & Ch.\ \ref{ch:resources} \\
ML-MCTDH & Multi-Layer Multiconfigurational Time-Dependent Hartree & Ch.\ \ref{ch:Introduction} \\
MP2 & Second-Order M{\o}ller–Plesset Perturbation Theory & App.\ \ref{app:symmetry_appendix} \\
MPI & Message Passing Interface & Ch.\ \ref{ch:model_ext} \\
MPS & Matrix Product State & Ch.\ \ref{ch:Introduction} \\
MQB & Mixed Qubit-Boson & Ch.\ \ref{ch:Introduction} \\
MRCI & Multireference Configuration Interaction & Ch.\ \ref{ch:Introduction} \\
MSW & Mikheyev--Smirnov--Wolfenstein (effect) & Ch.\ \ref{ch:Application} \\
MZ & Mach--Zehnder (interferometer) & Ch.\ \ref{ch:scattering} \\
%
%
NACME & Nonadiabatic Coupling Matrix Element & Ch.\ \ref{ch:Application} \\
NA-QMD & Nonadiabatic Quantum Molecular Dynamics & Ch.\ \ref{ch:Introduction} \\
NCSM & No-Core Shell Model & Ch.\ \ref{ch:Application} \\
NISQ & Noisy Intermediate-Scale Quantum & Ch.\ \ref{ch:Introduction} \\
NP & Nondeterministic Polynomial (complexity class) & Ch.\ \ref{ch:Introduction} \\
%
%
OGE & One-Gluon Exchange & Ch.\ \ref{ch:Application} \\
OPA & One-Photon Absorption (spectrum) & App.\ \ref{app:Classical_runtime} \\
OTOC & Out-of-Time-Ordered Correlator & Ch.\ \ref{ch:Introduction} \\
%
%
PES & Potential Energy Surface & Ch.\ \ref{ch:Introduction} \\
PG & Phase Gradient (method) & Ch.\ \ref{ch:resources} \\
pJT & Pseudo-Jahn--Teller (coupling) & App.\ \ref{app:symmetry_appendix} \\
PMNS & Pontecorvo--Maki--Nakagawa--Sakata (matrix) & Ch.\ \ref{ch:Application} \\
POTFIT & Potential Fitting (algorithm for SOP expansion) & App.\ \ref{app:Classical_runtime} \\
PSPACE & Polynomial Space (complexity class) & Ch.\ \ref{ch:Introduction} \\
%
%
QAOA & Quantum Approximate Optimization Algorithm & Ch.\ \ref{ch:Introduction} \\
QAQC & Quantum-Assisted Quantum Compiling & Ch.\ \ref{ch:marcus} \\
QEC & Quantum Error Correction & Ch.\ \ref{ch:Introduction} \\
QED & Quantum Electrodynamics & Ch.\ \ref{ch:Application} \\
QFT & Quantum Fourier Transform & Ch.\ \ref{ch:wavepacket_prop} \\
QKSD & Quantum Krylov Subspace Diagonalization & Ch.\ \ref{ch:Introduction} \\
QMD & Quantum Molecular Dynamics & Ch.\ \ref{ch:smooth_potentials} \\
QNM & Quasi-Normal Mode & Ch.\ \ref{ch:Application} \\
QPE & Quantum Phase Estimation & Ch.\ \ref{ch:Application} \\
QROAM & Quantum Read-Only Access Memory & Ch.\ \ref{ch:Application} \\
QROM & Quantum Read-Only Memory & Ch.\ \ref{ch:Introduction} \\
QSP & Quantum Signal Processing & Ch.\ \ref{ch:Introduction}, \ref{ch:comparison}\\
QSVT & Quantum Singular Value Transformation & Ch.\ \ref{ch:Introduction}, \ref{ch:comparison} \\
QVC & Quadratic Vibronic Coupling & Ch.\ \ref{ch:model_ext} \\
%
%
REQWIEM & Real-time Evolution of Quantum Wavepackets In Explicit Modularity & Ch.\ \ref{ch:Introduction} \\
RI & Resolution of the Identity & Ch.\ \ref{ch:smooth_potentials} \\
RI-CC2 & Resolution-of-Identity Coupled Cluster, 2nd order & Ch.\ \ref{ch:smooth_potentials} \\
%
%
SCC & Similarity-Constrained Coupled Cluster & Ch.\ \ref{ch:Introduction} \\
SC-IVR & Semiclassical Initial Value Representation & Ch.\ \ref{ch:Introduction} \\
SDK & Software Development Kit & Ch.\ \ref{ch:smooth_potentials} \\
SIL & Short Iterative Lanczos (integrator) & App.\ \ref{app:Classical_runtime} \\
SIMD & Single Instruction, Multiple Data & Ch.\ \ref{ch:Application} \\
SOP & Sum-of-Products (Hamiltonian form) & App.\ \ref{app:Classical_runtime} \\
SP & Single-Particle & Ch.\ \ref{ch:Introduction} \\
SPF & Single-Particle Function & Ch.\ \ref{ch:Introduction} \\
SPH & Smoothed Particle Hydrodynamics & Ch.\ \ref{ch:Application} \\
SU & Special Unitary (group) & Ch.\ \ref{ch:Introduction} \\
SVD & Singular Value Decomposition & Ch.\ \ref{ch:Introduction} \\
SWAP & SWAP gate & Ch.\ \ref{ch:model_ext} \\
%
%
TBF & Trajectory Basis Function & Ch.\ \ref{ch:Introduction} \\
TD-CC & Time-Dependent Coupled Cluster & Ch.\ \ref{ch:Introduction} \\
TD-CCSD & Time-Dependent Coupled Cluster Singles and Doubles & Ch.\ \ref{ch:Introduction} \\
TD-CCSDT & Time-Dependent Coupled Cluster Singles, Doubles, and Triples & Ch.\ \ref{ch:Introduction} \\
TDDFT & Time-Dependent Density Functional Theory & Ch.\ \ref{ch:smooth_potentials} \\
TD-DMRG & Time-Dependent Density Matrix Renormalization Group & Ch.\ \ref{ch:Introduction} \\
T-depth & Circuit depth measured in T-gate layers & Ch.\ \ref{ch:resources} \\
TDSE & Time-Dependent Schr{\"o}dinger Equation & Ch.\ \ref{ch:Introduction} \\
TDVP & Time-Dependent Variational Principle & Ch.\ \ref{ch:Introduction} \\
TEBD & Time-Evolving Block Decimation & Ch.\ \ref{ch:Introduction} \\
T-gate & Non-Clifford $\pi/8$ rotation gate & Ch.\ \ref{ch:resources} \\
TODD & T-Optimization by Diagonal Decomposition & App.\ \ref{app:tgate_pipeline} \\
TTNS & Tree Tensor Network State & App.\ \ref{app:Classical_runtime} \\
%
%
UCC & Unitary Coupled Cluster & Ch.\ \ref{ch:wavepacket_prop} \\
UCR & Uniformly Controlled Rotation & Ch.\ \ref{ch:model_ext} \\
UCU & Uniformly Controlled Unitary & Ch.\ \ref{ch:Application} \\  
USC & Ultrastrong Coupling & Ch.\ \ref{ch:Application} \\
%
%
VarQITE & Variational Quantum Imaginary Time Evolution & Ch.\ \ref{ch:wavepacket_prop} \\
VFF & Variational Fast Forwarding & Ch.\ \ref{ch:marcus} \\
VQE & Variational Quantum Eigensolver & Ch.\ \ref{ch:Introduction} \\
VUV & Vacuum Ultraviolet & App.\ \ref{app:grid_convergence} \\
%
%
WD & Wigner $D$-matrix & Ch.\ \ref{ch:Application} \\
WH-LCU & Wigner--Haagerup Linear Combination of Unitaries & Ch.\ \ref{ch:comparison} \\
WINGS & Wavepacket Initialization on Neighboring Grid States & Ch.\ \ref{ch:wavepacket_prop} \\
WS & Wigner--Seitz (cell) & Ch.\ \ref{ch:Application} \\
%
%
XOR & Exclusive OR & Ch.\ \ref{ch:Introduction} \\
%
%
ZPE & Zero-Point Energy & Ch.\ \ref{ch:smooth_potentials} \\
ZX & ZX-calculus & App.\ \ref{app:tgate_pipeline} \\
\end{longtable}

\chapter{Glossary of Technical Terms}
\label{app:glossary}

This appendix defines key technical terms, physical concepts, and quantum-computing
jargon used throughout this dissertation.  Terms are listed alphabetically; the
rightmost column indicates the chapter of first appearance.  Acronyms and
initialisms are catalogued separately in Appendix~\ref{app:acronyms}.

\vspace{1em}

\begin{longtable}{@{} p{0.22\textwidth} p{0.58\textwidth} l @{}}
\toprule
\textbf{Term} & \textbf{Definition} & \textbf{First Use} \\
\midrule
\endfirsthead
\toprule
\textbf{Term} & \textbf{Definition} & \textbf{First Use} \\
\midrule
\endhead
\midrule
\multicolumn{3}{r}{\textit{Continued on next page\ldots}} \\
\endfoot
\bottomrule
\endlastfoot
\textbf{Ab initio multiple spawning} & A method that represents the nuclear wavefunction as a swarm of Gaussian basis functions (trajectories) that can ``spawn'' new functions at nonadiabatic coupling regions. & Ch.\ \ref{ch:Introduction} \\[3pt]
\textbf{Adiabatic representation} & Molecular quantum representation in which the electronic basis states are instantaneous eigenstates of the electronic Hamiltonian at each nuclear geometry. Off-diagonal couplings then arise through the nuclear kinetic energy operator. & Ch.\ \ref{ch:Introduction} \\[3pt]
\textbf{Ancilla qubit} & An auxiliary qubit used as temporary workspace during a quantum computation, typically initialized to $|0\rangle$ and disentangled (uncomputed) before measurement. & Ch.\ \ref{ch:wavepacket_prop} \\[3pt]
\textbf{Anharmonicity} & Deviation of a potential energy surface from quadratic (harmonic) form, leading to unequally spaced energy levels and mode--mode coupling. & Ch.\ \ref{ch:Introduction} \\[3pt]
\textbf{Autocorrelation function} & The overlap $C(t) = \langle \psi(0)|\psi(t)\rangle$ between an initial and time-evolved wavepacket; its Fourier transform gives the energy spectrum. & Ch.\ \ref{ch:Introduction} \\[3pt]
\textbf{Block encoding} & A technique for embedding a non-unitary matrix (such as a Hamiltonian) into the upper-left block of a larger unitary, enabling its use in quantum algorithms. & Ch.\ \ref{ch:Introduction} \\[3pt]
\textbf{Bogoliubov transformation} & A canonical transformation mixing particle creation and annihilation operators to diagonalize a quadratic Hamiltonian, used in superconductivity and Bose--Einstein condensation. & Ch.\ \ref{ch:Application} \\[3pt]
\textbf{Born--Oppenheimer approximation} & The assumption that electronic and nuclear motions decouple due to the large mass ratio, allowing the total wavefunction to be written as a product of electronic and nuclear parts. & Ch.\ \ref{ch:Introduction} \\[3pt]
\textbf{Branching space} & The two-dimensional subspace of nuclear coordinates (gradient-difference and derivative-coupling vectors) that lifts the degeneracy at a conical intersection. & Ch.\ \ref{ch:model_ext} \\[3pt]
\textbf{Channel register} & The set of qubits encoding the discrete electronic (or internal) state label in a coupled-channel quantum simulation. & Ch.\ \ref{ch:marcus} \\[3pt]
\textbf{Circuit depth} & The number of sequential time steps (layers) in a quantum circuit, where gates in the same layer act on disjoint qubits and execute in parallel. & Ch.\ \ref{ch:Introduction} \\[3pt]
\textbf{Clebsch--Gordan coefficient} & Expansion coefficients for coupling two angular momentum eigenstates into a state of definite total angular momentum; appear in partial-wave decompositions. & Ch.\ \ref{ch:Application} \\[3pt]
\textbf{Clifford gate} & A quantum gate from the Clifford group (Hadamard, CNOT, S-gate, etc.) that maps Pauli operators to Pauli operators; these gates can be implemented transversally in many error-correcting codes. & Ch.\ \ref{ch:resources} \\[3pt]
\textbf{Conical intersection} & A molecular geometry at which two or more electronic potential energy surfaces become exactly degenerate, forming a cone-shaped crossing that mediates ultrafast nonadiabatic transitions. & Ch.\ \ref{ch:Introduction} \\[3pt]
\textbf{Cooper pair} & A bound pair of electrons with opposite spin and momentum in a superconductor, formed via phonon-mediated attraction; the microscopic basis of BCS superconductivity. & Ch.\ \ref{ch:Application} \\[3pt]
\textbf{Coupled-channel Hamiltonian} & A Hamiltonian of the form $H = T \otimes I\_C + \sum\_{ij} V\_{ij}(\mathbf{x})|i\rangle\langle j|$ coupling spatial dynamics to internal-state transitions. & Ch.\ \ref{ch:model_ext} \\[3pt]
\textbf{Coupling mode} & A vibrational mode that introduces off-diagonal coupling between electronic states, directly mediating nonadiabatic transitions. & Ch.\ \ref{ch:model_ext} \\[3pt]
\textbf{Density matrix renormalization group} & A variational algorithm that optimizes a matrix product state representation of the ground state by iteratively sweeping through lattice sites. & Ch.\ \ref{ch:Introduction} \\[3pt]
\textbf{Diabatic representation} & Molecular quantum representation in which the electronic basis states are chosen to vary smoothly with nuclear coordinates, transferring coupling from the kinetic operator into smooth potential-energy matrix elements. & Ch.\ \ref{ch:Introduction} \\[3pt]
\textbf{Discrete variable representation} & A grid-based method for quantum dynamics in which the wavefunction is represented at quadrature points and the potential operator is diagonal. & Ch.\ \ref{ch:Application} \\[3pt]
\textbf{Distorted-wave Born approximation} & A perturbative scattering method that uses distorted (non-plane-wave) reference wavefunctions to improve accuracy for strong short-range potentials. & Ch.\ \ref{ch:Application} \\[3pt]
\textbf{Duschinsky rotation} & A unitary mixing of normal-mode coordinates between two electronic states, arising when equilibrium geometries and/or force constants differ. & Ch.\ \ref{ch:model_ext} \\[3pt]
\textbf{Eigenstate} & A quantum state that, when acted on by a given operator, returns the same state multiplied by a scalar (its eigenvalue). & Ch.\ \ref{ch:Introduction} \\[3pt]
\textbf{Fan-out protocol} & A circuit motif that copies a channel-register state to ancillary qubits so that independent vibrational-mode operations can be applied in parallel across channels. & Ch.\ \ref{ch:model_ext} \\[3pt]
\textbf{Fault-tolerant quantum computing} & Quantum computation using error-correcting codes and encoded (logical) qubits so that arbitrarily long computations can be performed reliably despite physical gate errors. & Ch.\ \ref{ch:Introduction} \\[3pt]
\textbf{Fidelity} & A measure of closeness between two quantum states or operations, ranging from 0 (orthogonal) to 1 (identical); $F = |\langle\psi|\phi\rangle|^2$ for pure states. & Ch.\ \ref{ch:wavepacket_prop} \\[3pt]
\textbf{Finite basis representation} & Representation of a quantum state in a truncated set of basis functions (e.g., harmonic oscillator eigenstates), in which the kinetic operator is typically diagonal. & Ch.\ \ref{ch:Application} \\[3pt]
\textbf{Flavor mixing matrix} & A unitary matrix (e.g., PMNS for neutrinos, CKM for quarks) relating mass eigenstates to flavor eigenstates. & Ch.\ \ref{ch:Application} \\[3pt]
\textbf{Franck--Condon factor} & The squared overlap integral between vibrational wavefunctions of two electronic states, governing the intensity of vibronic transitions. & Ch.\ \ref{ch:marcus} \\[3pt]
\textbf{Gate count} & The total number of elementary quantum gates in a circuit; a primary metric for quantum resource estimation. & Ch.\ \ref{ch:Introduction} \\[3pt]
\textbf{Gray code} & An ordering of binary strings in which successive entries differ in exactly one bit; used to implement multi-controlled rotations with minimal CNOT overhead. & Ch.\ \ref{ch:model_ext} \\[3pt]
\textbf{Grid resolution} & The number of discrete spatial points per dimension used to represent the wavefunction; must be sufficient to resolve the shortest de Broglie wavelength. & Ch.\ \ref{ch:wavepacket_prop} \\[3pt]
\textbf{Hamiltonian} & The quantum-mechanical operator whose eigenvalues are the system's energy levels and whose structure governs time evolution via the Schr\"odinger equation. & Ch.\ \ref{ch:Introduction} \\[3pt]
\textbf{Harmonic oscillator} & A system with a quadratic potential $V(x)=\tfrac{1}{2}k x^2$; the simplest model for molecular vibrations, with equally spaced energy levels. & Ch.\ \ref{ch:wavepacket_prop} \\[3pt]
\textbf{Hermitian operator} & A linear operator equal to its own conjugate transpose ($H = H^\dagger$); observables in quantum mechanics are represented by Hermitian operators. & Ch.\ \ref{ch:Application} \\[3pt]
\textbf{High-harmonic generation} & A nonlinear optical process in which an intense laser field drives electron recombination with its parent ion, emitting photons at integer multiples of the driving frequency. & Ch.\ \ref{ch:Application} \\[3pt]
\textbf{Hilbert space} & The complete vector space of quantum states for a given system, equipped with an inner product that defines probabilities and expectation values. & Ch.\ \ref{ch:Introduction} \\[3pt]
\textbf{Irreducible representation} & A matrix representation of a symmetry group that cannot be reduced to smaller block-diagonal form; labels the transformation properties of electronic states and vibrational modes. & Ch.\ \ref{ch:model_ext} \\[3pt]
\textbf{Lennard-Jones potential} & An interatomic pair potential $V(r) = 4\varepsilon[(\sigma/r)^{12} - (\sigma/r)^{6}]$ combining short-range repulsion with long-range van der Waals attraction. & Ch.\ \ref{ch:scattering} \\[3pt]
\textbf{Logical qubit} & A qubit encoded redundantly across many physical qubits using a quantum error-correcting code, enabling fault-tolerant operations. & Ch.\ \ref{ch:resources} \\[3pt]
\textbf{Magic state distillation} & A protocol that consumes many noisy copies of a non-Clifford resource state (magic state) to produce fewer high-fidelity copies, enabling fault-tolerant T-gates. & Ch.\ \ref{ch:resources} \\[3pt]
\textbf{Marcus model} & A model for electron transfer in which two coupled harmonic (parabolic) diabatic surfaces cross, with the transfer rate governed by the electronic coupling and reorganization energy. & Ch.\ \ref{ch:marcus} \\[3pt]
\textbf{Matrix product state} & A one-dimensional tensor network ansatz that represents a quantum state as a chain of contracted matrices, with bond dimension controlling the entanglement capacity. & Ch.\ \ref{ch:Introduction} \\[3pt]
\textbf{Morse potential} & An anharmonic interatomic potential $V(r) = D\_e(1-e^{-a(r-r\_e)})^2$ that models bond stretching with correct dissociation behavior, unlike the harmonic approximation. & Ch.\ \ref{ch:smooth_potentials} \\[3pt]
\textbf{MCTDH} & A variational method for quantum nuclear dynamics that expands the wavefunction in a time-dependent basis of single-particle functions, enabling treatment of moderate-dimensional systems. & Ch.\ \ref{ch:Introduction} \\[3pt]
\textbf{Multilinear expansion} & Expansion of a smooth multivariable function on a discrete grid into a hierarchy of terms (constant, linear, bilinear, trilinear, \ldots) enabling systematic circuit construction. & Ch.\ \ref{ch:smooth_potentials} \\[3pt]
\textbf{Neutrino oscillation} & Quantum-mechanical phenomenon in which neutrinos change flavor (electron, muon, tau) during propagation, arising from misalignment between mass and flavor eigenstates. & Ch.\ \ref{ch:Application} \\[3pt]
\textbf{Nonadiabatic dynamics} & Molecular dynamics in which the Born--Oppenheimer approximation breaks down and nuclear motion is coupled to transitions between two or more electronic states. & Ch.\ \ref{ch:Introduction} \\[3pt]
\textbf{Oracle} & A black-box unitary subroutine assumed to implement a specific function; oracle-based algorithms treat it as a given primitive whose internal structure is unspecified. & Ch.\ \ref{ch:Introduction} \\[3pt]
\textbf{Oracle-free algorithm} & A quantum algorithm that constructs all unitary operations explicitly from elementary gates, without relying on an abstract oracle subroutine. & Ch.\ \ref{ch:Introduction} \\[3pt]
\textbf{Partial wave} & A component of the scattering wavefunction with definite angular momentum quantum number $\ell$, used to decompose the total cross-section into angular-momentum contributions. & Ch.\ \ref{ch:scattering} \\[3pt]
\textbf{Phase gradient method} & A quantum arithmetic technique that adds or multiplies numbers stored in a quantum register using phase rotations rather than explicit binary adders. & Ch.\ \ref{ch:resources} \\[3pt]
\textbf{Phase shift} & The angular displacement acquired by a partial wave due to scattering relative to the free-particle solution; encodes the strength and sign of the interaction. & Ch.\ \ref{ch:scattering} \\[3pt]
\textbf{Point group} & The set of all spatial symmetry operations (rotations, reflections, inversions) that leave a molecule's nuclear framework invariant. & Ch.\ \ref{ch:model_ext} \\[3pt]
\textbf{Population dynamics} & The time-dependent probability of occupying each electronic or internal state, obtained by tracing the total density matrix over spatial degrees of freedom. & Ch.\ \ref{ch:Introduction} \\[3pt]
\textbf{Position register} & The set of qubits encoding the discretized nuclear coordinate(s) on a spatial grid in a first-quantized simulation. & Ch.\ \ref{ch:marcus} \\[3pt]
\textbf{Potential energy surface} & A hypersurface giving the electronic energy as a function of nuclear coordinates, governing nuclear motion within a single electronic state. & Ch.\ \ref{ch:Introduction} \\[3pt]
\textbf{Product formula} & A general term for Trotter--Suzuki-type decompositions that approximate $e^{-i(A+B)t}$ as a product of exponentials $e^{-iAt}e^{-iBt}$ with controlled error. & Ch.\ \ref{ch:wavepacket_prop} \\[3pt]
\textbf{Quantum advantage} & The regime in which a quantum computer outperforms the best known classical algorithm for a given problem, typically by an exponential or large polynomial factor. & Ch.\ \ref{ch:Introduction} \\[3pt]
\textbf{Quantum Fourier transform} & The quantum-circuit implementation of the discrete Fourier transform, mapping a computational-basis state to an equal superposition with phase-encoded amplitudes. & Ch.\ \ref{ch:wavepacket_prop} \\[3pt]
\textbf{Quantum gate} & An elementary unitary operation on one or more qubits; the building block of quantum circuits, analogous to classical logic gates. & Ch.\ \ref{ch:Introduction} \\[3pt]
\textbf{Quantum Krylov subspace diagonalization} & A quantum algorithm that builds a subspace of time-evolved states $\{e^{-iHt\_k}|\psi\_0\rangle\}$ and diagonalizes $H$ within it, accessing excited states without variational optimization. & Ch.\ \ref{ch:Introduction} \\[3pt]
\textbf{Quantum read-only memory} & A quantum data structure that loads classically precomputed data into a quantum register via a tree of controlled-SWAP gates; its depth scales logarithmically in the table size. & Ch.\ \ref{ch:Introduction} \\[3pt]
\textbf{Quantum scars} & Non-ergodic quantum eigenstates with anomalously high amplitude along unstable classical periodic orbits, violating the eigenstate thermalization hypothesis. & Ch.\ \ref{ch:Introduction} \\[3pt]
\textbf{Quasiparticle} & An emergent excitation in a many-body system that behaves like a modified particle with renormalized mass, charge, or other properties. & Ch.\ \ref{ch:Application} \\[3pt]
\textbf{Qubit} & The fundamental unit of quantum information, a two-level quantum system whose state is a superposition $\alpha|0\rangle + \beta|1\rangle$. & Ch.\ \ref{ch:Introduction} \\[3pt]
\textbf{Qubitization} & An optimal Hamiltonian simulation technique based on block encoding that achieves query complexity linear in time and logarithmic in inverse error. & Ch.\ \ref{ch:Introduction} \\[3pt]
\textbf{Rabi oscillation} & Coherent oscillation of a two-level quantum system driven by a resonant perturbation, with frequency proportional to the coupling strength. & Ch.\ \ref{ch:Application} \\[3pt]
\textbf{Recurrence spectrum} & The time-autocorrelation function $\langle\psi(0)|\psi(t)\rangle$ of a propagated wavepacket, whose Fourier transform yields the absorption or emission spectrum. & Ch.\ \ref{ch:Introduction} \\[3pt]
\textbf{Reorganization energy} & In Marcus theory, the energy cost to distort the nuclear framework from the equilibrium geometry of one diabatic state to that of the other, without changing electronic state. & Ch.\ \ref{ch:marcus} \\[3pt]
\textbf{Resource estimation} & The process of counting the quantum gates, qubits, circuit depth, and other resources required to implement a quantum algorithm at a target precision. & Ch.\ \ref{ch:resources} \\[3pt]
\textbf{Ross--Selinger Synthesis} & Optimal ancilla-free decomposition of $R_z$ rotations into Clifford$+T$ gates. & Ch.\ \ref{ch:resources}, \ref{ch:comparison} \\
\textbf{Rotation synthesis} & The compilation of an arbitrary single-qubit rotation into a sequence of Clifford+T gates with a specified approximation error, typically using the Ross--Selinger or related algorithms. & Ch.\ \ref{ch:resources} \\[3pt]
\textbf{S-matrix} & The scattering matrix relating incoming and outgoing asymptotic states in a collision; its elements encode all transition amplitudes between channels. & Ch.\ \ref{ch:scattering} \\[3pt]
\textbf{Scattering cross-section} & A measure of the effective target area presented by a particle to an incoming projectile, encoding the probability of a particular collision outcome. & Ch.\ \ref{ch:scattering} \\[3pt]
\textbf{Smoothed particle hydrodynamics} & A mesh-free Lagrangian method for fluid dynamics that represents the continuum by a set of particles with smoothed density kernels. & Ch.\ \ref{ch:Application} \\[3pt]
\textbf{Spectral Range Penalty} & The QSP polynomial degree scales with the Hamiltonian spectral range $\alpha$, which grows as $4^n$ for grid-based kinetic operators. & Ch.\ \ref{ch:comparison} \\
\textbf{Split-operator method} & A propagation scheme that alternately applies kinetic and potential operators in the position and momentum bases, connected by Fourier transforms. & Ch.\ \ref{ch:wavepacket_prop} \\[3pt]
\textbf{Strang splitting} & A second-order symmetric operator-splitting scheme: $e^{-iHt} \approx e^{-iV t/2}\,e^{-iT t}\,e^{-iV t/2}$, widely used in split-operator wavepacket propagation. & Ch.\ \ref{ch:wavepacket_prop} \\[3pt]
\textbf{Surface code} & A topological quantum error-correcting code defined on a two-dimensional lattice of physical qubits, widely considered the leading candidate for fault-tolerant quantum computing. & Ch.\ \ref{ch:resources} \\[3pt]
\textbf{Surface hopping} & A mixed quantum--classical method in which nuclei follow classical trajectories on a single electronic surface, with stochastic hops between surfaces at nonadiabatic coupling regions. & Ch.\ \ref{ch:Introduction} \\[3pt]
\textbf{T-count} & The number of T-gates (or their equivalents) in a quantum circuit; the dominant cost metric in fault-tolerant architectures. & Ch.\ \ref{ch:resources} \\[3pt]
\textbf{T-depth} & The number of sequential layers containing at least one T-gate; determines the time cost of a fault-tolerant circuit when T-gates cannot be parallelized. & Ch.\ \ref{ch:resources} \\[3pt]
\textbf{T-gate} & The single-qubit $\pi/8$ phase gate $\mathrm{diag}(1, e^{i\pi/4})$; a non-Clifford gate whose count dominates the cost of fault-tolerant circuits. & Ch.\ \ref{ch:resources} \\[3pt]
\textbf{Tensor network} & A structured decomposition of a high-dimensional tensor (wavefunction) into a network of lower-order tensors, enabling efficient representation of entangled many-body states. & Ch.\ \ref{ch:Introduction} \\[3pt]
\textbf{Toffoli Floor} & The irreducible Toffoli gate cost in oracle-based methods, independent of rotation-synthesis precision. & Ch.\ \ref{ch:comparison} \\
\textbf{Trotter error} & The approximation error introduced by replacing the exact time-evolution operator with a finite product of partial exponentials (Trotter--Suzuki formula). & Ch.\ \ref{ch:wavepacket_prop} \\[3pt]
\textbf{Trotter--Dyson Duality} & Near conical intersections, Trotter error vanishes while perturbative truncation order diverges and QSP cost remains anchored to the gap-independent spectral range, giving a one-sided structural advantage to split-operator methods. & Ch.\ \ref{ch:comparison} \\
\textbf{Trotterization} & Approximation of the time-evolution operator $e^{-iHt}$ by a product of simpler exponentials, one for each term in the Hamiltonian, with error controlled by step size. & Ch.\ \ref{ch:wavepacket_prop} \\[3pt]
\textbf{Tuning mode} & A vibrational mode that modulates the energy gap between electronic states (shifts potentials along the diagonal), influencing the crossing geometry. & Ch.\ \ref{ch:model_ext} \\[3pt]
\textbf{Ultrastrong coupling} & A light--matter coupling regime where the interaction energy is a non-negligible fraction ($\gtrsim 10\%$) of the bare mode frequency, invalidating the rotating-wave approximation. & Ch.\ \ref{ch:Application} \\[3pt]
\textbf{Uniformly controlled rotation} & A quantum gate that applies a different rotation angle to a target qubit conditioned on the state of all control qubits; the key primitive for encoding position-dependent potentials. & Ch.\ \ref{ch:wavepacket_prop} \\[3pt]
\textbf{Unitary operator} & A linear operator $U$ satisfying $U^\dagger U = I$, guaranteeing norm preservation; all quantum gates and time-evolution operators are unitary. & Ch.\ \ref{ch:Introduction} \\[3pt]
\textbf{Variational compression} & A hybrid quantum--classical technique that compresses a many-Trotter-step circuit into a shallower parameterized ansatz optimized to match the full evolution. & Ch.\ \ref{ch:marcus} \\[3pt]
\textbf{Variational quantum eigensolver} & A hybrid quantum--classical algorithm that uses a parameterized quantum circuit (ansatz) and classical optimization to find the ground-state energy of a Hamiltonian. & Ch.\ \ref{ch:Introduction} \\[3pt]
\textbf{Vibronic coupling} & Interaction between vibrational (nuclear) and electronic degrees of freedom that drives transitions between electronic states through nuclear displacement. & Ch.\ \ref{ch:Introduction} \\[3pt]
\textbf{Wavepacket} & A spatially localized quantum state constructed as a superposition of plane waves or basis functions, whose time evolution encodes dynamical observables. & Ch.\ \ref{ch:Introduction} \\[3pt]
\textbf{Wigner $D$-matrix} & Matrix elements of the rotation operator in the angular-momentum basis, used to transform between body-fixed and space-fixed reference frames. & Ch.\ \ref{ch:Application} \\[3pt]
\textbf{Work register} & Auxiliary qubits used for intermediate computations (e.g., arithmetic, phase kickback) that are uncomputed before the end of a Trotter step. & Ch.\ \ref{ch:marcus} \\[3pt]
\textbf{XOR-class decomposition} & A partitioning of off-diagonal coupling pairs $(i,j)$ by the value $i \oplus j$, enabling parallel application of all couplings sharing the same XOR label. & Ch.\ \ref{ch:model_ext} \\[3pt]
\end{longtable}

\end{document}